%% file: main.tex
\documentclass[12pt,twoside,longbibliography]{mitthesis}
\usepackage{lgrind}
\usepackage{cmap}
\usepackage[T1]{fontenc}
\usepackage{color}
\usepackage{graphicx}
\usepackage{amsfonts}
\usepackage{amsmath}
\usepackage{amssymb}
\usepackage{capt-of}
\usepackage{url}
\usepackage{slashed}
\usepackage{epstopdf}
\usepackage{xspace}
\usepackage{multirow}
\usepackage{array}
\usepackage{makecell}
\usepackage{cite}
\usepackage{tabularx}
\usepackage{booktabs}
\usepackage{afterpage}
\usepackage{latexsym}
\usepackage{cancel}
\usepackage[utf8]{inputenc}
\usepackage[english]{babel}
\usepackage[framemethod=TikZ]{mdframed}
\usepackage[colorlinks=true
,urlcolor=blue
,anchorcolor=blue
,citecolor=blue
,filecolor=blue
,linkcolor=red
,menucolor=blue
,linktocpage=true
,pdfproducer=medialab
,pdfa=true
]{hyperref}
\newcommand{\es}[2] {\begin{equation} \label{#1} \begin{split} #2 \end{split} \end{equation}}
\newcolumntype{C}[1]{>{\centering\let\newline\\\arraybackslash\hspace{0pt}}m{#1}}

\newcommand{\be}{\begin{equation}}
\newcommand{\ee}{\end{equation}}
\newcommand{\bea}{\begin{eqnarray}}
\newcommand{\eea}{\end{eqnarray}}
\newcommand{\nn}{\nonumber \\ }
\def\dbar{{\mathchar'26\mkern-12mu d}}

\begin{document}

\include{cover}

\pagestyle{plain} 
\include{contents}
\include{intro}
\include{gevexcess}
\include{fermigg}
\include{dmdecay}
\include{cascsig}
\include{casclim}
\include{dm1loop}
\include{conc}

\appendix
\include{gevexcess-app}
\include{fermigg-app}
\include{dmdecay-app}
\include{cascsig-app}
\include{casclim-app}
\include{dm1loop-app}
\bibliography{main}

\end{document}

%% file: cover.tex
\title{Listening to the Universe through Indirect Detection}

\author{Nicholas Llewellyn Rodd}
       \prevdegrees{M.Sc., University of Melbourne (2012) \\
                    B.Sc. and LL.B., University of Melbourne (2010)}
\department{Department of Physics}

\degree{Doctor of Philosophy}

\degreemonth{June}
\degreeyear{2018}
\thesisdate{April 20, 2018}

\supervisor{Tracy R. Slatyer}{Assistant Professor of Physics}

\chairman{Scott A. Hughes}{Interim Associate Department Head of Physics}

\maketitle

\cleardoublepage
\setcounter{savepage}{\thepage}
\begin{abstractpage}
\input{abstract}
\end{abstractpage}

\cleardoublepage

\section*{Acknowledgments}

Life and science are fundamentally collaborative.
In this sense I feel that only having my name listed as an author for this thesis is misleading.
The work I undertook in grad school, culminating in this thesis, would not have been possible without the people around me.
For this reason, here I want to take the time to thank the following people for far more than just useful discussions and comments, but for helping me grow as a physicist and a person over the last five years.

My best decision and piece of good fortune during grad school was to have Tracy Slatyer as my advisor.
Tracy had me working on research almost immediately after we met at the MIT open house.
Although I knew little about dark matter and nothing about the {\it Fermi} telescope, she involved me in a project that turned into the galactic center analysis presented in this thesis.
This was undoubtedly one of the most exciting projects I've ever worked on, thinking that we may actually be looking at a signal emanating from dark matter (as you will see in the thesis, we were not).
During this period we would meet several times a week, and Tracy with near infinite patience brought me to a level where I could contribute to that paper.
Over time Tracy slowly encouraged me to become more independent, but in a manner where I hardly noticed it was happening, leaving me feeling confident about life after grad school.
Beyond this Tracy is a truly incredible academic, teacher, and person.
While this may be common knowledge, to have such a mentor over the last five years has been a privilege, and I cannot thank her enough.
So if you are reading this thanks again Tracy!

After Tracy, the person who had the most influence on my academic experience at MIT was Ben Safdi. 
Ben, who arrived at MIT as a postdoc in my second year, quickly went about setting us straight on the galactic center excess, demonstrating that it was most likely coming from unresolved point sources, not dark matter.
This work was extremely impressive, and it was clear in my mind Ben was someone I wanted to work with.
So shortly after we started working on a project.
After several twists and turns this became the galaxy group analysis presented in this thesis, and was just the first of many projects we undertook.
Throughout this time Ben acted almost as a second advisor to me, teaching me many new ways to look at problems, and emphasising again and again the importance of Monte Carlo!
In his role as pseudo advisor, Ben has been an incredible mentor.
But beyond this he is a remarkable academic, and his ability to generate interesting research questions and directions is an inspiration.
Our collaboration was one of my absolute highlights of grad school, and I look forward to our ongoing work in the future.

Although Tracy and Ben played the largest role, my academic experience at grad school has been shaped by a much larger number of people.
MIT, and in particular the CTP, has been the ideal grad school environment.
I have had the good fortune to work with all of the MIT phenomenology faculty, which in addition to Tracy includes Iain Stewart and Jesse Thaler.
Jesse and Iain are giants of the field.
Being able to see how they approach problems and how they dragged me through my own has taught me a lot.
They were incredibly generous with their time both when it came to research, but also more generally in discussing career advice. 
The CTP itself was also incredibly welcoming.
In my early years, older students were often looking out for me, and really made me feel welcome.
Other highlights have included morning coffee discussions with Yotam and the resource provided by Tracy's other grad students Hongwan, Chih-Liang, and Patrick.
Non-academic aspects of life in the CTP were always smooth, and as I know this does not just happen on its own, I have a lot of appreciation for the hard work of Scott Morley, Joyce Berggren, Charles Suggs, and Cathy Modica.
My research has enormously benefited from the many collaborators I've had both at MIT and elsewhere.
I have learnt from each and every one of them, but in particular I want to thank long time collaborator Sid Mishra-Sharma for saving our 2mass2furious project by initiating a full rewrite of our code, Grigory Ovanesyan for dragging me through my first box integral, and Josh Foster for patiently reminding me many times of basic probability. 
Finally, I also owe a det of gratitude to my physics advisors during my masters in Australia.
Archil Kobakhidze, Elisabetta Barberio, and in particular Ray Volkas set me off on the path I am now on and have continued to provide useful advice at times during grad school.

I've been fortunate enough to have a fantastic experience in the US over the last five years, an experience which has been due to more than just science, and is largely thanks to the number of great friends I've made during this time.
In particular I want to thank my next door office mate Nikhil for endless amusing discussions and pointing out the many logical flaws in Journey's ``Don't Stop Believin'\hspace{0.05cm}'' (who are the streetlight people?), Erik for always motivating me to go the gym and take full advantage of nights out, and my other housemates for the good times shared at 401, especially Darius and Ryan.
Beyond this I am fortunate enough to have stayed connected with so many friends back in Australia, which is one of the reasons I always enjoy heading home.

Finally I want to thank the people who have played the greatest role in getting me to my present position: my family Annabelle, Robyn, David, and Helen.
In particular I want to single out my mother.
From a young age, her encouragement and support have always driven me forwards, and no one in my life has played a larger part in my development than her.
But every member of my family has played an important role in supporting me, even though I'm now living on the other side of the world, and I'm endlessly grateful for it.
If any one of them had been missing I cannot imagine achieving half of what I have.
For the providing this same support I am also extremely grateful to my partner Felicia.
Meeting you has been the best thing to happen to me during this period.
You provide me strength and support, whilst making every day brighter.
I've cherished our time together so far and cannot wait for the next chapter to unfold in California.

%% file: abstract.tex
Indirect detection is the search for the particle nature of dark matter with astrophysical probes.
Manifestly, it exists right at the intersection of particle physics and astrophysics, and the discovery potential for dark matter can be greatly extended using insights from both disciplines.
This thesis provides an exploration of this philosophy.
On the one hand, I will show how astrophysical observations of dark matter, through its gravitational interaction, can be exploited to determine the most promising locations on the sky to observe a particle dark matter signal.
On the other, I demonstrate that refined theoretical calculations of the expected dark matter interactions can be used disentangle signals from astrophysical backgrounds.
Both of these approaches will be discussed in the context of general searches, but also applied to the case of an excess of photons observed at the center of the Milky Way.
This galactic center excess represents both the challenges and joys of indirect detection.
Initially thought to be a signal of annihilating dark matter at the center of our own galaxy, it now appears more likely to be associated with a population of millisecond pulsars.
Yet these pulsars were completely unanticipated, and highlight that indirect detection can lead to many new insights about the universe, hopefully one day including the particle nature of dark matter. 

%% file: contents.tex
\tableofcontents

%% file: intro.tex
\chapter{Introduction}

There is an enormous body of evidence pointing to the existence of dark matter all around us,\footnote{The evidence for dark matter is due to a body of theoretical, numerical, and experimental work conducted over decades by large fractions of the physics and astronomy community.
The range of scales over which dark matter's influence has been observed is staggering. 
The effects of dark matter stretch from its influence on our local region in the Milky Way, to the role it played in creating structures in the early universe, the imprints of which are left in the cosmic microwave background.
A recent review of the history of dark matter and the different threads of evidence pointing to its existence can be found in~\cite{Bertone:2016nfn}.}
and this evidence is entirely consistent with dark matter being a new fundamental particle.
Yet we are almost completely in the dark as to the basic properties of this particle if it exists.
For example, the mass of the dark matter particle and whether it experiences interactions with itself or the standard model beyond gravity are completely unknown, although limits exist.
Answering these questions, beyond resolving the question of what makes up 85\% of the mass in our universe, would have
 profound implications for both particle physics and astrophysics.
Famously, the standard model of particle physics does not contain a dark matter candidate.\footnote{The neutrino, being electrically neutral and having only feeble interactions with the rest of the standard model, is the only possible candidate.
Yet as neutrinos obey Fermi-Dirac statistics, there is a limit on the number density that can be packed into a given structure like a dark matter halo~\cite{Tremaine:1979we}.
This, combined with existing constraints on the smallness of the neutrino mass, forbid neutrinos from making up an $\mathcal{O}(1)$ fraction of the observed dark matter.}
In this sense, any insight as to the particle nature of dark matter would open a window into physics beyond the standard model.
In addition, insights into dark matter self-interactions, as well as the interactions between dark matter and the standard model could prove important ingredients towards a deeper understanding of how structure formed in the universe, both at the cosmological and galactic scales.

In short there are plenty of reasons to want a deeper understanding of dark matter.
In pursuit of this goal, three detection paradigms have emerged, all of which require the dark matter to have some coupling to the standard model.\footnote{Another possibility is to search for dark matter self-interactions, which could leave fingerprints on structures throughout the universe, as many of these are dark matter dominated. 
A comprehensive discussion of this approach can be found in~\cite{Buckley:2017ijx}.}
One strategy is to look for the production of dark matter at a collider.
At the Large Hadron Collider, for example, dark matter could be produced in a proton-proton collision, which we depict schematically as $p p \to {\rm DM}\, {\rm DM}$.
Of course, dark matter is famously hard to detect, so this would not result in an event we could actually see at the experiment.
If, however, one of the initial state protons emitted an observable particle such as a jet, weak boson, or photon, that we collectively denote $X$, then the process would become $p p \to {\rm DM}\, {\rm DM} + X$.
By looking for this single $X$ particle, and a large amount of missing energy associated with the fact we cannot see the dark matter particles, one can effectively search for various dark matter candidates.
This strategy is generically referred to as a mono-$X$ search, and for a recent review of the collider approach, see e.g.~\cite{Kahlhoefer:2017dnp}.
The second strategy is to search for the signatures of a dark matter particle scattering with the standard model, through an interaction of the form ${\rm SM}\, {\rm DM} \to {\rm SM}\, {\rm DM}$.
Such a scattering would cause the standard model particle to recoil, and if such an effect were detected it would be a direct indication of the influence of dark matter particles.
This approach, referred to as direct detection, has developed into a small industry, setting incredibly strong limits on the rate at which such scattering can occur.
A review of this approach can be found in, e.g.~\cite{Undagoitia:2015gya}.

The final paradigm, which represents the focus of this thesis, is referred to as indirect detection, and will be introduced in the following section.
Before proceeding, we note that often when referring to dark matter as a particle in this thesis, there will be an implicit assumption that it is a particle with a mass not too different to the particles in the standard model, the canonical example being an electroweak scale, $\mathcal{O}(100~{\rm GeV})$, supersymmetric weakly interactive massive particle (WIMP), see e.g.~\cite{Giudice:1998xp,Randall:1998uk}.
Nevertheless, we mention in passing that it is possible the dark matter could be an extremely light boson, potentially as light as $10^{-22}$ eV~\cite{Hui:2016ltb}.
This mass range includes theoretically well motivated particles such as the QCD axion~\cite{Peccei:1977hh,Peccei:1977ur,Weinberg:1977ma,Wilczek:1977pj}, which in addition to solving the strong CP problem is a viable dark matter candidate~\cite{Preskill:1982cy,Abbott:1982af,Dine:1982ah}. 
For such particles it is often more useful to think of the dark matter as a coherent field, rather than individual particles, just as how it is convenient to move from photons to waves when describing the electromagnetic field at lower energies.
This leads to a modification for the search strategies.
In recent years there has been a resurgence of efforts to search for the axion, and during grad school I contributed to this effort by introducing an analysis framework for the direct detection of axions~\cite{Foster:2017hbq}, although it will not be discussed further here.

\section{Introduction to Indirect Detection}

\begin{figure}[t!]
\begin{center}
\includegraphics[width=3.0in]{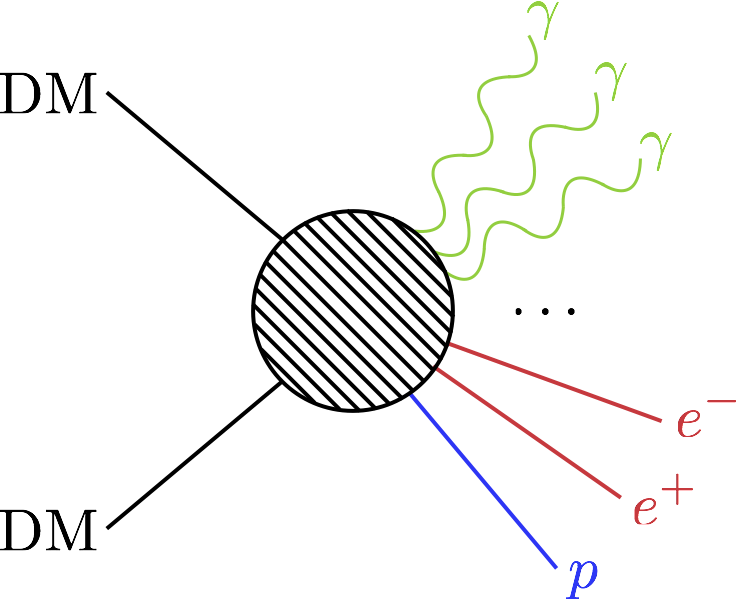}
\end{center}
\vspace{-0.5cm}
\caption{A schematic depiction of dark matter annihilation to standard model final states.
}
\label{fig:dmannihilation}
\end{figure}

If dark matter is a particle that has a fundamental interaction with the standard model, then it is possible that it could annihilate or decay into standard model final states.
This possibility, first suggested in 1978~\cite{Gunn:1978gr,Stecker:1978du}, is the dark matter analogue of familiar processes in the standard model, such as electron-positron annihilation to photons $e^+ e^- \to \gamma \gamma$ or muon decay $\mu^- \to e^- \bar{\nu}_e \nu_{\mu}$.
We can represent the dark sector analogues schematically as 
\begin{equation}\begin{aligned}
{\rm DM}\,\,\,{\rm DM} &\longrightarrow {\rm SM}~{\rm particles}\,, \\
{\rm DM} &\longrightarrow {\rm SM}~{\rm particles}\,.
\label{eq:DMschematic}
\end{aligned}\end{equation}
In both cases the identity of the standard model (SM) particles in the final state depends on the model and the mass of the dark matter (DM), as if it is too light certain states become kinematically inaccessible.
For annihilations, the standard WIMP picture is that the dark matter is its own antiparticle, allowing this process to occur, although if this is not the case, then the process above represents a dark matter particle antiparticle annihilation.
A schematic depiction of the annihilation case is presented in Fig.~\ref{fig:dmannihilation}.

Such interactions appear generically in a large class of dark matter models.
The same interactions that can give rise to dark matter annihilations can play an important role in the early universe, as at temperatures well above the dark matter mass they can keep the dark matter and standard model in thermal equilibrium.
Then, using our detailed understanding of the thermal history of the universe, this process leads to a prediction for the resultant dark matter mass fraction in the universe, which is a well measured observable.
Famously, if the dark matter mass and cross section both occur at the electroweak scale, we can exactly explain the observed dark matter density, a phenomena known as the WIMP miracle.
In fact for a wide range of masses, an electroweak cross section of $\langle \sigma v \rangle \approx 3 \times 10^{-26}~{\rm cm}^3/{\rm s}$ is required in order to obtain the observed abundance of dark matter.
We refer elsewhere for a detailed review of these points, see e.g.~\cite{Slatyer:2017sev}, however for our purposes this indicates a particular cross section value as an important target for indirect detection searches.
For the case of dark matter decay, for this particle to make up the dark matter of our universe, we expect its lifetime to be much larger than the age of the universe, which is $\sim$$10^{17}~{\rm s}$.
Considering the type of particle interactions that could induce such a decay, values for the dark matter lifetime of $10^{26}~{\rm s}$ or larger are well motivated, although we put the details aside for now as this will be discussed in detail in Chapter~\ref{chap:dmdecay} of this thesis.
The important point at this stage, is that these values provide a benchmark for experimental searches for these effects.

The central idea of indirect detection is that if these processes are occurring throughout the universe, then the standard model final states could be detectable.
As a simple example, if the final states are photons, then the universe should be illuminated by these processes in regions of high dark matter density.
Already at this stage, the two challenges of indirect detection can be identified.
On the one hand astrophysical inputs are required to determine what these regions of high dark matter density are, essentially identifying where we should look on the sky.
On the other, particle physics dictates the result of the processes in~\eqref{eq:DMschematic}; in particular it sets what types of final states we should see in a detector, and at what energies they will be observed.
In the next section we will make this precise and derive some basic results for indirect detection that will be used throughout this thesis.

\section{Deriving the Fundamental Observables}

The goal of this section is to calculate carefully what flux of particles from dark matter annihilations and decays would be predicted to arrive at a detector on Earth.
To simplify the calculation we will imagine that the standard model final states are photons, which will effectively free stream directly from the site of their production to the detector.
This should be contrasted with charged final states, such as electrons and positrons, that due to the magnetic fields that permeate the Milky Way and universe more generally take a much more complicated path to a detector. 
Although this diffusion of charged final states can be approximately accounted for, within this thesis we will focus almost exclusively on photon detectors and so the current discussion will suffice.

\begin{figure}[t!]
\begin{center}
\includegraphics[width=5.5in]{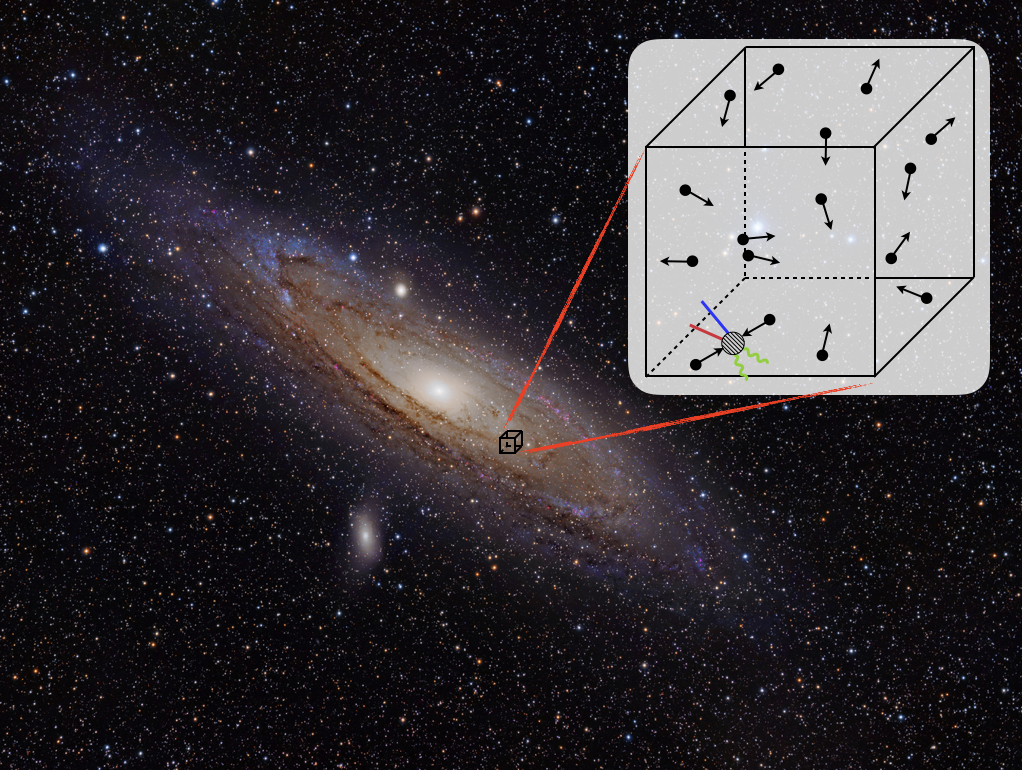}
\end{center}
\vspace{-0.5cm}
\caption{A cartoon depiction of the framework used to derive the expected flux at an experiment due to dark matter annihilation.
The starting point to this argument is to consider $N$ dark matter particles in a box of volume $V$, and consider the rate at which annihilations occur in this box.
Here we are considering the dark matter as its own antiparticle, and thus any of these particles can annihilate with any other.
The background image is of the Andromeda galaxy, and is taken from~\cite{AdamEvans}.
}
\label{fig:IDVolume}
\end{figure}

To begin with let us consider the case of dark matter annihilation to photons.
The ultimate goal is to derive the flux of photons deposited on a detector at Earth due to these annihilations, but as a starting point we will derive the rate at which annihilations are occurring in some arbitrary volume in the universe.
To this end, imagine we had the configuration shown in Fig.~\ref{fig:IDVolume}: a box of volume $V$ uniformly filled with a large number $N$ of identical dark matter particles, which are their own antiparticle.
This last condition means any particle can annihilate with any other, and is chosen as it is a common feature in dark matter models, although the case where the particle and antiparticle are distinguishable is a straightforward generalisation.
Now consider one of these particles and let us move to a frame where it is at rest.
We will consider this particle to be a target for the interaction leading to the annihilation.
The size of the target is set by the cross sectional area, $\sigma$.
In this frame the remaining $N-1$ particles will form an incident flux, and if one intersects the target's cross section it will initiate the annihilation.
The number density of the particles contributing to the flux is
\be
n_{\rm DM} \equiv (N-1)/V \approx N/V\,,
\ee
where we assumed the number of dark matter particles is large.
In terms of this the incident flux density of particles is $v n_{\rm DM}$, where $v$ represents the relative velocity between the particles as we are in the target's rest frame.
For the time being we imagine this velocity is fixed for all particles, but we will consider the more realistic case where it is drawn from a distribution shortly.
Combining this with the target area, $\sigma$, the rate at which an annihilation with this one particle would occur is simply the target area combined with the incident flux, or $\sigma v n_{\rm DM}$.
To determine the rate of annihilations in the whole box, we repeat this exercise but letting each particle take a turn as the target, which enhances the rate by $N$.
But this enhancement includes a double counting.
To see this, if two of the particles were labelled $a$ and $b$, we have counted the case where $a$ is a target hit by $b$, and where $b$ is a target hit by $a$.
More generally the number of pairs we can make from $N$ particles is $N(N-1)/2$, not $N(N-1)$ as our naive initial counting suggested.
Consequently the rate at which annihilations occur in this box is given by $\frac{1}{2} N \sigma v n_{\rm DM}$.
Then to remove any reference to the box, which was just a calculational tool, we instead re-express this as the rate of annihilations per unit volume, $\frac{1}{2} \sigma v n_{\rm DM}^2$.
As a final step, we address the fact that realistically the relative velocities will be drawn from a distribution which we should average over.
This then allows us to write the number of annihilations per unit volume per unit time as
\be
\frac{dN_{\rm ann.}}{dV dt} = \frac{1}{2} \langle \sigma v \rangle\,n_{\rm DM}^2\,.
\ee
In this expression the $\langle\,\cdot\,\rangle$ indicates an averaging over the velocity distribution, and we have also accounted for the fact that the cross section can in general depend on the relative velocity.

The above expression contains the intuitive fact that the higher the number density of dark matter particles, the more often the particles will find each other, and therefore the larger the annihilation rate will be.
Nevertheless, the rate depends on the number density of dark matter particles at a specific location in the universe, which is not something we can experimentally observe at present.
Instead our best tool is gravitational probes of dark matter, and gravitational effects are sensitive to the mass density $\rho_{\rm DM} = m_{\rm DM} n_{\rm DM}$.
Here the dark matter mass, $m_{\rm DM}$, has to be viewed as an input from the particle physics side.
To account for this, it is convenient to rewrite the above expression as
\be
\frac{dN_{\rm ann.}}{dV dt} = \frac{\langle \sigma v \rangle}{2 m_{\rm DM}^2}\,\rho_{\rm DM}^2\,.
\label{eq:annpervol}
\ee

The expression in~\eqref{eq:annpervol} achieves our goal of giving the rate of annihilations per volume at some point in the universe.
Next we want to determine the incident flux and spectrum of photons resulting from these annihilations.
For this we need a particle physics input, which is the spectrum of photons per annihilation describing the schematic process in \eqref{eq:DMschematic}, denoted $dN_{\gamma}/dE$.\footnote{We note another common convention is to consider the annihilation spectrum per dark matter particle rather than per annihilation, which will reduce spectra by a factor of two compared to those presented in this thesis.}
The spectrum is a function of energy itself, and can be defined as $dN_{\gamma}/dE(E)$ giving the number of photons in the energy range $[E,E+dE]$.
Further, the total number of dark matter particles expected from an annihilation can be readily determined from the spectrum as
\be
N_{\gamma} = \int_0^{E_{\rm max}} dE\, \frac{dN_{\gamma}}{dE}\,,
\ee
where $E_{\rm max}$ is the maximum photon energy allowed by the kinematics of the process, so $E_{\rm max} = m_{\rm DM}$ for annihilation and $m_{\rm DM}/2$ for decay.
Note $N_{\gamma}$ need not be an integer, as the annihilation process is dictated by quantum mechanics and hence is intrinsically probabilistic.
Instead the actual number of particles emerging from a given annihilation will be a draw from a Poisson distribution with mean $N_{\gamma}$.
The shape of the spectrum is dictated by at what energies it is most probable to emit a photon, and this probability is determined by quantum field theory.
Accordingly, it can be extracted from the cross section using\footnote{A similar result for decays holds if the cross section is replaced by the decay rate, $\langle \sigma v \rangle \to \Gamma$.
This point is expanded upon in App.~\ref{app:casclim-app}.}
\be
\frac{1}{N_{\gamma}} \frac{dN_{\gamma}}{dE} = \frac{1}{\langle \sigma v \rangle} \frac{d \langle \sigma v \rangle}{dE}\,.
\ee
The spectrum is highly model dependent and will be discussed extensively in this thesis, but to provide a concrete example, consider the particularly simple case where dark matter annihilates to two photons, ${\rm DM}\,{\rm DM} \to \gamma \gamma$.
In the center of mass frame for the collision, this spectrum takes the form
\be
\frac{dN_{\gamma}}{dE} = 2 \delta \left(E-m_{\rm DM} \right)\,,
\label{eq:simple2photonsspec}
\ee
such that there are two photons produced, and their energy is fixed to be the dark matter mass by the simple $2 \to 2$ kinematics.
Returning to the case of a general spectrum, combining this with the annihilation rate in~\eqref{eq:annpervol}, the number of photons per unit volume and per unit energy produced by annihilations is then
\be
\frac{dN_{\gamma}}{dE dV dt} = \frac{\langle \sigma v \rangle}{2 m_{\rm DM}^2} \frac{dN_{\gamma}}{dE}\,\rho_{\rm DM}^2\,.
\label{eq:dNdEdVdt}
\ee

At this stage we just know the rate at which photons are being injected into the universe.
If we want to detect this effect, the quantity of interest is the number of these photons incident on a detector at Earth.\footnote{We note that detection at Earth is not the only way to determine the impact of these processes.
If they have been occurring throughout the history of the universe, then their impact can be seen elsewhere, for example through perturbations to the cosmic microwave background.
This is a powerful probe and can be used to constrain annihilation~\cite{Slatyer:2015jla,Slatyer:2015kla}, decay~\cite{Slatyer:2016qyl}, and even contributions to processes such as reionization~\cite{Liu:2016cnk}.}
On average, the photons produced will disperse isotropically out over a sphere.
If the proper distance between the volume element under consideration and the telescope is $s$, then by the time the photons reach the Earth they are spread over an area $4 \pi s^2$.
Imagining that we have a detector with a differential effective area $dA$,\footnote{The effective area is an efficiency corrected notion of detector area.
The larger the collection area of the telescope, the more photons will be detected following the argument in the text.
Nevertheless any realistic experiment will not perfectly detect every incident photon, and instead will only do so with some efficiency.
The effective area is a way of quantifying this, and operates such that if you have a telescope of area 2~m$^2$ with a 50\% efficiency, then the effective area is 1~m$^2$.
In general the efficiency and hence effective area will vary with energy, and additional details such as the incident angle of the photon on the detector and where it hit, although we will put these complications aside for the present discussion.} then only $dA/(4 \pi s^2)$ of the photons produced will be detected, and we have to downweight the number of photons produced as given in~\eqref{eq:dNdEdVdt} by this factor.
Doing so, we arrive at\footnote{By writing the same energy on both sides of this expression we have implicitly assumed the emitted photon energy is equal to the detected value.
In general this will not be true.
For one example, a photon emitted a cosmological distance from the detector will be redshifted during its propagation.
Another example would be if the center of mass frame of the annihilation differs from the detector rest frame, the energy will be shifted.
This effect would cause an initial $\delta$ function line to be smeared out by the velocity dispersion of the dark matter.
We put these caveats aside for now, but where relevant will address them in the main body of this thesis.}
\be
\frac{dN_{\gamma}}{dE dV dA dt} = \frac{\langle \sigma v \rangle}{8 \pi m_{\rm DM}^2} \frac{dN_{\gamma}}{dE}\,\frac{\rho_{\rm DM}^2}{s^2}\,.
\label{eq:dNdEdVdAdt}
\ee
It is convenient at this point to define the notion of differential photon flux incident on the detector, as
\be
d \Phi_{\gamma} \equiv \frac{d N_{\gamma}}{dA dt}\,,
\ee
which has units of photons per effective area per time.\footnote{There are two common variants of flux used in indirect detection: 1. particles per unit effective area per unit time; and 2. particles per unit effective area per unit time per unit solid angle on the sky.
Our definition here corresponds to the former, and note that the difference between them is that the first definition is the integrated version of the second over the full sky.
We will not address this issue further, although we emphasize caution is required as the two naively differ by a factor $\mathcal{O}(4\pi)$, and errors due to confusing the two exist in the literature.
A careful description of the units appeared in one of the works I completed during grad school, see App. A of~\cite{Lisanti:2017qoz}.}
It includes all the experimental quantities, such as telescope size, efficiency, and observation time; increasing any of these leads to more collected photons.
This definition allows us to rewrite~\eqref{eq:dNdEdVdAdt} as
\be
\frac{d \Phi_{\gamma}}{dE} = \frac{\langle \sigma v \rangle}{8 \pi m_{\rm DM}^2} \frac{dN_{\gamma}}{dE}\,\frac{\rho_{\rm DM}^2}{s^2} dV\,.
\label{eq:dPhidEdV}
\ee

This last expression represents the differential energy flux produced by dark matter annihilations from a differential volume element $dV$ a distance $s$ away.
But of course there is a lot of dark matter out there in the universe, all of which can contribute photons at the detector.
To account for this we will want to integrate over some volume of dark matter, accounting for the fact that dark matter is not distributed homogeneously, but rather often collapsed into objects like the Milky Way halo.
We achieve this by making $\rho_{\rm DM}$ position dependent.
It is then convenient to perform this integral in a spherical coordinate system centered on the Earth, so that we can write $dV = s^2 ds d\Omega$.
From here note that by observing different regions of the celestial sphere, we can restrict the patch of solid angle we look at, but we cannot isolate a specific radial scale in general.
Incident photons could have come from a dark matter annihilation 1 mm or 1 Gpc from the detector and we could not distinguish them.
As such we need to integrate over all distances.
With this in mind, let us say we observe a region of solid angle $\Sigma$, which could be the full sky or a one degree circle around the galactic center for example, then the total detected energy flux is simply the integrated version of~\eqref{eq:dPhidEdV}, and is given by
\be
\frac{d \Phi_{\gamma}}{dE} = \frac{\langle \sigma v \rangle}{8 \pi m_{\rm DM}^2} \frac{dN_{\gamma}}{dE}\,\int_0^{\infty} ds \int_{\Sigma} d\Omega\,\rho_{\rm DM}^2(s, \Omega)\,.
\label{eq:IDfluxfinal}
\ee

This expression achieves our goal of expressing the photon flux arriving at an experiment on Earth due to dark matter annihilations.
Integrated over the energy range of the telescope, we can determine the expected flux, which we can turn into an expected number of observed photons when combined with the experimental parameters, specified via the detector effective area and observation time.
Yet predicting this flux is entirely dependent upon our ability to determine the various quantities appearing on the right hand side of~\eqref{eq:IDfluxfinal}.
This is the central challenge of indirect detection and the focus of the work in this thesis.

Observe that the various terms appearing in~\eqref{eq:IDfluxfinal} have factorized into quantities dictated by particle physics---the cross section, mass, and spectrum---and those fixed by astrophysics---the dark matter density.\footnote{In some models of dark matter, this factorization is not always exact.
For example, if the cross section has a large velocity dependence, which can happen in models with Sommerfeld enhancement of the annihilation~\cite{Hisano:2003ec,Hisano:2004ds,Cirelli:2007xd,ArkaniHamed:2008qn,Blum:2016nrz}, then the result is dependent on the velocity distribution of dark matter within $\Sigma$, which is also determined by astrophysics.
We will not consider such cases in this thesis, although see~\cite{Boddy:2017vpe,Boddy:2018qur}, for some recent work in this direction.}
Motivated by this, it is common to rewrite the expression in the following manifestly factorized form
\begin{equation}\begin{aligned}
\frac{d \Phi_{\gamma}}{dE} &= \frac{d\Phi_{\gamma}^{\rm PP}}{dE} \times J_{\Sigma}\,, \\
\frac{d\Phi_{\gamma}^{\rm PP}}{dE} &\equiv \frac{\langle \sigma v \rangle}{8 \pi m_{\rm DM}^2} \frac{dN_{\gamma}}{dE}\,, \\
J_{\Sigma} &\equiv \int_0^{\infty} ds \int_{\Sigma} d\Omega\,\rho_{\rm DM}^2(s, \Omega)\,.
\label{eq:fullannihilation}
\end{aligned}\end{equation}
The quantities on the final two lines are referred to as the particle physics factor, and the $J$-factor respectively.

For the case of decaying dark matter, an analogous argument holds, which we sketch out below.
Consider again $N$ particles in a box of volume $V$, and assume now these particles have a lifetime $\tau$.
The probability that one of these particles has decayed after a time $t$ is given by the cumulative distribution function $1-e^{-t/\tau}$.
Accordingly if we have $N$ particles undergoing the same process, then the expected number remaining after a time $t$ is $N(1-e^{-t/\tau})$, and rate of these decays is then the time derivative of this quantity.
Thus the rate of decays per unit volume is given by
\be
\frac{dN_{\rm dec.}}{dV dt} = \frac{e^{-t/\tau}}{\tau} n_{\rm DM}\,.
\ee
Recall for the dark matter to still be around, we require $\tau \gg t_{\rm universe}$, and indeed many models predict the lifetime to be many orders of magnitude longer than the age of the universe.
As such, $e^{-t/\tau} \approx 1$ is a very good approximation, and again moving to the more convenient mass density, we have
\be
\frac{dN_{\rm dec.}}{dV dt} = \frac{1}{m_{\rm DM} \tau} \rho_{\rm DM}\,.
\ee
From here the argument is completely analogous to the annihilation case, and we arrive at the factorized expression
\begin{equation}\begin{aligned}
\frac{d \Phi_{\gamma}}{dE} &= \frac{d\Phi_{\gamma}^{\rm PP}}{dE} \times D_{\Sigma}\,, \\
\frac{d\Phi_{\gamma}^{\rm PP}}{dE} &\equiv \frac{1}{4 \pi m_{\rm DM} \tau} \frac{dN_{\gamma}}{dE}\,, \\
D_{\Sigma} &\equiv \int_0^{\infty} ds \int_{\Sigma} d\Omega\,\rho_{\rm DM}(s, \Omega)\,.
\label{eq:fulldecay}
\end{aligned}\end{equation}

Both the particle physics and astrophysical factors are different in this case, with arguably the most striking distinction occurring in the astrophysical dependence.
For annihilation we have $J$, which depends on $\rho_{\rm DM}^2$, whereas $D$ is related to $\rho_{\rm DM}$.
Due to this, at first order the $D$-factor is just sensitive to the total dark matter mass of the object being observed, whereas the $J$-factor has a more complicated dependence on the substructure of the dark matter within an object, and in particular contributions from locally overdense regions can be strongly enhanced due to the $\rho_{\rm DM}^2$ scaling.
This implies immediately that the optimal targets of observation could well differ between annihilation and decay, and this is a topic explored in this thesis.

\section{Simple Scaling Estimates}
    
In the main work presented in this thesis we will consider very explicit forms of the various indirect detection quantities derived in the previous section.
Before going into this however, here we consider simplified versions of these expressions and consider their scaling with the various parameters.
Further, we will consider how to use these scalings to estimate the reach of indirect detection experiments.

To facilitate this estimate, consider the particularly simple case of annihilation or decay to two photons, as then we can use the simplified form of the spectrum provided in~\eqref{eq:simple2photonsspec} to simplify the results in~\eqref{eq:fullannihilation} and~\eqref{eq:fulldecay}.
Doing so, we have
\begin{equation}\begin{aligned}
\frac{d \Phi_{\gamma}^{\rm ann.}}{dE} &= \frac{\langle \sigma v \rangle J_{\Sigma}}{4 \pi m_{\rm DM}^2} \delta \left( E - m_{\rm DM} \right)\,, \\
\frac{d \Phi_{\gamma}^{\rm dec.}}{dE} &= \frac{D_{\Sigma}}{2 \pi m_{\rm DM} \tau} \delta \left( E - m_{\rm DM} \right)\,.
\end{aligned}\end{equation}
Further, if we assume that the detector effective area $\mathcal{E}$ and the observation time $T$ are independent of energy and $\Sigma$, we can determine the expected number of photons from the above as
\begin{equation}\begin{aligned}
\left. N_{\gamma}^{\rm ann.} \right|_{E=m_{\rm DM}} &= \frac{\langle \sigma v \rangle J_{\Sigma} \mathcal{E} T}{4 \pi m_{\rm DM}^2}\,, \\
\left. N_{\gamma}^{\rm dec.} \right|_{E=m_{\rm DM}} &= \frac{D_{\Sigma} \mathcal{E} T}{2 \pi m_{\rm DM} \tau} \,,
\label{eq:DMsigcounts}
\end{aligned}\end{equation}
where we have noted that this only holds for $E=m_{\rm DM}$, otherwise the detector will see zero photons.

\begin{figure}[t!]
\begin{center}
\includegraphics[width=5.5in]{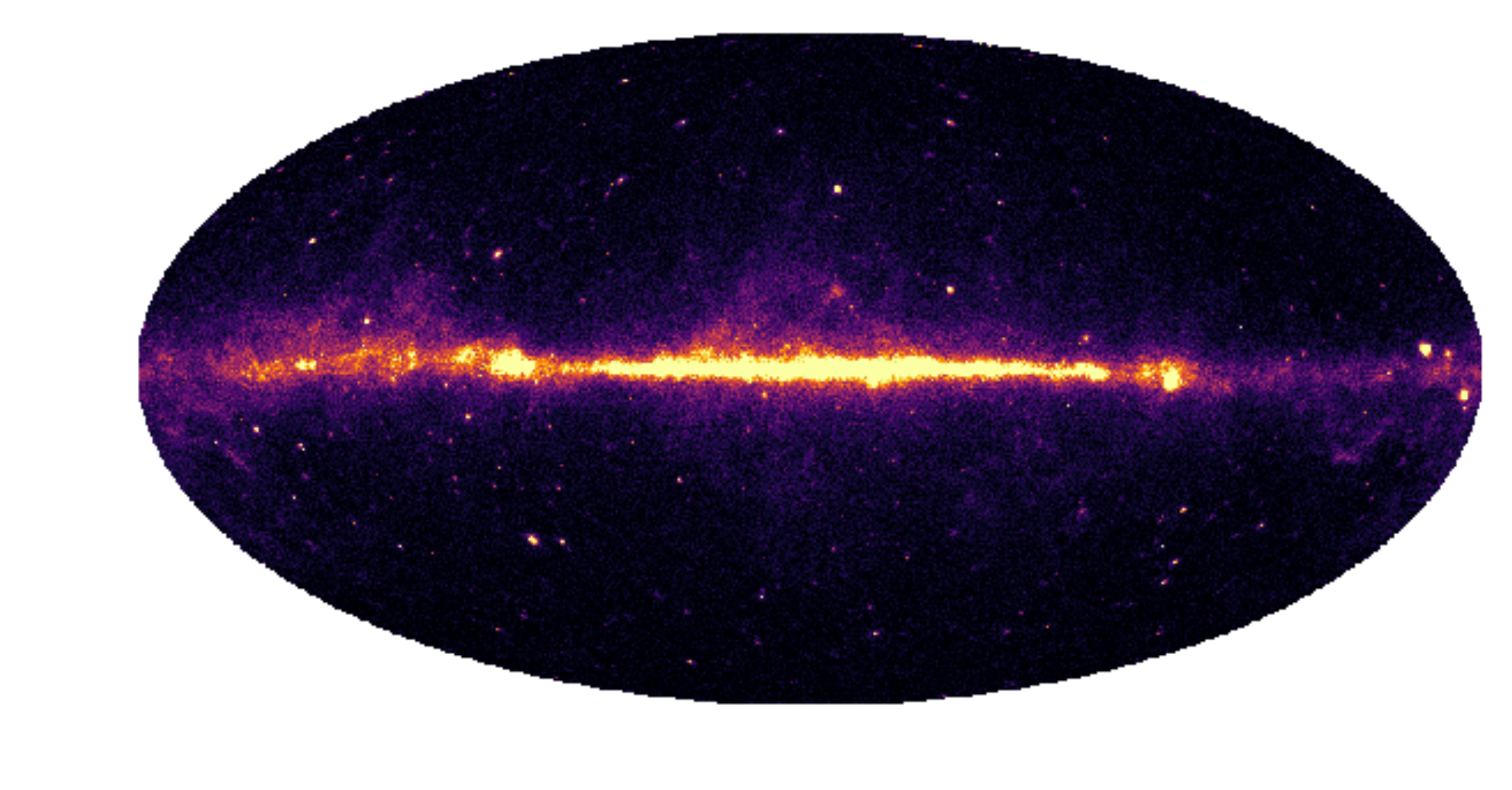}
\end{center}
\vspace{-0.5cm}
\caption{The gamma-ray sky as observed by the \textit{Fermi} Space Telescope.
Much of the work in this thesis is devoted to searching for the imprints of dark matter in this dataset.
In detail, the data represents photons collected from 200 MeV -- 2 TeV between August 4, 2008 through July 7, 2016.
In addition we are only showing the photons with the highest quality angular reconstruction.
The image represents a Mollweide projection of the celestial sphere in galactic coordinates, with the center of the Milky Way in the middle of the image.
}
\label{fig:Fermidata}
\end{figure}

At this point we can already estimate what type of experiment we might need to observe these effects.
To this end, consider the case of a 100 GeV dark matter candidate, which annihilates with the canonical thermal relic cross section $\langle \sigma v \rangle = 3 \times 10^{-26}$ cm$^3$/s.
For our observation, we take the Andromeda galaxy, which is expected to be the brightest extragalactic source of dark matter annihilation, with $J_{\Sigma} \approx 5 \times 10^{19}~ {\rm GeV}^2/{\rm cm}^5\cdot {\rm sr}$~\cite{Lisanti:2017qlb}.
To detect this, we clearly need to see at least one photon in our detector, and to have a better chance imagine we wanted to observe 40.
Then inverting the relation in~\eqref{eq:DMsigcounts}, we obtain
\be
\mathcal{E} T \approx \pi \times 10^{12}~{\rm cm}^2 \cdot {\rm s}\,.
\ee
The photons produced from this annihilation will be exactly at 100 GeV.
This is a particularly challenging energy to observe gamma-rays, as they interact strongly with the Earth's atmosphere, initiating a shower of particles.
At TeV and higher energies these showers become large enough that they can be observed on the planet's surface.
Yet at 100 GeV this is challenging, and to obtain a significant flux a satellite based experiment is required.
Given the costs and challenges associated with getting experiments into orbit, a telescope with a collection area of $\sim$1 m$^2$, or $10^4$ cm$^2$, is close to as large as we could expect the effective area of such an instrument to be.
Such an experiment would then have to observe Andromeda for
\be
T \approx \pi \times 10^8~{\rm s} \approx 10~{\rm years}\,.
\ee
This is a significant, although not unimaginable time scale.
Further, by combining observations of different targets with comparable values for $J_{\Sigma}$, one could hope to build up a similar amount of dark matter flux.
Excitingly, the \textit{Fermi} gamma-ray space telescope exactly fits the criteria described above.
Launched on June 11, 2008, it has almost exactly 10 years of data, and has a collection area of approximately 1 m$^2$.
This indicates that the ideal dataset for detecting electroweak scale dark matter annihilation has already been collected, and explains why data from the \textit{Fermi} satellite will feature heavily in this thesis. 
The dataset collected by \textit{Fermi} is shown in Fig.~\ref{fig:Fermidata}, which is a picture of the sky in gamma-rays.

Returning to our estimate for the scalings of indirect detection, we need to address the reality that the universe emits photons at almost all energies due to non-dark matter related phenomena.
The challenge then is to find a hint for a dark matter signal on top of this background, and for this we need an estimate for its contribution.
The background contribution will vary with energy, however generically follows a power law ($dN/dE \sim E^{-n}$), with various breaks corresponding to different physical phenomena.
Depending on the energy range of interest, the physical processes responsible for generating these photons varies.
In this thesis we will be primarily focussed on gamma-rays, which are photons with energy higher than $\sim$MeV.
At these energies, astrophysical photons emerge from non-thermal processes.
The dominant contribution arises from cosmic-ray proton collisions with interstellar hydrogen, which leads to a $p-p$ collision, much like the Large Hadron Collider.
In such processes, neutral pions are produced copiously, which leads to photon production through their decay $\pi^0 \to \gamma \gamma$.
Generically, we expect the initial cosmic-ray protons to have an energy spectrum which is a power law, scaling as $E^{-2}$, as a result of Fermi shock acceleration~\cite{Fermi:1949ee}, one of the dominant astrophysical mechanisms for accelerating charged particles to high energies.
The photons produced from the subsequent $p-p$ collisons will be expected to have a softer spectrum than the initial protons, but as a first order estimate, we can take the gamma-ray spectrum to also have a generic $E^{-2}$ scaling.\footnote{More realistically, the spectra associated with various astrophysical sources are not perfectly described by a power law, and where they are the index can deviate from $-2$.
For example, the \textit{Fermi} gamma-ray telescope has estimated the isotropic gamma-ray spectrum to scale as $\sim$$E^{-2.3}$, but with an exponential cutoff on the spectrum at several hundred GeV~\cite{Ackermann:2014usa}.
Further, the dominant $p-p$ galactic contribution, has an even softer spectrum of $\sim$$E^{-2.7}$, although sources with harder spectra exist, such as the \textit{Fermi} bubbles~\cite{Su:2010qj}.
As such, the background model used here is only a rough approximation, but is sufficient for the simple scaling arguments presented.}
Accordingly around these energies we expect the approximate scaling
\be
\frac{\Phi_{\gamma}^{\rm bkg.}}{dE} = A \left( \frac{E}{1~{\rm GeV}} \right)^{-2}\,,
\ee
where $A$ is an energy independent constant that determines the flux received at 1 GeV.
Assuming we can look away from the plane of the Milky Way, then a large contribution to the background comes from the position independent isotropic emission, and \textit{Fermi} measurements~\cite{Ackermann:2014usa} estimate this to have amplitude
\be
A \approx \Omega_{\Sigma}\,10^{-6}~{\rm photons/cm}^2{\rm /s/GeV}\,.
\ee
The amount of flux arriving from isotropic emission is of course dependent upon the size of the region considered, which we denote $\Omega_{\Sigma}$ in the above expression.
Now to compare to our line search, we are interested in the background flux near a particular dark matter mass.
For the sake of estimating the sensitivity to GeV scale dark matter, if we approximate the energy resolution as $\approx$1 GeV, then we can estimate the number of photons via
\be
N_{\gamma}^{\rm bkg.} = \frac{10^{-6}\,\Omega_{\Sigma} \mathcal{E} T}{m_{\rm DM}^2}\,.
\label{eq:DMbkgcounts}
\ee
where we now require $m_{\rm DM}$ is measured in GeV.
Of course we emphasize that this is a crude estimate of the actual gamma-ray background, for at least two reasons.
Firstly, as suggested above the background emission is often softer than $E^{-2}$, although not dramatically.
Secondly, generically the highly non-isotropic gamma-ray emission associated with sources within the Milky Way has a significant contribution, so this usually must be accounted for.
This second point is evident in Fig.~\ref{fig:Fermidata}, where the \textit{Fermi} dataset is brightest along the plane of the Milky Way. 
Nevertheless these complications should not significantly impact the simple estimates we are seeking here.

So now from \eqref{eq:DMsigcounts} and \eqref{eq:DMbkgcounts} we have a model for the expected signal and background contributions to the photon flux at an experiment.
We want to combine the two of these, which we will refer to as the signal counts $S$ and background counts $B$, to determine the scaling of the indirect detection sensitivity.
To this end we need a statistical, and more specifically a likelihood, framework.
Focussing on higher energies, say X-ray and above, the experiments are effectively counting the number of incident photons.
This implies the correct likelihood framework is the Poisson likelihood, where we have a predicted mean $S+B$. 
When performing analyses in the remainder of this thesis, this is the approach we will use, but for the following simple estimate we will assume we have a large enough number of photons that the Gaussian likelihood with mean $S+B$ and standard deviation $\sqrt{S+B}$ provides a good approximation.
In detail, if we observe $d$ photons, then we can write the likelihood as
\be
\mathcal{L}(d|S,B) = \frac{1}{\sqrt{2 \pi (S+B)}} \exp \left[ -\frac{\left( d-S-B\right)^2}{2(S+B)} \right]\,.
\ee
To test for the discovery of dark matter, a convenient test statistic to define is twice the log ratio of a hypothesis with and without dark matter, specifically
\be
{\rm TS} = 2 \left[ \ln \mathcal{L}(d|S,B) - \ln \mathcal{L}(d|0,B) \right]\,.
\ee
Substituting in the Gaussian form of the likelihood, we have
\be
{\rm TS} = -\frac{\left( d-S-B\right)^2}{S+B} + \frac{\left( d-B\right)^2}{B} - \ln \left( 1 + \frac{S}{B} \right)\,,
\ee
which is of course again a function of the signal and background models, as well as the data.
To calibrate our expectations, imagine that the data is actually perfectly described by a model with the background and signal, i.e. $d$ is a Poisson draw from $S+B$.
To determine our expected reach in this scenario, we can use the Asimov analysis framework~\cite{Cowan:2010js}, where we obtain the asymptotic expectation for TS under many experimental realisations by using $d=S+B$.
If so, then denoting the asymptotic TS as $\widetilde{\rm TS}$, we have
\be
\widetilde{\rm TS} = \frac{S^2}{B} - \ln \left( 1 + \frac{S}{B} \right) \approx \frac{S(S-1)}{B} \approx \frac{S^2}{B}\,.
\ee
In the second step above, we assumed that $S+B \approx B$, namely that the background will always be much larger than the signal.
Given that we have not discovered dark matter using these techniques, this is often a good approximation.\footnote{An exception to this can occur for line searches, where the entire dark matter flux is very localised in energy, whereas the background is not.
As such, if the energy resolution of the detector is good enough one can potentially achieve $S > B$.
This point will not impact the thrust of our main scaling arguments, however, so we put it aside.}
In the third step, we used the fact that even though $B \gg S$, we still want significantly more than one photons from dark matter to have a chance of discovering it, so $S \gg 1$.
Now in the case where we can ignore the look elsewhere effect, we can relate the ${\rm TS}$ to the local significance for discovery, $\sigma$, according to $\sqrt{\widetilde{\rm TS}} = \sigma$.
Requiring a 5$\sigma$ significance discovery then fixes
\be
S = 5\sqrt{B}\,.
\ee

Taking this result and returning to our expected signal and background counts from \eqref{eq:DMsigcounts} and \eqref{eq:DMbkgcounts}, our expected reach for the cross section and lifetime of dark matter will scale as
\begin{equation}\begin{aligned}
\langle \sigma v \rangle &\sim \frac{m_{\rm DM}}{J_{\Sigma}} \sqrt{\frac{\Omega_{\Sigma}}{\mathcal{E} T}}\,, \\
\tau^{-1} &\sim \frac{1}{D_{\Sigma}} \sqrt{\frac{\Omega_{\Sigma}}{\mathcal{E} T}}\,.
\label{eq:IDnaivescaling}
\end{aligned}\end{equation}
A number of basic scalings for indirect detection can be seen in these results.
As would be expected, increasing the $J$ or $D$-factor, or similarly the effective area or observation time, allow us to reach smaller cross sections and inverse lifetimes, as does reducing the background.
Interestingly, we see that finding better targets for observation, namely finding objects with better astrophysics factors, has a larger impact than improvements on the other parameters.
Note also that sensitivity to the annihilation cross section degrades with increasing dark matter mass, as highlighted by the presence of $m_{\rm DM}$ in the above expression, whilst the lifetime sensitivity is mass independent.
This basic scaling is common in indirect detection results, and is usually described as originating from the following heuristic argument.
Due to gravitational probes, we can determine the amount of dark matter mass in an object.
As we increase the mass of the individual dark matter particles, we must reduce their number density $n_{\rm DM}$.
As annihilation is dependent upon $n_{\rm DM}^2$ and decay $n_{\rm DM}$, this alone leads to a reduction of sensitivity in the two cases as $1/m_{\rm DM}^2$ and $1/m_{\rm DM}$ respectively.
Yet in both cases, the higher mass increases the power injected per event by $m_{\rm DM}$.
The combination of the two effects reproduces the scaling in~\eqref{eq:IDnaivescaling}.
Yet from the derivation of that result, we can see that alternative assumptions about the shape of the signal spectrum or the background can lead to variations in the scaling with mass.

We have reached the limit of what we can achieve with the rough scaling arguments presented above.
Within this thesis we will not only refine such arguments, but more importantly go beyond them to extend the discovery potential for dark matter in indirect detection through novel analysis strategies and refined theoretical predictions.

\newpage
\section{Organization of this Thesis}

As we have seen, the indirect detection flux factorizes into a contribution from astrophysics and particle physics.
In the same fashion, this thesis and much of my work during grad school approximately factorize down the same line.
The first half of the thesis, Chapters~\ref{chap:gevexcess},~\ref{chap:fermigg}, and~\ref{chap:dmdecay}, will focus on how we can search for evidence of dark matter, or set limits in its absence, by considering promising astrophysical targets.
The second half, Chapters~\ref{chap:cascsig},~\ref{chap:casclim}, and~\ref{chap:dm1loop}, will turn to refining the particle physics predictions, and demonstrating how these can enhance our understanding of how dark matter might first appear in the sky.
Note that for each of the substantive chapters in the main text there is an associated appendix where many of the technical details appear.

In more detail, the first half will be further subdivided into three parts.
In the first of these, presented in Chapter~\ref{chap:gevexcess} we provide a taste of what is considered the ultimate aim of indirect detection, analysis of a putative dark matter signal.
This signal is an excess of gamma-rays observed by the \textit{Fermi} telescope near the galactic center, and as such is commonly referred to as the galactic center excess (GCE).
The excess was observed almost as soon as the \textit{Fermi} data became publicly available~\cite{Goodenough:2009gk}, and then was followed up in a number of studies~\cite{Hooper:2010mq,Boyarsky:2010dr,Hooper:2011ti,Abazajian:2012pn,Gordon:2013vta,Abazajian:2014fta}.
That it was seen so quickly is consistent with a dark matter interpretation, as the galactic center is the location on the sky with the largest $J_{\Sigma}$.
My first contribution to the GCE anomaly came in~\cite{Daylan:2014rsa} and this represents the contents of Chapter~\ref{chap:gevexcess}.
In that work we demonstrated that the excess satisfies many properties you would expect for dark matter, such as being far more spherically symmetric than the expected background contributions.
This work generated a lot of excitement, as it gave further indication that this excess was due to dark matter.
Nevertheless it was later realised, due to the application of a novel statistical framework~\cite{Lee:2015fea} and a wavelet based technique~\cite{Bartels:2015aea}, that in fact the excess looks to be coming from a population of point sources.
The novel statistical technique is known as the non-Poissonian template fit, and my work has included applying this method to further study the GCE and in particular its spectrum at high energies~\cite{Linden:2016rcf}, and also in making the method into a publicly available code~\cite{Mishra-Sharma:2016gis}. 
Dark matter is not expected to have point-source-like spatial morphology, and thus the leading hypothesis is that GCE is due to an unresolved population of point sources, which are most likely millisecond pulsars, see e.g.~\cite{Brandt:2015ula}, although there is much ongoing work to fully understand this excess, examples of which include~\cite{Macias:2016nev,Bartels:2017vsx,Balaji:2018rwz,Bartels:2018eyb}.
Attempts have been made to find evidence for the existence of these millisecond pulsars amongst the resolved point sources \textit{Fermi} has seen~\cite{Fermi-LAT:2017yoi}, although as my collaborators and I have demonstrated such searches are at present unable to say anything definitive~\cite{Bartels:2017xba}.
Another basic challenge is presented by the fact that if the excess was due to dark matter, then we may have expected to see a signal from other regions with a large $J_{\Sigma}$, such as the Milky Way dwarf spheroidal galaxies.
Searches in the dwarfs, however, have not seen a similar excess~\cite{Ackermann:2015zua,Fermi-LAT:2016uux}, and in~\cite{Keeley:2017fbz} my collaborators and I quantified the existing tension between these two measurements.
As such, it is unlikely the GCE is associated with dark matter, but Chapter~\ref{chap:gevexcess} represents a study of the sort one would perform in the presence of an excess.

In Chapter~\ref{chap:fermigg} we look beyond the galactic center, and indeed beyond our own Milky Way, in search of extragalactic signals of dark matter annihilation.
As mentioned above, the largest expected regions of $J_{\Sigma}$ beyond the galactic center are associated with structures within the Milky Way, in particular dwarf spheroidal galaxies.
Non-observation of a dark matter signal in these objects leads to some of the strongest constraints on the annihilation cross section~\cite{Ackermann:2015zua,Fermi-LAT:2016uux}.
The question explored in Chapter~\ref{chap:fermigg}, is whether extragalactic observations can compete with the dwarf searches, and indeed we will demonstrate that they can.
The intuition is that even though extragalactic objects are much further away, they can be significantly more massive.
For example, the mass of the Milky Way is $\approx 10^{12}~M_{\odot}$, whereas the Virgo galaxy group has a substantially larger mass of $\approx 4 \times 10^{14}~M_{\odot}$.
The aim is to exploit this additional mass, combined with the fact there are an extraordinarily large number of galaxies and galaxy clusters outside the Milky Way, to compensate for the additional distance.
This Chapter represents work published in~\cite{Lisanti:2017qlb}, which appeared with a companion paper~\cite{Lisanti:2017qoz}, where my collaborators and I extensively validated our methods on $N$-body simulations.

Moving beyond annihilation, in Chapter~\ref{chap:dmdecay} we consider how to search for dark matter decay.
In this chapter, based on~\cite{Cohen:2016uyg}, my collaborators and I used data from the \textit{Fermi} telescope to set the strongest constraints on the dark matter lifetime over almost six orders of magnitude from a GeV to almost a PeV.
That we set limits is an indication that no clear signs of an excess was observed, although the methods we introduce allow for some of the deepest searches ever performed.
Further these methods have application beyond \textit{Fermi}, and as an example I worked with the HAWC collaboration to apply our methods to their instrument~\cite{Abeysekara:2017jxs}.

Starting in Chapter~\ref{chap:cascsig} we pivot to the particle physics side of indirect detection.
This chapter, based on work appearing in~\cite{Elor:2015tva}, should be considered as the particle physics side of exploring a potential excess, again in the context of the GCE.
As mentioned, the GCE was for a time considered a very promising candidate for a signal from annihilating dark matter, and as such generated a great deal of excitement and work in the particle physics community.
The central idea was to try and determine what type of particle physics interaction could give rise to the specific spectrum \textit{Fermi} was seeing, basically trying to determine exactly what was going on in Fig.~\ref{fig:dmannihilation}.
An enormous number of proposals were put forward, and the work presented in Chapter~\ref{chap:cascsig} focussed on trying to organize the space of models in a convenient way.
In that work we demonstrated that many complicated dark matter models, where there can be a lot of structure in the blob of Fig.~\ref{fig:dmannihilation}, can be well approximated using relativistic kinematics.
For example, if instead of having ${\rm DM}\,{\rm DM} \to {\rm SM}\,{\rm SM}$, the dark matter annihilated to an intermediate state particle $\phi$, then the process is now a cascade annihilation: ${\rm DM}\,{\rm DM} \to \phi \phi$, followed by two copies of $\phi \to {\rm SM}\,{\rm SM}$.
The spectrum obtained in this more complicated case can actually be derived form the earlier spectrum by use of relativistic kinematics, and in this way starting from simple models we can generate the expectation for more complicated scenarios straightforwardly, even when many cascades occur in the dark sector.
This formed a framework that allowed for a broad consideration of the type of models that could explain the GCE, and represents the contents of this chapter.

Chapter~\ref{chap:casclim} builds off the insights of cascade annihilations derived in Chapter~\ref{chap:cascsig} to consider how this general framework can be turned to set more model-independent constraints on dark matter.
In this chapter, based on work appearing in~\cite{Elor:2015bho}, we demonstrated how the usual constraints on dark matter annihilation arising from measurements of photons from \textit{Fermi}, electrons, positrons, and antiprotons at AMS-02, and observations of the cosmic microwave background by \textit{Planck}, are modified when a more complicated dark sector is considered.
This work significantly extends the use of the standard published limits, and allows theorists to more easily convert those results into ones applicable to more complicated dark matter models.

Chapter~\ref{chap:dm1loop} represents the final substantive topic of the thesis, and is devoted to a detailed study of the physics involved in a specific dark matter annihilation. 
For this purpose we focus on a particular model for dark matter, the supersymmetric wino, and perform a full one-loop calculation for ${\rm DM}\,{\rm DM} \to \gamma \gamma$ cross section in this theory.
The calculation demonstrates many of the complications that can arise when a given model is considered in detail, for example the Sommerfeld enhancement and the resummation of large logarithms are both relevant and included.
This chapter represents the work that appeared in~\cite{Ovanesyan:2016vkk}.
This work was recently followed up by my collaborators and I in~\cite{Baumgart:2017nsr}, where we showed that contributions from other final states such as $W^+ W^- \gamma$ can play an important role in the photon spectrum near the dark matter mass.
For any realistic instrument with imperfect energy resolution, such effects are impossible to disentangle from the pure $\gamma \gamma$ contribution and thus we showed they can significantly modify the experimental expectation.

The focus of my research at grad school has been on the two faces of indirect detection as described above.
Nevertheless there are two projects I have completed that fall outside this general program.
The first of these related to exploiting novel jet algorithms on data collected at LHCb to uncover the splitting functions of heavy quarks, in particular the charm $c$ and bottom $b$ quarks~\cite{Ilten:2017rbd}.
The second, which was mentioned above, related to an analysis framework for axion direct detection~\cite{Foster:2017hbq}, which is a method for searching for dark matter that is much lighter than what we consider in the remainder of this thesis.

%% file: gevexcess.tex
\chapter{The Characterization of the Gamma-Ray Signal from the Central Milky Way: A Case for Annihilating Dark Matter}\label{chap:gevexcess}

\section{Introduction}\label{intro}

Weakly interacting massive particles (WIMPs) are a leading class of candidates for the dark matter of our universe.
If the dark matter consists of such particles, then their annihilations are predicted to produce potentially observable fluxes of energetic particles, including gamma rays, cosmic rays, and neutrinos.
Of particular interest are gamma rays from the region of the Galactic Center which, due to its proximity and high dark matter density, is expected to be the brightest source of dark matter annihilation products on the sky, hundreds of times brighter than the most promising dwarf spheroidal galaxies.

Over the past few years, several groups analyzing data from the \textit{Fermi} Gamma-Ray Space Telescope have reported the detection of a gamma-ray signal from the inner few degrees around the Galactic Center (corresponding to a region several hundred parsecs in radius), with a spectrum and angular distribution compatible with that anticipated from annihilating dark matter particles~\cite{Goodenough:2009gk,Hooper:2010mq,Boyarsky:2010dr,Hooper:2011ti,Abazajian:2012pn,Gordon:2013vta,Abazajian:2014fta}.
More recently, this signal was shown to also be present throughout the larger Inner Galaxy region, extending kiloparsecs from the center of the Milky Way~\cite{Hooper:2013rwa,Huang:2013pda}.
While the spectrum and morphology of the Galactic Center and Inner Galaxy signals have been shown to be compatible with that predicted from the annihilations of an approximately 30-40 GeV WIMP annihilating to quarks (or a $\sim$7-10 GeV WIMP annihilating significantly to tau leptons), other explanations have also been proposed.
In particular, it has been argued that if our galaxy's central stellar cluster contains several thousand unresolved millisecond pulsars, they might be able to account for the emission observed from the Galactic Center~\cite{Hooper:2010mq,Abazajian:2010zy,Hooper:2011ti,Abazajian:2012pn,Gordon:2013vta,Abazajian:2014fta}.
The realization that this signal extends well beyond the boundaries of the central stellar cluster~\cite{Hooper:2013rwa,Huang:2013pda} disfavors such interpretations, however.
In particular, pulsar population models capable of producing the observed emission from the Inner Galaxy invariably predict that \textit{Fermi} should have resolved a much greater number of such objects.
Accounting for this constraint, Ref.~\cite{Hooper:2013nhl} concluded that no more than $\sim$5-10\% of the anomalous gamma-ray emission from the Inner Galaxy can originate from pulsars.
Furthermore, while it has been suggested that the Galactic Center signal might result from cosmic-ray interactions with gas~\cite{Hooper:2010mq,Hooper:2011ti,Abazajian:2012pn,Gordon:2013vta}, the analyses of Refs.~\cite{Linden:2012iv} and~\cite{Macias:2013vya} find that measured distributions of gas provide a poor fit to the morphology of the observed signal.
It also appears implausible that such processes could account for the more spatially extended emission observed from throughout the Inner Galaxy.

In this study, we revisit the anomalous gamma-ray emission from the Galactic Center and the Inner Galaxy regions and scrutinize the \textit{Fermi} data in an effort to constrain and characterize this signal more definitively, with the ultimate goal being to confidently determine its origin.
One way in which we expand upon previous work is by selecting photons based on the value of the \textit{Fermi} event parameter CTBCORE.
Through the application of this cut, we select only those events with more reliable directional reconstruction, allowing us to better separate the various gamma-ray components, and to better limit the degree to which emission from the Galactic Disk leaks into the regions studied in our Inner Galaxy analysis.
We produce a new and robust determination of the spectrum and morphology of the Inner Galaxy and the Galactic Center signals.
We go on to apply a number of tests to this data, and determine that the anomalous emission in question agrees well with that predicted from the annihilations of a 36-51 GeV WIMP annihilating mostly to $b$ quarks (or a somewhat lower mass WIMP if its annihilations proceed to first or second generation quarks).
Our results now appear to disfavor the previously considered 7-10 GeV mass window in which the dark matter annihilates significantly to tau leptons~\cite{Hooper:2010mq,Hooper:2011ti,Gordon:2013vta,Hooper:2013rwa,Abazajian:2014fta} (the analysis of Ref.~\cite{Gordon:2013vta} also disfavored this scenario).
The morphology of the signal is consistent with spherical symmetry, and strongly disfavors any significant elongation along the Galactic Plane.
The emission decreases with the distance to the Galactic Center at a rate consistent with a dark matter halo profile which scales as $\rho \propto r^{-\gamma}$, with $\gamma \approx 1.1-1.3$.
The signal can be identified out to angles of $\simeq 10^{\circ}$ from the Galactic Center, beyond which systematic uncertainties related to the Galactic diffuse model become significant.
The annihilation cross section required to normalize the observed signal is $\sigma v \sim 10^{-26}$ cm$^3$/s, in good agreement with that predicted for dark matter in the form of a simple thermal relic.

The remainder of this chapter is structured as follows.
In the following section, we review the calculation of the spectrum and angular distribution of gamma rays predicted from annihilating dark matter.
In Sec.~\ref{ctbcore}, we describe the event selection used in our analysis, including the application of cuts on the \textit{Fermi} event parameter CTBCORE.
In Secs.~\ref{inner} and~\ref{center}, we describe our analyses of the Inner Galaxy and Galactic Center regions, respectively.
In each of these analyses, we observe a significant gamma-ray excess, with a spectrum and morphology in good agreement with that predicted from annihilating dark matter.
We further investigate the angular distribution of this emission in Sec.~\ref{morphology}, and discuss the dark matter interpretation of this signal in Sec.~\ref{darkmatter}.
In Sec.~\ref{discussion} we discuss the implications of these observations, and offer predictions for other upcoming observations.
Finally, we summarize our results and conclusions in Sec.~\ref{summary}.
In the associated appendix of this chapter, we include supplemental material intended for those interested in further details of our analysis.

\begin{figure}[t!]
\begin{center}
\includegraphics[width=6.0in]{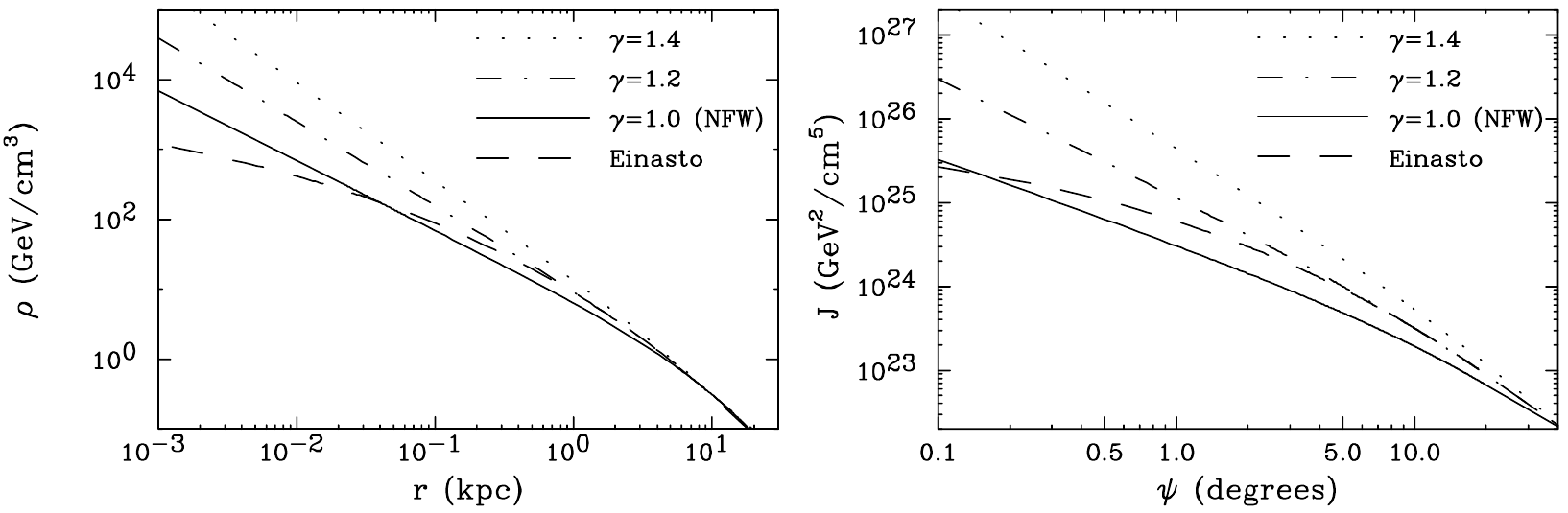}
\end{center}
\vspace{-0.5cm}
\caption{Left frame: The dark matter density as a function of the distance to the Galactic Center, for several halo profiles, each normalized such that $\rho=0.4$ GeV/cm$^3$ at $r=8.5$ kpc.
Right frame: The line-of-sight integral of the density squared, as defined in Eq.~\ref{J}, for the same set of halo profiles, as a function of the angular distance from the Galactic Center, $\psi$.
}
\label{profiles}
\end{figure}

\section{Gamma Rays From Dark Matter Annihilations in the Halo of the Milky Way}\label{intro2}

Dark matter searches using gamma-ray telescopes have a number of advantages over other indirect detection strategies.
Unlike signals associated with cosmic rays (electrons, positrons, antiprotons, etc), gamma rays are not deflected by magnetic fields.
Furthermore, gamma-ray energy losses are negligible on galactic scales.
As a result, gamma-ray telescopes can potentially acquire both spectral and spatial information, unmolested by astrophysical effects.

The flux of gamma rays generated by annihilating dark matter particles, as a function of the direction observed, $\psi$, is given by:
\begin{equation}
\Phi (E_{\gamma}, \psi) = \frac{\sigma v}{8 \pi  m_{X}^2}  \frac{\mathrm{d}N_{\gamma}}{\mathrm{d}E_\gamma}  \,  \int_{\text{los}} \rho^2(r ) \,\mathrm{d}l, 
\label{basic}
\end{equation}
where $m_X$ is the mass of the dark matter particle, $\sigma v$ is the annihilation cross section (times the relative velocity of the particles), $dN_{\gamma}/dE_{\gamma}$ is the gamma-ray spectrum produced per annihilation, and the integral of the density squared is performed over the line-of-sight (los).
Although N-body simulations lead us to expect dark matter halos to exhibit some degree of triaxiality (see~\cite{Kuhlen:2007ku} and references therein), the Milky Way's dark matter distribution is generally assumed to be approximately spherically symmetric, allowing us to describe the density as a function of only the distance from the Galactic Center, $r$.
Throughout this study, we will consider dark matter distributions described by a generalized Navarro-Frenk-White (NFW) halo profile~\cite{Navarro:1995iw,Navarro:1996gj}:
\begin{equation}
\rho( r)= \rho_0 \frac{(r/r_s)^{-\gamma}}{(1 + r/r_s)^{3-\gamma}}.
\label{gennfw}
\end{equation}  
Throughout this chapter, we adopt a scale radius of $r_s =20$ kpc, and select $\rho_0$ such that the local dark matter density (at $8.5\,{\rm kpc}$ from the Galactic Center) is $0.4$ GeV/cm$^3$, consistent with dynamical constraints~\cite{Iocco:2011jz,Catena:2009mf}.
Although dark matter-only simulations generally favor inner slopes near the canonical NFW value ($\gamma=1$)~\cite{Navarro:2008kc,Diemand:2008in}, baryonic effects are expected to have a non-negligible impact on the dark matter distribution within the inner $\sim$10 kiloparsecs of the Milky Way~\cite{Fry:1985tp,Ryden:1987ska,Gnedin:2011uj,Gnedin:2004cx,Governato:2012fa,Kuhlen:2012qw,Weinberg:2001gm,Weinberg:2006ps,Sellwood:2002vb,Valenzuela:2002np,Colin:2005rr}.
The magnitude and direction of such baryonic effects, however, are currently a topic of debate.
With this in mind, we remain agnostic as to the value of the inner slope, and take $\gamma$ to be a free parameter.

\begin{figure}[t!]
\begin{center}
\includegraphics[width=2.8in]{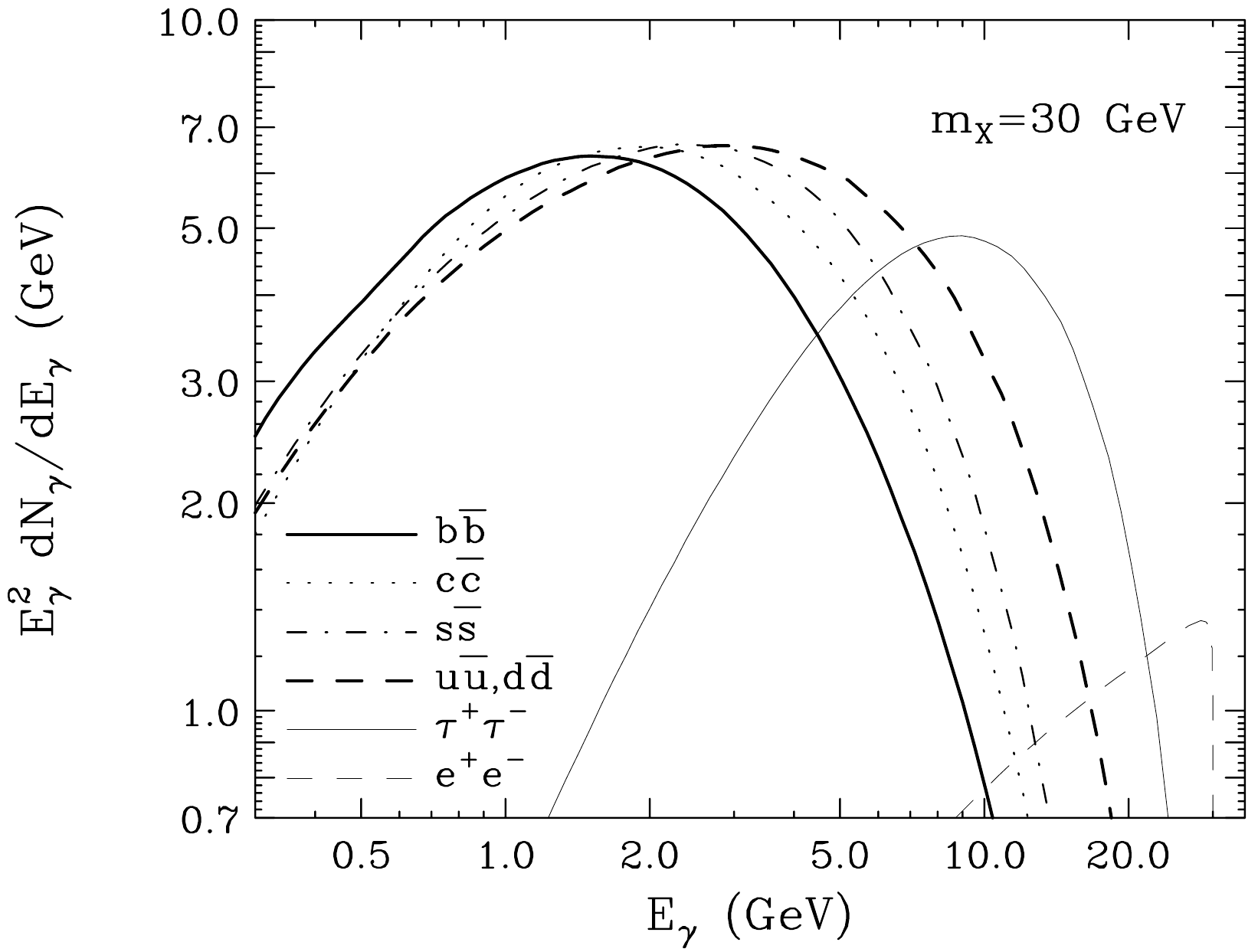}
\hspace{0.1in}
\includegraphics[width=2.8in]{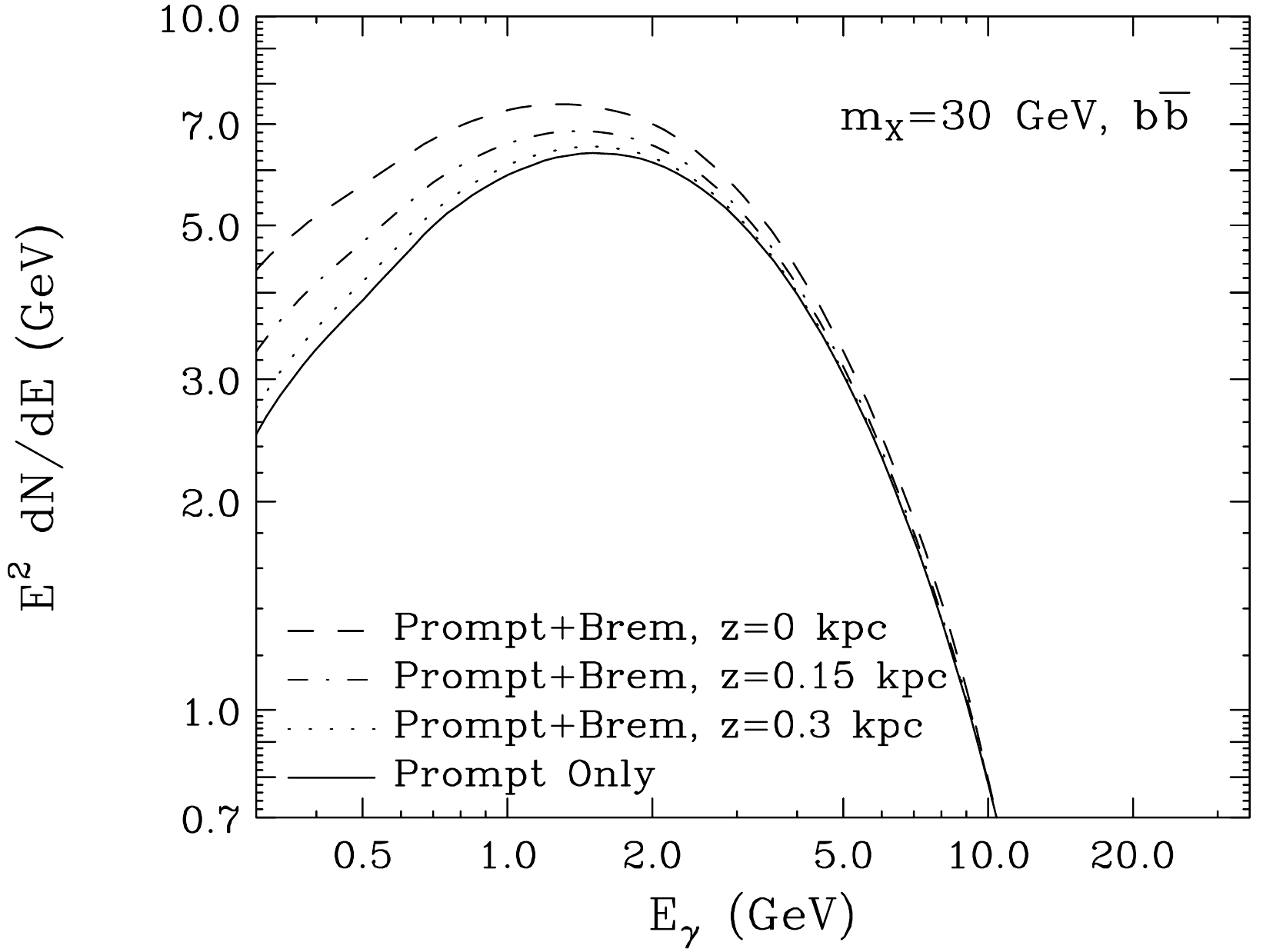}
\end{center}
\vspace{-0.5cm}
\caption{Left frame: The spectrum of gamma rays produced per dark matter annihilation for a 30 GeV WIMP mass and a variety of annihilation channels.
Right frame: An estimate for the bremsstrahlung emission from the electrons produced in dark matter annihilations taking place near the Galactic Center, for the case of a 30 GeV WIMP annihilating to $b\bar{b}$.
At $|z| \lesssim 0.3$ kpc ($|b| \lesssim 2^{\circ}$) and at energies below $\sim$1-2 GeV, bremsstrahlung could potentially contribute non-negligibly.
See text for details.
}
\label{dnde}
\end{figure}

In the left frame of Fig.~\ref{profiles}, we plot the density of dark matter as a function of $r$ for several choices of the halo profile.
Along with generalized NFW profiles using three values of the inner slope ($\gamma$=1.0, 1.2, 1.4), we also show for comparison the results for an Einasto profile (with $\alpha=0.17$)~\cite{Springel:2008cc}.
In the right frame, we plot the value of the integral in Eq.~\ref{basic} for the same halo profiles, denoted by the quantity, $J(\psi)$:
\begin{equation}
J(\psi)= \int_{\text{los}} \rho^2( r) \, dl,
\label{J}
\end{equation}
where $\psi$ is the angle observed away from the Galactic Center.
In the NFW case (with $\gamma=1$), for example, the value of $J$ averaged over the inner degree around the Galactic Center exceeds that of the most promising dwarf spheroidal galaxies by a factor of $\sim$$50$~\cite{Ackermann:2013yva}.
If the Milky Way's dark matter halo is contracted by baryons or is otherwise steeper than predicted by NFW, this ratio could easily be $\sim$$10^3$ or greater.

The spectrum of gamma rays produced per dark matter annihilation, $dN_{\gamma}/dE_{\gamma}$, depends on the mass of the dark matter particle and on the types of particles produced in this process.
In the left frame of Fig.~\ref{dnde}, we plot $dN_{\gamma}/dE_{\gamma}$ for the case of a 30 GeV WIMP mass, and for a variety of annihilation channels (as calculated using PYTHIA~\cite{Sjostrand:2006za}, except for the $e^+ e^-$ case, for which the final state radiation was calculated analytically~\cite{Bergstrom:2004cy,Birkedal:2005ep}).
In each case, a distinctive bump-like feature appears, although at different energies and with different widths, depending on the final state.

In addition to prompt gamma rays, dark matter annihilations can produce electrons and positrons which subsequently generate gamma rays via inverse Compton and bremsstrahlung processes.
For dark matter annihilations taking place near the Galactic Plane, the low-energy gamma-ray spectrum can receive a non-negligible contribution from bremsstrahlung.
In the right frame of Fig.~\ref{dnde}, we plot the gamma-ray spectrum from dark matter (per annihilation), including an estimate for the bremsstrahlung contribution.
In estimating the contribution from bremsstrahlung, we neglect diffusion, but otherwise follow the calculation of Ref.~\cite{Cirelli:2013mqa}.
In particular, we consider representative values of $\langle B \rangle =10 \, \mu$G for the magnetic field, and 10 eV$/$cm$^3$ for the radiation density throughout the region of the Galactic Center.
For the distribution of gas, we adopt a density of 10 particles per cm$^3$ near the Galactic Plane ($z=0$), with a dependence on $z$ given by $\exp(-|z|/0.15 \, {\rm kpc})$.
Within $\sim$$1^{\circ}$--\,$2^{\circ}$ of the Galactic Plane, we find that bremsstrahlung could potentially contribute non-negligibly to the low energy ($\lesssim$\,1--2 GeV) gamma-ray spectrum from annihilating dark matter.

\section{Making Higher Resolution Gamma-Ray Maps with CTBCORE}\label{ctbcore}

In most analyses of \textit{Fermi} data, one makes use of all of the events within a given class (Transient, Source, Clean, or Ultraclean).
Each of these event classes reflects a different trade-off between the effective area and the efficiency of cosmic-ray rejection.
Higher quality event classes also allow for somewhat greater angular resolution (as quantified by the point spread function, PSF).
The optimal choice of event class for a given analysis depends on the nature of the signal and background in question.
The Ultraclean event class, for example, is well suited to the study of large angular regions, and to situations where the analysis is sensitive to spectral features that might be caused by cosmic ray backgrounds.
The Transient event class, in contrast, is best suited for analyses of short duration events, with little background.
Searches for dark matter annihilation products from the Milky Way's halo significantly benefit from the high background rejection and angular resolution of the Ultraclean class and thus can potentially fall into the former category.

\begin{figure}[t!]
\begin{center}
\includegraphics[width=6.0in]{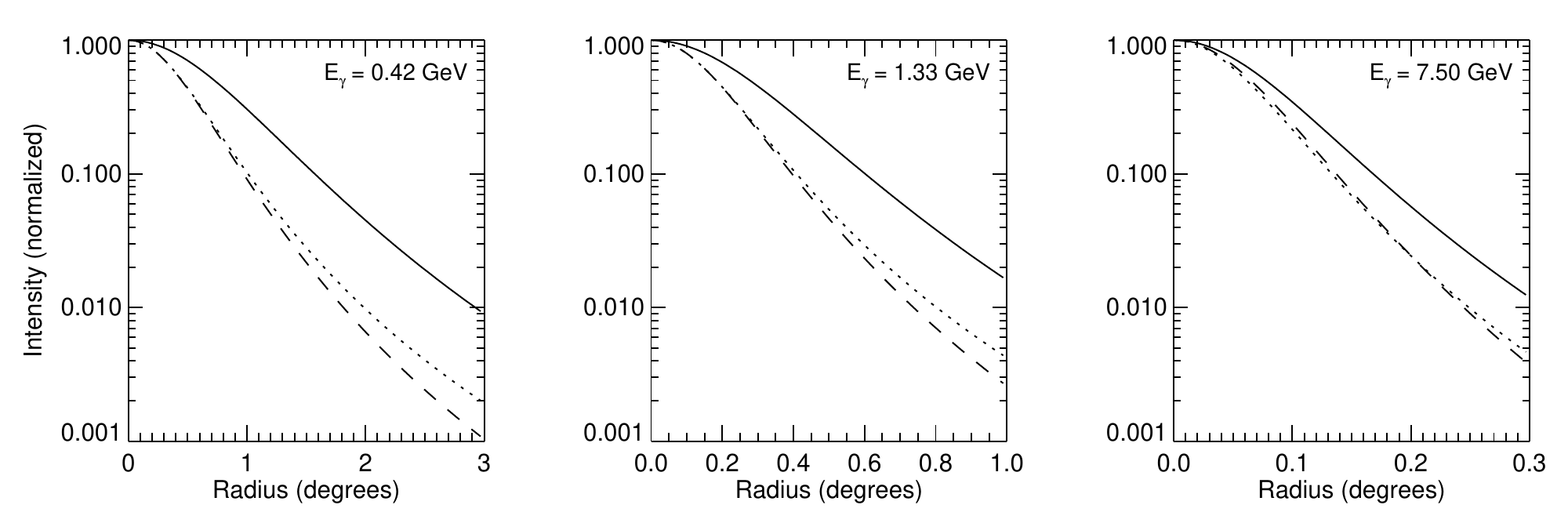}
\end{center}
\vspace{-0.5cm}
\caption{The point spread function (PSF) of the \textit{Fermi} Gamma-Ray Space Telescope, for front-converting, Ultraclean class events.
The solid lines represent the PSF for the full dataset, using the \textit{Fermi} Collaboration's default cuts on the parameter CTBCORE.
The dotted and dashed lines, in contrast, denote the PSFs for the top two quartiles (Q2) and top quartile (Q1) of these events, respectively, as ranked by CTBCORE.
See text for details.
}
\label{psf}
\end{figure}

As a part of event reconstruction, the \textit{Fermi} Collaboration estimates the accuracy of the reconstructed direction of each event.
Inefficiencies and inactive regions within the detector reduce the quality of 
the information available for certain events.
Factors such as whether an event is front-converting or back-converting, whether there are multiple tracks that can be combined into a vertex, and the amount of energy deposited into the calorimeter each impact the reliability of the reconstructed direction~\cite{Atwood:2009ez}.

In their most recent public data releases, the \textit{Fermi} Collaboration has begun to include a greater body of information about each event, including a value for the parameter CTBCORE, which quantifies the reliability of the directional reconstruction.
By selecting only events with a high value of CTBCORE, one can reduce the tails of the PSF, although at the expense of effective area~\cite{Atwood:2009ez}.

For this study, we have created a set of new event classes by increasing the CTBCORE cut from the default values used by the \textit{Fermi} Collaboration.
To accomplish this, we divided all front-converting, Ultraclean events (Pass 7, Reprocessed) into quartiles, ranked by CTBCORE.
Those events in the top quartile make up the event class Q1, while those in the top two quartiles make up Q2, etc.
For each new event class, we calibrate the on-orbit PSF~\cite{Ackermann:2012kna,Ackermann:2013yma} using the Geminga pulsar.
Taking advantage of Geminga's pulsation, we remove the background by taking the difference between the on-phase and off-phase images.
We fit the PSF in each energy bin by a single King function, and smooth the overall PSF with energy.
We also rescale \textit{Fermi}'s effective area according to the fraction of events that are removed by the CTBCORE cut, as a function of energy and incidence angle.

These cuts on CTBCORE have a substantial impact on \textit{Fermi}'s PSF, especially at low energies.
In Fig.~\ref{psf}, we show the PSF for front-converting, Ultraclean events, at three representative energies, for different cuts on CTBCORE (all events, Q2, and Q1).

Such a cut can be used to mitigate the leakage of astrophysical emission from the Galactic Plane and point sources into our regions of interest.
This leakage is most problematic at low energies, where the PSF is quite broad and where the CTBCORE cut has the greatest impact.
These new event classes and their characterization are further detailed in \cite{Portillo:2014ena}, and accompanied by a data release of all-sky maps 
for each class, and the instrument response function files necessary for use with the \textit{Fermi} Science Tools.

Throughout the remainder of this study, we will employ the Q2 event class by default, corresponding to the top 50\% (by CTBCORE) of \textit{Fermi}'s front-converting, Ultraclean photons, to maximize event quality.
We select Q2 rather than Q1 to improve statistics, since as demonstrated in Fig.~\ref{psf}, the angular resolution improvement in moving from Q2 to Q1 is minimal.
In Appendix \ref{app:consistency} we demonstrate that our results are stable upon removing the CTBCORE cut (thus doubling the dataset), or expanding the dataset to include lower-quality events.\footnote{An earlier version of this work found a number of apparent peculiarities in the results without the CTBCORE cut that were removed on applying the cut.
However, we now attribute those peculiarities to an incorrect smoothing of the diffuse background model.
When the background model is smoothed correctly, we find results that are much more stable to the choice of CTBCORE cut, and closely resemble the results previously obtained with Q2 events.
Accordingly, the CTBCORE cut appears to be effective at separating signal from poorly-modeled background emission, but is less necessary when the background is well-modeled.}

\section{The Inner Galaxy}\label{inner}

In this section, we follow the procedure previously pursued in Ref.~\cite{Hooper:2013rwa} (see also Refs.~\cite{Dobler:2009xz,Su:2010qj}) to study the gamma-ray emission from the Inner Galaxy.
We use the term ``Inner Galaxy'' to denote the region of the sky that lies within several tens of degrees around the Galactic Center, excepting the Galactic Plane itself ($|b|<1^{\circ}$), which we mask in this portion of our analysis.

\begin{figure}[t!]
\begin{center}
\includegraphics[width=2.8in]{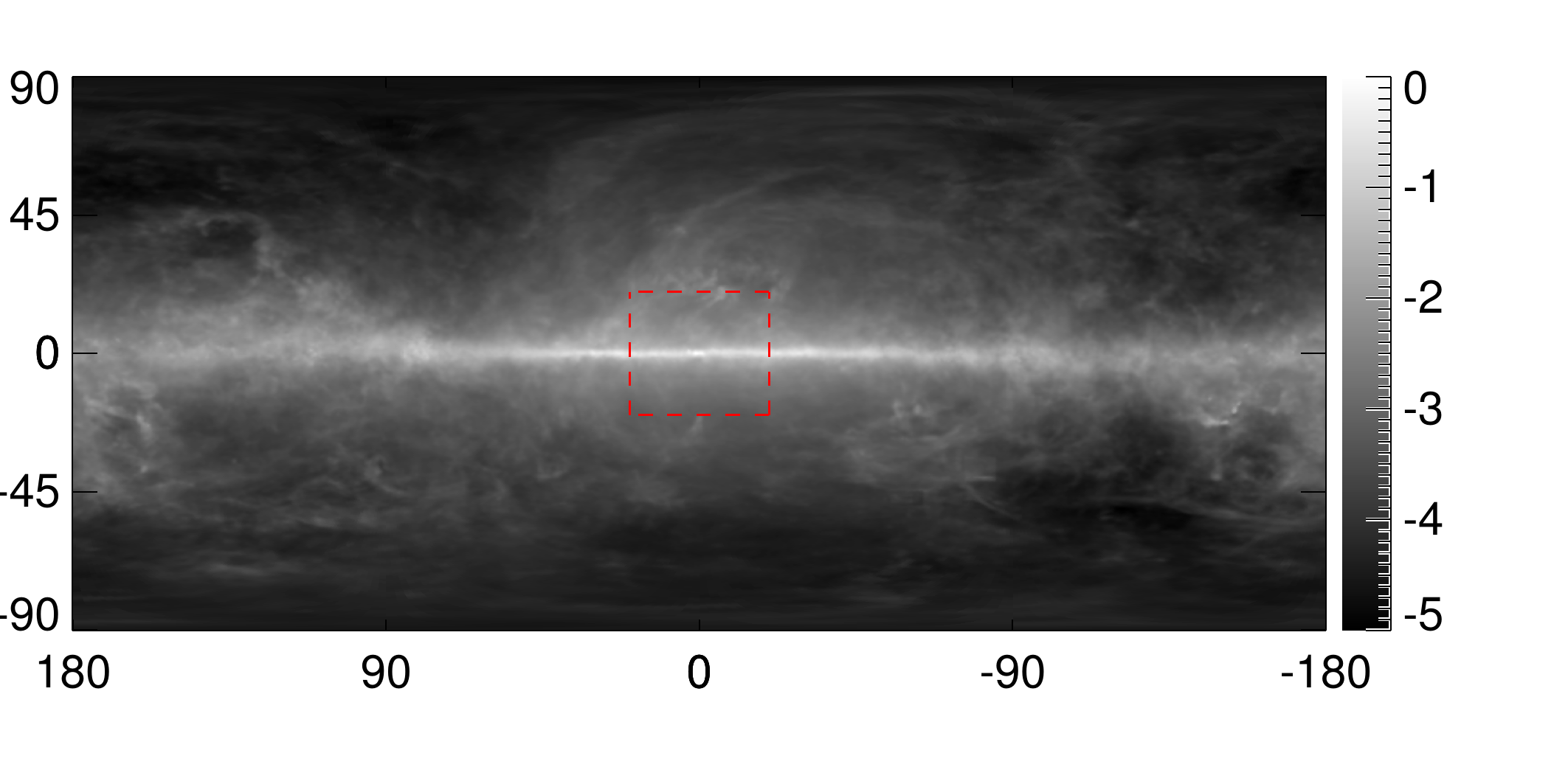}
\includegraphics[width=2.8in]{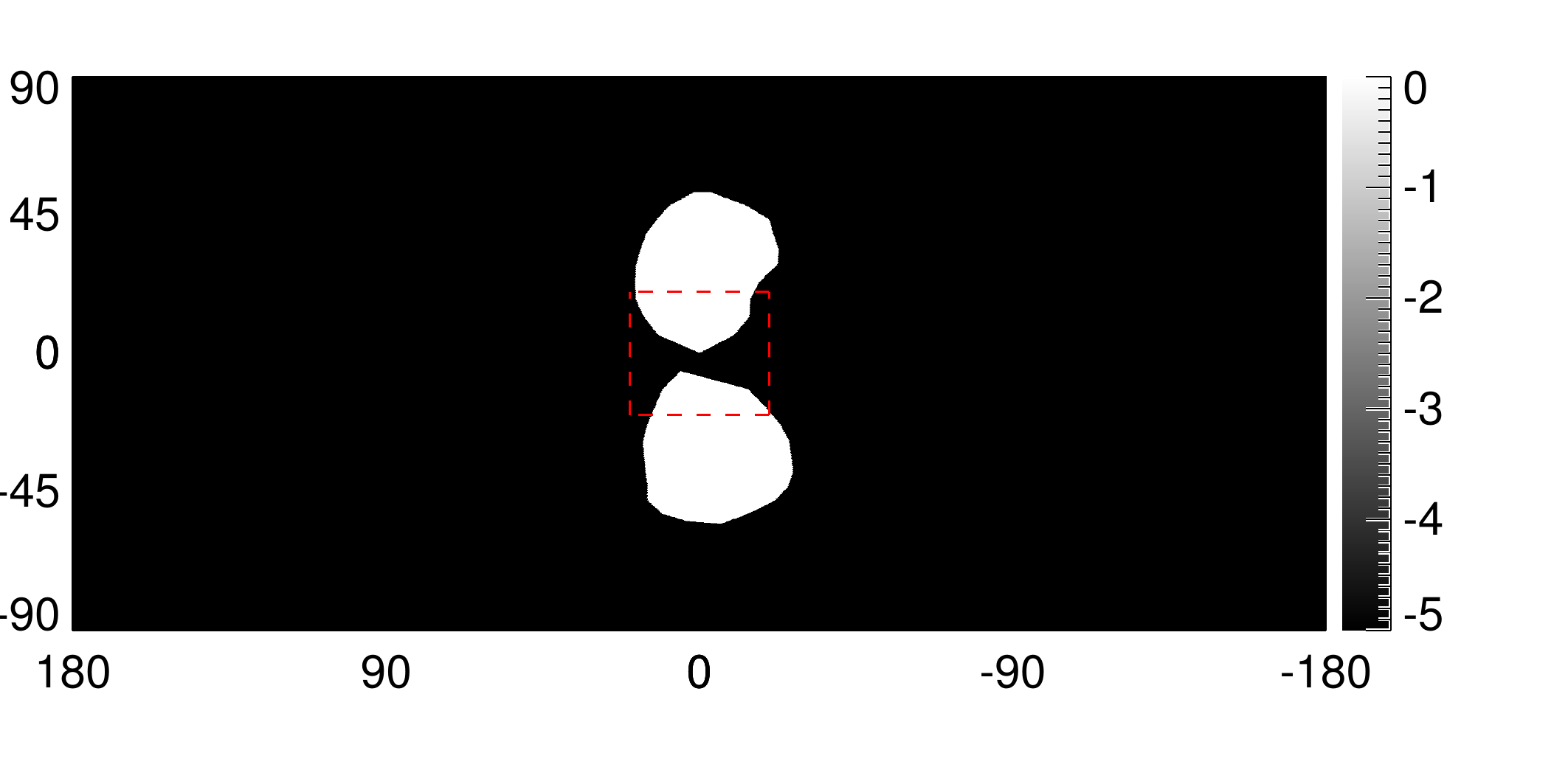}
\vspace{-0.1in}
\includegraphics[width=2.8in]{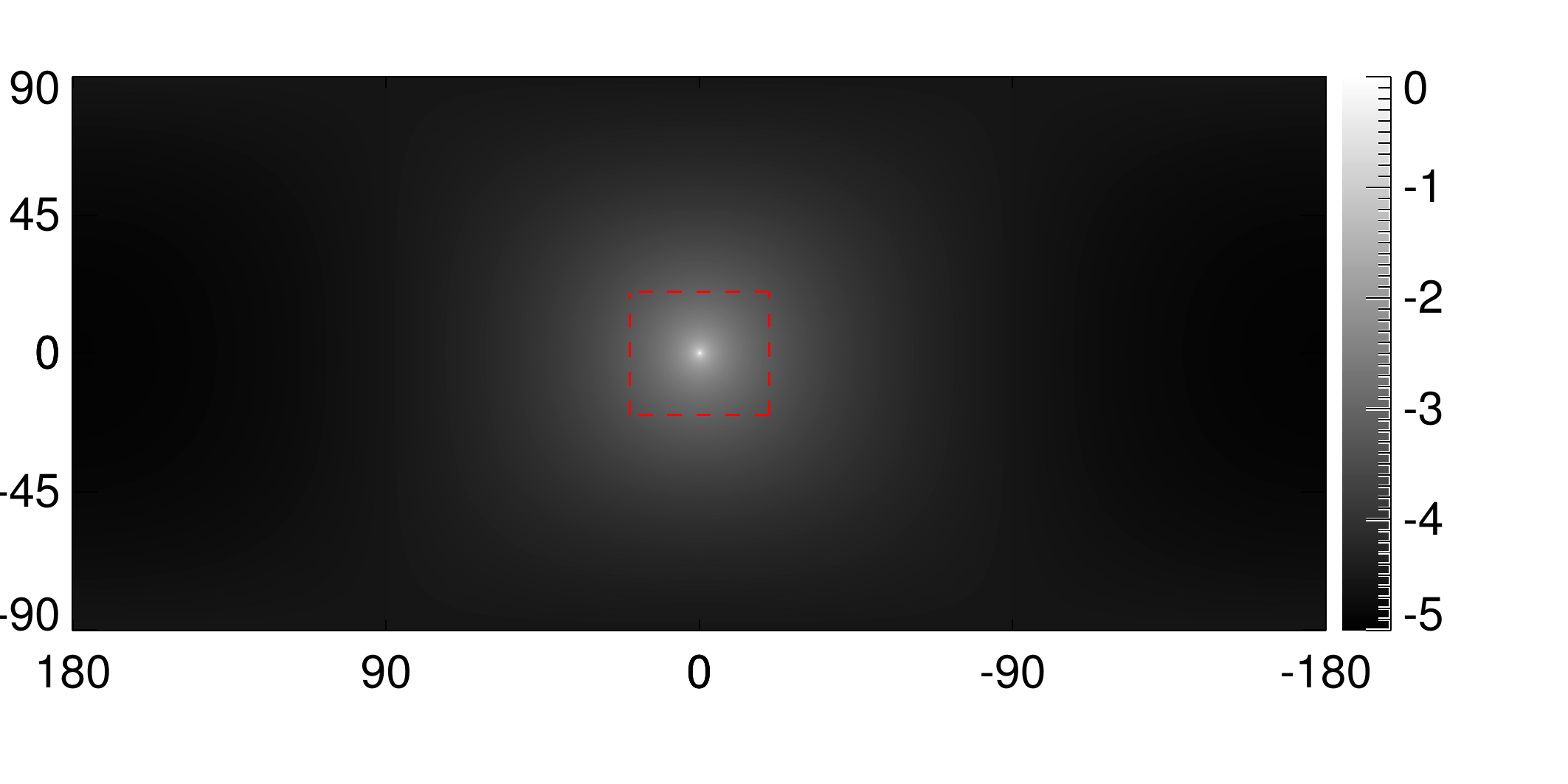}
\end{center}
\vspace{-0.5cm}
\caption{The spatial templates (in galactic coordinates) for the Galactic diffuse model (upper left), the \textit{Fermi} bubbles (upper right), and dark matter annihilation products (lower), as used in our Inner Galaxy analysis.
The scale is logarithmic (base 10), normalized to the brightest point in each map.
The diffuse model template is shown as evaluated at 1 GeV, and the dark matter template corresponds to a generalized NFW profile with an inner slope of $\gamma=1.18$.
Red dashed lines indicate the boundaries of our standard Region of Interest (we also mask bright point sources and the region of the Galactic plane with $|b| < 1^\circ$).
}
\label{templates}
\end{figure}

Throughout our analysis, we make use of the Pass 7 (V15) reprocessed data taken between August 4, 2008 and December 5, 2013, using only front-converting, Ultraclean class events which pass the Q2 CTBCORE cut as described in Sec.~\ref{ctbcore}.
We also apply standard cuts to ensure data quality (zenith angle $<100^{\circ}$, instrumental rocking angle $<52^{\circ}$, \texttt{DATA\_QUAL} = 1, \texttt{LAT\_CONFIG}=1).
Using this data set, we have generated a series of maps of the gamma-ray sky binned in energy, with 30 logarithmically spaced energy bins spanning the range from 0.3-300 GeV.
For the analyses presented in this chapter, by default we restrict to energies 50 GeV and lower to ensure numerical stability of the fit.
We apply the point source subtraction method described in Ref.~\cite{Su:2010qj}, updated to employ the 2FGL catalogue, and masking out the 300 brightest and most variable sources at a mask radius corresponding to $95\%$ containment.
We then perform a pixel-based maximum likelihood analysis on the map, fitting the data in each energy bin to a sum of spatial templates.
These templates consist of: 1) the \emph{Fermi} Collaboration \texttt{p6v11} Galactic diffuse model (which we refer to as the \texttt{p6v11} diffuse model),\footnote{Unlike more recently released Galactic diffuse models, the \texttt{p6v11} diffuse model does not implicitly include a component corresponding to the \textit{Fermi} Bubbles.
By using this model, we are free to fit the \textit{Fermi} Bubbles component independently.
See Appendix \ref{app:diffuse} for a discussion of the impact of varying the diffuse model.} 2) an isotropic map, intended to account for the extragalactic gamma-ray background and residual cosmic-ray contamination, and 3) a uniform-brightness spatial template coincident with the features known as the \textit{Fermi} Bubbles, as described in Ref.~\cite{Su:2010qj}.
In addition to these three background templates, we include an additional dark matter template, motivated by the hypothesis that the previously reported gamma-ray excess originates from annihilating dark matter.
In particular, our dark matter template is taken to be proportional to the line-of-sight integral of the dark matter density squared, $J(\psi)$, for a generalized NFW density profile (see Eqs.~\ref{gennfw}--\ref{J}).
The spatial morphology of the Galactic diffuse model (as evaluated at 1 GeV), \textit{Fermi} Bubbles, and dark matter templates are each shown in Fig.~\ref{templates}.

We smooth the Galactic diffuse model template to match the data using the \texttt{gtsrcmaps} routine in the \textit{Fermi} Science Tools, to ensure that the tails of the PSF are properly taken into account.\footnote{We checked the impact of smoothing the diffuse model with a Gaussian and found no significant impact on our results.} 
Because the Galactic diffuse model template is much brighter than the other contributions in the region of interest, relatively small errors in its smoothing could potentially bias our results.
However, the other templates are much fainter, and so we simply perform a Gaussian smoothing, with a FWHM matched to the FWHM of the \emph{Fermi} PSF at the minimum energy for the bin (since most of the counts are close to this minimum energy).
In all cases, when using CTBCORE data, we employ the appropriate (narrower) PSF, as derived in \cite{Portillo:2014ena}.

\begin{figure}[t!]
\begin{center}
\includegraphics[width=3in]{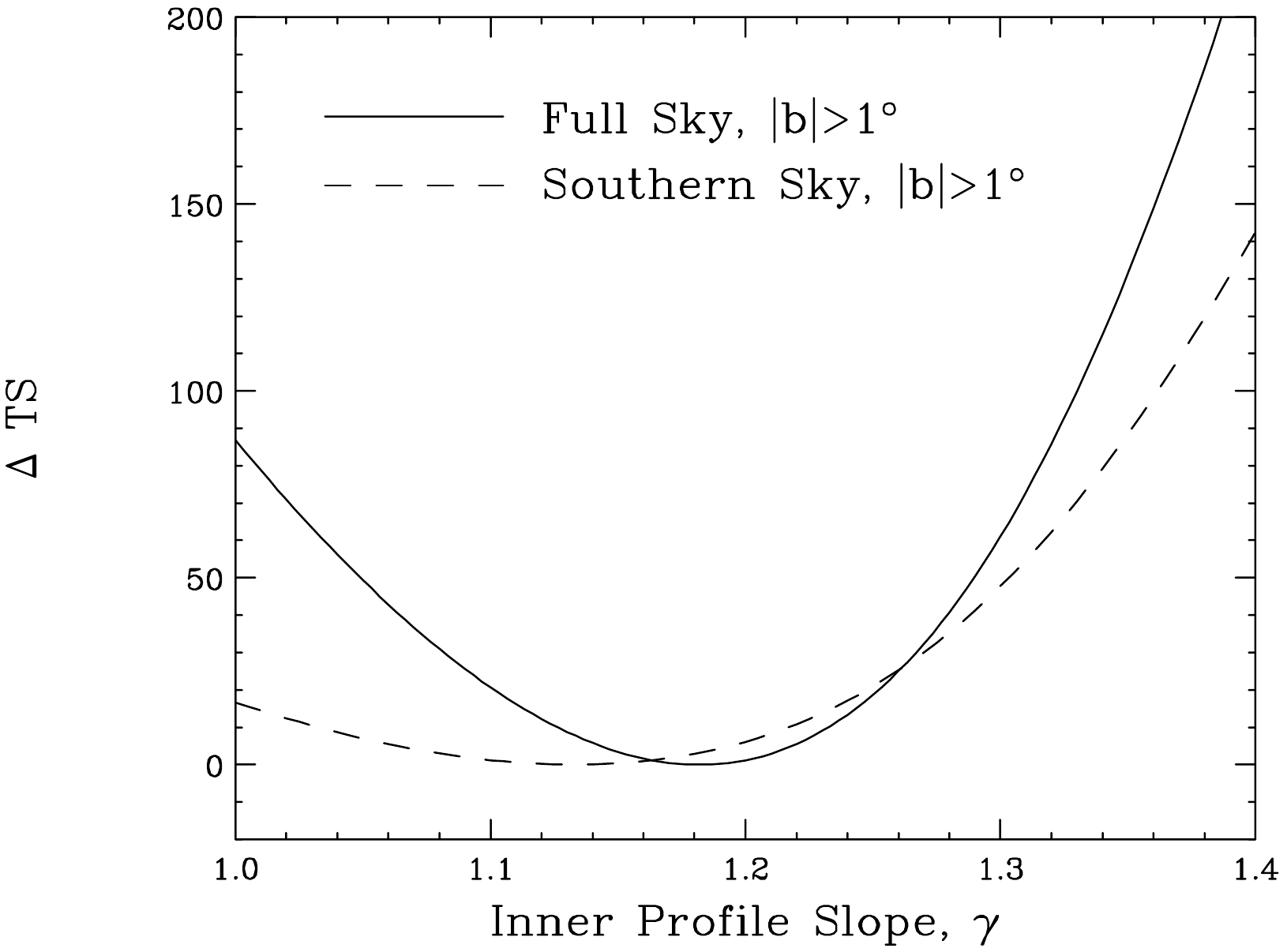}
\end{center}
\vspace{-0.5cm}
\caption{The variation in the quantity $-2 \Delta \ln {\mathcal L}$ (referred to as TS) extracted from the likelihood fit, as a function of the inner slope of the dark matter halo profile, $\gamma$.
All values are relative to the result for the best-fit (highest TS) template, and positive values thus indicate a reduction in TS.
Results are shown using gamma-ray data from the full sky (solid line) and only the southern sky (dashed line).
Unlike in the analysis of Ref.~\cite{Hooper:2013rwa}, we do not find any large north-south asymmetry in the preferred value of $\gamma$.
}
\label{innerslope}
\end{figure}

By default, we employ a Region of Interest (ROI) of $|\ell| < 20^\circ$, $1^\circ < |b| < 20^\circ$.
An earlier version of this work used the full sky (with the plane masked at 1 degree) as the default ROI; we find that restricting to a smaller ROI alleviates oversubtraction in the inner Galaxy and improves the stability of our results.\footnote{This approach was in part inspired by the work presented in Ref.~\cite{Calore:2014xka}.} Thus we present ``baseline'' results for the smaller region, but show the impact of changing the ROI in Appendix \ref{app:consistency}, and in selected figures in the main text.
Where we refer to the ``full sky'' analysis the Galactic plane is masked for $|b| < 1^\circ$ unless noted otherwise.

As found in previous studies~\cite{Hooper:2013rwa,Huang:2013pda}, the inclusion of the dark matter template dramatically improves the quality of the fit to the \textit{Fermi} data.
For the best-fit spectrum and halo profile, we find that the inclusion of the dark matter template improves the formal fit by TS$\equiv -2 \Delta \ln \mathcal{L} \simeq 1100$ (here TS stands for ``test statistic'').
This dark matter template has 22 degrees of freedom, corresponding to its normalization in each of the 22 energy bins below 50 GeV.
A naive translation from TS to $p$-value results in an apparent statistical preference greater than 30$\sigma$; however, when considering this enormous statistical significance, one should keep in mind that in addition to statistical errors there is a degree of unavoidable and unaccounted-for systematic error.
Neither model (with or without a dark matter component) is a ``good fit'' in the sense of describing the sky to the level of Poisson noise.
That being said, the data do very strongly prefer the presence of a gamma-ray component with a morphology similar to that predicted from annihilating dark matter (see Appendices \ref{app:consistency}-\ref{app:gc} for further details).

\begin{figure}[t!]
\begin{center}
\includegraphics[width=2.8in]{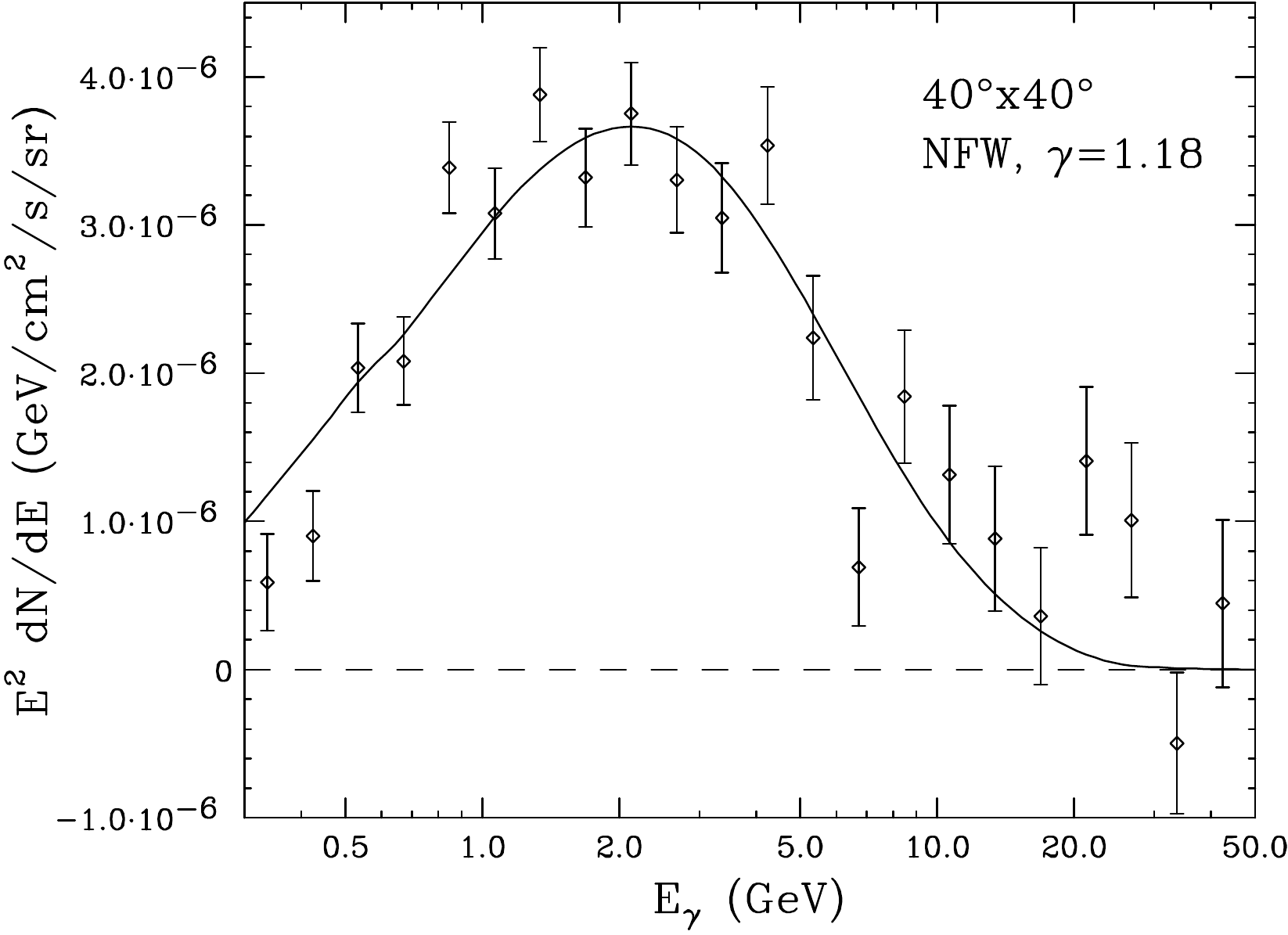}\hspace{0.1in}
\includegraphics[width=2.8in]{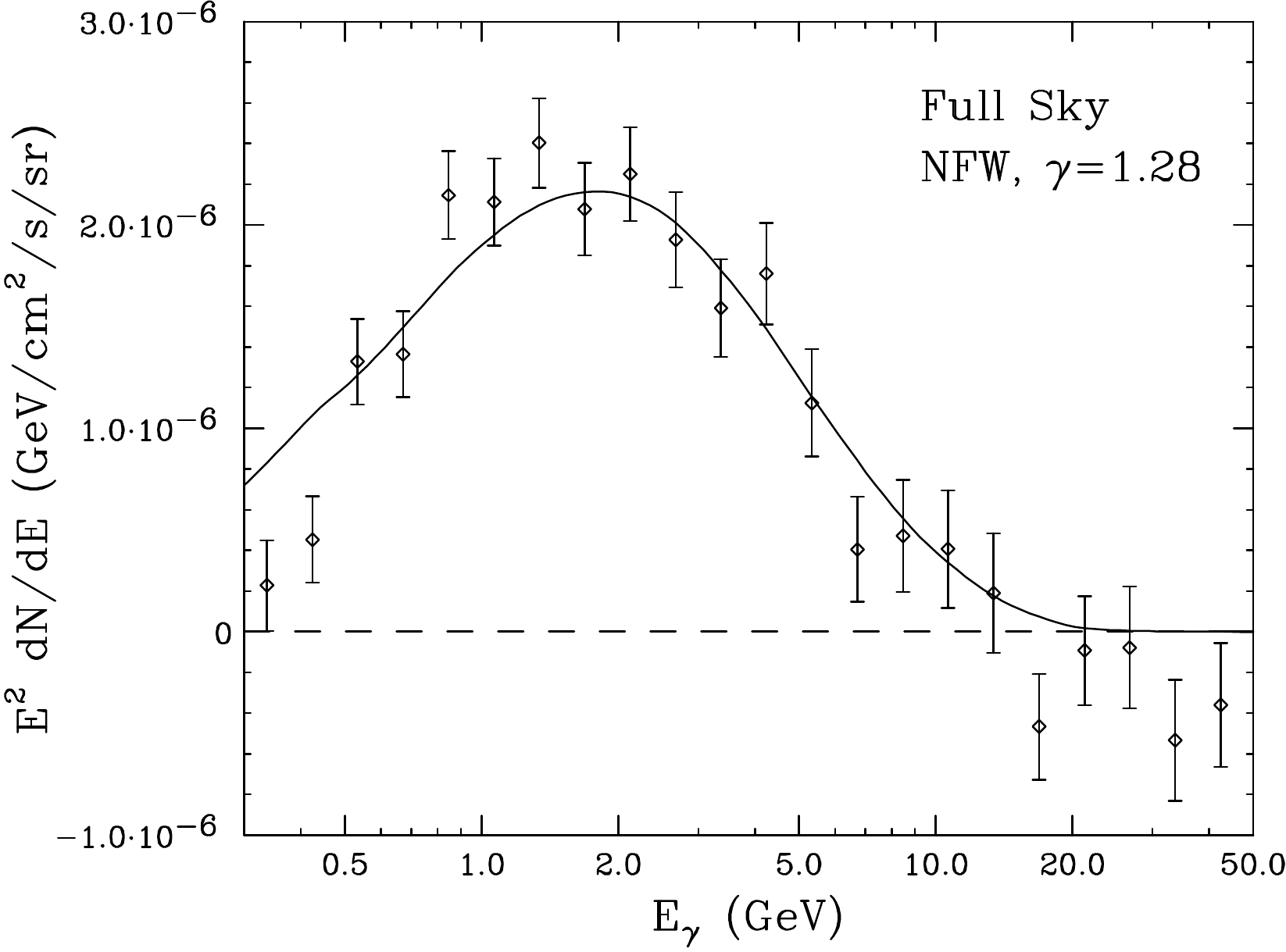}
\end{center}
\vspace{-0.5cm}
\caption{Left frame:  The spectrum of the dark matter component, extracted from a fit in our standard ROI ($1^\circ < |b| < 20^\circ$, $|l| < 20^\circ$) for a template corresponding to a generalized NFW halo profile with an inner slope of $\gamma=1.18$ (normalized to the flux at an angle of 5$^{\circ}$ from the Galactic Center).
Shown for comparison (solid line) is the spectrum predicted from a 43.0 GeV dark matter particle annihilating to $b\bar{b}$ with a cross section of $\sigma v = 2.25\times 10^{-26}$ cm$^3$/s $\, \times \, [(0.4 \,{\rm GeV}/{\rm cm}^3)/\rho_{\rm local}]^2$.
Right frame: as left frame, but for a full-sky ROI ($|b| > 1^\circ$), with $\gamma=1.28$; shown for comparison (solid line) is the spectrum predicted from a 36.
6 GeV dark matter particle annihilating to $b\bar{b}$ with a cross section of $\sigma v = 0.75\times 10^{-26}$ cm$^3$/s $\, \times \, [(0.4 \,{\rm GeV}/{\rm cm}^3)/\rho_{\rm local}]^2$.
}
\label{innerspec}
\end{figure}

\begin{figure}[t!]
\begin{center}
\includegraphics[width=6.0in]{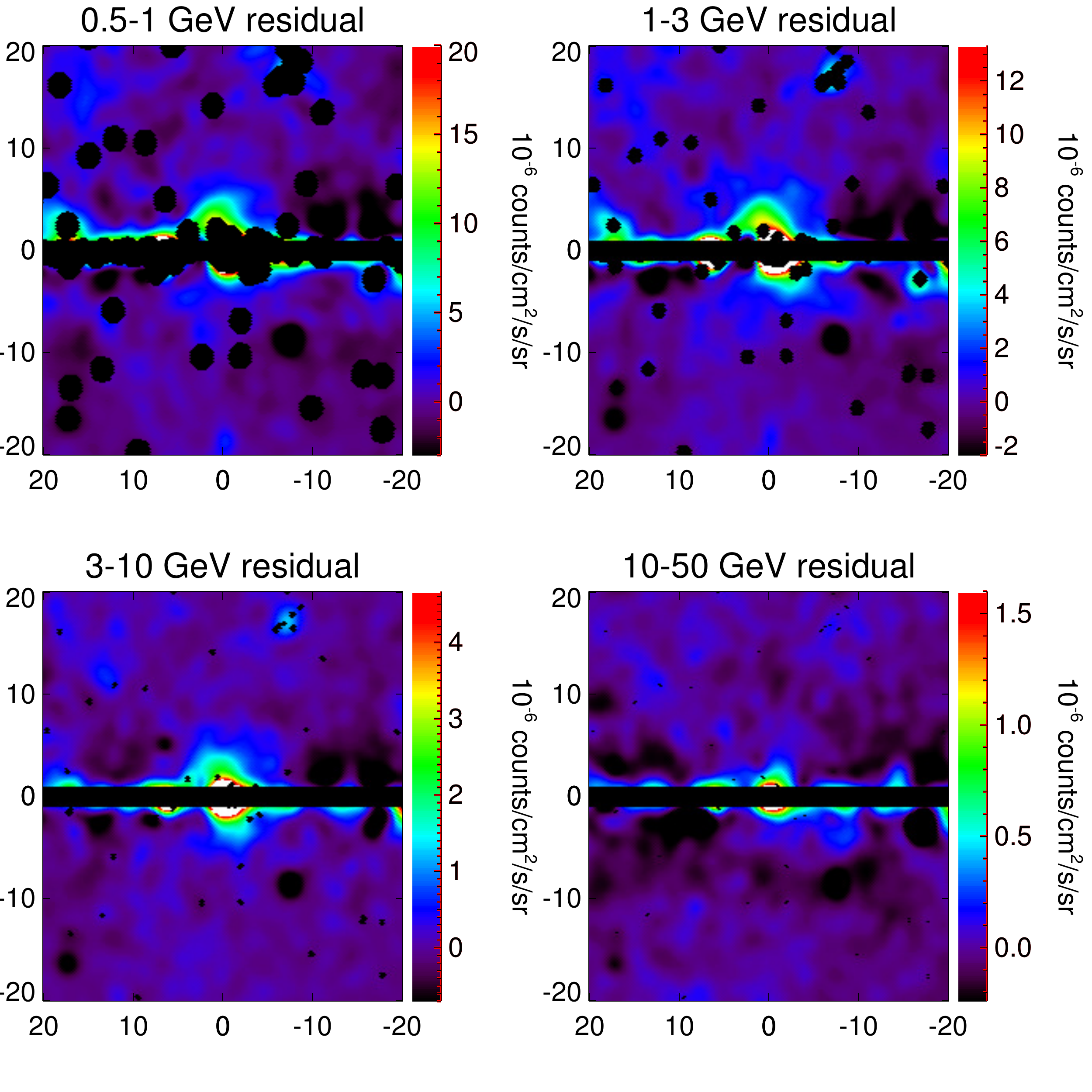}
\vspace{-1.5cm}
\end{center}
\caption{Intensity maps (in galactic coordinates) after subtracting the point source model and best-fit Galactic diffuse model, \textit{Fermi} bubbles, and isotropic templates.
Template coefficients are obtained from the fit including these three templates and a $\gamma=1.3$ DM-like template.
Masked pixels are indicated in black.
All maps have been smoothed to a common PSF of 2 degrees for display, before masking (the corresponding masks have \emph{not} been smoothed; they reflect the actual masks used in the analysis).
At energies between $\sim$0.5-10 GeV ({\it i.e.}~in the first three frames), the dark-matter-like emission is clearly visible around the Galactic Center.
}
\label{innerresidual}
\end{figure}

As in Ref.~\cite{Hooper:2013rwa}, we vary the value of the inner slope of the generalized NFW profile, $\gamma$, and compare the change in the log-likelihood, $\Delta \ln \mathcal{L}$, between the resulting fits in order to determine the preferred range for the value of $\gamma$.\footnote{Throughout, we describe the improvement in $-2 \Delta \ln \mathcal{L}$ induced by inclusion of a specific template as the ``test statistic'' or TS for that template.} 
The results of this exercise are shown in Fig.~\ref{innerslope}.
We find that our default ROI has a best-fit value of $\gamma=1.18$, consistent with previous studies of the inner Galaxy (which did not employ any additional cuts on CTBCORE) that preferred an inner slope of $\gamma \simeq 1.2$~\cite{Hooper:2013rwa}.
Fitting over the full sky, we find a preference for a slightly steeper value of $\gamma \simeq 1.28$.
These results are quite stable to our mask of the Galactic plane; masking the region with $|b| < 2^\circ$ changes the preferred value to $\gamma=1.25$ in our default ROI, and $\gamma=1.29$ over the whole sky.
In contrast to Ref.~\cite{Hooper:2013rwa}, we find no significant difference in the slope preferred by the fit over the standard ROI, and by a fit only over the southern half ($b<0$) of the ROI (we also find no significant difference between the fit over the full sky and the southern half of the full sky).
This can be seen directly from Fig.~\ref{innerslope}, where the full-sky and southern-sky fits for the same level of masking are found to favor quite similar values of $\gamma$ (the southern sky distribution is broader than that for the full sky simply due to the difference in the number of photons).
The best-fit values for gamma, from fits in the southern half of the standard ROI and the southern half of the full sky, are 1.13 and 1.26 respectively.

In Fig.~\ref{innerspec}, we show the spectrum of the emission correlated with the dark matter template in the default ROI and full-sky analysis, for their respective best-fit values of $\gamma=1.18$ and 1.28.\footnote{A  comparison between the two ROIs with $\gamma$ held constant is presented in Appendix \ref{app:consistency}.} 
While no significant emission is absorbed by this template at energies above $\sim$10 GeV, a bright and robust component is present at lower energies, peaking near $\sim$1-3 GeV.
Relative to the analysis of Ref.~\cite{Hooper:2013rwa} (which used an incorrectly smoothed diffuse model), our spectrum is in both cases significantly harder at energies below 1 GeV, rendering it more consistent with that extracted at higher latitudes (see Appendix A).\footnote{An earlier version of this work found this improvement only in the presence of the CTBCORE cut; we now find this hardening independent of the CTBCORE cut.} 
Shown for comparison (as a solid line) is the spectrum predicted from (left panel) a 43.0 GeV dark matter particle annihilating to  $b\bar{b}$ with a cross section of $\sigma v = 2.25 \times 10^{-26}$ cm$^3$/s $\, \times \, [(0.4 \,{\rm GeV}/{\rm cm}^3)/\rho_{\rm local}]^2$, and (right panel) a 36.6 GeV dark matter particle annihilating to $b\bar{b}$ with a cross section of $\sigma v = 0.75 \times 10^{-26}$ cm$^3$/s $\, \times \, [(0.4 \,{\rm GeV}/{\rm cm}^3)/\rho_{\rm local}]^2$.
The spectra extracted for this component are in moderately good agreement with the predictions of the dark matter models, yielding fits of $\chi^2=44$ and $64$ over the 22 error bars between 0.3 and 50 GeV.
We emphasize that these uncertainties (and the resulting $\chi^2$ values) are purely statistical, and there are significant systematic uncertainties which are not accounted for here (see the discussion in the appendices).
We also note that the spectral shape of the dark matter template is quite robust to variations in $\gamma$, within the range where good fits are obtained (see Appendix \ref{app:consistency}).

In Fig.~\ref{innerresidual}, we plot the maps of the gamma-ray sky in four energy ranges after subtracting the best-fit diffuse model, \textit{Fermi} Bubbles, and isotropic templates.
In the 0.5-1 GeV, 1-3 GeV, and 3-10 GeV maps, the dark-matter-like emission is clearly visible in the region surrounding the Galactic Center.
Much less central emission is visible at 10-50 GeV, where the dark matter component is absent, or at least significantly less bright.

We note that the \texttt{p6v11} diffuse model, like all other diffuse models created by the \emph{Fermi} Collaboration, was designed for point source subtraction rather than study of extended sources or large-scale diffuse excesses.
Accordingly, the \emph{Fermi} Collaboration does not recommend any standard background model for extended diffuse analyses, stating that the approach for such studies should be determined and tested on a case-by-case basis.\footnote{\url{http://fermi.gsfc.nasa.gov/ssc/data/analysis/LAT_caveats.html}.} The model also inherits fundamental limitations from the \texttt{GALPROP} code\footnote{\texttt{GALPROP} is publicly available at \url{http://galprop.stanford.edu}.} \cite{Strong:1998pw, Strong:1999sv, Strong:2007nh} used to compute the distribution of cosmic rays (for example, this code treats the Galaxy as axisymmetric).
Finally, the \texttt{p6v11} diffuse model was created based on a much earlier \emph{Fermi} dataset than the one employed in this work, with a different event selection.
It was fitted to the data assuming (a) no dark matter component, and (b) a set of instrument response functions that have since been superseded.
The \texttt{p6v11} diffuse model itself is a physical model for the gamma-ray emission, and is not convolved with those original instrument response functions.
However, there is no guarantee that it would still yield the best fit to our updated and modified dataset if the same analysis to be repeated, due to both increased statistics, and low-level systematic errors in the instrument response functions (that differ between our analysis and the original fit of the \texttt{p6v11} diffuse model to the data).
More fundamentally, in the absence of an accurate model for the cosmic ray distribution in the inner Galaxy, any diffuse model we construct will have difficult-to-gauge systematic differences from the data.
It is not unexpected that -- as mentioned above -- none of our models provide formally good fits to the data, to the level of Poisson noise.

Acknowledging these caveats, we proceed using the \texttt{p6v11} diffuse model, motivated by the results of Ref.~\cite{Hooper:2013rwa}, that demonstrated that (in a study of the inner Galaxy) consistent results were obtained using the \texttt{p6v11} diffuse model and a data-driven model for the diffuse emission based on the 0.5-1 GeV energy band (where the excess appears to be relatively faint).
By letting the model components float independently in each energy bin, as was done in Ref.~\cite{Hooper:2013rwa}, we remove any systematic effect from a mismodeled energy-dependent effective area in the original construction of the \texttt{p6v11} diffuse model.

Our approach in this work is to test the robustness of the excess to variations in the background modeling, rather than trying to construct the best possible background model.
In Appendix \ref{app:diffuse} we explore the effects of changing to a later \emph{Fermi} Collaboration diffuse model, and the impact of adding an additional template tracing the interstellar dust.
The dust map serves as a template for the gas distribution, and by allowing it to float freely we allow for mismodeling of the gas-correlated component in the \texttt{p6v11} diffuse model.
Both of these modifications can change the details of the extracted spectrum for the excess, but its presence and general shape (peaked at 1-3 GeV) appear fairly robust.
Only when a template sculpted to be near-spherical and sharply peaked toward the Galactic Center is added to the ``background'' model do we find substantial degeneracy with the excess at low energies, and a resulting shift in the peak of its spectrum to higher energies.\footnote{After the original appearance of this work, an independent study demonstrated that considering a wide range of \texttt{GALPROP}-based models also does not significantly alter the qualitative features of the excess \cite{Calore:2014xka}.} 
We also perform the tests of fitting in different sub-regions (Appendix \ref{app:consistency}), as one would expect the systematics due to mis-subtraction of the diffuse background to differ over the sky.
Finally, as we will show in the next section, the features of the excess discussed in this work are also reproduced in the Galactic Center, where we would again expect the systematic errors due to mismodeled diffuse emission to be different from those present at higher latitudes.

\section{The Galactic Center}\label{center}

In this section, we describe our analysis of the \textit{Fermi} data from the region of the Galactic Center, defined as $|b|<5^{\circ}$, $|l|<5^{\circ}$.
We make use of the same Pass 7 data set, with Q2 cuts on CTBCORE, as described in the previous section.
We performed a binned likelihood analysis to this data set using the \textit{Fermi} tool \texttt{gtlike}, dividing the region into 200$\times$200 spatial bins (each $0.05^{\circ}\times0.05^{\circ}$), and 12 logarithmically-spaced energy bins between 0.316-10.0 GeV.
Included in the fit is a model for the Galactic diffuse emission, supplemented by a model spatially tracing the observed 20 cm emission~\cite{Law:2008uk}, a model for the isotropic gamma-ray background, and all gamma-ray sources listed in the 2FGL catalog~\cite{Fermi-LAT:2011yjw}, as well as the two additional point sources described in Ref.~\cite{YusefZadeh:2012nh}.
We allow the flux and spectral shape of all high-significance ($\sqrt{{\rm TS}}>25$) 2FGL sources located within $7^{\circ}$ of the Galactic Center to vary.
For somewhat more distant or lower significance sources ($\psi=7^{\circ}-8^{\circ}$ and $\sqrt{{\rm TS}}>25$, $\psi=2^{\circ}-7^{\circ}$ and $\sqrt{{\rm TS}}=10-25$, or $\psi<2^{\circ}$ and any TS), we adopt the best-fit spectral shape as presented in the 2FGL catalog, but allow the overall normalization to float.
We additionally allow the spectrum and normalization of the two new sources from Ref.~\cite{YusefZadeh:2012nh}, the 20 cm template, and the extended sources W28 and W30~\cite{Fermi-LAT:2011yjw} to float.
We fix the emission from all other sources to the best-fit 2FGL values.

\begin{figure}[t!]
\begin{center}
\includegraphics[width=2.8in]{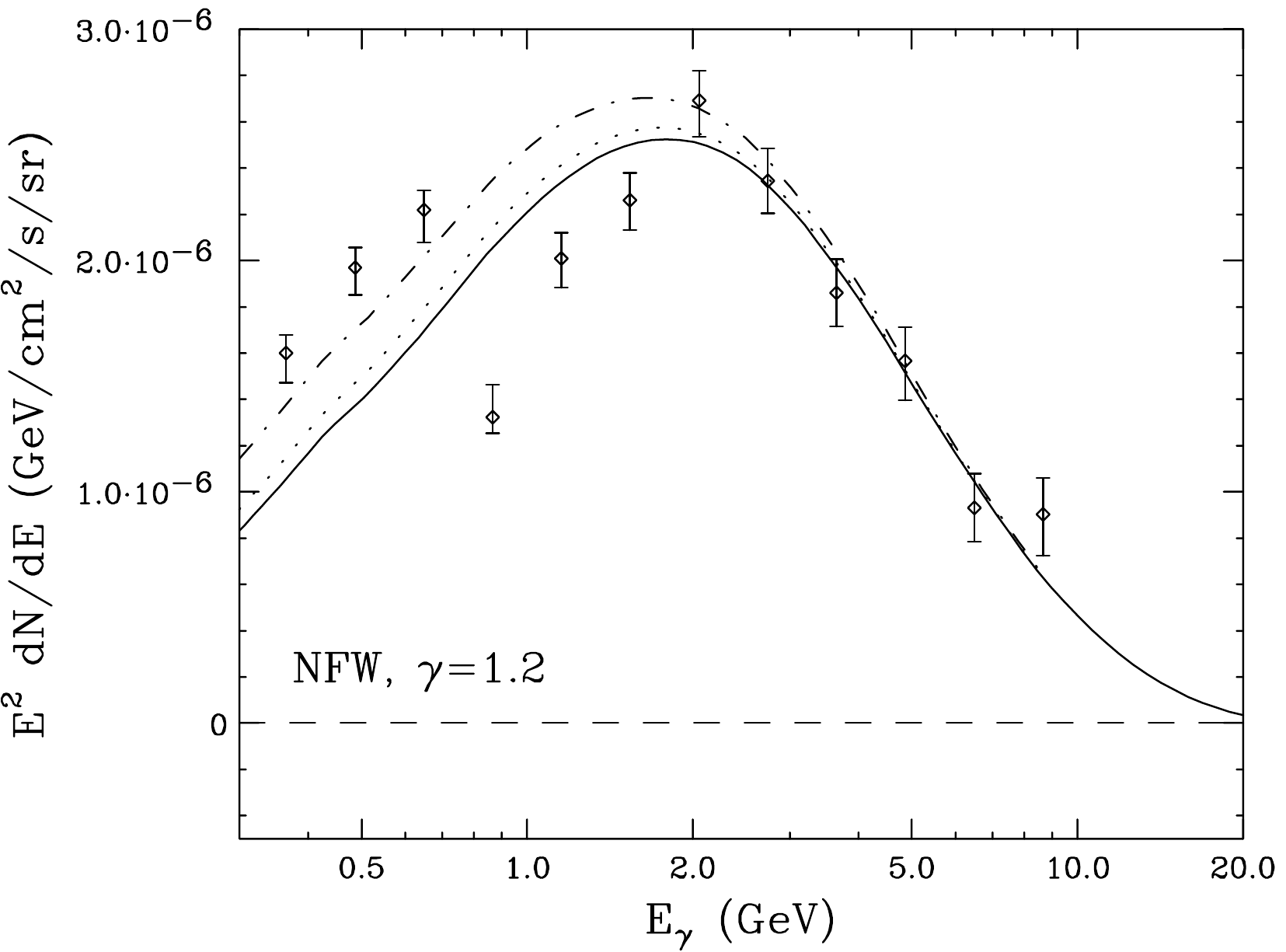}
\hspace{0.1in}
\includegraphics[width=2.8in]{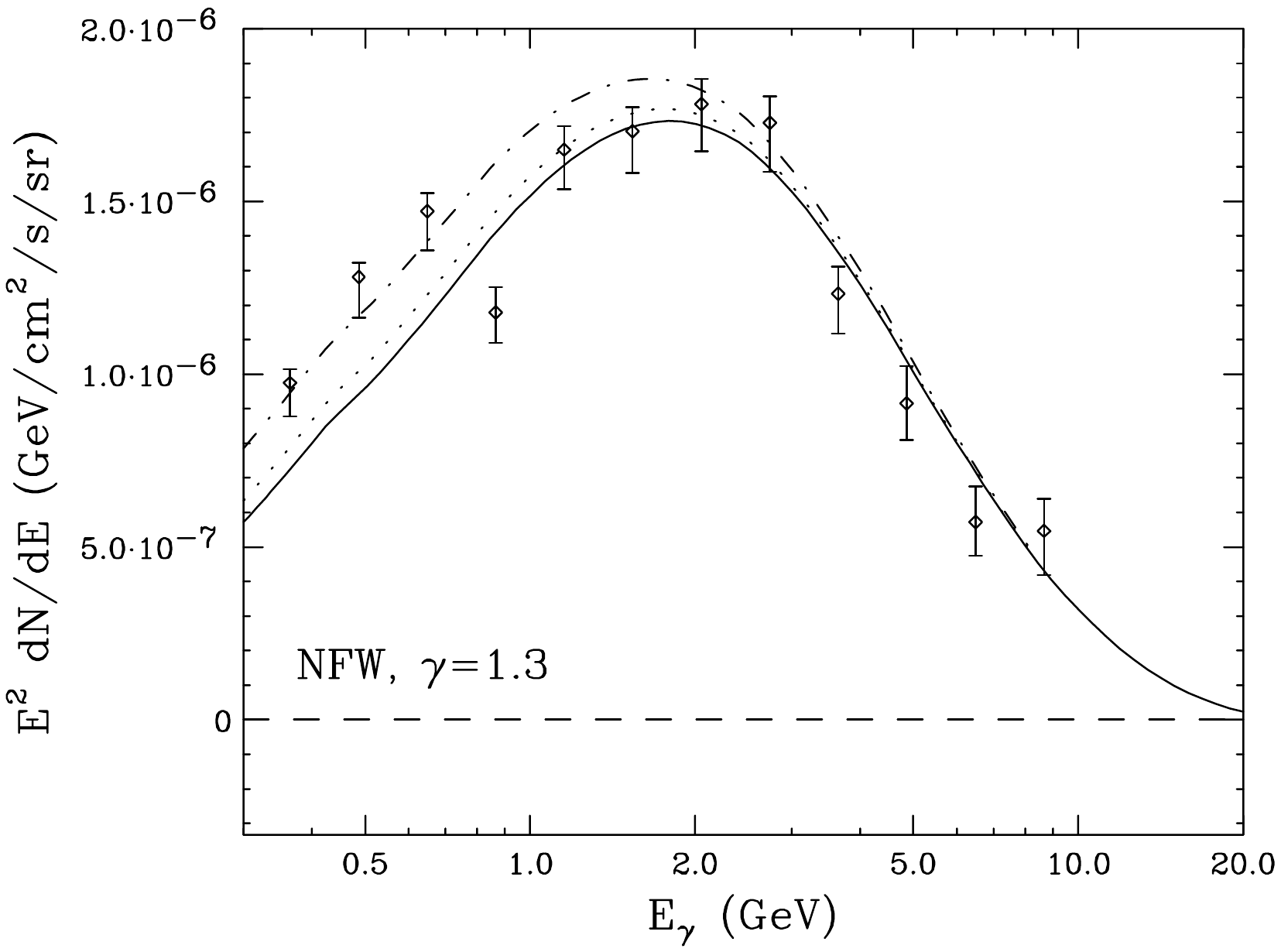}
\end{center}
\vspace{-0.5cm}
\caption{The spectrum of the dark matter component derived in our Galactic Center analysis, for a template corresponding to an NFW halo profile with an inner slope of $\gamma=1.2$ (left) or 1.3 (right), normalized to the flux at an angle of 5$^{\circ}$ from the Galactic Center.
We caution that significant and difficult to estimate systematic uncertainties exist in this determination, especially at energies below $\sim$1 GeV.
Shown for comparison (solid line) is the spectrum predicted from a 35.25 GeV dark matter particle annihilating to $b\bar{b}$ with a cross section of $\sigma v = 1.21\times 10^{-26}$ cm$^3$/s $\, \times \, [(0.4 \,{\rm GeV}/{\rm cm}^3)/\rho_{\rm local}]^2$ (left) or $\sigma v = 0.56\times 10^{-26}$ cm$^3$/s $\, \times \, [(0.4 \,{\rm GeV}/{\rm cm}^3)/\rho_{\rm local}]^2$ (right).
The dot-dash and dotted curves include an estimated contribution from bremsstrahlung, as shown in the right frame of Fig.~\ref{dnde}.
}
\label{timspec}
\end{figure}

\begin{figure}[t!]
\begin{center}
\includegraphics[width=3.0in]{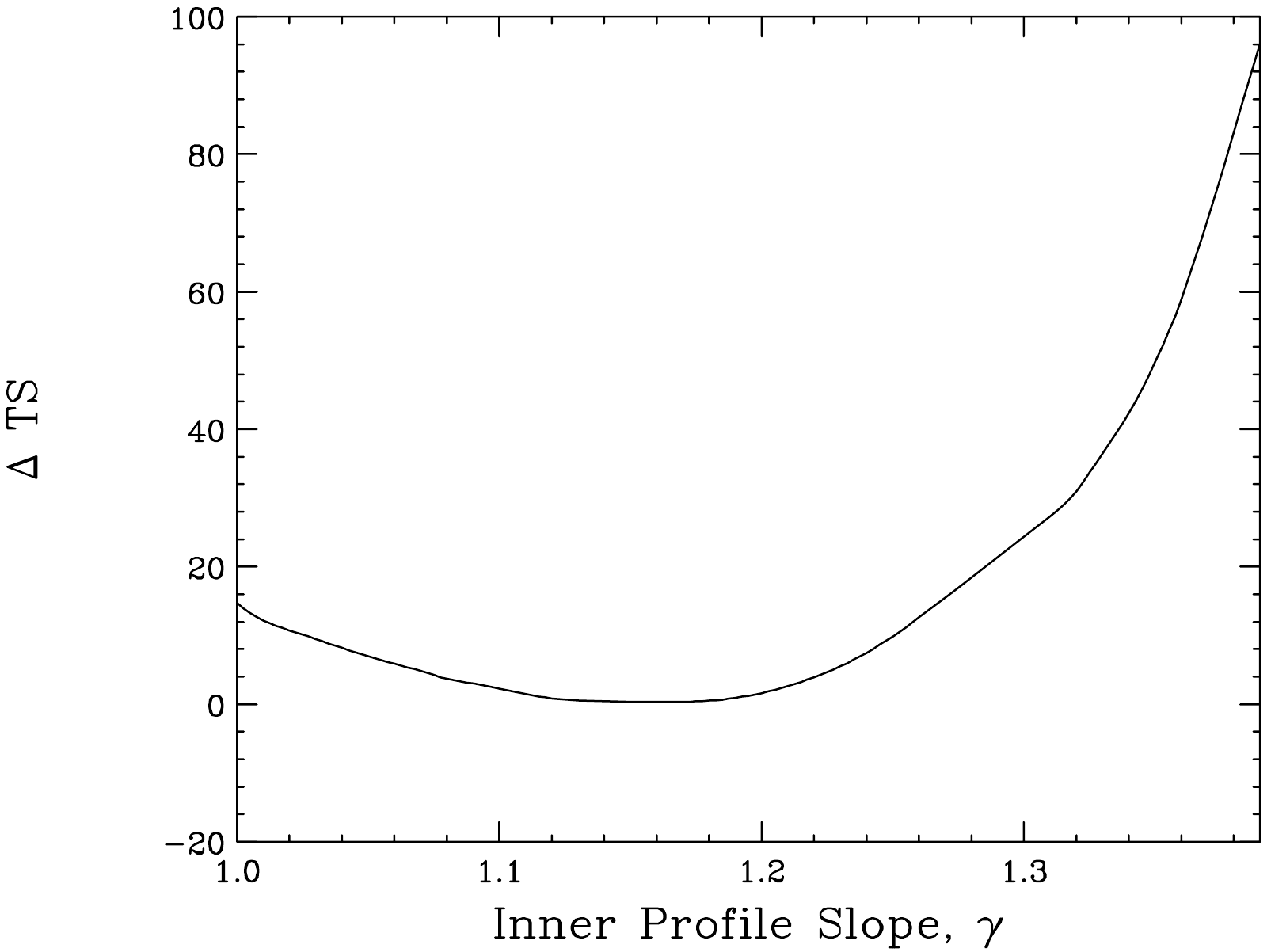}
\end{center}
\vspace{-0.5cm}
\caption{The change in TS for the dark matter template as a function of the inner slope of the dark matter halo profile, $\gamma$, as found in our Galactic Center likelihood analysis.
All values are relative to the result for the best-fit (highest TS) template, and positive values thus indicate a reduction in TS.
The best-fit value is very similar to that found in our analysis of the larger Inner Galaxy region (in the default ROI), favoring $\gamma \sim 1.17$ (compared to $\gamma \simeq 1.18$ in the Inner Galaxy analysis).
}
\label{timslope}
\end{figure}

For the Galactic diffuse emission, we adopt the model \texttt{gal\_2yearp7v6\_v0}.
Although an updated Galactic diffuse model has recently been released by the \textit{Fermi} Collaboration, that model includes additional empirically fitted features at scales greater than 2$^{\circ}$, and therefore is not recommended for studies of extended gamma-ray emission.
We use this ``\texttt{p7v6}  model''  in preference to the \texttt{p6v11} diffuse model for the Galactic Center analysis because of its finer binning and improved treatment of the Galactic plane and point sources.\footnote{\url{http://fermi.gsfc.nasa.gov/ssc/data/access/lat/Model_details/Pass7_galactic.html}} 
These factors are much more important in this region than in the inner Galaxy.
The disadvantage of the \texttt{p7v6} model for diffuse emission analyses, relative to its \texttt{p6v11} counterpart, is its inclusion of fixed, non-physical templates for large-scale residuals.
However, the impact of these templates is small in the bright Galactic Center region.

For the isotropic component, we adopt the model of Ref.~\cite{Abdo:2010nz}.
We allow the overall normalization of the Galactic diffuse and isotropic emission to freely vary.
In our fits, we found that the isotropic component prefers a normalization that is considerably brighter than the extragalactic gamma-ray background.
In order to account for this additional isotropic emission in our region of interest, we attempted simulations in which we allowed the spectrum of the isotropic component to vary, but found this to have a negligible impact on the fit.

\begin{figure}[t!]
\begin{center}
\includegraphics[width=6.0in]{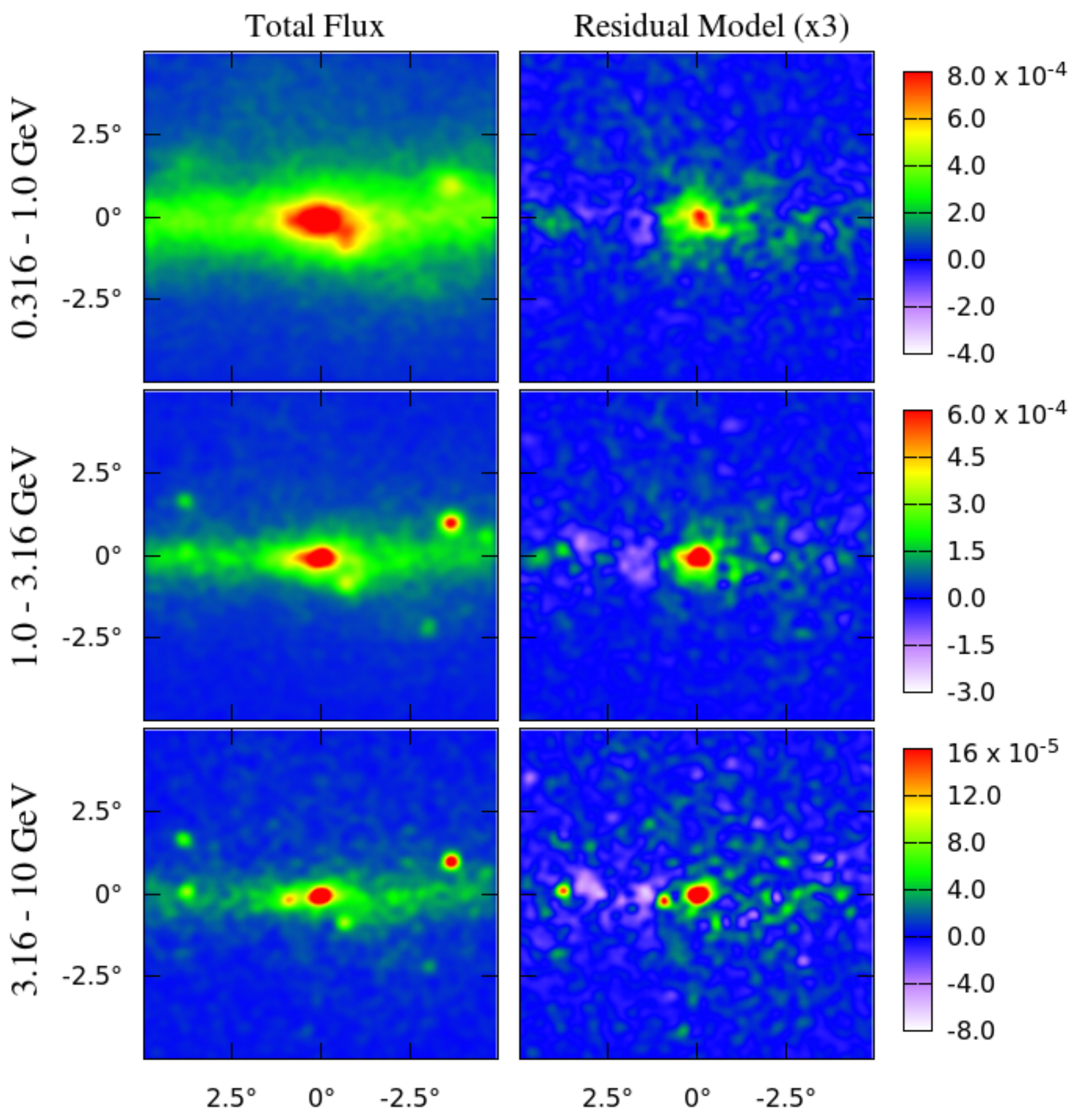}
\end{center}
\vspace{-0.5cm}
\caption{The raw gamma-ray maps (left) and the residual maps after subtracting the best-fit Galactic diffuse model, 20 cm template, point sources, and isotropic template (right), in units of photons/cm$^2$/s/sr.
The right frames clearly contain a significant central and spatially extended excess, peaking at $\sim$1-3 GeV.
Results are shown in galactic coordinates, and all maps have been smoothed by a 0.25$^\circ$~Gaussian.
}
\label{dmmap}
\end{figure}

In addition to these astrophysical components, we include a spatially extended model in our fits motivated by the possibility of annihilating dark matter.
The morphology of this component is again taken to follow the line-of-sight integral of the square of the dark matter density, as described in Sec.~\ref{intro2}.
We adopt a generalized NFW profile centered around the location of Sgr A$^*$ ($b=-0.04608^{\circ}$, $l=-0.05578^{\circ}$~\cite{YusefZadeh:1999bh}), and allow the inner slope ($\gamma$) and overall normalization (set by the annihilation cross section) to freely float.

In Figs.~\ref{timspec} and~\ref{timslope}, we show the main results of our Galactic Center likelihood analysis.
In Fig.~\ref{timslope}, we plot the change of the log-likelihood of our fit as a function of the inner slope of the halo profile, $\gamma$.
For our best-fit value of $\gamma=1.17$, the inclusion of the dark matter component (with two degrees of freedom corresponding to the normalization of spectrum based on the best-fit dark matter mass) can improve the overall fit with TS $\simeq 300$, corresponding to a statistical preference for such a component at the level of $\sim$17$\sigma$.
In Fig.~\ref{timspec}, we show the spectrum of the dark-matter-like component, for values of $\gamma=1.2$ (left frame) and $\gamma=1.3$ (right frame).
Shown for comparison is the spectrum predicted from a 35.25 GeV WIMP annihilating to $b\bar{b}$.
The solid line represents the contribution from prompt emission, whereas the dot-dashed and dotted lines also include an estimate for the contribution from bremsstrahlung (for the $z=0.15$ and 0.3 kpc cases, as shown in the right frame of Fig.~\ref{dnde}, respectively).
The normalizations of the Galactic Center and Inner Galaxy signals are compatible (see Figs.~\ref{innerspec} and ~\ref{timspec}), although the details of this comparison depend on the precise morphology that is adopted.

We note that the \textit{Fermi} tool \texttt{gtlike} determines the quality of the fit assuming a given spectral shape for the dark matter template, but does not generally provide a model-independent spectrum for this or other components.
In order to make a model-independent determination of the dark matter component's spectrum, we adopt the following procedure.
First, assuming a seed spectrum for the dark matter component, the normalization and spectral shape of the various astrophysical components are each varied and set to their best-fit values.
Then, the fit is performed again, allowing the spectrum of the dark matter component to vary in each energy bin.
The resultant dark matter spectrum is then taken to be the new seed, and this procedure is repeated iteratively until convergence is reached.

In Fig.~\ref{dmmap}, we plot the gamma-ray count maps of the Galactic Center region.
In the left frames, we show the raw maps, while in the right frames we have subtracted the best-fit contributions from each component in the fit except for that corresponding to the dark matter template (the Galactic diffuse model, 20 cm template, point sources, and isotropic template).
In each frame, the map has been smoothed by a 0.25$^{\circ}$ Gaussian (0.59$^{\circ}$ full-width-half-maximum).
The excess emission is clearly present in the right frames, and most evidently in the 1.0-3.16 GeV range, where the signal is most significant.

The slope favored by our Galactic Center analysis ($\gamma \simeq 1.04$--1.24) is very similar to that found in the Inner Galaxy analysis ($\gamma$~$\simeq$~1.15-1.22).
Our results are also broadly consistent with those of the recent analysis of Ref.~\cite{Abazajian:2014fta}, which studied a smaller region of the sky ($|b|<3.5^{\circ}$, $|l|<3.5^{\circ}$), and found a preference for $\gamma \simeq 1.12 \pm 0.05$.
We discuss this question further in Sec.~\ref{morphology}.

As mentioned above, in addition to the Galactic diffuse model, we include a spatial template in our Galactic Center fit with a morphology tracing the 20 cm (1.5 GHz) map of Ref.~\cite{Law:2008uk}.
This map is dominated by synchrotron emission, and thus traces a convolution of the distribution of cosmic-ray electrons and magnetic fields in the region.
As cosmic-ray electrons also generate gamma rays via bremsstrahlung and inverse Compton processes, the inclusion of the 20 cm template in our fit is intended to better account for these sources of gamma rays.
And although the Galactic diffuse model already includes contributions from bremsstrahlung and inverse Compton emission, the inclusion of this additional template allows for more flexibility in the fit.
In actuality, however, we find that this template has only a marginal impact on the results of our fit, absorbing some of the low energy emission that (without the 20 cm template) would have been associated with our dark matter template.

\begin{figure}[t!]
\begin{center}
\includegraphics[width=2.8in]{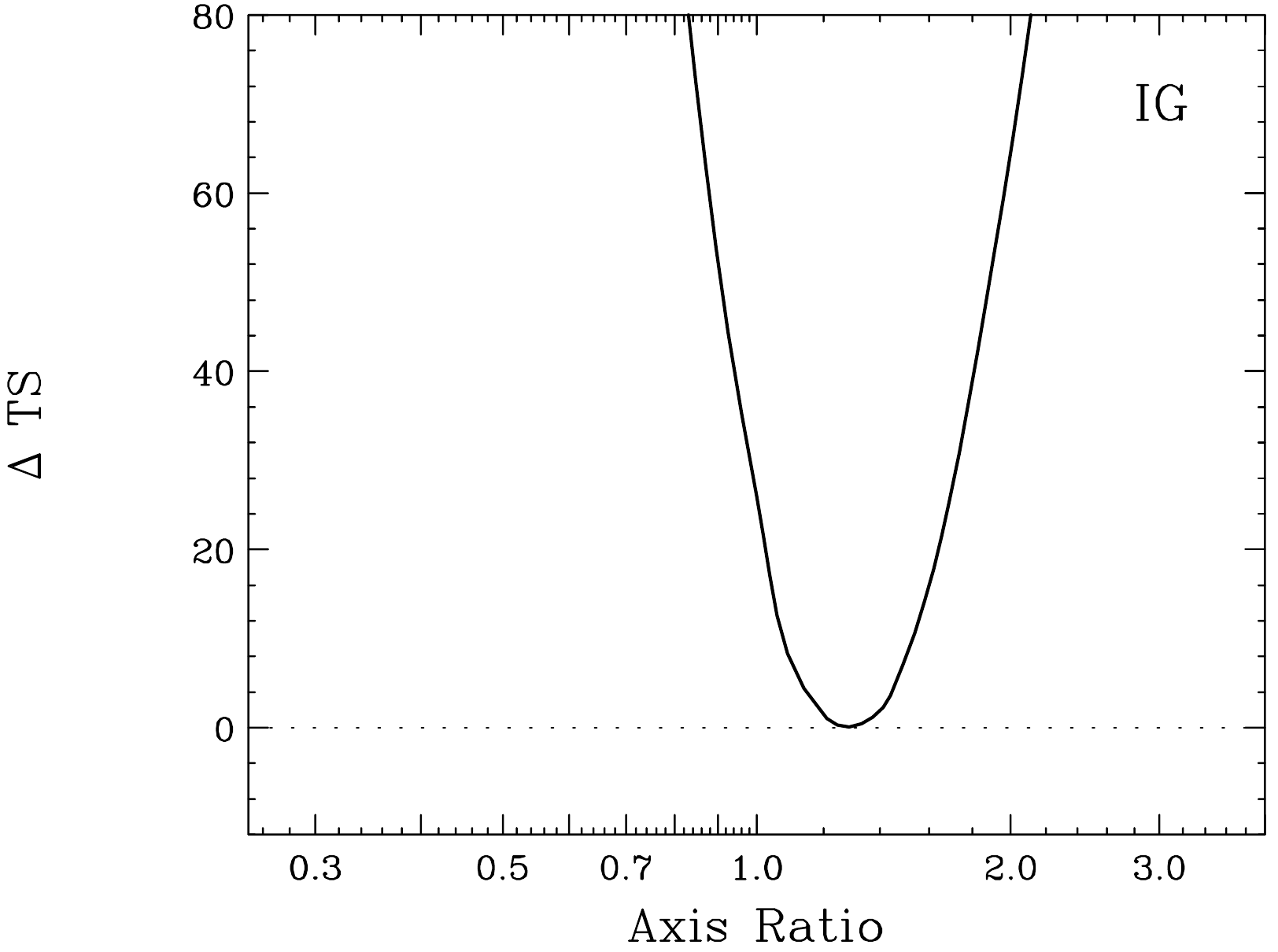}
\includegraphics[width=2.8in]{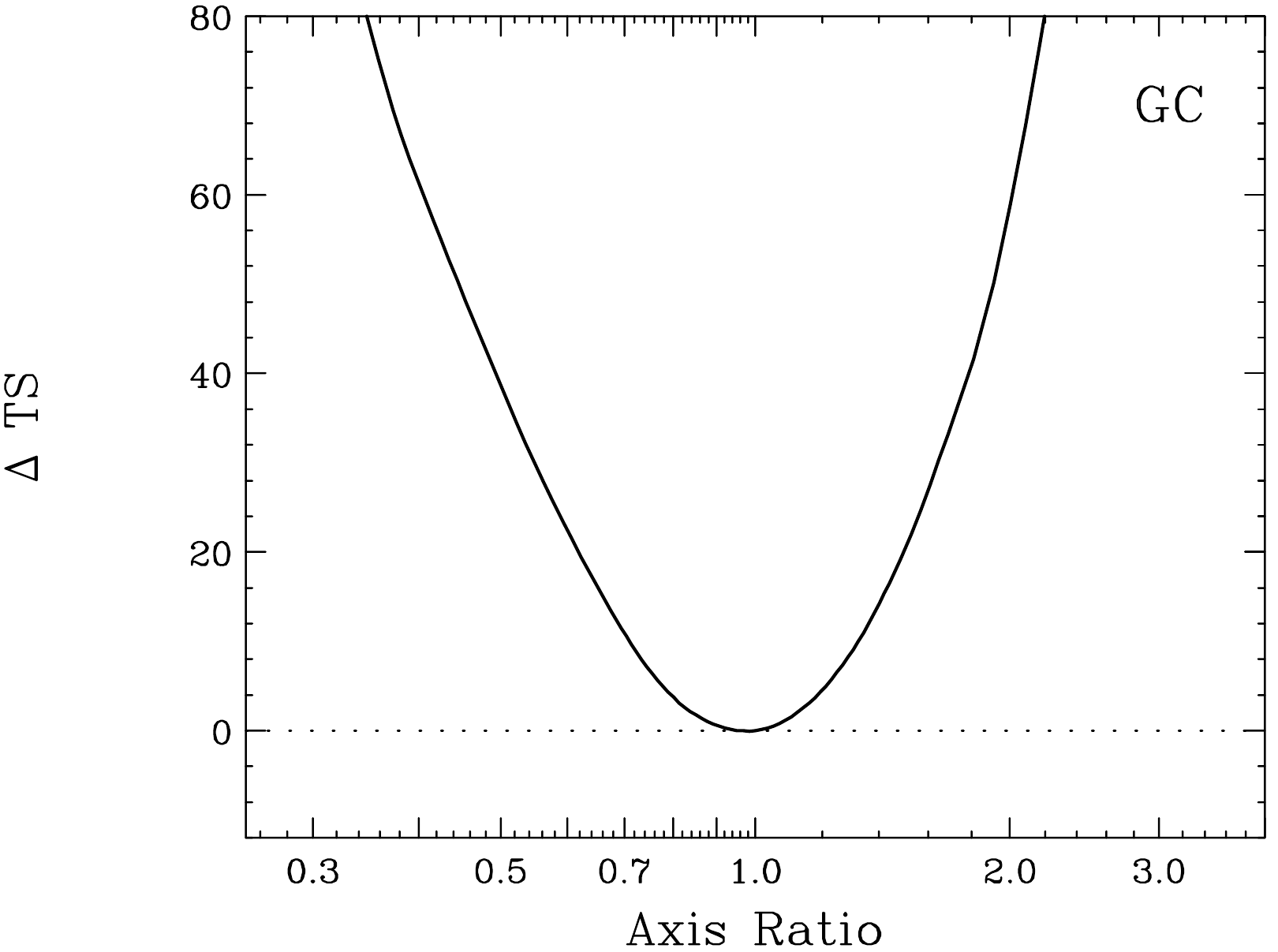}
\end{center}
\vspace{-0.5cm}
\caption{The variation in TS for the dark matter template, as performed in Sec.~\ref{inner}'s Inner Galaxy analysis (left frame) and Sec.~\ref{center}'s Galactic Center analysis (right frame), when breaking our assumption of spherical symmetry for the dark matter template.
All values shown are relative to the choice of axis ratio with the highest TS; positive values thus indicate a reduction in TS.
The axis ratio is defined such that values less than one are elongated along the Galactic Plane, whereas values greater than one are elongated with Galactic latitude.
The fit strongly prefers a morphology for the anomalous component that is approximately spherically symmetric, with an axis ratio near unity.
}
\label{asymmetry}
\end{figure}

\begin{figure}
\begin{center}
\includegraphics[width=3.695in]{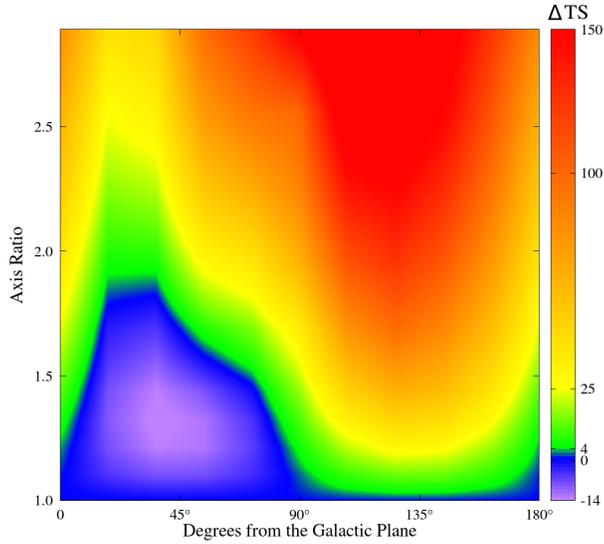}
\end{center}
\vspace{-0.5cm}
\caption{The change in the quality of the fit in our Galactic Center analysis, for a dark matter template that is elongated along an arbitrary orientation (x-axis) and with an arbitrary axis ratio (y-axis).
As shown in Fig.~\ref{asymmetry}, the fit worsens if the this template is significantly stretched either along or perpendicular to the direction of the Galactic Plane (corresponding to $0^{\circ}$ or $90^{\circ}$ on the x-axis, respectively).
A mild statistical preference, however, is found for a morphology with an axis ratio of $\sim$1.3-1.4 elongated along an axis rotated $\sim$35$^{\circ}$ clockwise from the Galactic Plane.
}
\label{alldirections}
\end{figure}

\begin{figure}
\begin{center}
\includegraphics[width=3.695in]{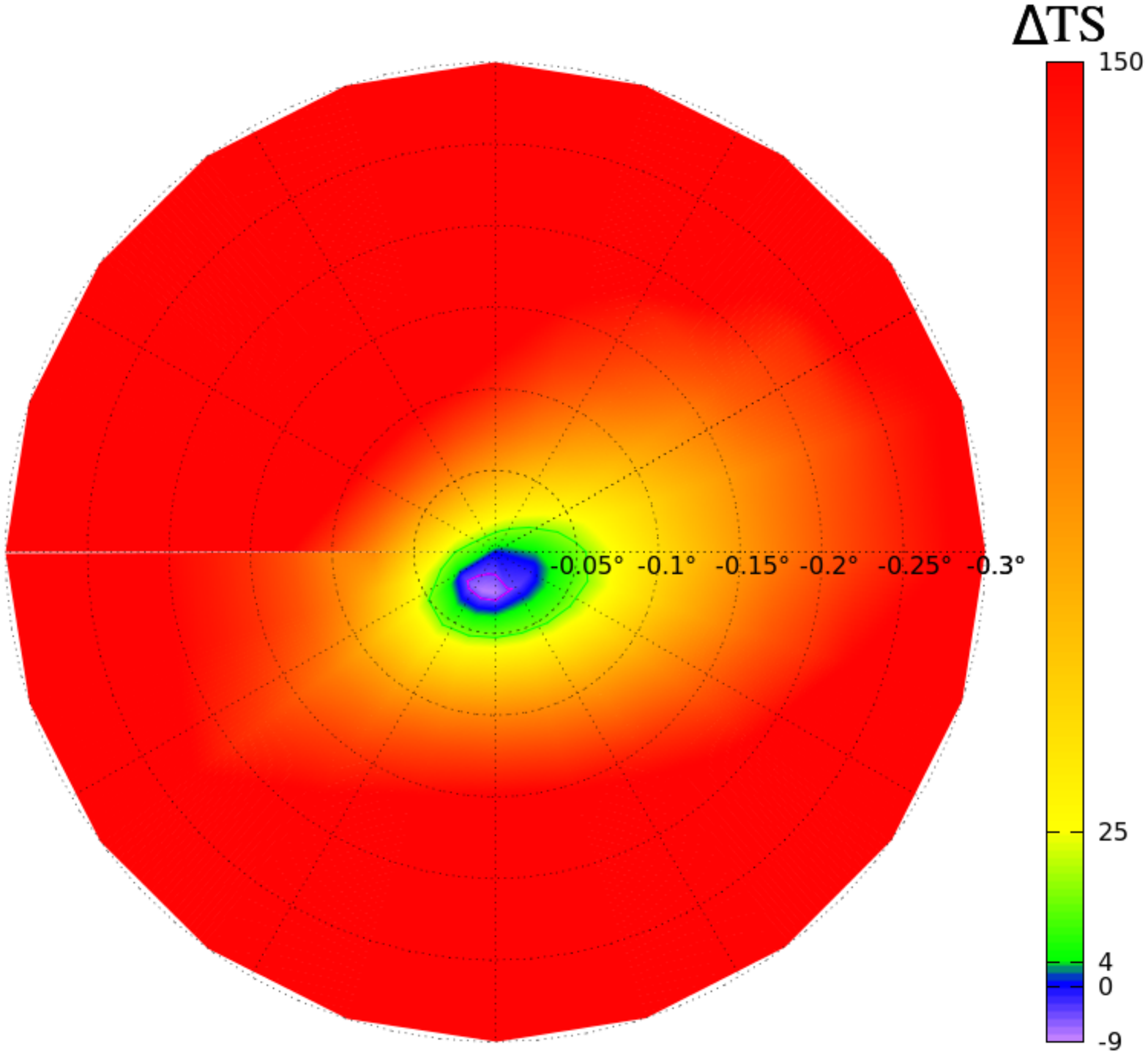}
\end{center}
\vspace{-0.5cm}
\caption{To test whether the excess emission is centered around the dynamical center of the Milky Way (Sgr A$^*$), we plot the change in the TS associated with the dark matter template found in our Galactic Center analysis, as a function of the center of the template.
Positive values correspond to a worse fit (lower TS).
The fit clearly prefers this template to be centered within $\sim$$0.05^{\circ}$ of the location of Sgr A$^*$.
} 
\label{offset}
\end{figure}

\section{Further Constraining the Morphology of the \\ Anomalous Gamma-Ray Emission}\label{morphology}

In the previous two sections, we showed that the gamma-ray emission observed from the regions of the Inner Galaxy and Galactic Center is significantly better fit when we include an additional component with an angular distribution that follows that predicted from annihilating dark matter.
In particular, our fits favor a morphology for this component that follows the square of a generalized NFW halo profile with an inner slope of $\gamma \simeq 1.1-1.3$.
Implicit in those fits, however, was the assumption that the angular distribution of the anomalous emission is spherically symmetric with respect to the dynamical center of the Milky Way.
In this section, we challenge this assumption and test whether other morphologies might provide a better fit to the observed emission.

We begin by considering templates which are elongated either along or perpendicular to the direction of the Galactic Plane.
In Fig.~\ref{asymmetry}, we plot the change in the TS of the Inner Galaxy (left) and Galactic Center (right) fits with such an asymmetric template, relative to the case of spherical symmetry.
The axis ratio is defined such that values less than unity are elongated in the direction of the Galactic Plane, while values greater than one are preferentially extended perpendicular to the plane.
The profile slope averaged over all orientations is taken to be $\gamma=1.2$ in both cases.
From this figure, it is clear that the gamma-ray excess in the Galactic Center prefers to be fit by an approximately spherically symmetric distribution, and disfavors any axis ratio which departs from unity by more than approximately 20\%.
In the Inner Galaxy there is a preference for a stretch \emph{perpendicular} to the plane, with an axis ratio of $\sim 1.3$.
As we will discuss in Appendix \ref{app:consistency}, however, there are reasons to believe this may be due to the oversubtraction of the diffuse model along the plane, and this result is especially sensitive to the choice of ROI.

In Fig.~\ref{alldirections}, we generalize this approach within our Galactic Center analysis to test morphologies that are not only elongated along or perpendicular to the Galactic Plane, but along any arbitrary orientation.
Again, we find that that the quality of the fit worsens if the the template is significantly elongated either along or perpendicular to the direction of the Galactic Plane.
A mild statistical preference is found, however, for a morphology with an axis ratio of $\sim$1.3-1.4 elongated along an axis rotated $\sim$35$^{\circ}$ clockwise from the Galactic Plane in galactic coordinates.\footnote{We define a ``clockwise'' rotation such that a 90$^\circ$ rotation turns +l into +b.} 
While this may be a statistical fluctuation, or the product of imperfect background templates, it could also potentially reflect a degree of triaxiality in the underlying dark matter distribution.

We have also tested whether the excess emission is, in fact, centered around the dynamical center of the Milky Way (Sgr A$^*$), as we have thus far assumed.
In Fig.~\ref{offset}, we plot the change in TS of the dark-matter-motivated template, as found in our Galactic Center analysis, when we vary the center of the template.
The fit clearly prefers this template to be centered within $\sim$$0.05^{\circ}$ of the location of Sgr A$^*$.
 
We smooth the ring templates to a $1^{\circ}$ Gaussian (full-width-half-max), and fit the normalization of each ring template independently.
Instead of allowing the spectrum of the ring templates to each vary freely (which would have introduced an untenable number of free parameters), we fix their spectral shape to that found for the dark matter component in the single template fit.
We also break the template associated with the \textit{Fermi} Bubbles into five templates, in 10$^{\circ}$ latitude slices (each with the same spectrum, but with independent normalizations).

The dark-matter-like emission is clearly and consistently present in each ring template out to $\sim$$10^{\circ}$, beyond which systematic and statistical limitations make such determinations difficult.
For comparison, we also show the predictions for a generalized NFW profile with $\gamma=1.4$ (after appropriate smoothing).
While this value for the profile slope is steeper that that found in Secs~\ref{inner} and~\ref{center}, we caution that systematic uncertainties associated with the diffuse model template may be biasing this fit toward somewhat steeper values of $\gamma$ (we discuss this question further in Appendix \ref{app:consistency}, in the context of the increased values of $\gamma$ found for larger ROIs).
It is also plausible that the dark matter slope could vary with distance from the Galactic Center, for example as exhibited by an Einasto profile~\cite{Springel:2008cc}.

\begin{figure}[t!]
\begin{center}
\includegraphics[width=3.0in,trim=1cm 11cm 1cm 0]{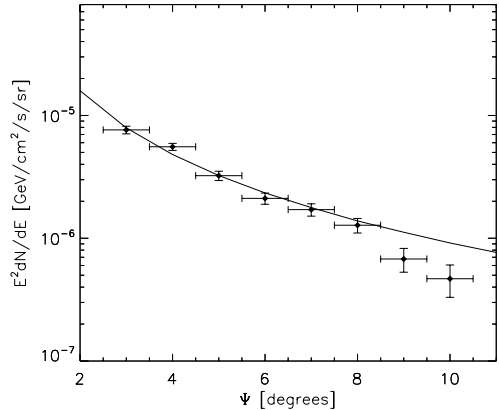}
\end{center}
\vspace{-0.5cm}
\caption{To constrain the degree to which the gamma-ray excess is spatially extended, we have repeated our Inner Galaxy analysis, replacing the dark matter template with a series of concentric ring templates centered around the Galactic Center.
The dark-matter-like emission is clearly and consistently present in each ring template out to $\sim$$10^{\circ}$, beyond which systematic and statistical limitations make such determinations difficult.
For comparison, we also show the predictions for a generalized NFW profile with $\gamma=1.3$.
The spectrum of the rings is held fixed at that of Fig.\ref{innerspec}, and the fluxes displayed in the plot correspond to an energy of 2.67 GeV.
}
\label{ringfit}
\end{figure}

An important question to address is to what degree the gamma-ray excess is spatially extended, and over what range of angles from the Galactic Center can it be detected? To address this issue, we have repeated our Inner Galaxy analysis, replacing the dark matter template with 8 concentric, rotationally symmetric ring templates, each 1$^{\circ}$ wide, and centered around the Galactic Center.
However instead of allowing the spectrum of the ring templates to each vary freely (which would have introduced an untenable number of free parameters), we fix their spectral shape between 0.3 GeV - 30 GeV to that found for the dark matter component in the single template fit.
By floating the ring coefficients with a fixed spectral dependence, we obtain another handle on the spatial extent and morphology of the excess.
In order to be self-consistent we inherit the background modeling and ROI from the Inner Galaxy analysis (except that we mask the plane for $|b| < 2^\circ$ rather than $|b| < 1^\circ$) and fix the spectra of all the other templates to the best fit values from the Inner Galaxy fit.
We also break the template associated with the \textit{Fermi} Bubbles into two sub-templates, in 10$^{\circ}$ latitude slices (each with the same spectrum, but with independent normalizations).
We smooth the templates to the \textit{Fermi} PSF.

The results of this fit are shown in Fig.~\ref{ringfit}.
The dark-matter-like emission is clearly and consistently present in each ring template out to $\sim 10^{\circ}$, beyond which systematic and statistical limitations make such determinations difficult.
In order to compare the radial dependence with that expected from a generalized NFW profile, we weight the properly smoothed NFW squared/projected template with each ring to obtain ring coefficients expected from an ideal NFW distribution.
We then perform a minimum $\chi^2$ fit on the data-driven ring coefficients taking as the template the coefficients obtained from an NFW profile with $\gamma=1.3$.
We exclude the two outermost outlier ring coefficients from this fit in order to avoid systematic bias on the preferred $\gamma$ value.
Since the ring templates spatially overlap upon smoothing, we take into account the correlated errors of the maximum likelihood fit, which add to the spectral errors in quadrature.
We show an interpolation of the best fit NFW ring coefficients with the solid line on the same figure.

We caution that systematic uncertainties associated with the diffuse model template may be biasing this fit toward somewhat steeper values of $\gamma$ (we discuss this question further in Appendix \ref{app:consistency}, in the context of the increased values of $\gamma$ found for larger ROIs).
It is also plausible that the dark matter slope could vary with distance from the Galactic Center, for example as exhibited by an Einasto profile~\cite{Springel:2008cc}.

To address the same question within the context of our Galactic Center analysis, we have re-performed our fit using dark matter templates which are based on density profiles which are set to zero beyond a given radius.
 We find that templates corresponding to density profiles set to zero outside of 800 pc (600 pc, 400 pc) provide a fit that is worse relative to that found using an untruncated template at the level of $\Delta$ TS=10.7 (57.6, 108, respectively).

We have also tested our Galactic Center fit to see if a cored dark matter profile could also provide a good fit to the data.
We find, however, that the inclusion of even a fairly small core is disfavored.
Marginalizing over the inner slope of the dark matter profile, we find that flattening the density profile within a radius of 10 pc (30 pc, 50 pc, 70 pc, 90 pc) worsens the overall fit by $\Delta$ TS=3.6 (12.2, 22.4, 30.6, 39.2, respectively).
The fit thus strongly disfavors any dark matter profile with a core larger than a few tens of parsecs.

Lastly, we confirm that the morphology of the anomalous emission does not significantly vary with energy.
If we fit the inner slope of the dark matter template in our Inner Galaxy analysis one energy bin at a time, we find a similar value of $\gamma\sim$1.1-1.3 for all bins between 0.7 and 13 GeV.
At energies $\sim 0.5$ GeV and lower, the fit prefers somewhat steeper slopes ($\gamma \sim 1.6$ or higher) and a corresponding spectrum with a very soft spectral index, probably reflecting contamination from the Galactic Plane.
At energies above $\sim 13$ GeV, the fit again tends to prefers a steeper profile.

The results of this section indicate that the gamma-ray excess exhibits a morphology which is both approximately spherically symmetric and steeply falling (yet detectable) over two orders of magnitude in galactocentric distance (between $\sim$20 pc and $\sim$2 kpc from Sgr A*).
This result is to be expected if the emission is produced by annihilating dark matter particles, but is not anticipated for any proposed astrophysical mechanisms or sources of this emission.

\begin{figure}[t!]
\begin{center}
\includegraphics[width=2.8in]{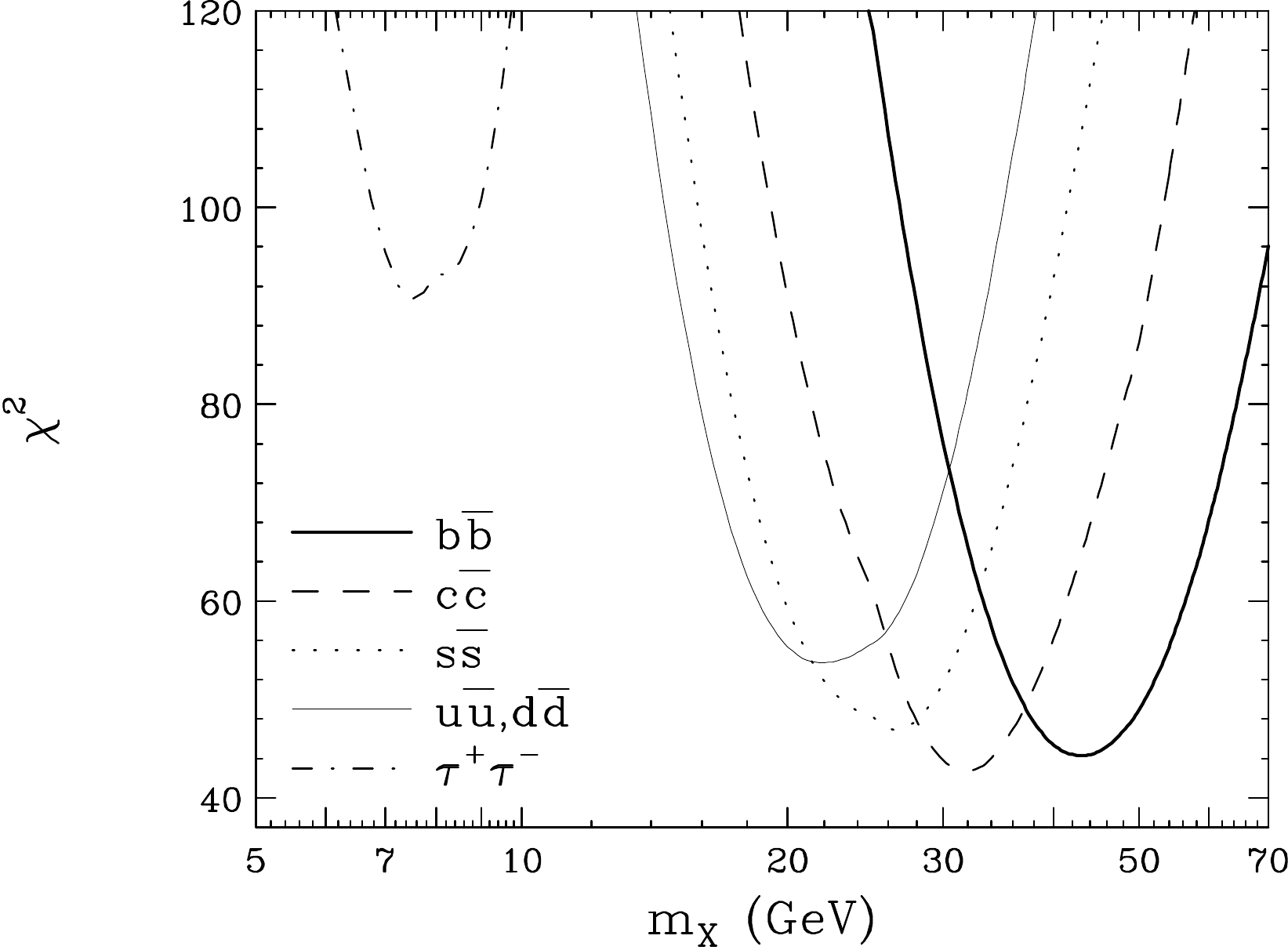}
\hspace{0.15in}
\includegraphics[width=2.8in]{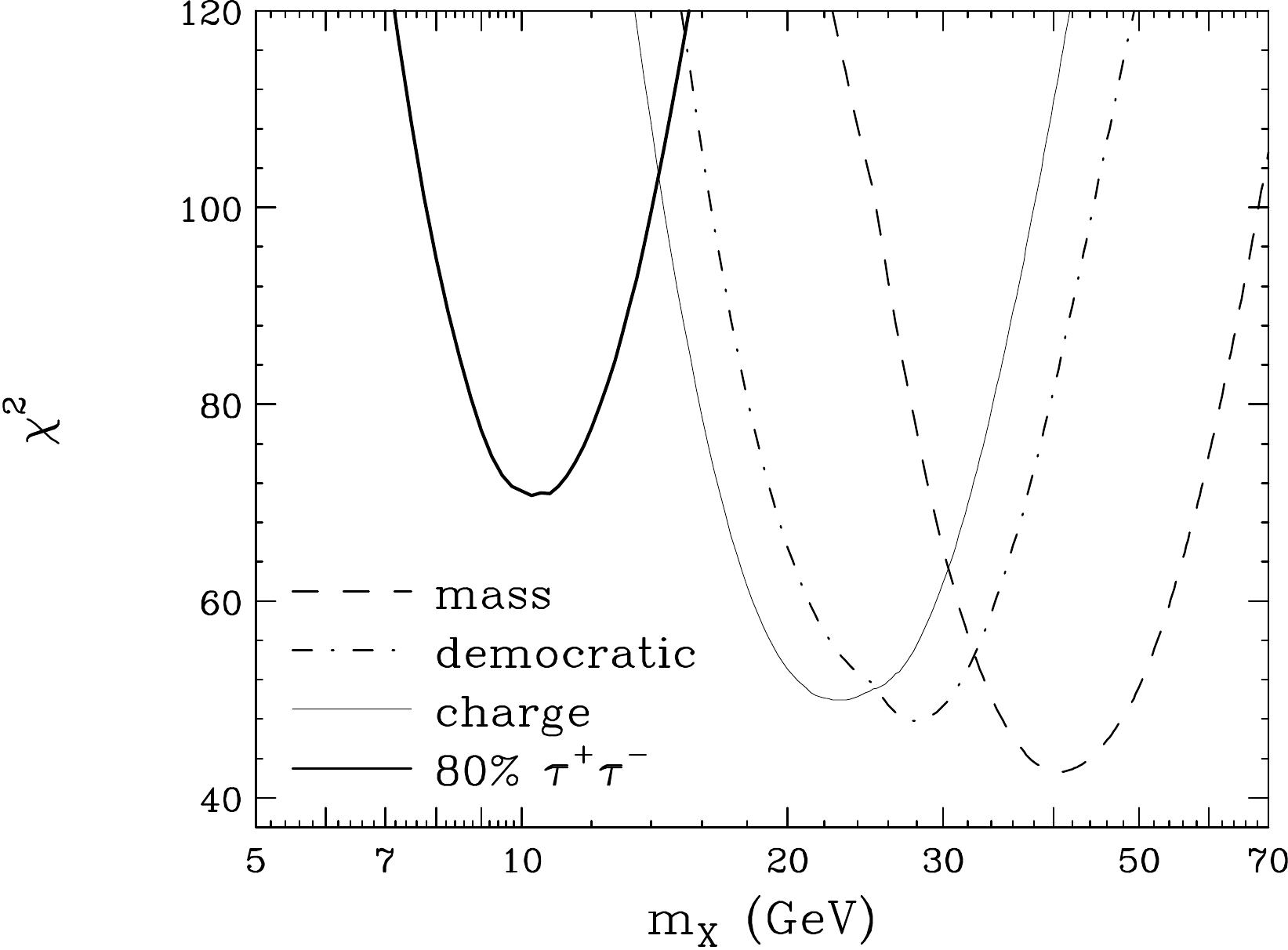}
\end{center}
\vspace{-0.5cm}
\caption{The quality of the fit ($\chi^2$, over 25-1 degrees-of-freedom) for various annihilating dark matter models to the spectrum of the anomalous gamma-ray emission from the Inner Galaxy (as shown in the left frame of Fig.~\ref{innerspec}) as a function of mass, and marginalized over the value of the annihilation cross section.
In the left frame, we show results for dark matter particles which annihilate uniquely to $b\bar{b}$, $c\bar{c}$, $s\bar{s}$, light quarks ($u\bar{u}$ and/or $d\bar{d}$), or $\tau^+ \tau^-$.
In the right frame, we consider models in which the dark matter annihilates to a combination of channels, with cross sections proportional to the square of the mass of the final state particles, the square of the charge of the final state particles, democratically to all kinematically accessible Standard Model fermions, or 80\% to $\tau^+ \tau^-$ and 20\% to $b\bar{b}$.
The best fits are found for dark matter particles with masses in the range of $\sim$20-60 GeV and which annihilate mostly to quarks.
}
\label{chisq}
\end{figure}

\section{Implications for Dark Matter}\label{darkmatter}

In this section, we use the results of the previous sections to constrain the characteristics of the dark matter particle species potentially responsible for the observed gamma-ray excess.

We begin by fitting various dark matter models to the spectrum of the gamma-ray excess as found in our Inner Galaxy analysis (as shown in the left frame of Fig.~\ref{innerspec}).
In Fig.~\ref{chisq}, we plot the quality of this fit ($\chi^2$) as a function of the WIMP mass, for a number of dark matter annihilation channels (or combination of channels), marginalized over the value of the annihilation cross section.
Given that this fit is performed over 22-1 degrees-of-freedom, a goodness-of-fit with a $p$-value of 0.05 (95\% CL) corresponds to a $\chi^2$ of approximately 36.8.
Given the systematic uncertainties associated with the choice of background templates, we take any value of $\chi^2 \lesssim 50$ to constitute a reasonably ``good fit'' to the Inner Galaxy spectrum.
Good fits are found for dark matter that annihilates to bottom, strange, or charm quarks.
The fits are slightly worse for annihilations to light quarks, or to combinations of fermions proportional to the square of the mass of the final state, the square of the charge of the final state, or equally to all fermionic degrees of freedom (democratic).
In the light mass region ($m_X$$\sim$7-10 GeV) motivated by various direct detection anomalies~\cite{Aalseth:2010vx,Aalseth:2011wp,Agnese:2013rvf,Angloher:2011uu,Bernabei:2008yi,Bernabei:2010mq}, the best fit we find is for annihilations which proceed mostly to $\tau^+ \tau^-$, with an additional small fraction to quarks, such as $b\bar{b}$.
Even this scenario, however, provides a somewhat poor fit, significantly worse that that found for heavier ($m_X\sim 20-60$ GeV) dark matter particles annihilating mostly to quarks.

\begin{figure}[t!]
\begin{center}
\includegraphics[width=2.8in]{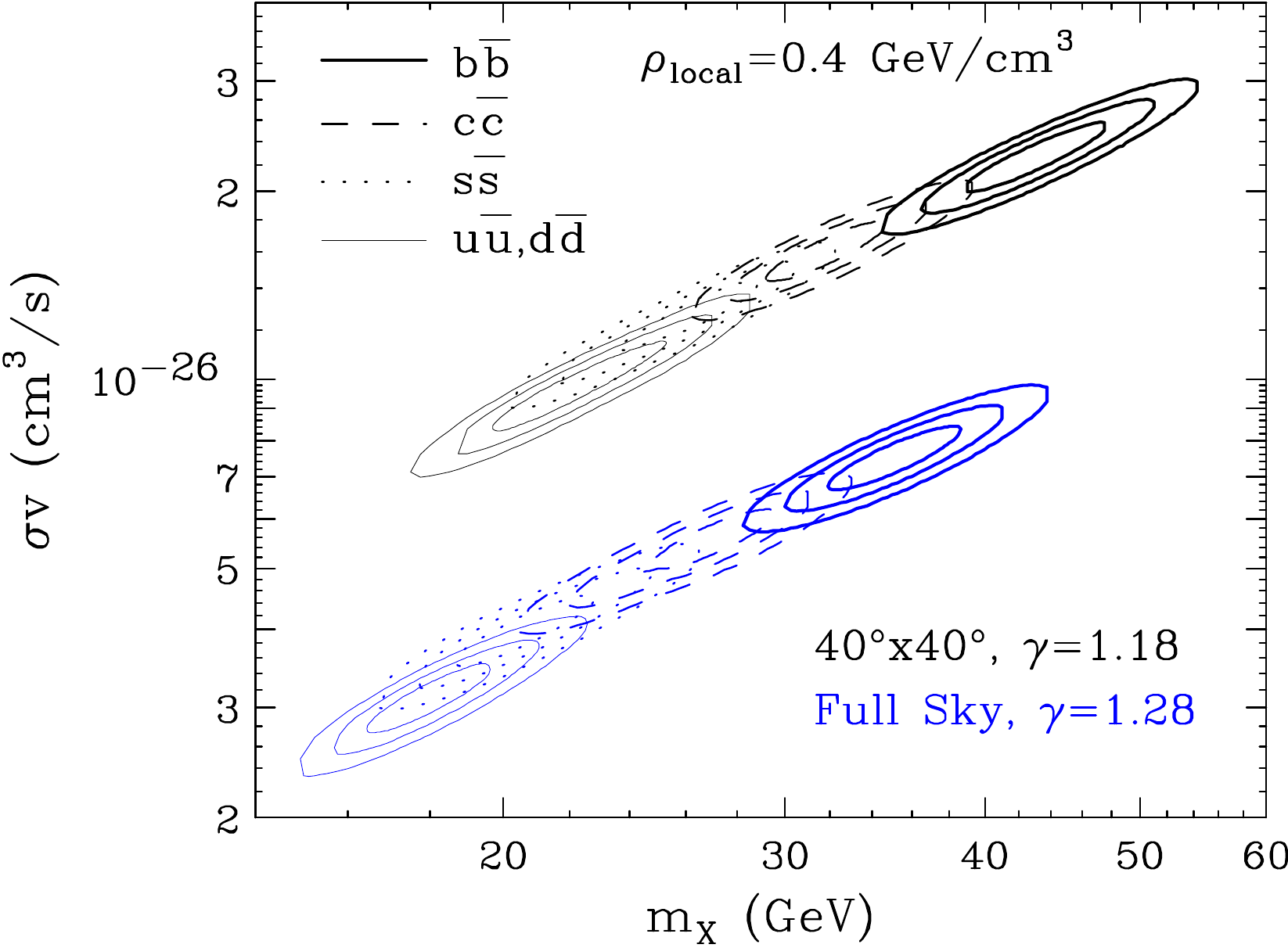}
\hspace{0.15in}
\includegraphics[width=2.8in]{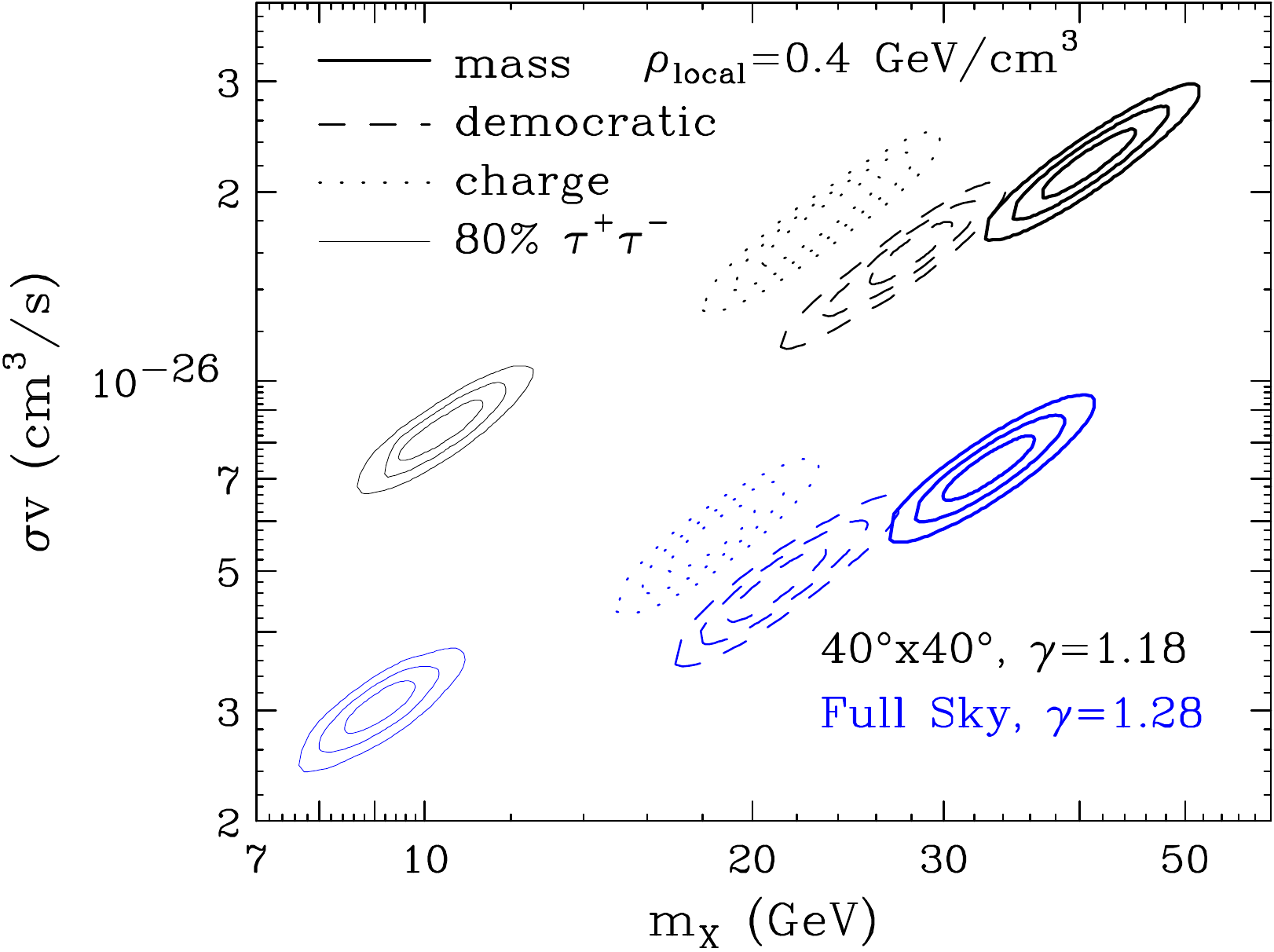}
\end{center}
\vspace{-0.5cm}
\caption{The range of the dark matter mass and annihilation cross section required to fit the gamma-ray spectrum observed from the Inner Galaxy, for a variety of annihilation channels or combination of channels (see Fig.~\ref{chisq}).
We show results for our standard ROI (black) and as fit over the full sky (blue).
The observed gamma-ray spectrum is generally best fit by dark matter particles with a mass of $\sim$20-50 GeV and that annihilate to quarks with a cross section of $\sigma v\sim 10^{-26}$ cm$^3$/s.
Note that the cross-section for each model is computed for the best-fit slope $\gamma$ in that ROI and the assumed dark matter densities at 5$^\circ$ from the Galactic Center (where the signal is normalized) are different for different values of $\gamma$.
This is responsible for roughly half of the variation between the best-fit cross-sections.
Figures~\ref{fig:igresults} and~\ref{regioncompare2} show the impact of changing the ROI when holding the assumed DM density profile constant.
}
\label{regions}
\end{figure}

In Fig.~\ref{regions}, we show the regions of the dark matter mass-annihilation cross section plane that are best fit by the gamma-ray spectrum shown in Fig.~\ref{innerspec}.
For each annihilation channel (or combination of channels), the 1, 2 and 3$\sigma$ contours are shown around the best-fit point (corresponding to $\Delta \chi^2=2.30$, 6.18, and 11.83, respectively).
Again, in the left frame we show results for dark matter particles which annihilate entirely to a single final state, while the right frame considers instead combinations of final states.
Generally speaking, the best-fit models are those in which the dark matter annihilates to second or third generation quarks with a cross section of $\sigma v\sim 10^{-26}$ cm$^3$/s.\footnote{The cross sections shown in Fig.~\ref{regions} were normalized assuming a local dark matter density of 0.4 GeV/cm$^3$.
Although this value is near the center of the range preferred by the combination of dynamical and microlensing data (for $\gamma=1.18$), there are non-negligible uncertainties in this quantity.
The analysis of Ref.~\cite{Iocco:2011jz}, for example, finds a range of $\rho_{\rm local}=0.26-0.49$ GeV/cm$^3$ at the 2$\sigma$ level.
This range of densities corresponds to a potential rescaling of the y-axis of Fig.~\ref{regions} by up to a factor of 0.7-2.4.}

This range of values favored for the dark matter's annihilation cross section is quite interesting from the perspective of early universe cosmology.
For the mass range being considered here, a WIMP with an annihilation cross section of $\sigma v \simeq 2.2 \times 10^{-26}$ cm$^3$/s (as evaluated at the temperature of freeze-out) will freeze-out in the early universe with a relic abundance equal to the measured cosmological dark matter density (assuming the standard thermal history)~\cite{Steigman:2012nb}.
The dark matter annihilation cross section evaluated in the low-velocity limit (as is relevant for indirect searches), however, is slightly lower than the value at freeze-out in many models.
For a generic $s$-wave annihilation process, for example, one generally expects dark matter in the form of a thermal relic to annihilate at low-velocities with a cross section near $\sigma v_{v=0} \simeq (1-2) \times 10^{-26}$ cm$^3$/s, in good agreement with the range of values favored by the observed gamma-ray excess.

\begin{figure}[t!]
\begin{center}
\includegraphics[width=3.0in]{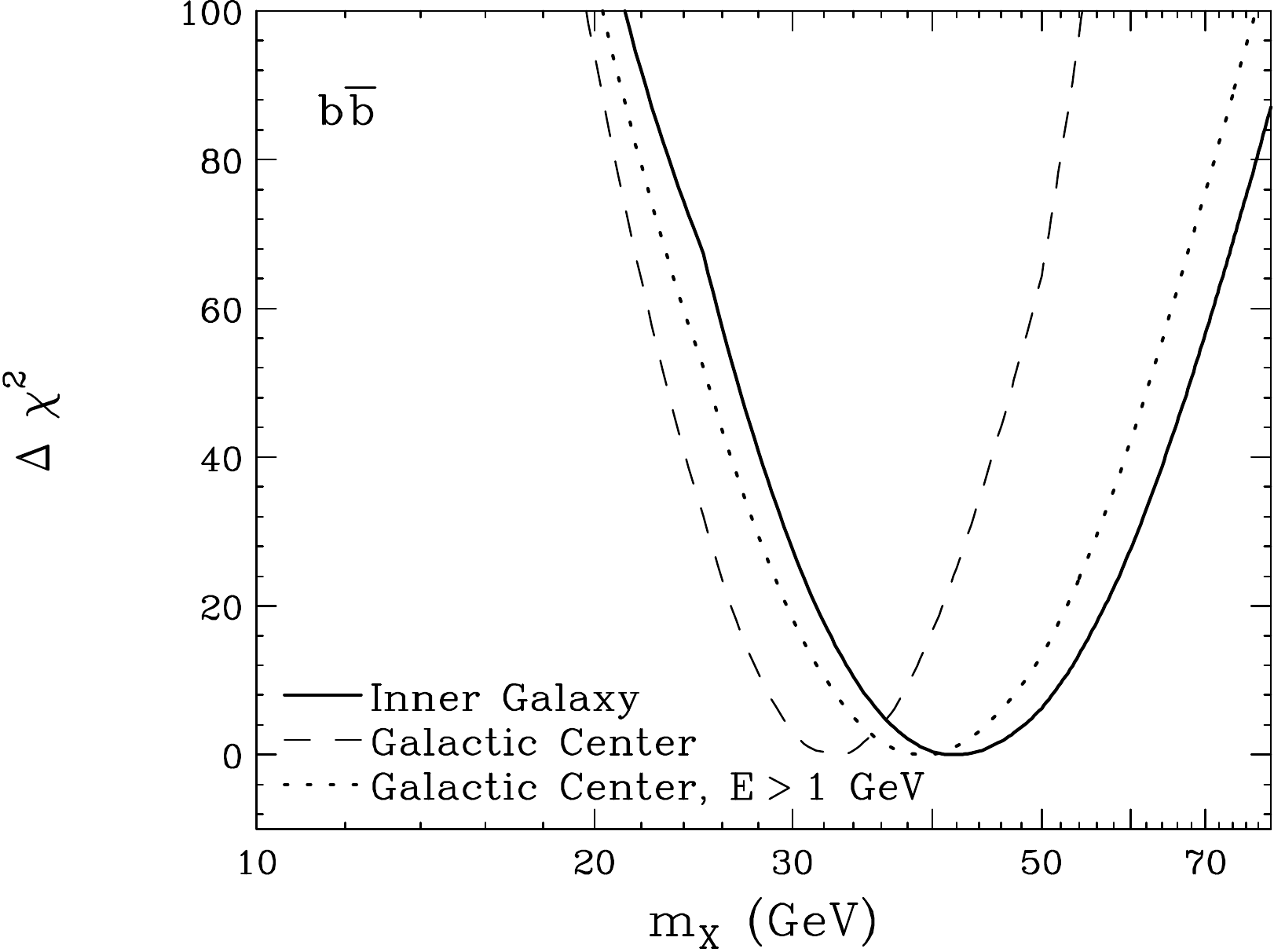}
\end{center}
\vspace{-0.5cm}
\caption{A comparison of the dark matter mass determination using the spectrum derived from our Inner Galaxy analysis (solid line) and using the spectrum derived from our Galactic Center analysis (dashed and dotted lines).
For each case shown, we have considered a profile with an inner slope of $\gamma=$1.2 and annihilations to $b\bar{b}$.
}
\label{compare}
\end{figure}

Thus far in this section, we have fit the predictions of various dark matter models to the gamma-ray spectrum derived from our Inner Galaxy analysis.
In Fig.~\ref{compare}, we compare the mass range best fit to the Inner Galaxy spectrum to that favored by our Galactic Center analysis.
Overall, these two analyses favor a similar range of dark matter masses and annihilation channels, although the Galactic Center spectrum does appear to be slightly softer, and thus prefers WIMP masses that are a few GeV lower than favored by the Inner Galaxy analysis.
This could, however, be the result of bremsstrahlung, which can soften the gamma-ray spectrum from dark matter in regions near the Galactic Plane (see Fig.~\ref{timspec} and the right frame of Fig.~\ref{dnde}).
Such emission could plausibly cause a $\sim$40-45 GeV WIMP, for example, to produce a gamma-ray spectrum along the Galactic Plane that resembles the prompt emission predicted from a $\sim$35-40 GeV WIMP.

\section{Discussion}\label{discussion}

In this chapter (and in previous studies~\cite{Goodenough:2009gk,Hooper:2010mq,Hooper:2011ti,Abazajian:2012pn,Gordon:2013vta,Hooper:2013rwa,Huang:2013pda,Abazajian:2014fta}), it has been shown that the gamma-ray excess observed from the Inner Galaxy and Galactic Center is compatible with that anticipated from annihilating dark matter particles.
This is not, however, the first time that an observational anomaly has been attributed to dark matter.
Signals observed by numerous experiments, including \textit{INTEGRAL}~\cite{Boehm:2003bt}, \textit{PAMELA}~\cite{Adriani:2008zr}, \textit{ATIC}~\cite{Chang:2008aa}, \textit{Fermi}~\cite{Weniger:2012tx,Su:2012ft}, \textit{WMAP}~\cite{Finkbeiner:2004us,Hooper:2007kb}, \textit{DAMA/LIBRA}~\cite{Bernabei:2008yi,Bernabei:2010mq}, \textit{CoGeNT}~\cite{Aalseth:2010vx,Aalseth:2011wp}, \textit{CDMS}~\cite{Agnese:2013rvf}, and \textit{CRESST}~\cite{Angloher:2011uu}, among others, have received a great deal of attention as possible detections of dark matter particles.
Most, if not all, of these signals, have nothing to do with dark matter, but instead result from some combination of astrophysical, environmental, and instrumental backgrounds (see e.g.~\cite{Hooper:2008kg, Profumo:2008ms, Dobler:2011rd, Dobler:2012ef, Ackermann:2013uma,Aalseth:2012if,Aprile:2012nq,Akerib:2013tjd}).
Given the frequency of such false alarms, we would be wise to apply a very high standard before concluding that any new signal is, in fact, the result of annihilating dark matter.

There are significant reasons to conclude, however, that the gamma-ray signal described in this chapter is more likely to be a detection of dark matter than any of the previously reported anomalies.
Firstly, this signal consists of a very large number of events, and has been detected with overwhelming statistical significance.
The excess consists of $\sim$$10^4$ gamma rays per square meter per year above 1 GeV (from within 10$^{\circ}$~of the Galactic Center).
Not only does this large number of events enable us to conclude with confidence that the signal is present, but it also allows us to determine its spectrum and morphology in some detail.
And as shown, the measured spectrum, angular distribution, and normalization of this emission does indeed match well with that expected from annihilating dark matter particles.

It is possible that a systematic mismodeling of the background could bias the extracted properties of the signal.
However, if this mismodeling is not itself fairly symmetric about the Galactic Center, it would be peculiar that the properties of the signal seem fairly consistent in different sub-regions of the ROI, and between the Galactic Center and the more extended inner Galaxy region.
Further, a mismodeling that is symmetric about the Galactic Center would be unexpected, as neither the data nor the background model possess this symmetry.
While it is possible that the true ``source of the excess'' and a background mismodeling could combine to yield an apparently spherically symmetric excess with a spectrum that does not appear to change significantly with position, even if neither component has such properties on its own, it would require a coincidence.
Accordingly, it seems likely that these observed properties reflect the actual nature of the signal.
(However, it is certainly possible that background mismodeling could induce subtle changes in the extracted spectrum or morphology or the signal.)

Secondly, the gamma-ray signal from annihilating dark matter can be calculated straightforwardly, and generally depends on only a few unknown parameters.
The morphology of this signal, in particular, depends only on the distribution of dark matter in the Inner Galaxy (as parameterized in our study by the inner slope, $\gamma$).
The spectral shape of the signal depends only on the mass of the dark matter particle and on what Standard Model particles are produced in its annihilations.
The Galactic gamma-ray signal from dark matter can thus be predicted relatively simply, in contrast to, {\it e.g}.,  dark matter searches using cosmic rays, where putative signals are affected by poorly constrained diffusion and energy-loss processes.
In other words, for the gamma-ray signal at hand, there are relatively few ``knobs to turn'', making it less likely that one would be able to mistakenly fit a well-measured astrophysical signal with that of an annihilating dark matter model.
(However, it is true that the spectrum of gamma-rays from measured millisecond pulsars closely resembles that arising from light DM annihilating into simple final states; we discuss this possible explanation below.)

Thirdly, we once again note that the signal described in this study can be explained by a very simple dark matter candidate, without any baroque or otherwise unexpected features.
After accounting for uncertainties in the overall mass of the Milky Way's dark matter halo profile~\cite{Iocco:2011jz}, our results favor dark matter particles with an annihilation cross section of $\sigma v = (0.4-6.6) \times 10^{-26}$ cm$^3$/s (for annihilations to $b\bar{b}$, see Fig.~\ref{regions}).
This range covers the long predicted value that is required of a thermal relic that freezes-out in the early universe with an abundance equal to the measured cosmological dark matter density ($2.2 \times 10^{-26}$ cm$^3$/s).
No substructure boost factors, Sommerfeld enhancements, or non-thermal histories are required.
Furthermore, it is not difficult to construct simple models in which a $\sim$30-50 GeV particle annihilates to quarks with the required cross section without violating constraints from direct detection experiments, colliders, or other indirect searches (for work related to particle physics models capable of accommodating this signal, see Refs.~\cite{Boehm:2014hva,Hardy:2014dea,Modak:2013jya,Huang:2013apa,Okada:2013bna,Hagiwara:2013qya,Buckley:2013sca,Anchordoqui:2013pta,Buckley:2011mm,Boucenna:2011hy,Marshall:2011mm,Zhu:2011dz,Buckley:2010ve,Logan:2010nw}).

And lastly, the dark matter interpretation of this signal is strengthened by the absence of well motivated alternatives.
There is no a priori reason to expect that any diffuse astrophysical emission processes would exhibit either the spectrum or the morphology of the observed signal.
In particular, the spherical symmetry of the observed emission with respect to the Galactic Center does not trace any combination of astrophysical components ({\it i.e.} radiation, gas, dust, star formation, etc.).
An energetic event at the Galactic Center might conceivably give rise to a spherically symmetric flux of cosmic rays, but the targets on which they scatter to produce gamma rays (gas, starlight) would \emph{not} share this symmetry.
Furthermore, the lack of any indication for a change in the spectrum with spatial position may argue against models where the signal photons are produced by scattering of electrons and positrons (whether originating from DM or astrophysical sources), as electrons lose energy rapidly as they diffuse through the Galactic halo.
In contrast, both the spherical symmetry and uniform spectrum of the signal are natural in the context of photons arising directly from DM annihilation, as the dark matter halo is inferred to be much more spherical than the disk (see e.g. \cite{Allgood:2005eu, Bernal:2014mmt} for discussions of general DM halos, and e.g. \cite{Law:2009yq, Law:2010pe} for studies of the Milky Way halo specifically).

The astrophysical interpretation most often discussed within the context of this signal is that it might originate from a large population of unresolved millisecond pulsars.
The millisecond pulsars observed within the Milky Way are largely located either within globular clusters or in or around the Galactic Disk (with an exponential scale height of $z_s \sim$~1 kpc~\cite{Gregoire:2013yta,Hooper:2013nhl}).
This pulsar population would lead to a diffuse gamma-ray signal that is highly elongated along the disk, and would be highly incompatible with the constraints described in Sec.~\ref{morphology}.
For example, the best-fit model of Ref.~\cite{Gregoire:2013yta}, which is based on the population of presently resolved gamma-ray millisecond pulsars, predicts a morphology for the diffuse gamma-ray emission exhibiting an axis ratio of $\sim$1-to-6.
Within $10^{\circ}$ of the Galactic Center, this model predicts that millisecond pulsars should account for $\sim$1\% of the observed diffuse emission, and less than $\sim$5-10\% of the signal described in this chapter.

To evade this conclusion, however, one could contemplate an additional (and less constrained) millisecond pulsar population associated with the Milky Way's central stellar cluster.
This scenario can be motivated by the fact that globular clusters are known to contain large numbers of millisecond pulsars, presumably as a consequence of their very high stellar densities.
If our galaxy's central stellar cluster contains a large number of millisecond pulsars with an extremely concentrated distribution (with a number density that scales approximately as $n_{\rm MSP} \propto r^{-2.4}$), those sources could plausibly account for much of the gamma-ray excess observed within the inner $\sim$1$^{\circ}$ around the Galactic Center~\cite{Hooper:2010mq,Abazajian:2010zy,Hooper:2011ti,Abazajian:2012pn,Gordon:2013vta,Abazajian:2014fta}.
However, it is more challenging to imagine that such a concentrated population could account for the more extended component of this excess, which we have shown to be present out to at least $\sim$$10^{\circ}$ from the Galactic Center.

If the number of faint millisecond pulsars required to generate the signal are present $\sim$$10^{\circ}$ ($\sim$1.5 kpc) north or south of the Galactic Center, and if they possess a similar luminosity function to other observed millisecond pulsars, a significant number of such sources should already have been resolved by \emph{Fermi} and appeared within the 2FGL catalog ~\cite{Fermi-LAT:2011yjw,Gregoire:2013yta,Hooper:2013nhl}.
The lack of such resolved sources strongly limits the abundance of millisecond pulsars in the region of the Inner Galaxy -- unless the signal originates from a novel population with an intrinsically faint luminosity function.\footnote{Since this work was posted to the arXiv, a toy model for a spherically symmetric pulsar population extending to high latitudes has been put forward \cite{Brandt:2015ula}; some evidence has also been presented in favor of a point source origin for the excess \cite{Lee:2015fea, Bartels:2015aea}.
Our arguments here cannot rule out a point source population with a different luminosity function and spatial distribution than the known/predicted pulsar populations we initially considered.}

\begin{figure}[t]
\begin{center}
\includegraphics[width=3.0in]{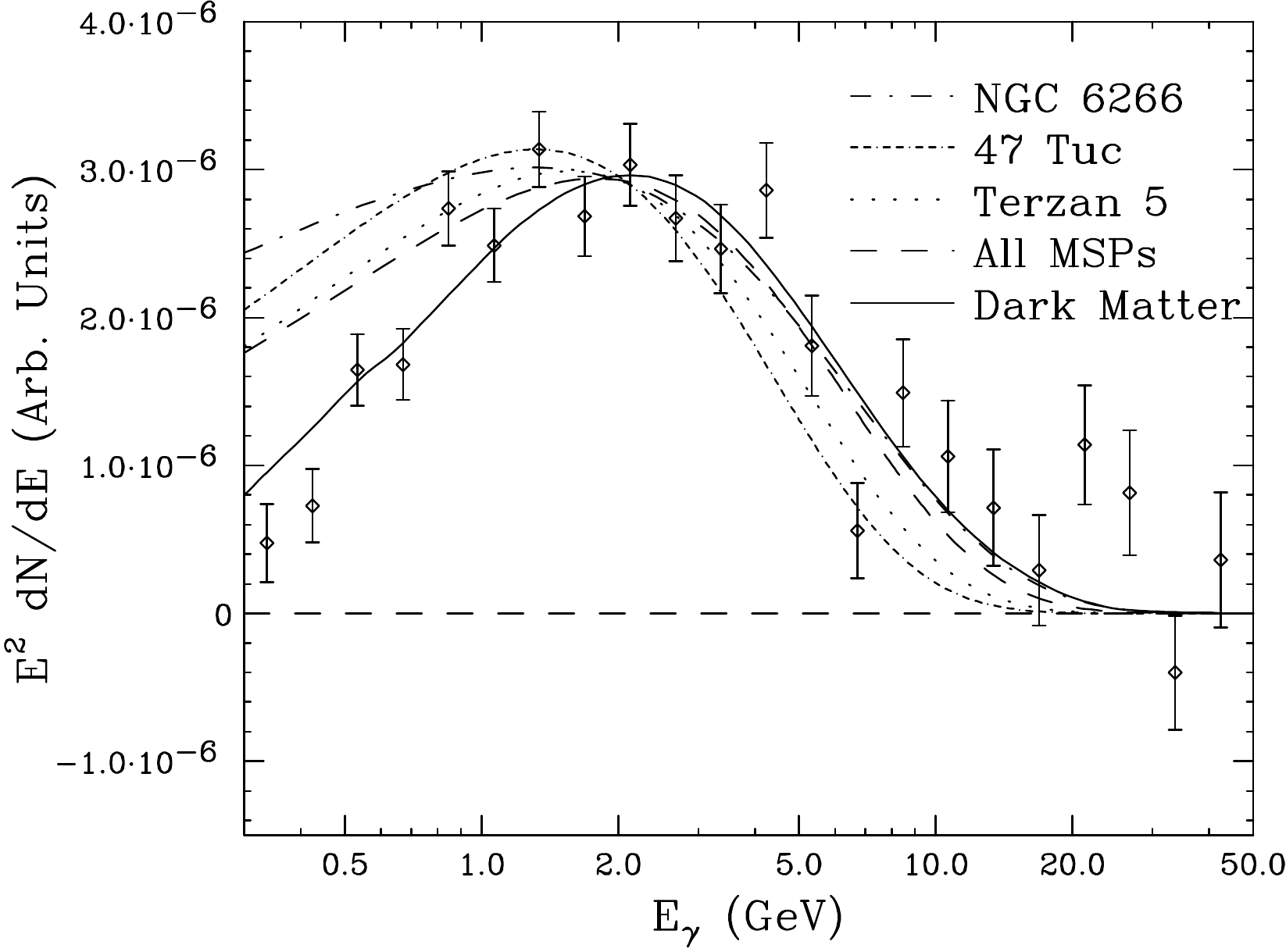}
\end{center}
\vspace{-0.5cm}
\caption{A comparison of the spectral shape of the gamma-ray excess described in this chapter (error bars) to that measured from a number of high-significance globular clusters (NGC 6266, 47 Tuc, and Terzan 5), and from the sum of all millisecond pulsars detected as individual point sources by \textit{Fermi}.
The gamma-ray spectrum measured from millisecond pulsars and from globular clusters (whose emission is believed to be dominated by millisecond pulsars) is consistently softer than that of the observed excess at energies below $\sim$1 GeV.
See text for details.
}
\label{pulsarspec}
\end{figure}

Furthermore, the shape of the gamma-ray spectrum observed from resolved millisecond pulsars and from globular clusters (whose emission is believed to be dominated by millisecond pulsars) appears to be somewhat softer than that of the gamma-ray excess observed from the Inner Galaxy.
n Fig.~\ref{pulsarspec}, we compare the spectral shape of the gamma-ray excess to that measured from a number of globular clusters, and from the sum of all resolved millisecond pulsars.
Here, we have selected the three highest significance globular clusters (NGC 6266, 47 Tuc, and Terzan 5), and plotted their best fit spectra as reported by the \textit{Fermi} Collaboration~\cite{collaboration:2010bb}.
For the emission from resolved millisecond pulsars, we include the 37 sources as described in Ref.~\cite{Hooper:2013nhl}.
Although each of these spectral shapes provides a reasonably good fit to the high-energy spectrum, they also each significantly exceed the amount of emission that is observed at energies below $\sim$1 GeV.
This comparison further disfavors millisecond pulsars as the source of the observed gamma-ray excess.
However, it is true that the low-energy spectrum is particularly prone to systematic errors in the background modeling, as the excess is faint at low energies; this is demonstrated in the appendices of this work (a similar effect was discussed in the context of the Galactic Center by \cite{Abazajian:2014fta}).
Thus, while the millisecond pulsar explanation may be disfavored by our preferred spectrum, it cannot be ruled out at high confidence on spectral grounds alone.

The near future offers encouraging prospects for detecting further evidence in support of a dark matter interpretation of this signal.
The dark matter mass and annihilation cross section implied by the gamma-ray excess is similar to \textit{Fermi}'s sensitivity from observations of dwarf spheroidal galaxies.
In fact, the \textit{Fermi} Collaboration has reported a modestly statistically significant excess ($\sim$2-3$\sigma$) in their search for annihilating dark matter particles in dwarf galaxies.
If interpreted as a detection of dark matter, this observation would imply a similar mass and cross section to that favored by our analysis~\cite{Ackermann:2013yva}.
A similar ($\sim$$3\sigma$) excess has also been reported from the direction of the Virgo Cluster~\cite{Han:2012uw,MaciasRamirez:2012mk}.
With the full dataset anticipated from \textit{Fermi}'s 10 year mission, it may be possible to make statistically significant detections of dark matter annihilation products from a few of the brightest dwarf galaxies, galaxy clusters, and perhaps nearby dark matter subhalos~\cite{Berlin:2013dva}.
Anticipated measurements of the cosmic-ray antiproton-to-proton ratio by \emph{AMS} may also be sensitive to annihilating dark matter with the characteristics implied by our analysis~\cite{Cirelli:2013hv,Fornengo:2013xda}.

\section{Summary and Conclusions}\label{summary}

In this study, we have revisited and scrutinized the gamma-ray emission from the central regions of the Milky Way, as measured by the \textit{Fermi} Gamma-Ray Space Telescope.
In doing so, we have confirmed a robust and highly statistically significant excess, with a spectrum and angular distribution that is in excellent agreement with that expected from annihilating dark matter.
The signal is distributed with approximate spherical symmetry around the Galactic Center, with a flux that falls off as $F_{\gamma} \propto r^{-(2.2-2.6)}$, implying a dark matter distribution of $\rho \propto r^{-\gamma}$, with $\gamma \simeq 1.1-1.3$.
The spectrum of the excess peaks at $\sim$1-3 GeV, and is well fit by 36-51 GeV dark matter particles annihilating to $b\bar{b}$.
The annihilation cross section required to normalize this signal is $\sigma v = (1.9-2.8) \times 10^{-26}$ cm$^3$/s (for a local dark matter density of 0.4 GeV/cm$^3$), in good agreement with the value predicted for a simple thermal relic.
Consequently, a dark matter particle with this cross section will freeze-out of thermal equilibrium in the early universe to yield an abundance approximately equal to the measured cosmological dark matter density (for the range of masses and cross sections favored for other annihilation channels, see Sec.~\ref{darkmatter}).

In addition to carrying out two different analyses (as described in Secs.~\ref{inner} and~\ref{center}), subject to different systematic uncertainties, we have applied a number of tests to our results in order to more stringently determine whether the characteristics of the observed excess are in fact robust and consistent with the signal predicted from annihilating dark matter.
These tests uniformly confirm that the signal is present throughout the Galactic Center and Inner Galaxy (extending out to angles of at least $10^{\circ}$ from the Galactic Center), without discernible spectral variation or significant departures from spherical symmetry.

At present, no known or proposed astrophysical diffuse emission mechanism naturally gives rise to these properties of the excess.
A population of several thousand millisecond pulsars could have plausibly been responsible for much of the anomalous emission observed from within the innermost $\sim1^{\circ}-2^{\circ}$ around the Galactic Center, but the extension of this signal into regions well beyond the confines of the central stellar cluster disfavors such objects as the primary source of this signal, unless the inner Galaxy hosts a new dense and approximately spherical pulsar population with an intrinsically fainter luminosity function than observed elsewhere in the Galaxy.
In light of these considerations, we consider annihilating dark matter particles to be the leading explanation for the origin of this signal, with potentially profound implications for cosmology and particle physics.

\textit{Note added: Since the completion of the work in this chapter, several analyses have demonstrated that the photon statistics of the excess are more consistent with a point source like population than smooth emission~\cite{Lee:2015fea,Bartels:2015aea}.
This poses a significant challenge for dark matter explanations of the excess, and the leading hypothesis at the time of writing is that the excess originates from a population of unresolved millisecond pulsars.
It remains true that such a population of pulsars would need to take on several surprising characteristics, although examples of how to generate these have been proposed, see e.g.~\cite{Brandt:2015ula}.}

%% file: fermigg.tex
\chapter{A Search for Dark Matter Annihilation in Galaxy Groups}\label{chap:fermigg}

\section{Introduction}

Weakly-interacting massive particles, which acquire their cosmological abundance through thermal freeze-out in the early universe, are leading candidates for dark matter (DM).  Such particles can annihilate into Standard Model states in the late universe, leading to striking gamma-ray signatures that can be detected with observatories such as the {\it Fermi} Large Area Telescope.  
Some of the strongest limits on the annihilation cross section have been set by searching for excess gamma-rays in the Milky Way's dwarf spheroidal satellite galaxies (dSphs)~\cite{Ackermann:2015zua,Fermi-LAT:2016uux}.  In this chapter, we present competitive constraints that are obtained using hundreds of galaxy groups within $z\lesssim0.03$. 

This work is complemented by a companion publication in which we describe the procedure for utilizing  galaxy group catalogs in searches for extragalactic DM~\cite{Lisanti:2017qoz}.  Previous attempts to search for DM outside the Local Group were broad in scope, but yielded weaker constraints than the dSph studies.  For example, limits on the annihilation rate were set by requiring that the DM-induced flux not overproduce the isotropic gamma-ray background~\cite{Ackermann:2015tah}.  These bounds could be improved by further resolving the contribution of sub-threshold point sources to the isotropic background~\cite{Zechlin:2016pme,Lisanti:2016jub}, or by  
looking at the auto-correlation spectrum~\cite{Ackermann:2012uf, Ackermann:2012uf,Ando:2006cr,Ando:2013ff}.  A separate approach involves cross-correlating~\cite{Xia:2011ax,Ando:2014aoa,Ando:2013xwa,Xia:2015wka,Regis:2015zka,Cuoco:2015rfa,Ando:2016ang} the {\it Fermi} data with galaxy-count maps constructed from, \emph{e.g.}, the Two Micron All-Sky Survey (2MASS)~\cite{Jarrett:2000me,Bilicki:2013sza}.  A positive cross-correlation was detected with 2MASS galaxy counts~\cite{Xia:2015wka}, which could arise from annihilating DM with mass $\sim$$10$--$100$~GeV and a near-thermal annihilation rate~\cite{Regis:2015zka}.  However, other source classes, such as misaligned Active Galactic Nuclei, could also explain the signal~\cite{Cuoco:2015rfa}.
  
An alternative to studying the full-sky imprint of extragalactic DM annihilation is to use individual galaxy clusters~\cite{Ackermann:2010rg, Ando:2012vu,Ackermann:2013iaq,Ackermann:2015fdi,Anderson:2015dpc,Rephaeli:2015nca,Ahnen:2016qkt,Liang:2016pvm,Adams:2016alz,Huang:2011xr}. Previous analyses along these lines have looked at  
  a small number of $\sim$$10^{14}$--$10^{15}$~M$_\odot$ X-ray--selected clusters. 
  Like the dSph searches, the cluster studies have the advantage that the expected signal is localized in the sky, which reduces the systematic uncertainties associated with modeling the foregrounds and unresolved extragalactic sources.  As we will show, however, the sensitivity to DM annihilation is enhanced---and is more robust---when a larger number of targets are included compared to previous studies.

 Our work aims to combine the best attributes of the cross-correlation and cluster studies to improve the search for extragalactic DM annihilation.  We use the galaxy group catalogs in Refs.~\cite{Tully:2015opa} and~\cite{2017ApJ...843...16K} (hereby T15 and T17, respectively), which contain accurate mass estimates for halos with mass greater than $\sim$$10^{12}$~M$_\odot$ and $z \lesssim 0.03$, to systematically determine the galaxy groups that are expected to yield the best limits on the annihilation rate.  The T15 catalog provides reliable redshift estimates in the range $0.01 \lesssim z \lesssim 0.03$, while the T17 catalog provides measured distances for nearby galaxies, $z \lesssim 0.01$, based on Ref.~\cite{Tully:2016ppz}. The T15 catalog was previously used for a gamma-ray line search~\cite{Adams:2016alz}, but our focus here is on the broader, and more challenging, class of continuum signatures.  We search for gamma-ray flux from these galaxy groups and interpret the null results as bounds on the annihilation cross section.   

\newpage
\section{Galaxy Group Selection} 
 
The observed gamma-ray flux from DM annihilation in an extragalactic halo is proportional to both the particle physics properties of the DM, as well as its astrophysical distribution:
\es{particle}{
\frac{d\Phi}{dE_{\gamma}} &= \left.  J \, \times \frac{\langle\sigma v\rangle}{8 \pi m_{\chi}^{2}} \, \, \sum_i \text{Br}_{i}\, \frac{dN_{i}}{dE'_{\gamma}} \right|_{E_{\gamma}' = (1 +z) E_{\gamma}}   \,,
}
with units of $[{\rm counts} \,\,{\rm cm}^{-2} \, {\rm s}^{-1} \, {\rm GeV}^{-1}]$.  Here, $E_\gamma$ is the gamma-ray energy, $\langle \sigma v \rangle$ is the annihilation cross section, $m_\chi$ is the DM mass, $\text{Br}_{i}$ is the branching fraction to the $i^\text{th}$ annihilation channel, and $z$ is the cosmological redshift.  The energy spectrum for each channel is described the function $dN_{i}/dE_{\gamma}$, which is modeled using PPPC4DMID~\cite{Cirelli:2010xx}.  The $J$-factor that appears in~Eq.~\ref{particle} encodes the astrophysical properties of the halo.  It is proportional to the line-of-sight integral of the squared DM density distribution, $\rho_\text{DM}$, and is written in full as 
\begin{equation}
J = \left(1+b_\text{sh}[M_\text{vir}] \right)  \int ds\,d \Omega \,\rho^{2}_\text{DM}(s,\Omega) \, ,
\label{eq:Jfactor}
\end{equation}
where $b_\text{sh}[M_\text{vir}]$ is the boost factor, which accounts for the enhancement due to substructure.  For an extragalactic halo, where the angular diameter distance $d_A[z]$ is much greater than the virial radius $r_\text{vir}$, the integral in Eq.~\ref{eq:Jfactor} scales as $M_{\rm vir} c_{\rm vir}^3\rho_c/d_A^2[z]$ for the Navarro-Frenk-White (NFW) density profile~\cite{Navarro:1996gj}.  Here, $M_\text{vir}$ is the virial mass, $\rho_c$ is the critical density, and $c_\text{vir}=r_\text{vir}/r_s$ is the virial concentration, with $r_s$ the scale radius.  We infer $c_\text{vir}$ using the concentration-mass relation from Ref.~\cite{Correa:2015dva}, which we update with the Planck 2015 cosmology~\cite{Ade:2015xua}.  
For a given mass and redshift, the concentration is modeled as a log-normal distribution with mean given by the concentration-mass relation.  We estimate the dispersion by matching to that observed in the \texttt{DarkSky-400} simulation for an equivalent $M_\text{vir}$~\cite{Lehmann:2015ioa}.  Typical dispersions range from $\sim$$0.14$--$0.19$ over the halo masses considered. 

The halo mass and redshift also determine the boost factor enhancement that arises from annihilation in DM substructure.  Accurately modeling the boost factor is challenging as it involves extrapolating the halo-mass function and concentration to masses smaller than can be resolved with current simulations.  Some previous analyses of extragalactic DM annihilation have estimated boost factors $\sim$$10^2$--$10^3$ for cluster-size halos (see, for example, Ref.~\cite{Gao:2011rf}) based on phenomenological extrapolations of the subhalo mass and concentration relations.  However, more recent studies indicate that the concentration-mass relation likely flattens at low masses~\cite{Anderhalden:2013wd,Ludlow:2013vxa,Correa:2015dva}, suppressing the enhancement. We use the model of Ref.~\cite{Bartels:2015uba}---specifically, the ``self-consistent'' model with $M_\text{min} = 10^{-6}$~M$_\odot$---which accounts for tidal stripping of bound subhalos and yields a modest boost $\sim$$5$ for $\sim$$10^{15}$~M$_\odot$ halos. Additionally, we model the boost factor as a multiplicative enhancement to the rate in our main analysis, though we consider the effect of possible spatial extension from the subhalo annihilation in the Supplementary Material. In particular, we find that modeling the boost component of the signal as tracing a subhalo population distributed as $\rho_\text{NFW}$ rather than $\rho^{2}_\text{NFW}$ degrades the upper limits obtained by almost an order of magnitude at higher masses $m_\chi \gtrsim 500$ GeV while strengthening the limit by a small $\mathcal O(1)$ factor at lower masses $m_\chi \lesssim 200$ GeV.

The halo masses and redshifts are taken from the galaxy group catalog T15~\cite{Tully:2015opa}, which is based on the 2MASS Redshift Survey (2MRS)~\cite{Crook:2006sw}, and T17~\cite{2017ApJ...843...16K}, which compiles an inventory of nearby galaxies and distances from several sources.  The catalogs provide group associations for these galaxies as well as mass estimates and uncertainties of the host halos, constructed from a luminosity-to-mass relation. The mass distribution is assumed to follow a log-normal distribution with uncertainty fixed at 1\% in log-space \cite{Lisanti:2017qoz}, which translates to typical absolute uncertainties of 25-40\%.\footnote{To translate, approximately, between log- and linear-space uncertainties for the mass, we may write $x = \log_{10} M_\text{vir}$, which implies that the linear-space fractional uncertainties are $\delta M_\text{vir} / M_\text{vir} \sim (\delta x / x) \log M_\text{vir}$. } This is conservative compared to the 20\% uncertainty estimate given in T15 due to their inference procedure. The halo centers are assumed to coincide with the locations of the brightest galaxy in the group.  We infer the $J$-factor using Eq.~\ref{eq:Jfactor} and calculate its uncertainty by propagating the errors on $M_\text{vir}$ and $c_\text{vir}$, which we take to be uncorrelated.  Note that we neglect the distance uncertainties, which are expected to be $\sim$5\%~\cite{Tully:2016ppz,2017ApJ...843...16K}, as they are subdominant compared to the uncertainties on mass and concentration.  We compile an initial list of nearby targets using the T17 catalog, supplementing these with the T15 catalog.  We exclude from T15 all groups with Local Sheet velocity $V_\text{LS} < 3000$~km s$^{-1}$ ($z \lesssim 0.01$) and $V_\text{LS} > 10,000$~km s$^{-1}$ ($z \gtrsim 0.03$), the former because of peculiar velocity contamination and the latter because of large uncertainties in halo mass estimation due to less luminous satellites.  When groups overlap between the two catalogs, we preferentially choose distance and mass measurements from T17.

\begin{table}[t!]
\footnotesize
\begin{center}
\begin{tabular}{cccccccccc}
\toprule
\Xhline{3\arrayrulewidth}
Name &   $\log_{10} J$  &  $\log_{10} M_\text{vir}$ &          $z \times 10^{3}$&        $\ell$ &        $b$ &  $\log_{10} c_\text{vir}$ &  $\theta_\text{s}$  &  $b_\text{sh}$   \\
 & {[GeV$^2$\,cm$^{-5}$\,sr]}& [$M_\odot$] &  & [deg] & [deg] & & [deg] &\\
\midrule
\hline
            NGC4472  &  19.11$\pm$0.35 &  14.6$\pm$0.14 &   3.58 &  283.94 &  74.52 &  0.80$\pm$0.18 &     1.16 &  4.53 \\
                  NGC0253 &  18.76$\pm$0.37 &  12.7$\pm$0.12 &   0.79 &   98.24 & -87.89 &  1.00$\pm$0.17 &     0.77 &  2.90 \\
                  NGC3031 &  18.58$\pm$0.36 &  12.6$\pm$0.12 &   0.83 &  141.88 &  40.87 &  1.02$\pm$0.17 &     0.64 &  2.76 \\
        NGC4696 &  18.34$\pm$0.35 &  14.6$\pm$0.14 &   8.44 &  302.22 &  21.65 &  0.80$\pm$0.18 &     0.48 &  4.50 \\
                  NGC1399 &  18.31$\pm$0.37 &  13.8$\pm$0.13 &   4.11 &  236.62 & -53.88 &  0.89$\pm$0.17 &     0.45 &  3.87 \\
\bottomrule
\Xhline{3\arrayrulewidth}
\end{tabular}
\end{center}
\caption{The top five halos included in the analysis, as ranked by inferred $J$-factor, including the boost factor.  For each group, we show the brightest central galaxy, as well as the virial mass, cosmological redshift, Galactic longitude $\ell$, Galactic latitude $b$, inferred virial concentration~\cite{Correa:2015dva}, angular extent, and boost factor~\cite{Bartels:2015uba}.  The angular extent is defined as $\theta_\text{s} \equiv \tan^{-1} (r_\text{s} / d_A[z])$, where $d_A[z]$ is the angular diameter distance and $r_\text{s}$ is the NFW scale radius.  Common names for NGC4472 and NGC4696 are Virgo and Centaurus, respectively. A complete table of the galaxy groups used in this analysis, as well as their associated properties, are provided at \url{https://github.com/bsafdi/DMCat}.
}
\label{Jtab}
\end{table}

The galaxy groups are ranked by their inferred $J$-factors, excluding any groups that lie within $|b| \leq 20^\circ$ to mitigate contamination from Galactic diffuse emission.  We require that halos do not overlap to within $2^\circ$ of each other, which is approximately the scale radius of the largest halos.  The exclusion procedure is applied sequentially starting with a halo list ranked by $J$-factor.  We manually exclude Andromeda, the brightest halo in the catalog, because its large angular size is not ideally suited to our analysis pipeline and requires careful individual study~\cite{Ackermann:2017nya}.  
As discussed later in this chapter, halos are also excluded if they show large residuals that are inconsistent with DM annihilation in the other groups in the sample.  Starting with the top 1000 halos, we end up with 495 halos that pass all these requirements.  Of the excluded halos, 276 are removed because they fall too close to the Galactic plane, 134 are removed by the $2^\circ$ proximity requirement, and 95 are removed because of the cut on large residuals. 

Table~\ref{Jtab} lists the top five galaxy groups included in the analysis, labeled by their central galaxy or common name, if one exists.  We provide the inferred $J$-factor including the boost factor, the halo mass, redshift, position in Galactic coordinates, inferred concentration, and boost factor.  Additionally, we show $\theta_\text{s} \equiv \tan^{-1} (r_\text{s} / d_A[z])$ to indicate the spatial extension of the halo.  We find that $\theta_\text{s}$ is typically between the 68\% and 95\% containment radius for emission associated with annihilation in the halos, without accounting for spread from the point-spread function (PSF).  For reference, Andromeda has $\theta_\text{s} \sim 2.57^\circ$.  

\section{Data Analaysis}

We analyze 413 weeks of Pass 8 {\it Fermi} data in the UltracleanVeto event class, from August 4, 2008 through July 7, 2016.  The data is binned in 26 logarithmically-spaced energy bins between 502~MeV and 251~GeV and spatially with a HEALPix pixelation~\cite{Gorski:2004by} with \texttt{nside}=128.\footnote{Our energy binning is constructed by taking 40 log-spaced bins between 200~MeV and 2~TeV and then removing the lowest four and highest ten bins, for reasons discussed in the companion paper~\cite{Lisanti:2017qoz}. }  The recommended set of quality cuts are applied to the data corresponding to zenith angle less than $90^\circ$, $\texttt{LAT\_CONFIG}=1$, and $\texttt{DATA\_QUAL}>0$.\footnote{\url{https://fermi.gsfc.nasa.gov/ssc/data/analysis/documentation/Cicerone/Cicerone_Data_Exploration/Data_preparation.html}.}  We also mask known large-scale structures~\cite{Lisanti:2017qoz}.

The template analysis that we perform using \texttt{NPTFit}~\cite{Mishra-Sharma:2016gis} is similar to that of previous dSph studies~\cite{Ackermann:2015zua,Fermi-LAT:2016uux}  and is detailed in our companion paper~\cite{Lisanti:2017qoz}.  We summarize the relevant points here.  Each region-of-interest (ROI), defined as the $10^\circ$ area surrounding each halo center, has its own likelihood.  In each energy bin, this  likelihood is the product, over all pixels, of the Poisson probability for the observed photon counts per pixel.  This probability depends on the mean expected counts per pixel, which depends on contributions from known astrophysical emission as well as a potential DM signal. Note that the likelihood is also multiplied by the appropriate log-normal distribution for $J$, which we treat as a single nuisance parameter for each halo and account for through the profile likelihood method.  

To model the expected counts per pixel, we include several templates in the analysis that trace the emission associated with: (i) the projected NFW-squared profile modeling the putative DM signal, (ii) the diffuse background, as described by the  {\it Fermi} \texttt{gll\_iem\_v06 (p8r2)} model, (iii) isotropic emission, (iv) the {\it Fermi} bubbles~\cite{Su:2010qj}, (v) 3FGL sources within $10^\circ$ to $18^\circ$ of the halo center, floated together after fixing their individual fluxes to the values predicted by the 3FGL catalog~\cite{Acero:2015hja}, and (vi) all individual 3FGL point sources within $10^{\circ}$ of the halo center.  Note that we do not model the contributions from annihilation in the smooth Milky Way halo because the brightest groups have peak flux significantly (approximately an order of magnitude for the groups in Tab.~\ref{Jtab}) over the foreground emission from Galactic annihilation and because we expect Galactic annihilation to be subsumed by the isotropic component.   

We assume that the best-fit normalizations (\emph{i.e.}, profiled values) of the astrophysical components, which we treat as nuisance parameters, do not vary appreciably with DM template normalization. This allows us to obtain the likelihood profile in a given ROI and energy bin by profiling over them in the presence of the DM template, then fixing the normalizations of the background components to the best-fit values and scanning over the DM intensity. We then obtain the total likelihood by taking the product of the individual likelihoods from each energy bin. In order to avoid degeneracies at low energies due to the large PSF, we only include the DM template when obtaining the best-fit background normalizations at energies above $\sim$$1$~GeV. At the end of this procedure, the likelihood is only a function of the DM template intensity, which can then be mapped onto a mass and cross section for a given annihilation channel. We emphasize that the assumptions described above have been thoroughly vetted in our companion paper~\cite{Lisanti:2017qoz}, where we show that this procedure is robust in the presence of a potential signal.

The final step of the analysis involves stacking the likelihoods from each ROI. The stacked log-likelihood, $\log \mathcal{L}$, is simply the sum of the log-likelihoods for each ROI.  It follows that the test statistic for data $d$ is defined as
 \begin{equation}\begin{aligned}
{\rm TS}(\mathcal{M}, \langle\sigma v\rangle, m_\chi) \equiv 2 &\left[ \log \mathcal{L}(d | \mathcal{M}, \langle\sigma v\rangle, m_\chi ) \right.\\
&\left.- \log \mathcal{L}(d | \mathcal{M}, \widehat{\langle\sigma v\rangle}, m_\chi ) \right]\,,
\label{eq:TSdef}
\end{aligned}\end{equation}
where $\widehat{\langle\sigma v\rangle}$ is the cross section that maximizes the likelihood for DM model $\mathcal{M}$.    The 95\% upper limit on the annihilation cross section is given by the value of $\langle\sigma v\rangle > \widehat{\langle \sigma v\rangle}$ where $\text{TS}=-2.71$.

Galaxy groups are expected to emit gamma-rays from standard cosmic-ray processes.  Using group catalogs to study gamma-ray emission from cosmic rays in these objects is an interesting study in its own right (see, {\it e.g.}, Ref.~\cite{Jeltema:2008vu,Huber:2013cia,Ackermann:2015fdi,Rephaeli:2015nca}), which we leave to future work.  For the purpose of the present analysis, however, we would like a way to remove groups with large residuals, likely arising from standard astrophysical processes in the clusters, to maintain maximum sensitivity to DM annihilation.  This requires care, however, as we must guarantee  that the procedure for removing halos does not remove a real signal, if one were present.  

We adopt the following algorithm to remove halos with large residuals that are inconsistent with DM annihilation in the other groups in the sample. A group is excluded if it meets two conditions. First, to ensure it is a statistically significant excess, we require twice the difference between the maximum log likelihood and the log likelihood with $\langle \sigma v \rangle = 0$ to be greater than 9 at any DM mass. This selects sources with large residuals at a given DM mass.  Second, the residuals must be strongly inconsistent with limits set by other galaxy groups. Specifically, the halo  must satisfy $\langle\sigma v\rangle_\text{best} > 10 \times \langle\sigma v\rangle^*_\text{lim}$, where $\langle\sigma v\rangle_\text{best}$ is the halo's best-fit cross section at \emph{any} mass and $\langle\sigma v\rangle^*_\text{lim}$ is the strongest limit out of all halos at the specified $m_\chi$. These conditions are designed to exclude galaxy groups where the gamma-ray emission is inconsistent with a DM origin.  This prescription has been extensively tested on mock data and, crucially, does not exclude injected signals~\cite{Lisanti:2017qoz}.

\newpage
\section{Results}

\begin{figure}[t!]
\centering
\includegraphics[width=0.45\textwidth]{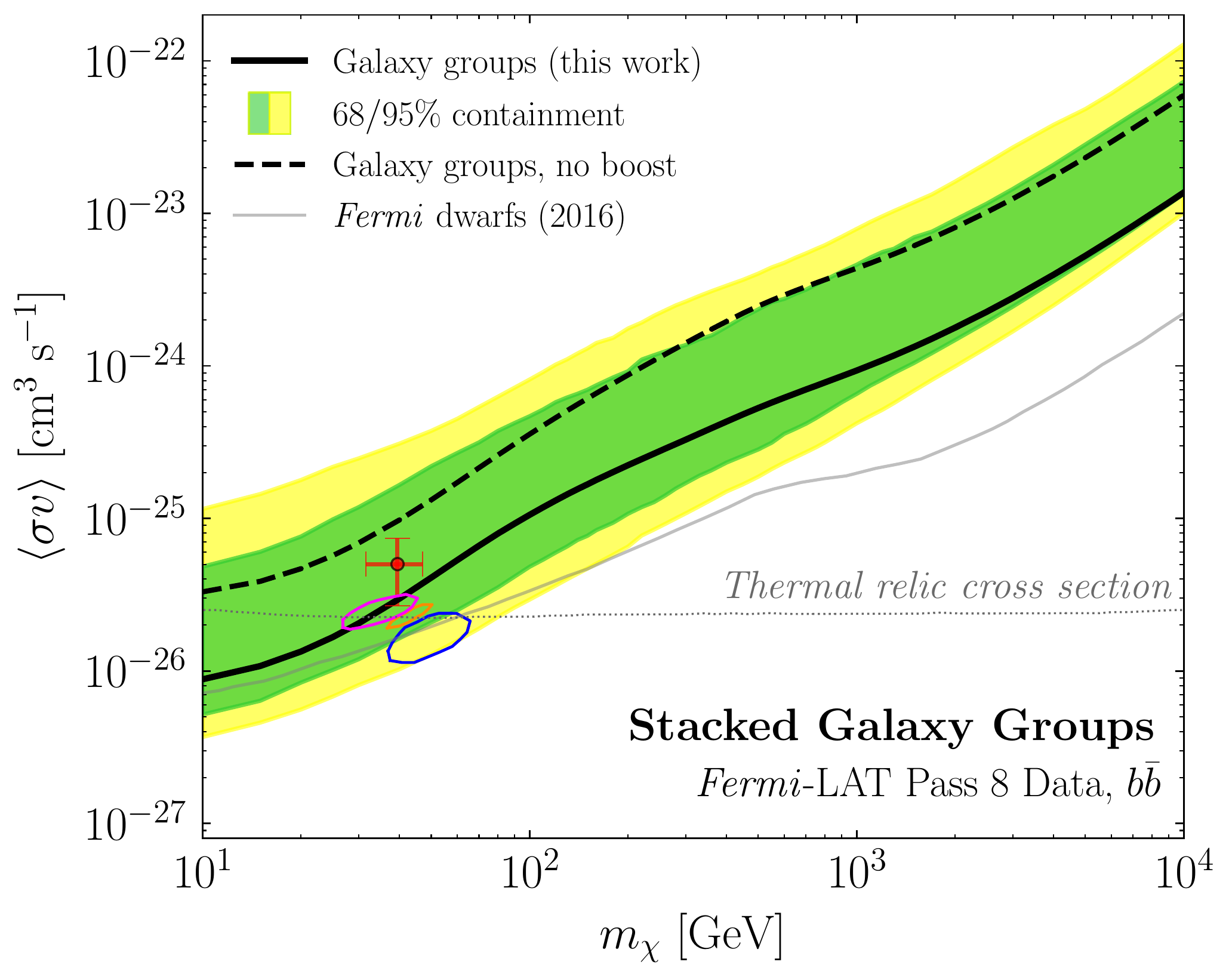} \hspace{4mm}
\includegraphics[width=0.45\textwidth]{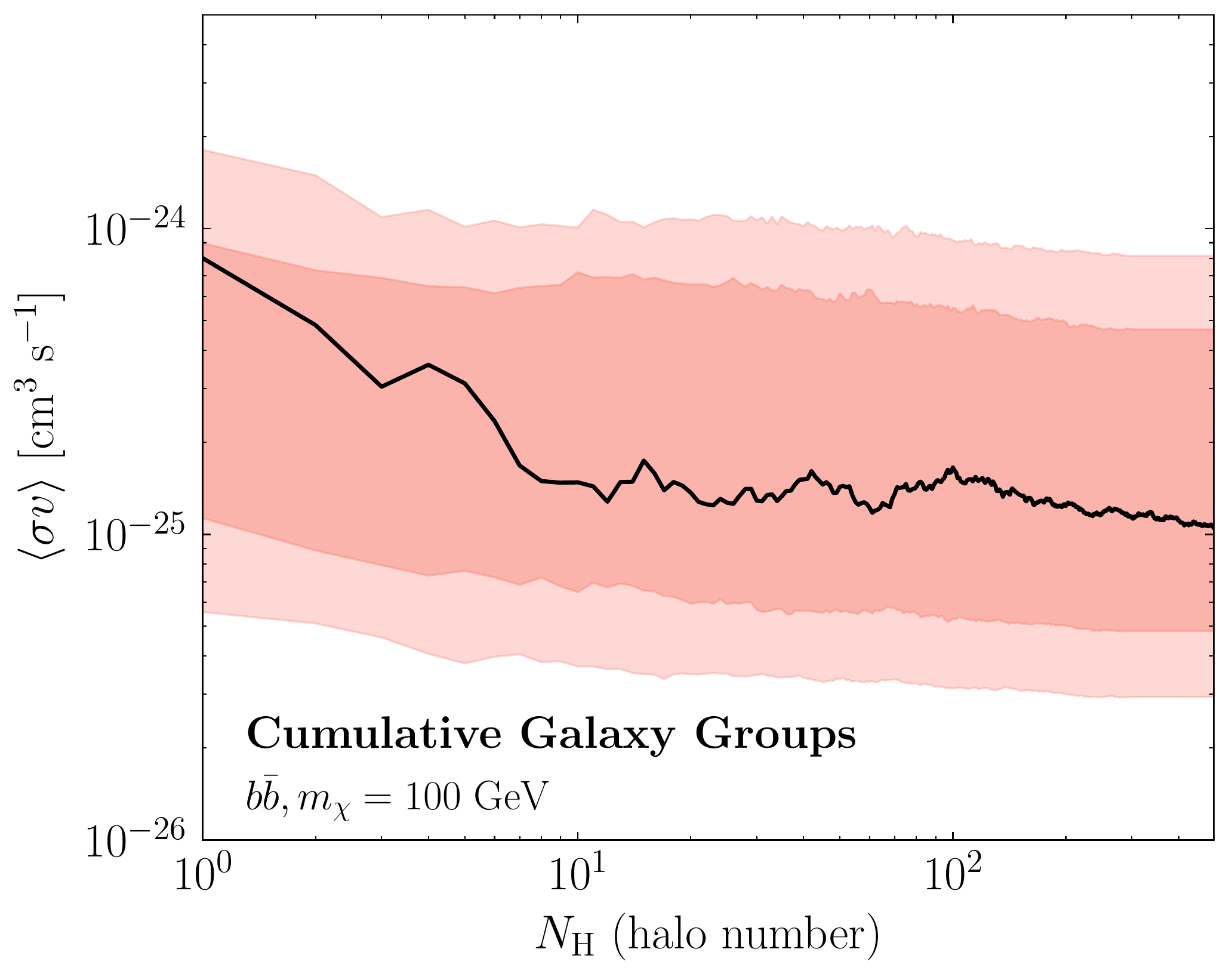}
\caption{(Left) The solid black line shows the 95\% confidence limit on the DM annihilation cross section, $\langle \sigma v \rangle$, as a function of the DM mass, $m_\chi$, for the $b \bar b$ final state, assuming the fiducial boost factor~\cite{Bartels:2015uba}. The containment regions are computed by performing the data analysis multiple times for random sky locations of the halos.  For comparison, the dashed black line shows the limit assuming no boost factor.  The {\it Fermi} dwarf limit is also shown, as well as the $2$$\sigma$ regions where DM may contribute to the Galactic Center Excess (see text for details).  The thermal relic cross section for a generic weakly interacting massive particle~\cite{Steigman:2012nb} is indicated by the thin dotted line.   (Right) The change in the limit for $m_\chi = 100$~GeV as a function of the number of halos that are included in the analysis, which are ranked in order of largest $J$-factor.  The result is compared to the expectation from random sky locations; the 68 and 95\% expectations from 200 random sky locations are indicated by the red bands.}
\label{fig:bounds}
\end{figure}

The left panel of Fig.~\ref{fig:bounds} illustrates the main results of the stacked analysis.  The solid black line represents the limit obtained for DM annihilating to a $b \bar b$ final state using the fiducial boost factor model~\cite{Bartels:2015uba}, while the dashed line  shows the limit without the boost factor enhancement.  To estimate the expected limit under the null hypothesis, we repeat the analysis by randomizing the locations of the halos on the sky 200 times, though still requiring they pass the selection cuts described above.  
The colored bands indicate the 68 and 95\% containment regions for the expected limit.  
The limit is consistent with the expectation under the null hypothesis.

The right panel of Fig.~\ref{fig:bounds} illustrates how the limits evolve for the $b \bar b$ final state with $m_\chi = 100$~GeV as an increasing number of halos are stacked.  We also show the expected 68\% and 95\% containment regions, which are obtained from the random sky locations.  As can be seen, no single halo dominates the bounds.  For example, removing Virgo, the brightest halo in the catalog, from the stacking has no significant effect on the limit.  Indeed, the inclusion of all 495 halos buys one an additional order of magnitude in the sensitivity reach.

The limit derived in this work is complementary to the published dSph bound~\cite{Ackermann:2015zua,Fermi-LAT:2016uux}, shown as the solid gray line in the left panel of Fig.~\ref{fig:bounds}. Given the large systematic uncertainties associated with the dwarf analyses (see~\emph{e.g.}, Ref.~\cite{Geringer-Sameth:2014qqa}), we stress the importance of using complementary targets and detection strategies to probe the same region of parameter space. Our limit also probes the parameter space that may explain the Galactic Center excess (GCE); the best-fit models are marked by the orange cross~\cite{Abazajian:2014fta}, blue~\cite{Gordon:2013vta}, red~\cite{Daylan:2014rsa}, and orange~\cite{Calore:2014xka} $2$$\sigma$ regions.  The GCE is a spherically symmetric excess of $\sim$GeV gamma-rays observed to arise from the center of the Milky Way~\cite{Goodenough:2009gk,Hooper:2010mq,TheFermi-LAT:2015kwa,Karwin:2016tsw}.  The GCE has received a considerable amount of attention because it can be explained by annihilating DM.  However, it can also be explained by more standard astrophysical sources; indeed, recent analyses have shown that the distribution of photons in this region of sky is more consistent with a population of unresolved point sources, such as millisecond pulsars, compared to smooth emission from DM~\cite{Lee:2015fea, Bartels:2015aea,Linden:2016rcf, Fermi-LAT:2017yoi}.  Because systematic uncertainties can be significant and hard to quantify in indirect searches for DM, it is crucial to have independent probes of the parameter space where DM can explain the GCE.  While our null findings do not exclude the DM interpretation of the GCE, their consistency with the dwarf bounds put it further in tension.  This does not, however, account for the fact that the systematics on the modeling of the Milky Way's density distribution can potentially alleviate the tension by changing the best-fit cross section for the GCE.   

\section{Conclusions}

This chapter presents the results of the first systematic search for annihilating DM in nearby galaxy groups.  We introduced and validated  a prescription to infer properties of DM halos associated with  these groups, thereby allowing us to build a map of DM annihilation in the local universe.  Using this map, we performed a stacked analysis of several hundred galaxy groups and obtained bounds that exclude thermal cross sections for DM  annihilating to $b \bar b$ with mass below $\sim$$30$~GeV, assuming a conservative boost factor model.  These limits are competitive with those obtained from the \emph{Fermi} dSph analyses and are in tension with the range of parameter space that can explain the GCE.  Moving forward, we plan to investigate the objects with gamma-ray excesses to see if they can be interpreted in the context of astrophysical emission.  In so doing, we can also develop more refined metrics for selecting the optimal galaxy groups for DM studies.    

%% file: dmdecay.tex
\chapter{Gamma-ray Constraints on Decaying Dark Matter and Implications for IceCube}\label{chap:dmdecay}

\section{Introduction}

\begin{figure}[t]
	\leavevmode
	\vspace{-.30cm}
	\begin{center}
        \includegraphics[width = 0.6 \columnwidth]{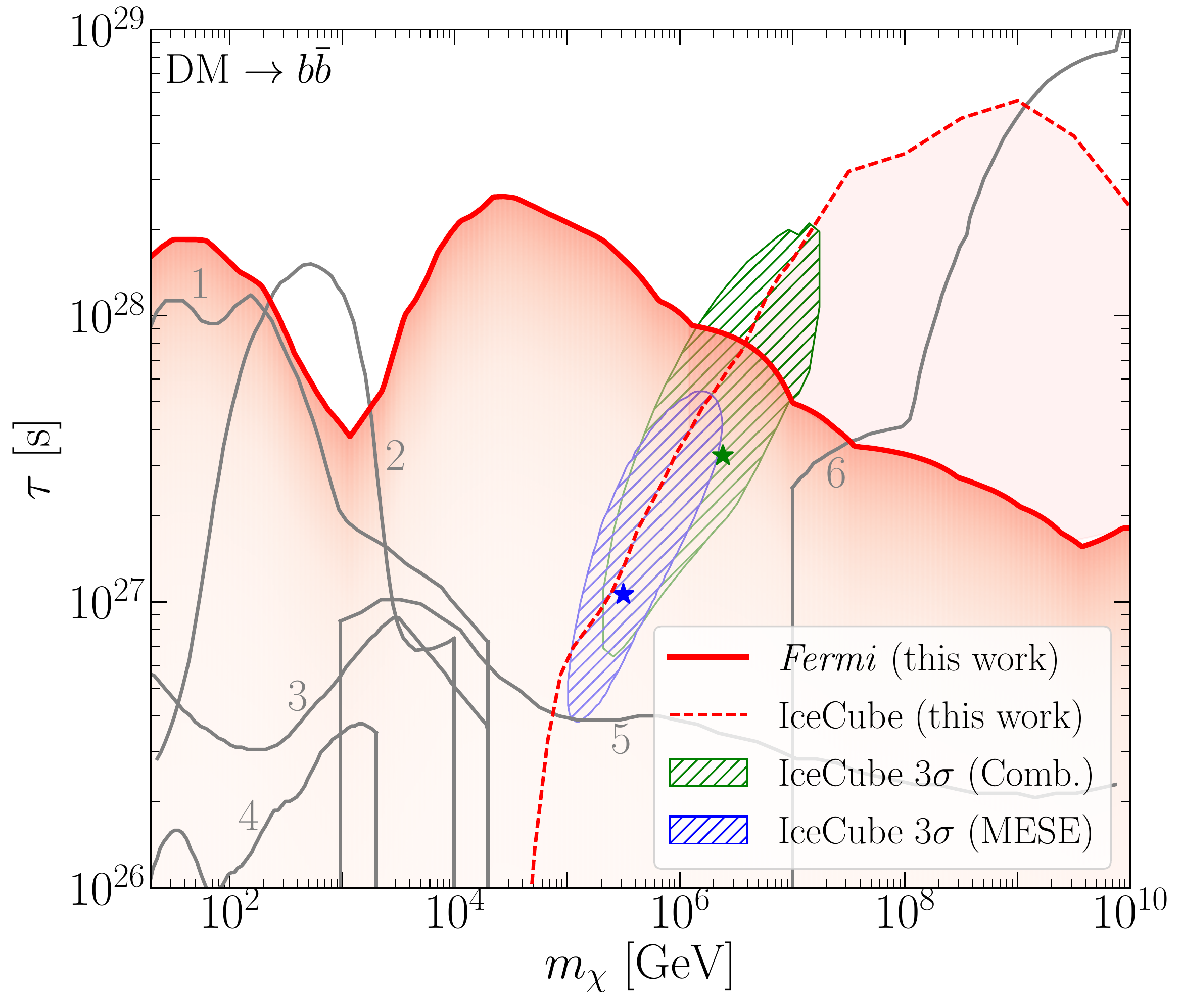}
	\end{center}
	\vspace{-.50cm}
	\caption{Limits 
	on DM decays to $b\, \bar{b}$, as compared to previously computed limits using data from {\it Fermi} (2,3,5), AMS-02 (1,4), and PAO/KASCADE/CASA-MIA  (6). 
	The hashed green (blue) region suggests parameter space where DM decay may provide a $\sim$$3 \sigma$ 
	improvement to the description of the combined maximum likelihood (MESE) IceCube neutrino flux.  The best-fit points, marked as stars, are in strong tension with our gamma-ray results.  The red dotted line provides a limit if we assume a combination of DM decay and astrophysical sources are responsible for the spectrum. 
	}
	\vspace{-0.15in}
	\label{Fig: bbResult}
\end{figure} 

A primary goal of the particle physics program is to discover the connection between dark matter~(DM) and the Standard Model~(SM).  While the DM is known to be stable over cosmological timescales, rare DM decays may give rise to observable signals in the spectrum of high-energy cosmic rays.  Such decays would be induced through operators involving both the dark sector and the SM.  In this chapter, we derive some of the strongest constraints to date on decaying DM for masses from $\sim$400 MeV to $\sim$$10^{7}$\,GeV by performing a dedicated analysis of \emph{Fermi} gamma-ray data from 200\,MeV to 2\,TeV.

The solid red line in Fig.~\ref{Fig: bbResult} gives an example of our constraint on the DM~($\chi$) lifetime, $\tau$, as a function of its mass, $m_\chi$, assuming the DM decays exclusively to a pair of bottom quarks. Our analysis includes three contributions to the photon spectrum:  (1) prompt emission, (2) gamma-rays that are up-scattered by primary electrons/positrons through inverse Compton (IC) within the Galaxy, and (3) extragalactic contributions. 

In addition to deriving some of the strongest limits on the DM lifetime across many DM decay channels, our results provide the first dedicated constraints on DM using the latest \emph{Fermi} data for $m_\chi \gtrsim 10$\,TeV.  To emphasize this point, we provide a comparison with other limits in Fig.~\ref{Fig: bbResult}.   The dashed red curve indicates our new estimate of the limits set by high-energy neutrino observations at the IceCube experiment~\cite{Aartsen:2013bka,Aartsen:2013jdh,Aartsen:2015knd,Aartsen:2015rwa}. Our IceCube constraint dominates in the range from $\sim$$10^7$ to $10^9$\,GeV.  

Constraints from previous studies are plotted as solid grey lines labeled from 1-6.  Curve 6 shows that for masses above $\sim$$10^9$~GeV, limits from null observations of ultra-high-energy gamma-rays at air shower experiments~\cite{Kalashev:2016cre}, such as the Pierre Auger Observatory (PAO)~\cite{Aab:2015bza}, KASCADE~\cite{Kang:2015gpa}, and CASA-MIA~\cite{Chantell:1997gs}, surpass our IceCube limits. Curves 2, 5, and 3 are from previous analyses of the extragalactic~\cite{Ando:2015qda,Murase:2012xs} and Galactic~\cite{Ackermann:2012rg} {\it Fermi} gamma-ray flux  (for related work see~\cite{Hutsi:2010ai,Cirelli:2012ut,Kalashev:2016xmy}).  Our results are less sensitive to astrophysical modeling than~\cite{Ando:2015qda}, which makes assumptions about the classes of sources and their spectra that contribute to the unresolved component of the extragalactic gamma-ray background.  We improve and extend beyond~\cite{Ackermann:2012rg,Murase:2012xs} in a number of ways: by including state-of-the-art modeling for cosmic-ray-induced gamma-ray emission in the Milky Way, a larger and cleaner data set, and a novel analysis technique that allows us to search for a combination of Galactic and extragalactic flux arising from DM decay. The limits labeled 1 and 4 in Fig.~\ref{Fig: bbResult} are from the AMS-02 antiproton~\cite{ams02pos,Giesen:2015ufa} and positron~\cite{Aguilar:2013qda,Ibarra:2013zia} measurements, respectively; these constraints are subject to considerable astrophysical uncertainties, due to the propagation of charged cosmic rays from their source to Earth.

An additional motivation for this chapter is the measurement of the so far unexplained high-energy neutrinos observed by the IceCube experiment~\cite{Aartsen:2013bka,Aartsen:2013jdh,Aartsen:2015knd,Aartsen:2015rwa}. If the DM has both a mass \mbox{$m_\chi \sim$} PeV and a long lifetime \mbox{$\tau \sim$ $10^{28}$} seconds, its decays could contribute to the upper end of the IceCube spectrum. These DM candidates would produce correlated cosmic-ray signals, yielding a broad spectrum of gamma rays with energies extending well into \emph{Fermi}'s energy range. Taking this correlation between neutrino and photon spectra into account enables us to constrain the DM interpretation of these neutrinos using the \emph{Fermi} data.  

Figure~\ref{Fig: bbResult} illustrates regions of parameter space where we fit a decaying DM spectrum to the high-energy neutrino flux at IceCube in hashed green. The corresponding region for the analysis of Ref.~\cite{Chianese:2016kpu} using lower-energy neutrinos is shown in blue. Clearly, much of the parameter space relevant for IceCube is disfavored by the gamma-ray limits; the best fit points (indicated by stars) are in strong tension with the {\it Fermi} observations.  We conclude that models where decaying DM could account for the entire astrophysical neutrino flux observed by IceCube are disfavored.  Furthermore, models where the neutrino flux results from a mix of decaying DM and astrophysical sources are strongly constrained.  

The rest of this chapter is organized as follows.  First, we discuss the various contributions to the gamma-ray flux resulting from DM decay.  Then, we give an overview of the data set and analysis techniques used in this chapter.  Next, we provide context for these limits by interpreting them as constraints on a concrete model (glueball DM), before concluding.

\section{The Gamma-ray flux}\label{sec:CalcGammaFlux}

Decaying DM contributes both a Galactic and extragalactic flux.  The Galactic contribution results primarily from prompt gamma-ray emission due to the decay itself, which is simulated with \textsc{Pythia} 8.219~\cite{Sjostrand:2006za,Sjostrand:2007gs,Sjostrand:2014zea} including electroweak showering~\cite{Christiansen:2014kba} (see~\emph{e.g.}~\cite{Kachelriess:2007aj,Regis:2008ij,Mack:2008wu,Bell:2008ey,Dent:2008qy,Borriello:2008gy,Bertone:2008xr,Bell:2008vx,Cirelli:2009vg,Kachelriess:2009zy,Ciafaloni:2010ti}).  

These effects can be the only source of photons for channels such as $\chi \rightarrow \nu \bar{\nu}$.

  In addition, the electrons and positrons from these decays IC scatter off of cosmic background radiation (CBR), producing gamma-rays (see \emph{e.g.}~\cite{Murase:2015gea,Esmaili:2015xpa}).  The prompt contribution follows the spatial morphology obtained from the line-of-sight (LOS) integral of the DM density, which we model with a Navarro-Frenk-White~(NFW) profile~\cite{Navarro:1995iw,Navarro:1996gj}, setting the local DM density \mbox{$\rho = 0.3~\text{GeV}/\text{cm}^3$}, and the scale radius \mbox{$r_s = 20$ kpc} (variations to the profile lead to similar results, see the associated appendix).  We only consider IC scattering off of the cosmic microwave background~(CMB), as scattering from integrated stellar radiation and the infrared background is expected to be sub-dominant, see the associated appendix.
 For scattering off of the CMB, the resulting gamma-ray morphology also follows the LOS integral of the DM density.  Importantly, as scattering off of the other radiation fields only increases the gamma-ray flux, neglecting these effects is conservative.  In the same spirit, we conservatively assume that the electrons and positrons lose energy due to synchrotron emission in a rather strong, uniform $B = 2.0$ $\mu$G magnetic field (see \emph{e.g.}~\cite{Mao:2012hx,Haverkorn:2014jka,Beck:2014pma}) and show variations in the associated appendix.

In addition to the Galactic fluxes, there is an essentially isotropic extragalactic contribution, arising from DM decays throughout the broader universe~\cite{Kribs:1996ac}. The extragalactic flux receives three important contributions: (1)~attenuated prompt emission; (2)~attenuated emission from IC of primary electrons and positrons; and~(3) emission from gamma-ray cascades. The cascade emission arises when an electron-positron pair is created by high-energy gamma rays scattering off of the CBR, inducing IC emission along with adiabatic energy loss.  We account for these effects following \cite{Murase:2012xs,Murase:2015gea}.

\section{Data Analysis}\label{sec:DataAnalysis}

We assess how well predicted Galactic~(NFW-correlated) and extragalactic~(isotropic) fluxes describe the data using the profile-likelihood method (see {\it e.g.}~\cite{Rolke:2004mj}), described in more detail in the associated appendix.
To this end, we perform a template fitting analysis (using \texttt{NPTFit}~\cite{Mishra-Sharma:2016gis}) with 413 weeks of {\it Fermi} Pass 8 data collected from August 4, 2008 to July 7, 2016.  We restrict to the UltracleanVeto event class; furthermore, we only use the top quartile of events as ranked by point-spread function (PSF).  The UltracleanVeto event class is used to minimize contamination from cosmic rays, while the PSF cut is imposed to mitigate effects from mis-modeling bright regions.   We bin the data in 40 logarithmically-spaced energy bins between $200$ MeV and $2$ TeV, and we apply the recommended quality cuts \texttt{DATA\_QUAL>0 \&\& LAT\_CONFIG==1} and zenith angle less than $90^\circ$.\footnote{See \url{http://fermi.gsfc.nasa.gov/ssc/data/analysis/documentation/Cicerone}.}  The data is binned spatially using a HEALPix~\cite{Gorski:2004by} pixelation with \texttt{nside}=128.

We constrain this data to a region of interest~(ROI) defined by Galactic latitude $|b| \geq 20^\circ$ within $45^\circ$ of the Galactic Center~(GC). The Galactic plane is masked in order to avoid issues related to mismodeling of diffuse emission in that region.  Similarly, we do not extend our region out further from the GC to avoid over-subtraction issues that may arise when fitting diffuse templates over large regions of the sky (see \emph{e.g.}~\cite{Daylan:2014rsa,Linden:2016rcf,Narayanan:2016nzy}). Finally we mask all point sources~(PSs) in the 3FGL PS catalog~\cite{Acero:2015hja} at their 95\% containment radius.

Using this restricted dataset, we then 
independently fit templates in each energy bin in order to construct a likelihood profile as a function of the extragalactic and Galactic flux. 
We separate our model parameters into those of interest  ${\bf \psi}$ and the nuisance parameters ${\bf \lambda}$.  The ${\bf \psi}$ include parameters for an isotropic template to account for the extragalactic emission, along with a template following a LOS-integrated NFW profile to model the Galactic emission.  Note that both the prompt and IC contribute to the same template, see the associated appendix for justification. The ${\bf \lambda}$ include parameters for the flux from diffuse emission within the Milky Way, flux from the {\it Fermi} bubbles, flux from isotropic emission that does not arise from DM decay (\emph{e.g.} emission from blazars and other extragalactic sources, along with misidentified cosmic rays), and flux from PSs, both Galactic and extragalactic, in the 3FGL PS catalog. Importantly, each spatial template is given a separate, uncorrelated degree of freedom in the northern and southern hemispheres, further alleviating over-subtraction. 

In our main analysis, we use the Pass 7 diffuse model \texttt{gal\_2yearp7v6\_v0} (\texttt{p7v6}) to account for diffuse emission in the Milky Way, coming from gas-correlated emission (mostly pion decay and bremsstrahlung from high-energy electrons), IC emission, and emission from large-scale structures such as the {\it Fermi} bubbles~\cite{Su:2010qj} and Loop~1~\cite{Casandjian:2009wq}.  Additionally, even though the {\it Fermi} bubbles are included to some extent in the \texttt{p7v6} model, we add an additional degree of freedom for the bubbles, following the uniform spatial template given in~\cite{Su:2010qj}.
  We add a single template for all 3FGL PSs based on the spectra in~\cite{Acero:2015hja}, though we emphasize again that all PSs are masked at 95\% containment.  See the associated appendix for variations of these choices.

Given the templates described above, we are able to construct 2-d log-likelihood profiles $\log p_i(d_i | \{I^i_\text{iso}, I^i_\text{NFW}\})$ as functions of the isotropic and NFW-correlated DM-induced emission $I^i_\text{iso}$ and  $I^i_\text{NFW}$, respectively, in each of the energy bins $i$.  Here, $d_i$ is the data in that energy bin, which simply consists of the number of counts in each pixel.  The likelihood profiles are given by maximizing the Poisson likelihood functions over the $\lambda$ parameters.

Any decaying DM model may be constrained from the set of likelihood profiles in each energy bin, which are provided as Supplementary Data~\cite{supp-data}. Concretely, given a DM model ${\cal M}$, the total log-likelihood $\log p(d | {\cal M},  \{\tau, m_\chi\})$ is simply the sum of the $\log p_i$, where the intensities in each energy bin are functions of the DM mass and lifetime.  The test-statistics (TS) used to constrain the model is twice the difference between the log-likelihood at a given $\tau$ and the value at $\tau = \infty$, where the DM contributes no flux.
The 95\% limit is given by $\text{TS} = -2.71$.

In order to compare our gamma-ray results to potential signals from IceCube, we determine the region of parameter space where DM may contribute to the observed high-energy neutrino flux.  We use the recent high-energy astrophysical neutrino spectrum measurement by the IceCube collaboration~\cite{Aartsen:2015knd}.  In that work, neutrino flux measurements from a combination of muon-track and shower data are given in 9 logarithmically-spaced energy bins between 10 TeV and 10 PeV, under the assumption of equal flavor ratios and an isotropic flux.\footnote{Constraints at high masses may be improved by incorporating recent results from~\cite{Aartsen:2016ngq}, which focused on neutrino events with energies greater than 10 PeV.}  We assume that DM decays are the only source of high-energy neutrino flux.  In Fig.~\ref{Fig: bbResult} (assuming the DM decays exclusively to $b \bar b$) we show the region where the DM model provides at least a 3$\sigma$ improvement over the null hypothesis of no high-energy flux at all.  The best-fit point is marked with a star.   The blue region in Fig.~\ref{Fig: bbResult} is the best-fit region~\cite{Chianese:2016kpu} for explaining an apparent excess in the 2-year medium energy starting event (MESE) IceCube data, which extends down to energies $\sim$1 TeV~\cite{Aartsen:2014muf}.   

The dashed red curve, on other other hand, shows the 95\% limit that we obtain on this DM channel under the assumption that astrophysical sources also contribute to the high-energy flux.  We parameterize the astrophysical flux by a power-law with an exponential cut-off, and we marginalize over the slope of the power-law, the normalization, and the cut-off in order to obtain a likelihood profile for the DM model, as a function of $\tau$ and $m_\chi$.  We emphasize that we allow the spectral index to float, as opposed to the analysis of \cite{Chianese:2016kpu}, which fixes the index equal to two.

\section{Interpretations}\label{sec:Interp}

In Fig.~\ref{Fig: bbResult}, we show our total constraint on the DM lifetime for a model where $\chi \to b\, \bar b$.  This result demonstrates tension in models where decaying DM explains or contributes to the astrophysical neutrino flux observed by IceCube.  PeV-scale decaying DM models have received attention recently (see \emph{e.g.}~\cite{Esmaili:2013gha,Feldstein:2013kka,Ema:2013nda,Zavala:2014dla,Bhattacharya:2014vwa,Higaki:2014dwa,Rott:2014kfa,Fong:2014bsa,Dudas:2014bca,Ema:2014ufa,Esmaili:2014rma,Murase:2015gea,Anchordoqui:2015lqa,Boucenna:2015tra,Ko:2015nma,Aisati:2015ova,Kistler:2015oae,Chianese:2016opp,Fiorentin:2016avj,Dev:2016qbd,DiBari:2016guw,Kalashev:2016cre,Chianese:2016smc}). 
In particular, while conventional astrophysical models such as those involving star-forming galaxies and galaxy clusters provide viable explanations for the neutrino data above 100\,TeV (see~\cite{Murase:2016gly} for a summary of recent ideas), the MESE data have been difficult to explain with conventional models~\cite{Murase:2015xka,Palladino:2016xsy}.
 Moreover, it is natural to expect heavy DM to slowly decay to the SM in a wide class of scenarios where, for example, the DM is stabilized through global symmetries in a hidden sector that are expected to be violated at the Planck scale or perhaps the scale of grand unification (the GUT scale).

From a purely data-driven point of view it is worthwhile to ask whether any set of SM final states may contribute significantly to or explain the IceCube data while being consistent with the gamma-ray constraints. 
In the associated appendix we provide limits on a variety of two-body SM final states.

 It is also important to interpret the bounds as constraints on the parameter space of UV models or gauge-invariant effective field theory~(EFT) realizations.  
If the decay is mediated by irrelevant operators, and given the long lifetimes we are probing, it is natural to assume very high cut-off scales $\Lambda$, such as the GUT scale $\sim$$10^{16} \text{\,GeV}$ or the Planck scale $m_\text{Pl} \simeq 2.4 \times 10^{18}$\,GeV.  We expect all gauge invariant operators connecting the dark sector to the SM to appear in the EFT suppressed by a scale $m_\text{Pl}$ or less (assuming no accidentally small coefficients and, perhaps, discrete global symmetries).  

\begin{figure}[t]
        \leavevmode
        \vspace{-.30cm}
        \begin{center}
        \includegraphics[width = 0.6 \columnwidth]{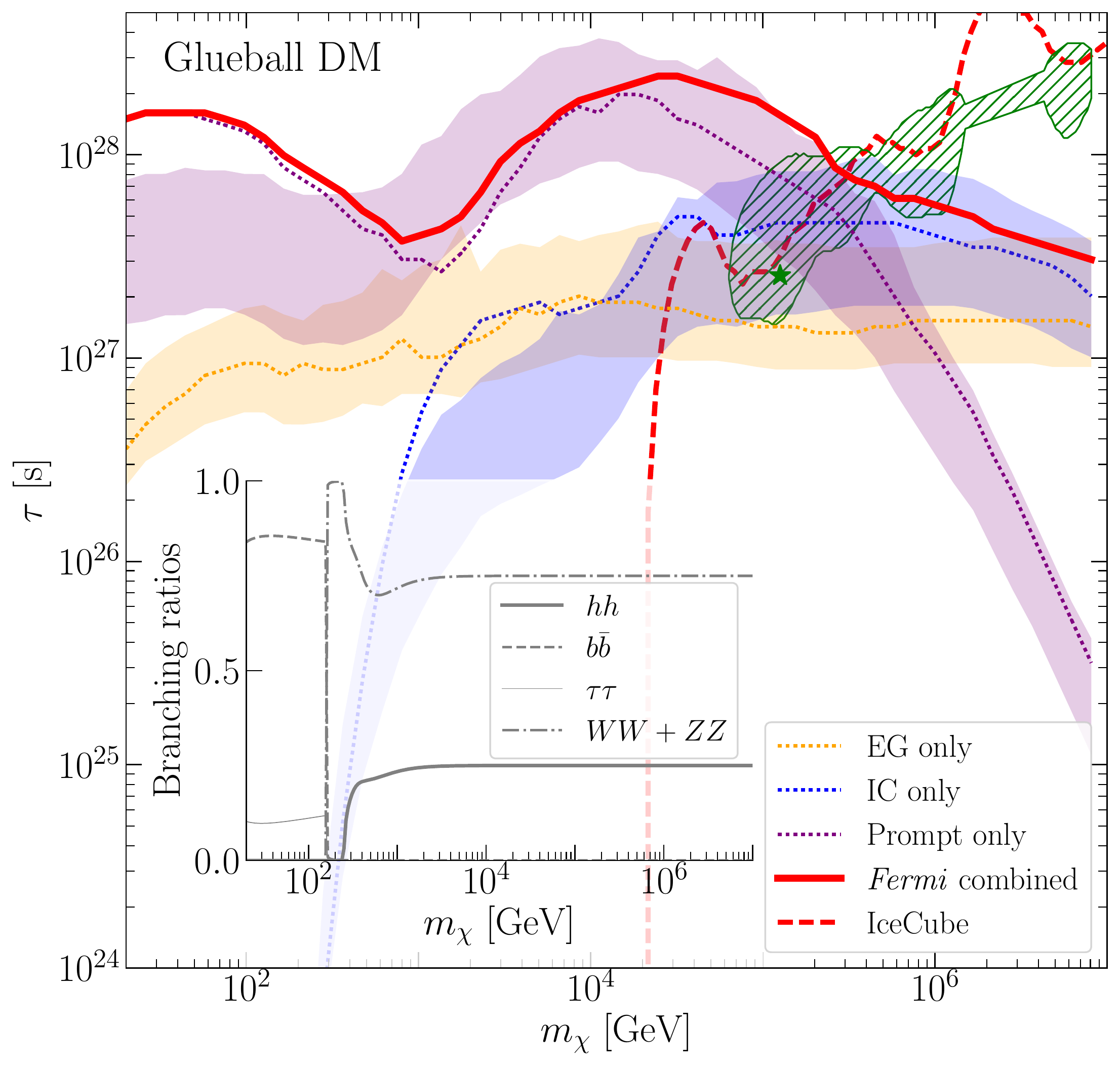}
        \end{center}
        \vspace{-.50cm}
        \caption{Limits
        on decaying glueball DM (see text for detals).
        We show limits obtained from prompt, IC, and EG emission only, along with the 95\% confidence window for the expectation of each limit from MC simulations.
         Furthermore, the parameter space where the IceCube data may be interpreted as a $\sim$$3\sigma$ hint for DM is shown in shaded green, with the best fit point represented by the star.
         (inset) The dominant glueball DM branching ratios.
        }
        \vspace{-0.15in}
        \label{Fig: glue}
\end{figure}

It is also interesting to consider 
models that could yield signals relevant for this analysis.  Many cases are explored in the associated appendix, and here we highlight one simple option:  a hidden sector that consists of a confining gauge theory, at scale $\Lambda_{D}$~\cite{Faraggi:2000pv}, without additional light matter. Hidden gauge sectors that decouple from the SM at high scales appear to be generic in many string constructions (see~\cite{Halverson:2016nfq} for a recent discussion).  Denoting 
the hidden-sector field strength as $G_{D \mu \nu}$, then the lowest dimensional operator connecting the hidden sector to the SM appears at dimension-6:  $\mathcal{L} \supset \lambda_D\, {G_{D \mu \nu}\, G_D^{ \mu \nu}\, |H|^2 / \Lambda^2 }$, where $\lambda_D$ is a dimensionless coupling constant, $\Lambda$ is the scale where this operator is generated, and $H$ the SM Higgs doublet.  The lightest $0^{++}$ glueball state in the hidden gauge theory is a simple DM candidate $\chi$, with $m_\chi \sim \Lambda_{D}$, though heavier, long-lived states may also play important roles (see \emph{e.g.}~\cite{Forestell:2016qhc}).  The lowest dimension EFT operator connecting $\chi$ to the SM is then $\sim \chi\, |H|^2 \,\Lambda_{D}^3 / \Lambda^2$.  Furthermore, $\Lambda_{D}\gtrsim100\,$MeV in order to avoid constraints on DM self-interactions~\cite{Boddy:2014yra}.

At masses comparable to and lower than the electroweak scale, the glueball decays primary to $b$ quarks through mixing with the SM Higgs, while at high masses the glueball decays predominantly to $W^\pm$, $Z^0$, and Higgs boson pairs (see the inset of Fig.~\ref{Fig: glue} for the dominant branching ratios).  In the high-mass limit, the lifetime is approximately 
 \es{tau_glue}{
\!\!\tau &\simeq 5 \cdot 10^{27}  \, \text{s} \left( { 3 \over N_D} {1 \over 4\, \pi  \lambda_D} \right)^2 \left( {\Lambda \over m_\text{Pl} } \right)^4 \left( {0.1  \, \text{PeV}  \over \Lambda_{D} } \right)^5 ,
 }
with $N_D$ the number of colors.  This is roughly the right lifetime to be relevant for the IceCube neutrino flux.    
 
   In Fig.~\ref{Fig: glue}, we show our constraint on this glueball model.  Using Eq.~\eqref{tau_glue}, these results suggest that models with $\Lambda_{D} \gtrsim 0.1 \, \, \text{PeV}$, $\lambda_D \gtrsim 1/(4 \pi)$, and $\Lambda = m_\text{Pl}$ are excluded.   As in Fig.~\ref{Fig: bbResult}, the shaded green is the region of parameter space where the model may contribute significantly to IceCube, and the dashed red line provides the limit we obtain from IceCube allowing for an astrophysical contribution to the flux.  As in the case of the $b\, \bar b$ final state, the gamma-ray limits derived in this chapter are in tension with the decaying-DM origin of the signal. 

Figure~\ref{Fig: glue} also illustrates the relative contribution of prompt, IC and extragalactic emissions to the total limit. The 95\% confidence interval is shown for each source, assuming background templates only, where the normalizations are fit to the data.
Across almost all of the mass range, and particularly at the highest masses, the limits obtained on the real data align with the expectations from MC.  In the statistics-dominated regime, we would expect the real-data limits to be consistent with those from MC, while in the systematics dominated regime the limits on real data may differ from those obtained from MC.  This is because the real data can have residuals coming from mis-modeling the background templates, and the overall goodness of fit may increase with flux from the NFW-correlated template, for example, even in the absence of DM.  Alternatively, the background templates may overpredict the flux at certain regions of the sky, leading to over-subtraction issues that could make the limits artificially strong.

\section{Discussion}\label{sec:Discussion}

In this chapter, we presented some of the strongest limits to date on decaying DM from a dedicated analysis of {\it Fermi} gamma-ray data incorporating spectral and spatial information, along with up-to-date modeling of diffuse emission in the Milky Way.  Our results disfavor a decaying DM explanation of the IceCube high-energy neutrino data.

There are several ways that our analysis could be expanded upon.
We have not attempted to characterize the spectral composition of the astrophysical contributions to the isotropic emission, which may strengthen our limits.  On the other hand, ideally, for a given, fixed decaying DM flux in the profile likelihood, we should marginalize not just over the normalization of the diffuse template but also over all of the individual components that go into making this template, such as IC emission and bremsstrahlung.

A variety of strategies beyond those described here have been used to constrain DM lifetimes (see \emph{e.g.}~\cite{Ibarra:2013cra} for a review).  These include gamma-ray line searches, such as those performed in~\cite{Abdo:2010nc,Vertongen:2011mu,Ackermann:2012qk,Ackermann:2013uma}, which are complementary to the constraints on broader energy emission given in this chapter. Limits from direct decay into neutrinos have also been considered \cite{Esmaili:2012us}. Less competitive limits  
have been set on DM decays resulting in broad energy deposition and nearby galaxies and galaxy clusters~\cite{Dugger:2010ys,Huang:2011xr}, large scale Galactic and extragalactic emission~\cite{Cirelli:2009dv,Zhang:2009ut,Zaharijas:2010ca,Ackermann:2012rg,Zaharijas:2012dr}, Milky Way Dwarfs~\cite{Aliu:2012ga,Baring:2015sza}, and the CMB~\cite{Slatyer:2016qyl}.  The upcoming Cherenkov Telescope Array~(CTA) experiment~\cite{Consortium:2010bc}
may have similar sensitivity as our results to DM masses $\sim$10 TeV~\cite{Pierre:2014tra}.  However, more work needs to be done in order to assess the potential for CTA to constrain or detect heavier, $\sim$PeV decaying DM.  
On the other hand, the High-Altitude Walter Cherenkov Observatory~(HAWC)~\cite{Abeysekara:2013tza} and air-shower experiments such as Tibet AS+MD~\cite{Sako:2009xa} will provide meaningful constraints on the Galactic diffuse gamma-ray emission. The constraints on DM lifetimes might be as stringent as ${10}^{27}-{10}^{28}$~s for PeV masses and hadronic channels, assuming no astrophysical emission is seen~\cite{Ahlers:2013xia,Murase:2015gea,Esmaili:2015xpa}.

Finally, we mention that our results also have implications for possible decaying DM interpretations (see {\it e.g.}~\cite{Cheng:2016slx}) of the positron~\cite{Aguilar:2013qda,Ting:2238506} and antiproton fluxes~\cite{ams02pos} measured by AMS-02. 
Recent measurements of the positron flux appear to exhibit a break at high masses that could indicate evidence for decaying DM to, for example, $e^+ \,e^-$ with $m_\chi \sim$ 1 TeV and $\tau \sim$ $10^{27} \text{ s}$.  However, our results appear to rule out the decaying DM interpretation of the positron flux for this and other final states.  For example, in the $e^+\, e^-$ case our limit for $m_\chi \sim$ 1 TeV DM is $\tau \gtrsim 5 \times 10^{28} \text{ s}$.

%% file: cascsig.tex
\chapter{Multi-Step Cascade Annihilations of Dark Matter and the Galactic Center Excess}\label{chap:cascsig}

\section{Introduction}

Over the past five years, numerous independent studies have confirmed a flux of few-GeV gamma rays from the inner Milky Way, steeply peaked toward the Galactic Center, that is not captured by models for the known diffuse backgrounds \cite{Goodenough:2009gk,Hooper:2010mq,Boyarsky:2010dr,Hooper:2011ti,Abazajian:2012pn,Hooper:2013rwa,Gordon:2013vta,Huang:2013pda,Abazajian:2014fta,Daylan:2014rsa,Calore:2014xka}. This ``Galactic Center excess'' (GCE), detected using public data from the {\it Fermi} Gamma-Ray Space Telescope, has a spatial morphology well described by the square of a generalized Navarro-Frenk-White (NFW) profile, projected along the line of sight. Furthermore, it is highly spherically symmetric,  centered on the Galactic Center (GC), and extends at least 10 degrees from the GC \cite{Daylan:2014rsa};\footnote{This analysis exploited improvements to the \emph{Fermi} point spread function as described in \cite{Portillo:2014ena}.} these conclusions remain unchanged when accounting for systematic uncertainties in the modeling of the diffuse backgrounds \cite{Calore:2014xka}. These spatial properties suggest the excess emission could arise from the annihilation of dark matter (DM) with an NFW-like density profile. Competing interpretations include a transient event at the GC producing high-energy cosmic rays that subsequently yield few-GeV gamma rays by scattering processes \cite{Petrovic:2014uda, Carlson:2014cwa}, or a population of many unresolved millisecond pulsars (MSPs) (e.g. \cite{Abazajian:2010zy, Gordon:2013vta}). However, these interpretations face significant challenges: it is unclear whether the proposed outflow models can match the spectrum and morphology of the excess \cite{LINDENTALK} (see also \cite{Macias:2013vya,Gordon:2014gya}), and estimates of the MSP population in the region of interest consistently underpredict the signal by an order of magnitude \cite{Hooper:2013nhl, Cholis:2014lta}.

Models where DM annihilates with a roughly thermal cross-section and has a mass of order several tens of GeV can readily account for the spectrum and size of the excess. However, when embedded in even a simplified DM model, there are often powerful constraints on these scenarios from direct detection and collider bounds (e.g. \cite{Alves:2014yha, Berlin:2014tja}). While UV-complete models where the DM annihilates directly to Standard Model (SM) particles do exist (e.g. \cite{Ipek:2014gua, Cheung:2014lqa, Gherghetta:2015ysa}), the constraints are much more easily evaded if the DM produces gamma-rays via a cascade process \cite{Pospelov:2007mp, Martin:2014sxa,Abdullah:2014lla,Ko:2014gha,Freytsis:2014sua}. In such scenarios, the DM is secluded in its own hidden dark sector, and first annihilates to other dark sector particles; these mediators subsequently decay into SM particles that produce gamma-rays.\footnote{Annihilation into the dark sector can also lead to a novel spatial distribution for the signal \cite{Rothstein:2009pm}, but the GCE favors a cuspy morphology, so in this chapter we assume all decays are prompt.}

The presence of an intermediate step between DM annihilation and the production of SM particles broadens the spectrum of SM particles produced, and consequently also broadens the resulting gamma-ray spectrum, unless the mediator is degenerate in mass with either the DM or  the total mass of the SM decay products. The gamma-ray multiplicity is increased by a factor of two, if each mediator decays into two SM particles, and the typical energy of the gamma-rays is reduced accordingly. Thus cascade models for the excess generically tend to accommodate:

\begin{itemize}
\item Higher DM masses,
\item Decays of the mediator to SM final states whose decays produce a more sharply peaked gamma-ray spectrum than favored by direct annihilation.
\end{itemize} 

In general, there may be more than one decay step within the dark sector; the dominant annihilation of the DM need not be to the lightest dark sector particle (e.g. \cite{Baumgart:2009tn, Nomura:2008ru}). If couplings within the dark sector are stronger than couplings between the sectors, dark sector particles will preferentially decay within the dark sector, with decays to the SM only occurring when no other states are available. Regardless of the model under consideration, in the absence of a mass degeneracy, each decay will increase the final gamma-ray multiplicity, decrease the typical gamma-ray energy, and broaden the spectrum (in the presence of a mass degeneracy only the first two effects will occur). Accordingly, long decay chains could potentially permit much heavier DM to explain the GCE, or favor decays to different SM states. In a sense, this description also characterizes the known decays of SM particles; final states whose decays produce gamma-rays through a lengthy cascade will generate a broader spectrum with a lower-energy peak, compared to final states that generate gamma-rays via a short cascade (we discuss this further in Sec. \ref{sec:results}).

It is this possibility of multi-step dark sector cascades that we explore in this chapter. For simplicity, we consider the case where all dark-sector particles involved in the cascade (except possibly the DM itself) are scalars - we briefly discuss the case of non-scalar mediators in Sec.~\ref{sec:generalcascade}. In this case, the results are largely independent of the details of the dark sector. The DM pair-annihilates into two scalar mediators which subsequently undergo a multi-step cascade in the dark sector, eventually producing a dark-sector state (with high multiplicity) that decays to the SM:
\be\begin{aligned}
\chi \chi \rightarrow \phi_n \phi_n &\rightarrow 2 \times \phi_{n-1} \phi_{n-1} \rightarrow . . . \\
 &\rightarrow 2^{n-1} \times \phi_1 \phi_1 \rightarrow 2^{n} \times f \bar{f}\,.
\label{eq:cascade}
\end{aligned}\ee
Here $f \bar{f}$ are SM lepton or quark pairs, which can subsequently decay; the decays shown above may also produce photons in the final step via final state radiation (FSR). By fitting the resulting photon spectrum to the GCE, we determine the allowed values of cross-section and DM mass for cascades with one to six steps, for a variety of SM final states. Provided that the masses of the particles at each step in the cascade are not near-degenerate, the final spectrum of gamma-rays becomes nearly independent of the exact masses at each step. This assumption is not limiting, as results for the quality of fit for the more general case of non-hierarchical cascades (with nearly-degenerate steps) can be simply extracted from results derived assuming a large hierarchy.

In Sec.~\ref{sec:methods} we outline the determination of the photon spectrum for an $n$-step cascade with specified SM final state, and discuss the procedure used to compare such a spectrum to the GCE. We present sample results of these fits in Sec.~\ref{sec:results} under certain assumptions. Section~\ref{sec:generalcascade} extends our results for general cascades, and contains our complete fit results. In Sec.~\ref{sec:signalsconstraints} we outline the existing experimental constraints a complete  model for the GCE via cascade decays would need to satisfy. We present our conclusions in Sec.~\ref{sec:conclusion}. In the appendices we provide additional details of our methodology and discuss some further model-dependent considerations. 

\section{Methodology}
\label{sec:methods}
The photon flux generated by the annihilations of self-conjugate DM\footnote{As discussed in Appendix~\ref{app:zerostep}, our results can be readily translated to the case of decays, although the steeply peaked morphology of the GCE disfavors this interpretation.} as a function of the direction observed in the sky, is given by:
\be
\Phi\left(E_\gamma, l, b \right) = \frac{\langle \sigma v \rangle}{8 \pi m_{\chi}^2} \frac{dN_\gamma}{dE_\gamma} J\left(l,b\right)\,,
\label{eq:flux}
\ee
where $\langle \sigma v \rangle$ is the thermally averaged annihilation cross-section, $m_\chi$ is the DM mass, and $dN_\gamma / dE_\gamma$ is the photon spectrum per DM annihilation, which has contributions from FSR and from the decay of the leptons or quarks and their subsequent hadronization products. The $J$-factor, the integral of DM density squared along the line-of-sight, is a function of the observed direction in the sky expressed in terms of Galactic coordinates $l$ and $b$:
\be
J\left(l,b\right) =   \int_{0}^{\infty}  \rho^2\left(\sqrt{s^2 + 2 r_{\odot} s  \cos l \cos b + r_{\odot}^2}  \right) ds\,,
\label{eq:Jfactor}
\ee
where $r_{\odot}\approx8.5$ kpc is the distance from the Sun to the Galactic Center, and  $s$ parametrizes the integral along the line-of-sight. We parameterize the DM density by a generalized NFW halo profile \cite{Navarro:1995iw,Navarro:1996gj}:
\be
\rho \left( r, \gamma \right) = \rho_0 \frac{(r/r_s)^{-\gamma}}{\left(1 + r/r_s \right)^{3-\gamma}}\,.
\ee
Here we use $r_s = 20$ kpc, $\rho_0 = 0.4$ GeV/cm$^3$ and $\gamma = 1.2$, following \cite{Calore:2014xka}, as we will compare our models to the data using the spectrum and covariance matrix determined by that work.

We focus on $n$-step cascades ending in $\phi_1 \rightarrow f \bar{f}$, where $f \bar{f}$ is a pair of electrons, muons, taus or $b$-quarks. Other SM final states are possible, of course, but these cases span the range from steeply peaked photon spectra close to the DM mass through to the lower-energy and broader spectra characteristic of annihilation to hadrons. In order to generate the cascade spectrum, we first start with the result from direct DM annihilation, which is equivalent to the spectrum from $\phi_1$ decay (in the $\phi_1$ rest frame) if the DM mass is half the $\phi_1$ mass. For the case of electrons or muons we determine this spectrum analytically using the results of \cite{Mardon:2009rc}, whilst for taus and $b$-quarks the results are simulated in \texttt{Pythia8} \cite{Sjostrand:2007gs}. We have relegated the details of calculating these spectra to Appendix~\ref{app:zerostep}.

We denote the spectrum obtained at this ``0th step'' by $dN_\gamma / dx_{0}$, where $x_0 = 2E_0 / m_1$, $m_1$ is the mass of $\phi_1$ and $E_0$ is the energy of the photon in the $\phi_1$ rest frame. The shape of the photon spectrum is determined by the identity of the final state particle $f$ and also the ratio $\epsilon_f = 2m_f / m_1$. In the limit where the decay of $\phi_1$ is dominated by a two-body final state (at least for the purposes of photon production), the photon spectrum converges to a constant shape (as a function of $x_0$) as $\epsilon_f \rightarrow 0$ and the $f \bar{f}$ become highly relativistic. However, final state radiation (FSR) and hadronization depend on the energy of the $f \bar{f}$ products of the $\phi_1$ decay in the $\phi_1$ rest frame, so in cases where these effects dominate, the dependence of the photon spectrum on $\epsilon_f$ is more complex.

\begin{figure}[t!]
\centering
\includegraphics[scale=0.75]{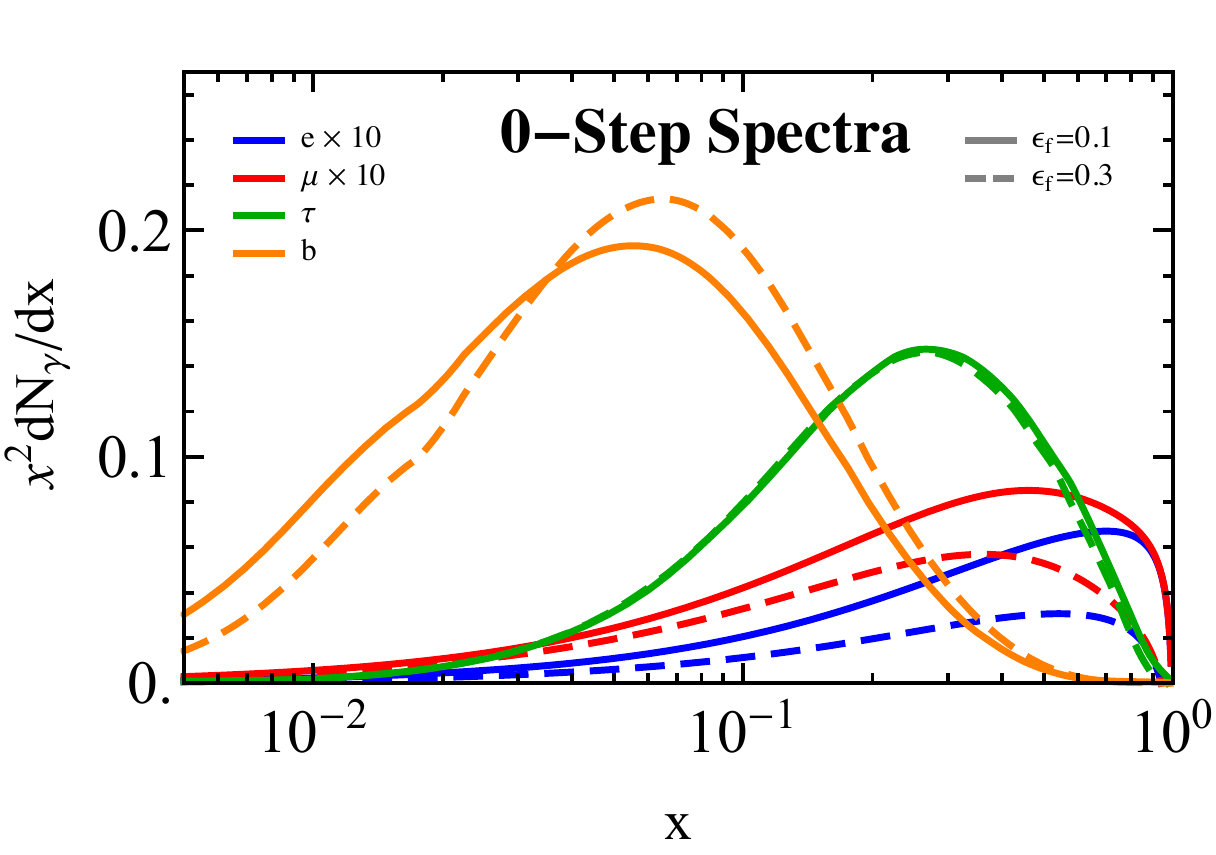}
\vspace{-0.2cm}
\caption{\footnotesize{0th step (direct annihilation) photon spectra $dN_\gamma/dx_{0}$ for $\phi_1$ decaying to $(e, \mu, \tau, b)$ in (blue, red, green, orange). Solid curves correspond to $\epsilon_f=0.1$, and dashed to $\epsilon_f=0.3$. The electron and muon spectra have been magnified by a factor of ten to appear comparable to the taus and $b$s.}}
\label{fig:0step-sig}
\end{figure}

In Fig.~\ref{fig:0step-sig} we show $dN_\gamma/dx_{0}$ per annihilation for the four different final states we considered, for $\epsilon_f = 0.1$ and $\epsilon_f = 0.3$. The photon spectra from electron and muon production are dominated by FSR, whereas for $b$-quarks fragmentation and hadronization are important. In the photon spectrum from taus, these effects are subdominant and so the impact of varying $\epsilon_f$ is minimal. Note that the spectrum for $b$-quarks is peaked at a significantly lower $x$, highlighting why models with this final state tend to accommodate higher DM masses.

Given the 0-step spectrum, determining the photon spectrum from an $n$-step cascade is particularly simple in the case of scalar mediators,\footnote{We discuss the case of vector mediators in Sec.~\ref{sec:generalcascade}.} where the calculation essentially reduces to Lorentz-boosting the photon spectrum up the ladder of particles appearing in the cascade. We review this calculation in Appendix~\ref{app:boost}. As observed in \cite{Mardon:2009rc}, in the case of large mass hierarchies between the steps in the cascade, the final photon spectrum can be simplified even further, as we now outline. 

\begin{figure}[t!]
\centering
\includegraphics[scale=0.7]{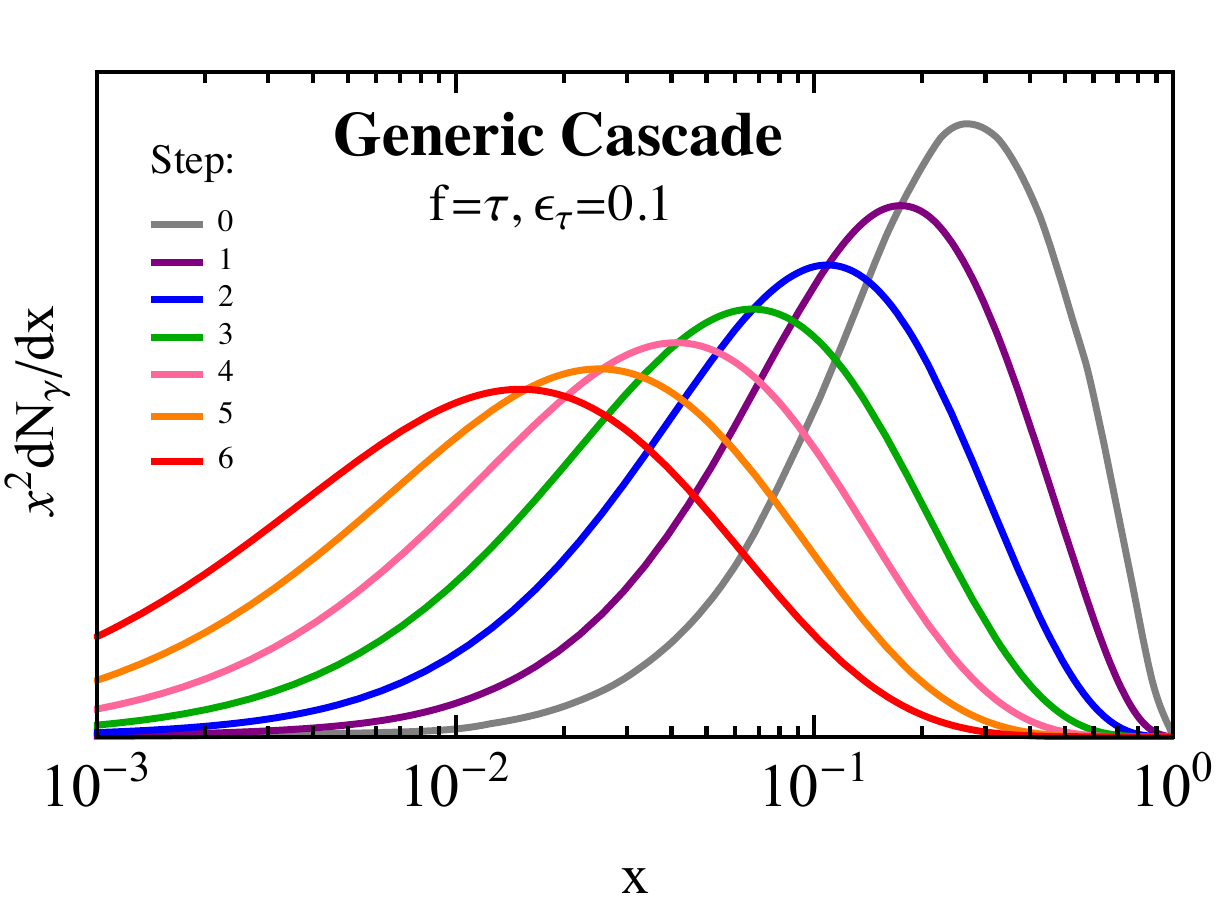}
\vspace{-0.2cm}
\caption{\footnotesize{An example photon spectrum from direct annihilation to taus (grey) and hierarchical cascades with $n$ = (1,2,3,4,5,6) steps, corresponding to (purple, blue, green, pink, orange, red) curves. The presence of each additional step in the cascade acts to broaden and soften the spectrum, and shift the peak to lower masses. All spectra are per annihilation.}}
\label{fig:GeneralSpectrum}
\end{figure}

Consider the $i$th step in the cascade, where the decay is $\phi_{i+1} \to \phi_i \phi_i$. Let us define $\epsilon_i = 2m_i/m_{i+1}$, and assume $\epsilon_i \ll 1$.\footnote{Note that the earlier-defined $\epsilon_f$ parameter does \emph{not} function in exactly the same way as these $\epsilon_i$ parameters: $\epsilon_f$ fully parameterizes the photon spectrum associated with production and decay of the SM particles, whereas the $\epsilon_i$ only describe Lorentz boosts.}  Suppose the photon spectrum from decay of a single $\phi_i$ (and the subsequent cascade), in the rest frame of the $\phi_i$ particle, is known and denoted by $dN_\gamma/dx_{i-1}$. Then, in the presence of a large mass hierarchy, the decay of $\phi_{i+1}$ produces two highly relativistic $\phi_i$ particles, each (in the rest frame of the $\phi_{i+1}$) carrying energy equal to $m_{i+1}/2 = m_i/\epsilon_i$. The photon spectrum in the rest frame of the $\phi_{i+1}$ is then given by a Lorentz boost (see Appendix~\ref{app:boost}), and in the limit $\epsilon_i \ll 1$ takes the simple form \cite{Mardon:2009rc}:
\be
\frac{dN_\gamma}{dx_i} = 2 \int_{x_i}^{1} \frac{dx_{i-1}}{x_{i-1}} \frac{dN_\gamma}{dx_{i-1}} + \mathcal{O}(\epsilon_i^2)\,.
\label{eq:boosteq-sig}
\ee
Here we have introduced the dimensionless variable $x_i = 2 E_i / m_{i+1}$, where $E_i$ is the photon energy in the $\phi_{i+1}$ rest frame. Following this, once we know the 0-step spectrum we can iteratively derive the $n$-step result. The error introduced by this assumption is $\mathcal{O}(\epsilon_i^2)$, as we quantify in Appendix~\ref{app:boost}. 

Beyond simplifying calculations, the large hierarchy approximation is also convenient for the following two reasons. Firstly in this limit, we can specify the shape of the spectrum simply by the identity of the final state $f$, the value of $\epsilon_f$, and finally the number of steps $n$. This is in contrast to the many possible parameters that could be present in a generic cascade. Secondly, as we will elaborate further in Sec.~\ref{sec:generalcascade}, it is also possible to read off the results for a generic hierarchy once we know the small $\epsilon_i$ result, making the assumption less limiting than it would initially appear. In particular in the limit when the masses become degenerate ($\epsilon_i \to 1$), the $\phi_i$'s are produced at rest. When they subsequently decay, there is no boost to the $\phi_{i+1}$ rest frame, and so an $n$-step cascade effectively reduces to a hierarchical $(n-1)$-step cascade, except for the additional final state multiplicity.  
 
The Galactic frame is approximately the rest frame of the (cold) DM; consequently, to determine the measured photon spectrum, we need to calculate the photon spectrum in the rest frame of the original DM particles. For an $n$-step cascade, this will involve $n$ such convolutions, starting from the $dN_\gamma/dx_0$ 0-step spectrum, where the highest mass scale in the cascade will be $m_{i=n} = 2 m_\chi$. Thus $x_{i=n} = E_n / m_{\chi}$, and the Galactic-frame photon spectrum will be $dN_\gamma/dx_n = m_{\chi} dN_\gamma / dE_n$. Fig.~\ref{fig:GeneralSpectrum} shows the resulting spectrum for a 0-6 step cascade in the case of final state taus with $\epsilon_{\tau}=0.1$. Each step in the cascade broadens out and softens the spectrum, and similar behaviour is seen for other final states.

In order to determine the favored parameter space, for a given choice of $f$, $\epsilon_f$, and number of steps in the cascade $n$, we vary $m_\chi$ and an overall normalization parameter $\eta$ (proportional to $\langle \sigma v \rangle /m_\chi^2$, as we will see below) and compare the model to the data using the spectrum and covariance matrix of \cite{Calore:2014xka}. In detail we calculate $\chi^2$ according to:
\be
\chi^2 = \sum_{ij} \left(\mathcal{N}_{i,\textrm{model}} - \mathcal{N}_{i,\textrm{data}} \right) C^{-1}_{ij} \left(\mathcal{N}_{j,\textrm{model}} - \mathcal{N}_{j,\textrm{data}} \right)\,,
\label{eq:chi}
\ee
where
\bea
&\mathcal{N}_{i,\rm{model}}& =  \left( \frac{\eta}{m_{\chi}}E_n^2\frac{dN}{dx_n}\right)_{i,\rm{model}} \\
&\mathcal{N}_{i,\rm{data}}& = \left(E^2 \frac{dN}{dE}\right)_{i,\rm{data}}
\label{eq:chi2}
\eea
and both model and data are expressed in units of GeV/cm$^2$/s/sr averaged over the region of interest. Here the $C^{-1}_{ij}$ are elements of the inverse covariance matrix, which together with the data points are taken from \cite{Calore:2014xka}. By Eq.~\ref{eq:flux}, the fitted normalization $\eta$ is related to the DM mass and the J-factor by:
\be
\langle \sigma v \rangle = \frac{8 \pi m_{\chi}^2 \eta} {J_{\rm norm}}\,.
\label{eq:xsecnorm}
\ee
For consistency with the spectrum normalization of \cite{Calore:2014xka} the J-factor is averaged over the ROI $|l| \leq 20^\circ$ and $2^\circ \leq |b| \leq 20^\circ$, so that:
\be\begin{aligned}
J_{\rm norm} &=  \int_{\rm ROI} d \Omega J\left(l,b\right)  / \int_{\rm ROI} d\Omega \\
&\sim 2.0618 \times 10^{23}~{\rm GeV}^2{\rm cm}^{-5}.
\label{eq:Jnorm}
\end{aligned}\ee 
(Note that $d\Omega = dl d\sin b$, not $dl d\cos b$, since $b$ measures the angle from the Galactic equator, not the north pole.)

\emph{Self-Consistency Requirements:} The procedure outlined above treats $m_\chi$ as a free parameter that can be adjusted to modify the 0-step spectrum; the fit only uses the shape of the spectrum provided by the 0-step result and the boost of Eq.~\ref{eq:boosteq-sig}. However, there is an additional condition required for a cascade scenario to be physically self-consistent: the mass hierarchy between the DM mass and the particles produced in the final state must be sufficiently large to accommodate the specified number of steps. Equivalently, there is a hard upper limit on the number of steps allowed, for a given DM mass and final state.

Recall that for an $n$-step cascade ending in a final state $f$, we defined $\epsilon_f = 2 m_f/m_1$, $\epsilon_1 = 2m_1/m_2$, $\epsilon_2 = 2m_2/m_3$ all the way up to $\epsilon_n=m_n/m_{\chi}$. Combining these, the DM mass is given in terms of $m_f$ and the $\epsilon$ factors by:
\be
m_\chi = 2^n \frac{m_f}{\epsilon_f  \epsilon_1  \epsilon_2  . . .  \epsilon_n}\,,
\label{eq:mDM-sig}
\ee
If the $\epsilon_i$ factors are allowed to float, we can still say that $0 < \epsilon_i \leq 1$ in all cases (since each decaying particle must have enough mass to provide the rest masses of the decay products), setting a strict lower bound on the DM mass of:
\be m_\chi \geq 2^n m_f/\epsilon_f\,. \label{eq:kinematic} \ee 
In the remainder of this article we refer to this bound as a ``self-consistency'' condition or defining ``kinematically allowed'' masses. For consistency with the assumption of hierarchical decays (i.e. $\epsilon_i \ll 1$), the true bound on $m_\chi$ will in general be somewhat stronger than this conservative estimate (although as we will discuss in Sec.~\ref{sec:generalcascade}, $\epsilon_i$ can become quite close to 1 before significantly modifying the fit relative to the $\epsilon_i \rightarrow 0$ case). 

\section{Results With the Assumption of Large Hierarchies}

Here we present the results from the fits performed using the procedure outlined in the previous section. Assuming hierarchical cascades, we perform fits for four different final states -- electrons, muons, taus, and $b$-quarks -- and fit over the photon energy range $0.5~\textrm{GeV} \leq E_{\gamma} \leq 300~\textrm{GeV}$.\footnote{By default, we omit the low energy data points with $0.3~\textrm{GeV} \leq E_{\gamma} \leq 0.5~\textrm{GeV}$, as in this region the spectrum suffers larger uncertainties under variations of the background modeling, and the preferred value of the NFW $\gamma$ parameter is not robust \cite{Daylan:2014rsa}. We have confirmed that including these low-energy data points has little impact on our results.} Later in this section we discuss the effects of cutting out high energy data points, and how the fits would change if we only considered statistical uncertainties. 

\begin{figure}[t!]
\centering
\begin{minipage}{.45\textwidth}
  \centering
  \includegraphics[scale=0.5]{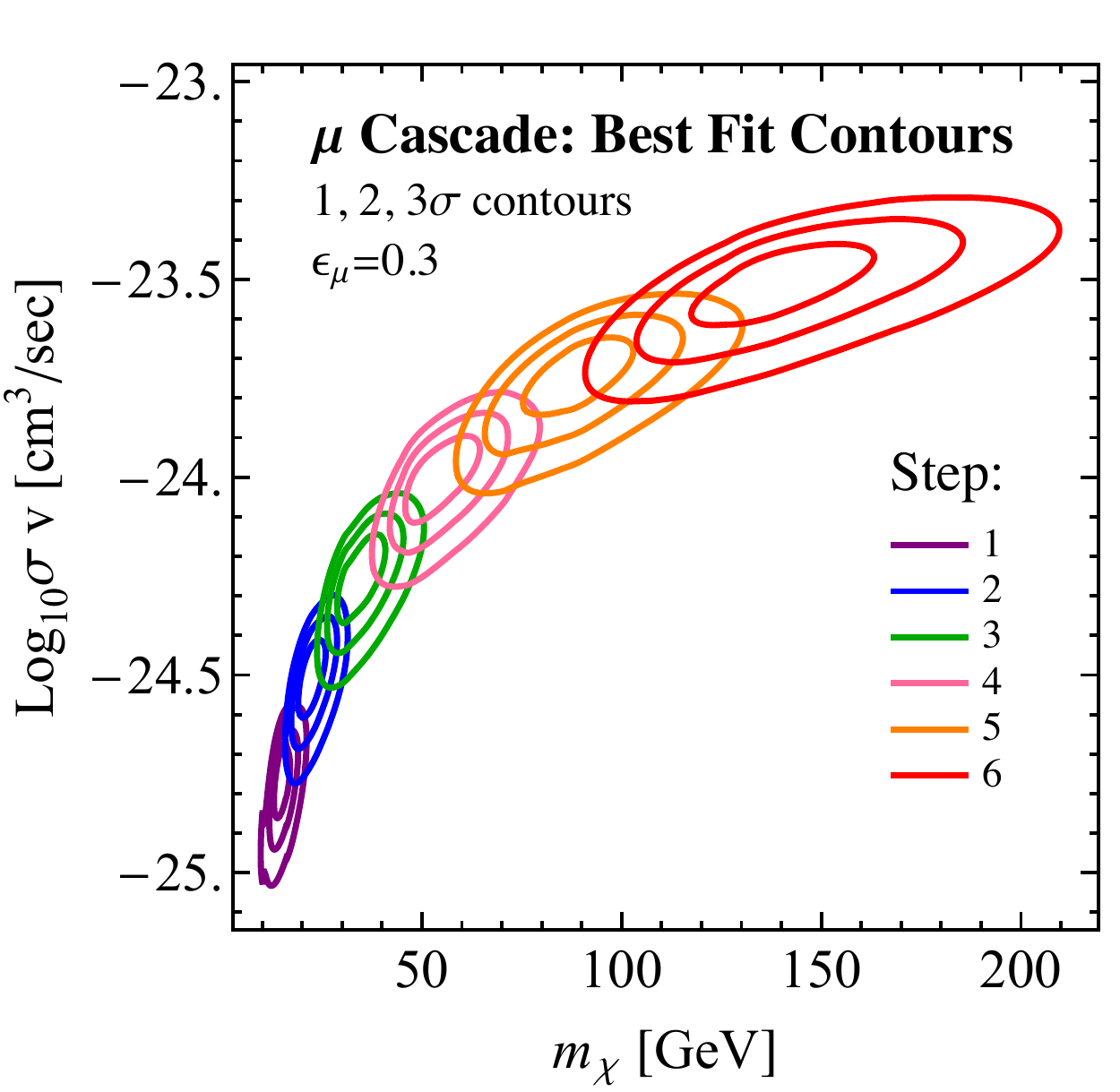}
  \vspace{-0.2cm}
  \captionof{figure}{\footnotesize{Contours of $\Delta \chi^2$ from the best-fit point (for a given step number $n$) corresponding to 1, 2 and 3$\sigma$ for final state $\mu$'s, with $\epsilon_{\mu} = 0.3$. The purple, blue, green, pink, orange and red colors correspond to $n =$ 1, 2, 3, 4, 5 and 6 steps in the cascades to final state $\mu$'s. Here we have fixed $\epsilon_{\mu} = 0.3$ and fit over the range 0.5 GeV $\leq E_\gamma \leq$ 300 GeV.}}
  \label{fig:TauChiPlot}
\end{minipage}
\hspace{0.4in}
\begin{minipage}{.45\textwidth}
\vspace{0.05in}
  \centering
  \includegraphics[scale=0.48]{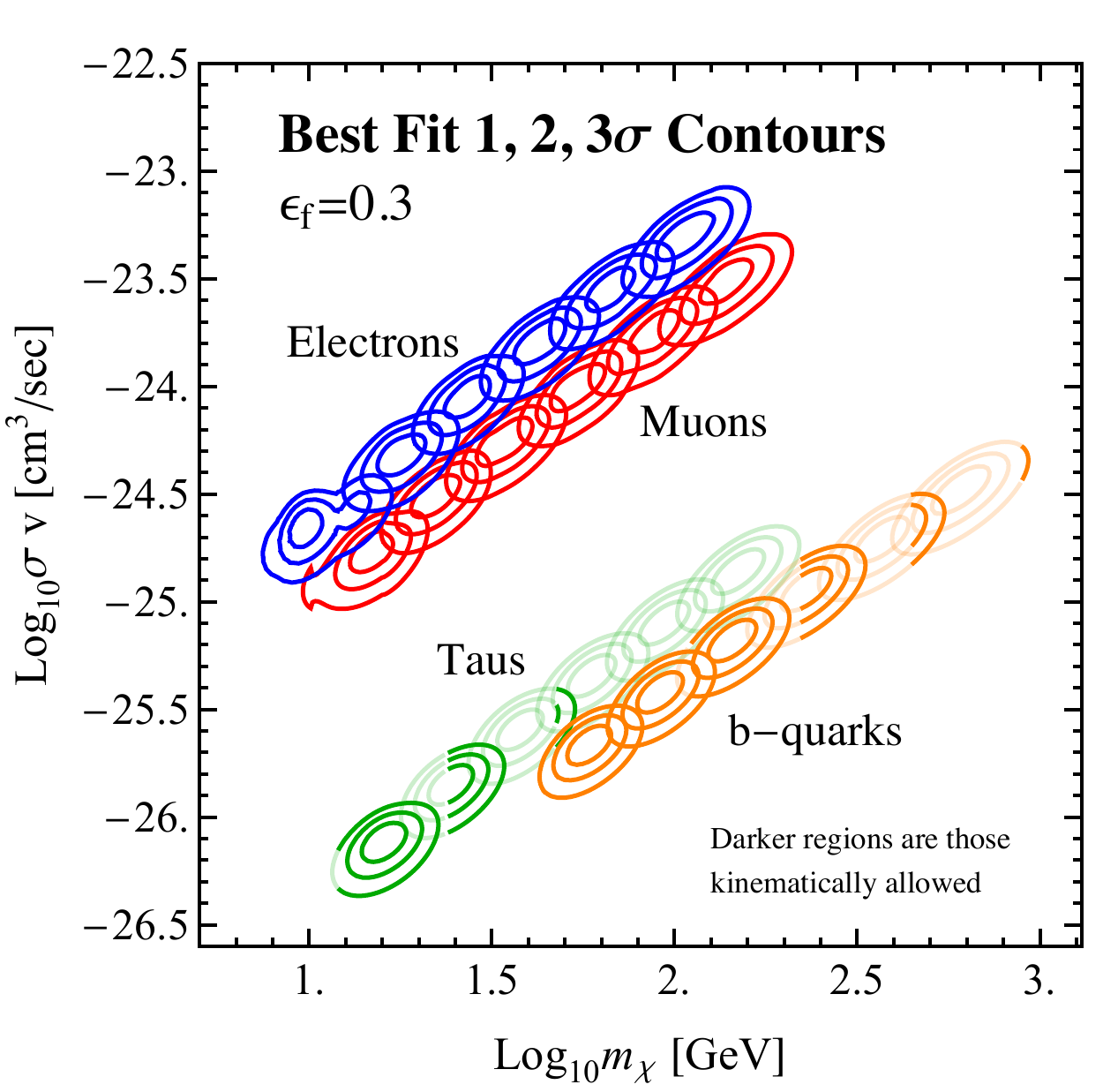}
  \vspace{-0.2cm}
  \captionof{figure}{\footnotesize{Contours of $\Delta \chi^2$ corresponding to 1, 2 and 3$\sigma$ for $n=1-6$ steps for $e$, $\mu$, $\tau$ and $b$ final states with $\epsilon_f=0.3$. The fit is performed over the range 0.5 GeV $\leq E_\gamma \leq$ 300 GeV. The best fit point of each step for all four final states follows a power law relation between $m_\chi$ and $\langle \sigma v \rangle$, with index $\sim 1.3$. Only the darker regions are kinematically allowed. See text for details.}}
  \label{fig:AllStatesLogChiPlot}
\end{minipage}
\end{figure}

In Fig.~\ref{fig:TauChiPlot} we show a sample result, in which we plot $\Delta \chi^2$ 1, 2 and 3$\sigma$ contours in $\left(m_{\chi}, \langle \sigma v \rangle \right)$ space for 1-6 step cascades ending in muons with $\epsilon_{\mu} = 0.3$. The trend in the best fit point for each step is as expected. Recall the generic behavior illustrated in Fig.~\ref{fig:GeneralSpectrum}; each progressive step in the cascade acts to reduce the height of the peak and shift it to lower masses. Therefore higher steps in the cascades will be better fit by larger DM mass and cross-section as is indeed the case in Fig.~\ref{fig:TauChiPlot}. The larger cross-section results from an interplay of effects as can be seen from Eq.~\ref{eq:xsecnorm}: an increased DM mass leads to a lower number density and hence a higher cross-section (scaling as $m_\chi^2$), but the increased power per annihilation implies a lower $\eta$ (adding a factor of $m_\chi^{-1}$), and finally the reduced height of the peak in the dimensionless spectrum for higher steps (as shown in Fig.~\ref{fig:GeneralSpectrum}) requires a larger $\eta$.

\label{sec:results}

\begin{figure}[t!]
\centering
\begin{tabular}{c}
\includegraphics[scale=0.47]{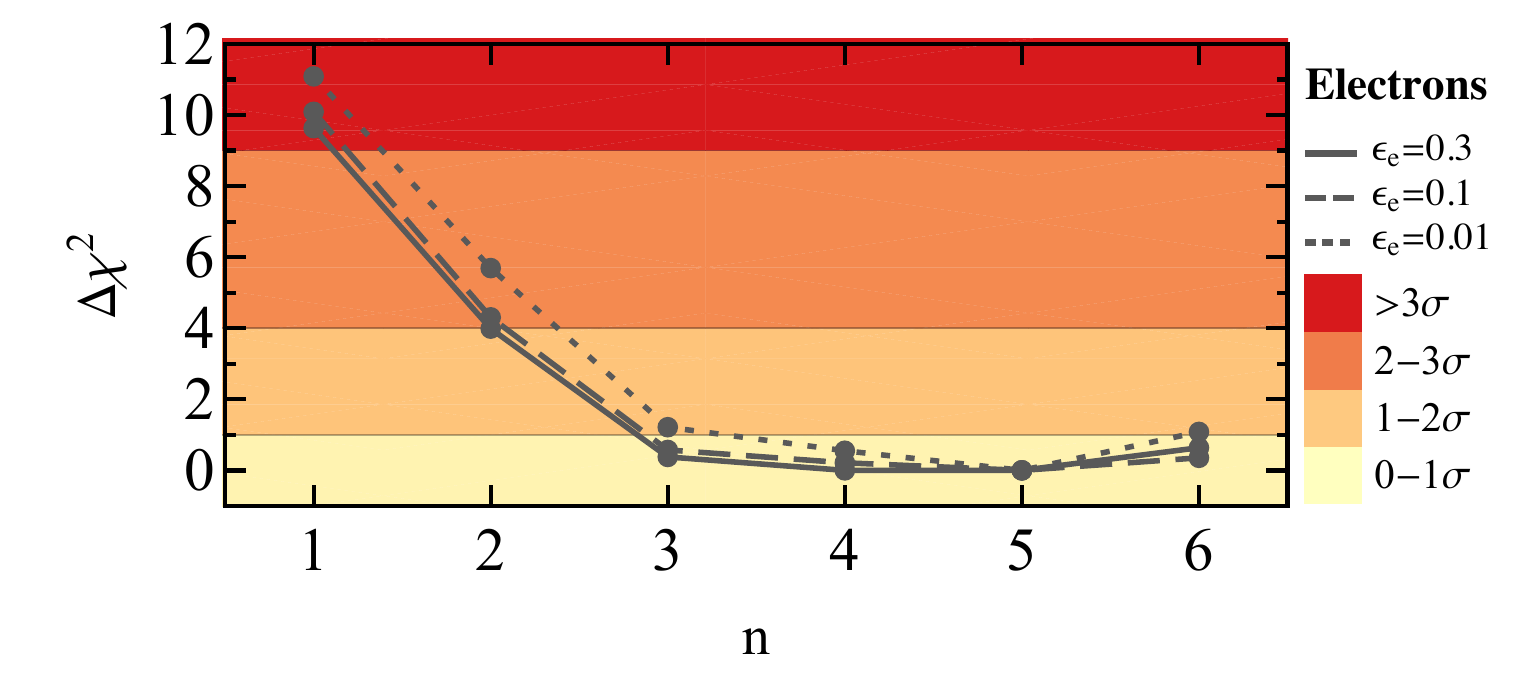} \hspace{0.1in}
\includegraphics[scale=0.47]{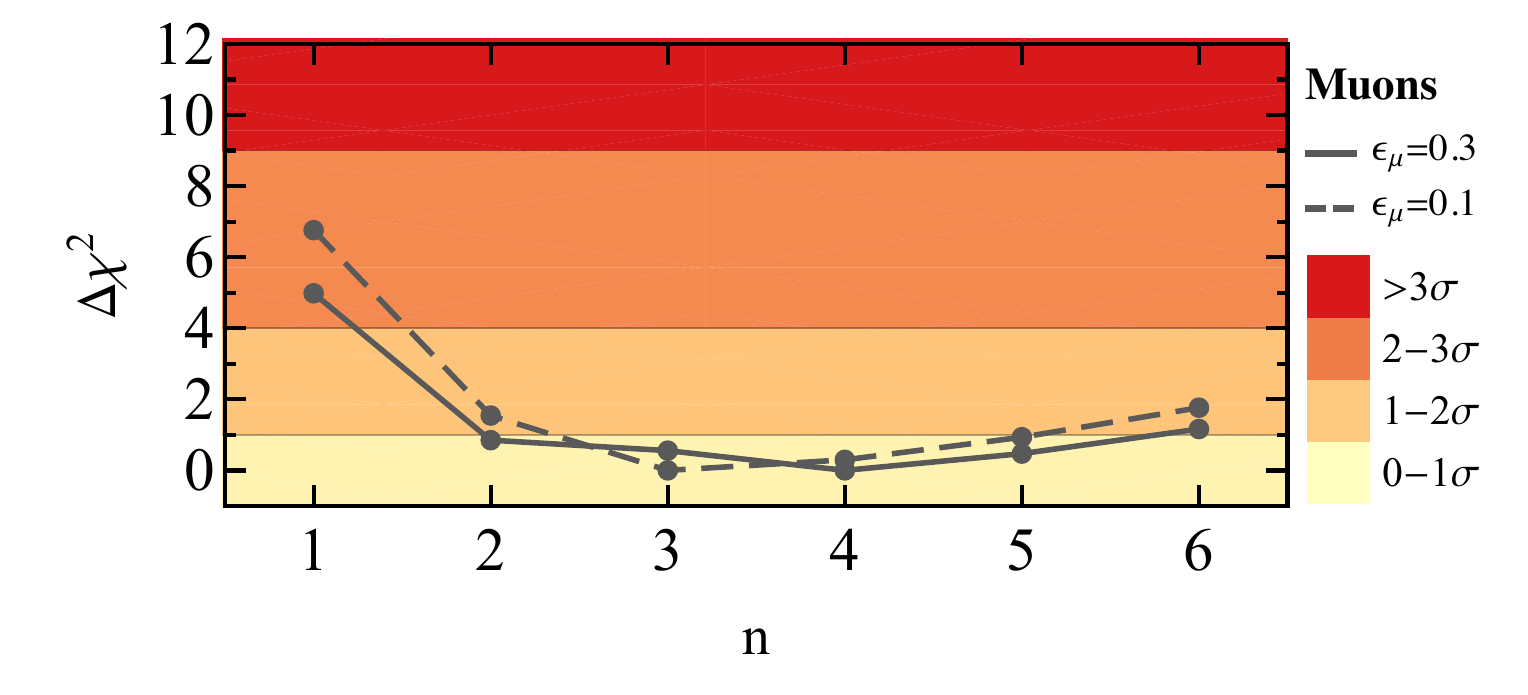}\\
\includegraphics[scale=0.47]{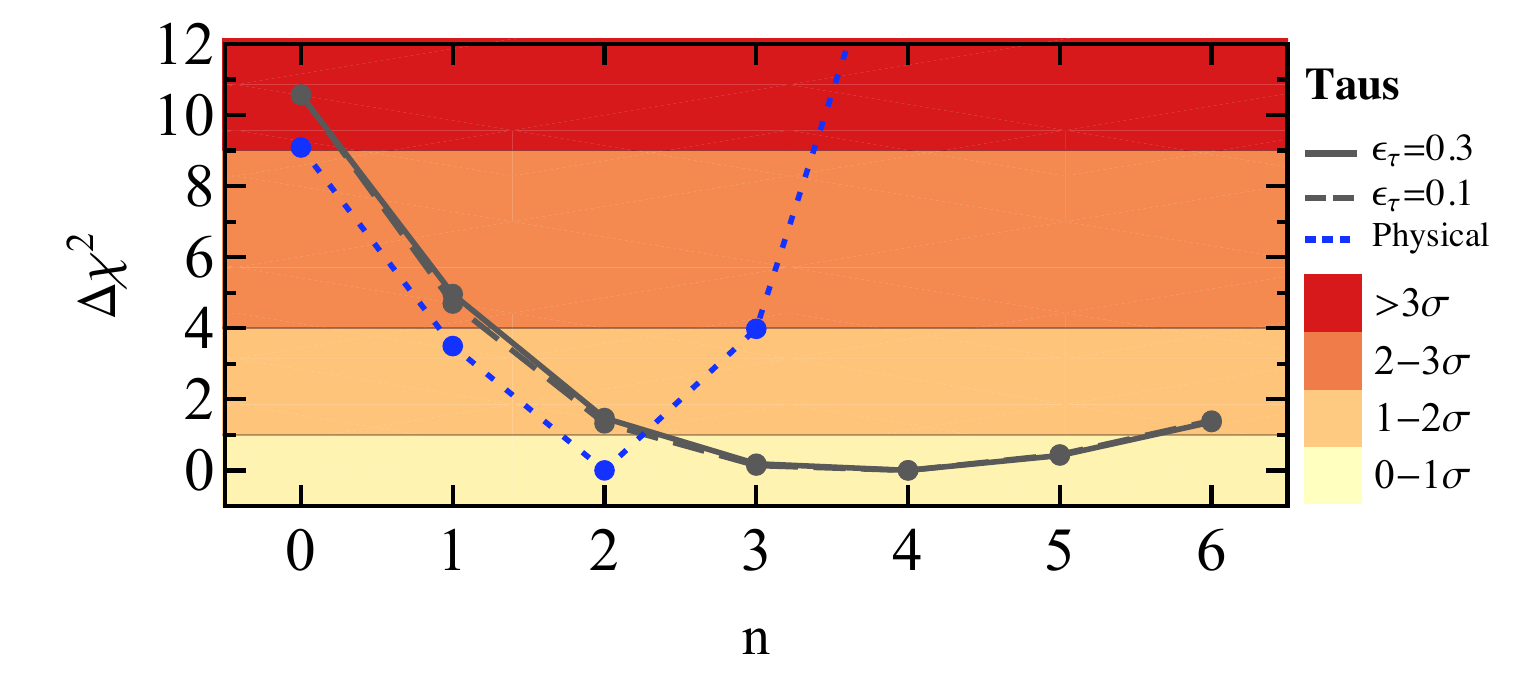} \hspace{0.1in}
\includegraphics[scale=0.47]{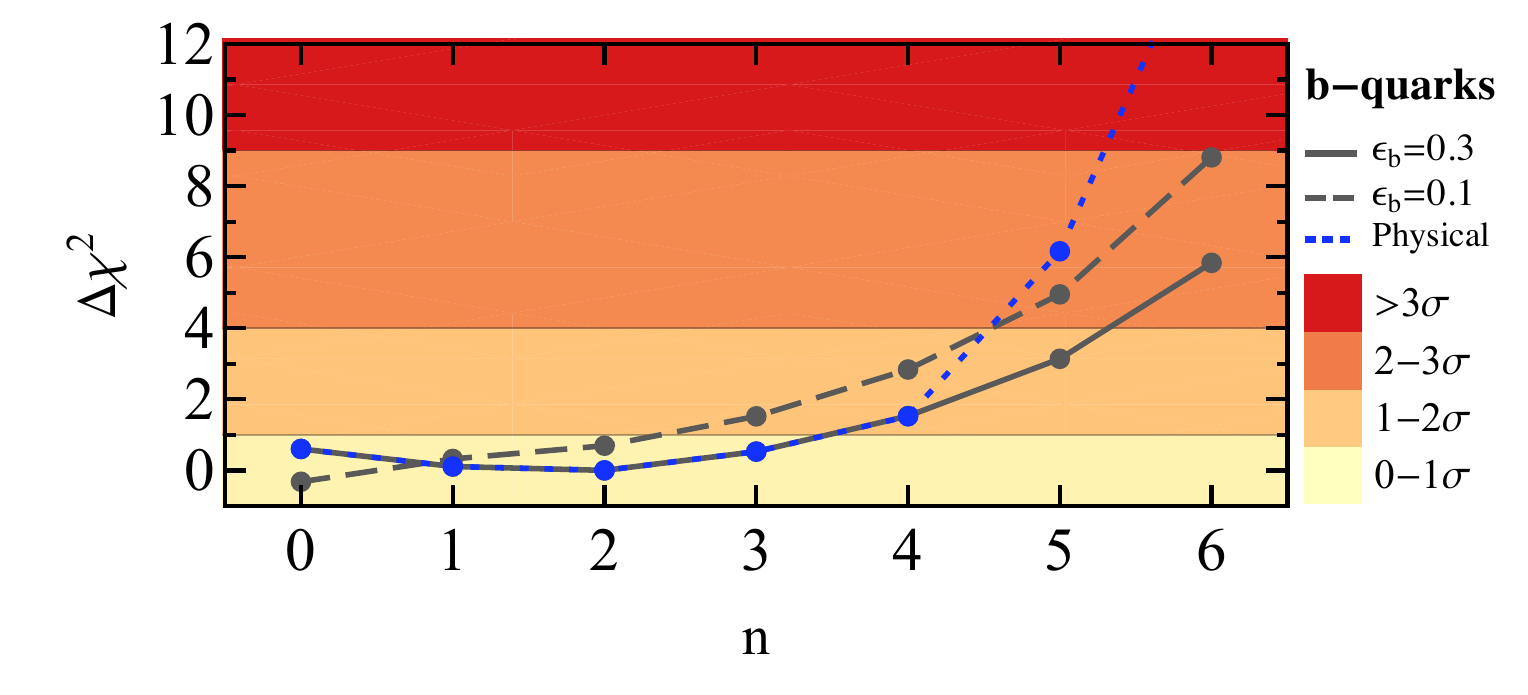}
\end{tabular}
\vspace{-0.2cm}
\caption{\footnotesize{Clockwise panels show the overall best fit for DM annihilating through an $n$-step cascade to electron, muon, $b$-quark and tau final states. The grey solid, dashed (and dotted) lines correspond to the $\Delta \chi^2$ between the best fit at that step, and the best fit for all $n$, for $\epsilon_{f} = 0.3,0.1$ (and $0.01$) respectively. In the case of tau and $b$-quark final states, the blue dotted curves, denoted `physical,' correspond to the case where only kinematically allowed  (self-consistent) masses are considered as per the discussion in Sec.~\ref{sec:methods} (we set $\epsilon_f = 0.3$ for these curves). Note that in the case of taus, the ``physical'' best-fit points for 0 and 1 steps have the same $\chi^2$ as the best-fit points when ``unphysical'' scenarios are allowed, but as the overall best fit is different (with higher $\chi^2$) their $\Delta \chi^2$ is lower. The shaded bands correspond to the quality of fit. 0-step results are not included for electrons and muons, as these fits are poor and have $\Delta \chi^2$ values well above the plotted $y$-axis. Electrons, muons and taus prefer longer 3-5 step cascades, whilst annihilations to $b$-quarks prefer shorter 0-2 step cascades. This is not surprising, since as has been already pointed out in the literature, $b$-quark final states are preferred for direct annihilations. Non-integer values of $n$ can be associated with cascades containing steps with one or more large $\epsilon_i$, as discussed in Sec.~\ref{sec:generalcascade}.}}
\label{fig:BestFit}
\end{figure}

\begin{figure}[t!]
\centering
\includegraphics[scale=0.65]{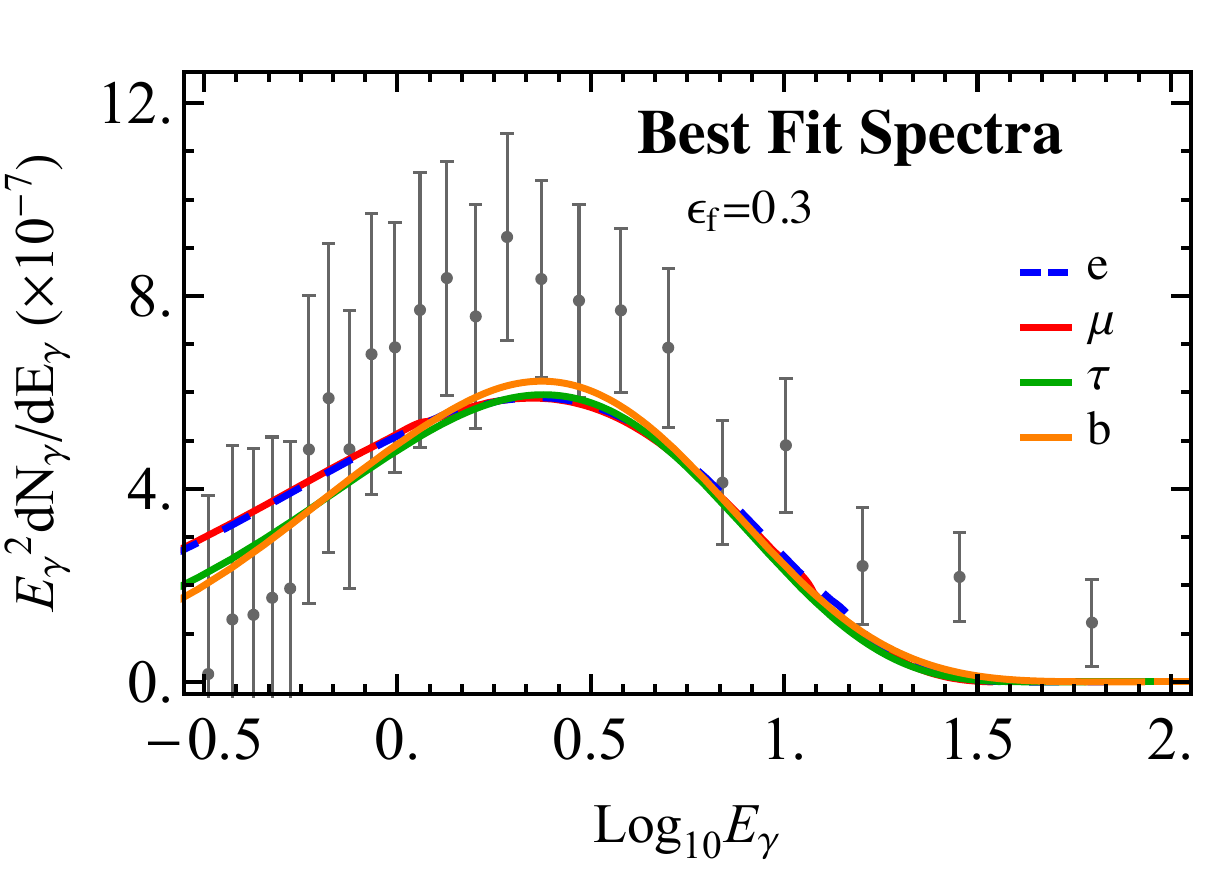}
\vspace{-0.2cm}
\caption{\footnotesize{The blue, red, green and orange curves correspond to the overall best fit spectrum for e, $\mu$, $\tau$ and $b$-quarks as determined from Fig.~\ref{fig:BestFit}. Overlaid are the data points and systematic errors from \cite{Calore:2014xka}. Note that due to correlations between energies, the best fit curves are not what would be naively expected if only statistical errors were present.}}
\label{fig:Spectrum0p3Full}
\end{figure}

In Fig.~\ref{fig:AllStatesLogChiPlot} we show the corresponding $\Delta \chi^2$ contours for electron, muon, tau, and $b$-quark final states, again fixing $\epsilon_f=0.3$. The best-fit mass and cross-section for each of the final states are empirically found to follow an approximate power law with $\langle \sigma v \rangle \propto m_\chi^{1.3}$. As discussed above we would expect $\langle \sigma v \rangle \propto m_\chi$ if the spectrum did not change in shape (simply being rescaled proportionally to $m_\chi$ to ensure energy conservation); the additional $m_\chi^{0.3}$ scaling factor reflects the change in shape of the spectrum.

As discussed above, for a given DM mass and final-state fermion with mass $m_f$, there is an absolute upper limit on the number of steps allowed in a cascade, since every step corresponds to a change in mass scale of at least a factor of 2. In Fig. \ref{fig:AllStatesLogChiPlot}, we show the contours if the limitation of Eq.~\ref{eq:kinematic} is \emph{ignored}, since this conveys information on the mass scale and number of steps at which the broadness of the spectrum best matches the data; however, the mass values that violate this condition and so do not represent a self-consistent physical scenario are shown in lighter shading. This issue is relevant for the heavier final-state fermions, taus and $b$-quarks, and particularly acute for taus. Finally note that the irregular shape of the contours for the one-step electrons and muons can be traced to the fact the 0-step FSR spectrum is both sharply peaked and has a kinematic edge, leading to a poor fit.

In Fig.~\ref{fig:BestFit} we show the $\Delta \chi^2$ values between the best fit at a given step number $n$ and the best fit overall, for each final state. We show results for both $\epsilon_f=0.3$ and $0.1$ in all cases, and also include $\epsilon_f=0.01$ for electrons. As expected the results do not depend strongly on $\epsilon_f$, especially in the case of taus, which is in accord with the results of Fig.~\ref{fig:0step-sig}. Note that the nominal overall best fit for the taus ($n=4$) falls into the kinematically disallowed (inconsistent) region; $n=4$ cannot be physically accommodated within 3$\sigma$ of its preferred DM mass. For this reason the results for taus and $b$-quarks were rerun allowing only self-consistent scenarios (in the sense of Eq.~\ref{eq:kinematic}); in these cases we obtain the results shown by the blue dotted curves in Fig.~\ref{fig:BestFit}. We summarize the best fit results for $\epsilon_f=0.3$ in Table~\ref{table:results0p3} and the 1$\sigma$ range as determined from Fig.~\ref{fig:BestFit} on these parameters in Table~\ref{table:results1sig}. 

\begin{table}[t!]\vspace{0.18in}
\begin{center}
\begin{tabular}{| c || c | c | c | c |}
    \hline
    Final State & $n$-step & $m_\chi$ (GeV)& $\sigma v$ ($\textrm{cm}^3/\textrm{sec}$) & $\chi^2$ \\ \hline \hline
    e & 5 & 67.2 & $2.9 \times 10^{-24}$ & 26.82 \\ \hline
    $\mu$ & 4 & 53.0 & $9.9 \times 10^{-25}$ & 26.94 \\ \hline
    $\tau_{\textrm{unphysical}}$ & 4 & 59.4 & $4.6 \times 10^{-26}$ & 24.13 \\ \hline
    $\tau_\textrm{physical}$ & 2 & 24.1 & $1.4 \times 10^{-26}$ & 25.59 \\ \hline
    $b$ & 2 &  91.2 & $ 3.9 \times 10^{-26}$ & 22.42 \\
    \hline
\end{tabular}
\end{center}
\vspace{-0.2cm}
\caption{Best fit to DM annihilations to various final states with $\epsilon_f = 0.3$. For the case of taus we show a best fit point if we include kinematically disallowed masses (unphysical) and also if we restrict ourselves to physical masses as discussed in Sec.~\ref{sec:methods}. Fits were performed over 20 degrees of freedom.}
\label{table:results0p3}
\end{table}

\begin{table}[t!]\vspace{0.2in}
\begin{center}
  \begin{tabular}{| c || c | c | c |}
    \hline
    Final State & $n$-step & $m_\chi$ (GeV)& $\sigma v$ ($\textrm{cm}^3/\textrm{sec}$) \\ \hline \hline
    e & 3-6 & 28-107 & $10^{-24.0}$-$10^{-23.3}$ \\ \hline
    $\mu$ &  2-5  & 22-89  & $10^{-24.5}$-$10^{-23.7}$ \\ \hline
    $\tau_{\textrm{unphysical}}$ & 3-5  & 37-94 & $10^{-25.6}$-$10^{-25.1}$ \\ \hline
    $\tau_\textrm{physical}$ & 2 & 24.1 & $10^{-25.8}$ \\ \hline
    $b$ & 0-3 & 40-150  & $10^{-25.8}$-$10^{-25.2} $ \\
    \hline
  \end{tabular}
\end{center}
\vspace{-0.2cm}
\caption{Range of parameters within 1$\sigma$ of the best fit step for $\epsilon_f=0.3$ for electrons, muons, taus and $b$-quarks. As in Table~\ref{table:results0p3} we show both physical and unphysical tau results.}
\label{table:results1sig}
\end{table}

\begin{figure}[t!]
\centering
\includegraphics[scale=0.7]{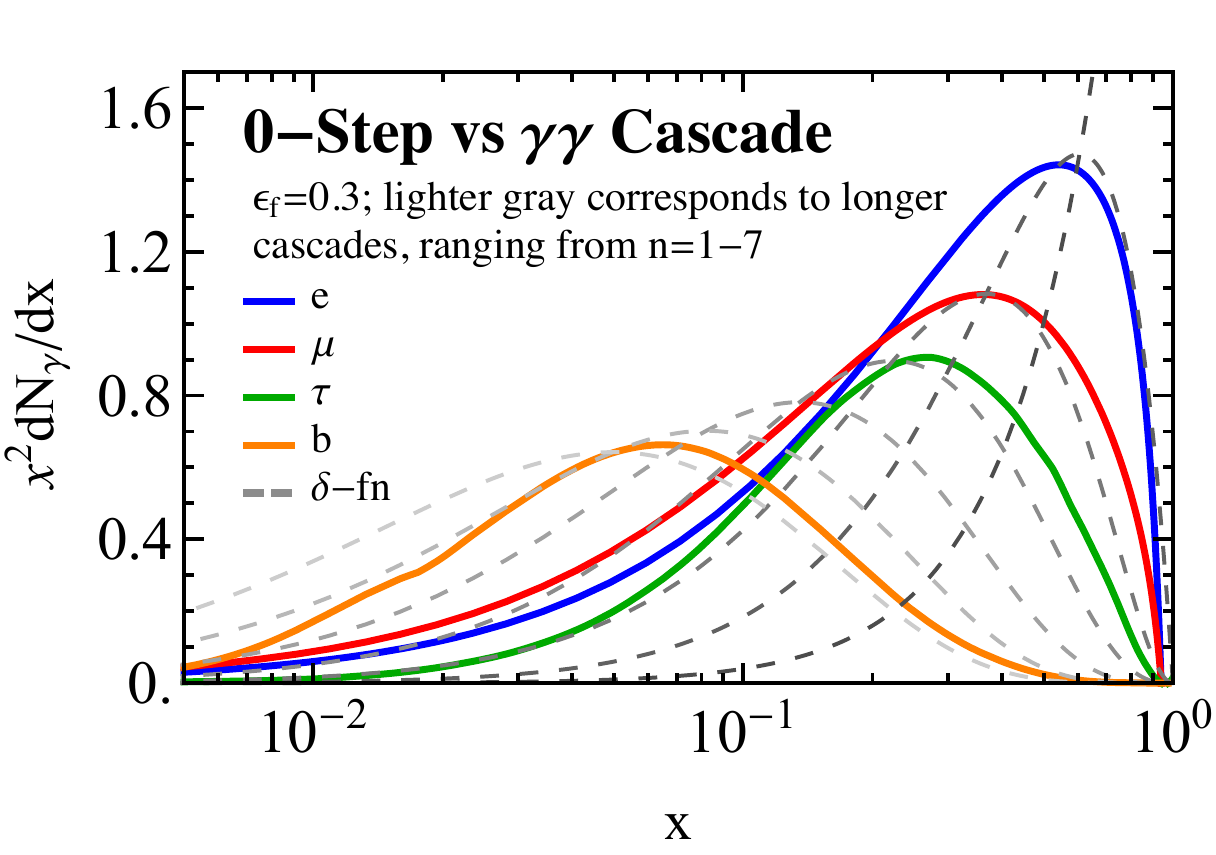}
\vspace{-0.2cm}
\caption{\footnotesize{The 0-step spectra for e, $\mu$, $\tau$ and $b$-quarks with $\epsilon_f=0.3$ are shown as the blue, red, green and orange curves. The dashed curves show the spectrum of a hierarchical $n$-step cascade that ends in $\phi_1 \to \gamma \gamma$ (a $\delta$-function in the $\phi_1$ rest frame) for $n=1-7$, with lighter curves corresponding to progressively longer cascades. In order to compare the shape of the spectra we have magnified the 0-step spectra by a factor of $470$, $190$, $6.2$ and $3.1$ for e, $\mu$, $\tau$ and $b$-quarks respectively. We see the electron spectrum is closest to a 2-3 step cascade ending in a $\delta$-function, muons and taus are closest to a 3-4 step cascade, whilst $b$-quarks most resemble 6-7.}}
\label{fig:DeltaCascade}
\end{figure}

In Fig.~\ref{fig:Spectrum0p3Full} we show the overall best fit spectrum for electron, muons, taus, and $b$-quarks with $\epsilon_{f} = 0.3$. Although the spectra for direct annihilation to these final states are quite different, after introducing the freedom to have multi-step cascades, a similar best fit spectrum is picked out in each case. 
To expand on this, we can compare the various 0-step spectra - as displayed in Fig.~\ref{fig:0step-sig} - to the result of a hierarchical $n$-step cascade that ends in $\phi_1 \to \gamma \gamma$. This comparison is shown in Fig.~\ref{fig:DeltaCascade}. The spectrum of photons from this process is just a $\delta$-function in the $\phi_1$ rest frame, and is in a sense the simplest possible photon spectrum. We find that the photon spectrum from direct annihilation to electrons is similar to that obtained by a 2-3 step cascade terminating in $\phi_1 \rightarrow \gamma \gamma$; for muons and taus the closest match is a 3-4 step cascade; and for $b$-quarks 6-7. Of course these correspondences are not exact -- for example, the $b$-quark spectrum is more complex than just applying Eq.~\ref{eq:boosteq-sig} to a $\delta$-function -- but they allow us to regard these 0-step spectra as arising approximately from a common ($\delta$-function) spectrum convolved with differing numbers of cascade steps. We can then intuit how many additional steps are required in each case, to bring the spectra to a similar shape. Combining these numbers with the preferred number of steps seen in Table~\ref{table:results0p3}, we find the GCE prefers a spectrum that can be roughly modeled as a $\delta$-function occurring at the endpoint of 7-9 cascade decays. In this sense it seems fits to the GCE prefer a cascade with a large number of steps, and that these can occur in the SM or dark sector.

Likewise, this general picture can approximately describe showers in the dark sector \cite{Freytsis:2014sua}. Such showers will effectively contain decay cascades of different lengths, but we find that the spectrum of \cite{Freytsis:2014sua} can be well described by a $\delta$-function $\phi_1 \rightarrow \gamma \gamma$ broadened by $\sim 3$ decay steps. The best-fit scenario found in that paper corresponds to a DM mass of $\sim 10$ GeV; this is consistent with the preferred mass for our 1-step electron case, which also corresponds to a $\delta$-function at the endpoint of a $\sim 3$-step cascade. A better fit to the data might therefore be obtained by combining such dark showering with a short dark-sector cascade. In Sec.~\ref{sec:generalcascade} we will return to this point, and discuss the sense in
which our results may be used to estimate the parameter space for dark shower models.

\subsection{Different Final States}
A few comments about the various final states are in order.

\textit{Electrons:} The photon spectrum from direct annihilations $\chi \chi \rightarrow e^+ e^-$ is sharply peaked. This tends to produce a worse fit to the GCE. As such we need several steps in the cascade in order to broaden the spectrum sufficiently to allow for a parameter space where a significantly improved fit is possible, and this is shown by the substantial decrease in the quality of fit at low $n$ in Fig.~\ref{fig:BestFit}. It should be noted that any model for the GCE with direct annihilation into electrons will likely be in severe tension with the data from AMS \cite{Bergstrom:2013jra}. This tension is likely to persist for at least the $n=1$ cascade, and possibly higher steps as well.\footnote{\footnotesize{Private communication, Wei Xue.}} As we go to higher-step cascades the spectrum broadens and the AMS bounds are expected to weaken, but the exact bounds should be worked out for any cascade scenario with a branching fraction to electrons. For the purposes of this chapter, we use the electron case as an example of a sharply peaked photon spectrum to demonstrate the impact of the spectral broadening, not necessarily as a realistic explanation for the excess. Similarly, constraints on DM annihilation from the cosmic microwave background (CMB) \cite{Ade:2015xua} are likely to rule out both the electron and muon favored regions shown in Fig.~\ref{fig:AllStatesLogChiPlot}, while leaving the $b$ and tau regions largely unconstrained. The figure of merit for CMB constraints is $\langle \sigma v \rangle/m_\chi$ \cite{Chen:2003gz, Padmanabhan:2005es}, up to an $\mathcal{O}(1)$ factor which is channel- and spectrum-dependent \cite{Slatyer:2009yq, Madhavacheril:2013cna}. As discussed above, for the best-fit regions (for hierarchical decays), this quantity scales as $\sim m_\chi^{0.3}$ as the number of steps increases; thus, we expect the constraint to become slightly stronger for longer cascades.

\textit{Muons:} In Fig.~\ref{fig:BestFit} we see that the muon final state spectrum has the same qualitative behavior as the electrons, and will be subject to similar constraints. This is unsurprising as the muon spectrum is quite similar to that from electrons, albeit with a less pronounced peak (see Fig.~\ref{fig:0step-sig}).

\textit{Taus:} As with other leptonic final states, taus also prefer multi-step cascades for the best fit. Note that the best fit point at 4 steps is in fact kinematically disallowed (inconsistent) as can be seen in Fig.~\ref{fig:AllStatesLogChiPlot} and as discussed in Sec.~\ref{sec:methods}. However, the best fit point after imposing the consistency condition, at 2 steps, is still a better fit than the high-step cases with electron and muon final states.

\textit{$b$-quarks:} DM annihilation to $b$-quarks is the preferred channel for direct annihilation identified in \cite{Daylan:2014rsa,Calore:2014xka}, where it already provides a good fit. Accordingly there is no need to broaden the spectrum with a large number of cascades -- however, as we will discuss in Sec.~\ref{sec:signalsconstraints}, even a short cascade can greatly alleviate constraints from colliders and direct searches (see also \cite{Martin:2014sxa,Abdullah:2014lla} and references therein). A cascade with several steps can still give an equally good or slightly better fit, and of course accommodates higher masses than for the case of direct annihilation. However, since the spectrum is already fairly broad, adding too many additional steps makes the fit worse, as shown in Fig.~\ref{fig:BestFit}. Accordingly, the DM mass cannot be pushed far above 100 GeV without significantly worsening the fit, at least in the context of hierarchical cascades.

\newpage
\subsection{Sensitivity to Systematics and Energy Cuts}

\begin{figure}[t!]
\centering
\includegraphics[scale=0.65]{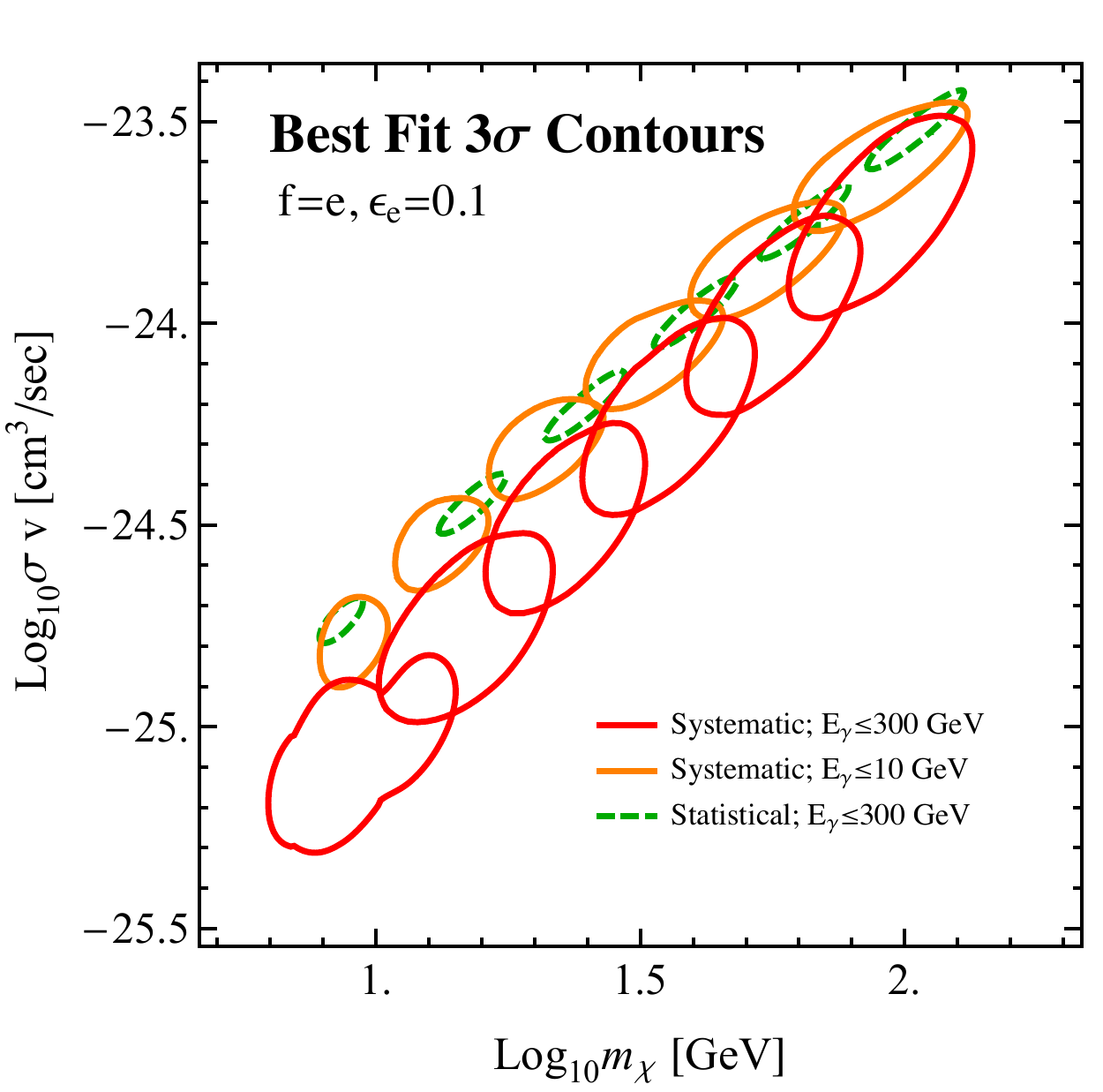}
\vspace{-0.2cm}
\caption{\footnotesize{The 3$\sigma$ contours for 1-6 step cascade annihilations to final state electrons with $\epsilon_e = 0.1$. Red contours correspond to fitting over the entire energy range $0.5~\textrm{GeV} \leq E_\gamma \leq 300~\textrm{GeV}$ with the full covariance matrix of \cite{Calore:2014xka}. Orange contours correspond to fitting with a cut on high energies $E_\gamma \leq 10~\textrm{GeV}$. Green contours correspond to a fit over the full energy range but with only the statistical errors of \cite{Calore:2014xka}.}}
\label{fig:Electron0p1StatsPlot}
\end{figure}

\begin{figure}[t!]
\centering
\begin{tabular}{c}
\hspace{0.04in}\includegraphics[scale=0.5]{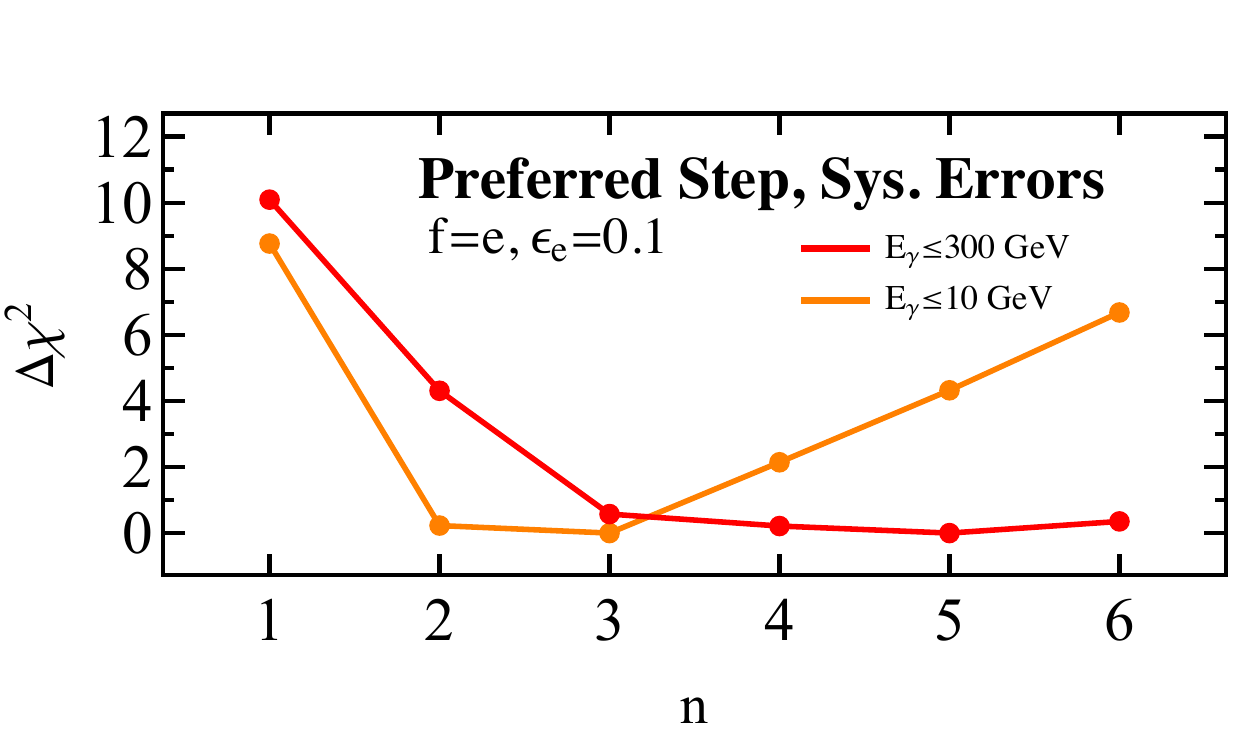} \hspace{0.16in}
\includegraphics[scale=0.51]{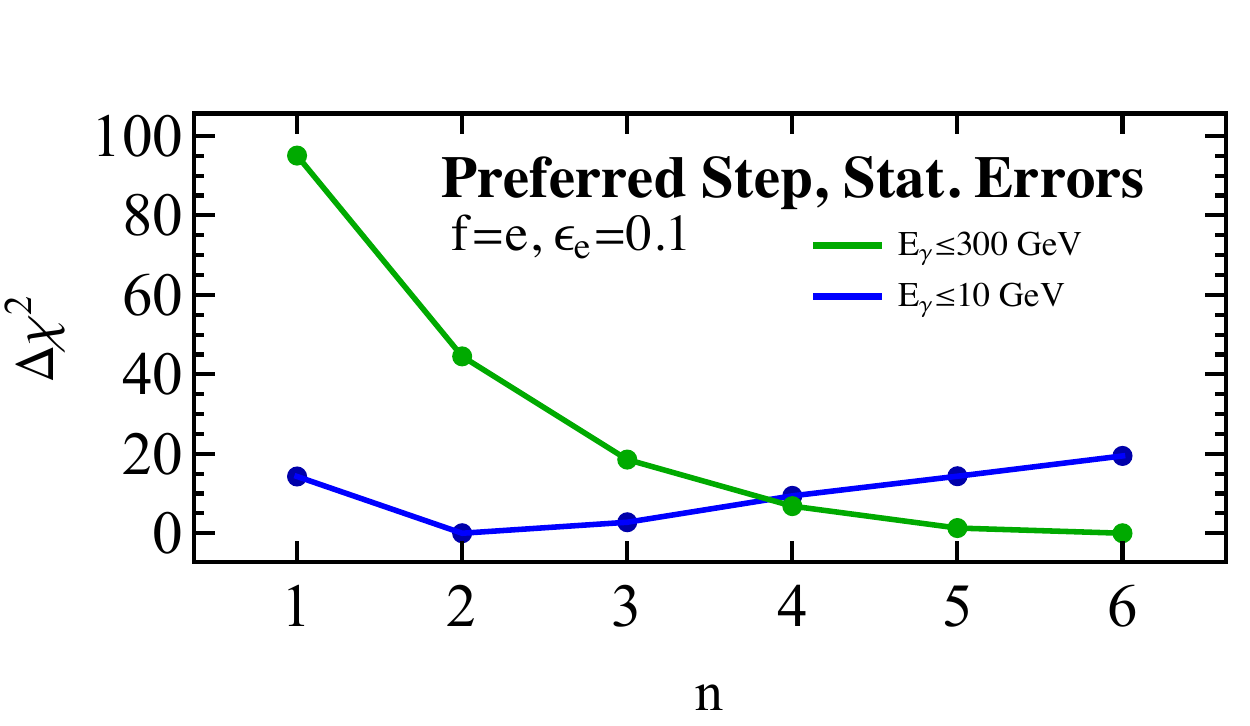} \\
\includegraphics[scale=0.535]{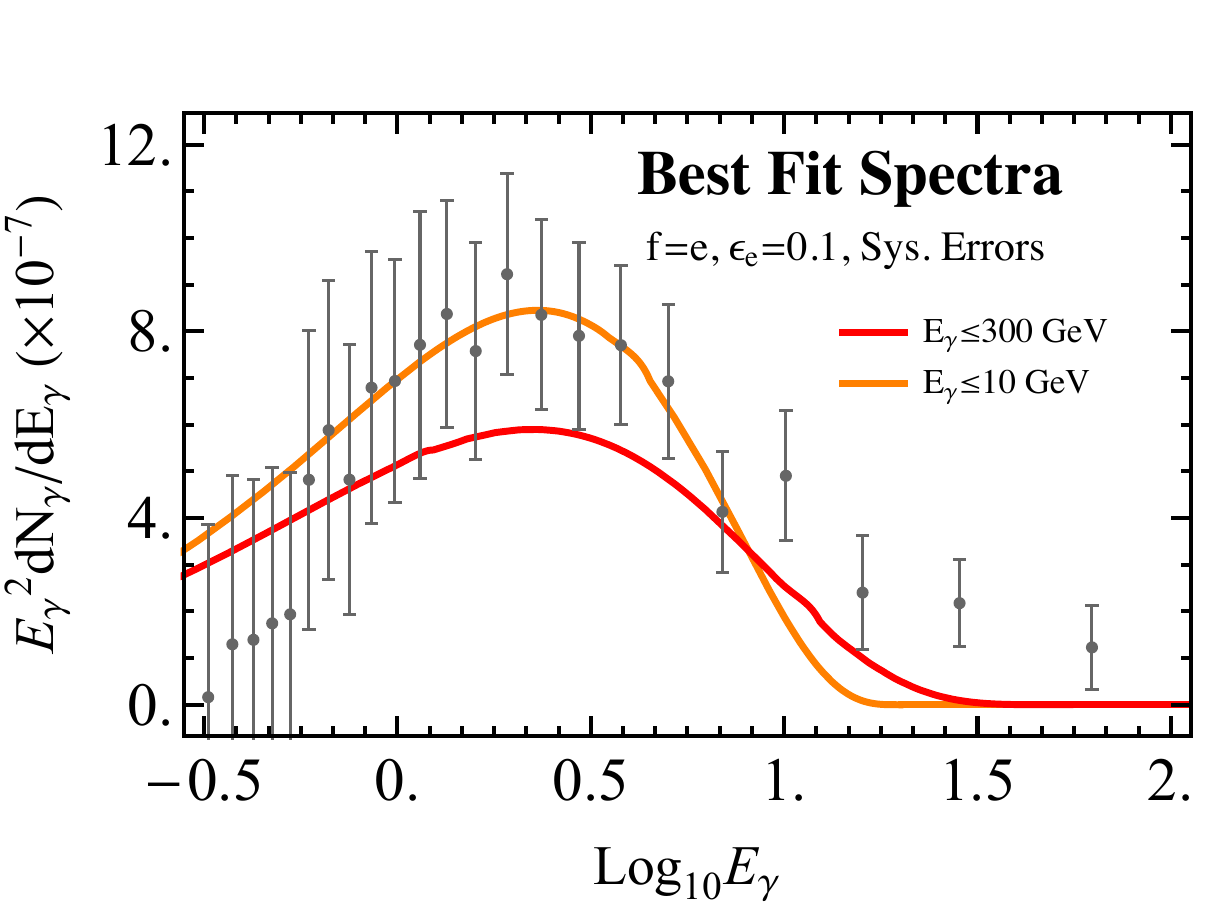} \hspace{0.15in}
\includegraphics[scale=0.535]{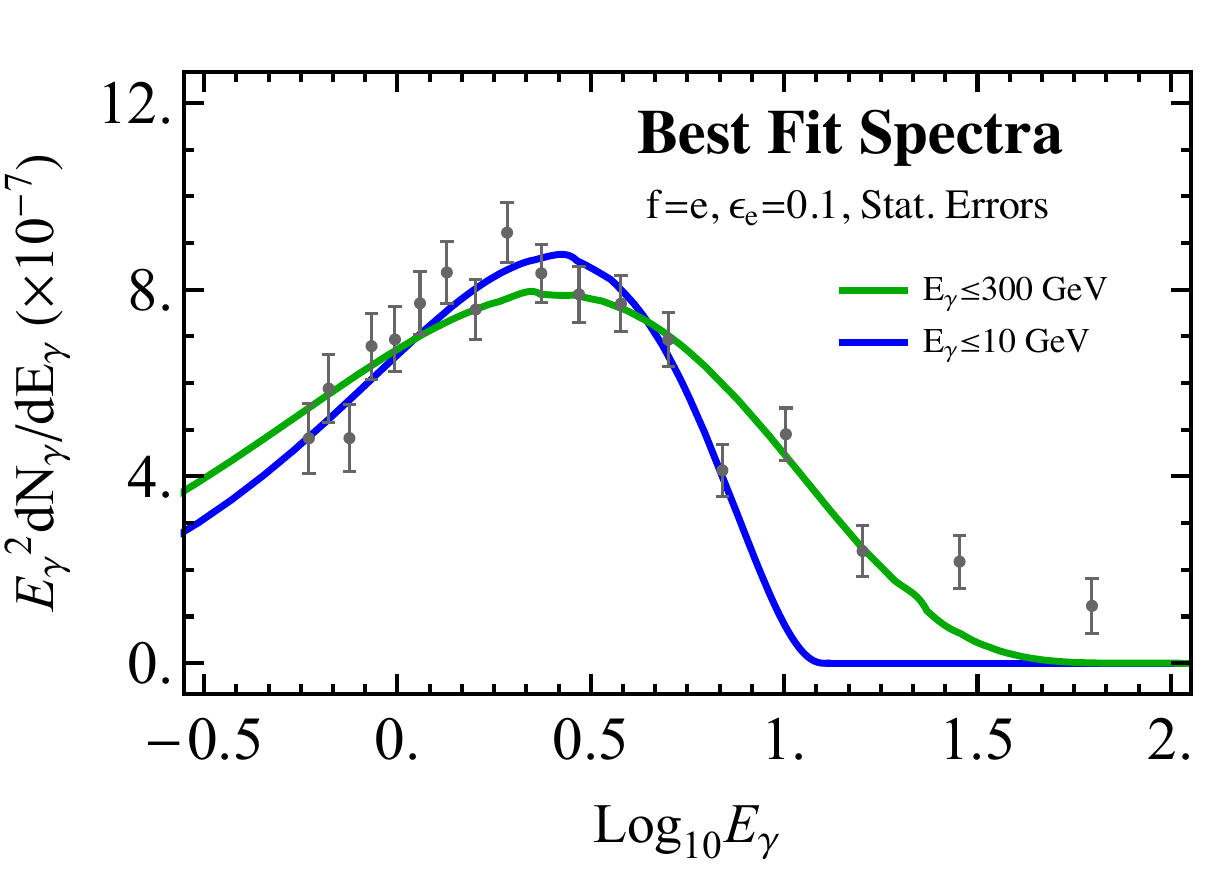}
\end{tabular}
\vspace{-0.2cm}
\caption{\footnotesize{Top Panels: We show the impact on the preferred number of steps when changing the energy range and error types considered. Each curve is for final state electrons with $\epsilon_e=0.1$. The left figure shows the use of systematic errors over the full and a restricted energy range ($E_\gamma \leq 10$ GeV) in red and orange respectively. The right figure is the equivalent for statistical errors, with the full energy range shown in green and the restricted in blue. Bottom panels: Here the best fit curves as determined from the top panels are shown with the appropriate data and errors from \cite{Calore:2014xka} overlaid, for the example case of the electron final state. The left panel shows the results for systematic errors, where the best fit point was $n=5$ for the full range (red curve) and $n=3$ for the restricted range (orange curve). The right panel shows the equivalent for statistical errors, where for the full range the $n=6$ curve is shown in green and for the restricted range the $n=2$ curve is in blue.}}
\label{fig:StatsDeltaChiandSpecPlot}
\end{figure}

In the results presented above we have fit the data of \cite{Calore:2014xka} to the photon spectrum from DM annihilations through multi-step cascades to various final states. The fit was performed over the energy range $0.5~\textrm{GeV} \leq E_{\gamma} \leq 300~\textrm{GeV}$. There is some evidence that the emission detected above 10 GeV may not share the same spatial profile as the main excess, suggesting a possible independent origin (for example, these high-energy data appear to prefer a morphology centered at negative $\ell$ and with a shallow spatial slope \cite{Calore:2014xka}), so we also test the impact of omitting the data above 10 GeV. Finally, we explore the impact of including only the statistical uncertainties of \cite{Calore:2014xka}, omitting systematic errors, to test the degree to which the constraints could improve with reduction in the systematic uncertainties.

We display the results of this study in Fig.~\ref{fig:Electron0p1StatsPlot}-\ref{fig:StatsDeltaChiandSpecPlot}, for the case of $n$-step cascade annihilations to final state electrons with $\epsilon_e = 0.1$. Annihilations to other final states generically display the same behavior as the energy range and error estimates are varied. Cutting out the high energy data points generically shifts the fit to prefer lower masses and narrower spectra, and therefore corresponds to cascades with fewer steps -- resembling a $\delta$-function at the endpoint of a 5-7 step cascade, rather than a 7-9 step cascade. At a fixed number of steps, the main impact of omitting the high-energy data points is to raise the preferred cross-section and shrink the contours. Understanding the high-energy data will thus be important in distinguishing quantitative models for the GeV excess. 

Fitting over statistical errors increases the actual $\chi^2$ values, and the rate at which $\chi^2$ increases away from its minimum (as expected), as demonstrated by the shrinking green contours of Fig.~\ref{fig:Electron0p1StatsPlot}. The overall preferred step in the cascade however is not dramatically affected, only changing by 0-1 steps, as shown in the top panels of Fig.~\ref{fig:StatsDeltaChiandSpecPlot} - we display the corresponding best fit spectra in the bottom panels. At a fixed number of steps, the preferred cross-section increases, becoming more similar to what we find when omitting the high energy points.

\section{Interpretation for General Cascades}
\label{sec:generalcascade}

\subsection{Relaxing the Assumption of Large Hierarchies}

The results displayed in the previous section were obtained assuming large mass hierarchies between each cascade step.  It is possible to recast these results to gain insight into the case of general $\epsilon_i$ values. To see this, consider the decay $\phi_{i+1} \rightarrow \phi_i \phi_i$. As previously discussed, in the limit when two mass scales become degenerate ($\epsilon_i \to 1$), an $n$-step cascade effectively reduces to an $(n-1)$-step cascade, except for the additional final state multiplicity.  Thus adding a degenerate step to a cascade is much simpler than adding one with a large hierarchy: we need only multiply the spectrum by two to account for the increased multiplicity, and halve the photon energy scale to account for the initial energy being spread between twice as many particles. (For completeness, we check analytically that the limit of $\epsilon_i\rightarrow 1$ has this behavior in Appendix~\ref{app:boost}.)

In light of this, an $n$-step cascade with one degenerate step and an $(n-1)$-step hierarchical cascade must provide equally good fits to the GCE, with the former preferring twice the annihilation cross-section and DM mass relative to the latter. The increased DM mass results from the halving of the energy scale, whilst to understand the cross-section we look back to Eq.~\ref{eq:xsecnorm}: adding the degenerate step doubles the photon multiplicity, which halves $\eta$ to compensate, but the doubling of the DM mass means overall the cross-section is increased by a factor of two. As such the results in Fig.~\ref{fig:AllStatesLogChiPlot} can be readily extended for additional degenerate steps. For each additional degenerate step on top of an initial hierarchical cascade (the degenerate step may occur anywhere in the cascade), the shape of the $\chi^2$ contours remains the same, but shifted upward by a factor of two in mass and cross-section. With a sufficiently large number of degenerate decays, the DM mass required to fit the GCE could be made arbitrarily high, although this would seem to require considerable fine-tuning. (A natural scenario in which one degenerate step arises due to a symmetry is discussed in \cite{Fan:2012gr}.)

\begin{figure}[t!]
\centering
\includegraphics[scale=0.75]{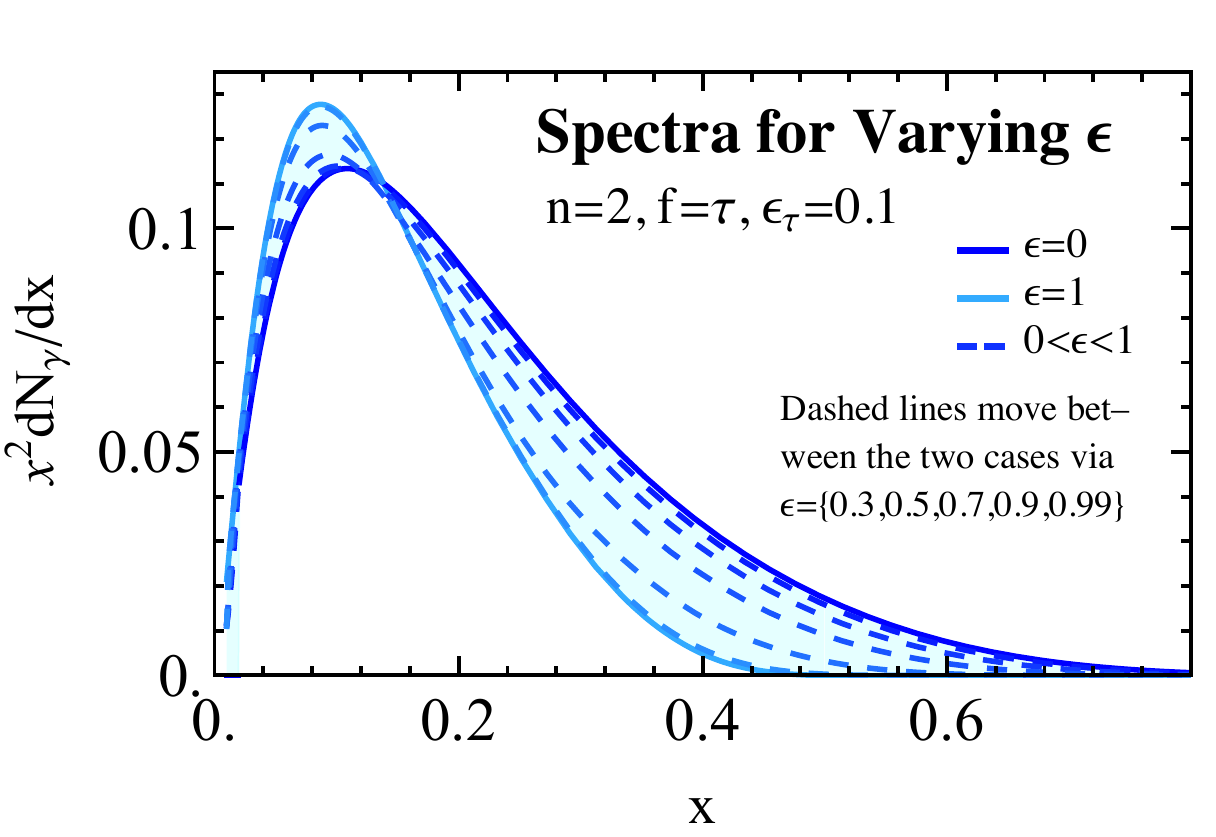}
\vspace{-0.2cm}
\caption{\footnotesize{The transition of the spectra between $\epsilon_2=0$ and $\epsilon_2=1$, calculated using Eq.~\ref{eq:fullepsilon}. The example case is a 2-step cascade with final state taus and $\epsilon_\tau = 0.1$. The dark blue is for $\epsilon=0$ and is what would result from the large hierarchies approximation. The $\epsilon=1$ case shown in light blue corresponds to a completely degenerate spectrum, and as such is equivalent to a shifted 1-step curve. In between these two, we show intermediate $\epsilon$ values as the dashed curves, specifically $\epsilon=\{0.3,0.5,0.7,0.9,0.99\}$. Note the rate of transition between the two cases is in keeping with the error in the large hierarchies case being of order $\mathcal{O}(\epsilon_i^2)$.}}
\label{fig:fullEpsilon0p1}
\end{figure}

\begin{figure}[t!]
\centering
\includegraphics[scale=0.6]{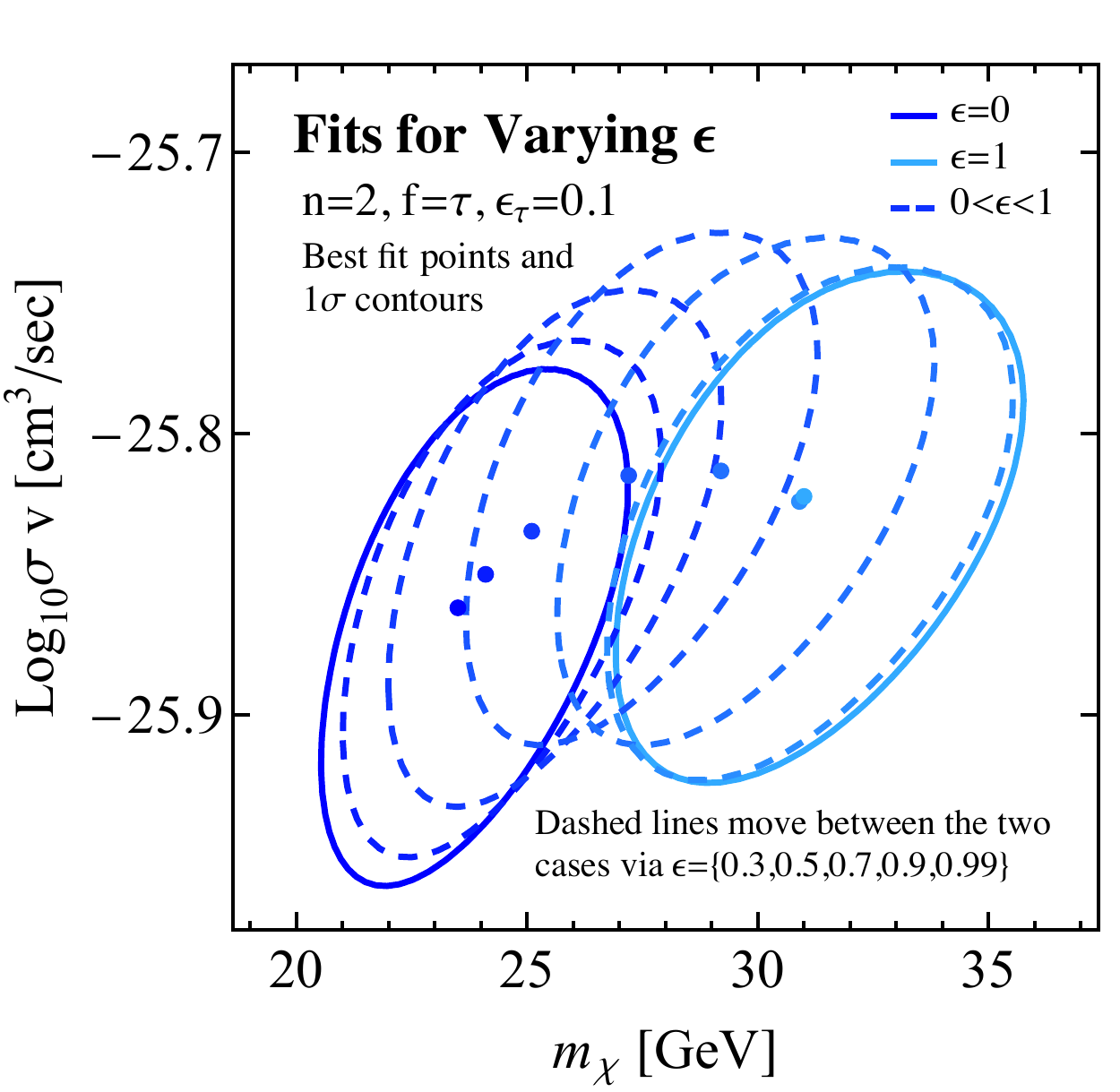}
\vspace{-0.2cm}
\caption{\footnotesize{The transition of the best fit point and 1$\sigma$ contours between $\epsilon_2=0$ and $\epsilon_2=1$, calculated using Eq.~\ref{eq:fullepsilon}. The example case is a 2-step cascade with final state taus and $\epsilon_\tau = 0.1$. The transition is between the $\epsilon=0$ case in dark blue and $\epsilon=1$ in light blue. The dashed curves map out the transition with intermediate values, specifically $\epsilon=\{0.3,0.5,0.7,0.9,0.99\}$.}}
\label{fig:fullEpsilonFits}
\end{figure}

Cascades with general values of $\epsilon_i$ in turn interpolate between the two simpler cases already considered, with small and large $\epsilon_i$. We give the general convolution formula in Appendix~\ref{app:boost}, and an example of how spectra evolve as a single $\epsilon_i$ shifts from 0 to 1 is shown in Fig.~\ref{fig:fullEpsilon0p1}. This interpolation provides an alternate interpretation for Fig.~\ref{fig:BestFit}: the $n$ on the $x$-axis of these plots can be thought of as representing the number of steps with a large hierarchy, rather than the total number of steps. If one of these steps becomes degenerate (while holding the total number of steps fixed), as previously discussed, we will move from $n$ to $n-1$ steps in terms of the spectral shape and hence quality of fit. Intermediate $\epsilon_i$ values will interpolate smoothly between these two cases. Thus for any arbitrary collection of hierarchical and degenerate steps, the quality of the fit and the location of the best-fit region in $m_\chi-\langle\sigma v\rangle$ parameter space can already be estimated from Figs.~\ref{fig:AllStatesLogChiPlot}-\ref{fig:BestFit}. A concrete example of the transition in preferred DM mass and cross-section is shown in Fig.~\ref{fig:fullEpsilonFits}, which corresponds to the variation of the spectrum shown in Fig.~\ref{fig:fullEpsilon0p1}. The curve plotted out by the best fit point for intermediate values of $\epsilon$ is not a straight line between the two extreme values, but does not deviate far from this. Similar behavior was seen for other final states and choice of degenerate step.

At a fixed DM mass, the perturbation to the $\epsilon_i=0$ photon spectrum evolves roughly as $\epsilon_i^2$ as $\epsilon_i$ varies from 0 to 1 (as discussed in Appendix~\ref{app:boost}); this behavior is shown in Fig.~\ref{fig:fullEpsilon0p1}, where the $\epsilon_2=0.3$ spectrum is almost indistinguishable from the $\epsilon_2=0$ spectrum, and $\epsilon_2=0.5$, $\epsilon_2=0.7$ and $\epsilon_2=0.9$ give spectra intermediate between the $\epsilon_2=0$ and $\epsilon_2=1$ cases. The perturbation to the best-fit $\chi^2$ will tend to increase even \emph{more} slowly than $\epsilon_i^2$, in the case where $\epsilon_i=0$ is a better fit than $\epsilon_i=1$, since the DM mass and cross-section can float to absorb changes in the spectrum and reduce the increase in $\chi^2$. In all examples tested the best-fit $\chi^2$ remains essentially unchanged from the $\epsilon_i=0$ case out to $\epsilon_i=0.7$.

\begin{figure}[t!]
\centering
\includegraphics[scale=0.65]{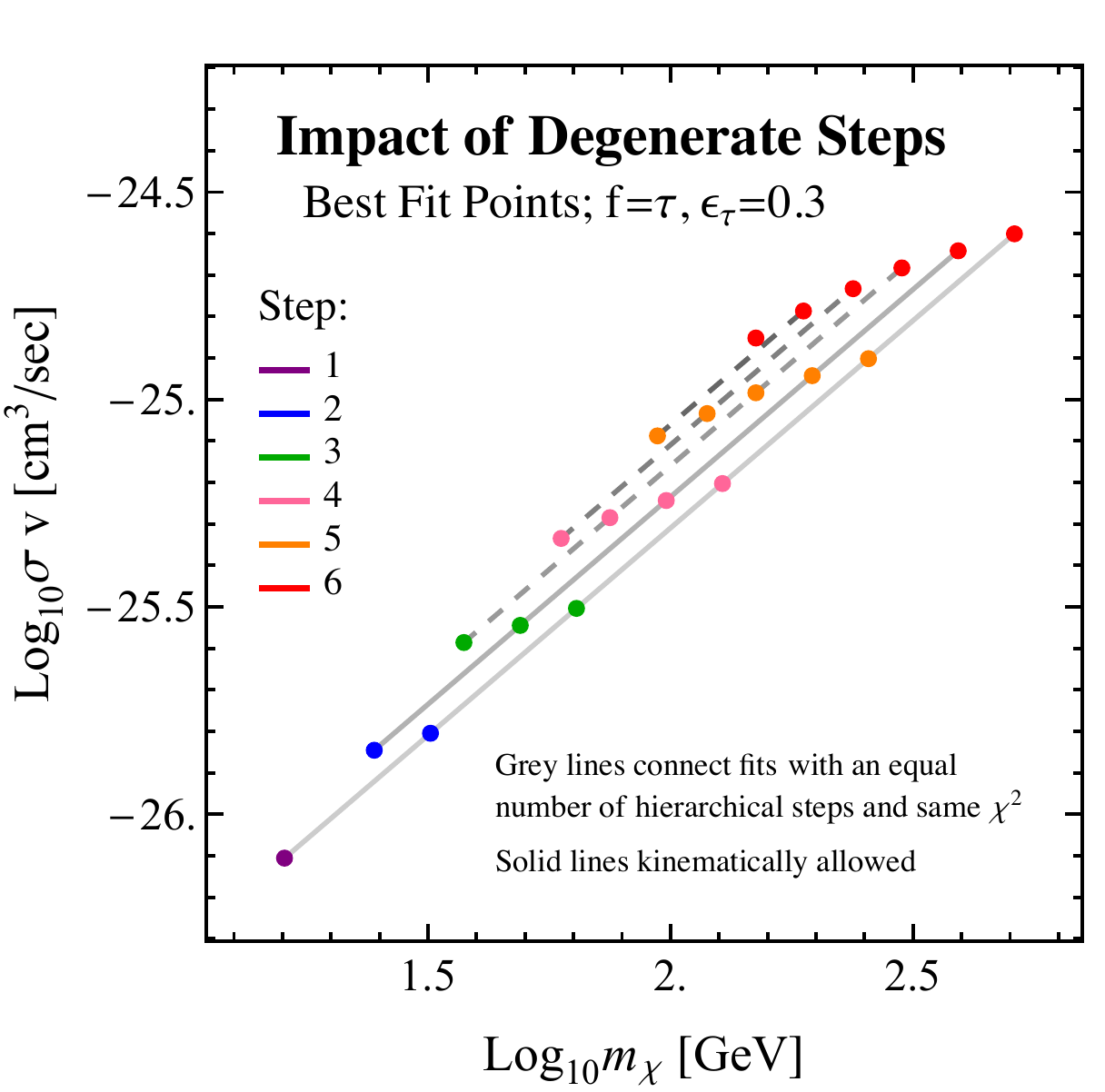}
\vspace{-0.2cm}
\caption{\footnotesize{The purple, blue, green, pink, orange and red points correspond to the best fit $(m_{\chi}, \sigma v)$ point for a total number of cascade steps (degenerate + hierarchical) $n$ = 1, 2, 3, 4, 5, 6 respectively; for annihilations to final state taus with $\epsilon_\tau = 0.3$. Points living on the same line have the same number of hierarchical steps and therefore result in equally good fits to the data. Points of the same color, but with progressively greater values of $(m_\chi, \sigma v)$, correspond to successively replacing hierarchical steps with degenerate steps, holding the number of total steps fixed. For the above case of taus only the one and two step hierarchical cascades are kinematically allowed as indicated in Fig.~\ref{fig:AllStatesLogChiPlot} (note that the kinematic constraint applies to lines as a whole, not individual points; see text), thus only points living on the solid lines are allowed as these lines correspond to cascades with one and two hierarchical steps respectively.}}
\label{fig:DegFit}
\end{figure}

In general a cascade with $n$ total steps, $n_d$ of which are degenerate ($n_d$ values of $\epsilon_i \rightarrow 1$) will have the same spectrum as a cascade with $(n-n_d)$ hierarchical steps with a factor of $2^{n_d}$ enhancement in mass and cross-section. This is illustrated in Fig.~\ref{fig:DegFit} for the case of decays to final state $\tau$'s with 1-6 total cascade steps. Relaxing the assumption of large hierarchies therefore results in a preferred triangular slice of parameter space, bounded by curves with $\langle \sigma v \rangle \propto m_\chi$ and $\langle \sigma v \rangle \propto m_\chi^{1.3}$. We can now understand the results of  Fig.~\ref{fig:BestFit}  as mapping out the variation in $\chi^2$ when moving between \emph{classes} of scenarios, each defined by a fixed number of hierarchical steps but containing scenarios with varying numbers of degenerate steps (each of these classes is represented by a line in Fig.~\ref{fig:DegFit}). Note also that the kinematic constraint Eq.~\ref{eq:mDM-sig} acts on classes rather than individual scenarios (since adding a degenerate step doubles the DM mass but increases the number of steps by 1, strengthening the constraint on DM mass by a factor of 2); if one scenario is disallowed the entire class is disallowed.

\begin{figure}[t!]
\centering
\begin{tabular}{c}
\includegraphics[scale=0.50]{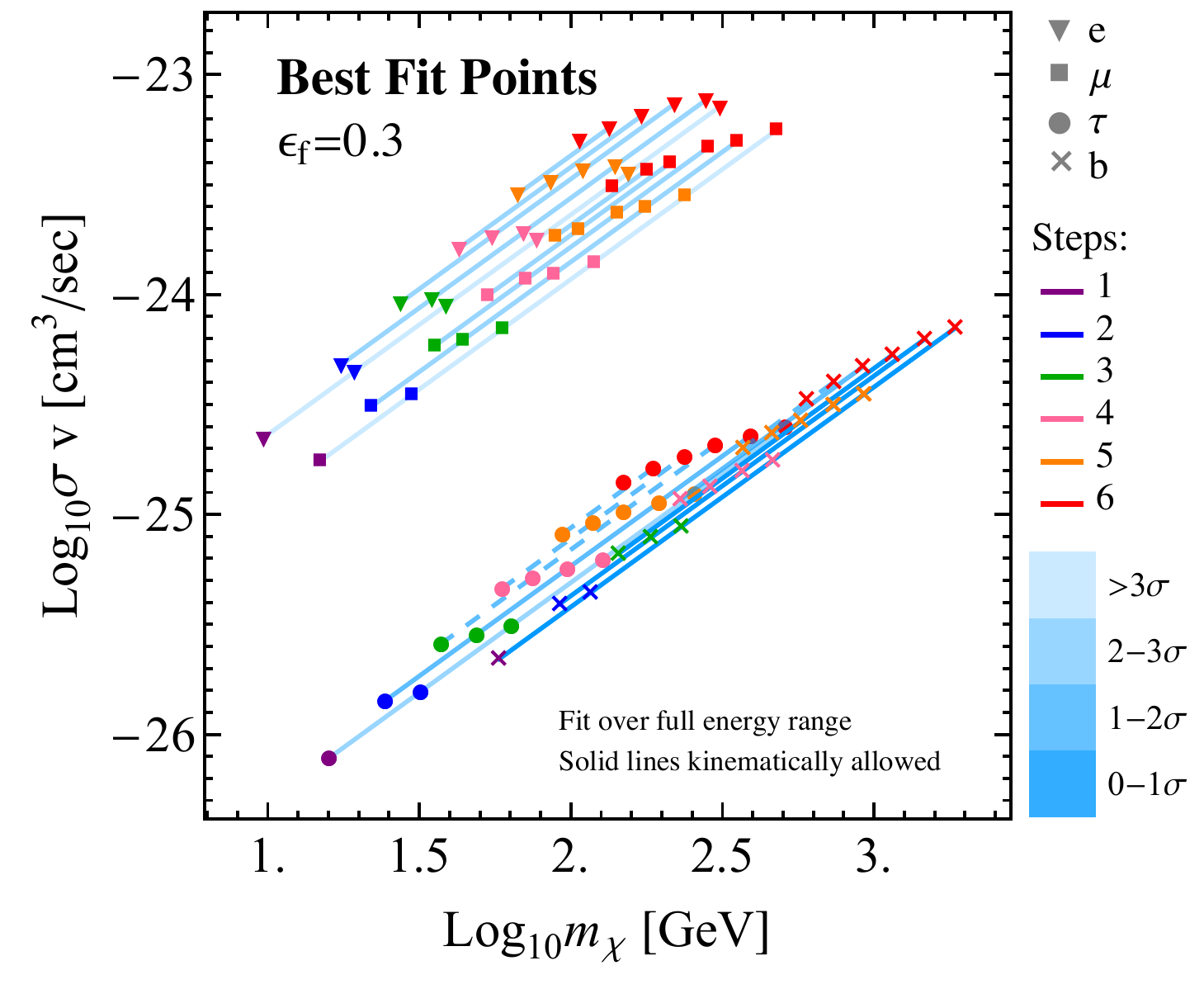}
\includegraphics[scale=0.50]{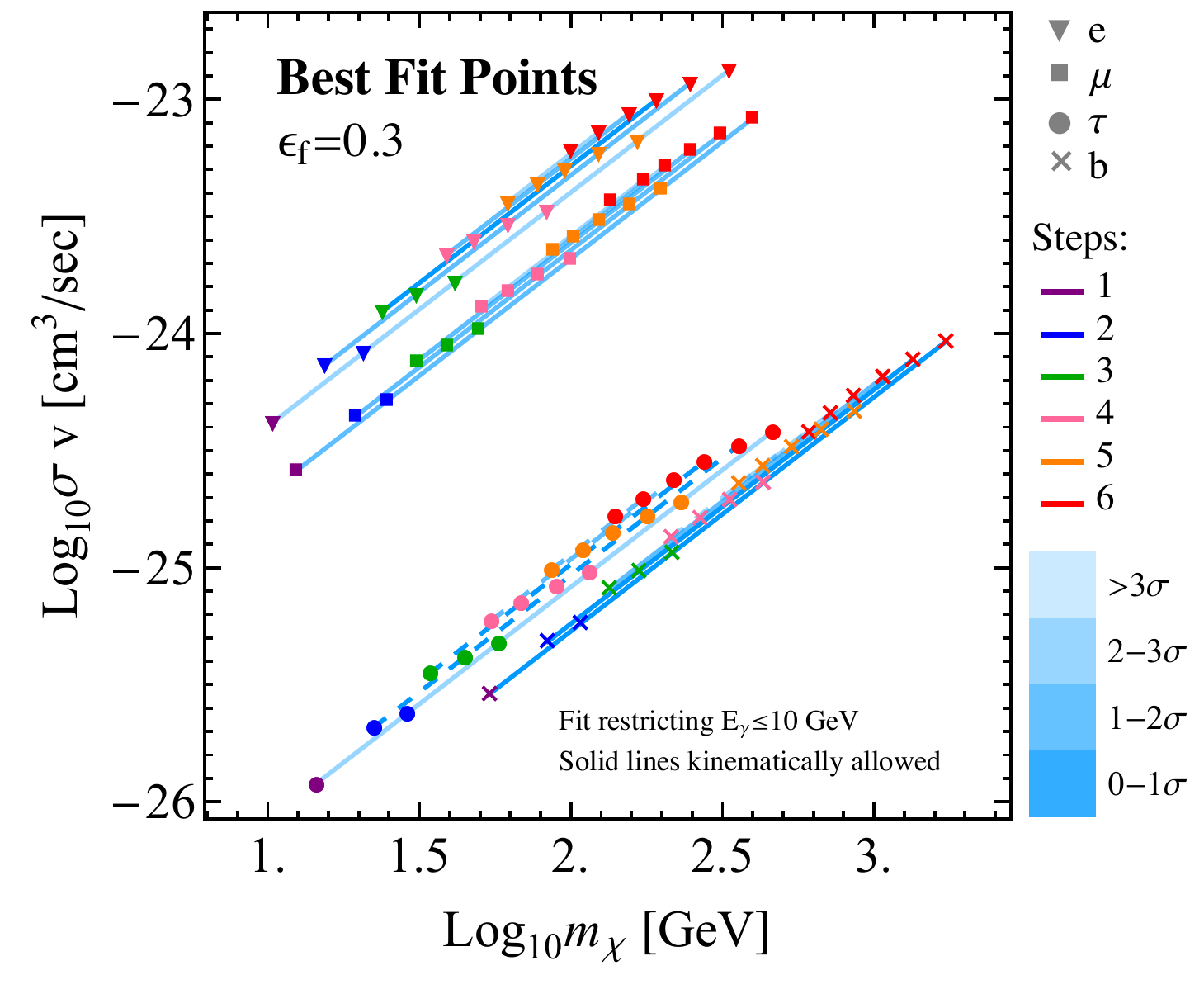}\\
\hspace{0.07in}\includegraphics[scale=0.45]{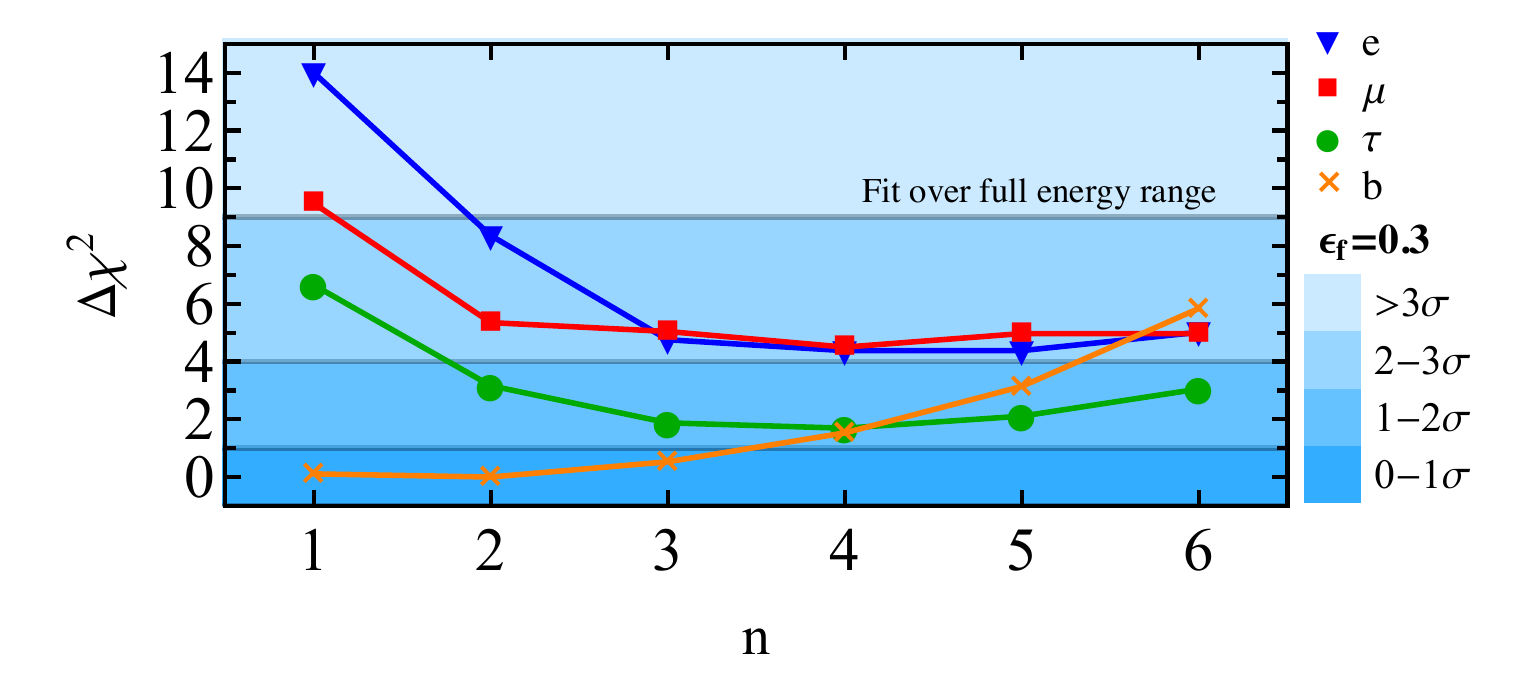}\hspace{0.13in}
\includegraphics[scale=0.45]{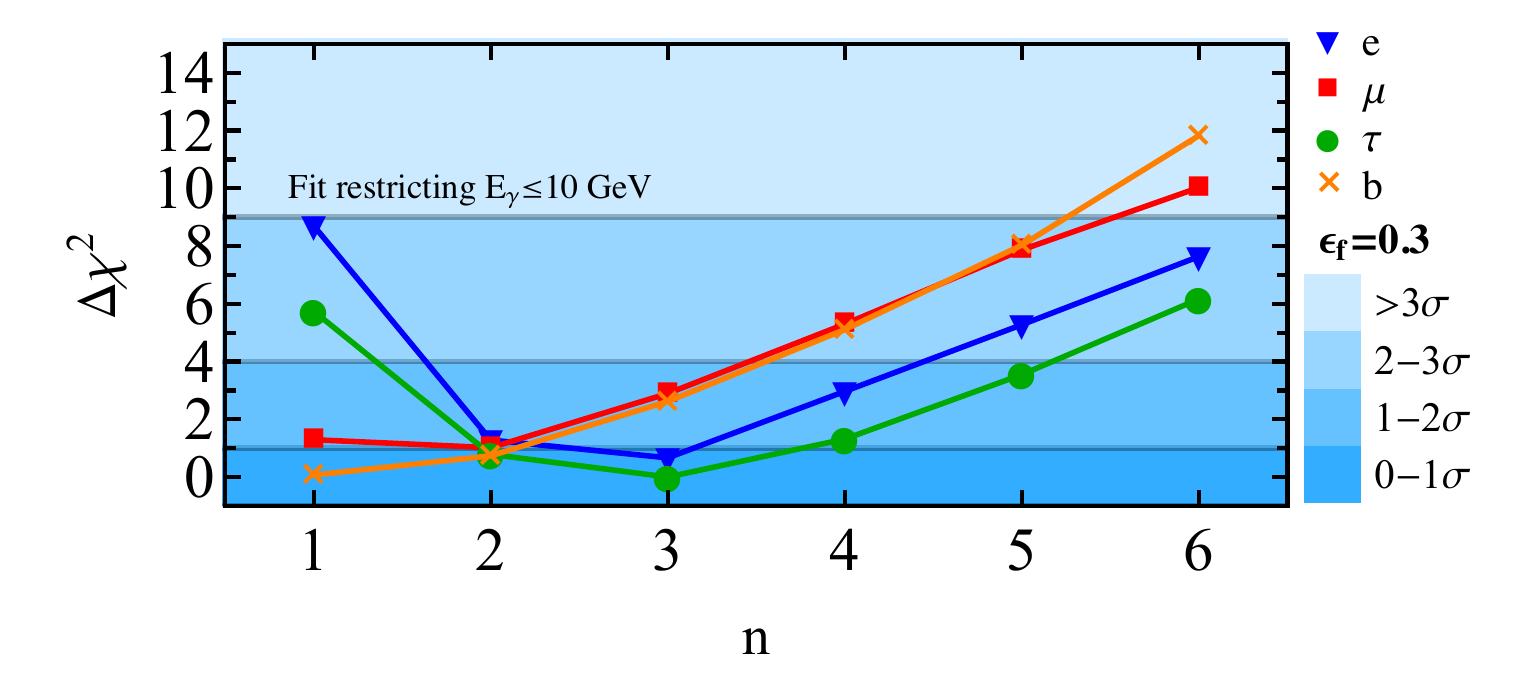}\\
\end{tabular}
\vspace{-0.2cm}
\caption{\footnotesize{Combined results of fits with $\epsilon_f = 0.3$ over the full energy range (left) or with a restriction $E_{\gamma} \leq 10$ GeV (right). Top panels: Best fit $(m_{\chi},\sigma v)$ for a cascade with 1-6 total (degenerate + hierarchical) steps ending in electrons, muons, taus of $b$-quarks. Points on the same line have the same number of hierarchical steps and therefore result in equally good fits to the data, following the discussion in Sec.~\ref{sec:generalcascade}. Points of the same color, but with sequentially greater values of $(m_{\chi},\sigma v)$, correspond to progressively replacing hierarchical steps with degenerate steps, holding the total number of steps fixed. The color of the lines indicate goodness of fit and only solid lines are kinematically allowed (as explained in see Sec.~\ref{sec:methods}). Bottom panels: Show the overall best fit for DM annihilation through an $n$-step hierarchical cascade to electron, muon, tau and $b$-quark final states. The curves show the $\Delta \chi^2$ of the best fit at that step and final state, as compared with best fit over all steps and final states. No restriction to physical kinematics is imposed, but where restrictions would apply can be inferred from the top panels. The shaded bands correspond to the quality of fit. For fits over the full energy range a fairly short cascade terminating in decay to $b$-quarks gives the preferred spectrum, whilst over the restricted energy range each final state can potentially provide approximately equally good fits.}}
\label{fig:CombinedResults}
\end{figure}

Fig.~\ref{fig:CombinedResults} summarizes our combined results. There, the top panels display the regions mapped out in the $\langle \sigma v \rangle-m_\chi$ plane by the best fit points involving 1-6 steps (either hierarchical or degenerate)  cascades to final state electrons, muons, taus and $b$-quarks. In the bottom panels, we indicate which hierarchical step and final state yield the best fit, and the comparative quality of fit for other combinations. We show all these results for fits over the full (left panels) and restricted (right panels) energy ranges. Additionally as shown in the top panels, electrons (taus) and muons ($b$-quarks) have some degree of overlap, especially once degenerate steps are included. The overlap of these regions is reduced when the high energy data points are excluded, as is clear by comparing the right and left panels.

\begin{figure}[t!]
\centering
\vspace{0.15in}
\begin{tabular}{c}
\includegraphics[scale=0.60]{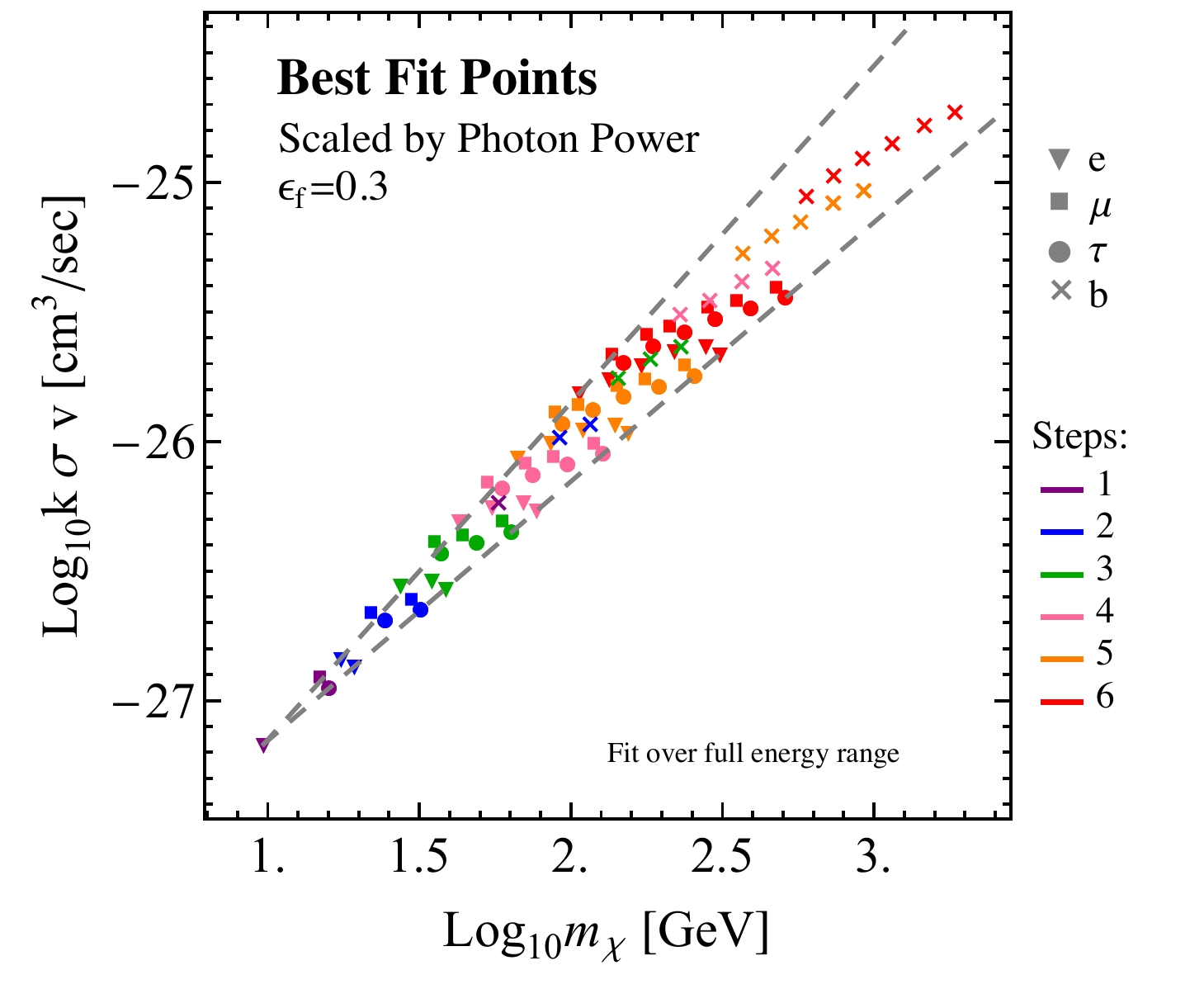}
\end{tabular}
\vspace{-0.2cm}
\caption{\footnotesize{Colored points indicate the best fits for different numbers of hierarchical and degenerate cascade steps, and different final states, as in Fig.~\ref{fig:CombinedResults}. However, here we rescale the cross-section by the fraction of power into photons $k$ for each final state ($3.0 \times 10^{-3}$, $7.0 \times 10^{-3}$, $0.14$ and $0.26$ for electrons, muons, taus and $b$-quarks respectively). All final states then pick out the same region of $(m_\chi, k \sigma v)$ parameter space. The dashed lines indicate curves with $k \langle \sigma v \rangle \propto m_\chi$ and $k \langle \sigma v \rangle \propto m_\chi^{1.3}$, chosen to originate from the lowest-mass point studied; these curves approximately bound the full parameter space of interest (see text).}}
\label{fig:PhotonPowerScaled}
\end{figure}

The positions of the triangular regions in Fig.~\ref{fig:CombinedResults} largely reflect the differing branching ratios to photons (rather than other stable SM particles) for the different final states. For each of the direct annihilation (0-step) spectra, we can compute a factor $k$, defined as the total energy in photons (per annihilation) as a fraction of $m_1 = 2 m_\chi$. For example, direct annihilation/decay to $\gamma \gamma$ would have $k=1$. For the final states we consider, we find $k=3.0 \times 10^{-3}$, $7.0 \times 10^{-3}$, $0.14$ and $0.26$ for electrons, muons, taus and $b$-quarks respectively. Final states with smaller $k$ will naturally require higher cross-sections in order to fit the signal. In Fig.~\ref{fig:PhotonPowerScaled} we show the results of Fig.~\ref{fig:CombinedResults} replotted in terms of $k \langle \sigma v \rangle$ and $m_\chi$: we see that once this factor is taken into account, all channels pick out essentially the \emph{same} triangular region of parameter space, bounded by curves with $k \langle \sigma v \rangle \propto m_\chi$ and $k \langle \sigma v \rangle \propto m_\chi^{1.3}$.

\textit{Incorporating dark showers:} This concordance between the different final states suggests that dark shower models may be expected to also inhabit this region. For instance, the authors of \cite{Freytsis:2014sua} find a preferred cross-section of $8\times 10^{-27}$ cm$^3$/s for their $SU(2)_V$ model, with a roughly $35\%$ branching ratio into stable dark sector baryons (with other decay channels ending in photons), and a preferred mass of $\sim 10$ GeV. At first glance this suggests a somewhat higher value for $k \langle \sigma v \rangle$ than the lower tip of the triangular region identified in Fig.~\ref{fig:PhotonPowerScaled}. However, \cite{Freytsis:2014sua} fits to a different spectrum for the GCE excess (taken from \cite{Daylan:2014rsa}), without a systematic uncertainty estimate, and assumes a lower local DM density (0.3 GeV/cm$^3$ rather than 0.4 GeV/cm$^3$).\footnote{Private communication, Dean Robinson.} In our analysis, omitting systematic errors (or removing high-energy data points) raises the preferred cross-section by a factor of $\sim 2$ (Fig.~\ref{fig:Electron0p1StatsPlot}), and likewise lowering the local DM density from 0.4 to 0.3 GeV/cm$^3$ would raise the required cross-section by a factor of $\sim 2$; the lower tip of our triangular region would then reside at $m_\chi \sim 10$ GeV and $k \langle \sigma v \rangle \sim 3 \times 10^{-27}$ cm$^3$/s, which seems roughly consistent with \cite{Freytsis:2014sua}.

\subsection{Models with Vector Mediators}

Thus far we have considered models of multi-step cascades through scalar mediators. However models in which the hidden sector mediators include vector, fermion or pseudo-scalar particles are at least as equally well motivated (e.g. \cite{Pospelov:2007mp} or the dark shower example discussed above \cite{Freytsis:2014sua}). In the case of vector or fermionic mediators the simple recursion formula Eq.~\ref{eq:boosteq-sig} will in general no longer hold, since the photon spectrum from the decay of mediators with spin need not in general be isotropic. The standard recursion formula will also break down if a decay is more than two-body, or if the decay is two-body but the decay products have different masses (although if the decay is strongly hierarchical the impact will be tiny), since these possibilities modify the Lorentz boost from the $\phi_i$ frame to the $\phi_{i+1}$ frame. Note this is different to having several possible decay chains with different branching ratios; in this case our analysis \emph{does} apply, and the final spectrum will simply be a linear combination of the spectra produced by the different decay chains.

Anisotropy of the photon spectrum is not in itself a sufficient condition for the recursion formula to break down. To modify the recursion, for some step $i$, the differential decay rate of $\phi_i$ must be a function of the angle $\theta$ between (1) the momenta of the decay products in the $\phi_i$ rest frame and (2) the boost direction from the $\phi_i$ rest frame to the $\phi_{i-1}$ rest frame. (Here we use $\phi_i$ to denote arbitrary mediators, independent of their spin.) Since the decays in the $\phi_i$ rest frame do not ``know'' about the $\phi_{i+1}$ frame, this sort of correlation is only possible if (1) the direction of the spin/polarization vector of the $\phi_i$ in its rest frame depends on the momentum with which it was produced in the $\phi_{i-1}$ rest frame, \emph{and} (2) the spectrum of the decay products of $\phi_i$ is a function of the angle between their momentum and the rest-frame spin/polarization vector of $\phi_i$. If only one of the two applies, averaging over the spin/polarization of $\phi_i$ will leave no $\theta$-dependence. However, both these properties will generically hold if $\phi_i$ is a vector: typically the decay of $\phi_{i-1}$ will prefer either longitudinally or transversely polarized vectors $\phi_i$, which will in turn decay with different angular distributions.

Let us consider the potential impact of such a $\theta$-dependence. For illustrative purposes, let us suppose that the \emph{photons} produced in the decays of $\phi_1$ (whether directly or by subsequent decays of the fermions) have essentially the same \emph{energy} spectrum as in the pure-scalar case, in the rest frame of the $\phi_1$. This assumption might fail if the spin of $\phi_1$ affects the correlations (if any) between the fermion spins, fermion momenta and photon momenta, but by making it we can isolate the impact of angular dependence in a single step of the cascade.

Consider a one step cascade $\chi \chi \rightarrow \phi_1 \phi_1$, $\phi_1 \rightarrow f \bar{f}$, where $\phi_1$ is a vector boson. Suppose the full spectrum of photons in the $\phi_1$ rest frame can be written as $\frac{dN}{dx_0} = f_{0}\left(y_0\right) dN/dx_0$, where $y_0 = \cos \theta_0$ and $dN/dx_0$ is the spectrum for the scalar mediator case $f_0 = 1$. Then the now familiar formula for the energy spectrum in the $\chi \chi$ center of mass frame is:
\be\begin{aligned}
\frac{dN_\gamma}{dx_1} &= 2 \int_{-1}^{1} d y_0 \int_{0}^1 dx_0 f_0\left(y_0\right) \frac{dN_\gamma}{dx_0}\\
& \delta\left( 2 x_1 - x_0 - y_0 x_0 \sqrt{1-\epsilon_1^2} \right) \\
&= 2 \int_{x_1}^1 \frac{dx_0}{x_0} f_0 \left( \frac{2x_1}{x_0} - 1 \right) \frac{dN_\gamma}{dx_0} + \mathcal{O}(\epsilon_1^2)\,.
\label{eq:Vectors}
\end{aligned}\ee
where we calculated the $y_0$ integral assuming $\epsilon_1 \ll 1$. Again we could extend this expression to an $n$-step cascade using the same formalism as in Appendix~\ref{app:boost}. The angular dependence at each step will in general be different depending on the model; we can parameterize this by specifying different functions $f_i\left(y_i\right)$ at each step. In the limit of small $\epsilon_i$ we find:
\be\begin{aligned}
 \frac{dN_\gamma}{dx_i} = 2 \int_{x_i}^1 \frac{dx_{i-1}}{x_{i-1}} f_{i-1}\left( \frac{2x_i}{x_{i-1}}-1 \right)\frac{dN}{dx_{i-1}} + \mathcal{O}(\epsilon_i^2)\,.
\label{eq:nStepVectors}
\end{aligned}\ee

\begin{figure}[t!]
\centering
\includegraphics[scale=0.6]{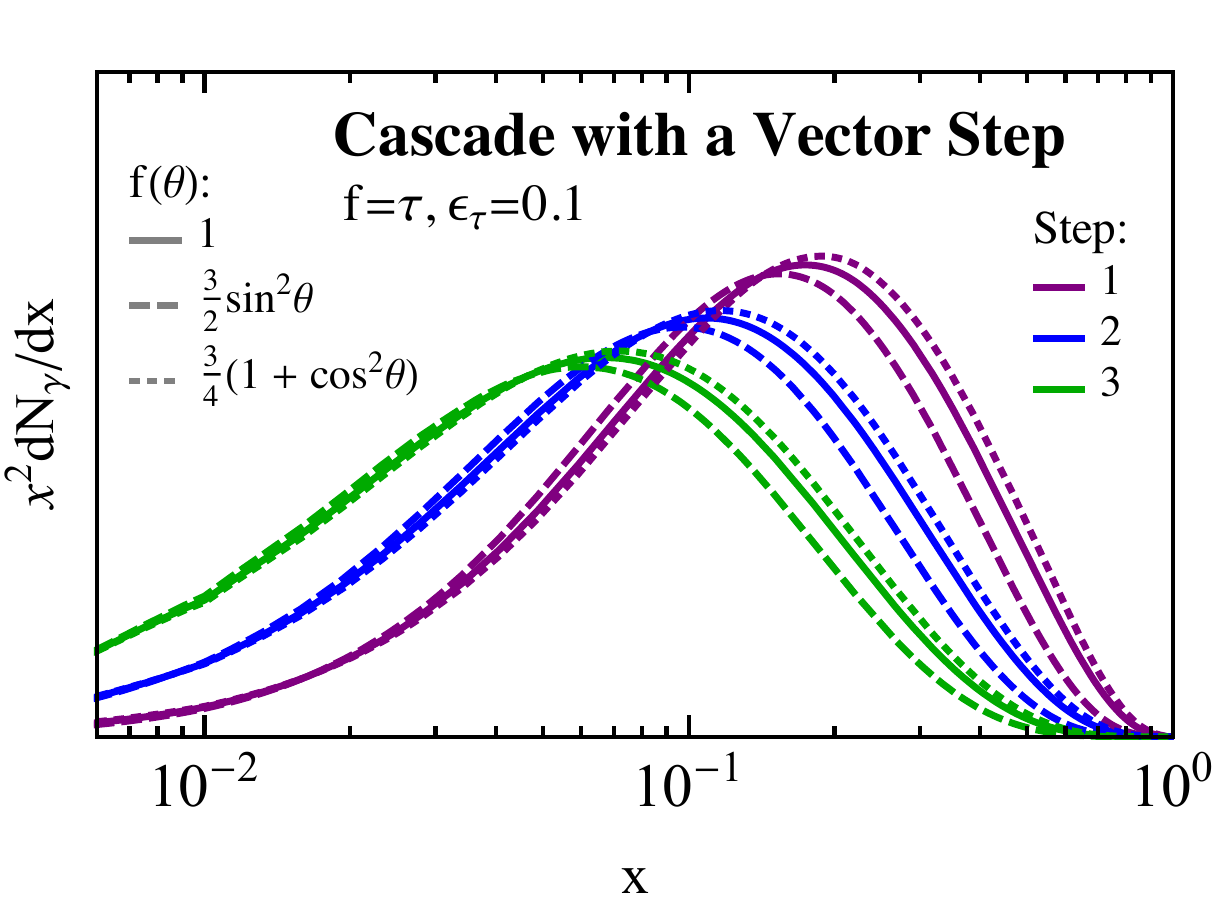}
\vspace{-0.2cm}
\caption{\footnotesize{Spectrum for a 1-3 step cascade with a vector mediator in the final step of the cascade $\phi_2 \rightarrow V V$, $V \rightarrow  f \bar{f}$. We consider three separate cases: $f(\theta) = 1$, $(3/4)(1+\cos^2 \theta)$, and $(3/2)\sin^2\theta$. The first of these is equivalent to a cascade with only intermediate scalars (and hence isotropic decays), the others correspond to common angular dependences (see text).}}
\label{fig:vector}
\end{figure}

A detailed study of the impact of vector or fermionic mediators is beyond the scope of this chapter; we leave it to future work. However, we can work out an explicit example motivated by the case where at the end of the cascade, a scalar/pseudoscalar resonance decays to two vectors which subsequently each decay into two fermions. This scenario has been studied in the context of Higgs decays \cite{Gao:2010qx}, furnishing results for a general resonance $X$ decaying to two identical vectors $VV$, which each in turn subsequently decay to $f \bar{f}$. (In our notation, the $V$ here would correspond to $\phi_1$ and $X$ to $\phi_{2}$.) The differential decay rate to fermions in this case is a linear combination of terms proportional to $\sin^2\theta$, $1 + \cos^2\theta$ and $\cos\theta$ (where $\theta$ is the angle defined above and in Appendix \ref{app:boost}), with coefficients depending on the axial and vector couplings of the fermions to the $V$, and the parity of the initial state $X$ \cite{Gao:2010qx}. In hierarchical decays of a scalar or pseudoscalar resonance to $VV$, where $V$ has vector (rather than axial vector) couplings to $f \bar{f}$, the dominant angular dependence is either $1 + \cos^2\theta$ or $\sin^2\theta$. For these specific (but common) angular dependences in the $\phi_1$ decay, we show the resulting changes to the photon spectrum in Fig. \ref{fig:vector}. The impact is modest, and so we expect our qualitative results should hold for more general cascades.

\section{Signals and Constraints}
\label{sec:signalsconstraints}

While we have remained agnostic regarding the choice of an actual model, we point out that any model with new light states in a dark sector that explains the GCE must also be consistent with the following experimental constraints:
\begin{itemize}
\item{Direct Detection: The coupling controlling $\sigma_{DD}$ must not be so large as to be in conflict with bounds from DM direct detection experiments \cite{Martin:2014sxa}.}
\item{Big Bang Nucleosynthesis (BBN): New light states must decay fast enough such that they do not spoil the predictions of BBN.}
\item{Collider constraints.}
\end{itemize}
These experimental constraints on a multi-step cascade will be very similar to those on a one-step cascade, with the key parameter being the coupling of the dark sector to the SM in both cases.

The simplest models that explain the GCE by direct DM annihilations to SM states are generally in conflict with direct detection bounds: the same coupling that must be small enough to avoid the LUX bound \cite{Akerib:2013tjd}, must also be large enough to explain the GCE with a thermal WIMP (note however that this conclusion is not inevitable; there are effective DM-SM couplings and simplified models that generically evade the bounds, e.g. \cite{Alves:2014yha, Berlin:2014tja}). As pointed out in \cite{Pospelov:2007mp,Martin:2014sxa,Abdullah:2014lla}, the addition of a dark sector with a single mediator allows for an explanation of the GCE while alleviating direct detection constraints. The reason is straightforward: any direct detection signal will be controlled by the coupling of the mediator to the SM, whereas the \emph{annihilation} rate is independent of this quantity, so the two can be tuned largely independently. We make this point more explicit in Appendix \ref{app:models}. Exactly the same property holds in models with expanded cascades, where the direct detection signal is controlled by the coupling between the dark sector and the SM; indeed, the direct detection signal may be suppressed even further if the coupling between the DM and the SM requires multiple mediators. If the couplings \emph{within} the dark sector are not highly suppressed, decays within the dark sector should in general proceed promptly (on timescales $\ll 1$ s), and so the constraint from BBN will primarily limit the coupling of the final mediator in the cascade to the SM. Accordingly, since it has been shown that for one-step cascades the constraint from BBN can be consistent with a null signal in direct detection experiments \cite{Martin:2014sxa}, the same should hold true for multi-step cascades (since in the multi-step case, the final step controls the coupling to the SM and hence provides the only relevant parameter for both BBN and direct detection). Collider bounds and limits from invisible decays of SM particles are also controlled by this final coupling, so can accordingly be dialled down in the same way as for one-step cascades, consistent with BBN bounds on the final coupling \cite{Martin:2014sxa}. A complex dark sector with multiple mediators could potentially give rise to interesting collider signatures (e.g. \cite{ArkaniHamed:2008qp, Baumgart:2009tn, Cheung:2009su}), but a detailed discussion is beyond the scope of this chapter.

\section{Conclusion}
\label{sec:conclusion}

We have laid out a general framework for characterizing the photon spectrum from multi-step decays within a secluded dark sector terminating in a decay to SM particles, and explored the ability of such a framework to produce the GeV gamma-ray excess observed in the central Milky Way. 

For any given SM final state, allowing multi-step decays expands the preferred region of $m_\chi-\langle \sigma v \rangle$ to a triangular region of parameter space, probed by cascades with different numbers of degenerate and hierarchical decays (where the decay products are slow-moving or relativistic, respectively), and bounded by curves with $\langle \sigma v \rangle \propto m_\chi$ and $\langle \sigma v \rangle \propto m_\chi^{1.3}$. Decays to different Standard model final steps correspond to different triangular regions in parameters space as shown in Fig.~\ref{fig:CombinedResults}. Large numbers of degenerate decays can raise the mass scale for the DM without bound, albeit at the cost of requiring a cross-section much higher than the thermal relic value and some degree of fine-tuning. Hierarchical decays broaden the photon spectrum, permitting a better fit to the data for SM final states that produce a sharply peaked photon spectrum; however, more than 4-5 hierarchical decays begin to reduce the quality of the fit even if the initial spectrum is very sharply peaked. In the absence of degenerate decays, the preferred mass range for the DM can then be constrained, and is consistently $\sim 20-150$ GeV across all channels; the corresponding cross-sections are close to the thermal relic value for tau and $b$-quark final states, and 1-2 orders of magnitude higher for $e$ and $\mu$ final states. Regardless of the final state, with the additional freedom of hierarchical decays the preferred spectrum tends to a similar shape, which can be approximated as the result of a cascade of 7-9 hierarchical decays terminating in a two-body $\gamma \gamma$ decay. We find that the best overall fits are still attained by DM annihilating to $b$-quarks (or other hadronic channels) with 0-2 hierarchical steps. 

Our preferred $\langle \sigma v \rangle-m_\chi$ regions are fairly insensitive to the details of the uncertainty analysis or the range of data points included. However, omitting high-energy data (above 10 GeV) substantially reduces the preferred number of hierarchical decay steps (from 4-5 to 2) for channels where the photon spectrum from direct annihilation is sharply peaked. There is currently disagreement between different analyses as to the high-energy photon spectrum associated with the excess; we do not take a position on this question, but note that its resolution may affect the range of dark-sector models that can provide viable explanations of the excess.

In this chapter we assumed that the directions of decay products in the rest frame of their progenitor are uncorrelated with the direction of the Lorentz boost to the rest frame of the previous progenitor particle in the sequence. Whilst always true for scalars, this may not hold for vector and fermionic mediators. We leave a more detailed discussion of concrete multi-step cascade models exploring these issues for future work.

%% file: casclim.tex
\chapter{Model-Independent Indirect Detection Constraints on Hidden Sector Dark Matter}\label{chap:casclim}

\section{Introduction}

\begin{figure}[t!]
\centering
  \includegraphics[scale=0.38]{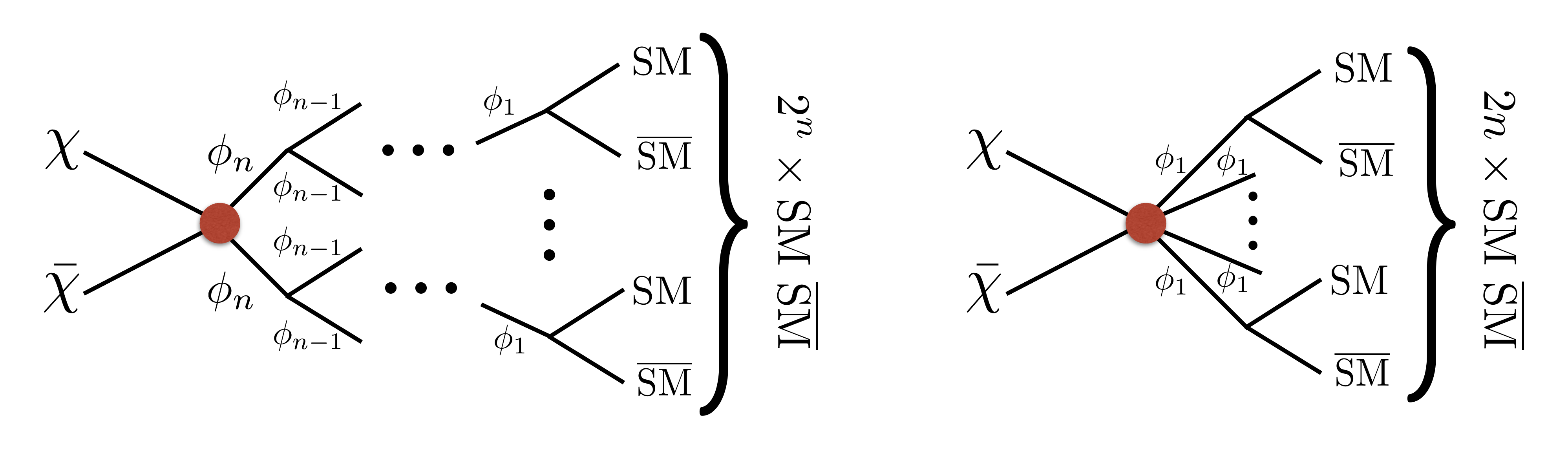}
\vspace{-0.7cm}
  \caption{\footnotesize{Left: Schematic diagram of a generic hidden sector cascade. The DM, secluded in its own hidden sector, first annihilates to a pair of hidden sector particles. These $\phi_n$ mediators subsequently decay to lighter particles in the hidden sector and finally to SM particles. Here we consider SM = $\{\gamma, e, \mu, \tau, b, H, W, g\}$ and $n =$ 0-6 step cascades (where $n =$ 0 refers to the usual case of direct annihilations). Right: An equivalent diagram depicting the case where the DM annihilates through an off-shell heavy mediator; effectively decaying to an $n$-body state in the hidden sector which then decays to SM particles.}}
  \label{fig:Cartoon}
\end{figure}

Indirect searches provide one of the best ways to probe the nature of dark matter (DM) beyond gravitational interactions. Through the observation of gamma rays, cosmic rays, and the anisotropies of the Cosmic Microwave Background (CMB), we may find a hint of DM annihilations to Standard Model (SM) particles. Many models have been proposed in which DM annihilates directly to a pair of SM particles through, for example, a Higgs~\cite{Patt:2006fw,MarchRussell:2008yu}, gauge boson~\cite{Dienes:1996zr}, axion~\cite{Nomura:2008ru}, or neutrino~\cite{Falkowski:2009yz}. Going beyond these simple models, we can consider scenarios in which DM is secluded in its own rich dark sector; such a setup is well motivated from top-down considerations (e.g. \cite{Essig:2013lka} and references therein). In such scenarios, the DM does not couple directly to SM particles (or such couplings are highly suppressed), but instead annihilates to unstable dark sector particles. These states may decay to SM particles or to other dark sector states, but eventually mediator particles that couple to the SM are produced. The mediators subsequently decay into SM particles, which in turn decay to stable and detectable photons, neutrinos, electrons, positrons, protons and/or antiprotons. We refer to this pattern as a ``cascade annihilation'' or simply ``cascade'', with a number of steps given by the number of distinct on-shell dark-sector states between the initial DM annihilation and the production of SM particles. We illustrate this setup schematically in Fig.~\ref{fig:Cartoon}.

Hidden Sector DM scenarios encompass a broad class of models. For instance models containing one light dark photon mediator \cite{Holdom:1985ag,ArkaniHamed:2008qn,Pospelov:2007mp}, generically give rise to one-step cascades decays; DM annihilates to two dark photons which decay to SM particles.  Multi-step cascades can occur naturally in hidden valley models \cite{Strassler:2006im,Han:2007ae}. In such models, production of the DM at terrestrial colliders and scattering in direct detection experiments can be generically suppressed by the small coupling between the dark and visible sectors. In contrast, indirect detection signals depend primarily on the annihilation rate of the dark matter to particles \emph{within} the dark sector; the small coupling between the sectors only suppresses the decay rate of the mediators to SM particles, which does not affect indirect searches provided the decay rate is small on astrophysical timescales (as in e.g. \cite{Rothstein:2009pm}). Thus cascade annihilations scenarios are often invoked to explain anomalies and suggest new DM signals.
For instance in \cite{Elor:2015tva,Hooper:2012cw,Ko:2014gha, Martin:2014sxa,Abdullah:2014lla, Berlin:2014pya,Cline:2014dwa,Liu:2014cma,Cline:2015qha} multi-step cascades were used to explain the apparent excess GeV gamma-rays identified in the central Milky Way \cite{Goodenough:2009gk,Hooper:2010mq,Boyarsky:2010dr,Hooper:2011ti,Abazajian:2012pn,Hooper:2013rwa,Gordon:2013vta,Huang:2013pda,Abazajian:2014fta,Daylan:2014rsa,Calore:2014xka, TheFermi-LAT:2015kwa}, while evading bounds from dark matter direct detection experiments.\footnote{There is recent evidence this excess may originate from a population of point-like objects, rather than DM \cite{Bartels:2015aea,Lee:2015fea}.} In general the injection of photons and other high energy secondary particles produced is constrained by a number of indirect searches. In particular we focus on:
\begin{itemize}
\item{Measurements of the CMB by {\it Planck} \cite{Ade:2015xua}}
\item{Bounds set by {\it Fermi} from DM searches in the Dwarf Spheroidal Galaxies of the Milky Way \cite{Ackermann:2015zua}}
\item{Measurements of the $e^+$ flux by AMS-02 \cite{Aguilar:2014mma,Accardo:2014lma}}
\end{itemize}

Constraints from the above three experiments can be parametrized model-independently for the case of direct DM annihilations (see for instance \cite{Cirelli:2008pk}), by classifying 
annihilations to all possible two-body SM final states, $\textrm{DM} + \textrm{DM} \rightarrow \textrm{SM} +\textrm{SM}$. For a given DM mass and final state, the spectra of secondary particles, is fixed independently of the form of the DM interaction and spin. Therefore constraints on DM annihilation rates are usually quoted in terms of the parameters relevant to the direct annihilation scenario, and do not encompass DM models embedded in a hidden sector.\footnote{\footnotesize{Signals and constraints for a class of 1-step hidden sector models were studied in~\cite{Mardon:2009gw}.}} Given the broad space of Hidden Sector DM models, it is essential to provide model-independent methods that cover the majority of model space. 

In the present chapter, we present DM mass dependent bounds on the DM cross section from the above three indirect detection experiments for DM annihilations via 0-6 step cascades to eight SM final states: $\gamma\gamma$, $e^+ e^-$, $\mu^+ \mu^-$, $\tau^+\tau^-$, $b \bar{b}$, $gg$, $W^+W^-$, and $h\bar{h}$. We remain agnostic about the details of the hidden sector, thus making our statements robust and model-independent.  Limits from the {\it Fermi} dwarfs and AMS-02 generally provide the strongest constraints on channels that are rich in photons and those that are not, respectively (although at sufficiently high masses, limits from H.E.S.S. \cite{Abramowski:2014tra} and VERITAS \cite{Zitzer:2015eqa} overtake those from \emph{Fermi}). While there may be arguably stronger photon bounds from the Galactic Center (e.g. \cite{Hooper:2012sr}) or galaxy clusters (e.g. \cite{Storm:2012ty}), these limits depend strongly on the assumed dark matter density profile and/or the degree of substructure. We include the CMB limits because they are robust and almost independent of the spectrum of the annihilation products; thus we expect them to be nearly unaffected by the transition from 2-body to multi-body SM final states.

Our results are presented in Fig.~\ref{fig:CMB0p3} - Fig.~\ref{fig:MasterLimits}, and our findings can be summarized as follows:
\begin{itemize}
\item{The {\it Planck} CMB bounds are robust and nearly model-independent varying by at most a factor of 1.5 over cascades with up to 6 steps for all final states.}
\item{For photon-rich final states (all states considered except electrons and muons), we find the dwarf limits yield the most sensitive robust constraint, and can be weakened or strengthened by about an order of magnitude or more as compared to the direct annihilation case. For high (low) DM masses and small (large) step number the dwarf bounds can be overtaken by the robust CMB bounds as the most limiting constraints.}
\item{For final states with few photons (electrons and muons), constraints from AMS-02 generally dominate the limits for low number of cascade steps. The limits can change by several orders of magnitude as compared to the direct case. As these weaken for higher DM masses and larger number of steps, CMB constraints become more important.}
\item{Taking the above three points into account we find that for a fixed DM mass and final state, the presence of a hidden sector can change the overall cross section constraints by up to an order of magnitude in either direction (although the effect can be much smaller).}
\end{itemize}
In addition to these constraints we also discuss how the bounds from multi-step cascades can be generalized to include the case of decays to $n$-body states in the dark sector. Finally as a supplement to this chapter we release code to generate the cascade spectrum. 

In Sec.~\ref{sec:Review} we review the procedure used in \cite{Elor:2015tva} to calculate the photon, electron and positron spectra from a multi-step cascade. Section~\ref{sec:Spec} contains a description of the SM final state spectra used. Then in Sec.~\ref{sec:nBody} we describe how results for multi-body decays can be estimated from our cascade results. Our main results are presented in Sec.~\ref{sec:CMB}-\ref{sec:AMS} where we show the model-independent bounds extracted from the CMB, dwarfs, and AMS-02 respectively. We discuss the interplay of the various experimental limit in Sec.~\ref{sec:gendis} and conclude in Sec.~\ref{sec:conclusion}. In the Appendices we describe the contents of the publicly available code, as well as additional details and cross-checks.

\section{Multi-Step Cascade Annihilations}
\label{sec:Review}

The multi-step cascade annihilation scenario is illustrated schematically in the left panel of Fig.~\ref{fig:Cartoon}. In this setup the DM pair annihilates into two scalar mediators (the case of non-scalar mediators was discussed in \cite{Elor:2015tva} where the conclusions proved to be relatively insensitive to choice of vector or scalar mediator\footnote{\footnotesize{A thorough investigation of possible exceptions to this result is left to future work.}}) which subsequently decay through a (possibly) multi-step cascade in the dark sector, eventually producing a dark-sector state (with high multiplicity) that decays to the SM. Schematically we have:
\be\begin{aligned}
\chi \chi \rightarrow \phi_n \phi_n &\rightarrow 2 \times \phi_{n-1} \phi_{n-1} \rightarrow . . . \\
 &\rightarrow 2^{n-1} \times \phi_1 \phi_1 \rightarrow 2^{n} \times \text{(SM final state)}\,.
\label{eq:cascade}
\end{aligned}\ee
Here $n$ is the number of steps as defined above.

A variation on this picture occurs when any of the heavy hidden sector mediators goes off-shell and can therefore be integrated out yielding an effective vertex, now with a multi-body decay in the hidden sector of the form $\phi_n \to m \phi_{n-1}$, with $m>2$. This possibility is illustrated schematically in the right panel of Fig.~\ref{fig:Cartoon} and for a 1-step cascade the analogue to Eq.~\ref{eq:cascade} would be:
\begin{equation}
\chi \chi \to n \times \phi \to 2n \times \text{(SM final state)}\,,
\label{eq:nbodysetup}
\end{equation}
from which the extension to higher step cascades should be clear. Naively this framework seems like it could give quite different results to iterated 2-body decays. Yet in both cases, the main effect is to distribute the energy of the annihilation among a larger number of particles, thus increasing the multiplicity of the SM final state, lowering the average energy of the annihilation products, and broadening their spectrum. Consequently, limits on such scenarios can be broadly understood in terms of the $n$-step cascade results. This again highlights the point emphasized in \cite{Elor:2015tva} that the simple framework of $n$-step 2-body scalar cascades can describe a wide class of models and in this sense provide a relatively model-independent framework.

In Eq.~\ref{eq:cascade} and Eq.~\ref{eq:nbodysetup} ``(SM final state)'' denotes the SM particles produced by a single decay of $\phi_1$, which in turn will (in general) subsequently decay to produce observable photons, neutrinos and charged stable particles. For example a SM final state may produce additional photons due to final state radiation (FSR) or the decay of neutral pions produced during hadronization. The mass ratio between $\phi_1$ and the sum of the masses of the SM particles in this state, which we denote $\epsilon_f$ ($\epsilon_f \equiv ( \sum m_\mathrm{SM}) / m_1)$, controls the level of FSR and hadronization, and so is a useful parameters for describing these decays; the details are discussed in \cite{Elor:2015tva}. When the SM particles are massless, the relevant parameter is instead just the mass of the $\phi_1$, which we denote interchangeably as $m_1$ or $m_\phi$.

The spectrum of particles in an intermediate step of a cascade may be obtained using the method discussed in \cite{Elor:2015tva}, which we briefly review in this section. Consider the ``ith step'' decay $\phi_{i+1} \to \phi_i \phi_i$. In the rest frame of  $\phi_{i+1}$ we will denote the spectrum of the subsequent photons, electrons or positrons as $dN / dx_{i}$, where $x_i = 2E_i / m_{i+1}$, $m_{i+1}$ is the mass of $\phi_{i+1}$ and $E_i$ is the energy of the photon, electron or positron in the $\phi_{i+1}$ rest frame. We define $\epsilon_i = 2m_i/m_{i+1}$, and will (by default) assume a large mass hierarchy between cascades steps such that $\epsilon_i \ll 1$. Assume that the spectrum in the rest frame of the $\phi_{i}$ particle is known and denoted by $dN/dx_{i-1}$. In the limit of a large mass hierarchy the decay of $\phi_{i+1}$ produces two highly relativistic $\phi_i$ particles, each (in the rest frame of the $\phi_{i+1}$) carrying energy equal to $m_{i+1}/2 = m_i/\epsilon_i$. The photon, electron, or positron spectrum per annihilation in the rest frame of the $\phi_{i+1}$ is then given by a Lorentz boost, and takes the simple form
\be
\frac{dN}{dx_i} = 2 \int_{x_i}^{1} \frac{dx_{i-1}}{x_{i-1}} \frac{dN}{dx_{i-1}} + \mathcal{O}(\epsilon_i^2)\,.
\label{eq:boosteq}
\ee

In this way, we can begin with a direct spectrum of $dN/dx_0$ from $\phi_1 \to {\rm SM~final~state}$ -- the details of which are described in the next section -- and generate a cascade spectrum inductively. By repeated application of this formula we can see that the presence of each additional step in a cascade acts to broaden and soften the spectrum, and shift the peak to lower masses. Importantly the shapes of these cascade spectra are very simple, being characterized by just three pieces of information: the number of steps $n$, the SM final state (often denoted $f$), and the value of $\epsilon_f$. Such cascades are independent of the details of each of the intermediate steps, within the large-hierarchy ($\epsilon_i \ll 1$) approximation, and as such are independent of the various $\epsilon_i$.\footnote{The order of the error in the large-hierarchy approximation is $\epsilon_i^2$; see \cite{Elor:2015tva} for more details.}

As pointed out in \cite{Elor:2015tva}, although the large-hierarchy approximation seems to discard information, the more general case can be recovered quite easily. To see this, consider the opposite limit where $\epsilon_i \rightarrow 1$, so that $2 m_i \approx m_{i+1}$. In this case, the rest frames of the $\phi_{i+1}$ and $\phi_i$ are the same, so no boost needs to be applied. As such, in this ``degenerate limit'', the final spectrum of annihilation products is the same as that for a hierarchical cascade with one fewer step, with half the initial dark matter mass and half the annihilation cross-section. The intermediate regime, where neither $\epsilon_i$ nor $1-\epsilon_i$ are particularly small, smoothly interpolates between these two cases. Thus by studying the parameter space of $(m_\chi, \langle \sigma v \rangle$, no. of steps) in the hierarchical limit, it is possible to quickly estimate results for a general cascade. 

Again this framework is more general than it might initially appear. For example, simple extensions where a $\phi_i$ decays to two $\phi_{i-1}$ with different masses will not change our results in the large-hierarchy limit, as those results are independent of the intermediate masses. Additionally, as pointed out in \cite{Elor:2015tva}, for larger $n$ our cascade scenarios can approximate models with hadronization in the dark sector (see e.g. \cite{Freytsis:2014sua,vonHarling:2012sz}), and additionally as we will show in Sec.~\ref{sec:nBody}, multi-body decays can also be approximately captured within this framework.

Note that the cascade scenario must be physically self-consistent: the mass hierarchy between the DM mass and the SM particles in the final state must be sufficiently large to accommodate the specified number of steps. In other words, there is a hard upper limit on the number of steps allowed, for a given DM mass and final state. In detail, for an $n$-step cascade ending in a final state consisting of two particles each with mass $m_f$, we defined $\epsilon_f = 2 m_f/m_1$, $\epsilon_1 = 2m_1/m_2$, $\epsilon_2 = 2m_2/m_3$ and so on until $\epsilon_n=m_n/m_{\chi}$. Combining these, the DM mass is given in terms of $m_f$ and the $\epsilon$ factors by:
\be
m_\chi = 2^n \frac{m_f}{\epsilon_f  \epsilon_1  \epsilon_2  . . .  \epsilon_n}\,.
\label{eq:mDM}
\ee
If the $\epsilon_i$ factors are allowed to float, we can still say that $0 < \epsilon_i \leq 1$ in all cases (since each decaying particle must have enough mass to provide the rest masses of the decay products), setting a strict lower bound on the DM mass of:
\be m_\chi \geq 2^n m_f/\epsilon_f\,. \label{eq:kinematic} \ee 
Where this limit is \emph{not} satisfied, the spectra should not be thought of as potentially representing a physical dark-sector scenario, but only as a parameterization for general spectral broadening. For the massless final states considered here (photons and gluons) $m_f=0$, but we can still derive a condition from the value of $m_{\phi}$, specifically:
\be m_\chi \geq 2^{n-1} m_{\phi}\,. \label{eq:kinematicmphi} \ee 

\section{Direct Spectra}
\label{sec:Spec}

\begin{figure}[t!]
\centering
\begin{tabular}{c}
\includegraphics[scale=0.38]{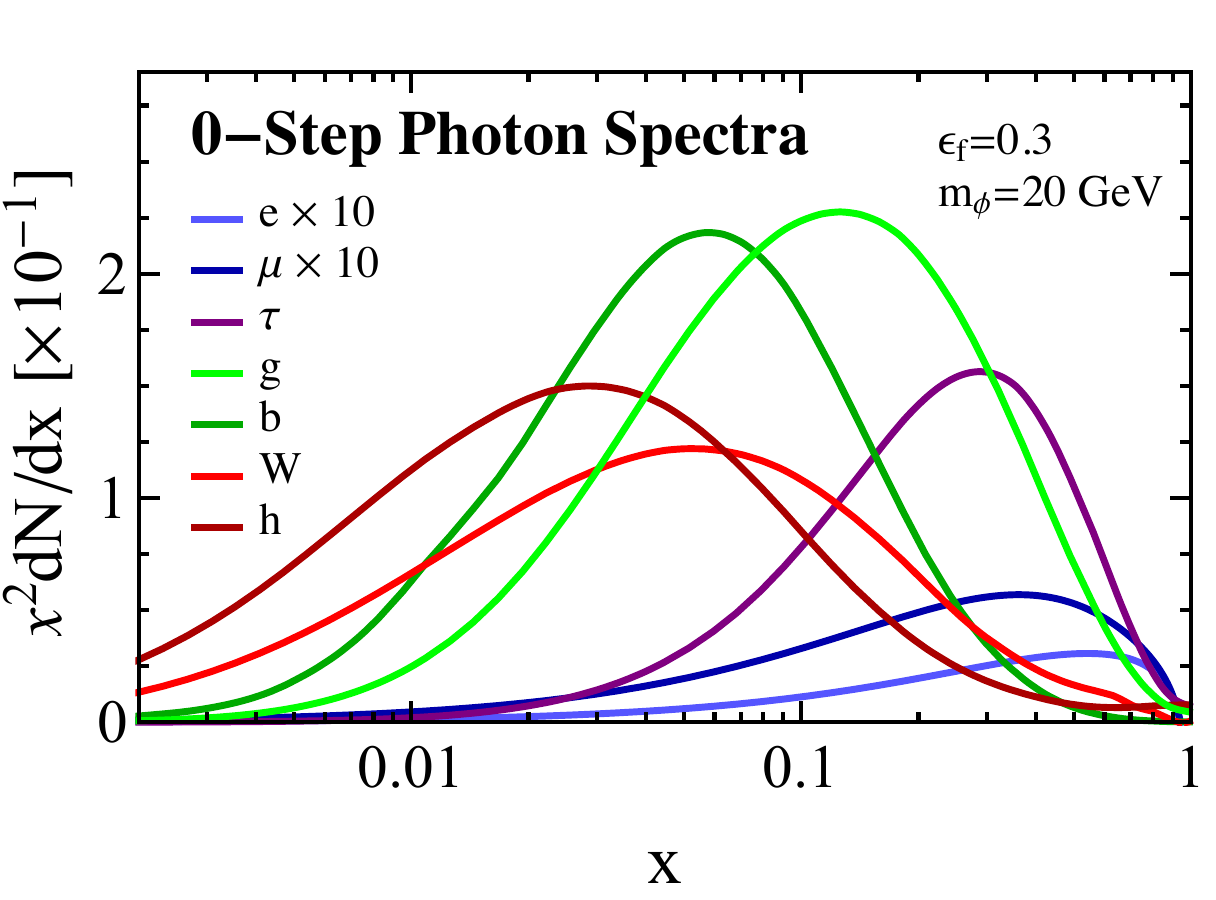} \hspace{0.05in}
\includegraphics[scale=0.38]{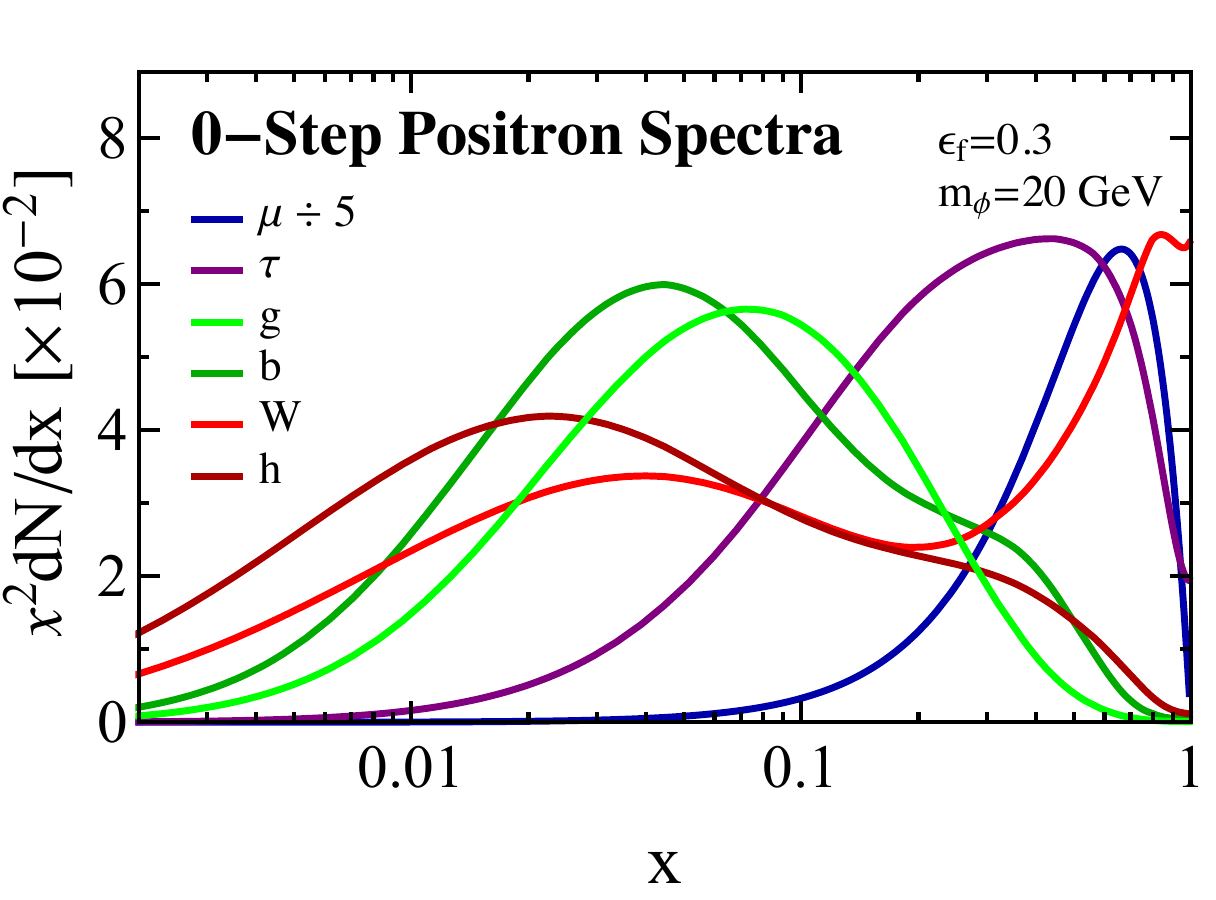} \hspace{0.05in}
\includegraphics[scale=0.38]{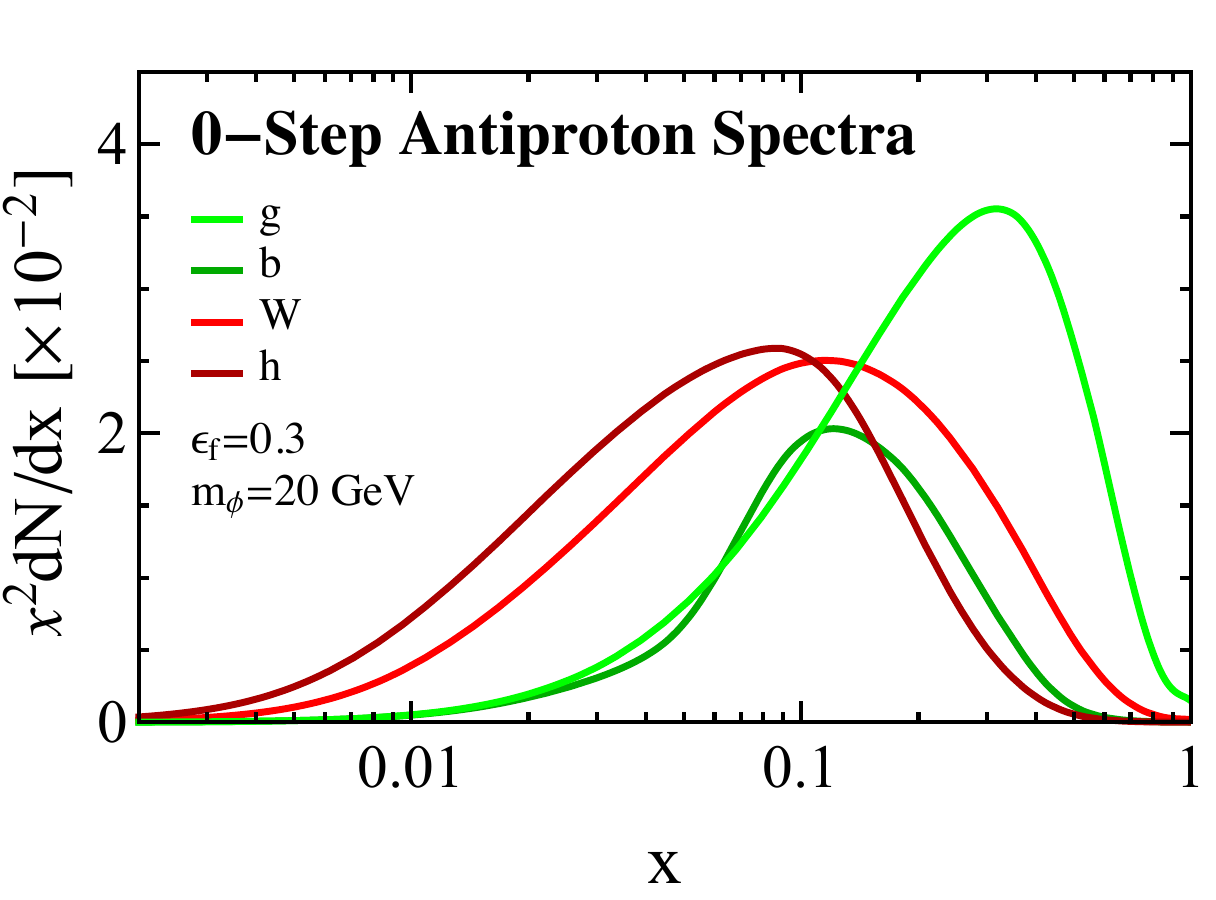}
\end{tabular}
\vspace{-0.2cm}
\caption{\footnotesize{The 0-step or direct photon (left), positron (center) or antiproton (right) spectrum for the various final states considered in this chapter. We have left out the $\gamma \gamma$ spectrum in the photon case and the electron spectrum in the positron case as both of these are $\delta$-functions. Where applicable spectra are plotted with $\epsilon_f=0.3$ or $m_{\phi}=20$ GeV in the case of gluons.
}
}
\label{fig:0step}
\end{figure}

Using the formalism outlined in the previous section, from a given ``direct'' spectrum we can straightforwardly generate an $n$-step cascade spectrum, to compare with various indirect searches. We outline the different SM final states considered for the direct (0-step) spectra in this section. To obtain limits using bounds from the dwarfs, CMB and AMS-02 we need the spectrum of photons, electrons and positrons, and so we determine the spectrum for these particles arising from the boosted decays of the following eight SM states: $\gamma \gamma$, $e^+ e^-$, $\mu^+ \mu^-$, $\tau^+ \tau^-$, $\bar{b} b$, $W^+ W^-$, $h\bar{h}$, and $gg$.\footnote{In our publicly released code we also provide the antiproton spectrum for $b$-quarks, $W$-bosons, Higgs and gluons.} We choose these states as a representative sample of possible spectra. For example decays of light quarks generally give signals similar to those of $b$-quarks and the $ZZ$ final state is similar to $W^+ W^-$.

As discussed in the previous section, many of the cascade spectra depend on the parameter $\epsilon_f = \sum m_\mathrm{SM} / m_1 = 2m_f / m_1$ (the final equality holds for all the processes we consider here). In the context of generating the direct (0-step) spectrum, we can imagine two analogous processes: either the direct annihilation $\chi \chi \to {\rm SM~final~state}$, in which case $\epsilon_f = m_f/m_{\chi}$, or the final step in a cascade annihilation, $\phi_1 \to {\rm SM~final~state}$, so that $\epsilon_f = 2m_f/m_1$ as stated. If the $({\rm SM~final~state})$ is a photon or a gluon, then clearly $\epsilon_f$ is no longer a useful parameter; instead $m_{\phi} = m_1$ (equivalent to $2 m_\chi$ in the case of direct annihilation) plays this role. For many spectra no such parameter is needed. For example the $\gamma \gamma$ photon spectrum, as well as the positron spectra from $e^+ e^-$ or $\mu^+ \mu^-$ final states, are independent of any such parameter, since they are either just $\delta$-functions or arise from decay rather than FSR or hadronization. 

In all but five cases, we use the results of the PPPC4DMID package \cite{Cirelli:2010xx} to produce the direct spectra (hereafter referred to simply as PPPC). The exceptions to this are:
\begin{itemize}
\item the $\gamma \gamma$ photon and $e^+ e^-$ electron or positron spectra, which are just $\delta$-functions, to a good approximation (we neglect the effect of FSR on the $e^+ e^-$ spectra in the case of annihilation/decay to $e^+ e^-$),
\item the spectra of photons produced in conjunction with the $e^+ e^-$ and $\mu^+ \mu^-$ final states, for which we use the analytic results of \cite{Mardon:2009rc,Elor:2015tva} and \cite{Kuno:1999jp,Mardon:2009rc,Elor:2015tva} respectively,
\item the spectrum of electrons or positrons from muon decay, where we use the analytic Michel spectrum \cite{Michel:1949qe,Mardon:2009rc}.
\end{itemize}

Finally we briefly comment on the $\epsilon_f$ or $m_{\phi}$ dependence of the various direct spectra as it is often useful in interpreting results, noting that \cite{Elor:2015tva} has a more detailed discussion of several cases for photon spectra. For photons produced from $e^+ e^-$ and $\mu^+ \mu^-$ final states, the spectra arise entirely from FSR and so are strongly dependent on $\epsilon_f$, increasing in flux and becoming more sharply peaked near the maximum energy as $\epsilon_f$ decreases. Similarly the photon spectrum produced from the $W$-boson final state, in addition to a broad continuum peaked at low $x$, acquires a sharp spike at high $x$ due to FSR when $\epsilon_f$ becomes small. The photon spectrum from the $b$-quarks final state broadens and moves its peak to smaller $x$ as $\epsilon_f$ decreases; the gluon spectrum behaves similarly as $m_{\phi}$ increases. Finally the photon spectra from $\tau^+ \tau^-$ and $\bar{h}h$ final states are largely independent of $\epsilon_f$.

The positron spectra produced from the Higgs and tau final states again show no real variation with $\epsilon_f$. For positrons the spectrum from the $W^+ W^-$ final state is also quite independent of $\epsilon_f$, whilst the $b$-quark and gluon spectra behave much as they did in the photon case. Lastly, for antiprotons, once more the spectra from Higgs and $W$-boson final states are independent of $\epsilon_f$, whilst now for decreasing $\epsilon_f$ (increasing $m_{\phi}$) the $b$-quark (gluon) spectrum increases in height without substantially changing the position of its peak.

\newpage
\section{Multi-Body Cascades}
\label{sec:nBody}

So far we have focused on cascades comprised of 2-body scalar decays. In this section we discuss the extension of this framework to the case of $n$-body cascades, schematically illustrated on the right of Fig.~\ref{fig:Cartoon}. As we will see, in the large hierarchies regime the $n$-body decays can be understood in terms of our existing 2-body results, again emphasizing the model-independence of our results. The explicit calculations and examples to help build intuition are provided in App.~\ref{app:multibody}. 

As explained in the introduction, a multi-body decay can arise if there is a heavy mediator in the cascade that has been integrated out. This can happen anywhere in a cascade, but here we restrict to a 1-step cascade of the form $\chi \chi \to n \times \phi \to 2n \times \text{(SM final state)}$ (c.f. Eq.~\ref{eq:cascade}). From here the extension to higher step cascades is intuitively clear, and in practice can be calculated using Eq.~\ref{eq:boosteq}. As shown in the appendix, an analogue of this equation can be derived for the multi-body case:
\begin{equation}
\frac{dN}{dx_1} = n(n-1)(n-2) \int_0^1 d \xi (1-\xi)^{n-3} \int_{x_1/\xi}^1 \frac{dx_0}{x_0} \frac{dN}{dx_0} + \mathcal{O}(\epsilon_1^2)
\label{eq:nbodyboosteq}
\end{equation}
where again $dN/dx_0$ represents the direct spectrum. Intuitively, the $dx_0$ integral accounts for the boosting of the decay products, just as in Eq.~\ref{eq:boosteq}, whilst the $\xi$ integral samples from the $n$-body phase space to give the correct degree of boosting.

\begin{figure}[t!]
\centering
  \includegraphics[scale=0.7]{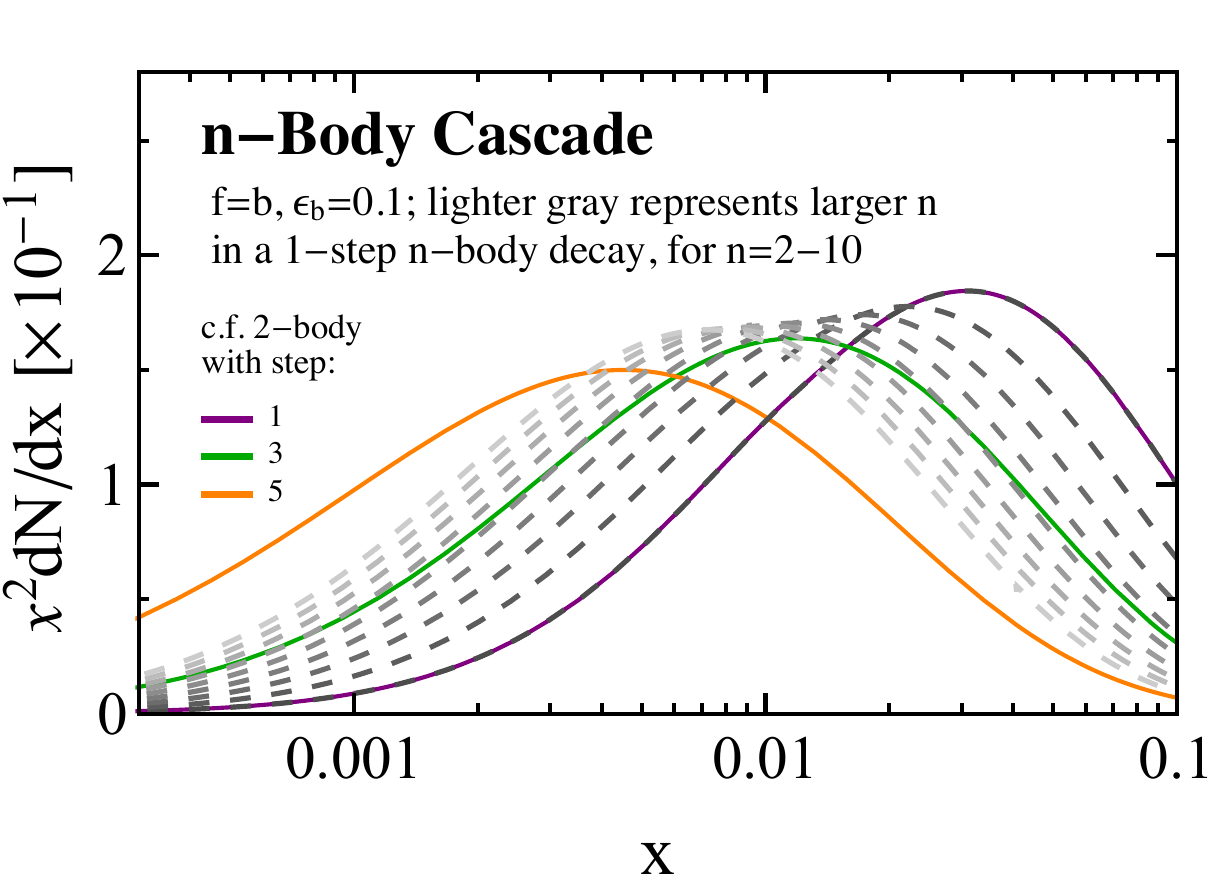}
  \vspace{-0.2cm}
  \caption{\footnotesize{The 1-step spectrum for an $n$-body cascade from a direct annihilation to $b\bar{b}$ with $\epsilon_b=0.1$ is shown as the dashed gray curves for $n=$ 2-10, where lighter curves correspond to larger $n$. In purple, green and orange we show a 2-body 1-step, 3-step and 5-step cascade spectrum respectively, for the same direct spectrum. These three curves outline the $n$-body results and show that the result of 1-step multi-body spectra should be encapsulated in the multi-step 2-body results.}}
  \label{fig:nbody}
\end{figure}

At first glance it appears that this formula could produce marked differences to our standard cascade framework, but as we show in Fig.~\ref{fig:nbody}, this is not the case. There we show the 1-step spectrum for an $n$-body cascade ending in annihilation into the SM state $b \bar{b}$ with $\epsilon_b=0.1$, for $n=$ 2-10. Overlaid is the $m$-step 2-body cascade for $m=$ 1,3,5. Of course when $n=$ 2 and $m=$ 1 these two are the same by definition. But increasing $n$ and increasing $m$ perturb the spectra in quite similar ways (albeit to different degrees, as expected since the multiplicities of final-state particles are not equal for $m=n$ with $n > 2$), and so we expect the observational signatures of multi-body decays to lie within the space mapped out by the simple cascade annihilations. An example of the constraints on multi-body decays, and how they lie within the band of cascade constraints, is given in Fig.~\ref{fig:DwarfdeltaNBody}, in App.~\ref{app:multibody}.

\section{CMB Bounds from $\textit{Planck}$}
\label{sec:CMB}

Dark matter annihilation during the cosmic dark ages can inject ionizing particles into the universe, modify the ionization history of the hydrogen and helium gas, and consequently perturb  the anisotropies of the CMB. Sensitive measurements of the CMB by $\textit{Planck}$ \cite{Ade:2015xua} (and previously WMAP and other experiments) can place quite model-independent limits on such energy injections, which when applied to dark matter annihilation are competitive with other indirect searches.

The figure of merit for CMB limits on dark matter annihilation is the parameter $p_\mathrm{ann} = f_\mathrm{eff} \langle \sigma v \rangle/m_\chi$, where $f_\textrm{eff}$ is an efficiency factor that depends on the spectrum of injected electrons and photons, and the other factors describe the total power injected by dark matter annihilation per unit time. In principle, different dark matter models could give rise to different patterns of anisotropies in the CMB -- but for WIMP models of dark matter that annihilate through $s$-wave processes, it has been shown \cite{Finkbeiner:2011dx} that the impact on the CMB is identical at the sub-percent level up to an overall rescaling by $p_\mathrm{ann}$ (related studies \cite{Galli:2011rz,Hutsi:2011vx,Giesen:2012rp} independently found that the signal was largely controlled by a single parameter). In \cite{Slatyer:2015jla}-\cite{Slatyer:2015kla} this result was generalized to include any class of DM models for which $\langle \sigma v \rangle$ can be treated as a constant during the cosmic dark ages, which is generically true for the models considered in the present chapter.  We use the results of \cite{Slatyer:2015jla} to compute $f_{\text{eff}}$ as a function of dark matter mass and annihilation channel. In particular, we compute positron and photon spectra for the case of direct annihilations using PPPC, and then convolve to find the resulting spectrum for an $n$-step cascade as discussed above. The spectrum of electrons is equal to that of positrons by the assumption of charge symmetry. We then integrate the resulting spectra over the $f_{\textrm{eff}}(E)$ curves provided in \cite{Slatyer:2015jla} to obtain the weighted $f_{\textrm{eff}}(m_\chi)$ for $n=0$-6 step cascades for dark matter annihilations to various final states:
\bea \nonumber
f_{\textrm{eff}}(m_\chi) = \frac{\int^{m_\chi}_0 E dE \left[  2 f_{\textrm{eff}}^{e^+ e^-} \left(\frac{dN}{dE}\right)_{e^+} + f_{\textrm{eff}}^{\gamma} \left(\frac{dN}{dE}\right)_{\gamma}  \right]}{2m_\chi}\,. \\
\eea
We neglect the contribution from protons and antiprotons, as for all the channels we consider, the fraction of power proceeding into these channels is rather small, and consequently including them should only slightly increase $f_\mathrm{eff}$ \cite{Weniger:2013hja}. The constraints we present are therefore somewhat conservative (they could be strengthened slightly by a careful treatment of protons and antiprotons). As discussed in \cite{Slatyer:2015kla}, we use the best-estimate curves suited for the ``3 keV'' baseline prescription, which are most appropriate for applying constraints derived by {\it{Planck}}.

The bound set on the annihilation parameter, $p_\mathrm{ann}$, from \textit{Planck} temperature and polarization data is taken to be \cite{Ade:2015xua}:
\bea
\frac{f_{\textrm{eff}} \langle \sigma v \rangle}{m_\chi} < 4.1 \times 10^{-28}~\textrm{cm}^3/\textrm{s}/\textrm{GeV}\,.
\label{eq:CMBbound}
\eea

\begin{figure}[htbp]
\begin{center}
\begin{tabular}{c}
\includegraphics[scale=0.64]{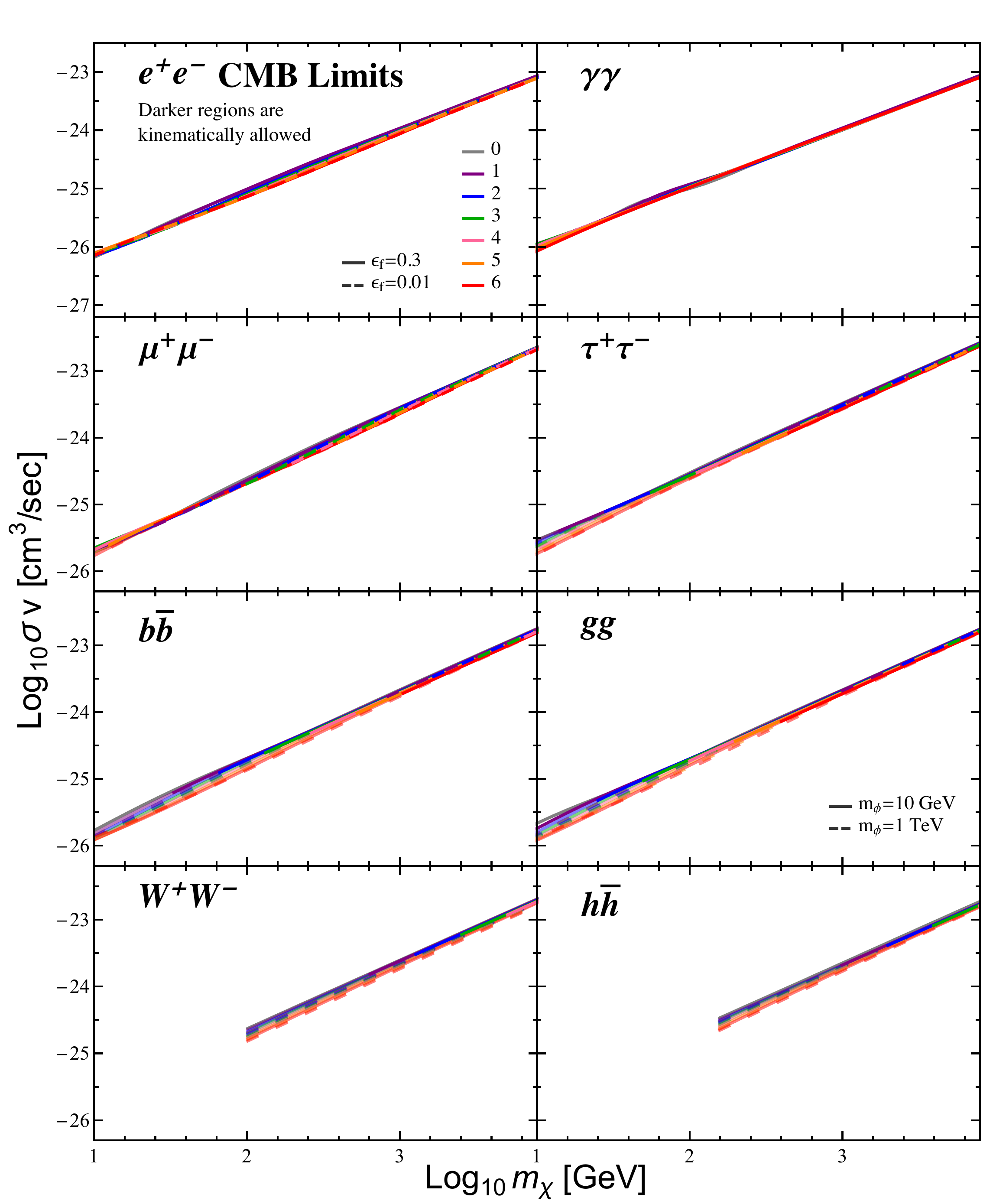}
\end{tabular}
\end{center}
\caption{\footnotesize{The bound on DM cross-section Eq.~\ref{eq:CMBbound}  for $n=$ 1-6 step cascade for various final states, with $\epsilon_f = 0.3$ (solid) and  $\epsilon_f = 0.01$ (dashed). The shaded out portions of the plot correspond to values of $m_\chi$ that are kinematically forbidden. As discussed above the number of steps does not affect the total power deposited by the DM annihilation per unit time. Therefore the constraints are insensitive to the number of steps as the only impact comes from the energy dependence of $f_{\textrm{eff}}$ and the broadening of the spectrum.}}
\label{fig:CMB0p3}
\end{figure}

In Fig.~\ref{fig:CMB0p3} we present our results for the bound on DM cross-section as a function of $m_\chi$ for various numbers of cascade steps and SM final states. We note that the number of steps does not affect the total power deposited by dark matter annihilation per unit time (at least in the simple scenario where all that power eventually goes into SM particles). Each additional step reduces the average energy of the final-state photons/positrons/electrons by a factor of 2, but simultaneously increases their multiplicity by a factor of 2. Thus the only possible impact on the constraints comes from the energy dependence of $f_\mathrm{eff}$, combined with the softening and broadening of the spectrum.

In accordance with our expectations, we find that the effect of the spectral broadening and softening is rather mild, typically changing the constraints by no more then 0.1-0.15 decades (corresponding to a factor of $\sim 1.5$). There is no general trend, in that constraints on these high-multiplicity final states may be either weaker or stronger than those pertaining to direct annihilation; this arises from the fact that $f_\mathrm{eff}$ is not a monotonic function of energy, so lowering the average energy of the injected particles may either increase or decrease the deposition efficiency. In general, $f_\mathrm{eff}$ and hence the upper bound on the ratio $\langle \sigma v \rangle/m_\chi$ varies less as a function of mass for higher-multiplicity final states (as expected, from the broader resulting spectrum), but this effect is very small. The choice of $\epsilon$ parameters, again, does not perturb the constraints outside this $\sim 0.15$-decade band. We refer the reader to the App.~\ref{app:CMBdetails} for additional details regarding the behavior of $f_{\textrm{eff}}$.

\section{Dwarf Limits from \textit{Fermi}}
\label{sec:Dwarfs}

The dwarf spheroidal galaxies of the Milky Way are expected to produce some of the brightest signals of DM annihilation on the sky. Whilst less intense than the emission expected from the galactic center, the dwarfs have the distinct advantage of an enormous reduction in the expected astrophysical background. These features make them ideal candidates for analysis with the data available from the \textit{Fermi} Gamma-Ray Space Telescope. Indeed the \textit{Fermi} Collaboration has set stringent limits on the dark matter annihilation cross-section using the dwarfs \cite{Ackermann:2015zua}, and together with the DES Collaboration have used 8 newly discovered dwarf satellites \cite{Bechtol:2015cbp,Koposov:2015cua} to set independent limits \cite{Drlica-Wagner:2015xua}. We note in passing that several groups have pointed out an apparent gamma-ray excess in the direction of one of the new dwarfs, Reticulum II \cite{Geringer-Sameth:2015lua,Drlica-Wagner:2015xua, Hooper:2015ula}, albeit with considerable variation as to its significance (with estimates ranging from $\sim 3\sigma$ to completely insignificant). We will not discuss this tentative excess here, other than to note as it appears roughly consistent with the emission coming from the GCE, the implications for dark sector cascades will be analogous to those discussed in \cite{Elor:2015tva}.

Here we focus on understanding how the presence of cascade annihilations can modify the limits obtained from these dwarf galaxies. In order to do this we use the publicly released bin-by-bin likelihoods provided for each of the dwarfs considered in \cite{Ackermann:2015zua}.\footnote{These results are available for download from \\ http://www-glast.stanford.edu/pub\_data/1048/} This analysis made use of 6 years of Pass 8 data and found no evidence for an excess over the expected background. Note the \textit{Fermi} collaboration produced an earlier analysis of the same dwarfs using 4 years of Pass 7 data in \cite{Ackermann:2013yva}. In App.~\ref{app:p7p8} we show that the results are similar between the two, but that the limits set using the newer analysis are usually about half an order of magnitude stronger.

Although \cite{Ackermann:2015zua} considered 25 dwarf galaxies, when setting limits they restricted this to 15, choosing a non-overlapping subset of dwarfs with kinematically determined $J$-factors. Specifically the 15 dwarfs considered were: Bootes I, Canes Venatici II, Carina, Coma Berenices, Draco, Fornax, Hercules, Leo II, Leo IV, Sculptor, Segue 1, Sextans, Ursa Major II, Ursa Minor, and Willman 1.

For a given dwarf \textit{Fermi} provides the likelihood curves as a function of the integrated energy flux in each of the energy bins considered in their analysis, covering the energy range from 500 MeV to 500 GeV. Thus to obtain the likelihood curves for our cascade models we need to firstly determine the integrated energy flux per bin. This will be a function of the DM mass $m_{\chi}$, annihilation cross-section $\langle \sigma v \rangle$, and shape of the cascade spectrum $dN/dx$ -- which itself depends on the number of cascade steps, the identity of the final state particle and possibly either $\epsilon_f$ or $m_{\phi}$. For an energy bin running from $E_{\rm min}$ to $E_{\rm max}$, the energy flux in GeV/cm$^2$/s is:
\begin{equation}
\Phi_E = \frac{\langle \sigma v \rangle}{8\pi m_{\chi}^2} \left[ \int_{E_{\rm min}}^{E_{\rm max}} E \frac{dN}{dE} dE \right] J_i\,,
\label{eq:Eflux}
\end{equation}
where $J_i$ is the $J$-factor appropriate for the individual dwarf $i$. Treating the energy bins as independent, we can simply multiply the likelihoods for the various bins to obtain the full likelihood for a given dwarf $i$: $\mathcal{L}_i\left({\boldsymbol \mu} \vert \mathcal{D}_i\right)$, which is a function of both the model parameters ${\boldsymbol \mu}$ and the data $\mathcal{D}_i$.  At a given mass and for a given channel (final state and number of cascade steps),  ${\boldsymbol \mu}$ just describes the annihilation cross-section $\langle \sigma v\rangle$. There is, however, one additional source of error that should be accounted for: the uncertainty in the $J$-factor. Following \cite{Ackermann:2015zua} we incorporate this as a nuisance parameter on the global likelihood, modifying the likelihood as follows:
\begin{equation}\begin{aligned}
&\tilde{\mathcal{L}}_i\left({\boldsymbol \mu}, J_i \vert \mathcal{D}_i\right) = \mathcal{L}_i\left({\boldsymbol \mu} \vert \mathcal{D}_i\right) \\&\times \frac{1}{\ln(10) J_i \sqrt{2\pi} \sigma_i} e^{-\left( \log_{10}(J_i) - \overline{\log_{10}(J_i)} \right)^2/2 \sigma_i^2} \,,
\label{eq:Jlike}
\end{aligned}\end{equation}
where for $\overline{\log_{10}(J_i)}$ and $\sigma_i$ we use the values provided in \cite{Ackermann:2015zua} for a Navarro-Frenk-White profile \cite{Navarro:1996gj}. This approach allows us to account for the $J$-factor uncertainties using the profile likelihood method \cite{Rolke:2004mj}. We obtain the full likelihood function by multiplying the likelihoods for each of the 15 dwarfs together.

Using this likelihood function, for a given DM mass and cascade spectrum we can then determine the 95\% confidence bound on the annihilation cross-section. We follow this procedure for cascade annihilations with 0-6 steps, for final state electrons, muons, taus, $b$-quarks, $W$-bosons, Higgses, photons and gluons, considering two different values of $\epsilon_f$ or $m_{\phi}$ where appropriate. 

\begin{figure}[htbp]
\begin{center}
\includegraphics[scale=0.64]{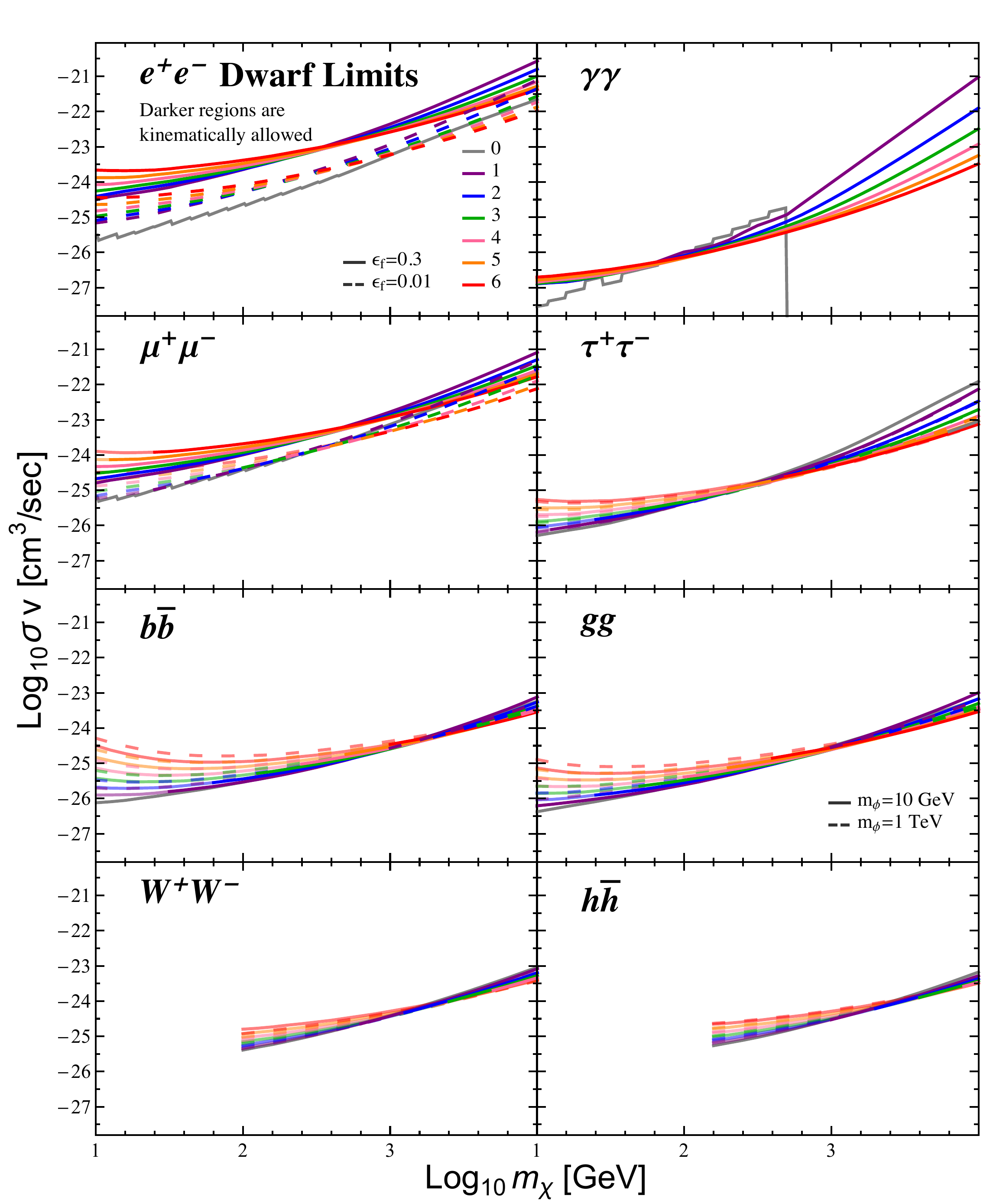}
\end{center}
\caption{\footnotesize{95\% confidence limits on dark matter cross-section for cascade models using the data from 15 dwarf spheroidal galaxies. Results are shown for the photon spectrum obtained from eight different final states: electrons, photons, muons, taus, $b$-quarks, gluons, $W$-bosons, and Higgs. In each case we show the results of a 0 (direct), or 1-6 step cascade. Additionally where it makes sense we show results for two different $\epsilon_f$ values, solid lines representing $0.3$  and dashed $0.01$. Note the $\gamma \gamma$ spectrum is independent of $\epsilon_f$, so we only show one set of limits there, and for the gluon spectrum the relevant parameter is instead $m_{\phi}$ and we show results for $10$ GeV in solid and $1$ TeV as dashed. Only the darker regions are kinematically allowed. See text for a discussion of the results.}}
\label{fig:DwarfLimits}
\end{figure}

Results are shown in Fig.~\ref{fig:DwarfLimits}. For the final states considered in \cite{Ackermann:2015zua}, our direct/0-step results are in agreement. Recall that there is a physical limitation on realizing a given cascade scenario set by $m_{\chi} \geq 2^n m_f/\epsilon_f$, as mentioned in Sec.~\ref{sec:Review}. The constraints corresponding to scenarios that satisfy this condition are indicated by darker lines, but we also show the limits for cases that do not satisfy this condition (and so cannot be physically realized as a cascade annihilation of the type we have considered), to demonstrate the effect of spectral broadening.

Before discussing results for each final state independently, there are a few generic features worth pointing out. Recall that higher-step cascades have a spectrum peaked at lower $x=E_{\gamma}/m_{\chi}$. Thus in order to produce emission at an equivalent energy, higher-step cascades require a larger DM mass, which in turn requires a larger cross-section to inject the same amount of power (as the DM number density scales inversely with the mass). Equivalently, at a fixed mass and cross-section, larger numbers of cascade steps will tend to produce a larger number of lower-energy photons; at low masses, some of these photons may lie outside the energy range of the \emph{Fermi} analysis, and the astrophysical backgrounds will also generally be larger at low energies. These factors tend to weaken the constraints, and indeed we see a systematic trend for weaker bounds with increasing $n$ for low-mass dark matter, for all channels.

Nevertheless this conclusion is not inevitable. Specific energy bins may allow stronger constraints than neighboring bins, purely due to statistical accidents; adding cascade steps smooths out such effects. The total number of emitted photons is increased with larger $n$ (albeit while preserving the total injected power). 

Most generically, if the DM mass is large, much of the spectrum may be above the 500 GeV cutoff of this analysis in the case of direct annihilation. In this case, adding cascade steps can strengthen the constraints by moving the photons into the range of sensitivity for the search. This effect is most pronounced, and occurs at the lowest DM masses, for final states with spectra peaked at large $x$ (electrons, muons, taus and photons): for softer direct-annihilation spectra, even at the heaviest masses tested, the peak of the spectrum does not move past 500 GeV. Inclusion of higher-energy data, e.g. from studies of the dwarf galaxies with VERITAS \cite{Zitzer:2015eqa}, would potentially strengthen the constraints at high DM masses, but for this reason we expect the improvement to be smaller for higher-step cascades.

Thus in general we see a weakening in the cascade constraints relative to the direct-annihilation case at low DM masses, and a strengthening at high DM masses, with the crossover point and the width of the band varying based on the SM final state. For some final states, the cascade constraints can be weaker or stronger than those for the direct-annihilation case by more than an order of magnitude. Let us now discuss the detailed results for each SM final state (shown in Fig.~\ref{fig:DwarfLimits}) separately:

\textit{Electrons:} the generic behaviors discussed above are clearly demonstrated in these results. There is also a striking difference between the results for direct and cascade annihilations. The photon spectrum in the direct case originates from FSR and is very sharply peaked (especially for small $\epsilon_f$); even a single cascade step will smooth out the spectrum and considerably change its shape. Further, the bounds are strongly dependent on the value of $\epsilon_f$, as the FSR photon spectrum diverges as $\epsilon_f \to 0$. As such, for smaller $\epsilon_f$ we expect stronger limits, and this is exactly what we observe. Nonetheless note that the position of the peak of the spectrum in $x$ is not strongly dependent on $\epsilon_f$, so we should expect the crossover behavior between different spectra mentioned above should happen at a similar location for different $\epsilon_f$ values and this is exactly what we observe. Finally note that the bumps in the direct spectrum are a result of the sharply peaked 0-step spectrum moving between energy bins. The width of these bumps is exactly the width of the energy bins in the data. As we move to cascade scenarios, the spectrum is smoothed out and the majority of the emission is no longer in a single bin, meaning these bumps vanish.

$\gamma\gamma$: the most noticeable feature here is the jagged direct spectrum. As the direct spectrum of $\gamma \gamma$ is just a $\delta$-function at the mass considered, these jumps are an extreme realization of the issue mentioned for the 0-step electron limits: we get a jump as the emission moves from one of the energy bins considered to the next. Of course physically the {\it Fermi} instrument has a finite energy resolution, which will act to smooth out such a sharp feature. To approximate this we smooth the 0-step by a Gaussian with a width set to 10\% of the energy value, yet this ultimately had little impact on the extracted limit. Note also that once the spike moves beyond 500 GeV, which occurs at roughly $\log_{10} m_{\chi} = 2.7$, the \textit{Fermi} data can no longer constrain this scenario so the limit completely drops off.

\textit{Muons:} the photon spectrum for the muon final state is very similar to that for the electron final state, except that it is slightly less dependent on $\epsilon_f$. The results here are accordingly very similar to those for the electron final state, except that the variations with $\epsilon_f$ are less pronounced.

\textit{Taus:} the fact that the tau spectrum is only weakly dependent on $\epsilon_f$ is clearly visible; otherwise only the generic behavior is apparent.

$b$-\textit{quarks:} There is a modest dependence on $\epsilon_f$, which does not change the qualitative results. The crossover  where the direct constraints become weaker than the cascade constraints occurs at a DM mass around 100 GeV. Due to the kinematic bounds, over the physically allowed region the variation in the band width is fairly modest varying by at most 0.4 decades.

\textit{Gluons:} the gluon spectrum behaves very similarly to the $b$-quark spectrum, if we swap decreasing $\epsilon_f$ for increasing $m_{\phi}$. As such the results are similar to those for $b$-quarks.

$W$-\textit{bosons:} firstly note that the kinematic edge in these results appears from the threshold requirement to have enough energy to create on-shell $W$'s. Other than this we see that the limits are somewhat stronger for smaller values of $\epsilon_f$, which is because the $W$ spectrum includes a small FSR component which is larger for smaller $\epsilon_f$. The width of the band of possible results is at most 0.7 decades. Again we also see a crossover where the direct constraints become weaker than the cascade constraints, here at roughly 500 GeV.

\textit{Higgs:} as with the $W$-bosons, our results again have a kinematic edge. Furthermore, like final state taus, the Higgs spectrum is only weakly dependent on $\epsilon_f$ and thus so are the results. As for the $W$ case, the width again has a maximum around 0.7 decades, whilst this time the direct crossover first happens at about a TeV. 

\section{Positron bounds from AMS-02}
\label{sec:AMS}

AMS-02 has recently released a precise measurement of cosmic ray electrons and positrons in the energy range of $\sim 1$ GeV to $\sim 500$ GeV~\cite{Aguilar:2014mma,Accardo:2014lma}.
The measured positron ratio exceeds the prediction of the standard cosmic ray 
propagation models at energies larger than $\sim 10$ GeV. There are many possible 
explanations for this rise in the positron ratio, including DM physics (although the annihilation scenario seems challenged by a range of other null results, e.g. \cite{Ade:2015xua}), nearby pulsars~\cite{Hooper:2008kg,Profumo:2008ms} 
or supernovae~\cite{Blasi:2009hv}.

The presence of an apparent large positron excess of unknown origin makes it challenging to set stringent limits on general DM annihilation scenarios. The situation is further complicated by the effects of solar modulation at energies 
below  $\sim 10$ GeV \cite{Gleeson:1968zza,Moskalenko:2001ya,Beischer:2009zz}, which modifies the cosmic ray flux in a charge-dependent manner and adds significant astrophysical uncertainties. However, the data do indicate that both the positron ratio (flux of $e^+$ divided by the flux of $e^+ + e^-$) and the fluxes of cosmic ray electrons and positrons are fairly smooth; there is no clear structure 
in the spectrum within the energy resolution of AMS-02. Accordingly, it is possible to set quite strong constraints on DM models that predict a sharp spectral feature in the positron spectrum (e.g. \cite{Bergstrom:2013jra,Hooper:2012gq,Ibarra:2013zia}).

As discussed in the previous sections, DM annihilation through multi-step cascades usually gives rise to a softer and broader
spectrum than direct annihilation to the SM states, generally leading to weaker
bounds from AMS-02. In this section, we study this effect quantitatively. We note that our goal here is to study the impact of these spectral changes, not to explore possible explanations for the rise in the positron fraction or systematic uncertainties in the modeling of the background or signal.

To set bounds on annihilating DM, we first need to parametrize or model 
the backgrounds. Here the backgrounds that we refer to are the astrophysical 
cosmic ray flux, plus some new smooth ingredient to account for the observed rise in the positron flux. Since we do not know the source of the new ingredient,  
polynomial functions of degree 6 are introduced to fit the AMS-02 positron flux data (the 6 degrees are employed to obtain a good $\chi^2$ fit to the data).
To derive the limits, we float the 6 parameters from the polynomial functions 
within 30$\%$ of the best fit values from the fit without DM, together with the DM annihilation cross-section. We check that increasing the range of allowed values for the background parameters does not weaken the constraints.

We derive limits from only the positron flux, as both the positron and electron backgrounds are required to float in the fit to the positron ratio. Such an analysis would require many additional free parameters, and is beyond the scope of the current chapter. As a cross-check we attempted a simplified fit to the positron ratio data (using AMS-02 measurements of the total $e^++e^-$ spectrum) and found constraints of comparable strength to those we present here.

The positron flux from DM annihilation is obtained by propagating the injected
positron spectrum using the public code DRAGON~\cite{Evoli:2008dv,2011ascl.soft06011M}. There are substantial
systematic uncertainties in the propagation of electrons and positrons in the 
galaxy, affecting diffusion, energy loss, convection, and solar modulation. 
In particular, accounting for uncertainties in the modeling of energy loss and solar modulation can significantly weaken the constraints on DM annihilation. 

Once electrons and positrons are injected into the halo, they will diffuse and lose
energy. Their number density $N_i$ evolves according to the following diffusion equation,
\begin{eqnarray}
   \frac{ \partial N_i} { \partial t} &=&
      \vec{\nabla} \cdot \left( D \vec{\nabla} - \vec{v}_c \right) N_i
      + \frac{\partial} { \partial p} \left ( \dot{p}
      -\frac{p}{3} \vec{\nabla} \cdot \vec{v_c} \right) N_i
   \nonumber \\
   &+& \frac{ \partial} { \partial p} p^2 D_{pp} \frac{\partial}{\partial p}
      \frac{ N_i} { p^2}  +   Q_i ( p, r, z)
   \nonumber \\
     &+&  \sum_{j>i} \beta n_\mathrm{gas} (r,z) \sigma_{ji} N_j
      - \beta n_\mathrm{gas} \sigma_i^{in}( E_k) N_i ~,
   \label{eqn::prop}
\end{eqnarray}
where $D$ is the spatial diffusion coefficient, depending on the spatial position
and energy. It is parametrized by the following form
\begin{equation}
   D ( \rho, r, z) = D_0 \mathrm{e}^{|z|/z_t} \left( \frac{ \rho} 
      {\rho_0} \right)^\delta~,
\end{equation}
where we assume the diffusion zone is axisymmetric, and use the cylindrical coordinate system $( r, z) $. Most of the electrons and positrons are trapped in the 
diffusion zone with thickness $2 z_t$. Here $\rho = p / (Ze)   $ is the rigidity of the charged
particle with $Z= 1$ for electron and positron. 
$D_0$ normalizes the diffusion at the 
rigidity $\rho_0 = 4$ GV. In Eq.~\ref{eqn::prop}, $v_c$
is the velocity of the convection winds; $\dot{p}$ accounts for the energy loss; 
$Q_i$ is the source of the cosmic ray, where DM is one kind of the source; 
$D_{pp}$ accounts for  the diffusion in the momentum space; the last two terms
in Eq.~\ref{eqn::prop} parameterize how the nuclei inelastic scattering with the gas to 
affects the number density of the cosmic rays. Although there are many parameters
in the diffusion equation, we do not simulate the backgrounds (instead modeling them with a polynomial function), which decreases the systematic uncertainties of the limit substantially.

\begin{figure}[t!]
\centering
\begin{tabular}{c}
\includegraphics[scale=0.6]{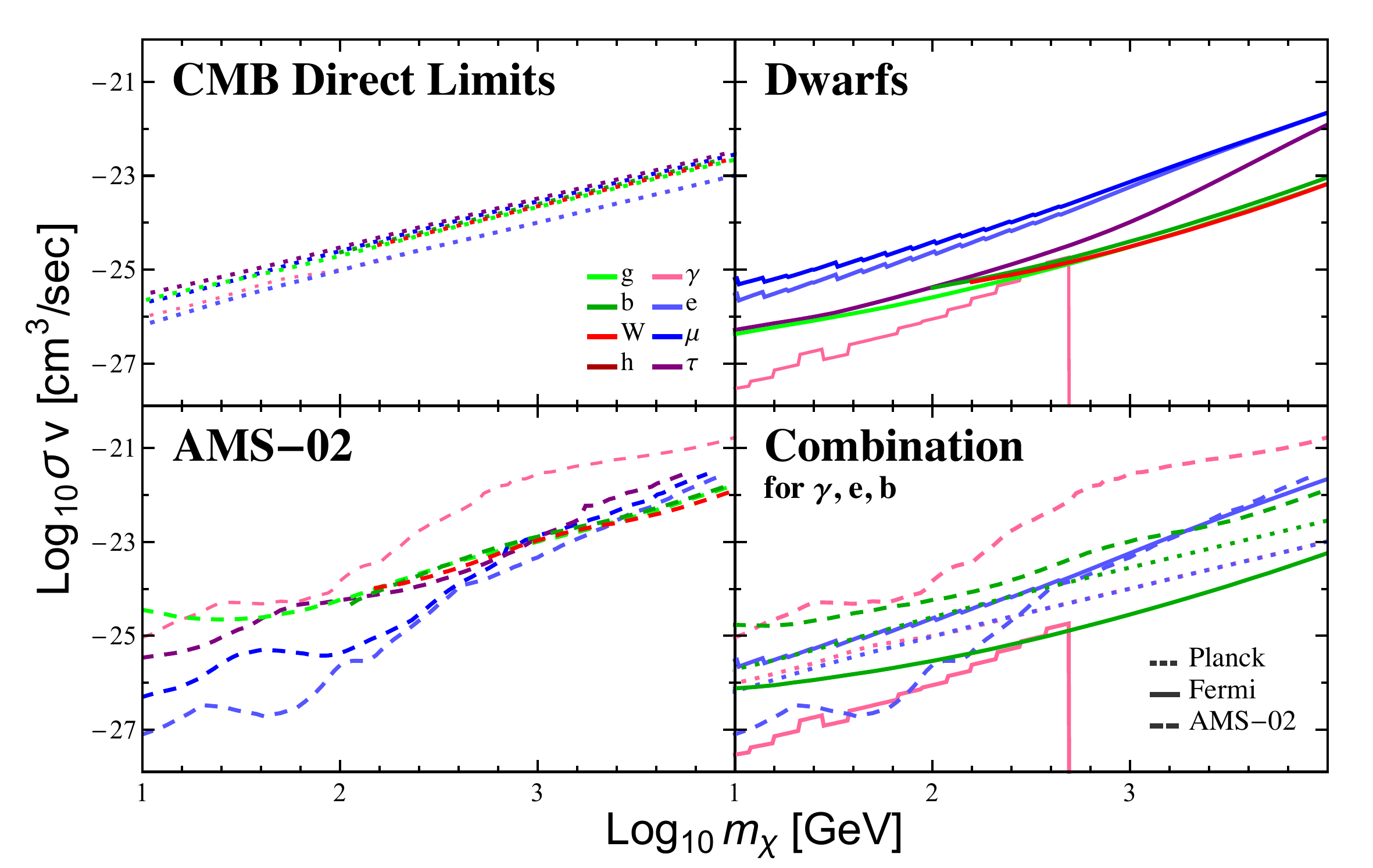}
\end{tabular}
\caption{\footnotesize{Constraints on $\langle \sigma v \rangle$ for the case of direct annihilations to photons, electrons, muons, taus, $b$-quarks, gluons, $W$'s and Higgs final states derived from CMB (top left), dwarfs (top right) and AMS-02 (bottom left). In the bottom right panel we overlay the constraints from all three experiments for the case of direct annihilations to final state photons, electrons and $b$-quarks.}}
\label{fig:DirectCombo}
\end{figure}

We use a specific model to propagate the electrons and positrons 
injected by DM annihilation~\cite{Evoli:2011id}. In this model, $D_0 = 2.7 \times 10^{28}$ cm$^2$/s, 
$z_t = 4$ kpc, $\delta = 0.6$ and we take the local density of DM to be 0.4 GeV/cm$^3$. We set the convection and diffusion in momentum 
space equal to zero, since they do not change the spectrum significantly in the 
energy range of interest~\cite{Strong:2007nh}. There are other propagation 
models with different diffusion terms or diffusion zone heights that can be employed 
here.
However, since the energy loss effect is dominant for the propagation of high 
energy leptons, we choose only one propagation model to derive the limits. While there may be remaining systematic effects due to the choice of propagation model, we reiterate that the purpose of this analysis is not to explore all the uncertainties in these constraints.

Cosmic-ray propagation is affected by the magnetic field, which determines both how the cosmic rays diffuse and their energy losses due to synchrotron radiation. The magnetic field is modeled by two components, one regular and one 
turbulent (irregular) ~\cite{Pshirkov:2011um, DiBernardo:2012zu}. 
\cite{Jaffe:2009hh} gives the constraints on these components. To be 
conservative, here we set the value of the magnetic field at the Sun to
$B_\odot \sim 8.9$ $\mu$G. With this magnetic field, the local radiation field
and magnetic field energy density is 3.1 eV/cm$^3$, which is 
close to (but somewhat higher than) the 2.6 eV/cm$^3$ value used for conservative constraints in \cite{Bergstrom:2013jra}. For this reason, 
the constraint we obtain for the direct annihilation is slightly weaker than even the conservative case studied in \cite{Bergstrom:2013jra}, as the energy loss rate for the positrons is higher. The main effect of changing the local energy density is to rescale all the constraint curves, with lesser effects on the variation of the constraint with DM mass and number of cascade steps. 

For cosmic rays with energy smaller than $10$ GeV, although there are 
many other parameters in the propagation model, 
we only consider the uncertainty from the solar modulation, which is modeled by
the modulation potential. The modulation potential $\phi$ in the range of 
$( 0, 1 )$ GeV is fixed by minimizing the $\chi^2$ to fit the AMS-02 data.  

\begin{figure}[htbp]
\begin{center}
\includegraphics[scale=0.64]{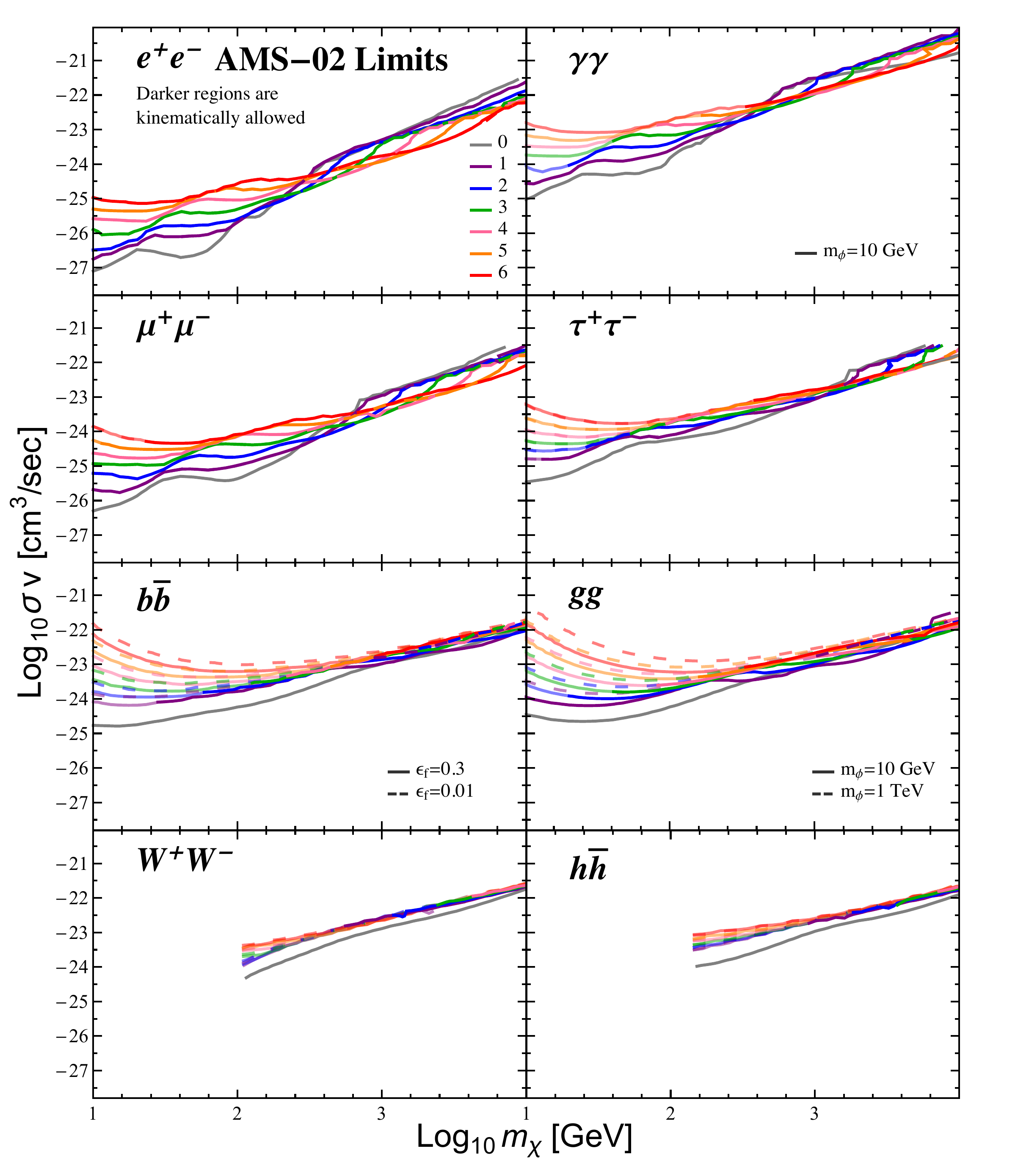}
\end{center}
\caption{\footnotesize{95\% confidence limits on DM cross-section for cascade
models. Details are similar to the previous two plots. The limits obtained are strongest for electron and muon final states, and generically we find that the addition of cascades steps can change the limits by up to several orders of magnitude.}}
\label{fig:AMSLimits}
\end{figure}

In summary, we derive the limits on DM annihilation by using AMS-02 positron 
flux starting from $1 \mathrm{GeV}$. The background is parametrized by a polynomial function of 6 degree, and
to derive the bounds we let the 6 parameters float within 30$\%$ of their best 
fit values. The diffusion model is employed here to propagate the DM positron flux,
and the solar modulation potential is allowed to float in the range  
$( 0, 1 )~\mathrm{GeV}$ when minimizing the likelihood function. The limits are summarized in Fig.~\ref{fig:AMSLimits}.

In general, similar to the dwarf galaxies, the constraints on cascade models can be substantially weaker than those on the direct-annihilation case for low DM masses (below $\sim 100$ GeV), by up to several orders of magnitude depending on the channel. This weakening likely arises from a combination of (a) positrons falling below the minimum energy of the search, and (b) broadening of the spectrum making it easier for the background model to compensate for a DM component. The effect can be up to two orders of magnitude in most channels at sufficiently low masses (the exceptions are the $W$, Higgs and $b$-quark final states where low DM masses are kinematically forbidden). At high masses, the bands of possible constraints are narrower, of order half an order of magnitude or less; for the electron, muon, tau and gamma final states the direct-annihilation constraints are systematically weaker than those for cascade scenarios. This is likely due to the cascade scenarios producing greater numbers of positrons in the energy range of the search, but may also be due to the hardening of the positron flux at high energies mimicking a hard signal from DM annihilation.

\section{General Discussion}
\label{sec:gendis}

\begin{figure}[htbp]
\begin{center}
\includegraphics[scale=0.64]{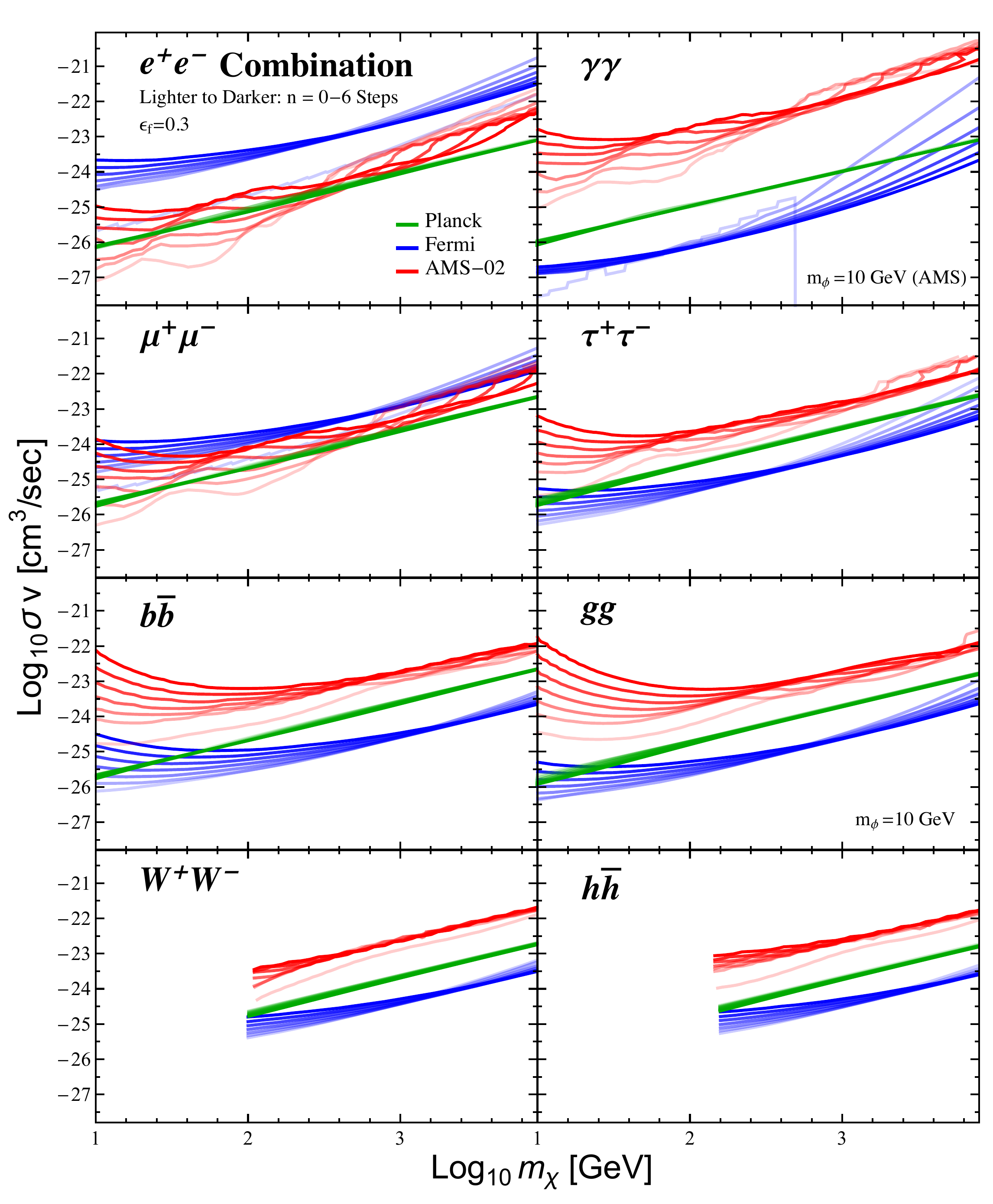}
\end{center}
\caption{\footnotesize{Overlaid constraints from the CMB (green), AMS-02 (red) and the $\it{Fermi}$ Dwarfs (blue) for $n =$ 0-6 step cascades (lighter to darker shading) for various SM final states. The CMB bounds are very weakly dependent on the number of cascade steps, while the AMS-02 and dwarf results change noticeably. The AMS-02 bounds are most constraining for electron and muon final states, and weaken rapidly as the positron spectrum broadens with increasing cascades steps. The dwarfs are generally most constraining for final states with a high multiplicity of photons.}}
\label{fig:MasterLimits}
\end{figure}

We summarize our main results in Fig.~\ref{fig:MasterLimits}, where we overlay the combined constraints from the three experiments as a function of DM mass for an $n =$ 0-6 step hidden sector cascade. Furthermore in Fig.~\ref{fig:DirectCombo} we show results just for the direct, or 0-step, annihilation, in order to highlight the interplay between the experiments. As discussed above the CMB constraint is fairly insensitive to the SM final state and number of cascade steps. The AMS-02 bounds, which are most constraining for direct annihilation for electron and muon final states, weaken rapidly at low masses as the positron spectrum broadens with increasing cascade steps, but for masses above a few hundred GeV are also fairly robust. The dwarfs are generally most constraining for final states with a high multiplicity of photons. However, for lepton-rich or photon-rich final states respectively, the CMB bounds can become more constraining than the AMS-02 or \emph{Fermi} limits for large numbers of cascade steps at low masses, or small numbers of cascade steps at high masses. We summarize the results the various SM final states below. 

\textit{Electrons:}  The spectrum of positron and photon spectrum is very sharply peaked in the case of direct DM annihilations to $e^+ e^-$. Thus AMS-02 places the most constraining bound for $n =$ 0-4 step cascades at low masses $m_{\chi} \lesssim 400 \mathrm{GeV}$. As the number of steps increases, $n > 3$, the spectrum smooths and broadens thereby weakening the AMS-02 bound so that the CMB bound becomes the most constraining. The CMB bounds are generically stronger at high DM masses, above a few hundred GeV. The dwarf limits are, in all cases, 1-2 orders of magnitude less constraining than the AMS-02 and CMB bounds. This is unsurprising given the dwarfs are only sensitive to the photon spectrum from the final state electrons, which represents only a small fraction of the available power per annihilation.

$\gamma\gamma$: The strongest constraints almost always arise from the \emph{Fermi} dwarfs, although at high DM masses and for small numbers of steps, the CMB bounds may be more stringent. However, in this case VERITAS or H.E.S.S dwarf searches may actually provide a stronger limit. For AMS-02 the positron spectrum is similar in shape to that of the electron channel; the photon generates a hard electron spectrum via Drell-Yan. Nonetheless this process is suppressed by a factor of $\alpha_e$ as well as phase space. Combining these, approximately two order magnitude suppression relative to electron case would be expected and is in fact observed.

\textit{Muons:} Recall that the spectrum of positrons and photons from DM annihilations to muons is similar to the corresponding spectra for the electron final state, except the photon spectrum is less dependent on $\epsilon_f$ and the positron spectrum is somewhat broader. For 0-2 cascade steps, the most stringent constraints are from AMS-02 at low masses, below a few hundred GeV. At higher step numbers (for all masses) and higher masses (for all cascade scenarios), the CMB limit becomes more restrictive.

\textit{Taus:} The tau final state is richer in photons than the other leptonic final states, and yields smoother and broader photon and positron spectra even in case of direct annihilation. Thus the bound from the dwarfs is more sensitive and constraining than AMS-02, and generally also stronger than the CMB limits. The exceptions are at low mass and large number of steps, or inversely high mass and a small number of steps, as in both cases the CMB bounds dominate the constraint.

$b$-\textit{quarks:} The direct spectrum for DM annihilations to $b\bar{b}$ is much softer then the previously discussed channels. So the \emph{Fermi} dwarf limits almost always provide the strongest constraint; for low masses and $n =$ 3-6 steps there is a region of parameter space where the CMB bounds appear to be more stringent, however this region is kinematically disallowed.

\textit{Gluons:} As previously discussed the gluon spectrum behaves very similarly to the $b$-quark spectrum, if we swap the decreasing $\epsilon_f$ for increasing $m_{\phi}$. As such the results are similar to those for $b\bar{b}$.

$W$-\textit{bosons:} For annihilation to $W$ final states the bounds are quite robust, with the dwarfs always setting the strongest limits. 

\textit{Higgs:} Annihilations to final state Higgses is similar to the $W$ case; the results are almost identical aside from the difference in the kinematic edge between the H and $W$ mass.

\section{Conclusion}
\label{sec:conclusion}

We have shown that results from current DM indirect searches can be extended to constrain a broad space of dark sector models. We summarize our main points below:
\begin{itemize}
\item{Photon rich final states are generally most constrained by bounds from the {\it Fermi} dwarfs.}
\item{Electron and muon final states are generally most constrained by AMS-02 (albeit subject to uncertainties in the propagation and background modeling).}
\item{The CMB bounds from {\it Planck}  are robust and insensitive to number of dark sector steps. As a result the CMB bounds may become the most limiting in certain cases where the AMS-02 or dwarf bounds weaken as a result of large $m_\chi$ or increasing number of dark sector steps.}
\item{We find that for a fixed DM mass and final state, the presence of a hidden sector can change the overall cross section constraints by up to an order of magnitude in either direction (although the effect can be much smaller).}
\end{itemize}
For hadronic SM final states ($b$-quarks, gluons, gauge bosons, Higgses), constraints from gamma-ray studies of dwarf galaxies generally remain the most limiting, and -- within the kinematically allowed region -- are generally fairly robust, although they can weaken at low masses and strengthen at high masses. More specifically for small but kinematically allowed masses the bound for final state gauge bosons, Higgses and $b$-quarks can weaken by about 0.1 decades. For the gluon final state, where very low DM masses are in principle possible, this bound can weaken by up to 1.1 decade; however a careful consideration of this regime would require taking into account the mass of the mediators, which may be comparable to $\Lambda_\mathrm{QCD}$. At high masses the bounds will strengthen by about 0.3-0.5 decades for the hadronic final states. 

The photon-rich tau and photon final states behave similarly, with the dwarf limits dominating the constraints except perhaps at very high masses (where it may be important to take constraints from VERITAS and H.E.S.S. into account). Adding extra cascade steps has little effect on the dwarf constraints on the photon final state at low masses (after the addition of the first cascade step, which weakens the limit by up to 0.8 decades), whereas for the tau final state it can weaken the bound by about 0.1 decades within the kinematically allowed regime.

For leptonic final states with few photons (electrons and muons), constraints from AMS-02 often appear to dominate the limits, but are quite sensitive to the number of cascade steps (as well as assumptions on the cosmic-ray propagation and local magnetic field; our limits are more conservative than others in the literature). At low masses (below a few hundred GeV), increasing the number of cascade steps can weaken the constraints by up to 2 orders of magnitude, at which point bounds from the CMB become more constraining. Above a few hundred GeV, however, adding more cascade steps tends to strengthen the constraints, so using results quoted for direct annihilation gives conservative bounds; the CMB limits are also generically stronger than the AMS-02 limits in this mass range. 

If a quick estimate of constraints is needed, the CMB limits almost always appear to be within an order of magnitude of the strongest limit, for the cases we have tested, and vary by at most a factor of 1.3 to 1.5 over cascades with up to 6 steps. 

The details of our code for $n$-step cascades which were used to produced our results are described in App.~\ref{app:fileoutline} and are available at: \url{http://web.mit.edu/lns/research/CascadeSpectra.html}.

%% file: dm1loop.tex
\chapter{The One-Loop Correction to Heavy Dark Matter Annihilation}\label{chap:dm1loop}

\section{Introduction}
\label{sec:int}

It is now well established that if dark matter (DM) is composed of TeV scale Weakly Interacting Massive Particles (WIMPs) then its present day annihilation rate is poorly described by the tree-level amplitude. Correcting this shortcoming is important for determining accurate theoretical predictions for existing and future indirect detection experiments focussing on the TeV mass range, such as H.E.S.S \cite{Hinton:2004eu,Abramowski:2013ax}, HAWC \cite{Sinnis:2004je,Harding:2015bua,Pretz:2015zja}, CTA \cite{Consortium:2010bc}, VERITAS \cite{Weekes:2001pd,Holder:2006gi,Geringer-Sameth:2013cxy}, and MAGIC \cite{FlixMolina:2005hv,Ahnen:2016qkx}.

The origin of the breakdown in the lowest order approximation can be traced to two independent effects. The first of these is the so called Sommerfeld enhancement: the large enhancement in the annihilation cross section when the initial states are subject to a long-range potential. In the case of WIMPs this potential is due to the exchange of electroweak gauge bosons and photons. This effect has been widely studied (see for example \cite{Hisano:2003ec,Hisano:2004ds,Cirelli:2007xd,ArkaniHamed:2008qn,Blum:2016nrz}) and can alter the cross section by as much as several orders of magnitude. The Sommerfeld enhancement is particularly important when the relative velocity of the annihilating DM particles is low, as it is thought to be in the present day Milky Way halo.

The second effect is due to large electroweak Sudakov logarithms of the heavy DM mass, $m_{\chi}$, over the electroweak scale, which enhance loop-level diagrams and cause a breakdown in the usual perturbative expansion. The origin of these large corrections can be traced to the fact that the initial state in the annihilation is not an electroweak gauge singlet, and that a particular $\gamma$ or $Z$ final state is selected, implying that the KLN theorem does not apply \cite{Ciafaloni:2000df,Ciafaloni:1998xg,Ciafaloni:1999ub,Chiu:2009mg}. While the importance of this effect for indirect detection has only been appreciated more recently (see for example \cite{Hryczuk:2011vi,Baumgart:2014vma,Bauer:2014ula,Ovanesyan:2014fwa,Baumgart:2014saa,Baumgart:2015bpa}), it must be accounted for, as it can induce $\mathcal{O}(1)$ changes to the cross section. Hryczuk and Iengo \cite{Hryczuk:2011vi} (hereafter HI) calculated the one-loop correction to the annihilation rate of heavy winos to $\gamma \gamma$ and $\gamma Z$, and found large corrections to the tree-level result, even after including a prescription for the Sommerfeld enhancement. These large corrections are symptomatic of the presence of large logarithms $\ln (2m_{\chi}/m_Z)$ and $\ln (2m_{\chi}/m_W)$, which can generally be resummed using effective field theory (EFT) techniques. This observation has been made by a number of authors who introduced EFTs to study a variety of models and final states. The list includes the case of exclusive annihilation into $\gamma$ or $Z$ final states for the standard fermionic wino \cite{Ovanesyan:2014fwa} and also a scalar version of the wino \cite{Bauer:2014ula}, as well as semi-inclusive annihilation into $\gamma + X$ for the wino \cite{Baumgart:2014vma,Baumgart:2014saa,Baumgart:2015bpa} and higgsino \cite{Baumgart:2015bpa}.

In principle the EFT calculations are systematically improvable to higher order and in a manner where the perturbative expansion is now under control. In order to fully demonstrate perturbative control has been regained, however, it is important to extend these works to higher order. To this end, in this chapter we extend the calculation of exclusive annihilation of the wino, which has already been calculated to next-to-leading logarithmic (NLL) accuracy \cite{Ovanesyan:2014fwa}. Doing so includes determining the one-loop correction in the full theory, as already considered in HI. Nonetheless the results in that reference were calculated numerically and are not in the form needed to extend the EFT calculation to higher order. As such, here we revisit that calculation and analytically determine high or DM-scale one-loop matching coefficients. We further calculate the low or electroweak-scale matching at one loop, thereby including the effects of finite gauge boson masses. Taken together these two effects extend the calculation to NLL$^{\prime} =~{\rm NLL}+\mathcal{O}(\alpha_2)$ one-loop corrections, where $\alpha_2 = g_2^2 /4\pi$ and $g_2$ is the SU(2)$_{\rm L}$ coupling. We estimate that our result reduces the perturbative uncertainty from Sudakov effects to $\mathcal{O}(1\%)$, improving on the NLL result where the uncertainty was $\mathcal{O}(5\%)$. Our calculation is complementary to the NLL$^{\prime}$ calculation for the scalar wino considered in \cite{Bauer:2014ula}, and where relevant we have cross checked our work against that reference. In Sec.~\ref{sec:NLL} we outline the EFT setup and review the NLL calculation. Then in Sec.~\ref{sec:1WC} we state the main results of this chapter, the one-loop high and low-scale matching, leaving the details of their calculation to App.~\ref{app:oneloopfull} and App.~\ref{app:lowscalematching} respectively. Detailed cross checks on the results are provided in App.~\ref{app:consistency} and App.~\ref{app:consistencylow}, whilst lengthy formulae are delayed till App.~\ref{app:PiFunctions}. We compare our analytic results to the numerical ones of HI in Sec.~\ref{sec:Comp} and then conclude in Sec.~\ref{sec:conclusion}.

\section{The EFT Framework}
\label{sec:NLL}

We begin by outlining the EFT framework for our calculation, and in doing so review the calculation of heavy DM annihilation to NLL, focussing on the treatment of the large logarithms that were partly responsible for the breakdown in the tree-level approximation. We choose the concrete model of pure wino DM -- the same as used in HI and \cite{Ovanesyan:2014fwa} -- to study these effects. Nevertheless we emphasize the point that the central aim is to quantify the effect of large logarithms which can occur in many models of heavy DM, rather than to better understand this particular model. Ultimately it would be satisfying to extend these results to DM with arbitrary charges under a general gauge group to make the analysis less model specific. This is possible for GeV scale DM indirect detection where the tree-level approximation is generally accurate (see for example \cite{Elor:2015bho,Elor:2015tva}). However, understanding the effects in a simple model is an important step towards this goal.

The model considered takes the DM to be a wino: an SU(2)$_{\rm L}$ triplet of Majorana fermions. As already highlighted, this is a simple example where both the Sommerfeld enhancement and large logarithms are important. Furthermore this model is of interest in its own right. Neutralino DM is generic in supersymmetric theories \cite{Giudice:1998xp,Randall:1998uk}; models of ``split supersymmetry’’ naturally accommodate wino-like DM close to the weak scale, while the scalar superpartners can be much heavier \cite{Wells:2004di,ArkaniHamed:2004fb,Giudice:2004tc}. DM transforming as an SU(2)$_{\rm L}$ triplet has been studied extensively in the literature, both within split-SUSY scenarios \cite{Arvanitaki:2012ps,ArkaniHamed:2012gw,Hall:2012zp} and more generally \cite{Cohen:2013ama,Cirelli:2007xd,Ciafaloni:2012gs}. The model augments the Standard Model (SM) Lagrangian with
\begin{equation}
\mathcal{L}_{\rm DM} = \frac{1}{2} {\rm Tr} \bar{\chi} \left( i \slashed{D} - M_{\chi} \right) \chi\,.
\label{eq:DMlagrangian}
\end{equation}
We take $M_{\chi} = m_{\chi} \mathbb{I}$, such that in the unbroken theory all the DM fermions have the same mass. After electroweak symmetry breaking, the three states $\chi^{1,2,3}$ break into a Majorana fermion $\chi^0$ and a Dirac fermion $\chi^+$. A small mass difference, $\delta m$, between these states is then generated radiatively, ensuring that $\chi^0$ makes up the observed stable DM. Note, however, that both the charged and neutral states will be included in the EFT.

An effective field theory for this model, NRDM-SCET, was introduced in \cite{Ovanesyan:2014fwa} and used to calculate the rates for the annihilation processes $\chi \chi \to ZZ, Z\gamma, \gamma \gamma$. Specifically the EFT generalizes soft-collinear effective theory (SCET) \cite{Bauer:2000ew,Bauer:2000yr,Bauer:2001ct,Bauer:2001yt} to include non-relativistic dark matter (NRDM) in the initial state. Schematically the calculation involves several steps. Firstly the full theory has to be matched onto the relevant NRDM-SCET$_{\rm EW}$ operators at the high scale of $\mu \simeq 2 m_{\chi}$. The qualifier EW indicates that this is a theory where electroweak degrees of freedom -- the $W$ and $Z$ bosons, top quark, and the Higgs -- are dynamical, as introduced in \cite{Chiu:2007yn,Chiu:2007dg,Chiu:2008vv,Chiu:2009mg,Chiu:2009ft}. These operators then need to be run down to the electroweak scale, $\mu \simeq m_Z$. At this low scale, we then match NRDM-SCET$_{\rm EW}$ onto a theory where the electroweak degrees of freedom are no longer dynamical, NRDM-SCET$_{\gamma}$. This matching accounts for the effects of electroweak symmetry breaking, such as the finite gauge boson masses. At this stage we can now calculate the low scale matrix elements which provide the Sommerfeld enhancement. We now briefly review each of these steps.

The first requirement is to match NRDM-SCET$_{\rm EW}$ and the full theory at the high scale $\mu_{m_{\chi}}$. The relevant operators in the EFT to describe DM annihilation have the following form:
\begin{equation}
O_r = \frac{1}{2} \left( \chi_v^{aT} i \sigma_2 \chi_v^b \right) \left( S_r^{abcd} \mathcal{B}_{n \perp}^{ic} \mathcal{B}_{\bar{n} \perp}^{jd} \right) i \epsilon^{ijk} (n - \bar{n})^k\,,
\label{eq:Ops}
\end{equation}
which is written in terms of the basic building blocks of the effective theory, and in the center of momentum frame we can define $v=(1,0,0,0)$, $n=(1,\hat{n})$, and $\bar{n}=(1,-\hat{n})$ where $\hat{n}$ is the direction of an outgoing gauge boson. In more detail $\chi_v^a$ is a non-relativistic two-component fermionic field of gauge index $a$ corresponding to the DM and $\mathcal{B}_{\bar{n},n}$ contain the outgoing (anti-)collinear gauge bosons $A_{\bar{n},n}^{\mu}$, which can be seen as
\begin{equation}
\mathcal{B}_{n \perp}^{\mu} = A^{\mu}_{n \perp} - \frac{k_{\perp}^{\mu}}{\bar{n} \cdot k} \bar{n} \cdot A_n^{\mu} + \ldots \,,
\label{eq:SCETB}
\end{equation}
where the higher order terms in this expression involve two or more collinear gauge fields. For $\mathcal{B}_{\bar{n} \perp}^{\mu}$ we simply interchange $n \leftrightarrow \bar{n}$. The full form of $\mathcal{B}_{n \perp}^{\mu}$ can be found in \cite{Bauer:2002nz}, and is collinear gauge invariant on its own. Finally the gauge index connection is encoded in $S_r^{abcd}$:
\begin{equation}\begin{aligned}
S_1^{abcd} &= \delta^{ab} ( \mathcal{S}_n^{ce} \mathcal{S}_{\bar{n}}^{de} )\,,\\
S_2^{abcd} &= ( \mathcal{S}_v^{ae} \mathcal{S}_{n}^{ce} ) ( \mathcal{S}_v^{bf} \mathcal{S}_{\bar{n}}^{df} )\,.
\label{eq:GaugeIndex}
\end{aligned}\end{equation}
These expressions are written in terms of adjoint Wilson lines of soft gauge bosons along some direction $n$, $\bar{n}$, or $v$; in position space the incoming Wilson line is
\begin{equation}
\mathcal{S}_v(x) = P \exp \left[ ig \int_{-\infty}^0 ds v \cdot A_v(x+ns) \right]\,,
\label{eq:SCETS}
\end{equation}
where the matrix $A^{bc}_v = -i f^{abc} A^a_v$ and for outgoing Wilson lines the integral runs from $0$ to $\infty$.

The fact there are only two possible forms of $S_r^{abcd}$ means there are only two relevant NRDM-SCET operators. An important requirement of the operators is that the incoming DM fields must be in an $s$-wave configuration. Then being a two-particle state of identical fermions, the initial state must be a spin singlet. If the annihilation was $p$-wave or higher, it would be suppressed by powers of the low DM velocity relative to these operators. The Wilson coefficients associated with these operators are determined by the matching. Calculating to NLL simply requires the tree-level result where $C_1(\mu_{m_{\chi}}) = - C_2(\mu_{m_{\chi}}) = - \pi \alpha_2(\mu_{m_{\chi}})/m_{\chi}$, where again $\alpha_2$ is the SU(2)$_{\rm L}$ fine structure constant. We extend this result to one loop in Sec.~\ref{sec:1WC}.

After matching, the next step is to evolve these operators down to the low scale, effectively resumming the large logarithms that caused a breakdown in the perturbative expansion of the coupling. This is done using the anomalous dimension matrix $\hat{\gamma}$ of the two operators (a matrix as the operators will in general mix during the running). In general the matrix can be broken into a diagonal collinear piece $\gamma_{W_T}$, and a non-diagonal soft contribution $\hat{\gamma}_S$, as
\begin{equation}
\hat{\gamma} = 2 \gamma_{W_T} \mathbb{I} + \hat{\gamma}_S\,.
\label{eq:anomdim}
\end{equation}
To NLL these results are given by \cite{Ovanesyan:2014fwa}:
\begin{equation}\begin{aligned}
\gamma_{W_T} &= \frac{\alpha_2}{4\pi} \Gamma_0^g \ln \frac{2m_{\chi}}{\mu} - \frac{\alpha_2}{4\pi} b_0 + \left( \frac{\alpha_2}{4\pi} \right)^2 \Gamma_1^g \ln \frac{2m_{\chi}}{\mu}\,, \\
\hat{\gamma}_S &= \frac{\alpha_2}{\pi} (1-i\pi) \begin{pmatrix} 2 & 1 \\ 0 & -1 \end{pmatrix} - \frac{2\alpha_2}{\pi} \begin{pmatrix} 1 & 0 \\ 0 & 1 \end{pmatrix}\,.
\label{eq:anomdimparts}
\end{aligned}\end{equation}
Here the collinear anomalous dimension has been written in terms of the SU(2)$_{\rm L}$ one-loop $\beta$-function, $b_0 = 19/6$, as well as the cusp anomalous dimensions, $\Gamma_0^g = 8$ and $\Gamma_1^g = 8 \left(\frac{70}{9} - \frac{2}{3} \pi^2 \right)$. Below the matching scale, the spin of the DM is no longer important. As such the anomalous dimension determined in \cite{Ovanesyan:2014fwa} for the fermionic wino should resum the same logarithms as those that appear in the scalar case considered in \cite{Bauer:2014ula}, and we have confirmed they agree.

We can then explicitly use the full anomalous dimension to evolve the operators as follows:
\begin{eqnarray}
\begin{bmatrix} C_{\pm}^X (\left\{ m_i \right\}) \vspace{0.1cm}\\ C_{0}^X (\left\{ m_i \right\}) \end{bmatrix} &= &e^{\hat{D}^X(\mu_Z,\left\{ m_i \right\}))} P \exp \left( \int_{\mu_{m_{\chi}}}^{\mu_Z} \frac{d\mu}{\mu} \hat{\gamma}(\mu, m_{\chi}) \right) \nonumber \\
&\times& \begin{bmatrix} C_1(\mu_{m_{\chi}},m_{\chi}) \\ C_2(\mu_{m_{\chi}},m_{\chi}) \end{bmatrix}\,, \label{eq:Running}
\end{eqnarray}
Let us carefully explain the origin and dependence of each of these terms. Starting from the right, $C_1$ and $C_2$ are the high-scale Wilson coefficients of the operators stated in Eq.~\eqref{eq:Ops}, resulting from a matching of the full theory onto NRDM-SCET$_{\rm EW}$. These only depend on the high scales, specifically $\mu_{m_{\chi}}$ and $m_{\chi}$. Next the anomalous dimension $\hat{\gamma}$ is also a high scale object, and so only depends on $m_{\chi}$ and now $\mu$ as it runs between the relevant scales. $\hat{D}^X$ is a factor accounting for the low-scale matching from NRDM-SCET$_{\rm EW}$ onto NRDM-SCET$_{\gamma}$ -- a theory where the electroweak modes have been integrated out, see \cite{Chiu:2007yn,Chiu:2007dg,Chiu:2008vv,Chiu:2009mg,Chiu:2009ft}. It is a matrix as soft gauge boson exchanges can mix the operators. Furthermore $\hat{D}^X$ is labelled by $X$ to denote its dependence on the specific final state considered, $\gamma \gamma$, $\gamma Z$ or $ZZ$. This object depends on the low-scale physics and so depends on $\mu_Z$ and all the masses in the problem, which we denote as $\{ m_i \}$. Finally on the left we have our final coefficients $C_{\pm}^X$ and $C_{0}^X$, which as explained below can be associated with the charged and neutral annihilation processes. In an all orders calculation of all terms in Eq.~\eqref{eq:Running}, the scale dependence would completely cancel on the right hand side, implying that $C_{\pm}^X$ and $C_{0}^X$ depend only on the mass scales in the problem and not $\mu_{m_{\chi}}$ or $\mu_Z$. Nevertheless at any finite perturbative order, the scale dependence does not cancel completely and so a residual dependence is induced in these coefficients. We will exploit this to estimate the uncertainty in our results associated with missing higher order terms.

As we are performing a resummed calculation, the order to which we calculate is defined in terms of the large electroweak logarithms we can resum. In general the structure of the logarithms can be written schematically as:
\begin{equation}
\ln \frac{C}{C^{\rm tree}} \sim \sum_{k=1}^{\infty} \left[ \vphantom{\alpha_2^k \ln^{k+1}} \right. \underbrace{\alpha_2^k \ln^{k+1}}_{\rm LL} + \underbrace{\alpha_2^k \ln^k}_{\rm NLL} + \underbrace{\alpha_2^k \ln^{k-1}}_{\rm NNLL} + \ldots \left. \vphantom{\alpha_2^k \ln^{k+1}} \right]\,,
\label{eq:NLLprimeDef}
\end{equation}
where since Sudakov logarithms exponentiate, we have defined the counting in terms of the log of the result. Furthermore all corrections are defined with respect to the tree level result $C^{\rm tree} \sim \mathcal{O}(\alpha_2)$, which is a convention we will follow throughout. With this definition of the counting, to perform the running in Eq.~\eqref{eq:Running} to NLL order, there are three effects that must be accounted for: 1. high-scale matching at tree level; 2. two-loop cusp and one-loop non-cusp anomalous dimensions; and 3. the low-scale matching at tree level, together with the rapidity renormalization group (\cite{Chiu:2012ir,Becher:2010tm}) at NLL. To extend this to NNLL all three of these need to be calculated to one order higher. In between these two is the NLL$^{\prime}$ result we present here, which involves determining both the high and low-scale matching at one loop. In terms Eq.~\eqref{eq:NLLprimeDef}, this amounts to determining the leading $k=1$ piece of the NNLL result. To the extent that $\mathcal{O}(\alpha_2)$ corrections are larger than those at $\mathcal{O}(\alpha^2_2 \ln (\mu_{m_{\chi}}^2/\mu_Z^2))$, the NLL$^{\prime}$ result is an improvement over NLL and more important than NNLL.

Before presenting the result of that calculation, however, it is worth emphasizing another advantage gained from the effective theory. In addition to allowing us to resum the Sudakov logarithms, the effective theory also allows this problem to be cleanly separated from the issue of low velocity Sommerfeld enhancement in the amplitude -- in NRDM-SCET there is a Sommerfeld-Sudakov factorization. At leading power the relevant SCET Lagrangian contains no interaction with the DM field. On the other hand NRDM does contain soft modes, which are responsible for running the couplings, however these modes do not couple the Sommerfeld potential to the hard interaction at leading power. Consequently matrix elements for the DM factorize from the elements of the states annihilated into. This allows for an all orders factorized formula for the DM annihilation amplitude in this theory:
\begin{equation}\begin{aligned}
\mathcal{M}_{\chi^0\chi^0 \to X} &= 4 \sqrt{2} m_{\chi} P_X \left[ s_{00} \left( \Sigma_1^X - \Sigma_2^X \right) + \sqrt{2}  s_{0\pm} \Sigma_1^X \right]\,, \\
\mathcal{M}_{\chi^+\chi^- \to X} &= 4 m_{\chi} P_X \left[ s_{\pm0} \left( \Sigma_1^X - \Sigma_2^X \right) + \sqrt{2}  s_{\pm\pm} \Sigma_1^X \right]\,.
\label{eq:Factorized}
\end{aligned}\end{equation}
Here $X$ can be $\gamma \gamma$, $\gamma Z$ or $ZZ$ and $P_{\gamma \gamma} = - e^2 \epsilon_{n \perp}^i \epsilon_{\bar{n} \perp}^j \epsilon^{ijk} \hat{n}^k/(2m_{\chi})$, whilst $P_{\gamma Z} = \cot \bar{\theta}_W P_{\gamma \gamma}$ and $P_{ZZ} = \cot^2 \bar{\theta}_W P_{\gamma \gamma}$, with $\bar{\theta}_W$ the $\overline{\rm MS}$ Weinberg angle. The key physics in this equation is that the contribution from Sommerfeld enhancement is captured in the terms $s_{ij}$, whilst the contribution from electroweak logarithms is in $\Sigma_i^X$; the two are manifestly factorized and can be calculated independently.

The focus of the present chapter is to extend the calculation of the Sudakov effects. In terms of the factorized result stated in Eq.~\eqref{eq:Factorized} this amounts to a modification of $\Sigma_i^X$. Explicitly, from there we can see that:
\begin{equation}\begin{aligned}
\left| \Sigma_1^X \right|^2 &= \frac{\sigma_{\chi^+ \chi^- \to X}^{\cancel{\rm SE}}}{\sigma_{\chi^+ \chi^- \to X}^{\rm tree}}\,, \\
\left| \Sigma_1^X - \Sigma_2^X \right|^2 &= \frac{\sigma_{\chi^0 \chi^0 \to X}^{\cancel{\rm SE}}}{\sigma_{\chi^+ \chi^- \to X}^{\rm tree}}\,,
\label{eq:SigmaDef}
\end{aligned}\end{equation}
where $\cancel{\rm SE}$ denotes a calculation where Sommerfeld Enhancement is intentionally left out. To be even more explicit, we can write these Sudakov effects in terms of the Wilson coefficients in Eq.~\eqref{eq:Running}. Specifically we have:
\begin{equation}\begin{aligned}
\Sigma_1^X &= \frac{C_{\pm}^X}{C_1^{\rm tree}}\,, \\
\Sigma_1^X - \Sigma_2^X &= \frac{C_{0}^X}{C_1^{\rm tree}}\,,
\label{eq:SigmaDefExplicit}
\end{aligned}\end{equation}
where as stated above $C_1^{\rm tree} = - \pi \alpha_2/m_{\chi}$.

\section{The One-Loop Correction}
\label{sec:1WC}

In terms of the formalism described in the previous section, we now state one of the main results of this chapter: the high-scale Wilson coefficients $C_r$ calculated to one loop as shown in Eq.~\eqref{eq:WilsonCoeff}. The details have been eschewed to App.~\ref{app:oneloopfull}. In short this calculation involves enumerating and evaluating the 25 one-loop diagrams that mediate $\chi^a \chi^b \to W^c W^d$ in the unbroken full theory and then matching this result onto the NRDM-SCET$_{\rm EW}$ operators. For example, we evaluate diagrams such as
\begin{center}
\includegraphics[height=0.15\columnwidth]{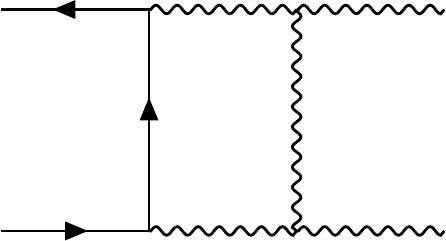} \hspace{0.1in}
\includegraphics[height=0.15\columnwidth]{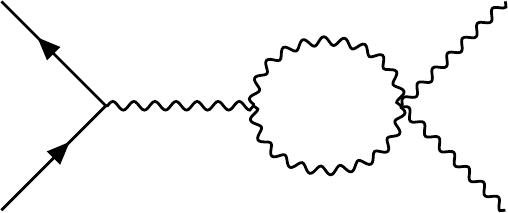}
\end{center}
and provide the analytic expression graph by graph. In addition we account for the counter term contribution, the change in the running of the coupling through the matching, and also ensure that the calculation maintains the Sudakov-Sommerfeld factorization. Combining all of these we find
\begin{equation}\begin{aligned}
C_1(\mu) &=- \frac{\pi\alpha_2(\mu)}{m_{\chi}} + \frac{\alpha_2(\mu)^2}{4m_{\chi}} \left[ 2 \ln^2 \frac{\mu^2}{4m_{\chi}^2} \right. \\
&\left. + 2 \ln \frac{\mu^2}{4m_{\chi}^2} + 2 i \pi \ln \frac{\mu^2}{4m_{\chi}^2} + 8 - \frac{11\pi^2}{6} \right] \,, \\
C_2(\mu) &=\frac{\pi\alpha_2(\mu)}{m_{\chi}} - \frac{\alpha_2(\mu)^2}{2m_{\chi}} \left[ \ln^2 \frac{\mu^2}{4m_{\chi}^2} \right. \\
&\hspace{0.318in}\left. + 3 \ln \frac{\mu^2}{4m_{\chi}^2} - i \pi \ln \frac{\mu^2}{4m_{\chi}^2} - \frac{5\pi^2}{12} \right]\,,
\label{eq:WilsonCoeff}
\end{aligned}\end{equation}
where here and throughout this section $\alpha_2(\mu)$ is the coupling defined below the scale of the DM mass, $m_{\chi}$. We explain this distinction carefully in App.~\ref{app:oneloopfull}. For each coefficient in Eq.~\eqref{eq:WilsonCoeff} the first term represents the tree-level contribution. A cross check on this result is provided in App.~\ref{app:consistency}, where we check that the $\mu$ dependence of this result properly cancels with that of the NLL resummation for the $\mathcal{O}(\alpha_2)$ corrections. The cancellation occurs between our result in Eq.~\eqref{eq:WilsonCoeff} and the running induced by the anomalous dimension stated in Eqs.~\eqref{eq:anomdim} and \eqref{eq:anomdimparts}; this can be seen clearly in Eq.~\eqref{eq:Running} as these are the only two objects that depend on $\mu_{m_{\chi}}$. As the anomalous dimension is independent of the DM spin, the logarithms appearing in our high-scale matching coefficients should also be, and indeed ours match those in the scalar calculation of \cite{Bauer:2014ula}. Of course the finite terms should not be, and are not, the same.

We next state the contribution from the low-scale matching. Unsurprisingly, as this effect accounts for electroweak symmetry breaking effects such as the gauge boson masses, it is in general dependent upon the identity of the final states. Again this is a matching calculation and involves evaluating diagrams that appear in SCET$_{\rm EW}$, but not SCET$_{\gamma}$, and we provide three examples below.
\begin{center}
\includegraphics[height=0.2\columnwidth]{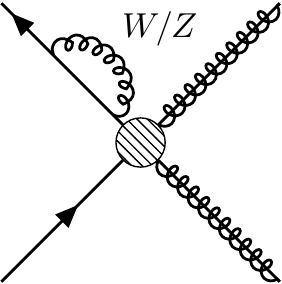} \hspace{0.1in}
\includegraphics[height=0.2\columnwidth]{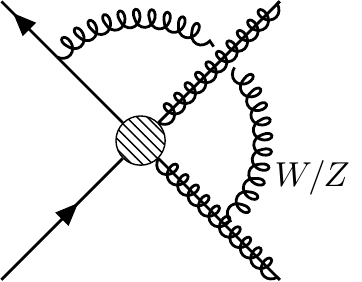} \hspace{-0.05 in}
\includegraphics[height=0.2\columnwidth]{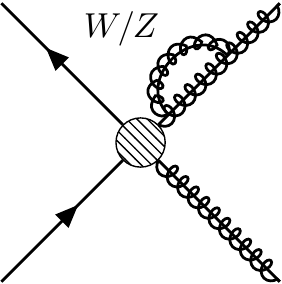}
\end{center}
A central difficulty in the calculation is accounting for the effects of electroweak symmetry breaking, see for example \cite{Denner:2016etu} for a recent discussion. In order to simplify this we make use of the general formalism for electroweak SCET of \cite{Chiu:2007yn,Chiu:2007dg,Chiu:2008vv,Chiu:2009mg,Chiu:2009ft}, which we have extended to include the case of non-relativistic external states.\footnote{This calculation can also be performed using the rapidity renormalization group \cite{Chiu:2012ir,Becher:2010tm}, but we will not use that formalism here.} We postpone the details to App.~\ref{app:lowscalematching}. The approach breaks the full low-scale matching into a soft and collinear component, which are the labels associated with the non-diagonal and diagonal contributions respectively, rather than the effective theory modes that give rise to them. This distinction is explained in detail in App.~\ref{app:lowscalematching}. In our case, $\hat{D}^X(\mu)$ in Eq.~\eqref{eq:Running} can be specified through
\begin{equation}
\exp \left[\hat{D}^X(\mu) \right] = \left[ \hat{D}_s(\mu) \right] \left[ \vphantom{\hat{D}} D_c^{\chi}(\mu) \mathbb{I} \right] \exp \left[ \sum_{i\in X} D_c^i(\mu) \mathbb{I} \right] \,,
\label{eq:lowbreakdown}
\end{equation}
where again $X$ can be $\gamma \gamma$, $\gamma Z$ or $ZZ$, $\hat{D}_s(\mu)$ is the non-diagonal soft contribution and a matrix as it mixes the operators, whilst $D_c^{\chi}(\mu)$ and $D_c^i(\mu)$ are the initial and final state diagonal contributions respectively. Note both $\hat{D}_s(\mu)$ and the identity matrix $\mathbb{I}$ are $2 \times 2$ matrices. The terms that are not exponentiated in Eq.~\eqref{eq:lowbreakdown} are only determined to $\mathcal{O}(\alpha_2)$, whereas the final state collinear contribution has its largest contribution resummed to all orders. Using this definition we find that the components of the soft matrix are (see App.~\ref{app:lowscalematching}):
\begin{eqnarray}
~{[\hat{D}_s]}_{11} &=& 1 + \frac{\alpha_2(\mu)}{2\pi} \left[ \ln \frac{m_W^2}{\mu^2} (1-2i\pi) + c_W^2 \ln \frac{m_Z^2}{\mu^2} \right]\,, \nn
{[\hat{D}_s]}_{12} &=& \frac{\alpha_2(\mu)}{2\pi} \ln \frac{m_W^2}{\mu^2} (1-i\pi)\,, \label{eq:LowSoft} \\
{[\hat{D}_s]}_{21} &=& 1 + \frac{\alpha_2(\mu)}{2\pi} \ln \frac{m_W^2}{\mu^2} (2-2i\pi)\,, \nn
{[\hat{D}_s]}_{22} &=& 1\,. \nonumber
\end{eqnarray}
Here and throughout we use the shorthand $c_W = \cos \bar{\theta}_W$ and $s_W = \sin \bar{\theta}_W$. Further, the collinear contributions can be written as:
\begin{equation}\begin{aligned}
D_c^{\chi}(\mu) = & 1 - \frac{\alpha_2(\mu)}{2\pi} \left[ \ln \frac{m_W^2}{\mu^2} + c_W^2 \ln \frac{m_Z^2}{\mu^2} \right]\,, \\
D_c^i(\mu) = &\frac{\alpha_2(\mu)}{2\pi} \left[ \ln \frac{m_W^2}{\mu^2} \ln \frac{4 m_{\chi}^2}{\mu^2} - \frac{1}{2} \ln^2 \frac{m_W^2}{\mu^2} \right. \\
&\hspace{0.5in}\left. - \ln \frac{m_W^2}{\mu^2} + c^i_1 \ln \frac{m_Z^2}{\mu^2} + c^i_2 \right]\,,
\label{eq:LowColinear}
\end{aligned}\end{equation}
where $i = Z$ or $\gamma$ and we have:
\begin{equation}\begin{aligned}
c^Z_1 &= \frac{5-24s_W^2-22s_W^4}{24c_W^2}\,, \\
c^{\gamma}_1 &= 1-\frac{47}{36}s_W^2\,,
\label{eq:LowColinearConsts1}
\end{aligned}\end{equation}
and 
\begin{equation}\begin{aligned}
c^Z_2 &= -1.5534 - 3.0892 i\,, \\
c^{\gamma}_2 &= -0.812092\,.
\label{eq:LowColinearConsts2}
\end{aligned}\end{equation}
Analytic expressions for these last results are provided in App.~\ref{app:lowscalematching} and App.~\ref{app:PiFunctions}, and we give numerical values here as the expressions are lengthy.

The $\mu$ dependence of the low-scale matching is demonstrated to cancel with that in our high-scale matching result when the running is turned off, the details being shown in App.~\ref{app:consistencylow}. We emphasize that this cross check involves not only the $\mu$ dependence of the objects in Eq.~\eqref{eq:lowbreakdown}, but also the $\mu$ dependence of the high-scale coefficients stated in Eq.~\eqref{eq:WilsonCoeff} and further the SM SU(2)$_{\rm L}$ and U(1)$_Y$ $\beta$-functions. The full $\mu$ cancellation is non-trivial -- it requires the interplay between each of these objects. This ultimately provides us with confidence in the results as stated. As a further check, our low-scale matching result does not depend on the spin of the DM. As such we should be again able to compare our result to the scalar case calculated in \cite{Bauer:2014ula}. In that work they only considered the $\gamma \gamma$ final state, and also neglected the impact of SM fermions. Restricting our calculation to the same assumptions, we confirm that the $\mu$ dependence in our result matches theirs.

Taking our results in combination, we can extend the NLL calculation to NLL$^{\prime}$. Of course we cannot show full NNLL results in the absence of the higher order anomalous dimension calculation, nevertheless the results we state here determine the cross section with perturbative uncertainties on the Sudakov effects reduced to the percent level. At $\mathcal{O}(\alpha_2^2)$,\footnote{Again note that all counting here is relative to the lowest order contribution, which occurs at $C^{\rm tree} \sim \mathcal{O}(\alpha_2)$. As such the absolute order of the terms in this sentence is $\mathcal{O}(\alpha_2^3)$.} our calculation accounts for all terms of the form $\alpha_2^2 \ln^4 (\mu_{m_{\chi}}^2/\mu_Z^2)$, $\alpha_2^2 \ln^3 (\mu_{m_{\chi}}^2/\mu_Z^2)$, and $\alpha_2^2 \ln^2 (\mu_{m_{\chi}}^2/\mu_Z^2)$. The first perturbative term we are missing at this order is $\alpha_2^2 \ln (\mu_{m_{\chi}}^2/\mu_Z^2)$. Taking $\mu_Z = m_Z$ and $m_{\chi}$ anywhere from $m_Z$ to $20$ TeV, we find the absence of these terms induces an uncertainty that is less than 1\%, demonstrating the claimed accuracy.

\begin{figure}[t!]
\centering
\begin{tabular}{c}
\includegraphics[scale=0.55]{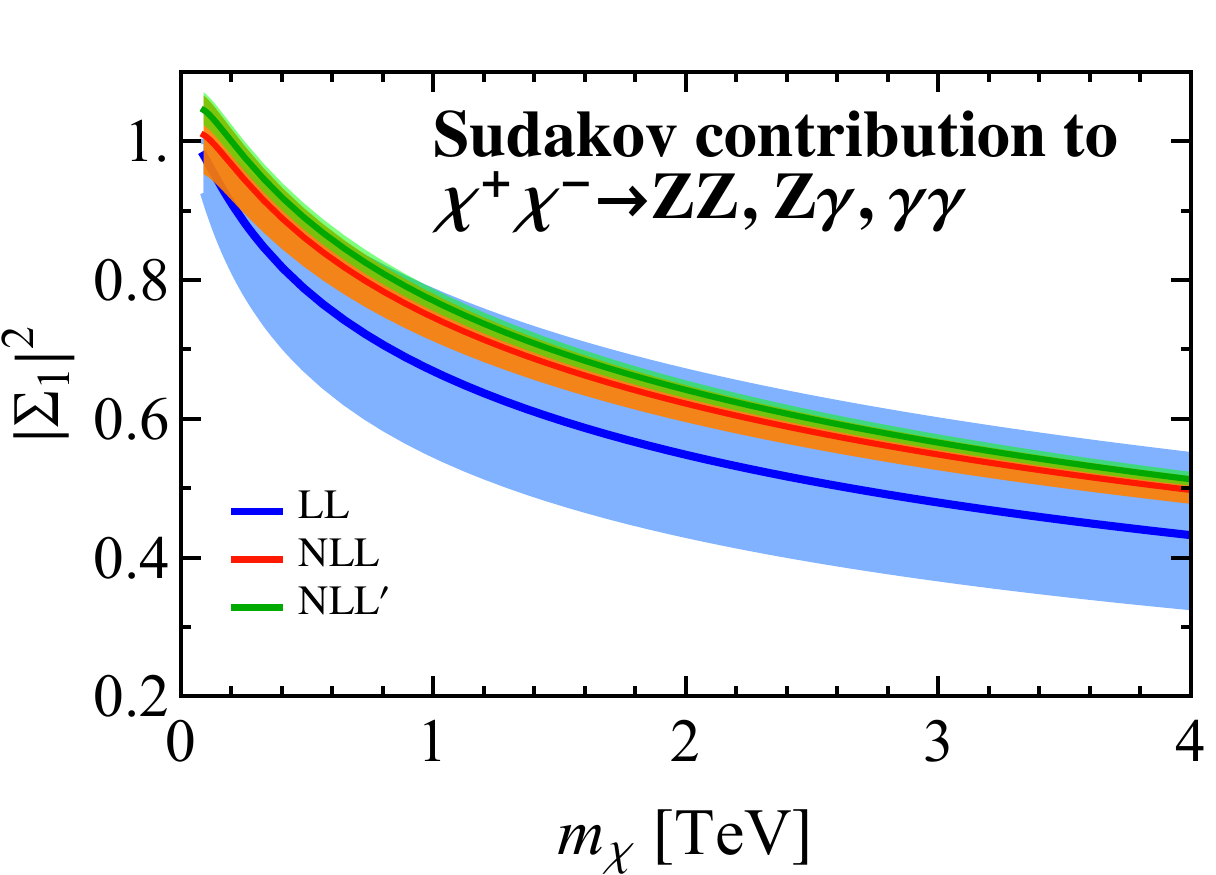} \hspace{0.12in}
\includegraphics[scale=0.565]{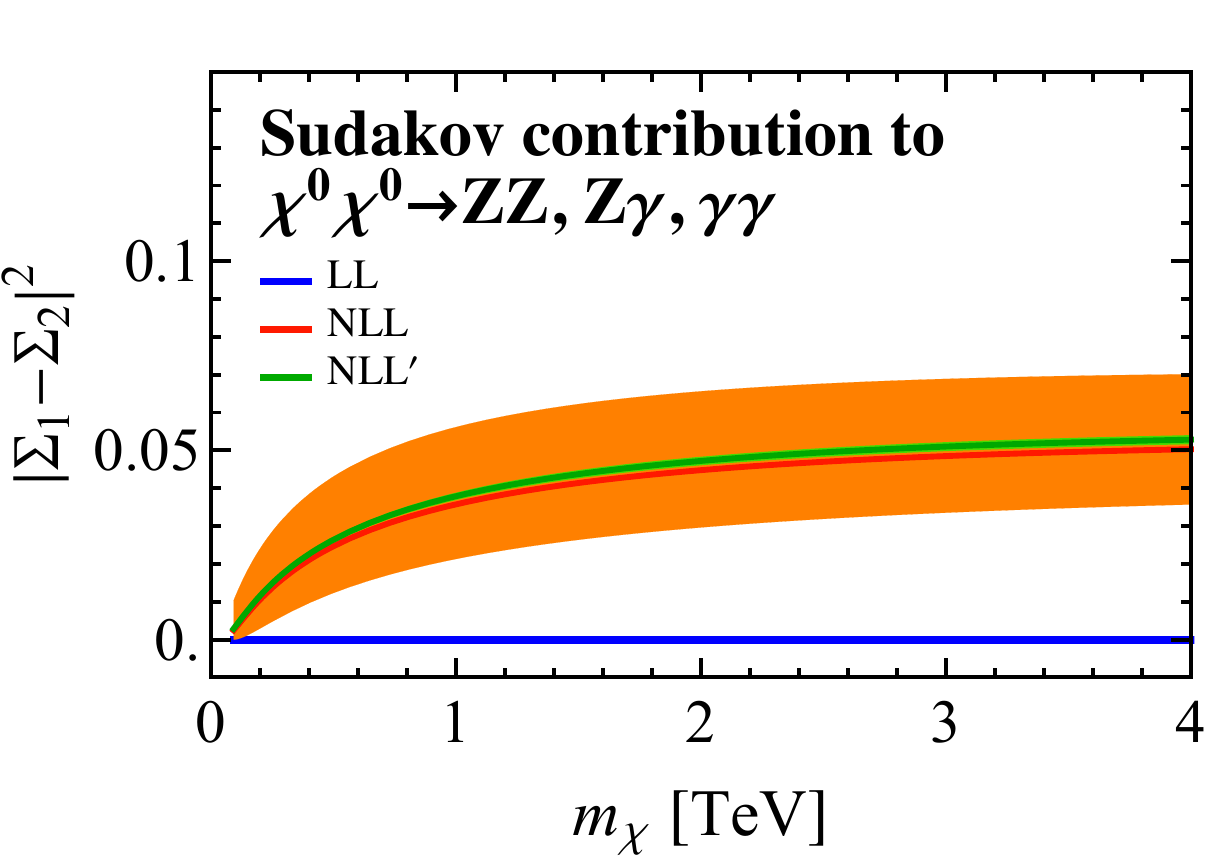}
\end{tabular}
\caption{\footnotesize{Here we show the NLL$^{\prime}$ electroweak corrections to the charged (left) and neutral (right) DM annihilations obtained by adding the one-loop high and low-scale corrections to the NLL result. The correction is in good agreement with the NLL calculation, whilst the scale uncertainties have been reduced. Bands are derived by varying the high scale between $m_{\chi}$ and $4 m_{\chi}$.}}
\label{fig:Nll1Loop}
\end{figure}

\begin{figure}[t!]
\centering
\begin{tabular}{c}
\includegraphics[scale=0.55]{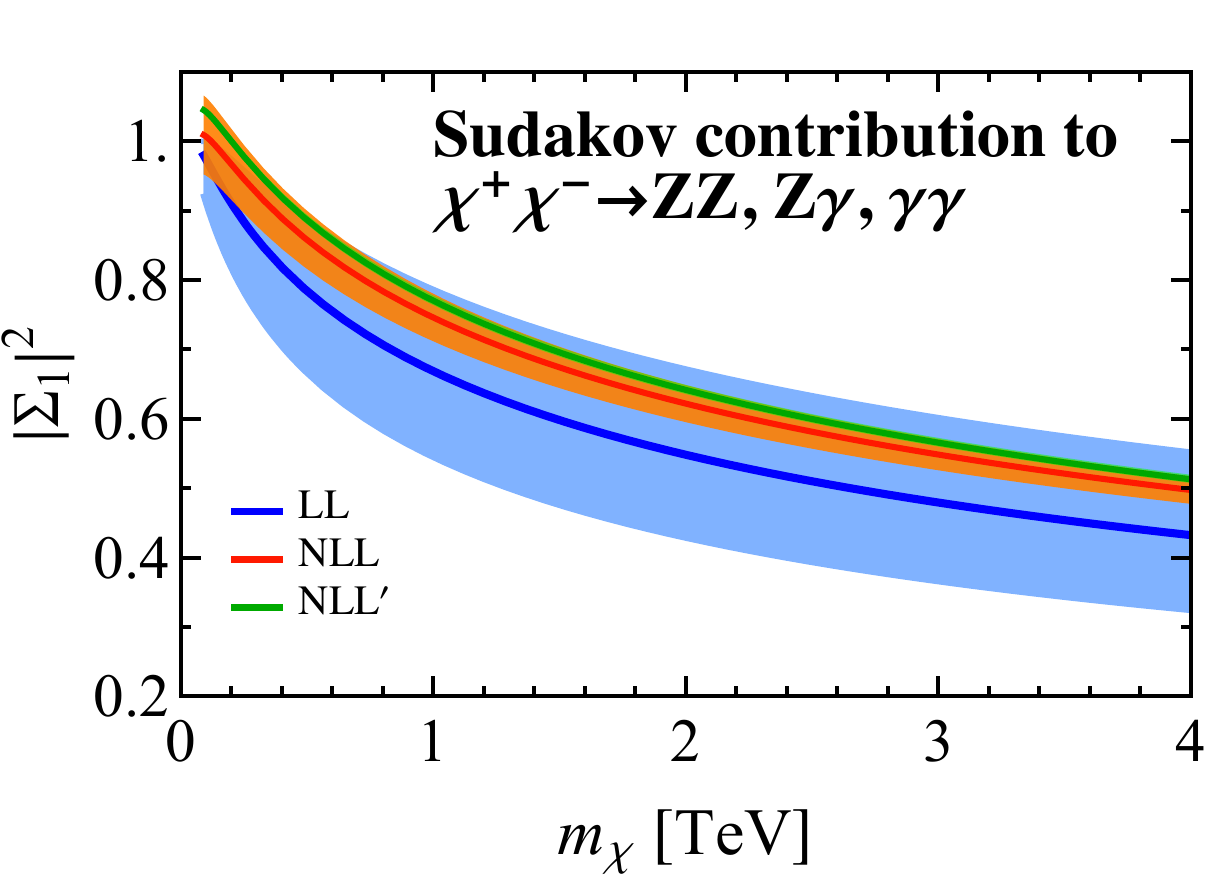} \hspace{0.12in}
\includegraphics[scale=0.565]{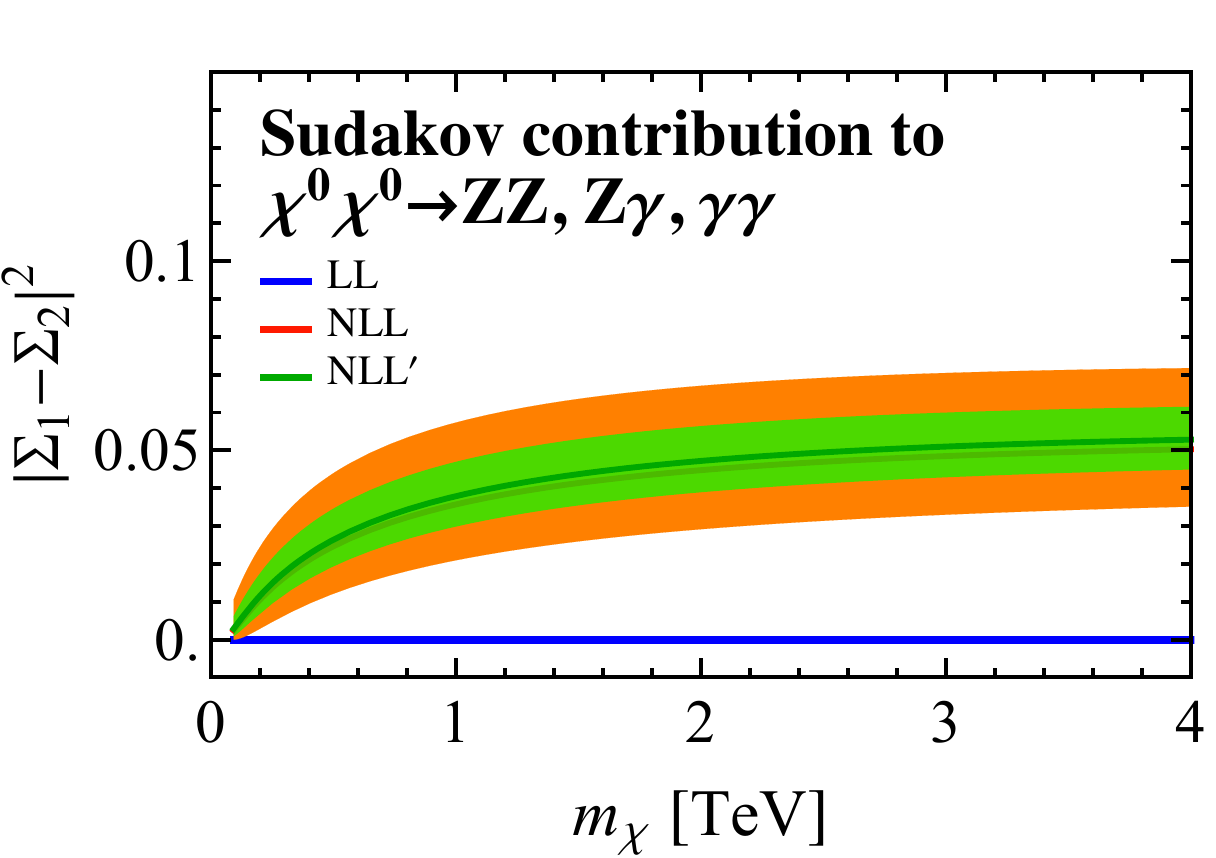}
\end{tabular}
\caption{\footnotesize{As for Fig.~\ref{fig:Nll1Loop}, but showing a variation in the low-scale matching between $m_Z/2$ and $2m_Z$, rather than a variation of the high-scale matching as shown there. As can be seen the NLL$^{\prime}$ contribution has reduced the low scale dependence in both cases, but is again consistent with the NLL result.}}
\label{fig:Nll1LoopLowScale}
\end{figure}

To combine the various results stated above into the cross section we take the factorized results in Eq.~\eqref{eq:Factorized}, and note that as the higher order Wilson coefficients have nothing to do with the Sommerfeld enhancement, their contribution is included in the $\Sigma$ terms as given explicitly in Eq.~\eqref{eq:SigmaDefExplicit}. We know that at tree level $s_{00} = s_{\pm \pm} = 1$ and $s_{0 \pm} = s_{\pm 0}=0$, implying that when the Sommerfeld enhancement can be ignored we can associate $|\Sigma_1|^2$ with the Sudakov contribution to $\chi^+ \chi^-$ annihilation and $|\Sigma_1 - \Sigma_2|^2$ with $\chi^0 \chi^0$.

\begin{figure}[t!]
\centering
\begin{tabular}{c}
\includegraphics[scale=0.46]{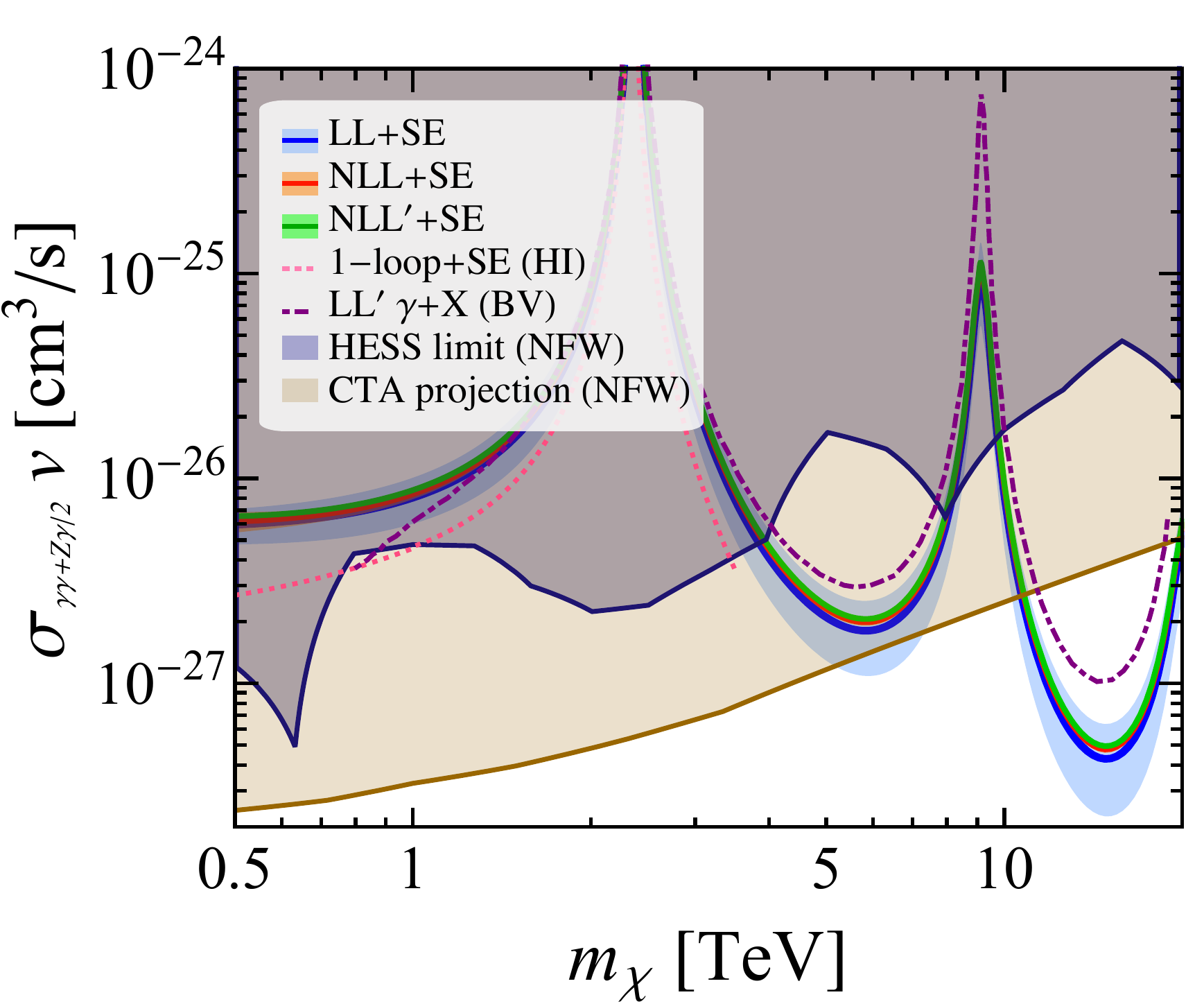}
\end{tabular}
\caption{\footnotesize{The impact of the NLL$^{\prime}$ result on the full cross section, which includes the Sommerfeld Enhancement (SE), is shown to be consistent with the lower orders result, suggesting the electroweak corrections are under control. Also shown is the rate for the semi-inclusive process $\gamma + X$ calculated to LL$^{\prime}$ in \cite{Baumgart:2015bpa}. In addition on this plot we show current bounds from H.E.S.S. and projected ones from CTA, determined assuming 5 hours of observation time. See text for details.}}
\label{fig:CrossSec}
\end{figure}

For this reason, in Fig.~\ref{fig:Nll1Loop} and Fig.~\ref{fig:Nll1LoopLowScale} we show the contributions to $|\Sigma_1|^2$ and $|\Sigma_1 - \Sigma_2|^2$ for LL, NLL and NLL$^{\prime}$. In both cases we see the addition of the one-loop corrections is completely consistent with the NLL results, suggesting that this approach has the Sudakov logarithms under control. In these plots we take a central value of $\mu_{m_{\chi}} = 2 m_{\chi}$ and $\mu_Z = m_Z$. In Fig.~\ref{fig:Nll1Loop} the bands are derived from varying the high-scale matching between $m_{\chi}$ and $4 m_{\chi}$. Recall that if we were able to calculate these quantities to all orders, they would be independent of $\mu$, and so varying these scales estimates the impact of missing higher order terms. For the $|\Sigma_1|^2$ NLL result, taking $\mu_{m_{\chi}} = 2 m_{\chi}$ is a minimum in the range varied over, so we symmetrise the uncertainties in order to indicate the range of uncertainty. Similarly in Fig.~\ref{fig:Nll1LoopLowScale} we show the equivalent plot, but here the bands are derived by varying the low scale $\mu_Z$ from $m_Z/2$ to $2m_Z$. Improving on the high and low-scale matching, as we have done here, should lead to a reduction in the scale uncertainty. In all four cases shown this is clearly visible and furthermore all results are still consistent with the NLL result within the uncertainty bands.

We can also take this result and determine the impact on the full DM annihilation cross section into line photons from $\gamma \gamma$ and $\gamma Z$ in this model, as we show in Fig.~\ref{fig:CrossSec}. We take the uncertainty on our final result to include the high and low-scale variations added in quadrature. For H.E.S.S. limits we use \cite{Abramowski:2013ax}, whilst for the CTA projection we assume 5 hours of observation time and use \cite{Cohen:2013ama,Bergstrom:2012vd}. For both we assume an NFW profile with a local DM density of 0.4 GeV/cm$^3$. We see again that our partial NLL$^{\prime}$ results are consistent with the NLL conclusions.\footnote{A digitized version of our cross section is available with the arXiv submission or upon request.} In this figure we also include the LL$^{\prime}$ result for the semi-inclusive process $\gamma+X$ taken from Fig.~7 of \cite{Baumgart:2015bpa}, denoted by (BV). The semi-inclusive result is above our line photon result, except at low DM masses. Note that this work does not show scale uncertainties, so the precise difference is hard to quantify numerically.

\section{Comparison to Earlier Work}
\label{sec:Comp}

\begin{figure}[t!]
\centering
\begin{tabular}{c}
\includegraphics[scale=0.55]{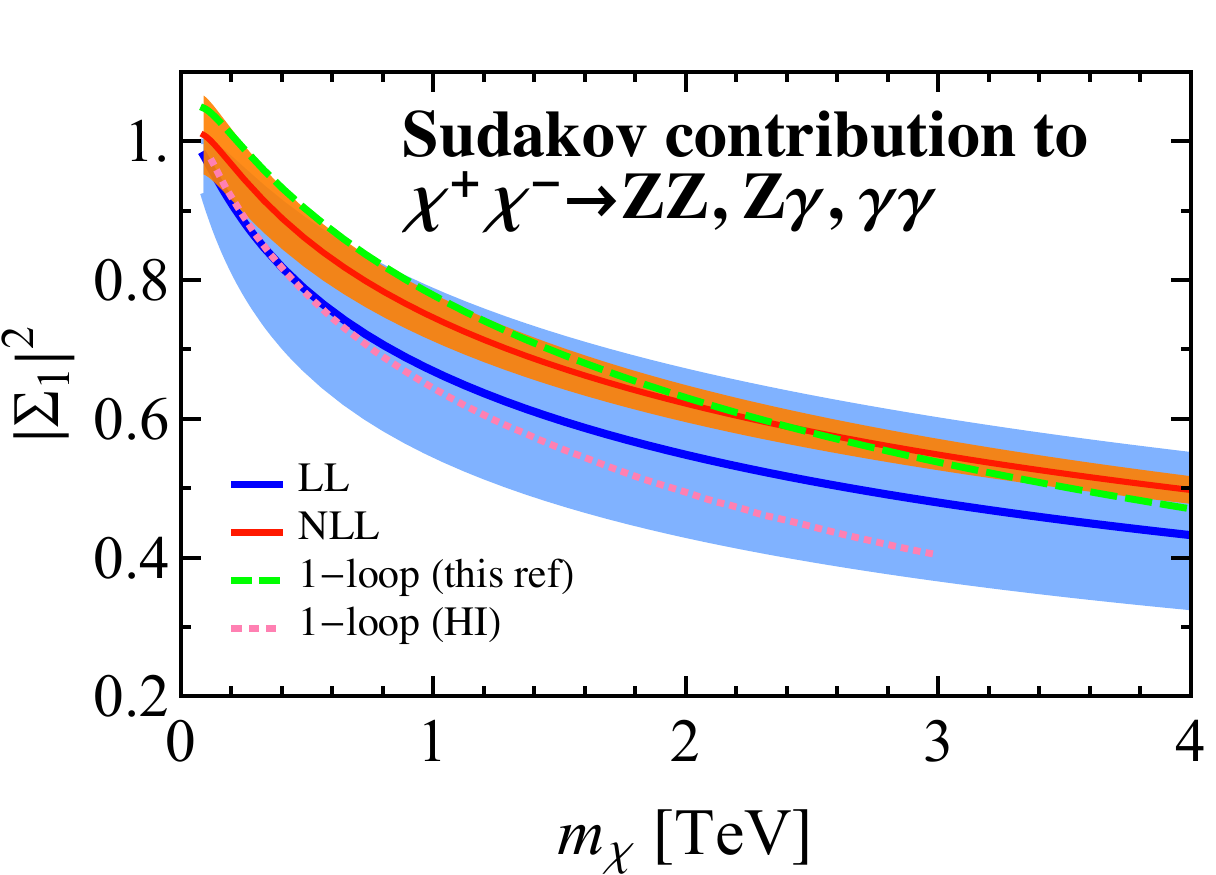} \hspace{0.12in}
\includegraphics[scale=0.565]{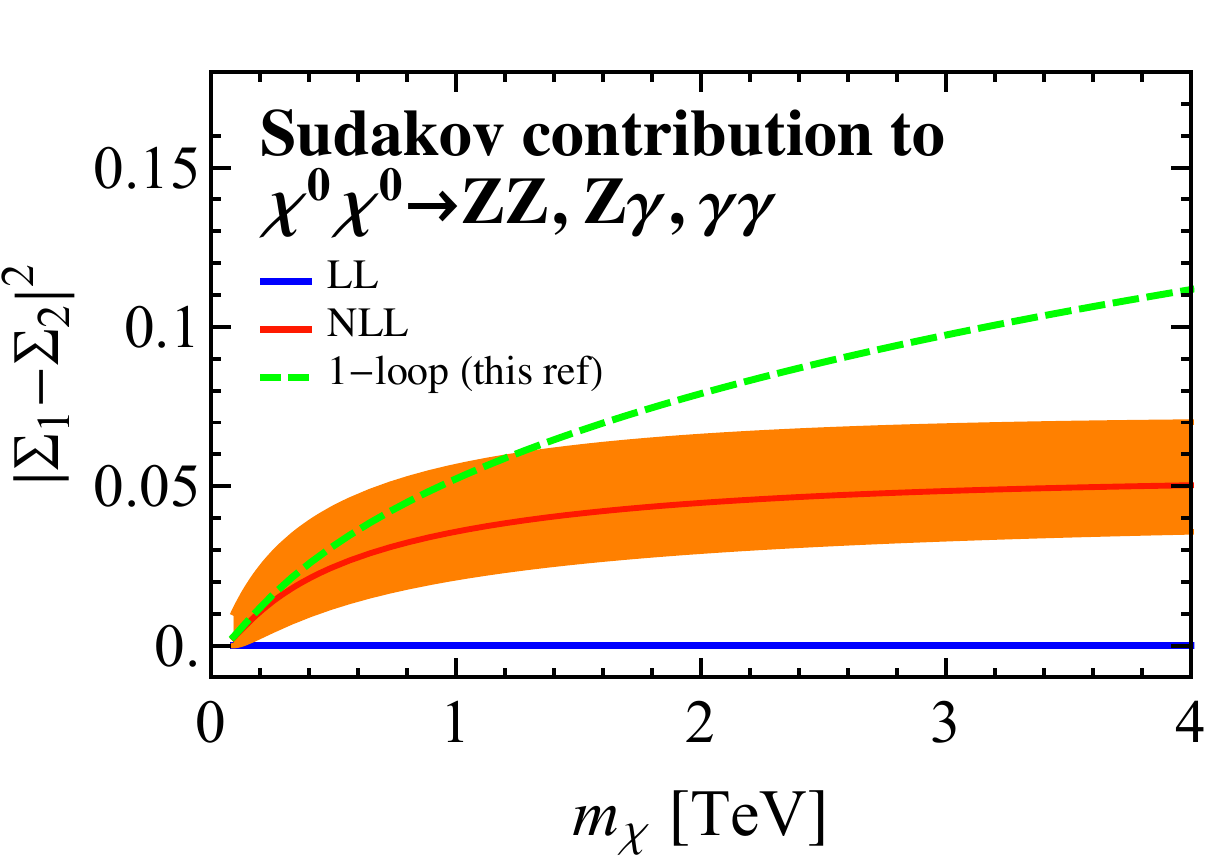}
\end{tabular}
\caption{\footnotesize{Similar to Fig.~\ref{fig:Nll1Loop}, but instead of the NLL$^{\prime}$ results we show our high and low-scale one-loop results including no running from the anomalous dimension. For the case of $\chi^+ \chi^-$ annihilation we further show the equivalent result of HI, taken from Fig.~11 of their work (which only extends up to 3 TeV and is the origin of the cutoff). There is evidently some discrepancy between the results. Note that at low masses where the Sudakov logarithms are not too large, our result is consistent with the NLL result as would be expected. See text for details.}}
\label{fig:Nllcf1loop}
\end{figure}

In addition to using our results from the previous section in conjunction with the running due to the anomalous dimension, we can also consider the case where we take our one-loop result in isolation. In this sense we should be able to reproduce the initial problem of large logarithms seen by HI. We show this in Fig.~\ref{fig:Nllcf1loop}, compared to the LL and NLL result. For $\Sigma_1$ our one-loop result is consistent with that from NLL, indicating the importance of the $\alpha_2 \ln^2 (\mu_{m_{\chi}}^2/\mu_Z^2)$ and $\alpha_2 \ln (\mu_{m_{\chi}}^2/\mu_Z^2)$ corrections to $C^{\rm tree}$. For $\Sigma_1 - \Sigma_2$, which starts at NLL, our one-loop result is only consistent with the NLL expression in the small $m_{\chi}$ region.

\begin{figure}[t!]
\centering
\begin{tabular}{c}
\includegraphics[scale=0.55]{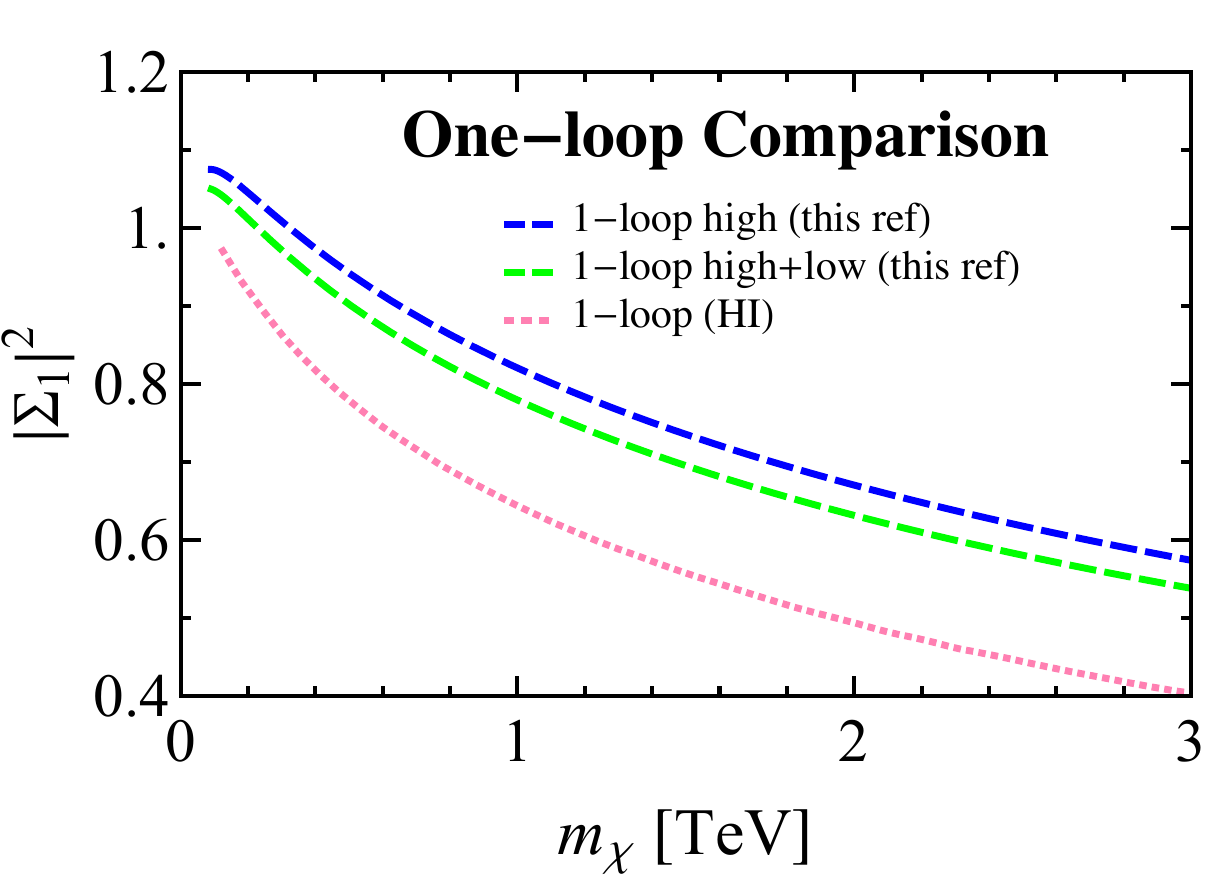} \hspace{0.12in}
\includegraphics[scale=0.55]{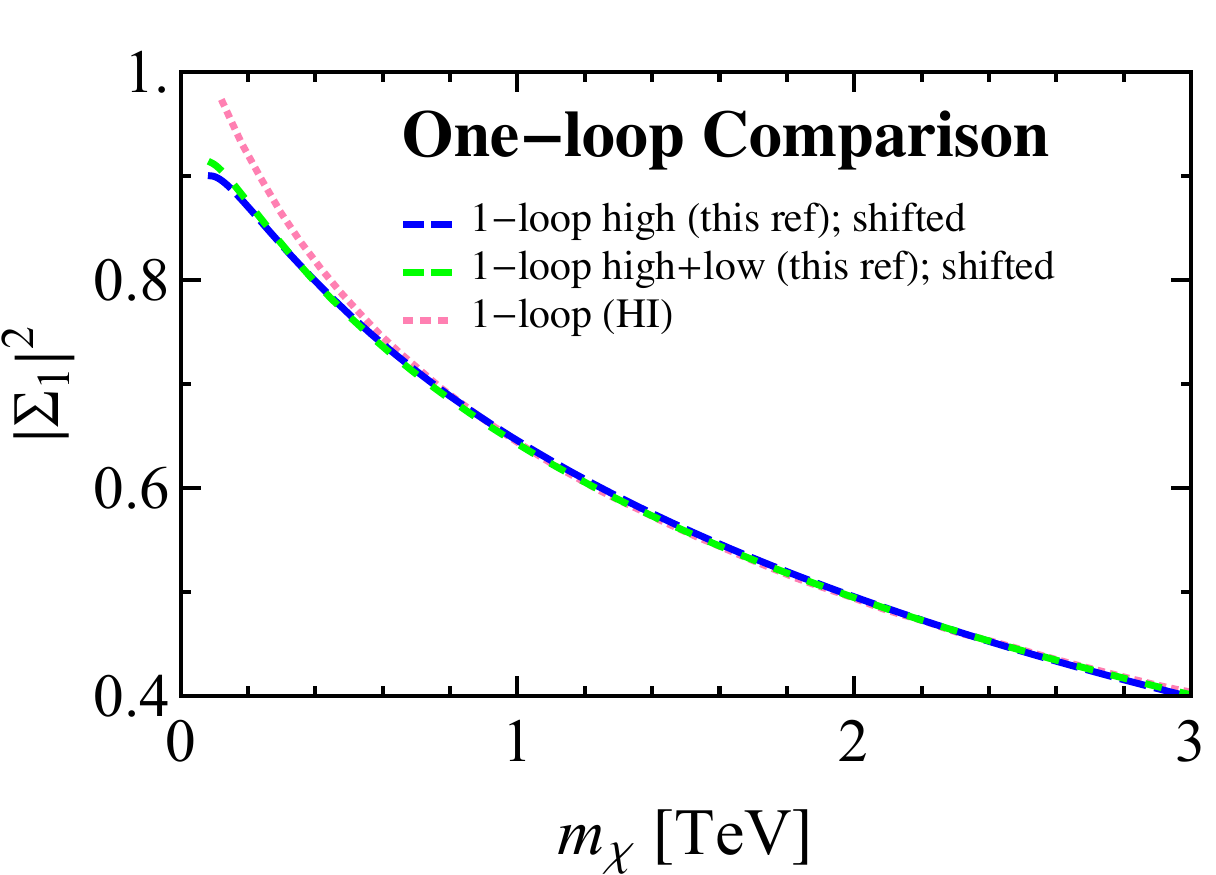}
\end{tabular}
\caption{\footnotesize{We show the result of HI for $|\Sigma_1|^2$ compared to two variations of our result. Firstly in the left panel we show our result with the high only or high and low-scale calculations compared to the result of HI, taken from Fig.~11 of their paper. In the right panel we take our results and shift them by a constant factor. The shifted results show that above around 1~TeV the shape of our result is in good agreement with HI, but the constant offset highlights there is tension.}}
\label{fig:UsVsHI}
\end{figure}

For the $|\Sigma_1|^2$ case we also show on that plot the equivalent curve for HI as extracted from Fig.~11 of their paper. From here it is clear that the qualitative shape of our results agrees with theirs but that there is some tension. This tension is already clear in Fig.~\ref{fig:CrossSec}, but in Fig.~\ref{fig:UsVsHI} we explore this difference in more detail. In the left panel we show the difference between their result and ours, showing our calculation with and without the low-scale matching included. Given the low-scale matching accounts for the electroweak masses, which were included in HI, we would expect including it to improve the agreement. This is seen, but it does not substantially relieve the tension.

To further explore the difference, in the right panel of Fig.~\ref{fig:UsVsHI} we take our results and shift them down by a constant: 0.175 for the high only result and 0.137 for the high and low combination. Such a constant offset could originate from a difference in $m_{\chi}$ independent terms between our result and HI. Unfortunately, however, a difference in such terms could originate from almost any of the graphs contributing to the result. Comparing our analytic expressions to the numerical results of HI we have been unable to pinpoint the exact location of the disagreement, although it is clear that we agree on the shape of the higher order corrections.

Despite the discrepancy between our result and that of HI, we emphasize that we have confidence in our result as stated. This confidence is derived from the non-trivial cross checks we have performed on our result. In detail, these are
\begin{itemize}
\item The cancellation in the $\mathcal{O}(\alpha_2)$ corrections of the $\mu_{m_{\chi}}$ dependence in our high-scale matching coefficients, stated in Eq.~\eqref{eq:WilsonCoeff}, with the high-scale dependence entering from the anomalous dimension, as stated in Eqs.~\eqref{eq:anomdim} and \eqref{eq:anomdimparts}. This cancellation is demonstrated in App.~\ref{app:consistency};
\item In the absence of running, the cancellation in the $\mathcal{O}(\alpha_2)$ corrections of the $\mu$ dependence between our high and low-scale results, where the latter is stated in Eqs.~\eqref{eq:lowbreakdown}, \eqref{eq:LowSoft}, \eqref{eq:LowColinear}, \eqref{eq:LowColinearConsts1}, and \eqref{eq:LowColinearConsts2}. This cancellation also depends on the SM SU(2)$_{\rm L}$ and U(1)$_Y$ $\beta$-functions and is shown in App.~\ref{app:consistencylow};
\item We have confirmed that the $\mu$ dependence in our low-scale result matches that in \cite{Bauer:2014ula}, when we reduce our calculation to make the same assumptions used in that work;
\item The form of the dominant $\mu$ independent terms in the low-scale matching are in agreement with the results of \cite{Chiu:2007yn,Chiu:2007dg,Chiu:2008vv,Chiu:2009mg,Chiu:2009ft}, as discussed in App.~\ref{app:lowscalematching}; and
\item We have confirmed that the framework used to calculate the low-scale matching for our non-relativistic initial state kinematics, reproduces the results of \cite{Chiu:2007yn,Chiu:2007dg,Chiu:2008vv,Chiu:2009mg,Chiu:2009ft} when we instead consider massless initial states as used in those references.
\end{itemize}

\section{Conclusion}
\label{sec:conclusion}

In this chapter we provide analytic expressions for the full one-loop corrections to heavy wino dark matter annihilation, allowing the systematic resummation of electroweak Sudakov logarithms to NLL$^{\prime}$ for the line cross section. We have compared our result to earlier numerical calculations of such effects, finding results similar in behaviour but quantitatively different. Our result is stated in a manner that can be straightforwardly extended to higher order, with our result already reducing the perturbative uncertainty from Sudakov effects on this process to $\mathcal{O}(1\%)$.

%% file: conc.tex
\chapter{Conclusions}

In this thesis I have shown just a taste of how insights from astrophysics and particle physics can help uncover the fingerprints of dark matter through indirect detection.
Yet in a very real sense there is work left to be done, as we remain in the dark as to the particle nature of dark matter.
Many of the techniques presented in this thesis can be further refined, and indeed in ongoing work I am already pursuing these directions.
Beyond this the data from recent and upcoming experiments like IceCube, HAWC, and CTA will provide new avenues for exploration.
Taken together I believe there is reason to be optimistic that a combination of improved techniques and experiments could in the near future uncover the first hints of dark matter shining back at us, and finally revealing its true nature.

%% file: gevexcess-app.tex
\chapter{A Gamma-Ray Signal in the Central Milky Way}

\section{Stability Under Modifications to the Analysis}\label{app:consistency}

\subsection{Changing the Region of Interest}\label{app:roi}

\begin{figure}
\begin{center}
\includegraphics[width=0.49\textwidth]{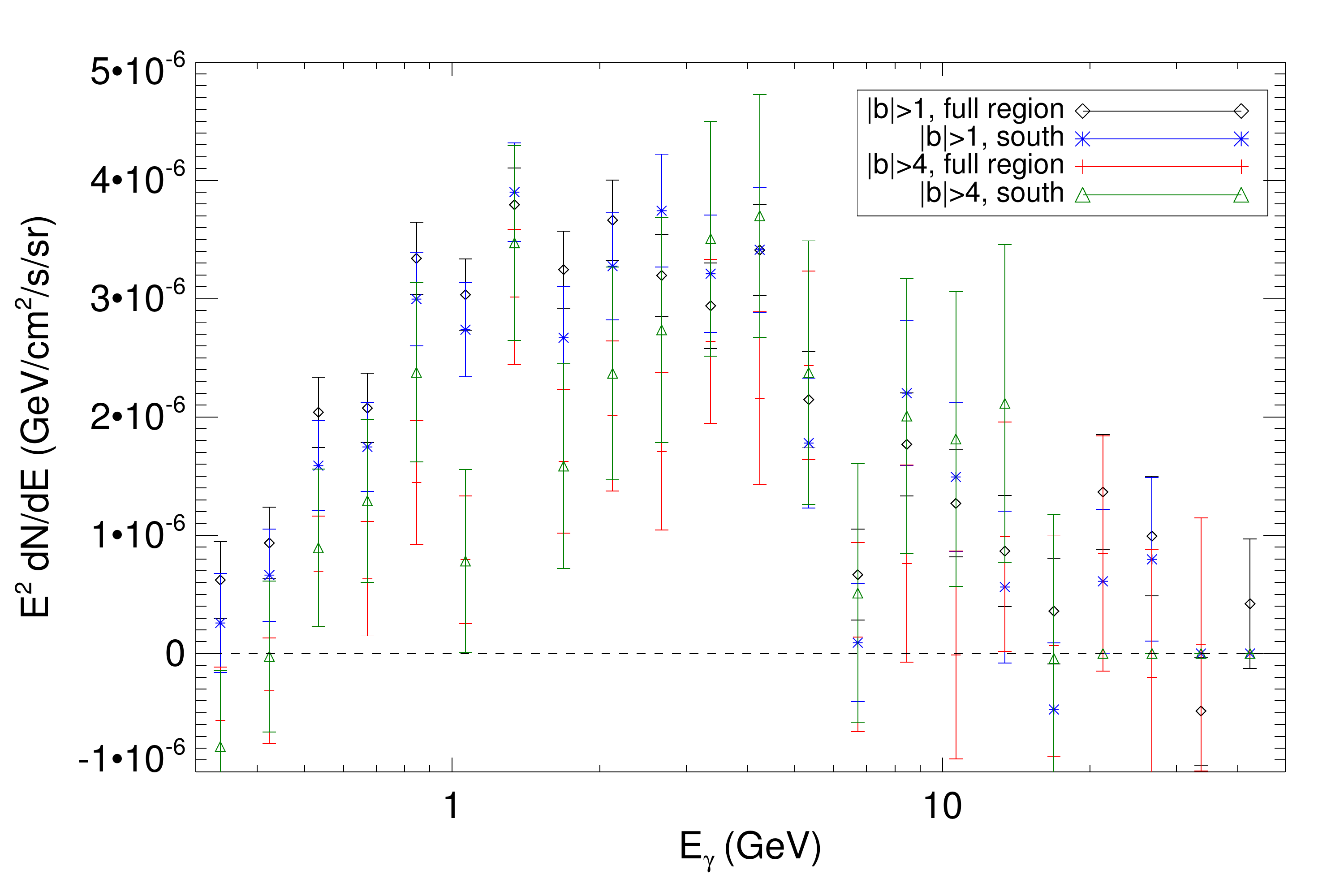}
\includegraphics[width=0.49\textwidth]{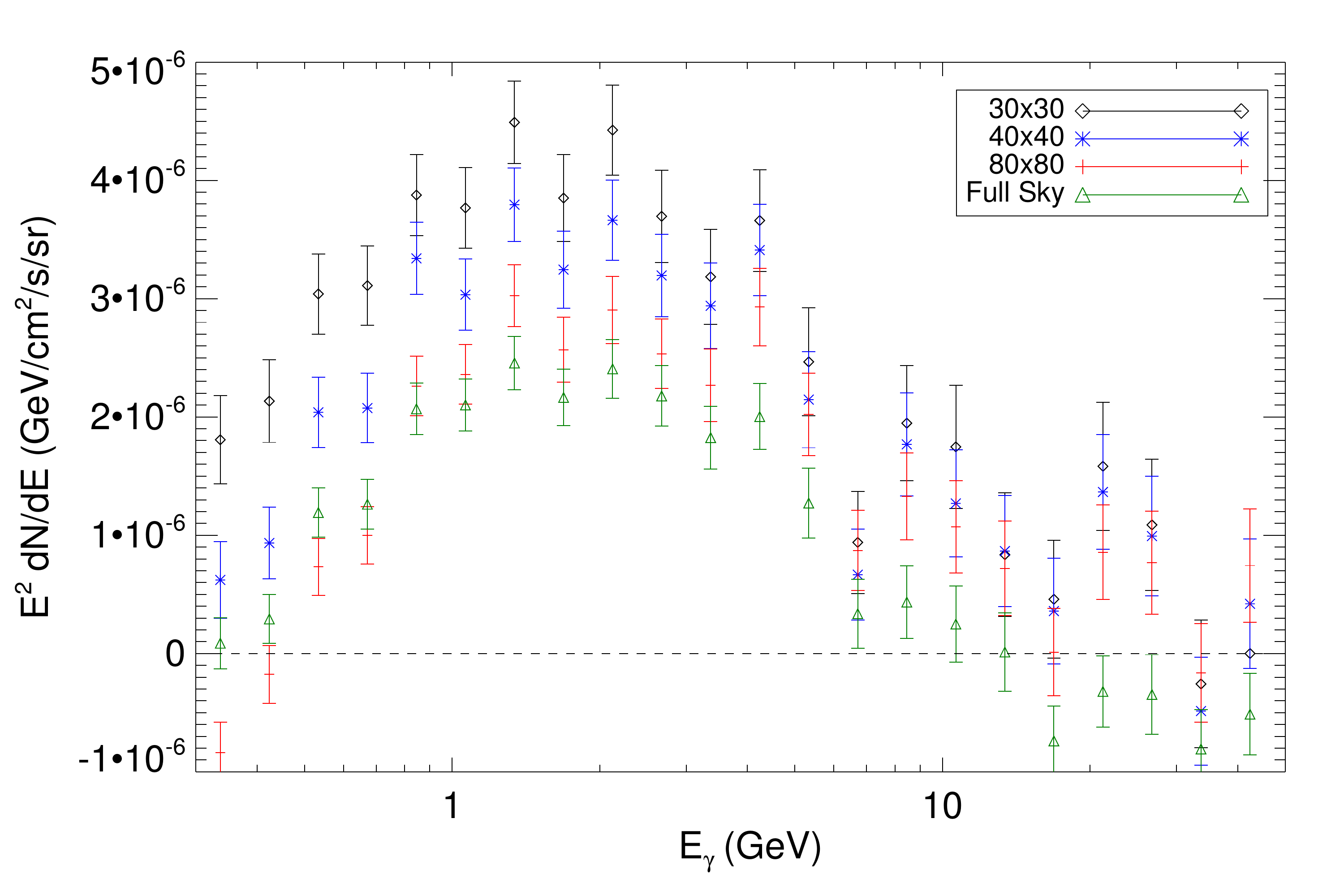}
\end{center}
\vspace{-0.5cm}
\caption{The spectrum of the dark matter template found in our Inner Galaxy analysis when performing the fit over different regions of the sky.
Using our standard ROI as a baseline, in the left panel we show variations of the Galactic plane mask and fits restricted to the southern sky, where backgrounds are typically somewhat lower, i.e. $|b| > 1^\circ$, $b < -1^\circ$, $|b| > 4^\circ$, and $b < -4^\circ$.
All fits employ a single template for the Bubbles, the \texttt{p6v11} \emph{Fermi} diffuse model, and a dark matter motivated signal template with an inner profile slope of $\gamma=1.2$.
In the right frame, we show the impact of varying the region over which the fit is performed.
All ROIs have $|b| > 1^\circ$; aside from this Galactic plane mask, the ROIs are $|b| < 15^\circ, |l| < 15^\circ$ (``$30\times30$''), $|b| < 20^\circ, |l| < 20^\circ$ (``$40\times40$'', standard ROI),  $|b| < 40^\circ, |l| < 40^\circ$ (``$80\times80$''), and the full sky.
}
\label{fig:igresults}
\end{figure}

In Fig.~\ref{fig:igresults}, we compare the spectrum correlated with the dark matter template (with $\gamma=1.2$) for variations of the ROI. In the left panel, we study different degrees of masking the Galactic Plane ($|b|>1^{\circ}$ and $|b|>4^{\circ}$), and the impact of performing the fit only in the southern sky (where the diffuse backgrounds are somewhat fainter) rather than in the full ROI. In the right panel, we show the impact of expanding or shrinking the ROI.

There is no evidence of asymmetry between the southern sky and the overall signal. Masking at $4^\circ$ gives rise to a similar spectral shape but a lower overall normalization than obtained with the $1^\circ$ mask, albeit with large error bars. As discussed in Sec.~\ref{morphology}, this may reflect a steepening of the spatial profile at larger distances from the Galactic Center, although the fainter emission at these larger radii is likely also more sensitive to mismodeling of the diffuse gamma-ray background.

\begin{figure}
\begin{center}
\includegraphics[width=0.45\textwidth]{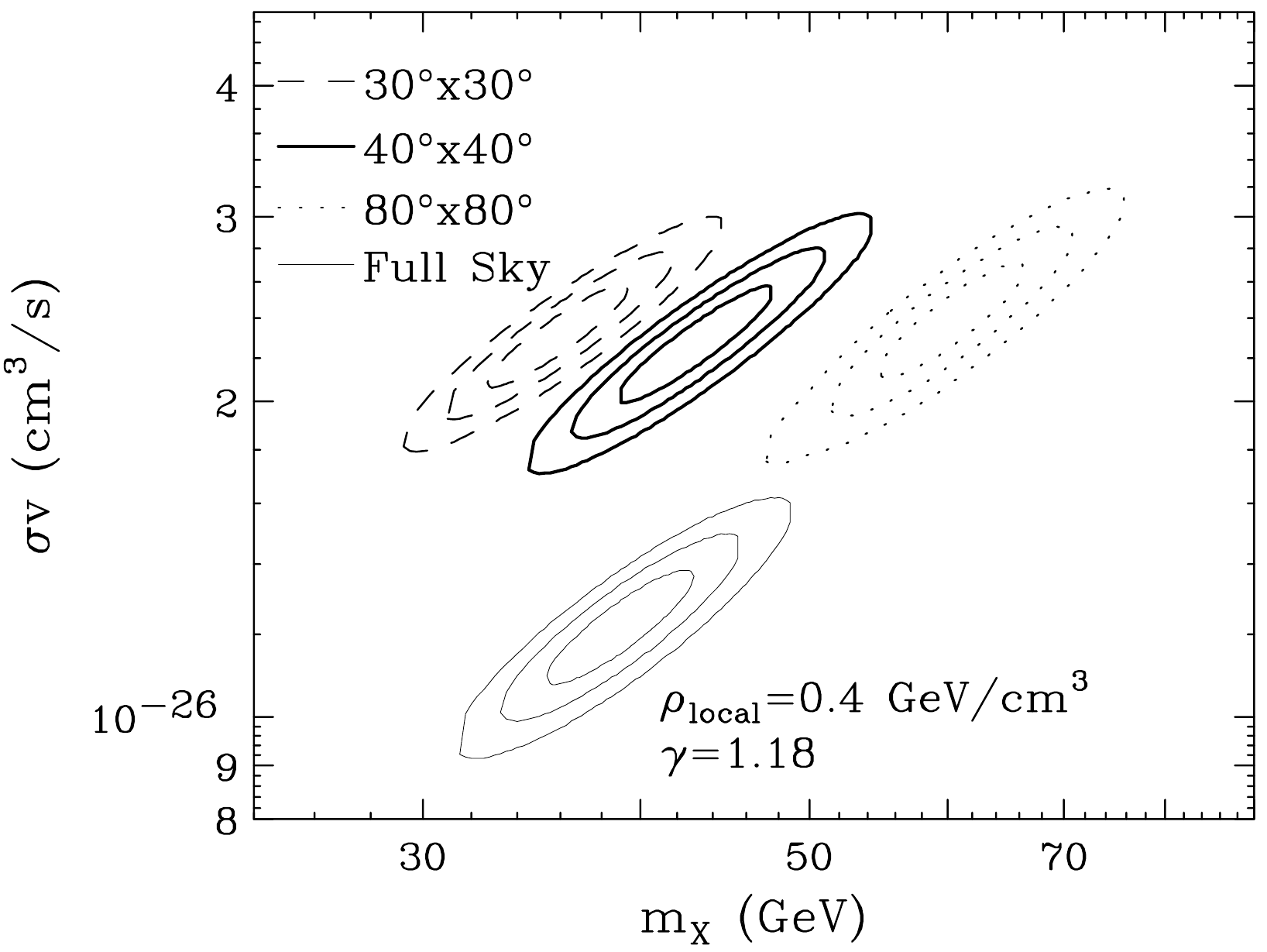}
\end{center}
\vspace{-0.5cm}
\caption{A comparison of the regions of the dark matter mass-annihilation cross section plane (for annihilations to $b\bar{b}$ and an inner slope of 1.18) best fit by the spectrum found in our default Inner Galaxy analysis (fit over the central $40^{\circ} \times 40^{\circ}$ region), to that found for fits to other ROIs.
See text for details.
}
\label{regioncompare2}
\end{figure}

Shrinking or expanding the size of the ROI also changes the height of the peak, while preserving a ``bump''-like spectrum that rises steeply at low energies and peaks around $\sim 2$ GeV. In general, larger ROIs give rise to lower normalizations for the signal. This effect appears to be driven by a higher normalization of the diffuse background model for larger ROIs; when the fit is confined to the inner Galaxy, the diffuse model prefers a lower coefficient than when fitted over the full sky, suggesting that the \texttt{p6v11} model has a tendency to overpredict the data in this region. This may also explain why larger ROIs prefer a somewhat steeper slope for the profile (higher $\gamma$); subtracting a larger background will lead to a greater relative decrease in the signal at large radii, where it is fainter. We also find evidence for substantial oversubtraction of the Galactic plane in larger ROIs, consistent with this hypothesis, as we will discuss in Appendix \ref{app:spherical}.

In Fig.~\ref{regioncompare2}, we show the regions of the dark matter mass-annihilation cross section plane favored by our fit, for several choices of the ROI (for annihilations to $b\bar{b}$ and an inner slope of 1.18). The degree of variation shown in this figure provides a measure of the systematic uncertainties involved in this determination; we see that the cross section is always very close to the thermal relic value, but the best-fit mass can shift substantially (from $\sim 35-60$ GeV). As previously, the contours are based on statistical errors only.

\subsection{Varying the Event Selection}\label{app:selections}

\begin{figure}
\begin{center}
\includegraphics[width=0.49\textwidth]{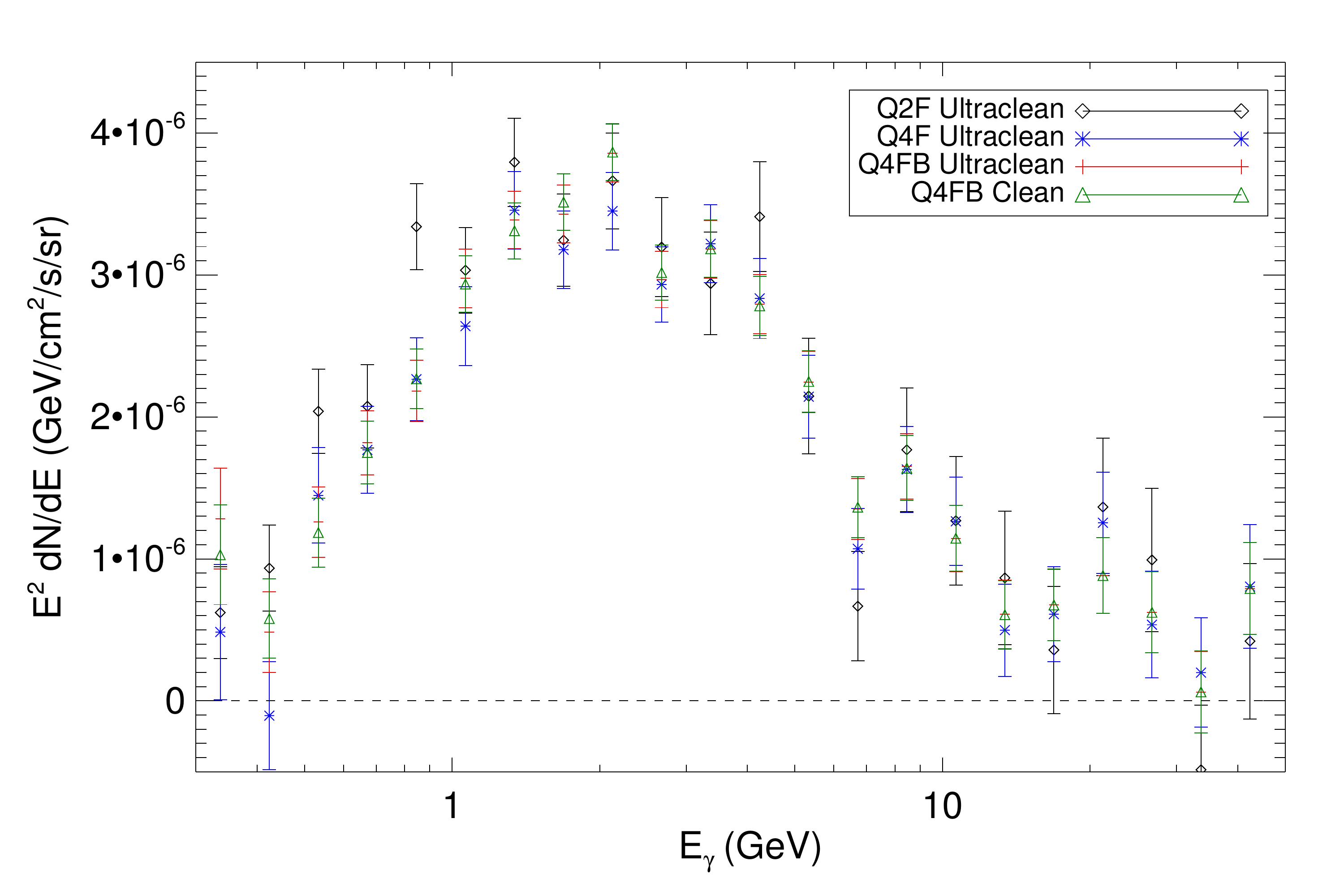}
\end{center}
\vspace{-0.5cm}
\caption{The spectrum of emission associated with the dark matter template, corresponding to a generalized NFW profile with an inner slope of $\gamma=1.2$, as performed for four different event selections. Black diamonds indicate the spectrum extracted from the usual fit. The blue stars, red crosses and green triangles represent the spectra extracted from repeating our analysis on datasets without a CTBCORE cut, for (respectively) ULTRACLEAN front-converting events, all ULTRACLEAN events, and all CLEAN events.}
\label{fig:eventselec}
\end{figure}

By default, we employ cuts on the CTBCORE parameter to improve angular resolution and minimize cross-leakage between the background and the signal. In an earlier version of this work, this resulted in a pronounced improvement in the consistency of the spectrum between different regions (in particular, in the hardness of the low-energy spectrum); however, this appears to have been due to a mismodeling of the background emission.\footnote{We suggested in that earlier work that the soft low-energy spectrum observed in the absence of a CTBCORE cut was likely due to contamination by mismodeled diffuse emission from the Galactic plane; our current results support that interpretation.} We now find that when the backgrounds are treated correctly, the spectrum has a consistent shape independent of the CTBCORE cut, and the significant changes in the tails of the point spread function (PSF) associated with a CTBCORE cut do not materially affect our results. Similarly, we find that our results are robust to the choice of ULTRACLEAN or CLEAN event selection, and to the inclusion or exclusion of back-converting events. We show the spectra extracted for several different event selections in Fig.~\ref{fig:eventselec}. Systematics associated with these choices are therefore unlikely to affect the observed excess.

\begin{figure}
\begin{center}
\includegraphics[width=0.4\textwidth]{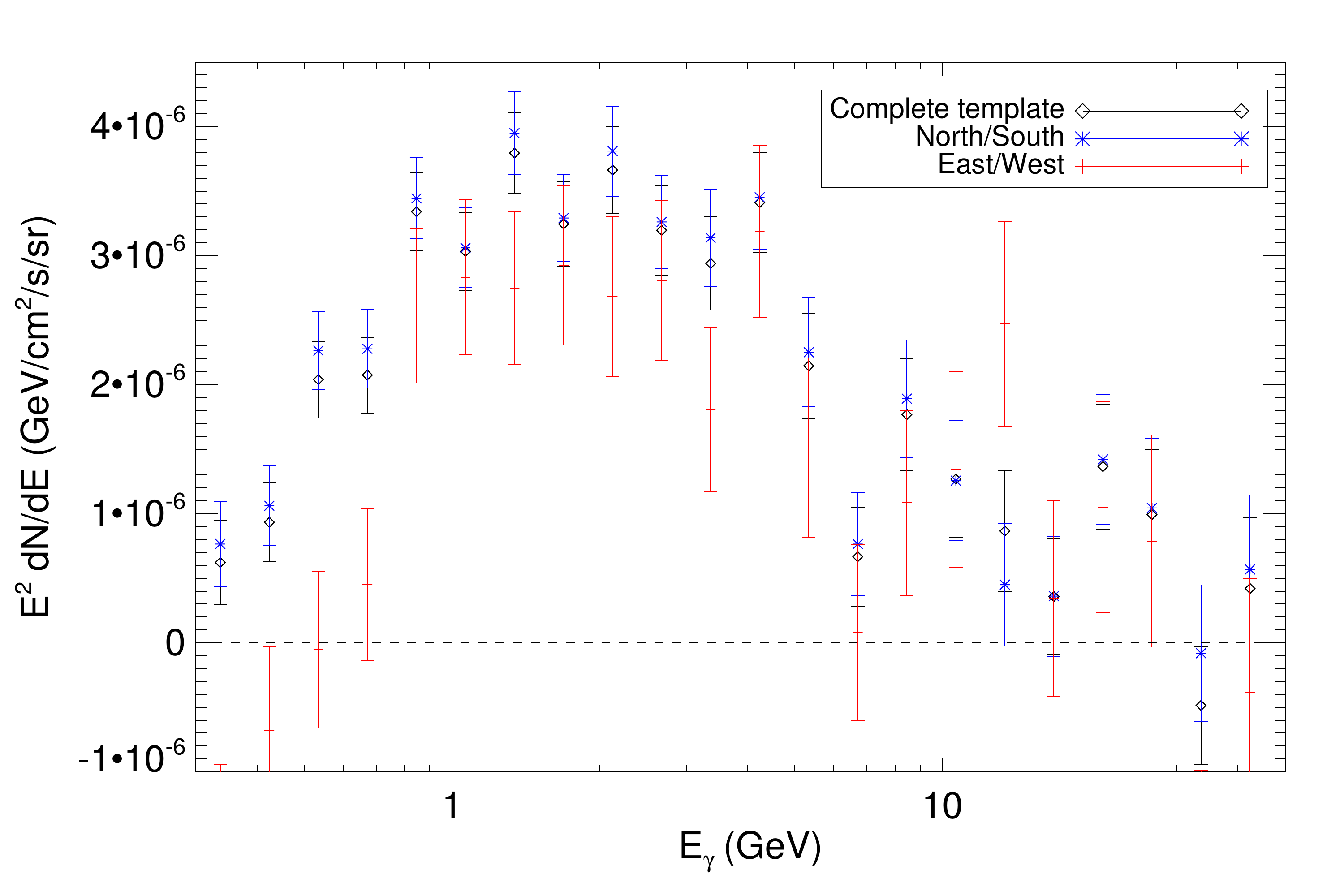} \\
\includegraphics[width=0.45\textwidth]{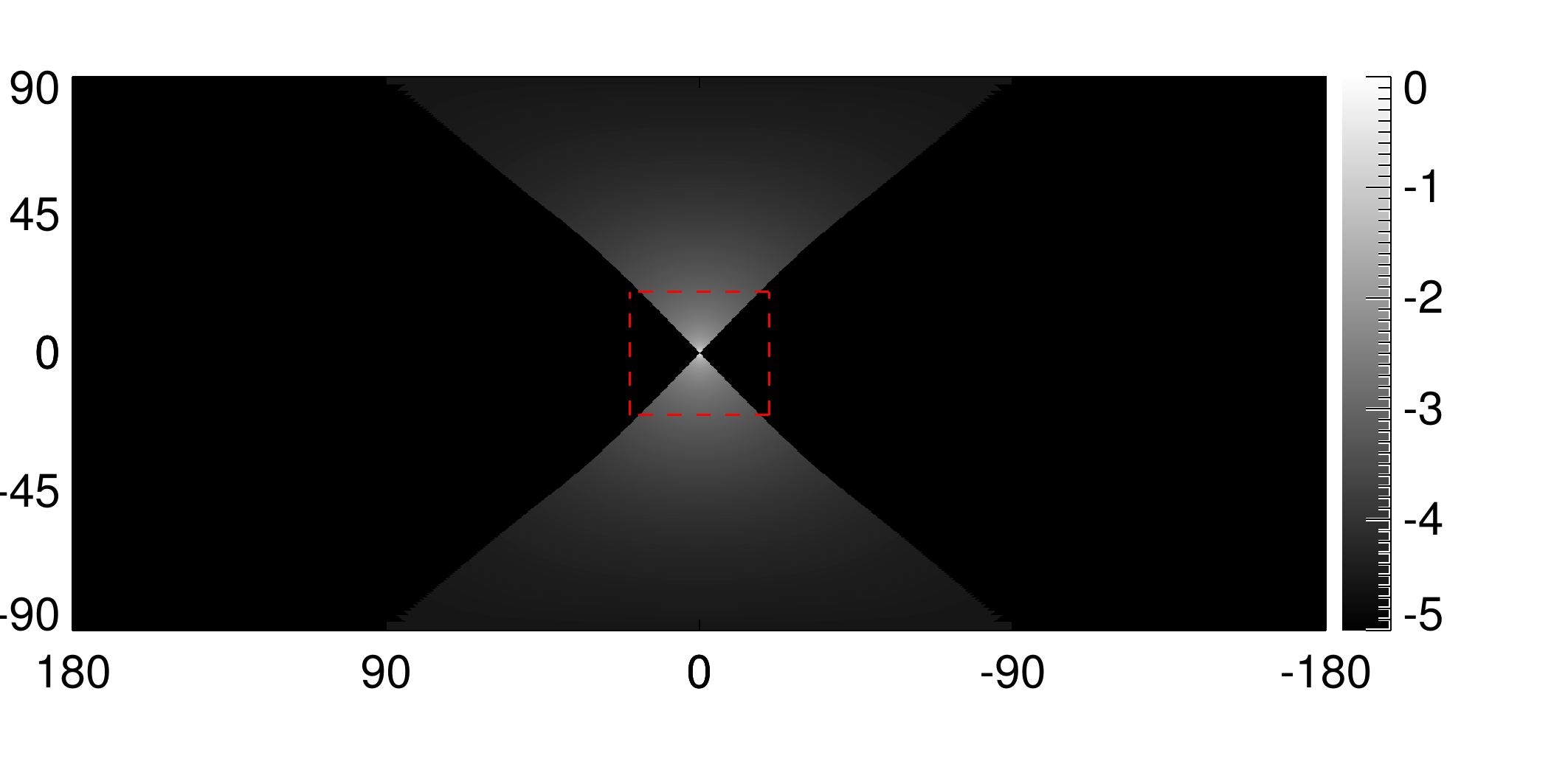}
\includegraphics[width=0.45\textwidth]{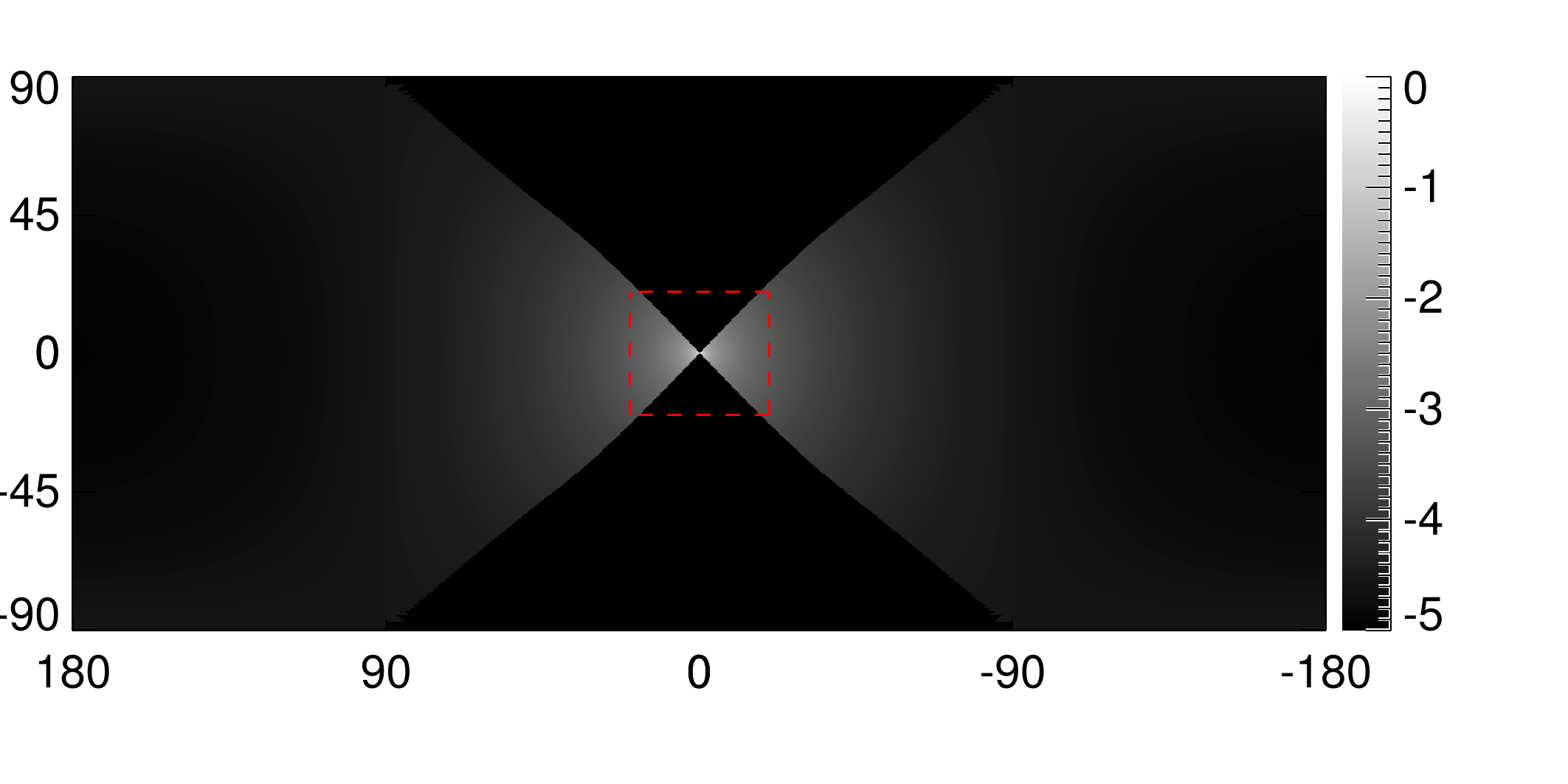}
\end{center}
\vspace{-1.1cm}
\caption{In the upper frame, we show the spectra of the emission associated with the dark matter template, corresponding to a generalized NFW profile with an inner slope of $\gamma=1.2$, as performed over three regions of the sky. Black diamonds indicate the spectrum extracted from the usual fit, whereas the blue stars and red crosses represent the spectra correlated with the parts of the template in which $|b| > |l|$ and $|b| < |l|$, respectively (when the two are allowed to vary independently). The corresponding spatial templates are shown in the lower row, in logarithmic (base 10) units, normalized to the brightest point in each map. Red dashed lines indicate the boundaries of our standard ROI.}
\label{fig:igspherical}
\vspace{-0.2cm}
\begin{center}
\includegraphics[width=0.4\textwidth]{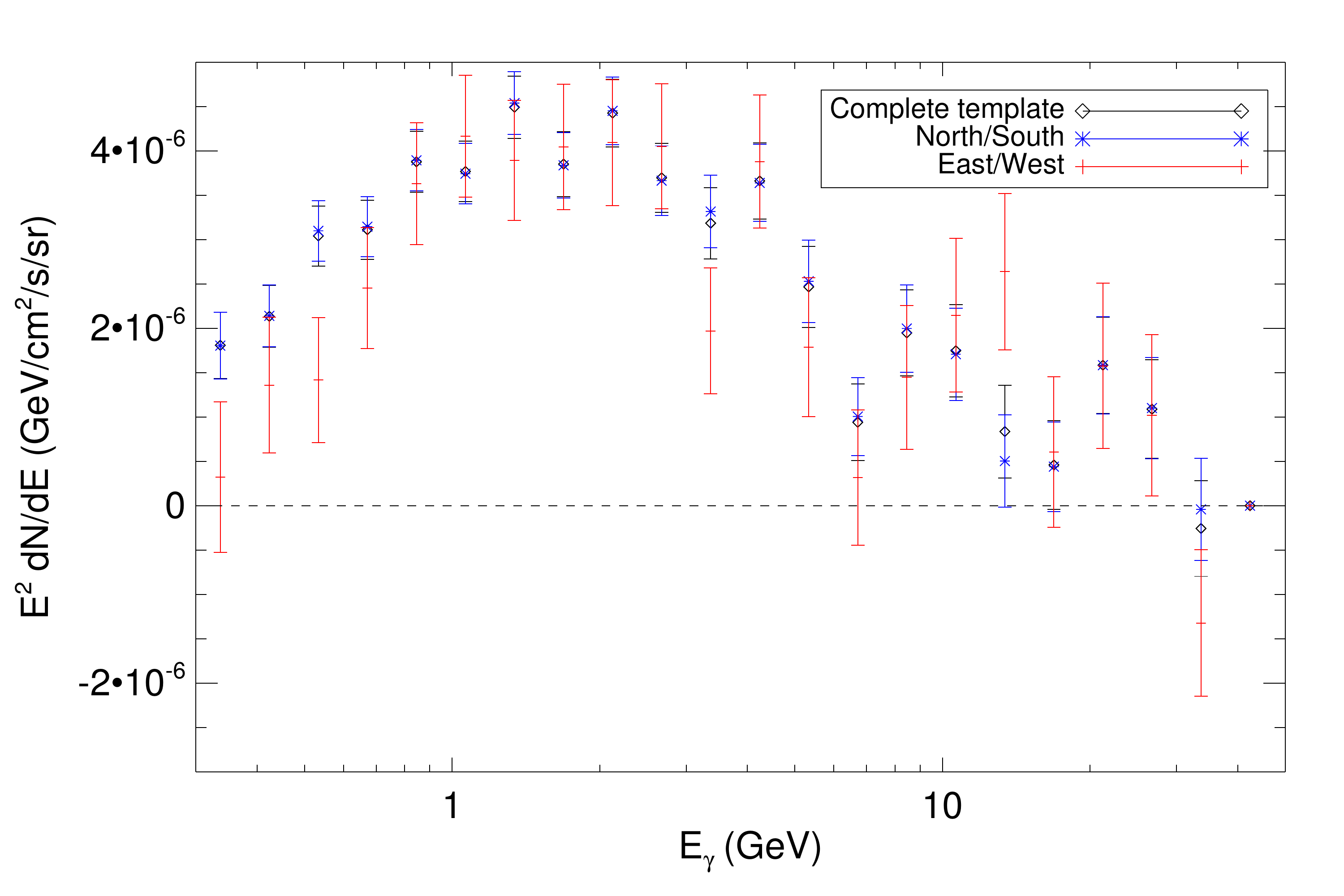}
\includegraphics[width=0.4\textwidth]{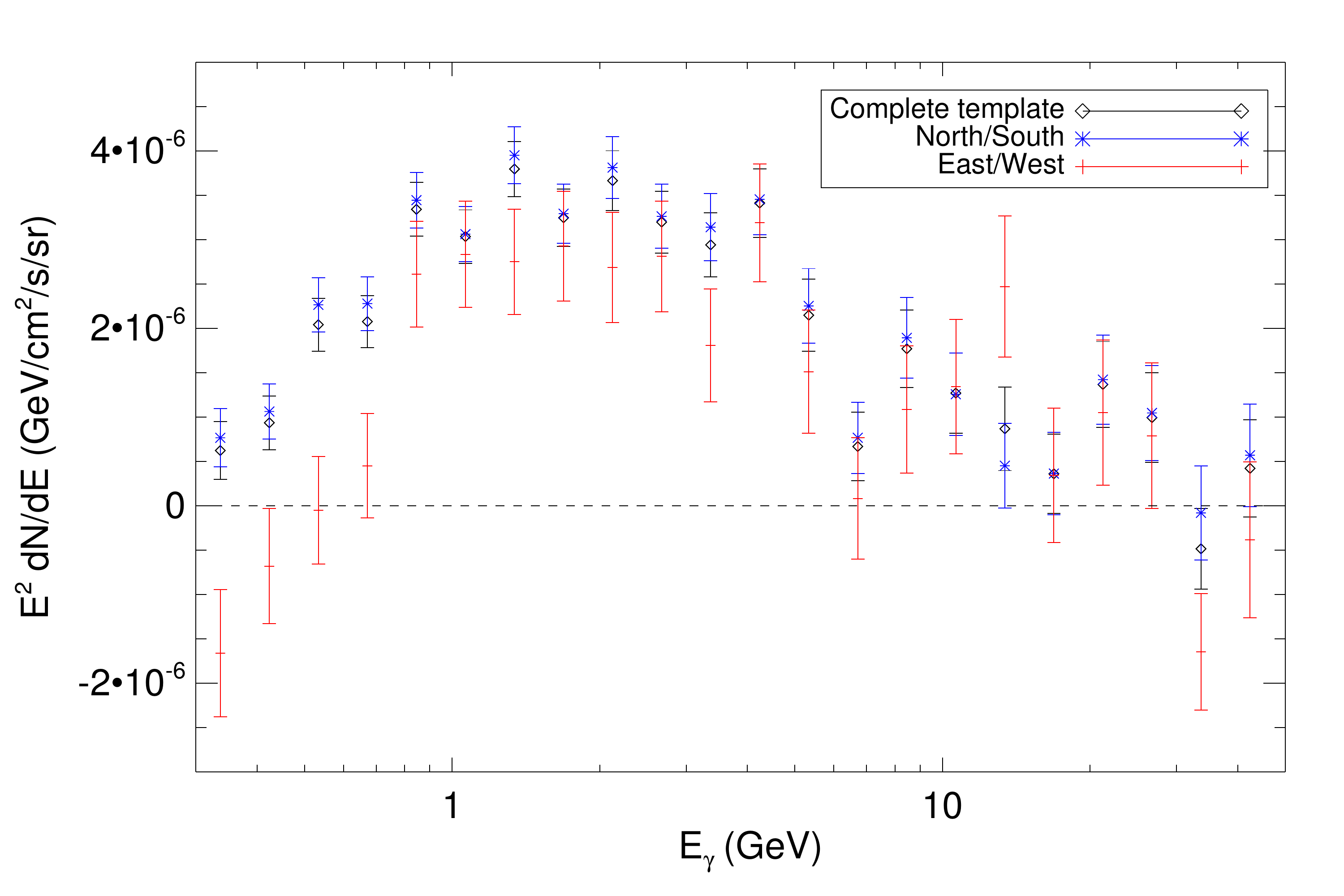} 
\\
\includegraphics[width=0.4\textwidth]{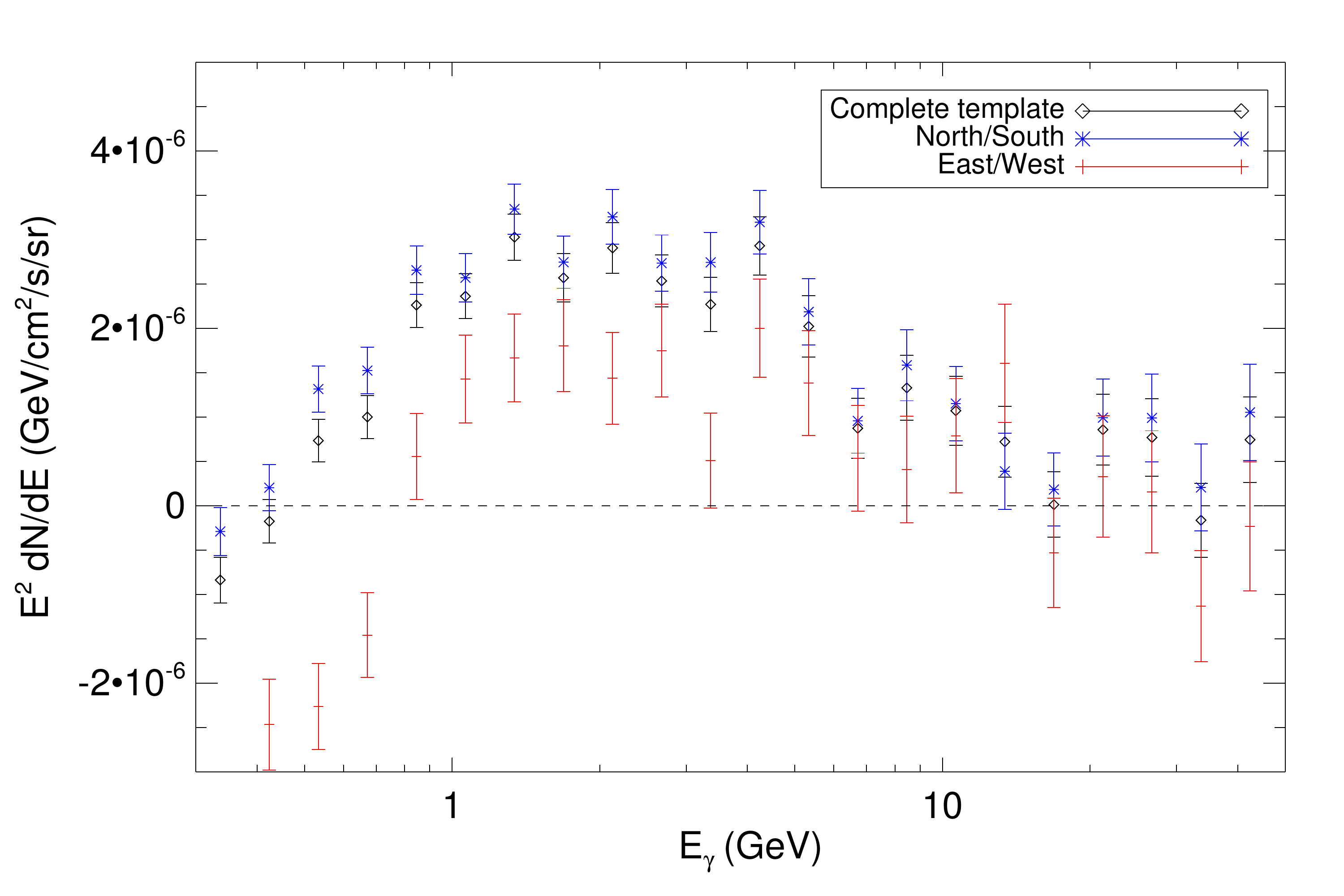}
\includegraphics[width=0.4\textwidth]{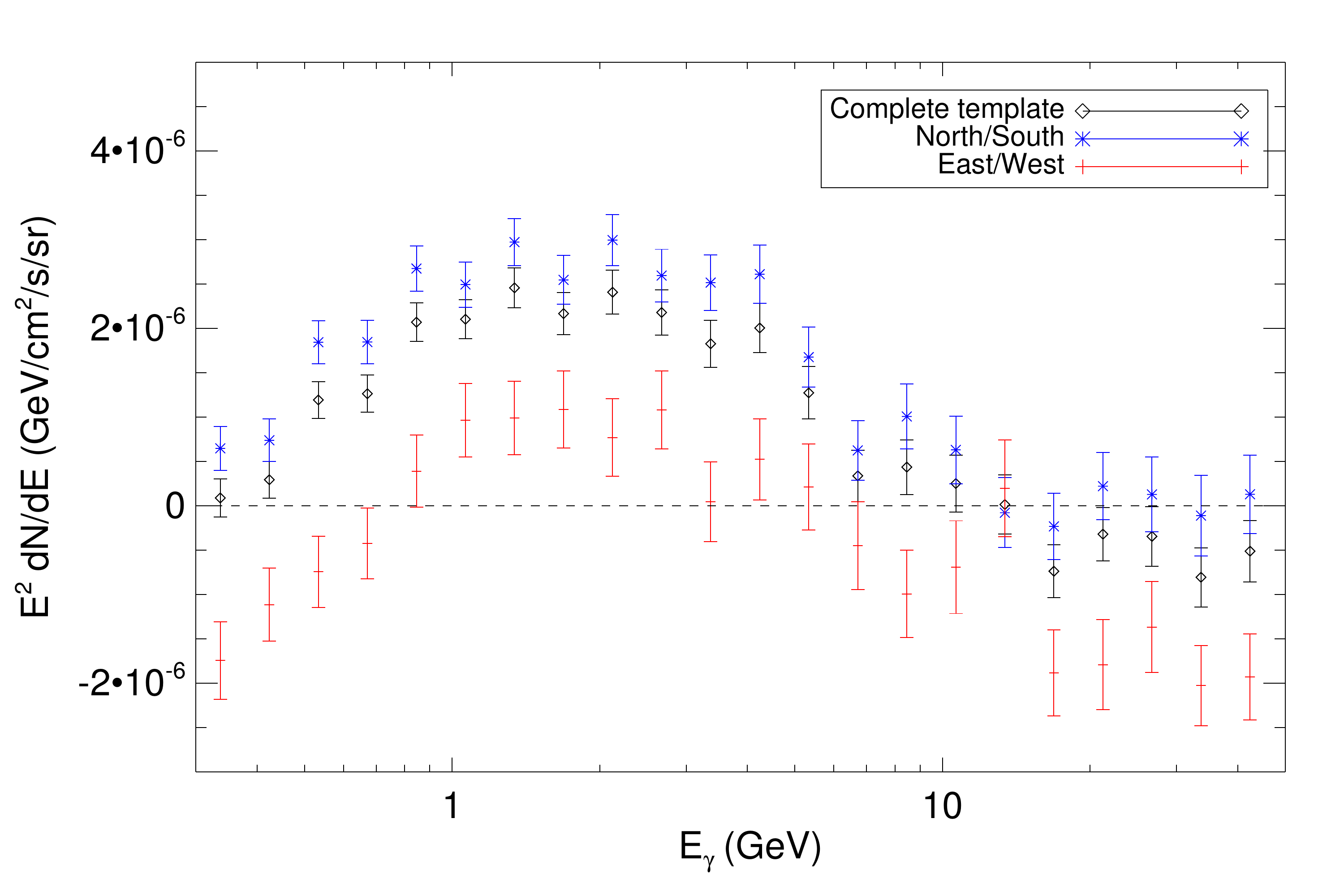}
\end{center}
\vspace{-1.0cm}
\caption{As the upper panel of Fig.~\ref{fig:igspherical}, but for ROIs given by (upper left frame) $|b|,|l| < 15^\circ$, (upper right frame) $|b|, |l| < 20^\circ$, (lower left frame) $|b|, |l| < 40^\circ$, (lower right frame) full sky. In all cases the Galactic plane is masked for $|b| < 1^\circ$. We attribute the lower emission in the East/West quadrants in the larger ROIs to oversubtraction by the Galactic diffuse model along the Galactic plane. The slope parameter for the dark matter template is set to $\gamma=1.2$ in all cases.}
\label{fig:nsewroi}
\end{figure}

\subsection{A Simplified Test of Elongation}\label{app:spherical}

Probing the morphology of the Inner Galaxy excess is complicated by the bright emission correlated with the Galactic Plane. In Ref.~\cite{Hooper:2013rwa}, it proved difficult to robustly determine whether any signal was present outside of the regions occupied by the \emph{Fermi} Bubbles, as the regions both close to the Galactic Center and outside of the Bubbles were dominated by the bright emission from the Galactic Plane. The improved analysis presented in this work mitigates this issue.

In addition to the detailed study of morphology described in Sec.~\ref{morphology}, we perform here a fit dividing the signal template into two independent templates, one with $|l| > |b|$ and the other with $|b| > |l|$. The former template favors the Galactic Plane, while the latter contains the \emph{Fermi} Bubbles. As previously, the fit also includes a single template for the Bubbles in addition to the \emph{Fermi} diffuse model and an isotropic offset. The extracted spectra of the signal templates are shown in Fig. \ref{fig:igspherical}. Both regions exhibit a clear spectral feature with broadly consistent shape and normalization, although the best-fit spectrum for the region with $|l| > |b|$ is generally slightly lower and has larger uncertainties. A lower normalization in these quadrants is expected, from the preference for a slight stretch perpendicular to the Galactic plane noted for the inner Galaxy in Sec.~\ref{morphology}.

As shown in Appendix \ref{app:roi}, the impact of the choice of ROI on the overall shape of the spectrum is modest. However, upon repeating this analysis in each of the ROIs, we find that the spectrum extracted from the quadrants lying along the Galactic plane ($|l| > |b|$) is much more sensitive to this choice. While a spectral ``bump'' peaked at $\sim 2$ GeV is always present, it appears to be superimposed on a negative offset which grows larger as the size of the ROI is increased. As discussed above, we believe this is due to oversubtraction along the plane by the Galactic diffuse model, which is most acute when the diffuse model normalization is determined by regions outside the inner Galaxy. We display this progression explicitly in Fig. \ref{fig:nsewroi}.

The relative heights of the spectra in the $|l| > |b|$ and $|b| > |l|$ regions are a reasonable proxy for sphericity of the signal; the former will be higher if the signal is elongated along the plane, and lower if the signal has perpendicular extension. Increased oversubtraction along the plane thus induces an apparent elongation of the signal perpendicular to the plane; we suspect this may be the origin of the apparent stretch perpendicular to the plane shown in Fig.~\ref{asymmetry}.

One might wonder whether this oversubtraction might give rise to apparent sphericity even if the true signal were elongated perpendicular to the plane. We argue that this is unlikely, as our results appear to converge to sphericity as the size of the ROI is reduced and the constraint on the normalization of the diffuse background is relaxed; the Galactic Center analysis, which includes the peak of the excess and the region where the signal-to-background ratio is largest, also prefers a spherical excess. 

We also performed the additional test of \emph{not} including any model for the point sources in the fit, allowing their flux to be absorbed by the NFW template. Since many point sources are clustered along the plane, over-subtracting them could bias the extracted morphology of the signal and hide an elongation along the plane. However, we found that even when no sources were subtracted, there was no ROI in which the spectrum extracted from the $|l| > |b|$ quadrants exceeded that for the $|b| > |l|$ quadrants.

\begin{figure}
\begin{center}
\includegraphics[width=0.49\textwidth]{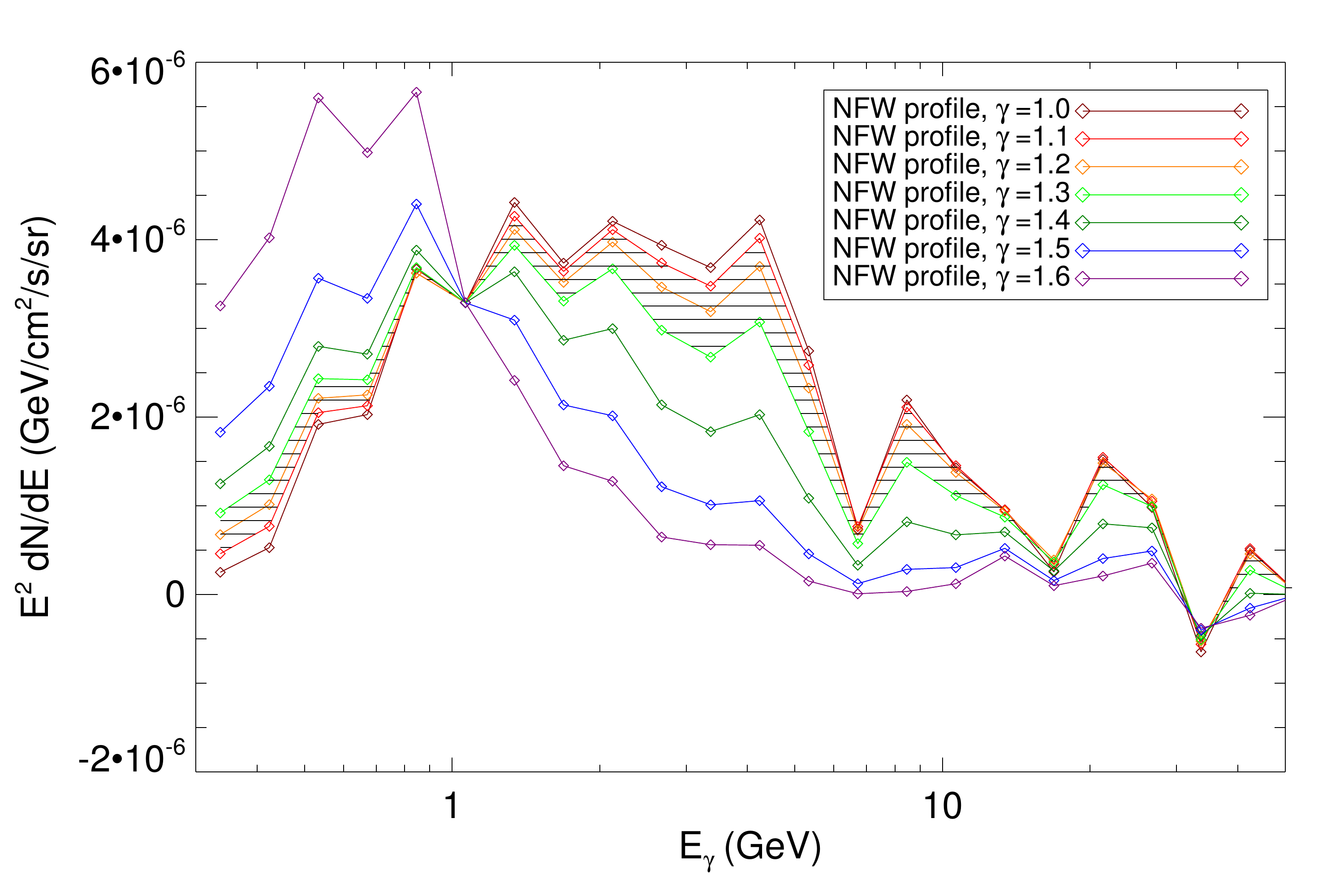}
\end{center}
\vspace{-0.5cm}
\caption{The central values of the spectra of the dark matter templates for different values of the dark matter profile's inner slope, $\gamma$. To better facilitate comparison, each curve has been rescaled to match the $\gamma=1.0$ curve at 1 GeV. All fits have been performed with the \texttt{p6v11} \emph{Fermi} diffuse model, a single flat template for the Bubbles, and the dark matter signal template. The region between the $\gamma=1.1$ and $\gamma=1.3$ lines, preferred by the fit, is cross-hatched. Error bars are not shown to avoid cluttering the plot. In this preferred range, the spectra are remarkably consistent. Allowing very high values of $\gamma$ seems to pick up a much softer spectrum, likely due to contamination by the Galactic plane, but these high values of $\gamma$ provide commensurately worse fits to the data.}
\label{fig:slopebias}
\end{figure}

\subsection{Sensitivity of the Spectral Shape to the Assumed Morphology}

In our main analyses, we have derived spectra for the component associated with the dark matter template assuming a dark matter density profile with a given inner slope, $\gamma$. One might ask, however, to what degree uncertainties in the morphology of the template might bias the spectral shape extracted from our analysis. In Fig.~\ref{fig:slopebias}, we plot the (central values of the) spectrum found for the dark matter template in our Inner Galaxy analysis, for a number of values of $\gamma$. The shapes of the spectra are quite consistent, within the range of slopes favored by our fits ($\gamma=1.1-1.3$); the extracted spectrum is not highly sensitive to the specified signal morphology. However, for $\gamma \gtrsim 1.5$ this statement is no longer true: higher values of $\gamma$ pick up a much softer spectrum, which we ascribe to contamination from the Galactic plane at the edge of the mask. Of course, such high values of $\gamma$ also have much worse TS.

\section{Modeling of Background Diffuse Emission in the Inner Galaxy}\label{app:diffuse}

\subsection{The \emph{Fermi} Bubbles}

\begin{figure}
\begin{center}
\includegraphics[width=0.49\textwidth]{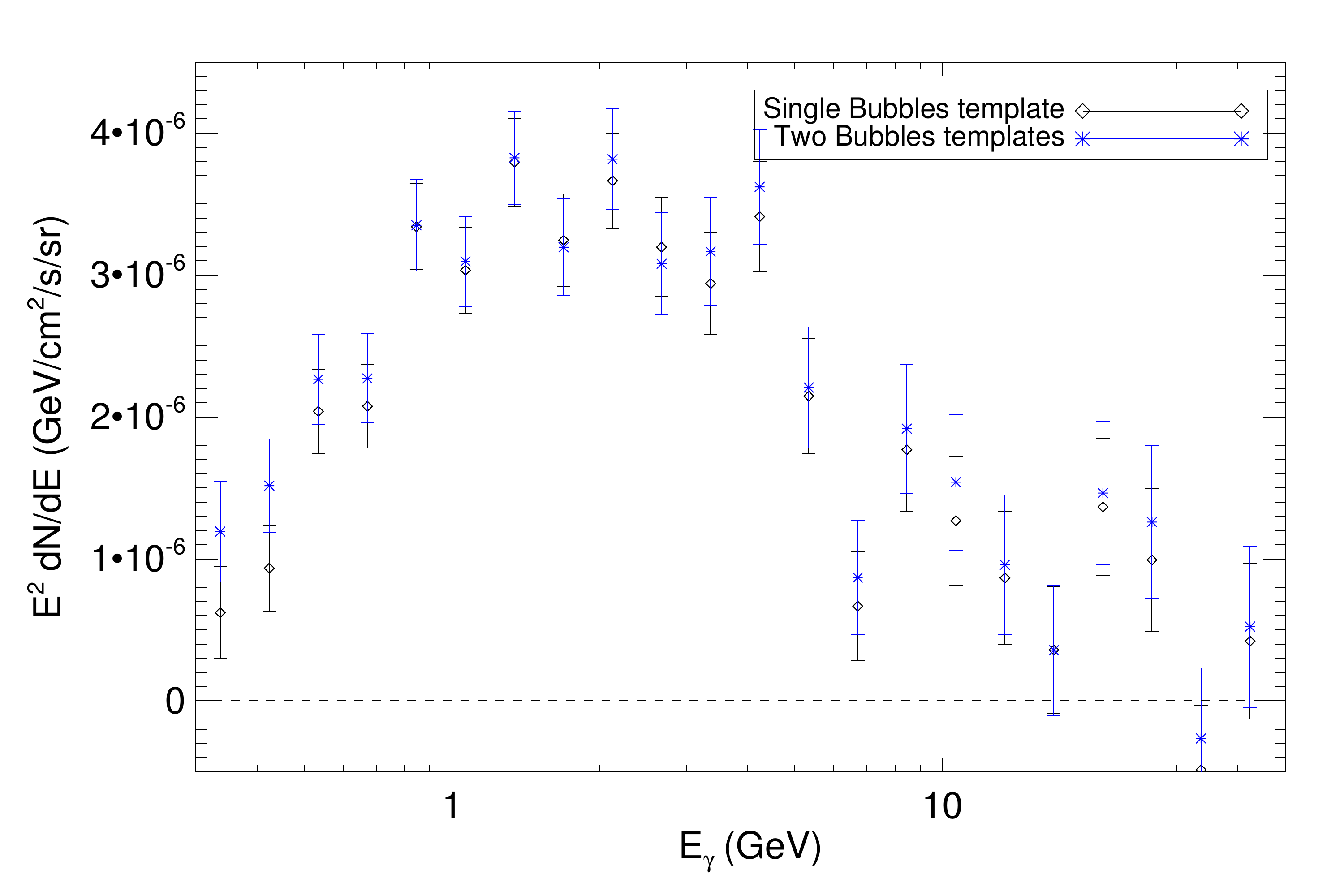}
\end{center}
\vspace{-0.5cm}
\caption{The spectrum of the emission correlated with a dark matter template, corresponding to a generalized NFW profile with an inner slope of $\gamma=1.2$, obtained by a fit containing either a single template for the \emph{Fermi} Bubbles (black diamonds) or two templates for 10-degree-wide slices in Galactic latitude through the Bubbles (blue stars). The latter allows the spectrum of the \emph{Fermi} Bubbles to vary somewhat with Galactic latitude (there are only two templates, in contrast to the five employed in \cite{Hooper:2013rwa}, because the ROI only extends to $\pm 20$ degrees). }
\label{fig:simplebubble}
\end{figure}

The fit described in Sec.~\ref{inner} is a simplified version of the analysis performed in Ref.~\cite{Hooper:2013rwa}, where the spectrum of the Bubbles was allowed to vary with latitude. From the results in Ref.~\cite{Hooper:2013rwa}, it appears that this freedom is not necessary -- the spectrum and normalization of the Bubbles varies only slightly with Galactic latitude.

It is straightforward to reintroduce this freedom, and we show in Fig.~\ref{fig:simplebubble} the spectrum correlated with the dark matter template if this is done. Above 0.5 GeV, the spectrum of the excess is not significantly altered by fixing the Bubbles to have a single spectrum; at low energies, reintroducing this freedom slightly raises the extracted spectrum for the dark matter template.

\begin{figure}
\begin{center}
\includegraphics[width=0.49\textwidth]{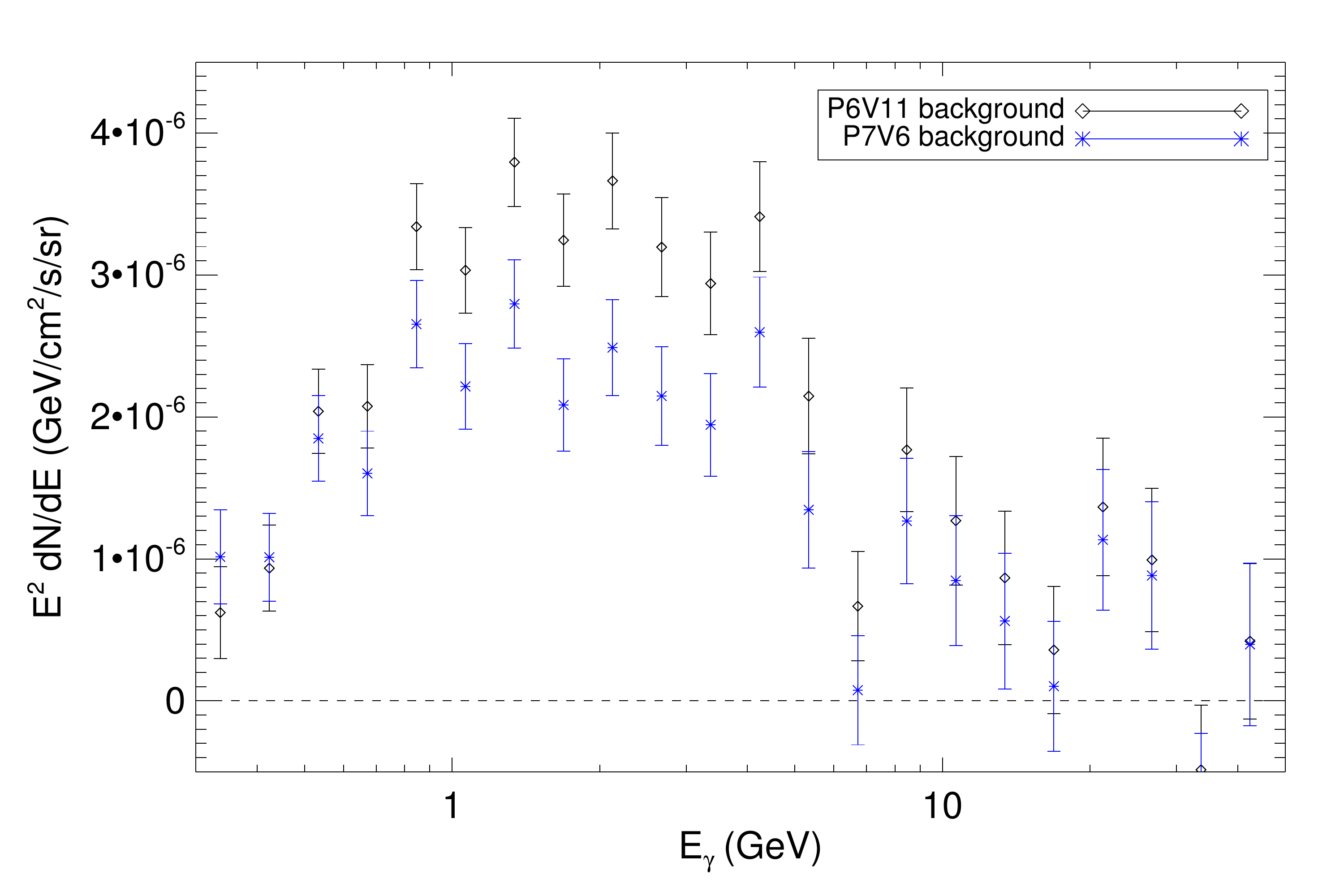}
\end{center}
\vspace{-0.5cm}
\caption{The spectra of the emission correlated with a dark matter template, corresponding to a generalized NFW profile with an inner slope of $\gamma=1.2$, with the background modeled by the \texttt{p6v11} diffuse model (black diamonds) or the \texttt{p7v6} diffuse model (blue stars). In both cases, the fit also contains an isotropic offset and a template for the \emph{Fermi} Bubbles.}
\label{fig:p7diffmodel}
\end{figure}

\subsection{The Choice of Diffuse Model}

Throughout our Inner Galaxy analysis, we employed the \texttt{p6v11} diffuse model released by the \textit{Fermi} Collaboration, rather than the more up-to-date \texttt{p7v6}  model. As noted earlier, this choice was made because the \texttt{p7v6} model contains artificial templates for the \emph{Fermi} Bubbles and other large-scale features (with fixed spectra), making it more difficult to interpret any residuals.

\begin{figure}
\begin{center}
\includegraphics[width=0.45\textwidth]{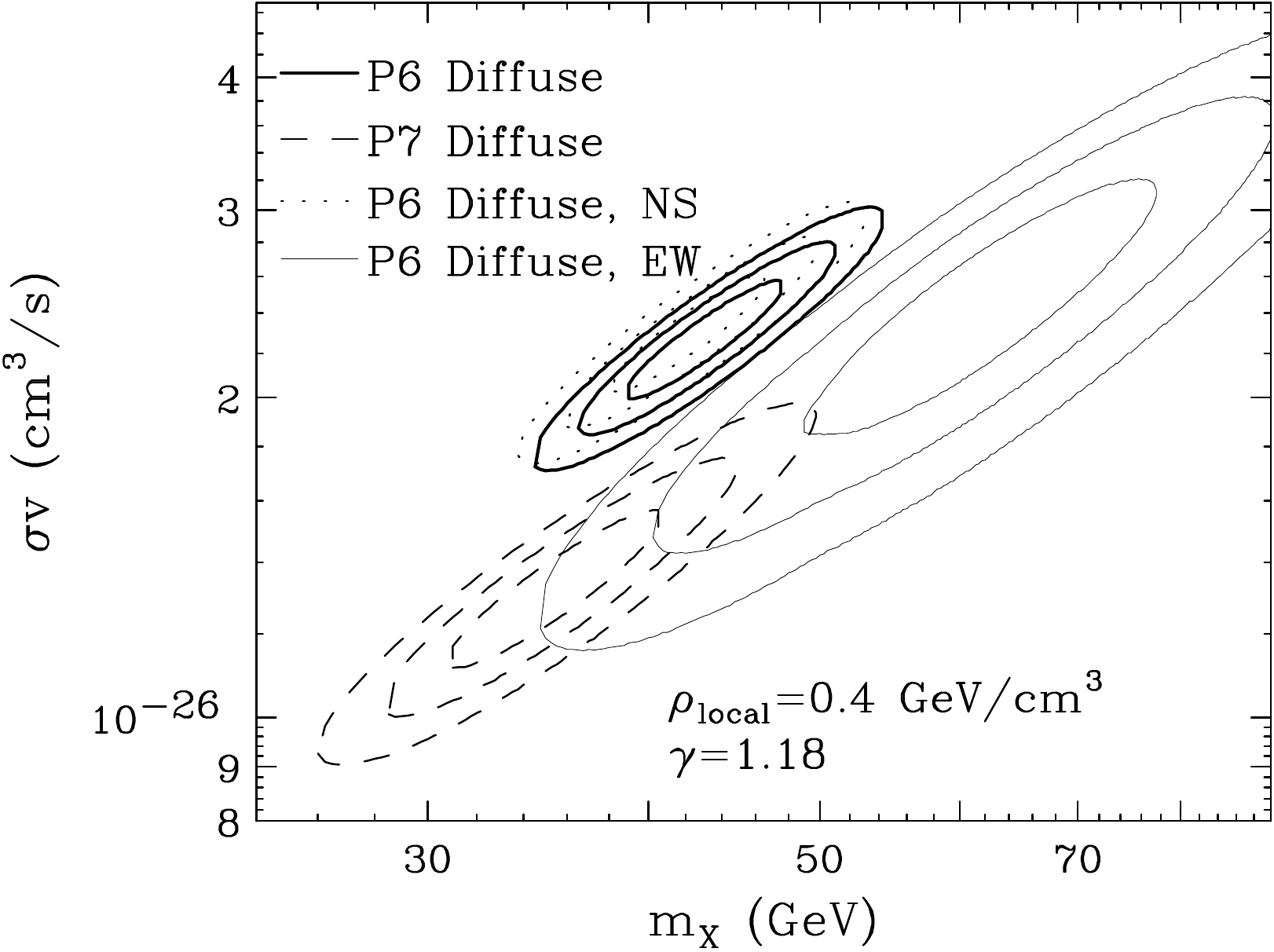}
\end{center}
\vspace{-0.5cm}
\caption{A comparison of the regions of the dark matter mass-annihilation cross section plane (for annihilations to $b\bar{b}$) best fit by the spectrum found in our default Inner Galaxy analysis (using the \texttt{p6v11} Galactic diffuse model, and fit over the standard ROI), to that found for the spectra shown in Figs.~\ref{fig:igspherical} and~\ref{fig:p7diffmodel}. See text for details.}
\label{regioncompare}
\end{figure}

Having shown that a single flat-luminosity template for the Bubbles is sufficient to capture their contribution without biasing the spectrum of the signal template, one might also employ the \texttt{p7v6} model in \emph{addition} to an independent template for the Bubbles, in order to absorb any deviations between the true spectrum of the Bubbles and their description in the model. Unfortunately, the template for the \emph{Fermi} Bubbles employed in constructing the \texttt{p7v6} diffuse model (which is not separately characterized from the overall Galactic diffuse emission) is different to the one employed in our analysis, especially in the regions close to the Galactic plane. Consequently, this approach gives rise to residuals correlated with the spatial differences between these templates. For this reason, we employ the \texttt{p6v11} diffuse model for our principal analysis. However, using the \texttt{p7v6} model does not quantitatively change our results, although the peak of the spectrum is somewhat lower (yielding results more comparable to that obtained from the full-sky ROI with the \texttt{p6v11} model). A direct comparison of these two results is shown in Fig.~\ref{fig:p7diffmodel}.

In Fig.~\ref{regioncompare}, we compare the regions of the dark matter mass-annihilation cross section plane (for annihilations to $b\bar{b}$) that are best fit by the spectrum found in our default Inner Galaxy analysis (using the \texttt{p6v11} Galactic diffuse model, and fit over the $|l| < 20^\circ$, $20^\circ > |b|>1^{\circ}$ ROI), to that found for the spectra shown in Figs.~\ref{fig:igspherical} and~\ref{fig:p7diffmodel}. The excess is still clearly present and consistent with a dark matter interpretation, and the qualitative results do not change with choice of diffuse model.

\begin{figure}
\begin{center}
\includegraphics[width=0.49\textwidth]{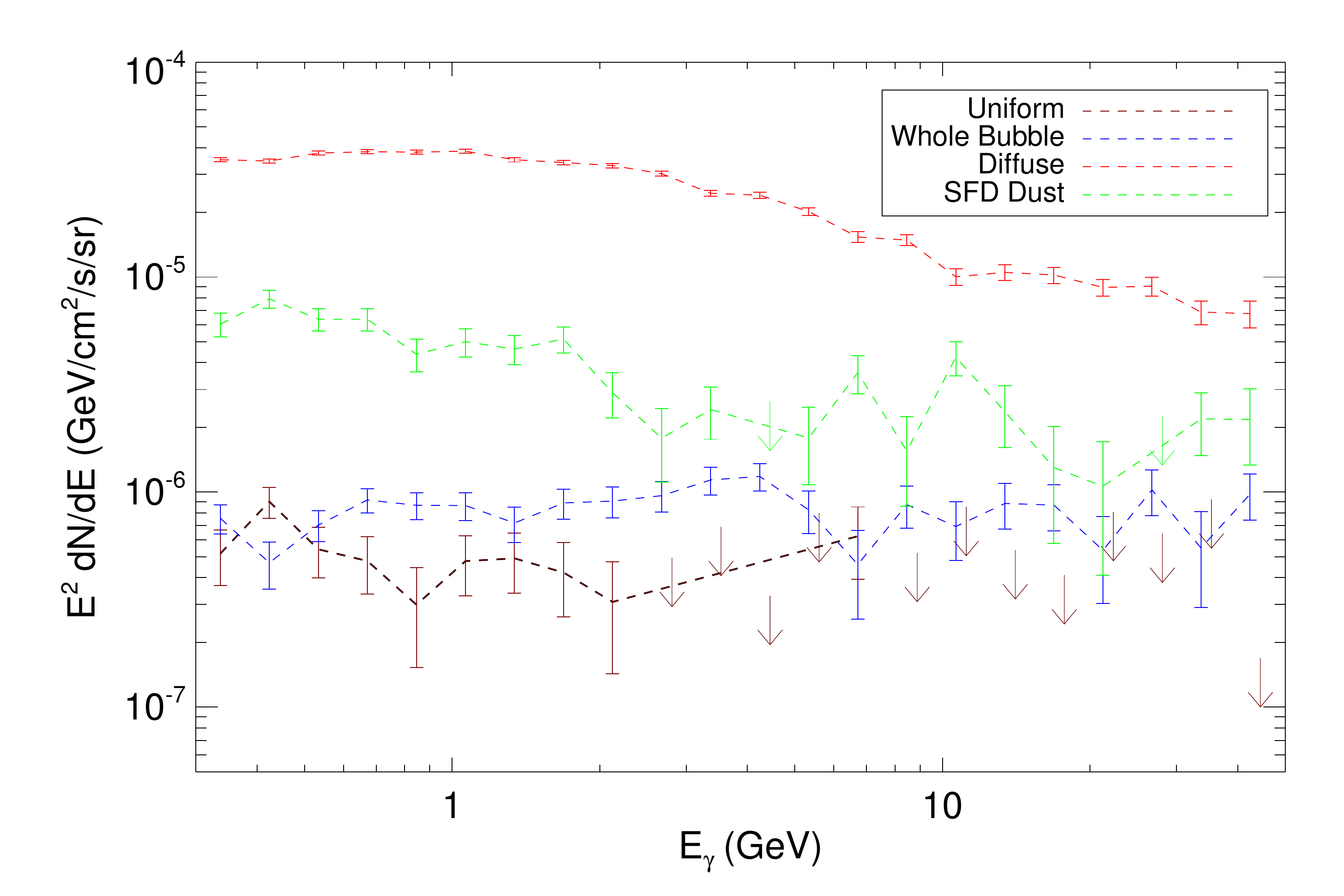}
\includegraphics[width=0.49\textwidth]{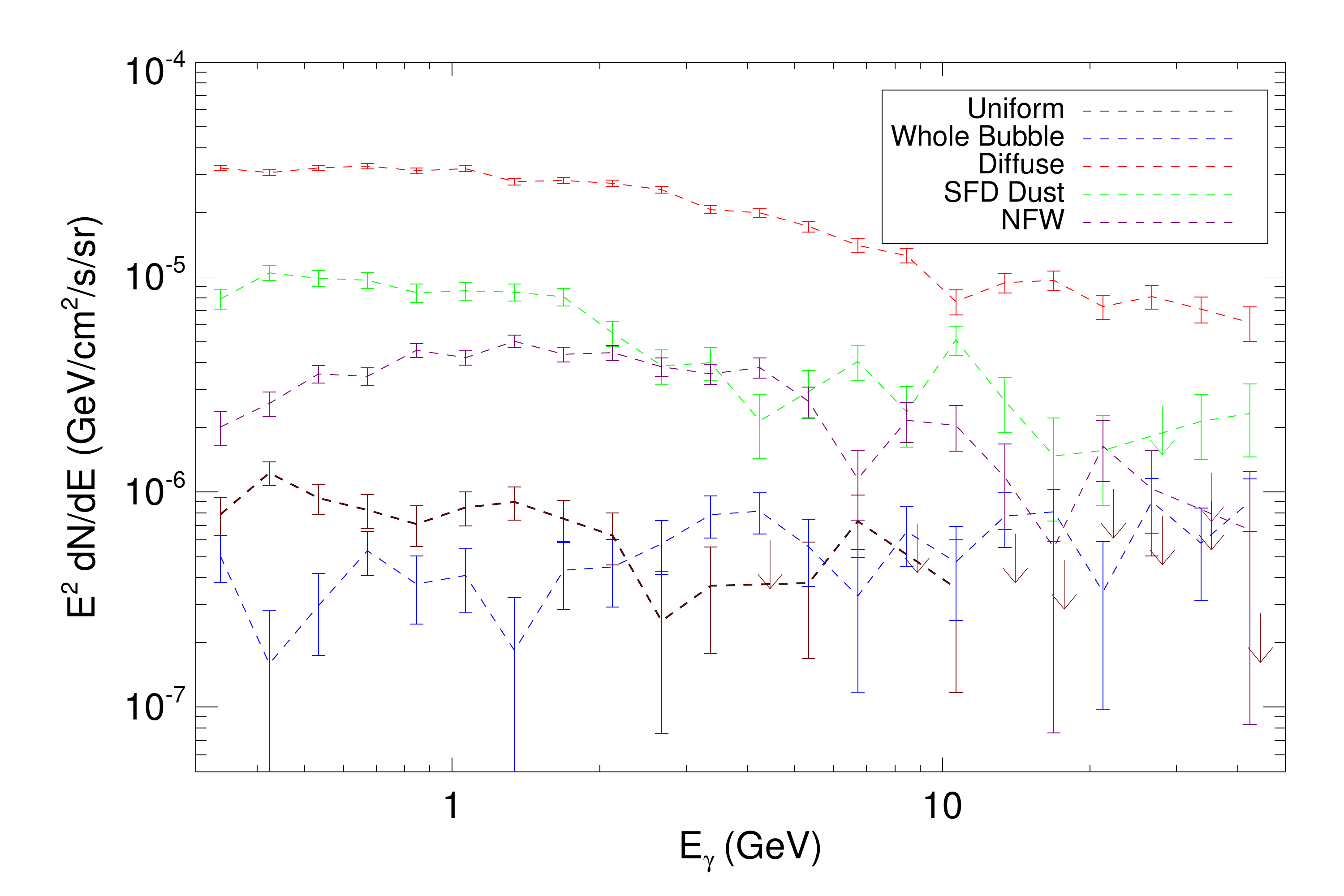}
\end{center}
\vspace{-0.5cm}
\caption{In the left frame, we show the spectra correlated with the various templates, from a fit with the usual backgrounds as well as the Schlegel-Finkbeiner-Davis (SFD) dust map, with the standard ROI. The right frame shows the results of the same fit, but also including a dark matter template with $\gamma=1.2$. The spectra for the dust map and diffuse model represent the average flux correlated with those templates outside the $|b| < 1^\circ$ mask and within $5^\circ$ of the Galactic Center.}
\label{fig:coeffsextraSFD}
\end{figure}

\subsection{Variation in the $\pi^0$ Contribution to the Galactic Diffuse Emission}\label{app:sfddust}

Although the spectrum of the observed excess does not appear to be consistent with gamma rays produced by interactions of proton cosmic rays with gas, one might wonder whether the \emph{difference} between the true spectrum and the model might give rise to an artificially peaked spectrum. While we fit the spectrum of emission correlated with the \emph{Fermi} diffuse model from the data, the model contains at least two principal emission components with quite different spectra (the gamma rays from the inverse Compton scattering of cosmic-ray electrons, and those from the interactions between cosmic-ray protons and gas), and their ratio is essentially fixed by our choice to use a single template for the diffuse Galactic emission (although we do allow for an arbitrary isotropic offset). Mismodeling of the cosmic-ray spectrum or density in the inner Galaxy could also give rise to residual differences between the data and model.

As a first step in exploring such issues, we consider relaxing the constraints on the background model by adding the Schlegel-Finkbeiner-Davis (SFD) map of interstellar dust \cite{Schlegel:1997yv} as an additional template. This dust map has previously been used effectively as a template for the gas-correlated gamma-ray emission~\cite{Dobler:2009xz,Su:2010qj}. By allowing its spectrum to vary independently of the \emph{Fermi} diffuse model, we hope to absorb systematic differences between the model and the data correlated with the gas. While the approximately spherical nature of the observed excess (see Sec.~\ref{morphology}) makes the dust template unlikely to absorb the majority of this signal, if the spectrum of the excess were to change drastically as a result of this new component, that could indicate a systematic uncertainty associated with the background modeling.

In Fig.~\ref{fig:coeffsextraSFD}, we show the results of a template fit using the three background templates described in Sec.~\ref{inner}, as well as the SFD dust map. The additional template improves the fit markedly, and absorbs significant emission across a broad range of energies. However, when the dark matter template is added, the fit still strongly prefers its presence and recovers the familiar spectrum with power peaked at $\sim$1-3 GeV.

\subsection{Modulating the $\pi^0$ Contribution}

The use of the SFD dust map as a tracer for the emission from cosmic-ray proton interactions with gas (producing neutral pions) is predicated on the assumption that the distribution of cosmic-ray protons is approximately spatially uniform. In this appendix, we demonstrate the robustness of the observed signal against the relaxation of this assumption. Specifically we consider an otherwise unmotivated modulation of the gas-correlated emission that seems most likely to be capable of mimicking the signal: the proton density at energies of a few tens of GeV increasing toward the Galactic Center in such a way as to produce the spatially concentrated spectral feature found in the data. Since the gas density is strongly correlated with the Galactic Disk while the signal appears to be quite spherically symmetric (see Sec.~\ref{morphology}), this would require the modulation from varying the cosmic-ray proton density to be aligned perpendicular to the Galactic Plane.

\begin{figure}
\begin{center}
\includegraphics[width=2.8in]{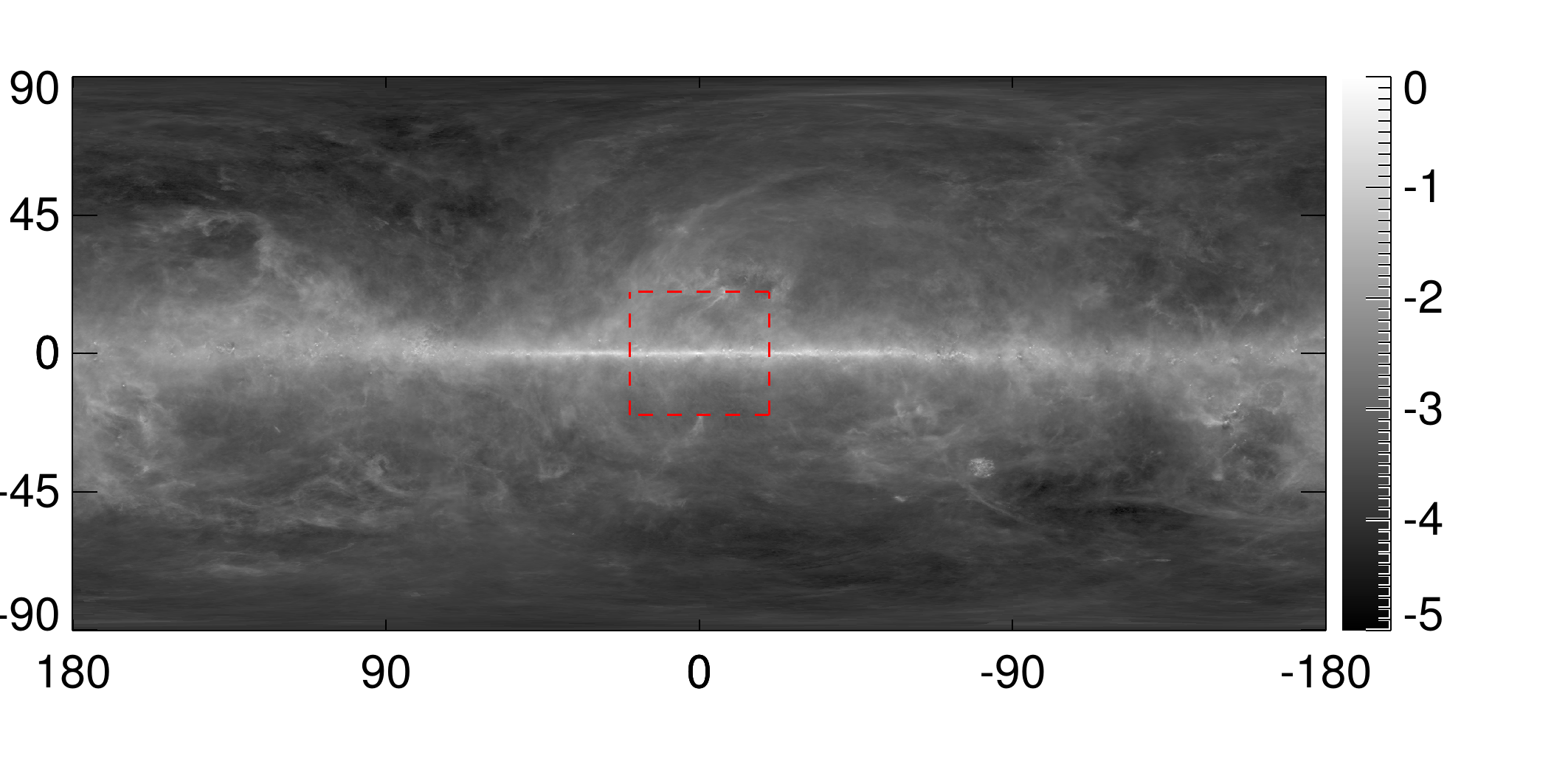}
\includegraphics[width=2.8in] {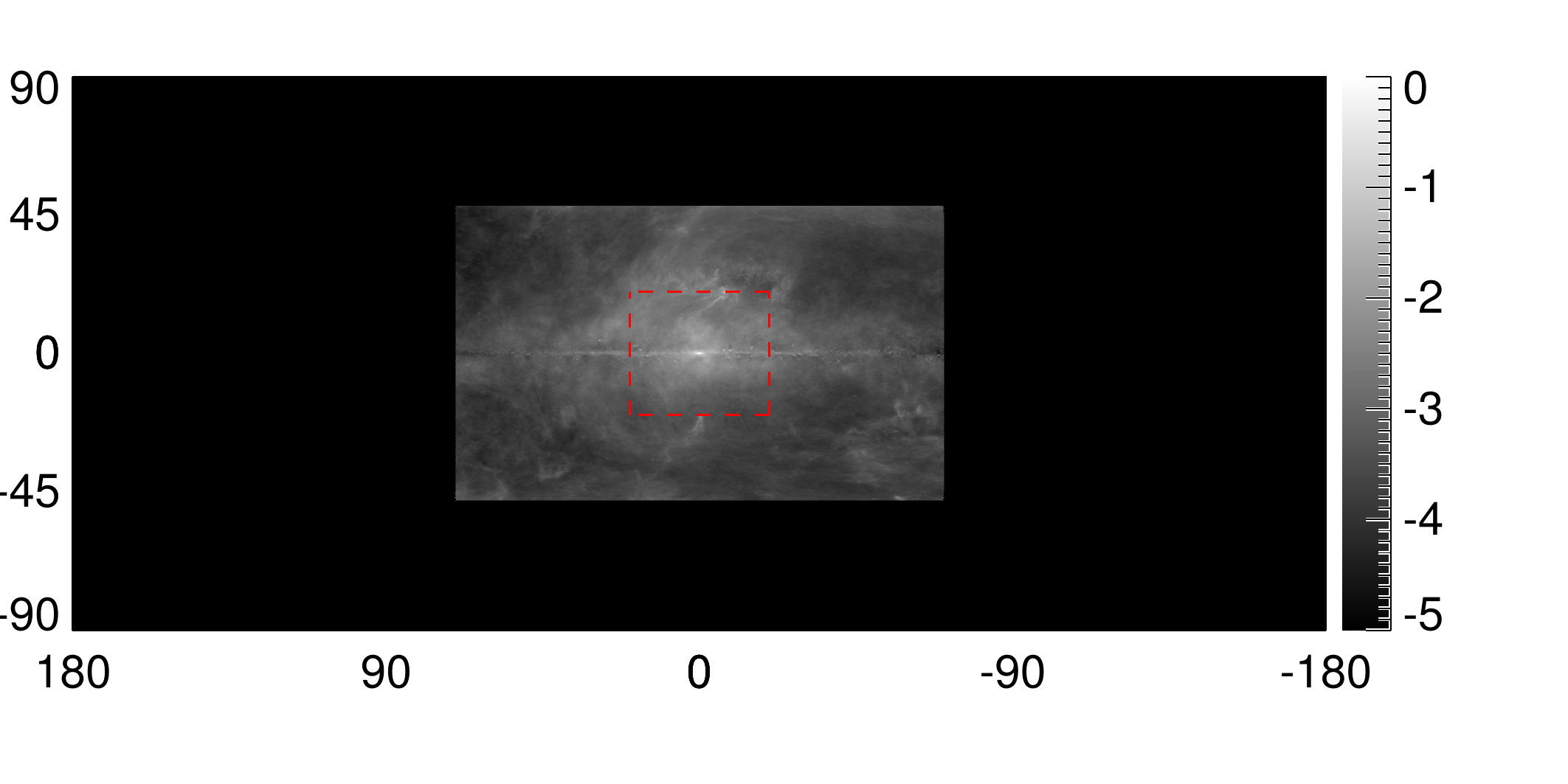}
\end{center}
\vspace{-0.5cm}
\caption{Left frame: The Schlegel-Finkbeiner-Davis dust map, used as a tracer for emission from proton-gas interactions. Right frame: An example of a dust-modulation template, created by multiplying the dust map by $f(r)/g(r)$, in the case where $f(r)$ is a projected squared NFW with $\gamma=0.8$. Red dashed lines indicate the boundaries of our standard ROI. All maps are given in logarithmic (base 10) units, normalized to the brightest point in each map. The modulated-dust template is artificially set to zero for $|b| > 45^{\circ}$ and $|l| > 70^{\circ}$, to avoid errors due to the denominator factor becoming small; as these bounds lie outside our ROI, they will not affect our results. See text for details.}
\label{fig:healcartdeltad}
\end{figure}

To this end we created additional templates of the form:
\begin{equation} {\rm Modulation} = ({\rm SFD~dust~map})\times \frac{f(r)}{g(r)}, \end{equation}
where $f(r)$ is a projected squared NFW template and $g(r)$ is a simple data-driven characterization of how the SFD dust map falls off with increasing galactic latitude and longitude. In this sense we have factored out how the dust map itself increases towards the Galactic Center and replaced this with a slope that matches a generalized NFW profile. Different modulations were generated by varying $f(r)$, which was done by choosing various values of the NFW inner slope, $\gamma$, from 0.5 to 2.0 in 0.1 increments. In order to determine $g(r)$, the dust map was binned in longitude and latitude and a rough functional form was chosen for each. For longitude, we analyzed the region with $|l|<70^{\circ}$, and fit the profile of the dust map with a Gaussian. For latitude, we considered $|b|<45^{\circ}$ and determined a best-fit using a combination of an exponential and linear function. These two best-fits were then multiplied to give $g(r)$. Each of the new templates were normalized such that the average value of all pixels with an angle between 4.9 and 5.1 degrees from the Galactic Center was set to unity. This was done in order to aid a comparison with the projected squared NFW template, which is normalized similarly. An example of the final template is shown in Fig.~\ref{fig:healcartdeltad}, which was created using an $f(r)$ with $\gamma=0.8$. 

Note that there is no particular physics motivation behind this choice of modulating function; we are attempting to create a dust-correlated map that mimics the observed signal as closely as possible, even if it is not physically reasonable. Since the dust map is integrated along the line of sight, the modulation we have performed is also not precisely equivalent to the effect of changing the cosmic ray density in the inner Galaxy -- this analysis serves as a test of correlation with the gas, but the modulation should \emph{not} be interpreted as a cosmic-ray density map.

\begin{figure}
\begin{center}
\includegraphics[width=2.8in]{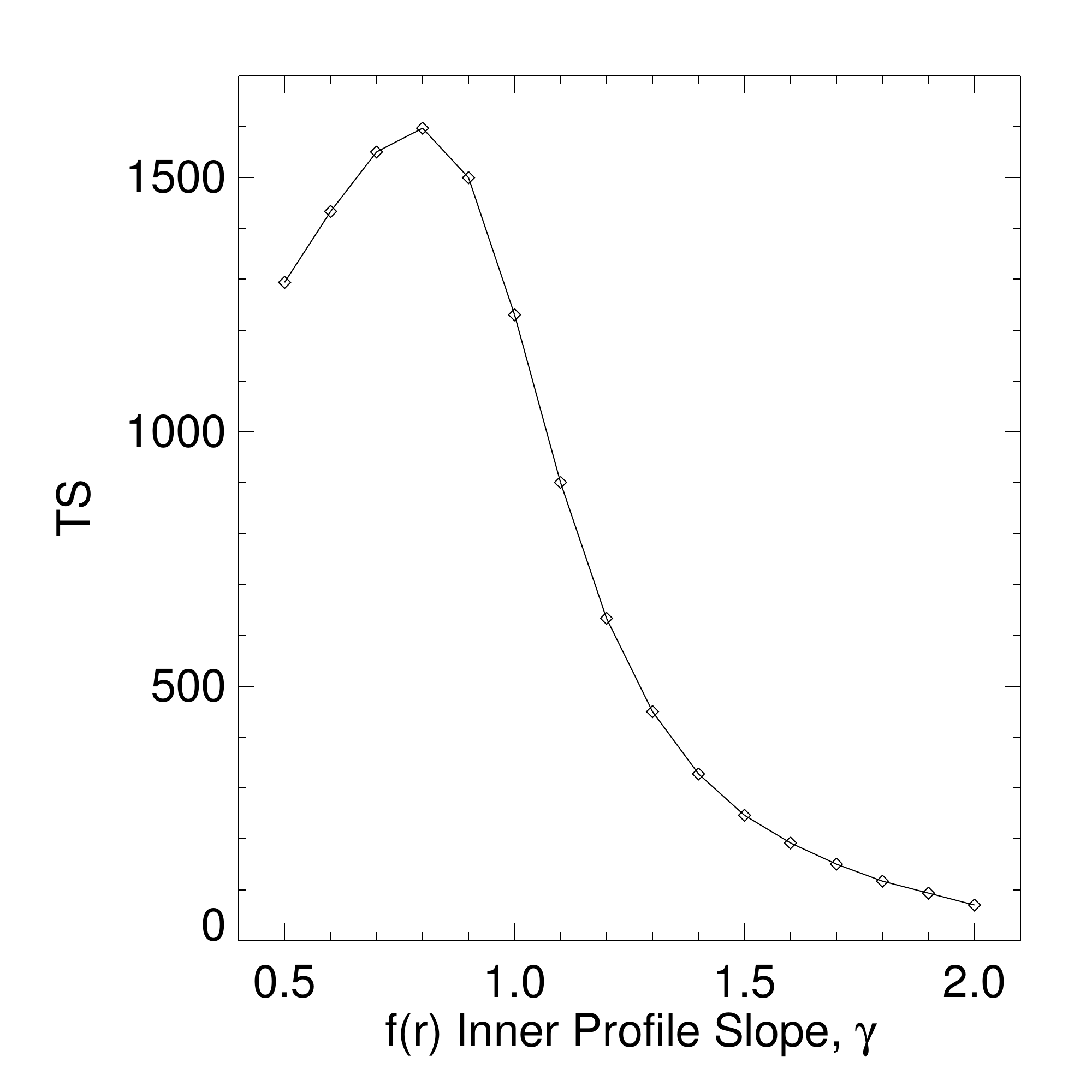}
\includegraphics[width=2.8in]{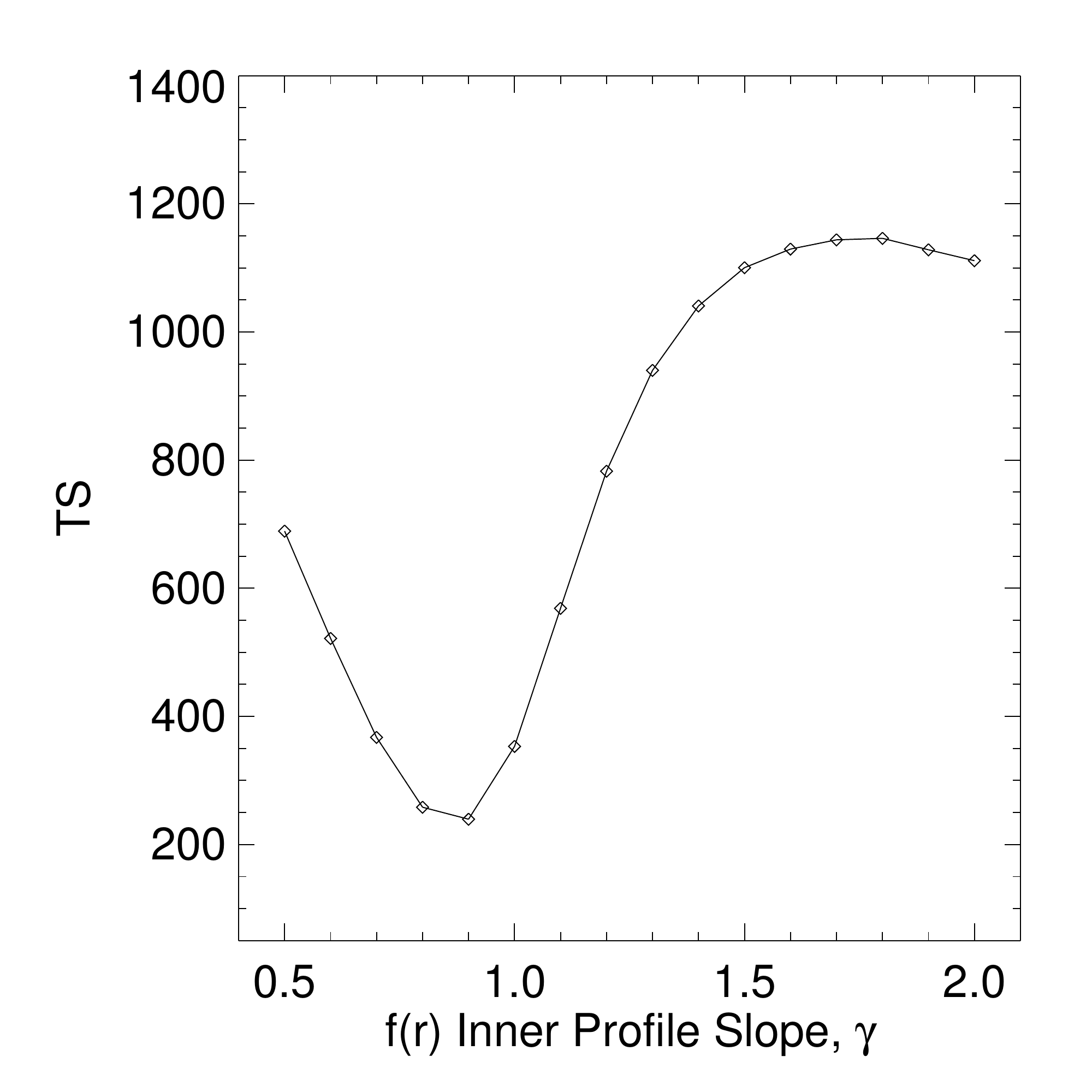}
\end{center}
\vspace{-0.5cm}
\caption{In the left frame, we plot the improvement in TS between the template fit performed using known backgrounds and a modulated Schlegel-Finkbeiner-Davis dust map (22 degrees of freedom, corresponding to the 22 energy bins), and the fit using only the known backgrounds, as a function of the inner profile slope $\gamma$ of the $f(r)$ template used in constructing the modulation. In the right frame, we show the improvement in TS when a $\gamma=1.18$ dark matter template is added to the previous fit, as a function of the inner profile slope $\gamma$ of the $f(r)$ template.}
\label{fig:chisqdeltad}
\end{figure}

\begin{figure}
\begin{center}
\includegraphics[width=0.49\textwidth]{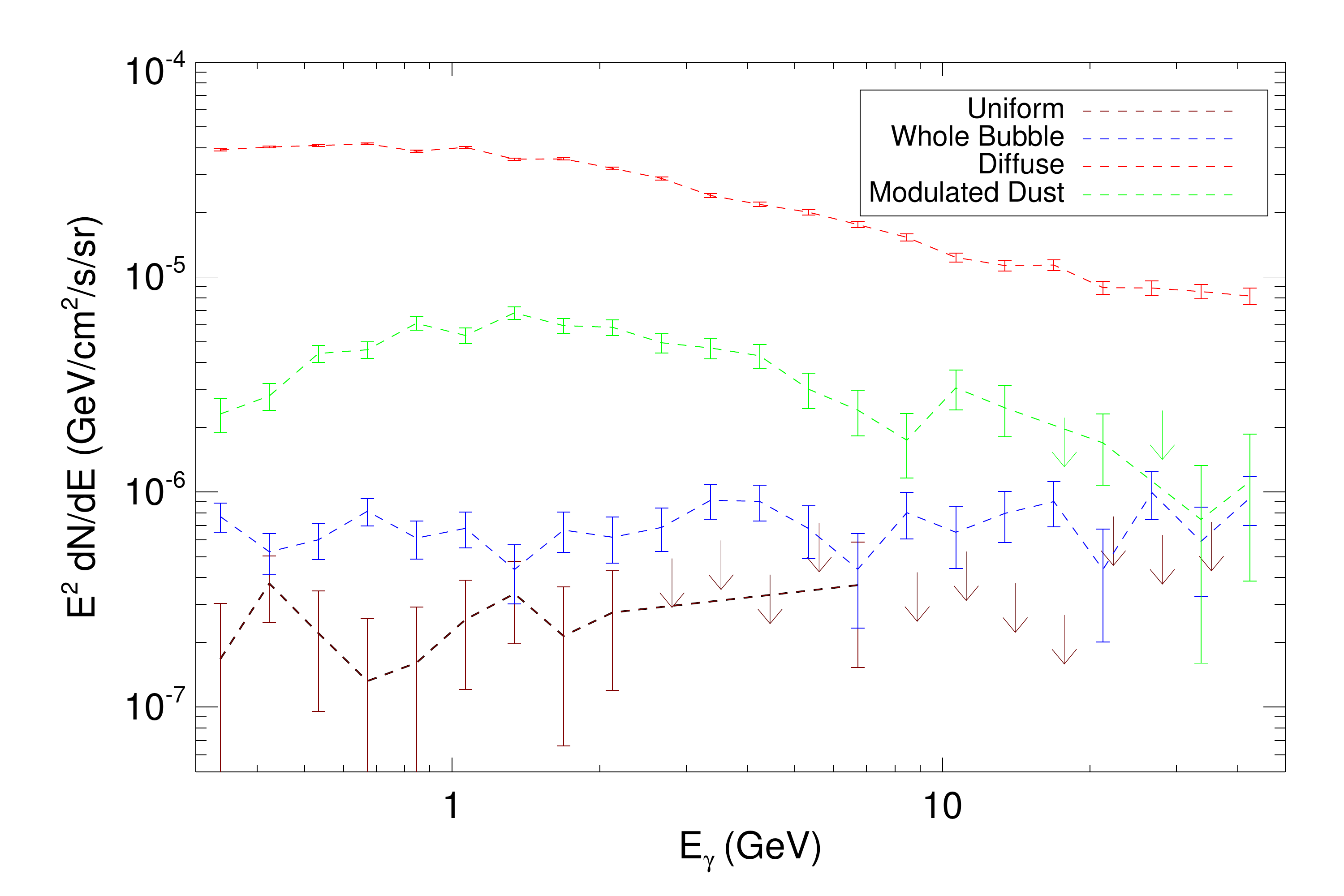}
\includegraphics[width=0.49\textwidth]{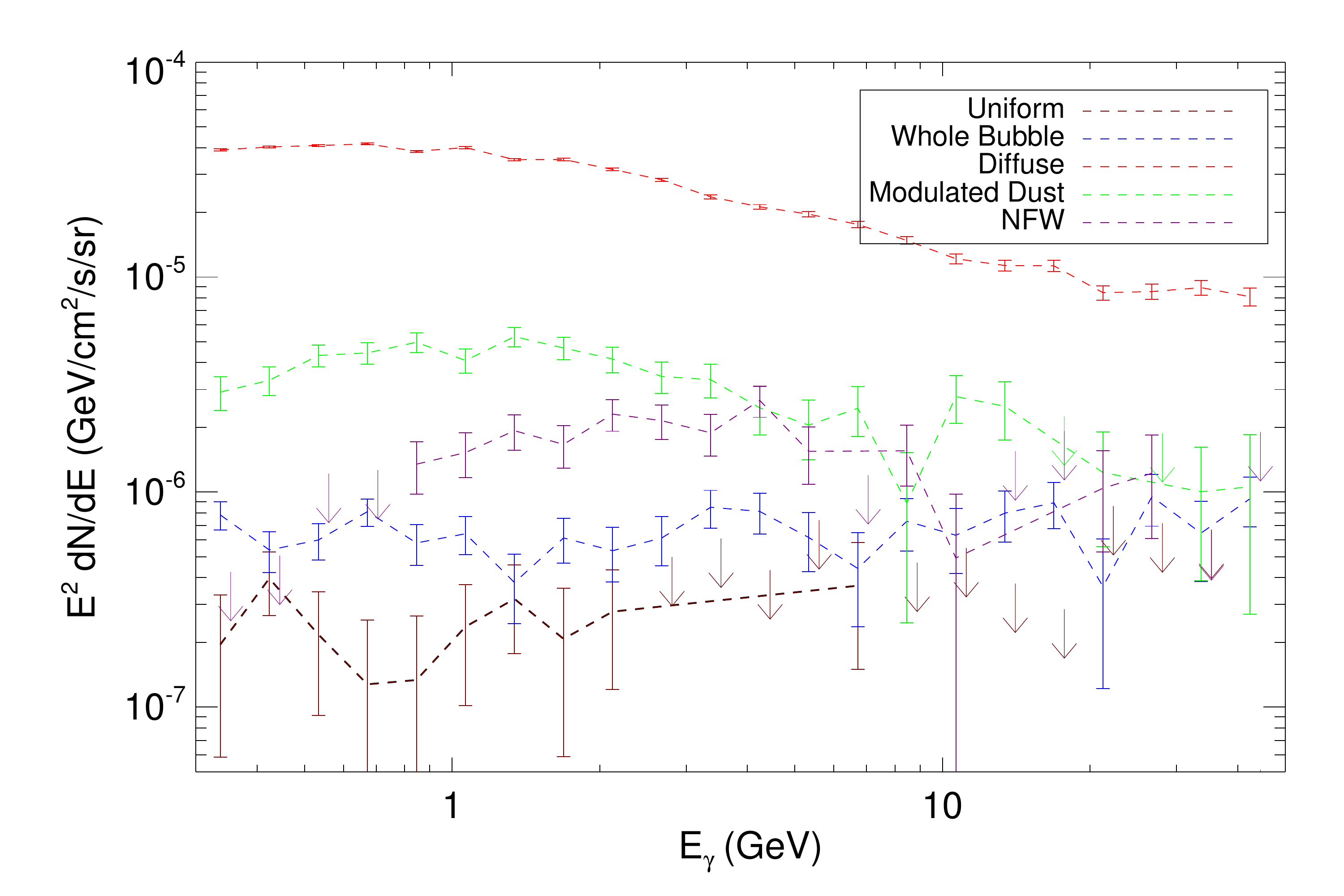}
\end{center}
\vspace{-0.5cm}
\caption{The left frame shows the spectra obtained from a template fit employing the standard backgrounds and a modulated dust template, choosing $f(r)$ with $\gamma=0.8$ (see text), in the standard ROI. In the right frame, we plot the coefficients from the same template fit, but with an additional $\gamma=1.18$ dark matter template included. The normalization of the spectrum for the modulated dust template is described in the text; the normalization of the diffuse model spectrum is as in Fig. \ref{fig:coeffsextraSFD}. Due to the large variation in the amplitudes of the different spectra, we use a log scale; where the central values are negative, we instead plot the $3\sigma$ upper limit in that bin.}
\label{fig:coeffsdeltad}
\end{figure}

Each of the modulated-dust templates was combined with the three background templates described in Sec.~\ref{inner} and run through the maximum likelihood analysis. The results can be seen in the left frame of Fig.~\ref{fig:chisqdeltad}. Generically, the modulated-dust template acquires an appreciable coefficient in a similar energy range to the observed excess. (This should not be surprising, as the modulated-dust templates have been designed to absorb the excess to the greatest degree possible.) The spectrum associated with the template fit using an $f(r)$ of $\gamma=0.8$, near where the $\chi^2$ was improved most, is shown in the left frame of Fig.~\ref{fig:coeffsdeltad}. Nevertheless, when a dark matter template was added to the analysis, there was always a substantial improvement in quality of the fit, as shown in the right frame of Fig.~\ref{fig:chisqdeltad} for a dark matter template with an inner slope of $\gamma=1.18$.

When the dark matter template and modulated dust map are added to the fit together, both acquire non-negligible coefficients, as shown in the right frame of Fig.~\ref{fig:coeffsdeltad}. The modulated dust map is correlated with a soft spectrum, similar to that of the diffuse model, while the dark matter template acquires power in the $\sim 1-5$ GeV range around the peak of the excess. The presence of the modulated dust map \emph{does} in this case substantially bias the extracted spectrum for the dark matter template -- this is not greatly surprising, as by construction the two templates are very similar in shape.

The observant reader may note that the TS of the best-fit modulated dust map is actually \emph{greater} than the TS for the dark matter template. However, it appears this may be due to the modulated dust map doing a better job of picking up unmodeled emission correlated with the \emph{dust}, rather than with the few-GeV excess. If the SFD dust map is added to the fit to provide an additional degree of freedom to the diffuse model, as described in App. \ref{app:sfddust}, the TS for the best fit dark matter template becomes 1748, compared to 1302 for the best fit modulated dust.

The above conclusions were checked to be robust against the choice of ROI and diffuse model; very similar results were found when the analysis was repeated using the full sky or the \texttt{p7v6} model.

It thus appears that a spatial modulation of the gas-correlated emission with coincidental similarities to a dark matter signal could significantly bias the extracted spectrum, but it is difficult (at least within the tests we have performed) to absorb the excess completely. The Galactic Center analysis also finds no evidence for correlation between the excess and known gas structures. Thus even if the $\pi^0$ background has been modeled incorrectly, this deficiency seems unlikely to provide an explanation for the observed signal.

\section{Modifications to the Point Source Modeling and Masking for the Inner Galaxy}
\label{app:pointsource}

As the point sources are concentrated along the Galactic disk and toward the Galactic Center, mismodeling of point sources might plausibly affect the extraction of the signal. To study the potential impact of mismodeling, and check the validity of our point source model, in the Inner Galaxy analysis, we perform the following independent tests:

\begin{itemize}
\item We allow the overall normalization of the point source model to float independently in each energy bin (the relative normalizations of different sources at the same energy are held fixed).
\item We halve or double the flux of all sources in the point source model, relative to the values given in the \texttt{2FGL} catalog.
\item We omit the point source model from the fit entirely.
\item We furthermore investigate the impact of our (fairly arbitrary) choice of mask radius, which is set at the $95\%$ containment radius of the (energy-dependent) PSF by default.
\end{itemize}

\begin{figure}
\begin{center}
\includegraphics[width=0.49\textwidth]{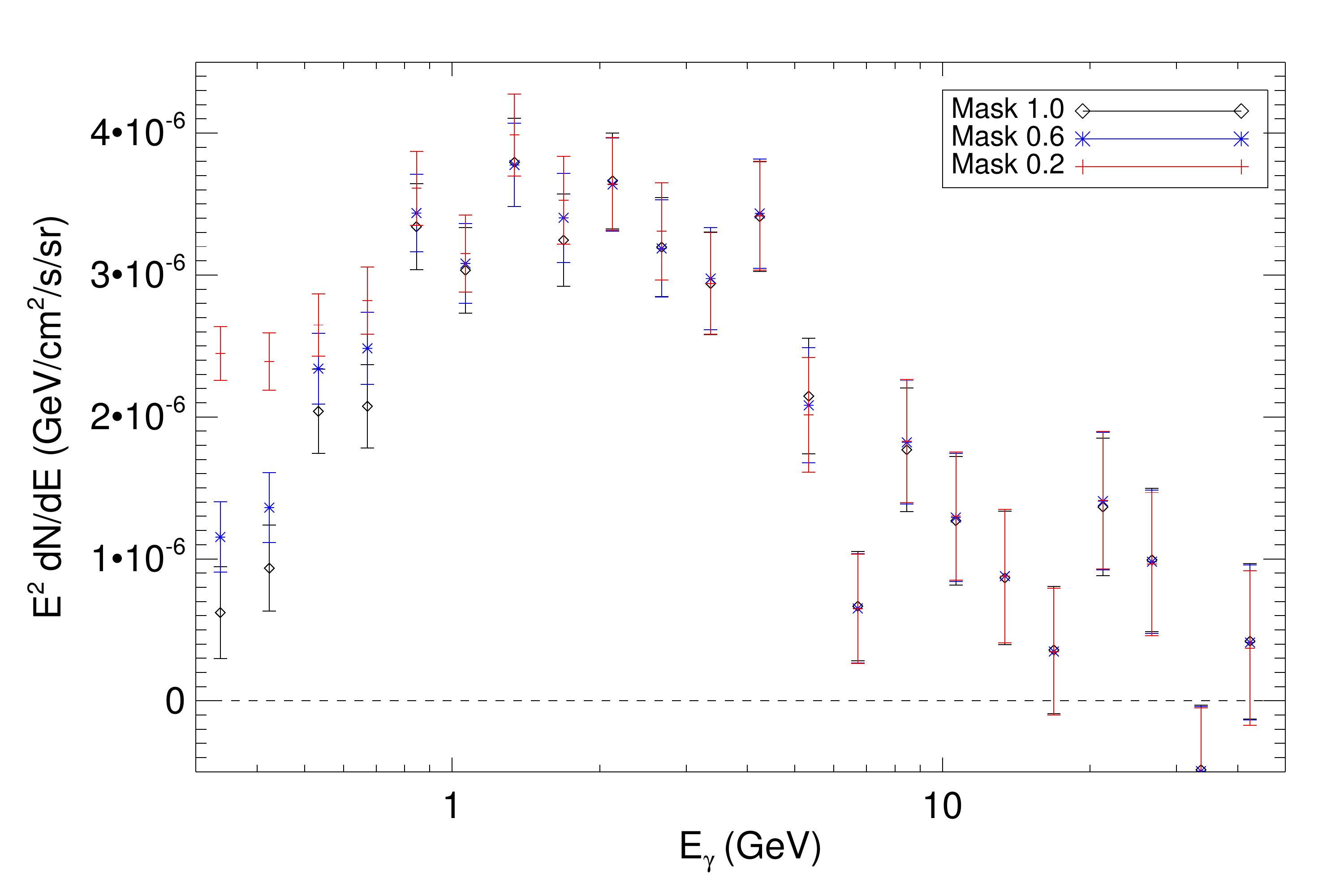}\\
\includegraphics[width=0.49\textwidth]{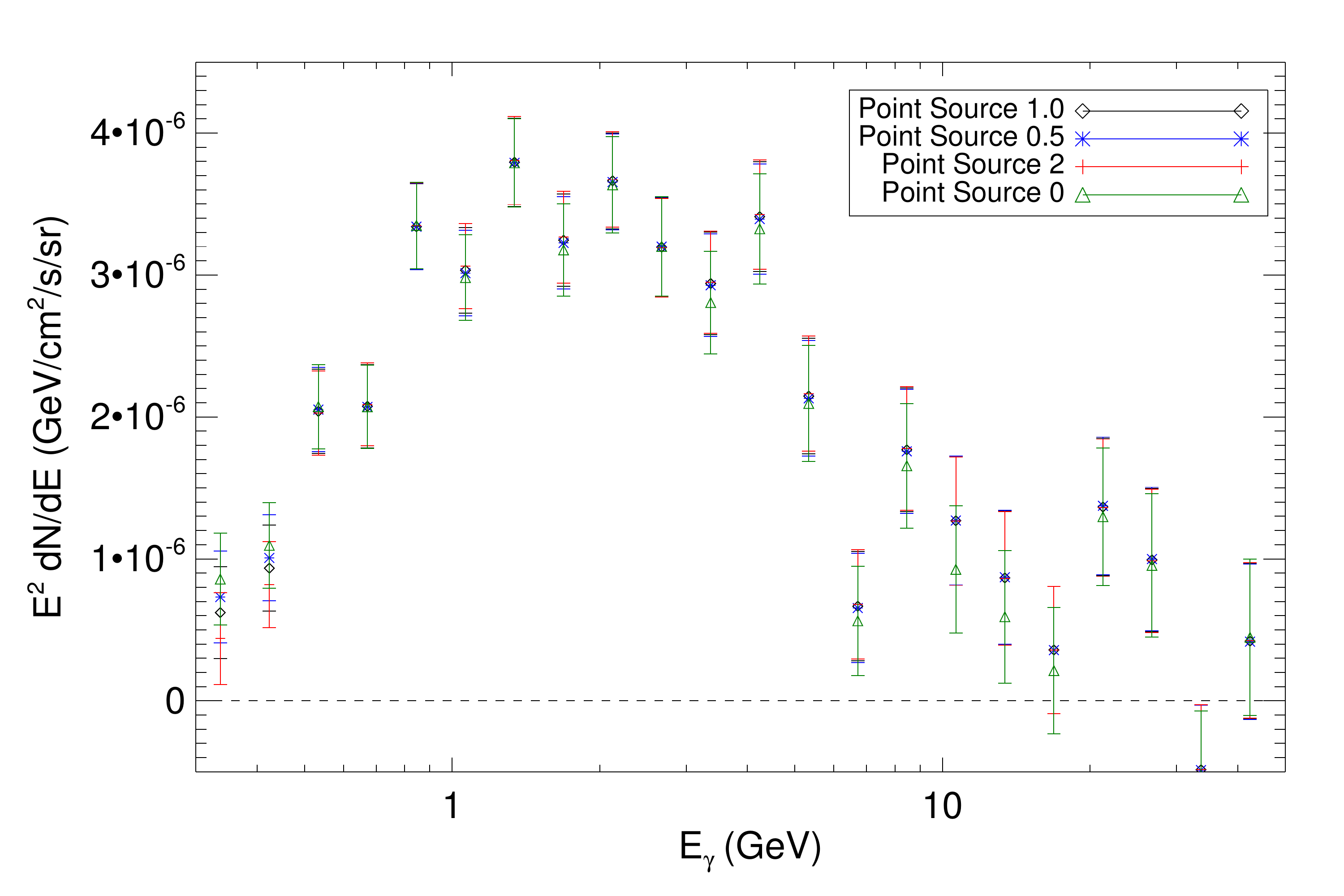}
\includegraphics[width=0.49\textwidth]{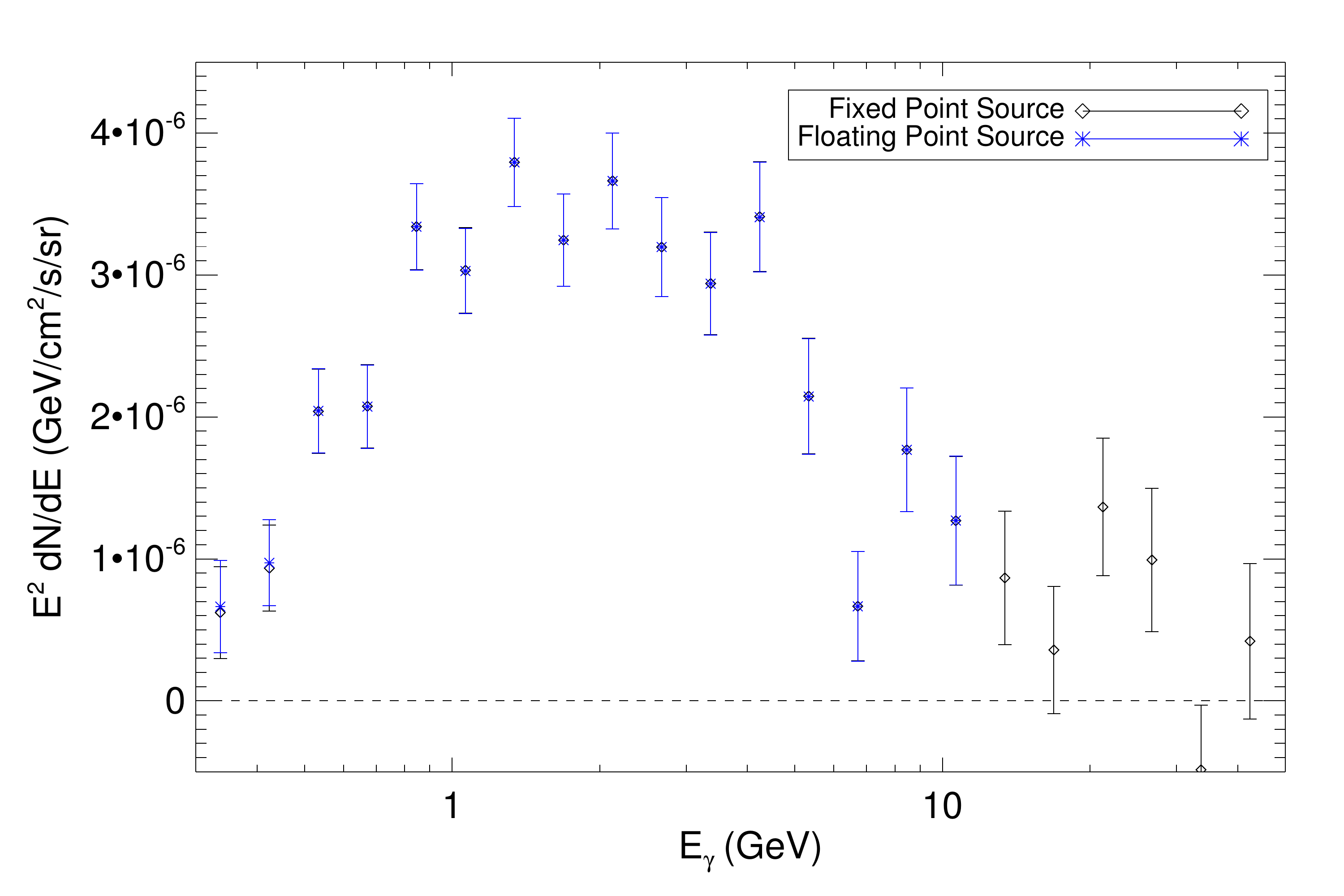}
\end{center}
\vspace{-0.5cm}
\caption{Upper panel: Here we show the impact of changing the point source mask radius, shrinking its original size from 1.0 to 0.6 and 0.2. We see that only at the lower energies is there any impact. Lower panel, left frame: We show the result of subtracting the point sources multiplied by a several values: 0, 0.5, 1 and 2. Lower panel, right frame: We show the difference of allowing the point source model to float at each energy as opposed to keeping it fixed. In the floating case we only perform the fit up to 10 GeV; beyond this point it becomes numerically unstable. Our NFW template has $\gamma=1.2$ for all fits.}
\label{fig:psmask}
\end{figure}

Plots showing the results of these various checks are found in Fig.~\ref{fig:psmask}. We find that the impact on the spectrum of even quite severe errors in the point source modeling (such as omitting it entirely or multiplying all source fluxes by a factor of two) is negligible, with the standard mask. Reducing the mask to a very small value has a greater effect, but is still only substantial at the lowest energies; we attribute the extra emission here to leakage from unmasked and poorly-subtracted bright sources.

\section{Shifting the Dark Matter Contribution Along the Plane}\label{app:shift}

\begin{figure}
\begin{center}
\includegraphics[width=0.49\textwidth]{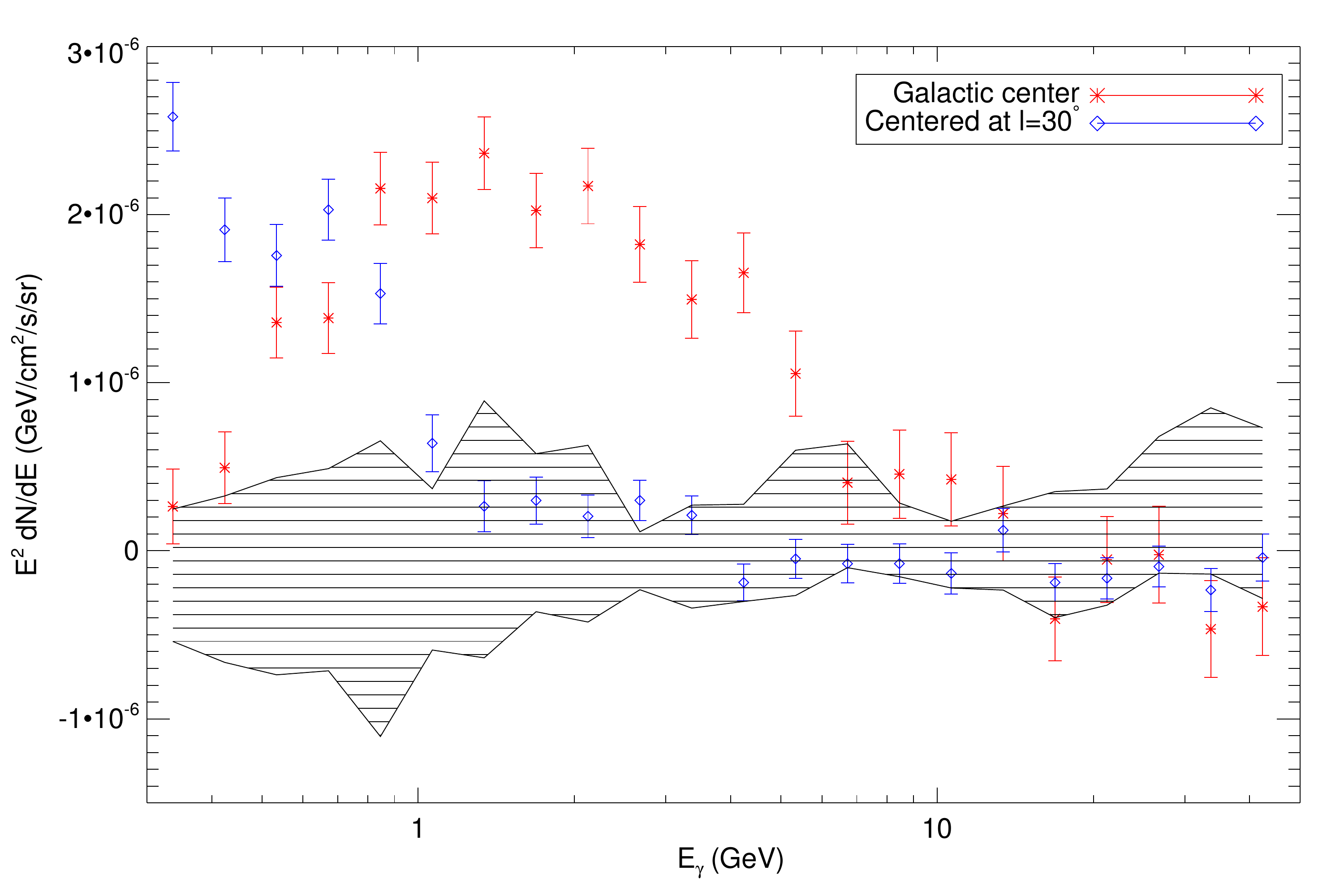}
\end{center}
\vspace{-0.5cm}
\caption{Red stars indicate the Galactic Center spectrum, whereas blue diamonds indicate the spectrum correlated with a DM-annihilation-like template (corresponding to an NFW profile with an inner slope $\gamma = 1.3$) centered at $b=0^\circ, l=30^\circ$, instead of at the Galactic Center. The band of horizontal lines indicates the spread of the best-fit spectra correlated with DM-annihilation-like templates shifted in $30^\circ$ increments along the Galactic plane: for the ten other cases sampled ($l=60^\circ, 90^\circ, ..., 330^\circ$), the emission correlated with the DM-annihilation-like template was nearly an order of magnitude below the Galactic Center excess at its peak, with no evidence of spectral similarity. In this case we perform the fit over the full-sky ROI (with an appropriate best-fit $\gamma$), rather than our standard ROI, to ensure stability of the fit and keep the fitted normalizations of the background templates similar over the different runs.}
\label{fig:planescan}
\end{figure}

The maps of Fig. 6 show residual bright structure along the Galactic plane. The presence of other bright excesses with the same spectrum along the disk, not simply in the Galactic Center, could favor astrophysical explanations for the signal. To test this possibility, we shift the DM-annihilation-like spatial template along the Galactic plane in 30$^\circ$ increments; as usual, the other templates in the fit are the \emph{Fermi} Bubbles, the diffuse model and an isotropic offset. For numerical stability and consistency of the background modeling, we perform these fits over the full sky rather than the standard ROI. All templates are normalized so that their spectra reflect the flux five degrees from their centers. For ten of the twelve points sampled, the emission correlated with this template is very small; the cross-hatched band in Fig. \ref{fig:planescan} shows the full range of the central values for these ten cases. For the point centered at $l=30^\circ$, there is substantial emission correlated with the template at energies below 1 GeV, but its spectrum is very soft, resembling the Galactic plane more than the excess at the Galactic Center. The last point is the Galactic Center.

We have performed the same test shifting the center of the DM-annihilation-like template in $5^\circ$ increments from $l=-30^\circ$ to $l=30^\circ$. The templates centered at $l = \pm 5^\circ$ absorb emission associated with the Galactic Center excess, albeit with lower amplitude; none of the other cases detect any excess of comparable size with a similar spectrum.

\newpage
\section{Variations to the Galactic Center Analysis}\label{app:gc}

In the default set of templates used in our Galactic Center analysis, we have employed astrophysical emission models which include several additional components that are not included within the official \textit{Fermi} diffuse models or source catalogs. These include the two point sources described in Ref.~\cite{YusefZadeh:2012nh} and a model tracing the 20 cm synchrotron emission. In models without a dark matter contribution, these structures are extremely significant; the addition of the 20 cm template is preferred with TS=130 (when fit with a broken power-law slope with 4 d.o.f), and the inclusion of the additional two point sources is favored with TS=15.9 and 59.3 (when the first is fit with a broken power-law with 4 d.o.f. the second with a simple power-law with 2 d.o.f).

Upon including the dark matter template in the fit, however, the significance of these additional components is lessened substantially. In this fit, the addition of the 20 cm template and the two new point sources is preferred at only TS=12.2, 21.8, and 14.6, respectively. Additionally, our best-fit models attribute extremely soft spectra to each of these sources. The 20 cm component has a hard spectrum at low energies but breaks to a spectral index of -3.3 above 0.6 GeV.  The spectral indices of the two point sources are -3.1 and -2.8, respectively. The total improvement in TS for the addition of these combined sources is 47.6.

\begin{figure}
\begin{center}
\includegraphics[width=0.46\textwidth]{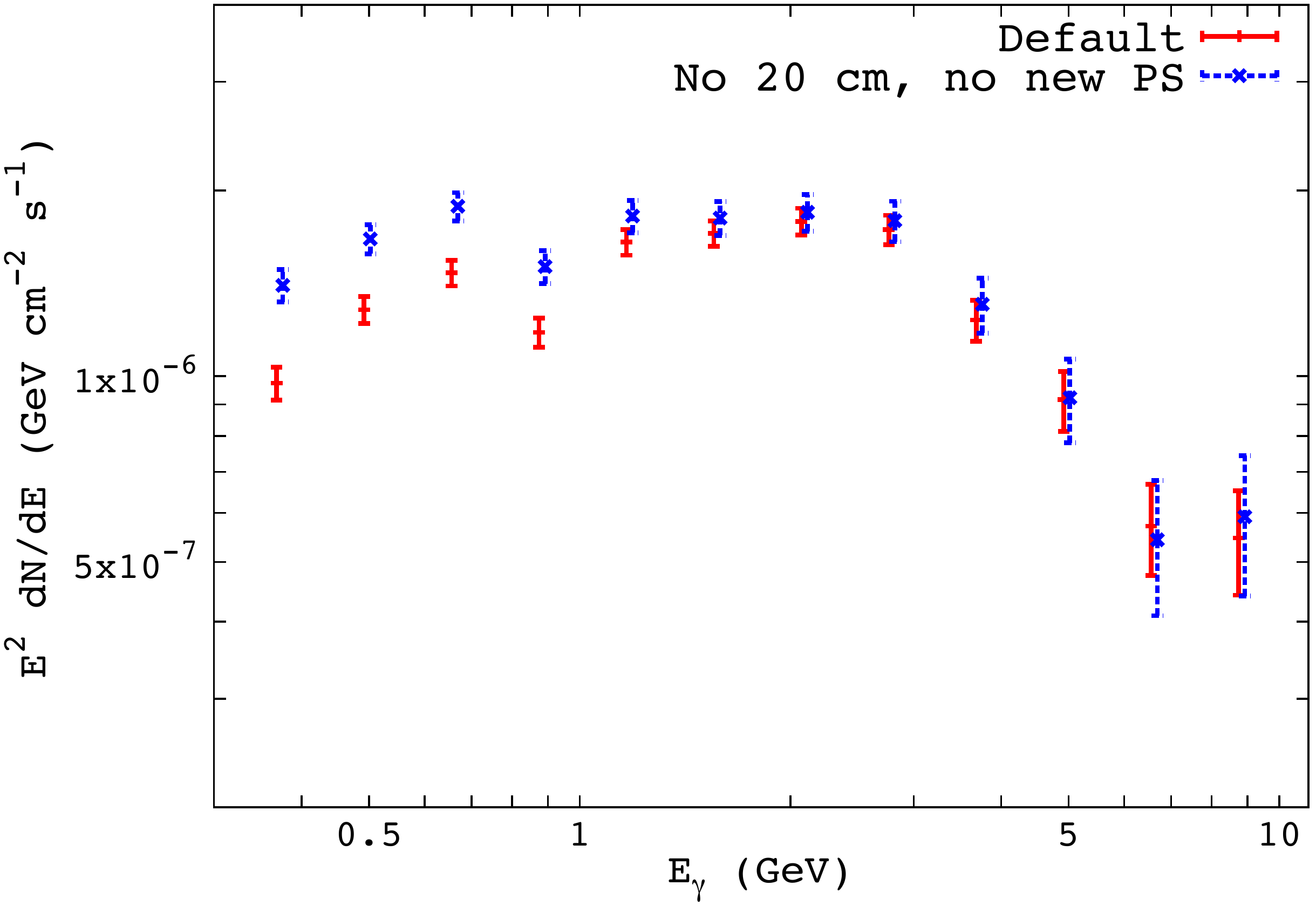}
\includegraphics[width=0.46\textwidth]{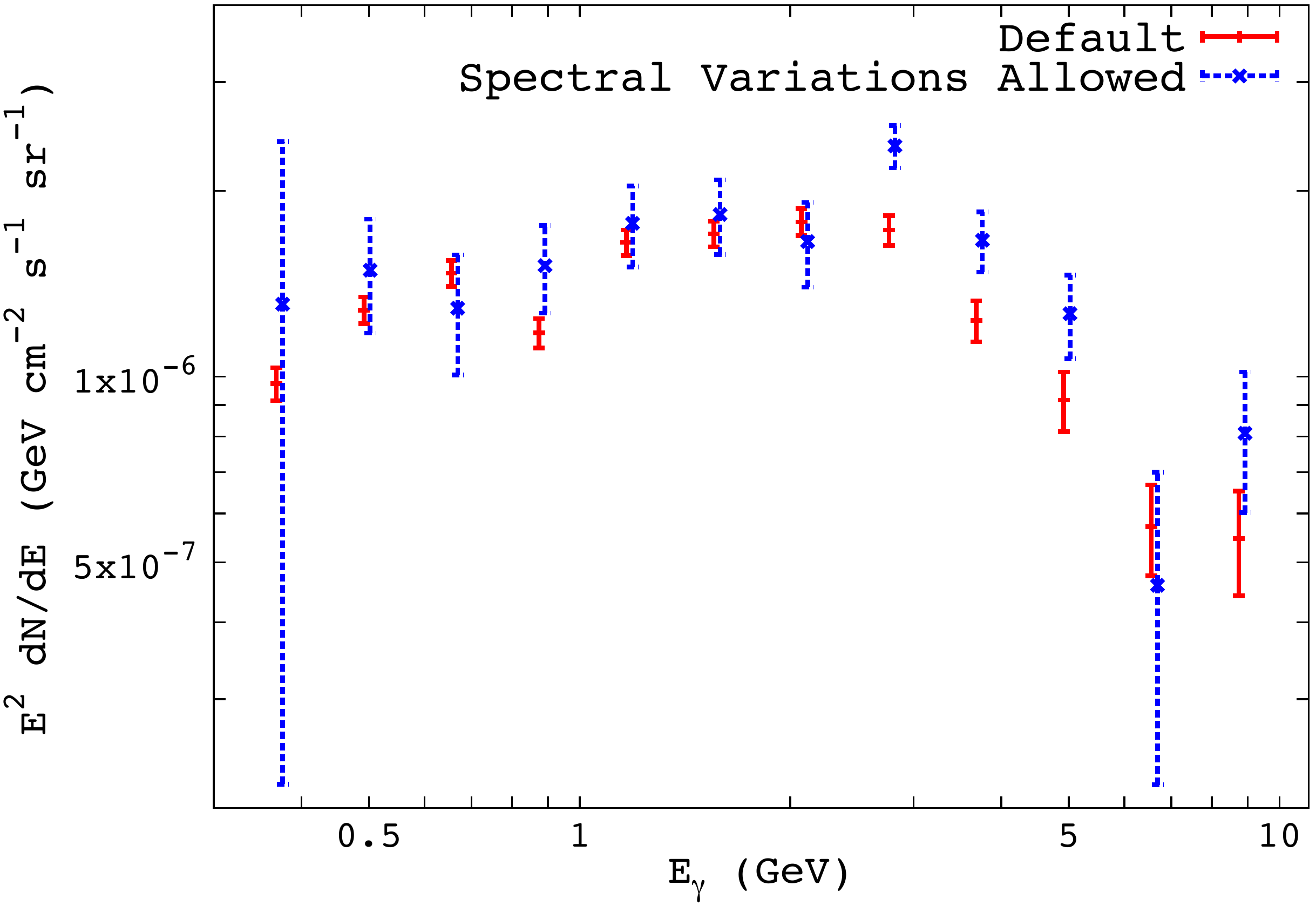}
\end{center}
\vspace{-0.5cm}
\caption{Left frame:  A comparison of the spectrum of the dark matter template found in our default Galactic Center analysis to that found when the 20 cm template and two additional point sources are not included in the fit (for $\gamma=1.3)$. The exclusion of these additional components from the fit leads to a softer spectrum at energies below $\sim$1 GeV, but does not influence the spectrum or intensity of the dark matter residual at higher energies.  Right frame: The spectrum of the dark matter template found in our Galactic Center analysis under our default assumptions, and when the flux in each energy bin is allowed to float freely for each of the isotropic, 20 cm, and Galactic diffuse components. Although the error bars become larger when this additional freedom is allowed, the residual excess remains and is robust across a wide range of energies. See text for details.}
\label{timappendix}
\end{figure}

In the upper frame of Fig.~\ref{timappendix}, we compare the spectrum of the dark matter template found in our default analysis to that found when the 20 cm template and two additional point sources are not included (for $\gamma=1.3)$. The exclusion of these additional components from the fit leads to a softer spectrum at energies below $\sim$1 GeV, but does not influence the spectrum or intensity of the dark matter residual at higher energies.

In our default Galactic Center analysis, the isotropic emission is taken to follow a power-law form, while the emission associated with the 20 cm template is allowed to follow a broken power-law, and the Galactic diffuse model adopts a spectrum as given by the model provided by the \textit{Fermi} Collaboration. As a test of the robustness of our results to these assumptions, we perform our fit once again, allowing the flux in each energy bin to float freely for each of the isotropic, 20 cm, and Galactic diffuse components.  In the lower frame of Fig.~\ref{timappendix}, we compare the spectrum extracted in this exercise to that found using our default assumptions. Although the error bars become larger, the residual excess is found to be robust across a wide range of energies. 

Lastly, to explore the possibility that the gas distribution as implicitly described by the diffuse model has a systematically biased radial distribution, we performed our fits after distorting the morphology of the diffuse model template such that it becomes brighter at a higher or lower rate as one approaches the center of the Galaxy. However, we found this variation to yield no significant improvement in our fits.

%% file: fermigg-app.tex
\chapter{Dark Matter in Galaxy Groups}

This appendix is organized as follows. First, we provide an extended description of the main analysis results presented in chapter~\ref{chap:fermigg}, including limits for different annihilation channels, injected signal tests, individual bounds on the top ten galaxy groups studied, and sky maps of the extragalactic DM halos. Secondly, we show how the results are affected by variations in the analysis procedure, focusing specifically on the halo selection criteria, data set type, foreground models, halo density and concentration, substructure boost, and the galaxy group catalog.  

\section{Extended results}
\label{sec:extended}

\begin{figure}[t!]
   \centering
    \includegraphics[width=.49\textwidth]{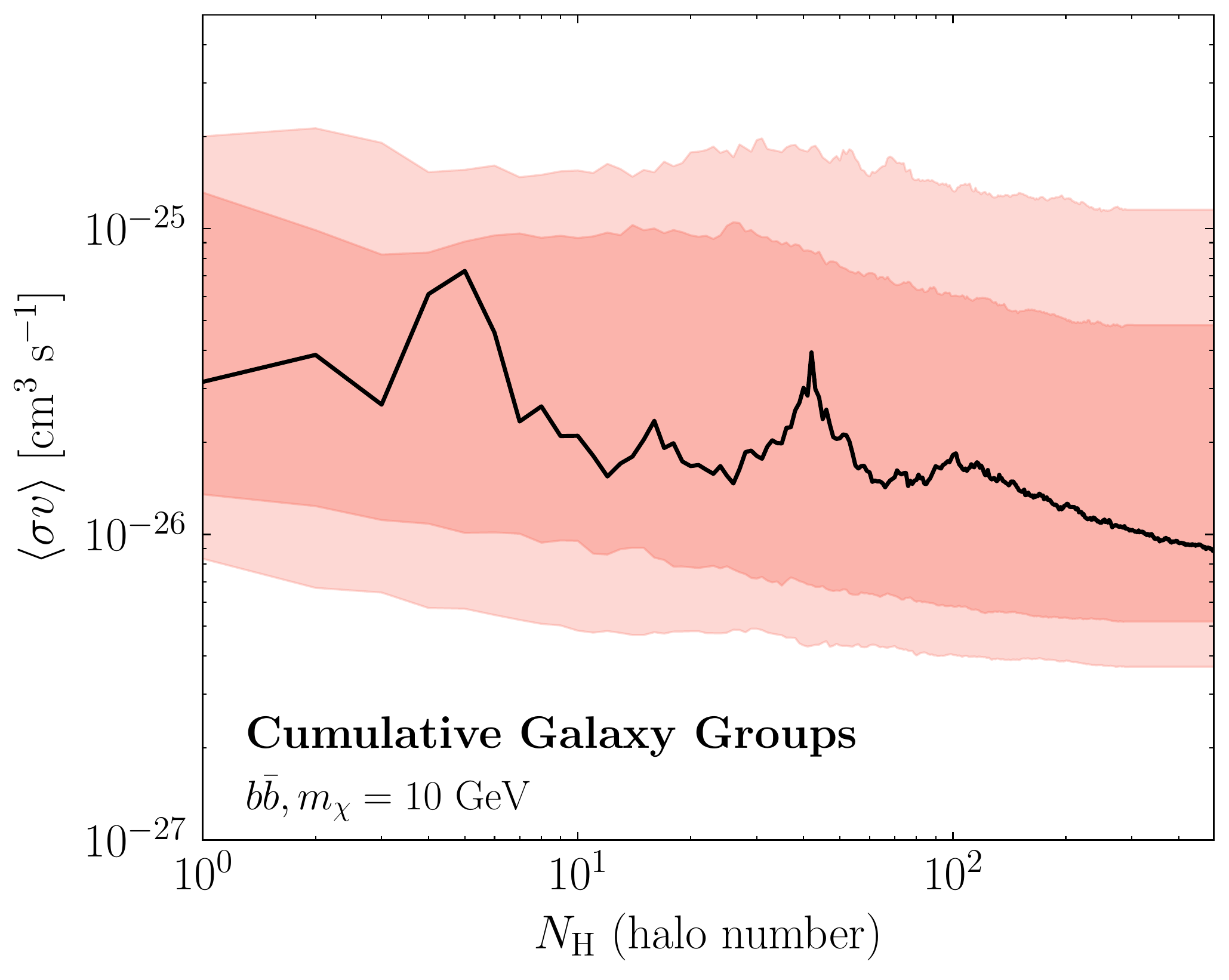}
    \includegraphics[width=.49\textwidth]{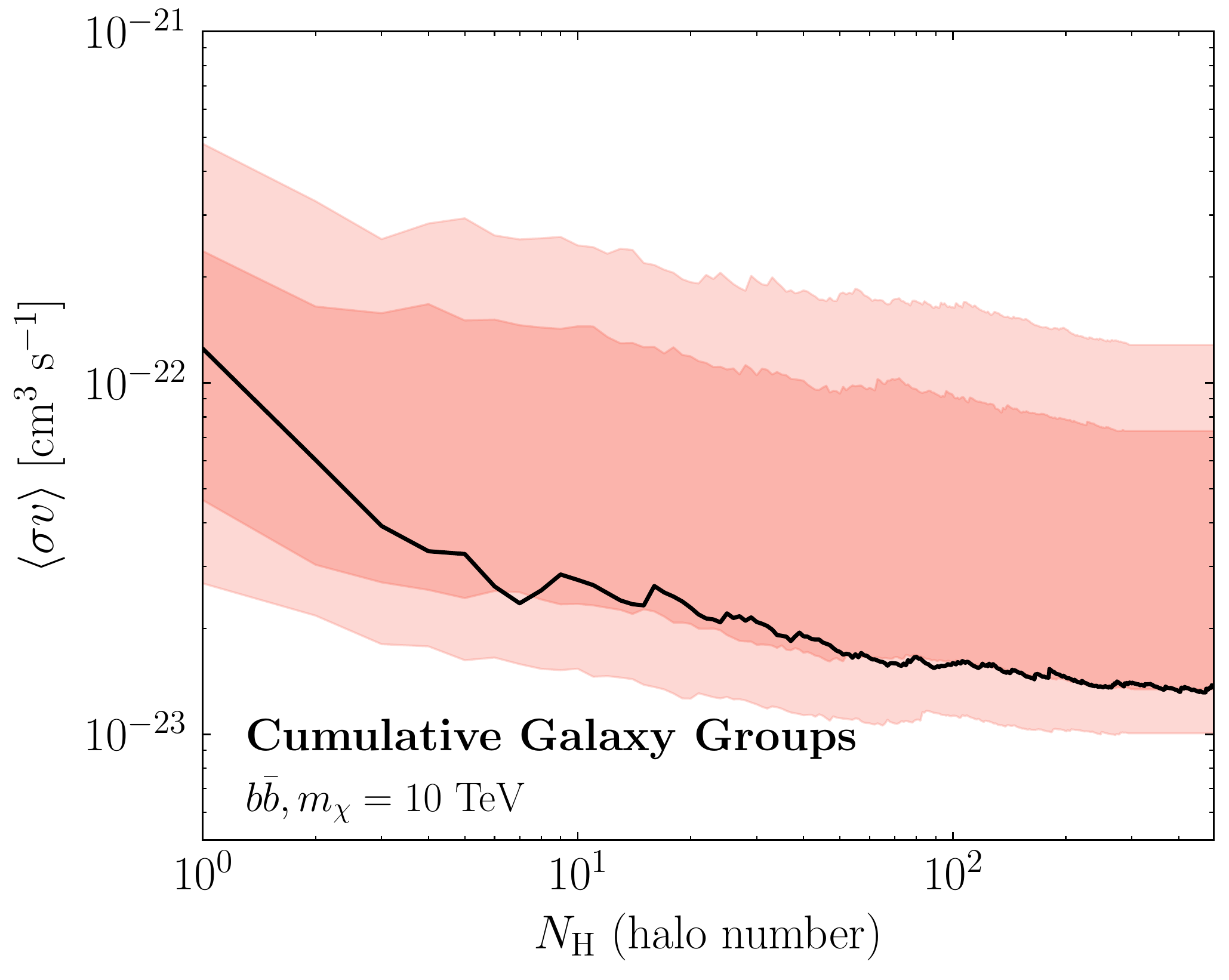}
   \caption{The change in the limit on the $b\bar{b}$ annihilation channel as a function of the number of halos included in the stacking, for $m_\chi= 10$~GeV (left) and 10~TeV (right). The 68 and 95\% expectations from 200 random sky locations are indicated by the red bands.}
   \label{fig:moreelephants}
\end{figure}

\noindent  {\bf The $\mathbf{b\bar{b}}$ Channel.}  
In the main Letter, the right panel of Fig.~\ref{fig:bounds}  demonstrates how the limit on the  $b\bar b$ annihilation cross section  depends on the number of halos included in the stacking, for the case where  $m_\chi = 100$~GeV. In Fig.~\ref{fig:moreelephants}, we show the corresponding plot for $m_\chi = 10$~GeV (left) and $10$~TeV (right).  As in the 100~GeV case, we see that no single halo dominates the bound and that stacking a large number of halos considerably improves the sensitivity.

\begin{figure}[t]
   \centering
    \includegraphics[width=0.49\textwidth]{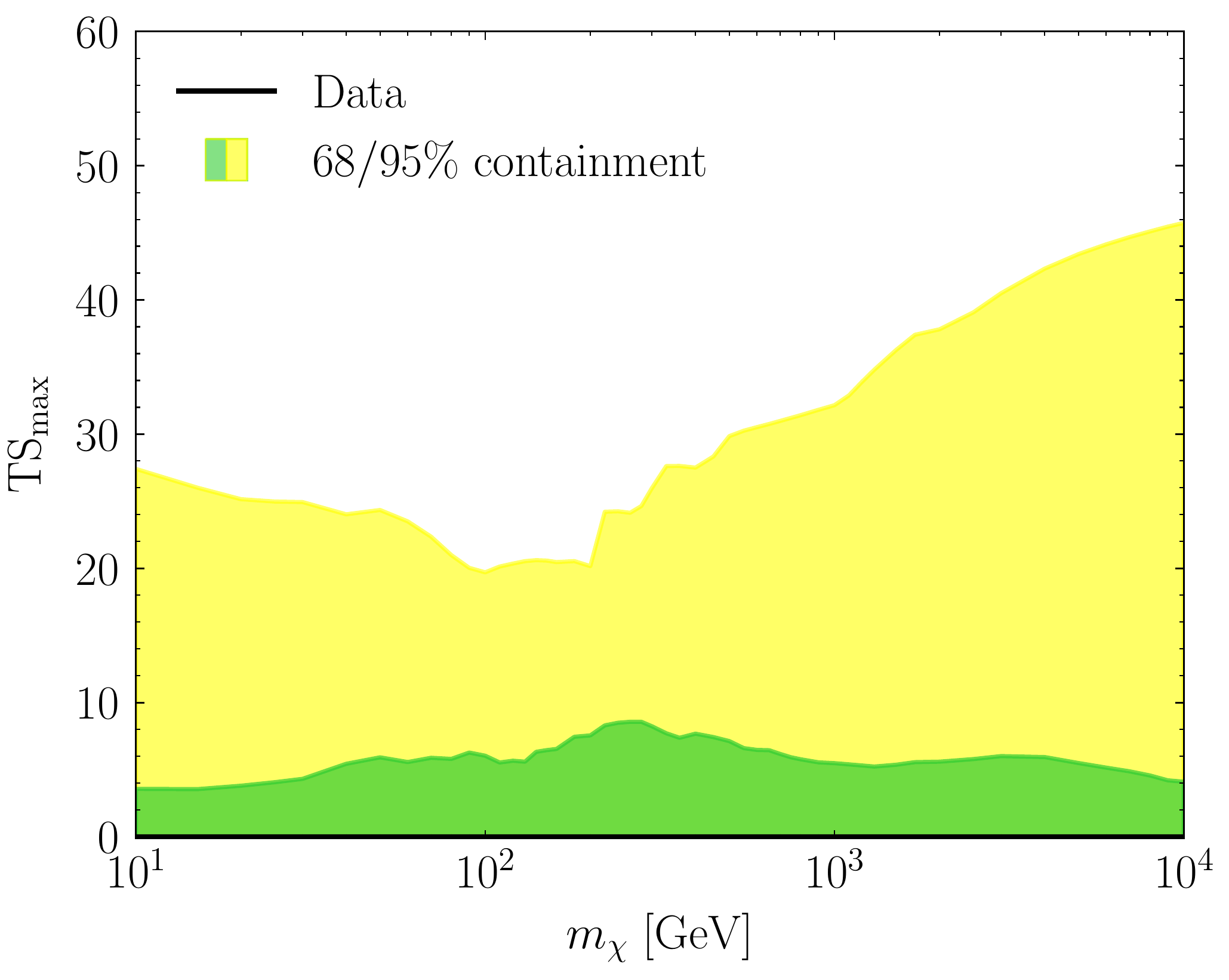}
   \includegraphics[width=.49\textwidth]{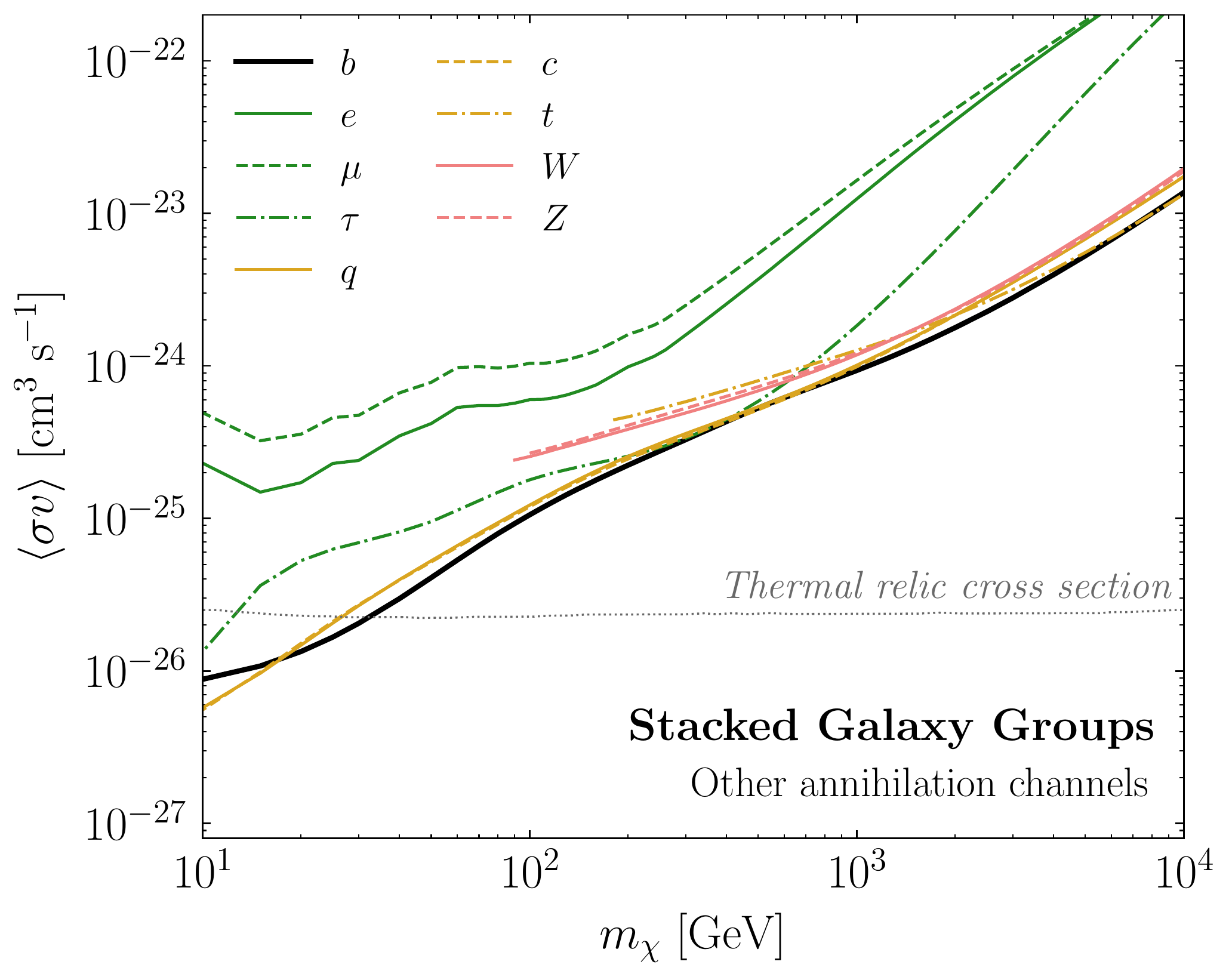}
   \caption{(Left) Maximum test statistic, TS$_\text{max}$, for the stacked analysis comparing the model with and without DM annihilating to $b \bar b$.  The green~(yellow) bands show the 68\%~(95\%) containment over multiple random sky locations.  (Right) The 95\% confidence limits on the DM annihilation cross section, as a function of the DM mass, for the Standard Model final states indicated in the legend.  These limits assume the fiducial boost factor taken from Ref.~\cite{Bartels:2015uba}.  Note that we neglect Inverse Compton emission and electromagnetic cascades, which can be relevant for the leptonic decay channels at high energies.}
   \label{fig:other_lims}
\end{figure}

The left panel of Fig.~\ref{fig:other_lims} shows the maximum test statistic, TS$_\text{max}$, recovered for the stacked analysis in the $b\bar{b}$ channel.  For a given data set $d$, we define the maximum test-statistic in preference for the DM model, relative to the null hypothesis without DM, as 
\es{maxTS}{
{\rm TS}_\text{max}(\mathcal{M}, m_{\rm \chi}) \equiv & \,2 \left[ \log \mathcal{L}(d | \mathcal{M}, \widehat{\langle\sigma v\rangle}, m_\chi ) - \log \mathcal{L}(d | \mathcal{M}, \langle\sigma v\rangle =0, m_\chi ) \right] \, ,
}
where $\widehat{\langle\sigma v\rangle}$ is the cross section that maximizes the likelihood for DM model $\mathcal{M}$.  The observed TS$_\text{max}$ is negligible at all masses and well-within the null expectation (green/yellow bands), consistent with the conclusion that we find no evidence for DM annihilation.  \vspace{0.1in}

\noindent  {\bf Other Annihilation Channels.}  
In general, DM may annihilate to a variety of Standard Model final states.  Figure~\ref{fig:other_lims} (right) interprets the results of the analysis in terms of limits on additional final states that also lead to continuum gamma-ray emission.  Final states that predominantly decay hadronically  ($W^+ W^-$, $ZZ$, $q \bar{q}$, $c \bar c$, $b \bar b$, $t \bar t$) give similar limits because their energy spectra are mostly set by boosted pion decay.  The leptonic channels ($e^+ e^-$, $\mu^+ \mu^-$) give weaker limits because gamma-rays predominantly arise from final-state radiation or, in the case of the muon, radiative decays.  The $\tau^+ \tau^-$ limit is intermediate because roughly 35\% of the $\tau$ decays are leptonic, while the remaining are hadronic.   Of course, the DM could annihilate into even more complicated final states than the two-body cases considered here and the results can be extended to these cases~\cite{Elor:2015tva,Elor:2015bho}.  Note that the limits we present for the leptonic final states are conservative, as they neglect Inverse Compton (IC) emission and electromagnetic cascades, which are likely important at high DM masses---see~\emph{e.g.}, Ref.~\cite{Cirelli:2009dv, Murase:2012xs}.  A more careful treatment of these final states requires modeling the magnetic field strength and energy loss mechanisms within the galaxy groups. 
\vspace{0.1in}

\begin{figure}[t]
   \centering
 	\includegraphics[width=.32\textwidth]{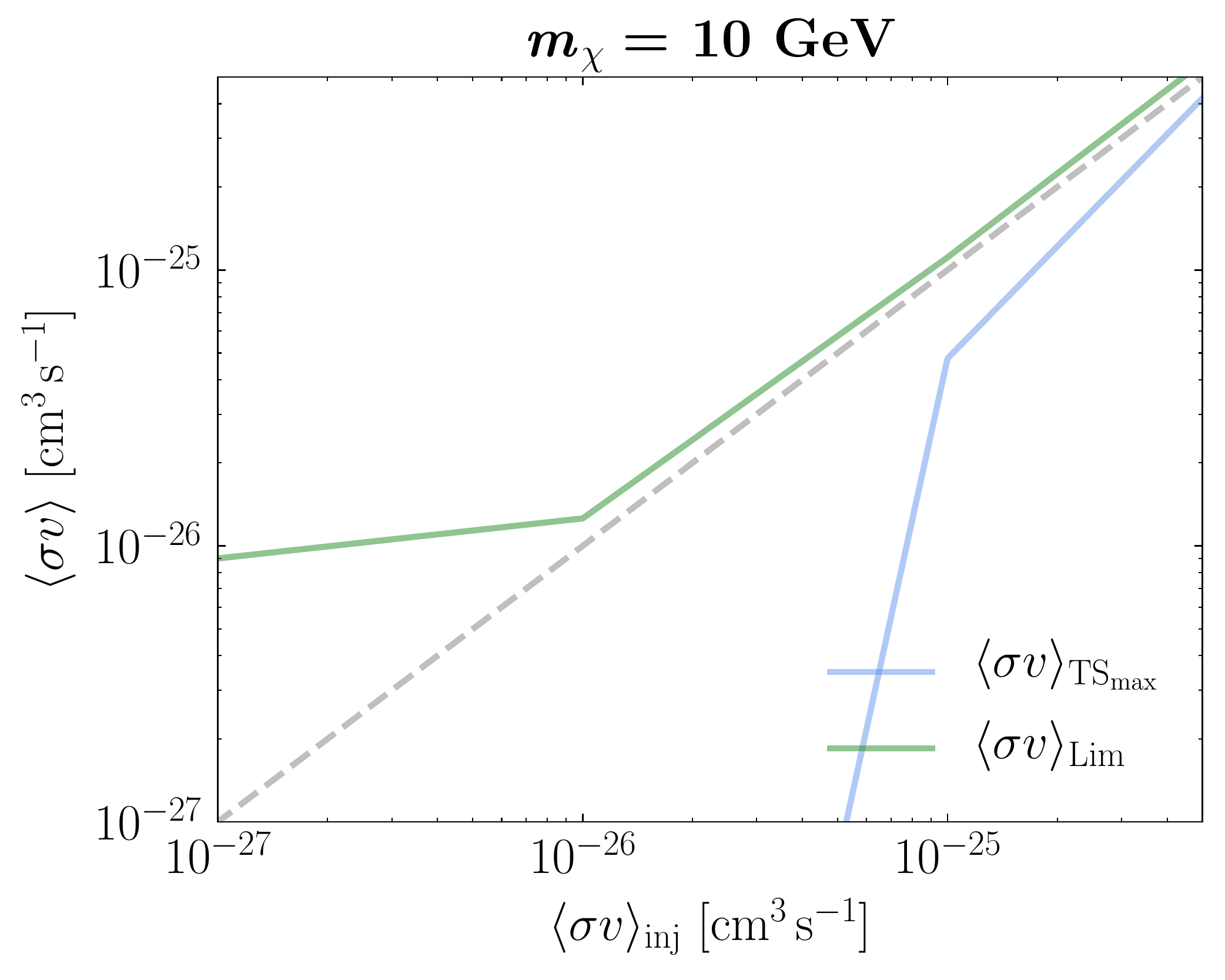}
	\includegraphics[width=.32\textwidth]{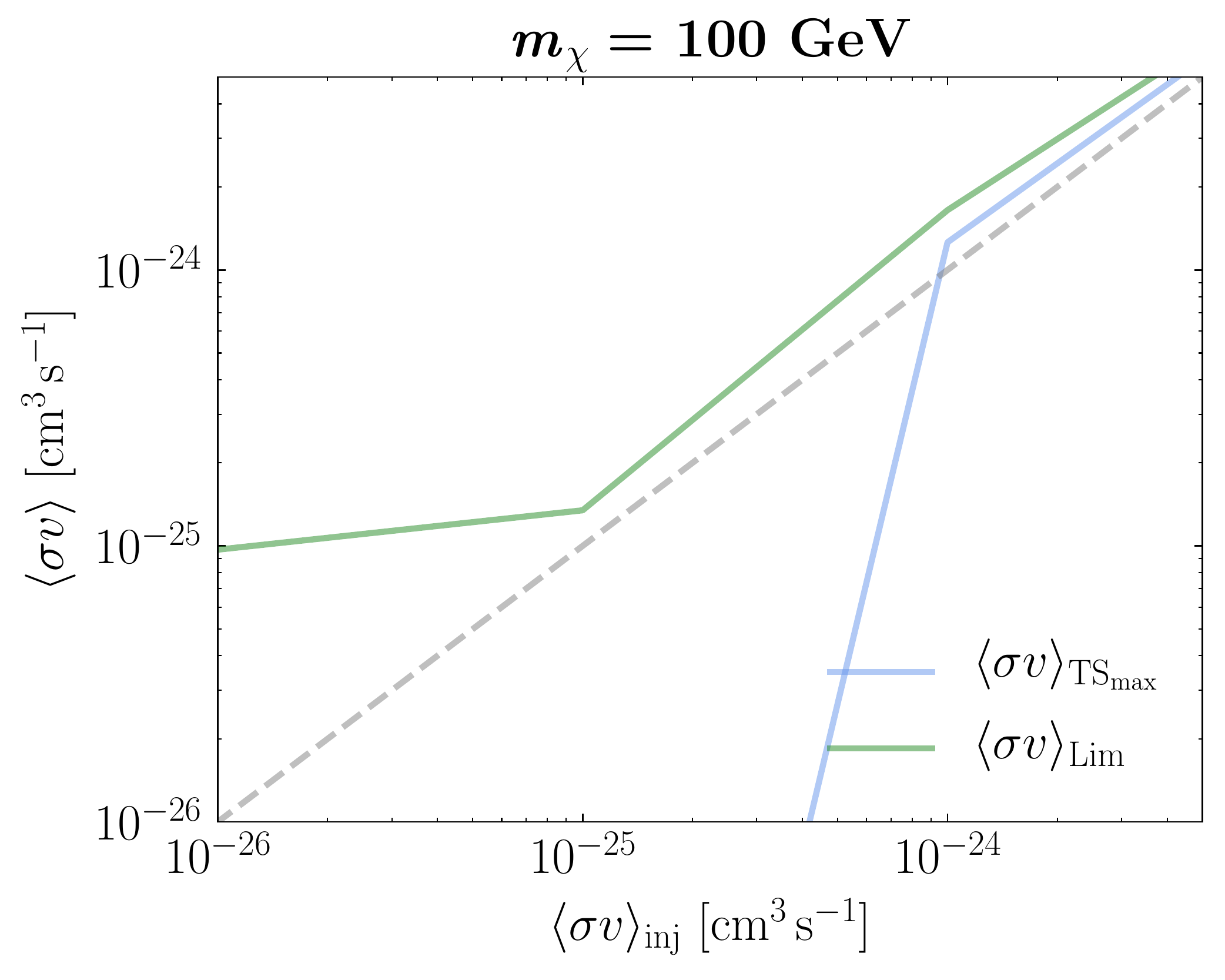}
	\includegraphics[width=.32\textwidth]{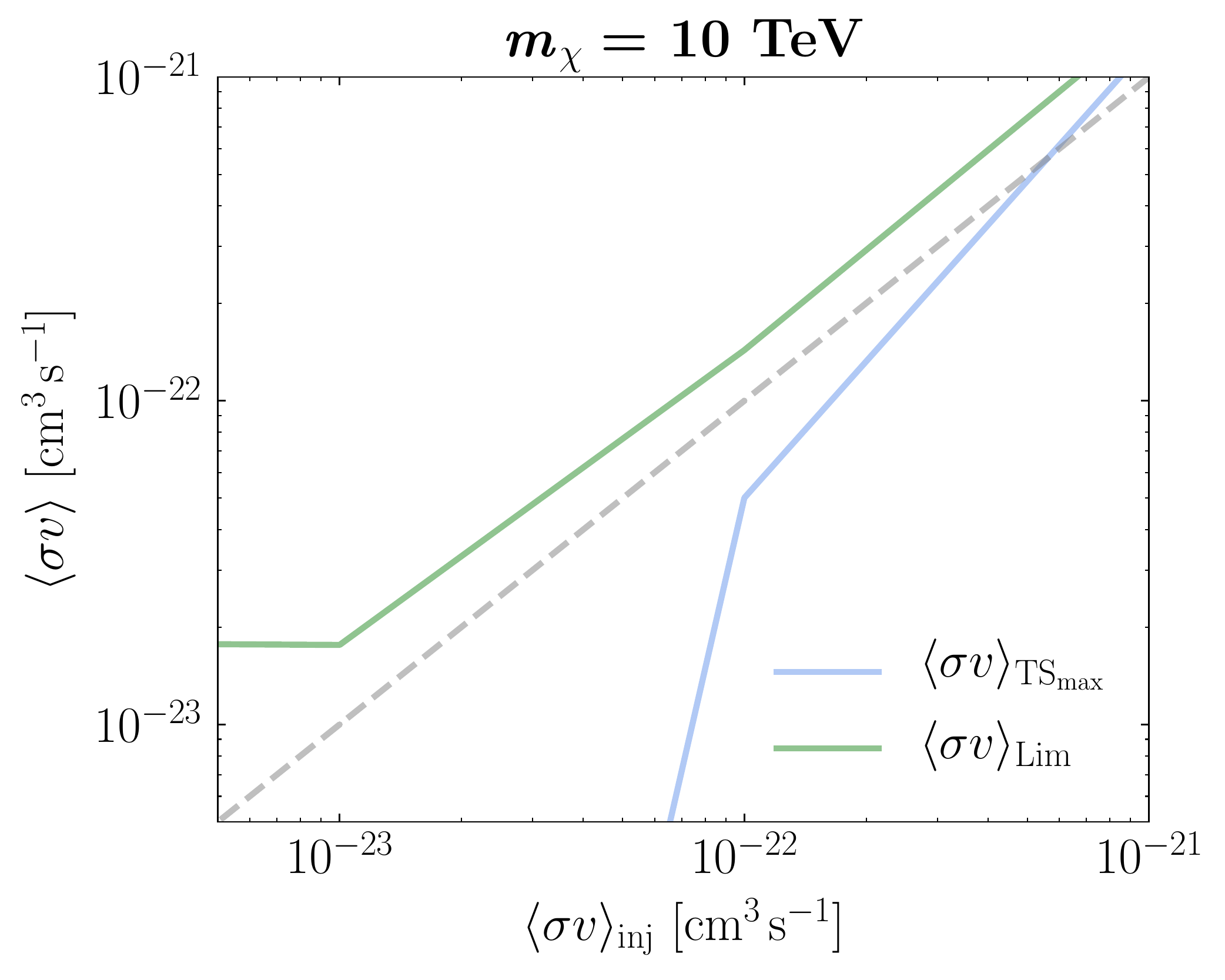} \\
	\includegraphics[width=.32\textwidth]{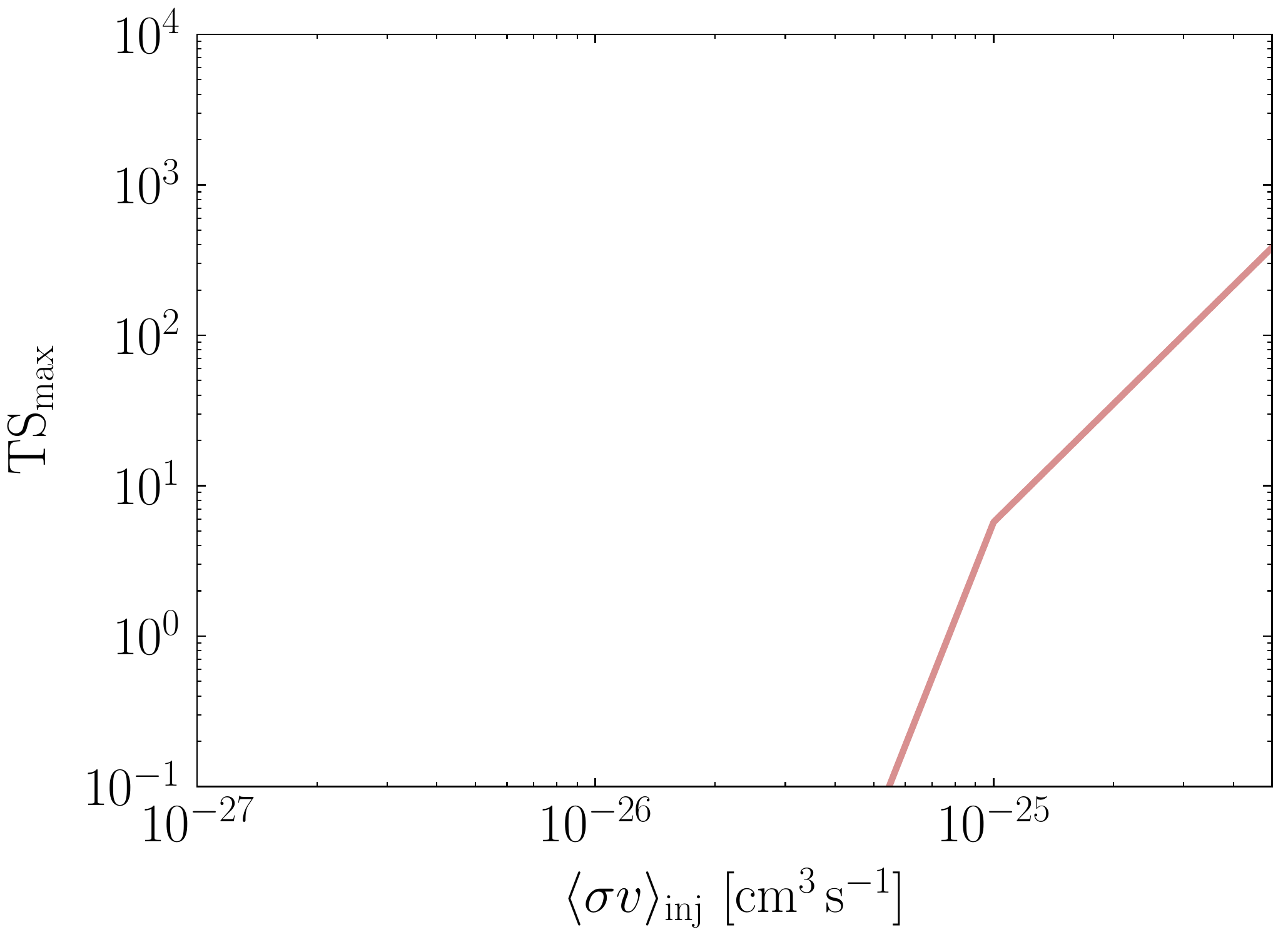}
	\includegraphics[width=.32\textwidth]{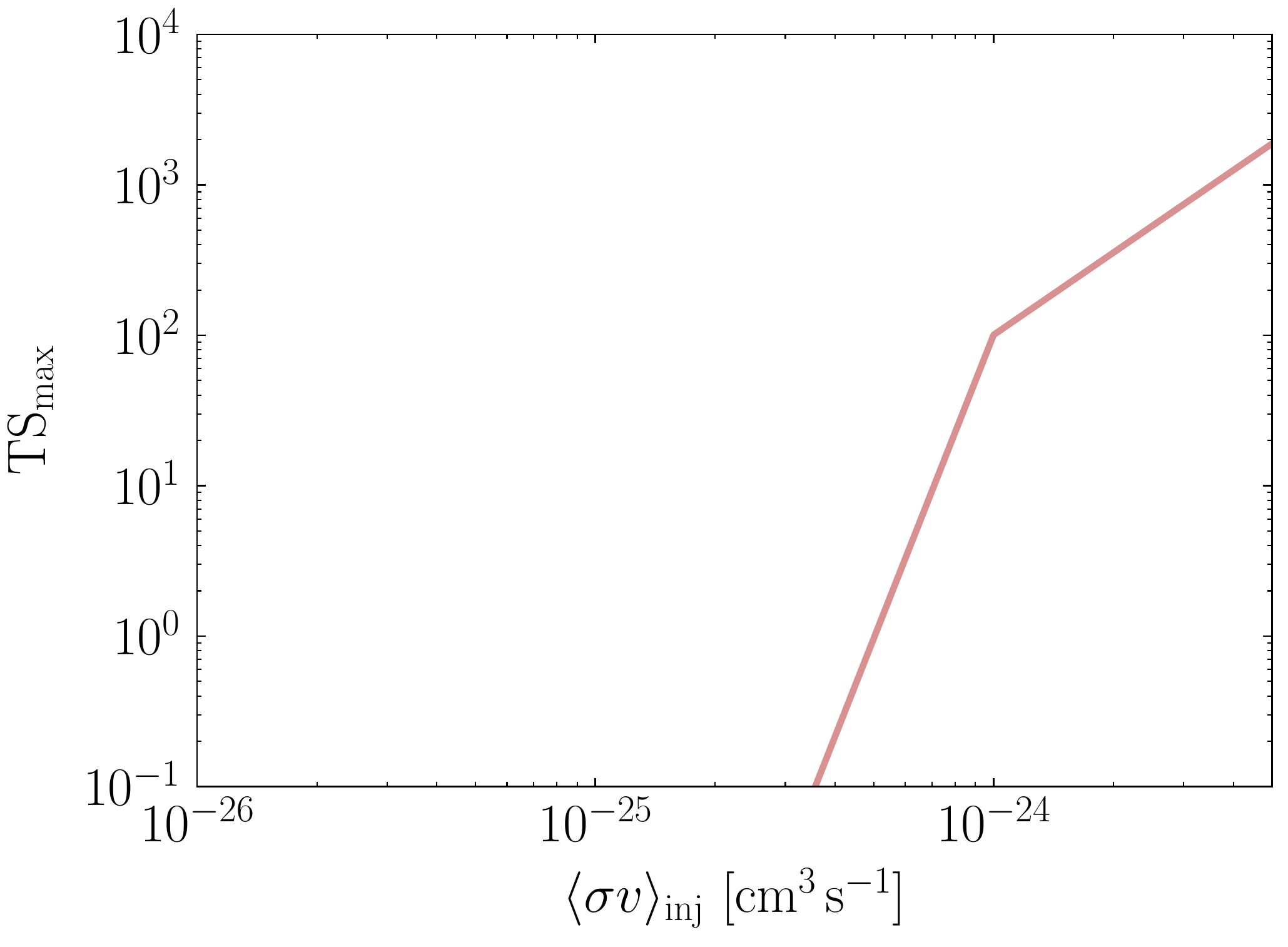}
	\includegraphics[width=.32\textwidth]{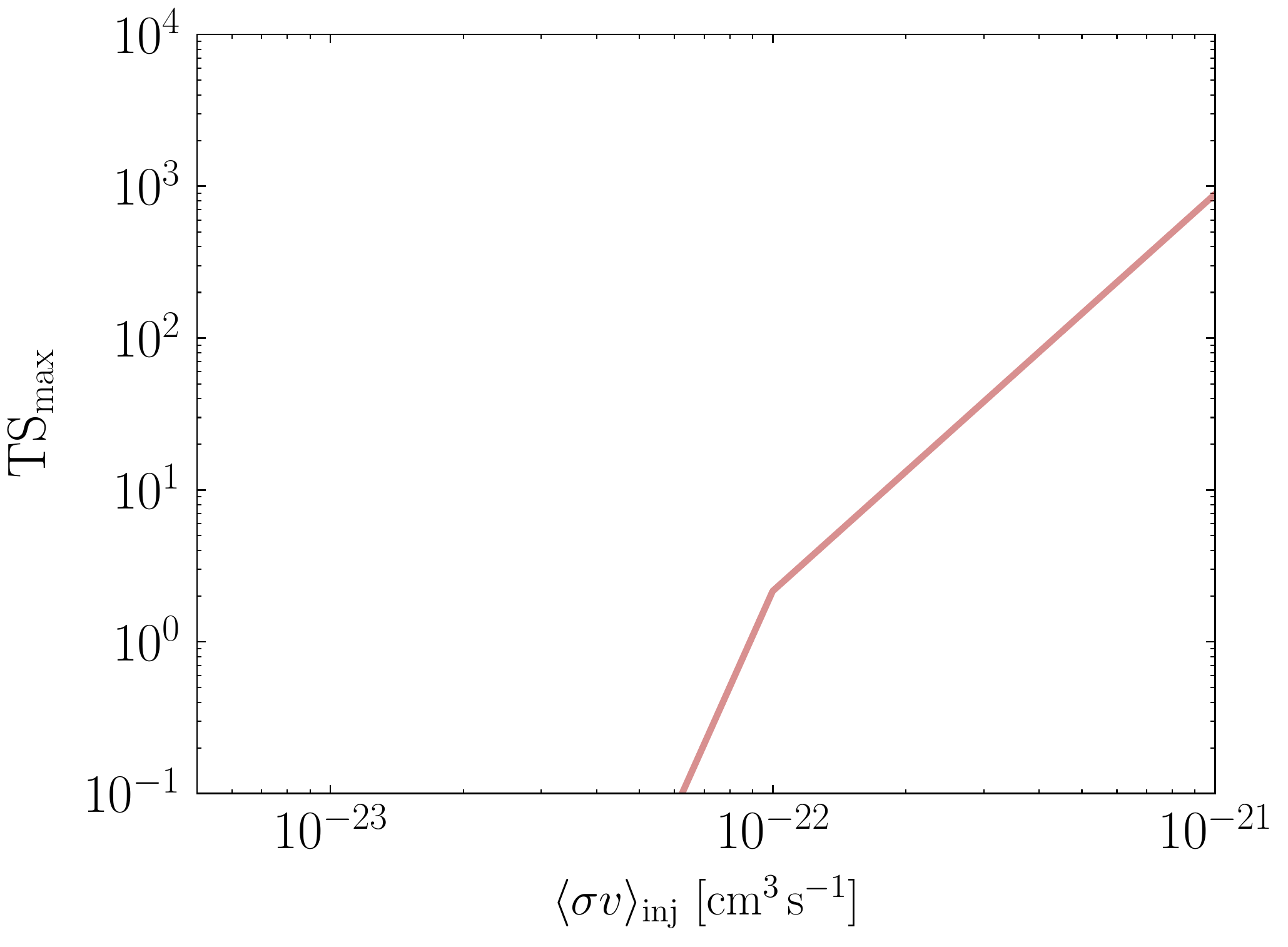}
   \caption{(Top) Recovered cross section at maxiumum test statistic, TS$_\text{max}$, (blue line) and limit (green line) obtained for various signals injected on top of the data. (Bottom) The maximum test statistic obtained at various injected cross section values. }
   \label{fig:injsig}
\end{figure}

\noindent  {\bf Injected Signal.} An important consistency requirement is to ensure that the limit-setting procedure does not exclude a putative DM signal. The likelihood procedure employed here was extensively vetted in our companion paper~\cite{Lisanti:2017qoz}, where we demonstrated that the limit never excludes an injected signal.  In Fig.~\ref{fig:injsig}, we demonstrate a data-driven version of this test. In detail, we inject a DM signal on top of the actual data set used in the main analysis, focusing on the case of DM annihilation to $b \bar{b}$ for a variety of cross sections and masses. We then apply the analysis pipeline to these maps.  The top panel of Fig.~\ref{fig:injsig} shows the recovered cross sections, as a function of the injected values.  The green line corresponds to the 95\% cross section limit, while the blue line shows the best-fit cross section.  Note that statistical uncertainties arising from DM annihilation photon counts are not significant here, as the dominant source of counts arises from the data itself. 
The columns correspond to 10, 100, and 10$^4$~GeV DM annihilating to $b \bar b$ (left, center, right, respectively).  The bottom row shows the maximum test statistic in favor of the model with DM as a function of the injected cross section.  The best-fit cross sections are only meaningful when the maximum test statistic is $\gtrsim 1$, implying evidence for DM annihilation.     
 We see that across all masses, the cross section limit  (green line) is always weaker than the injected value.  Additionally, the recovered cross section (blue line) closely approaches that of the injected signal as the significance of the DM excess  increases.   
\vspace{0.1in}

\noindent  {\bf Results for Individual Halos.}  Here, we explore the properties of the individual galaxy groups that are included in the stacked analysis.  These galaxy groups are taken from the catalogs in Ref.~\cite{Tully:2015opa} and~\cite{2017ApJ...843...16K}, which we refer to as T15 and T17, respectively.  Table~\ref{tab:tully_extended} lists the top thirty galaxy groups, ordered by the relative brightness of their inferred $J$-factor. Not all groups in this table are included in the stacking, as some of them satisfy one or more of the following conditions: 
\es{eq:selection}{
\begin{cases} 
    |b| \leq 20^\circ \, , \\
    \text{overlaps another halo to within}~2^\circ~\text{of its center} \, ,\\
    \text{TS}_\text{max} > 9\, \text{ and } (\sigma v)_\text{best} > 10 \times (\sigma v)^*_\text{lim} \, .
\end{cases}
}
Note that the overlap criteria is applied sequentially in order of increasing $J$-factor.
These selection criteria have been extensively studied on mock data in our companion paper~\cite{Lisanti:2017qoz} and have been verified to not exclude a potential DM signal, even on data as discussed above. Of the five halos with the largest $J$-factors that are excluded, Andromeda is removed because of its large angular extent, and the rest fail the latitude cut. 

The exclusion of Andromeda is not a result of the criteria in Eq.~\ref{eq:selection}, so some more justification is warranted. As can be seen in Table~\ref{tab:tully_extended}, the angular extent of Andromeda's  scale radius, $\theta_{s}$, is significantly larger than that of any other halo.  To justify $\theta_{s}$ as a proxy for angular extent of the emission, we calculate the 68\% (95\%) containment angle of the expected DM annihilation flux, without accounting for the PSF, and find 1.2$^{\circ}$ (4.4$^{\circ}$). This can be contrasted with the equivalent numbers for the next most important halo, Virgo, where the corresponding  68\% (95\%) containment angles are 0.5$^{\circ}$ (2.0$^{\circ}$). 
Because Andromeda is noticeably more extended beyond the \textit{Fermi} PSF, one must carefully model the spatial distribution of both the smooth DM component and the substructure.  Such a dedicated analysis of Andromeda was recently performed by the \emph{Fermi} collaboration~\cite{Ackermann:2017nya}.  Out of an abundance of caution, we remove Andromeda from the main joint analysis, but we do show how the limits change when Andromeda is included further below.

Figure~\ref{fig:individual_lims} shows the individual limits on the $b\bar{b}$ annihilation cross section for the top ten halos that pass the selection cuts and Fig.~\ref{fig:individual_maxts}  shows the maximum test statistic (TS$_\text{max}$), as a function of $m_\chi$, for these same halos. The green and yellow bands in Fig.~\ref{fig:individual_lims} and~\ref{fig:individual_maxts} represent the 68\% and 95\% containment regions obtained by randomly changing the sky location of each individual halo 200 times (subject to the selection criteria listed above). 
As is evident, the individual limits for the halos  are consistent with expectation under the null hypothesis---\emph{i.e.}, the black line falls within the green/yellow bands for each of these halos.  Some of these groups have been analyzed in previous cluster studies.  For example, the \emph{Fermi} Collaboration provided DM bounds for Virgo~\cite{Ackermann:2015fdi}; our limit is roughly consistent with theirs, and possibly a bit stronger, though an exact comparison is difficult to make due to differences in the data set and DM model assumptions.\footnote{Note that the $J$-factor in Ref.~\cite{Ackermann:2015fdi} is a factor of $4\pi$ too large.}

Figure~\ref{fig:individual_flux} provides the 95\% upper limits on the gamma-ray flux associated with the DM template for each of the top ten halos.  The upper limits are provided for 26 energy bins and compared to the expectations under the null hypothesis.  The upper limits are generally consistent with the expectations under the null hypothesis, though small systematic discrepancies do exist for a few halos, such as NGC3031, at high energies.  This could be due to subtle differences in the sky locations and angular extents between the objects of interest and the set of representative halos used to create the null hypothesis expectations. 

To demonstrate the case of a galaxy group with an excess, we show the TS$_\text{max}$ distribution and the limit for NGC6822 in Fig.~\ref{fig:maxTSoneobject}.  This object fails the selection criteria because it is too close to the Galactic plane. However, it also exhibits a TS$_\text{max}$ excess and, as expected, the limit is weaker than the expectation under the null hypothesis. \vspace{0.1in}

\noindent  {\bf Sky maps.} Fig.~\ref{fig:jfactor_maps} shows a Mollweide projection of all the $J$-factors inferred using the T15 and T17 catalogs,  smoothed at $2^\circ$ with a Gaussian kernel. The map is shown in Galactic coordinates with the Galactic Center at the origin. Looking beyond astrophysical sources, this is how an extragalactic DM signal might show up in the sky. Although this map has no masks added to it, a clear extinction is still visible along the Galactic plane. This originates from the incompleteness of the catalogs along the Galactic plane. 

In Fig.~\ref{fig:individual_skyrois}, we show the counts map in $20^\circ \times 20^\circ$ square regions around each of the top nine halos that pass the selection cuts.  For each map, we show all photons with energies above $\sim$500 MeV, indicate all {\it Fermi} 3FGL point sources with orange stars, and show the extent of $\theta_s$ with a dashed orange circle.  Given a DM signal, we would expect to see emission extend out to $\theta_s$ at the center of these images.

\begin{table*}[htb]
\footnotesize
\begin{tabular}{cccccccccc}
\toprule
\Xhline{3\arrayrulewidth}
Name &   $\log_{10} J$  &  $\log_{10} M_\text{vir}$ &          $z \times 10^{3}$&        $\ell$ &        $b$ &  $\log_{10} c_\text{vir}$  & $\theta_\text{s}$ &   $b_\text{sh}$ & TS$_\text{max}$ \\ 
\midrule
\hline
             Andromeda &  19.8$\pm$0.4 &  12.4$\pm$0.1 &   0.17 &  121.5 & -21.8 &  1.04$\pm$0.17 &     2.6 &  2.6 &   2.9 \\ 
         \begin{tabular}{c} NGC4472 \\ Virgo \end{tabular} &  19.1$\pm$0.4 &  14.6$\pm$0.1 &   3.58 &  283.9 &  74.5 &  0.80$\pm$0.18 &     1.2 &  4.5 &   1.0 \\ 
               NGC5128 &  18.9$\pm$0.4 &  12.9$\pm$0.1 &   0.82 &  307.9 &  17.1 &  0.99$\pm$0.17 &     0.9 &  3.1 &   0.0 \\ 
               NGC0253 &  18.8$\pm$0.4 &  12.7$\pm$0.1 &   0.79 &   98.2 & -87.9 &  1.00$\pm$0.17 &     0.8 &  2.9 &   0.6 \\ 
              Maffei 1 &  18.7$\pm$0.4 &  12.6$\pm$0.1 &   0.78 &  136.2 &  -0.4 &  1.01$\pm$0.17 &     0.7 &  2.8 &   7.3 \\ 
              NGC6822 &  18.6$\pm$0.4 &  10.7$\pm$0.1 &   0.11 &   25.3 & -18.4 &  1.17$\pm$0.17 &     0.8 &  1.7 &  16.7 \\ 
               NGC3031 &  18.6$\pm$0.4 &  12.6$\pm$0.1 &   0.83 &  141.9 &  40.9 &  1.02$\pm$0.17 &     0.7 &  2.8 &   0.0 \\ 
     \begin{tabular}{c} NGC4696 \\ Centaurus \end{tabular} &  18.3$\pm$0.4 &  14.6$\pm$0.1 &   8.44 &  302.2 &  21.7 &  0.80$\pm$0.18 &     0.5 &  4.5 &   6.6 \\ 
               NGC1399 &  18.3$\pm$0.4 &  13.8$\pm$0.1 &   4.11 &  236.6 & -53.9 &  0.89$\pm$0.17 &     0.5 &  3.9 &   0.7 \\ 
                IC0356 &  18.3$\pm$0.4 &  13.5$\pm$0.1 &   3.14 &  138.1 &  12.7 &  0.92$\pm$0.17 &     0.4 &  3.5 &   0.0 \\ 
               NGC4594 &  18.3$\pm$0.4 &  13.3$\pm$0.1 &   2.56 &  299.0 &  51.3 &  0.94$\pm$0.17 &     0.4 &  3.4 &   0.0 \\ 
  Norma &  18.2$\pm$0.3 &  15.1$\pm$0.2 &  17.07 &  325.3 &  -7.2 &  0.74$\pm$0.18 &     0.4 &  5.2 &   1.7 \\ 
               IC 1613 &  18.2$\pm$0.4 &  10.6$\pm$0.1 &   0.17 &  129.7 & -60.6 &  1.18$\pm$0.17 &     0.5 &  1.7 &   0.0 \\ 
      \begin{tabular}{c} NGC1275 \\ Perseus \end{tabular} &  18.1$\pm$0.3 &  15.0$\pm$0.2 &  17.62 &  150.6 & -13.3 &  0.75$\pm$0.18 &     0.4 &  5.2 &   0.0 \\ 
               NGC4736 &  18.1$\pm$0.4 &  12.2$\pm$0.1 &   1.00 &  124.8 &  75.8 &  1.05$\pm$0.17 &     0.4 &  2.6 &   0.9 \\ 
               NGC3627 &  18.1$\pm$0.4 &  13.0$\pm$0.1 &   2.20 &  241.5 &  64.4 &  0.98$\pm$0.17 &     0.4 &  3.2 &  27.2 \\ 
        \begin{tabular}{c} NGC1316 \\ Fornax \end{tabular} &  18.0$\pm$0.4 &  13.5$\pm$0.1 &   4.17 &  240.0 & -56.7 &  0.92$\pm$0.17 &     0.3 &  3.5 &   2.3 \\ 
               NGC5236 &  18.0$\pm$0.4 &  12.2$\pm$0.1 &   1.09 &  314.6 &  32.0 &  1.05$\pm$0.17 &     0.3 &  2.6 &  22.1 \\ 
                IC0342 &  18.0$\pm$0.4 &  11.8$\pm$0.1 &   0.73 &  138.5 &  10.7 &  1.09$\pm$0.17 &     0.3 &  2.3 &   1.9 \\ 
 Coma &  18.0$\pm$0.3 &  15.2$\pm$0.2 &  24.45 &   57.2 &  87.9 &  0.73$\pm$0.18 &     0.3 &  5.2 &  41.2 \\ 
               NGC4565 &  18.0$\pm$0.4 &  13.1$\pm$0.1 &   2.98 &  229.9 &  86.1 &  0.96$\pm$0.17 &     0.3 &  3.3 &   2.4 \\ 
         \begin{tabular}{c} NGC3311 \\ Hydra \end{tabular} &  18.0$\pm$0.3 &  14.4$\pm$0.1 &  10.87 &  269.6 &  26.4 &  0.82$\pm$0.17 &     0.3 &  4.3 &   0.1 \\ 
        \begin{tabular}{c} NGC1553 \\ Dorado \end{tabular} &  17.9$\pm$0.4 &  13.4$\pm$0.1 &   4.02 &  265.6 & -43.5 &  0.94$\pm$0.17 &     0.3 &  3.4 &   0.0 \\ 
               NGC3379 &  17.9$\pm$0.4 &  12.9$\pm$0.1 &   2.42 &  233.6 &  57.8 &  0.99$\pm$0.17 &     0.3 &  3.1 &   0.0 \\ 
               NGC5194 &  17.9$\pm$0.4 &  12.6$\pm$0.1 &   1.84 &  104.9 &  68.5 &  1.01$\pm$0.17 &     0.3 &  2.8 &   4.9 \\ 
            ESO097-013 &  17.9$\pm$0.4 &  11.6$\pm$0.1 &   0.60 &  311.3 &  -3.8 &  1.11$\pm$0.17 &     0.3 &  2.1 &  13.5 \\ 
               NGC4258 &  17.9$\pm$0.4 &  12.5$\pm$0.1 &   1.64 &  139.0 &  68.9 &  1.03$\pm$0.17 &     0.3 &  2.7 &   0.5 \\ 
               NGC1068 &  17.9$\pm$0.4 &  13.3$\pm$0.1 &   3.60 &  172.0 & -51.9 &  0.95$\pm$0.17 &     0.3 &  3.3 &   7.0 \\ 
               NGC4261 &  17.9$\pm$0.4 &  13.9$\pm$0.1 &   7.16 &  281.9 &  67.5 &  0.88$\pm$0.17 &     0.3 &  4.0 &  12.6 \\ 
               NGC4826 &  17.9$\pm$0.4 &  12.1$\pm$0.1 &   1.16 &  315.7 &  84.4 &  1.06$\pm$0.17 &     0.3 &  2.5 &   3.3 \\ 

         \bottomrule
\Xhline{3\arrayrulewidth}
\end{tabular}
\caption{The top thirty halos included from the T15~\cite{Tully:2015opa} and T17~\cite{2017ApJ...843...16K} catalogs, as ranked by inferred $J$-factor, which includes the boost factor.  For each group, we show the brightest central galaxy and the common name, if one exists, as well as the virial mass, cosmological redshift, Galactic coordinates, inferred concentration using Ref.~\cite{Correa:2015dva}, angular extension, boost factor using the fiducial model from Ref.~\cite{Bartels:2015uba}, and the maximum test statistic (TS$_\text{max}$) over all $m_\chi$ between the model with and without DM annihilating to $b \bar b$. 
A complete listing of all the halos used in this study is provided as Supplementary Data.
}
\label{tab:tully_extended}
\end{table*}

\afterpage{
\begin{figure}[p]
   \centering
   \includegraphics[width=0.95\textwidth]{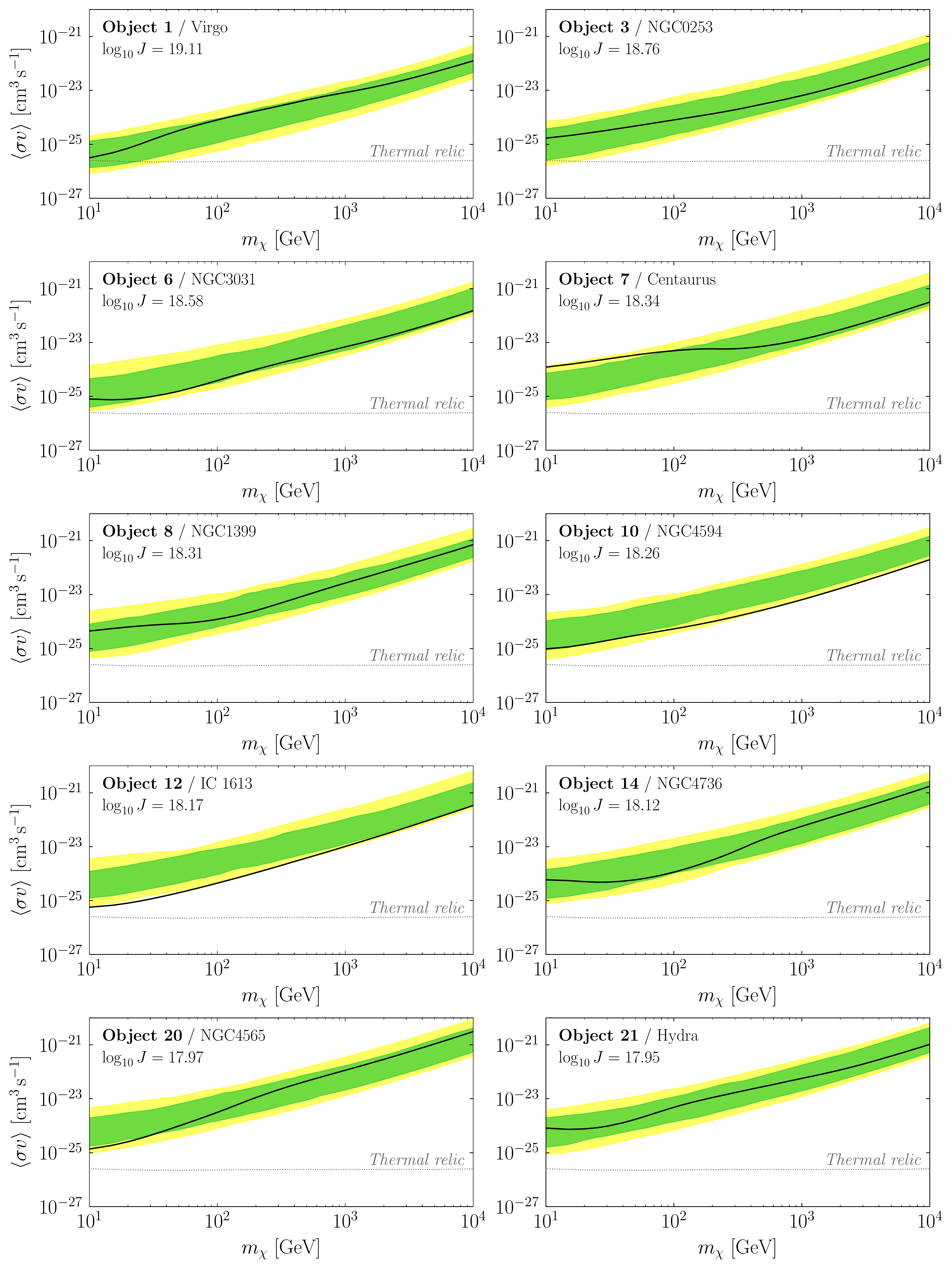}
   \caption{The 95\% confidence limit on the DM annihilation cross section to the $b \bar b$ final state for each of the top ten halos listed in Tab.~\ref{tab:tully_extended} that pass the selection cuts. For each halo, we show the 68\% and 95\% containment regions (green and yellow, respectively), which are obtained by placing the halo at 200 random sky locations.  The inferred ${J}$-factors, assuming the fiducial boost factor model~\cite{Bartels:2015uba}, are provided for each object.}
   \label{fig:individual_lims}
\end{figure}
\clearpage}

\afterpage{
\begin{figure}[htbp]
  \centering
   \includegraphics[width=0.95\textwidth]{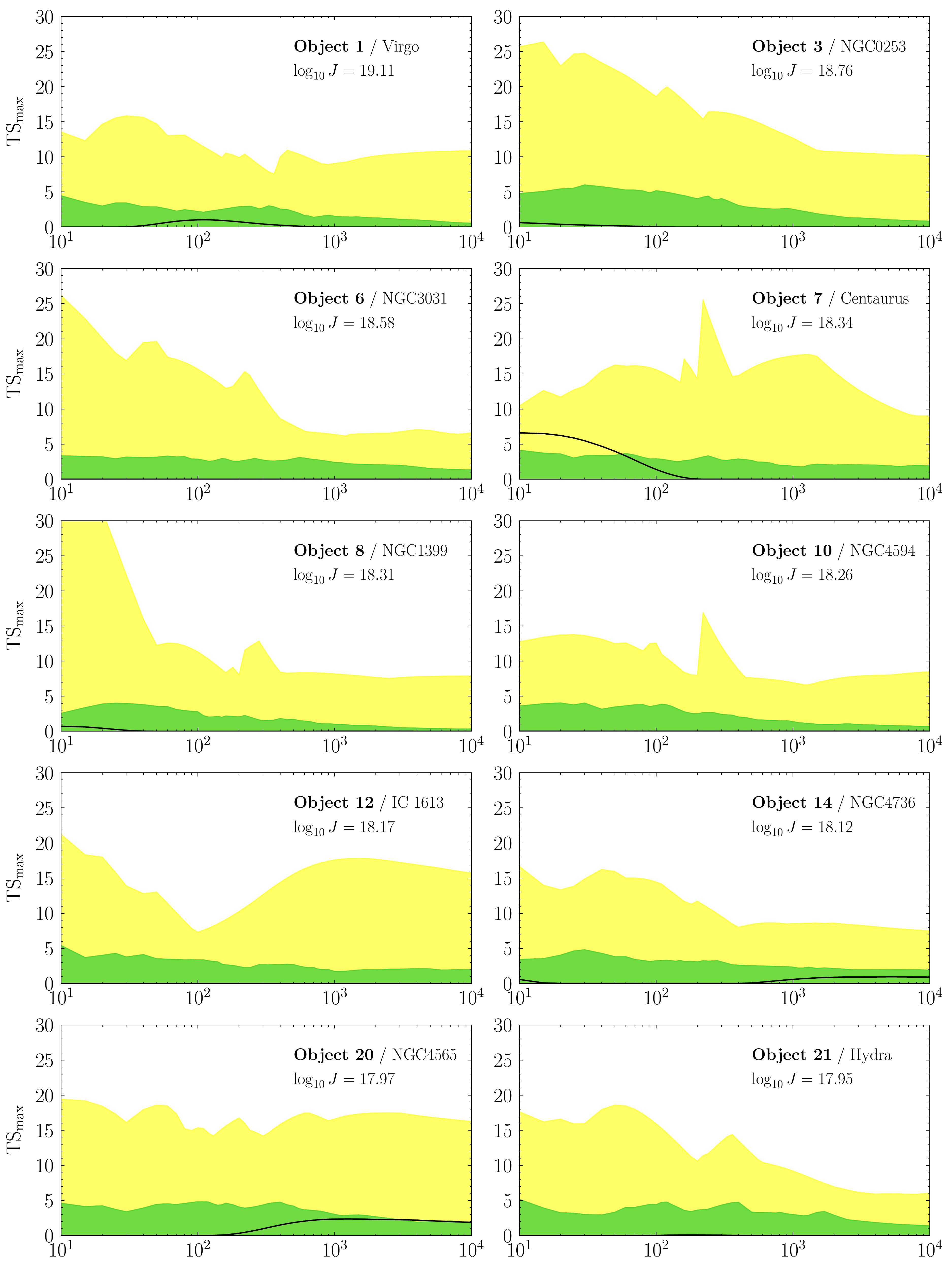}
  \caption{Same as Fig.~\ref{fig:individual_lims}, except showing the maximum test statistic (TS$_\text{max}$) for each individual halo, as a function of DM mass. These results correspond to the $b \bar b$ annihilation channel.}
   \label{fig:individual_maxts}
\end{figure}
\clearpage}

\afterpage{
\begin{figure}[htbp]
  \centering
   \includegraphics[width=0.95\textwidth]{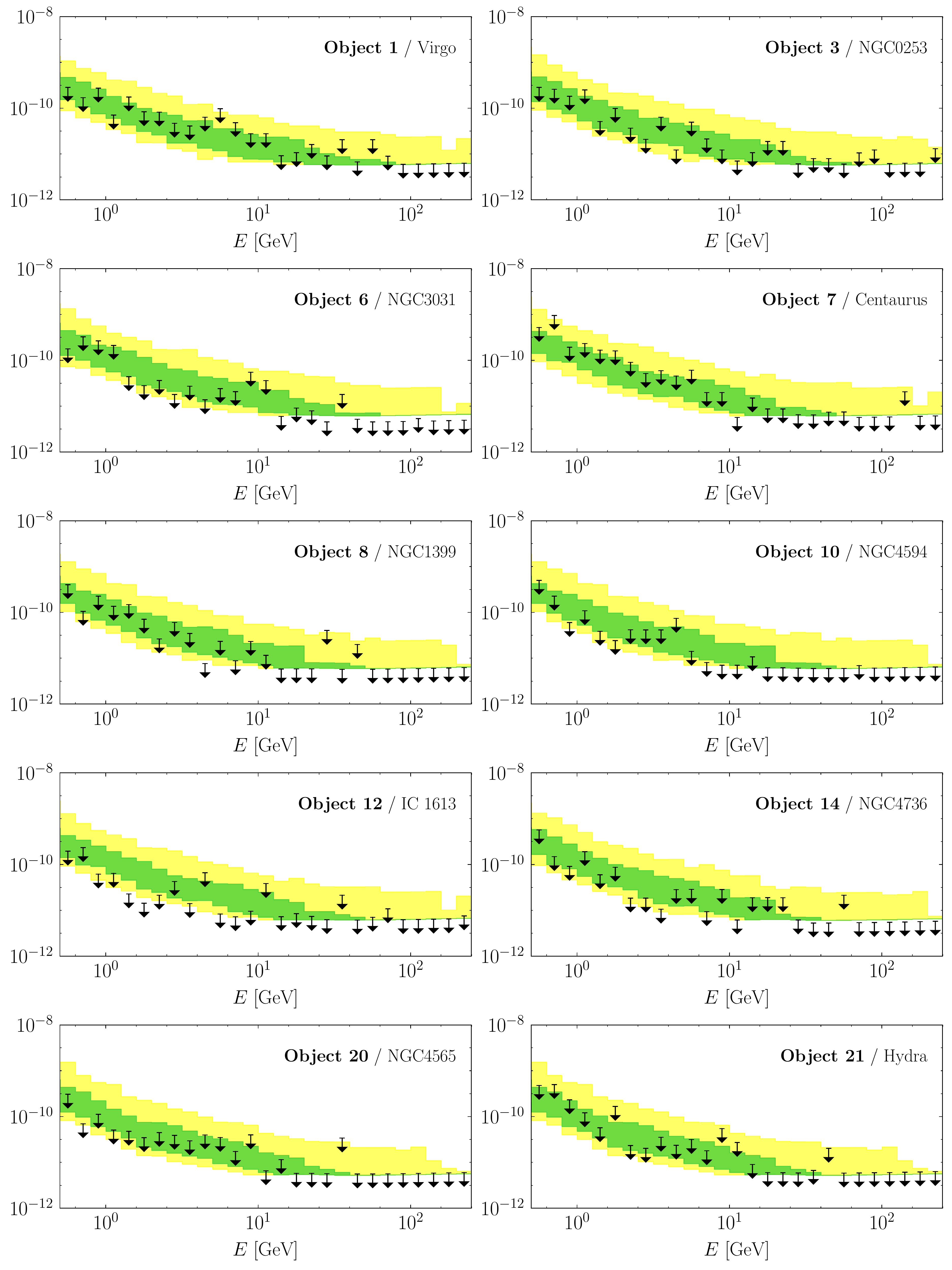}
  \caption{Same as Fig.~\ref{fig:individual_lims}, except showing the 95\% upper limit on the gamma-ray flux correlated with the DM annihilation profile in each halo.  We use 26 logarithmically spaced energy bins between 502~MeV and 251~GeV. 
  }
   \label{fig:individual_flux}
\end{figure}
\clearpage}

\afterpage{
\begin{figure}[htbp]
   \centering
   \includegraphics[width=0.95\textwidth]{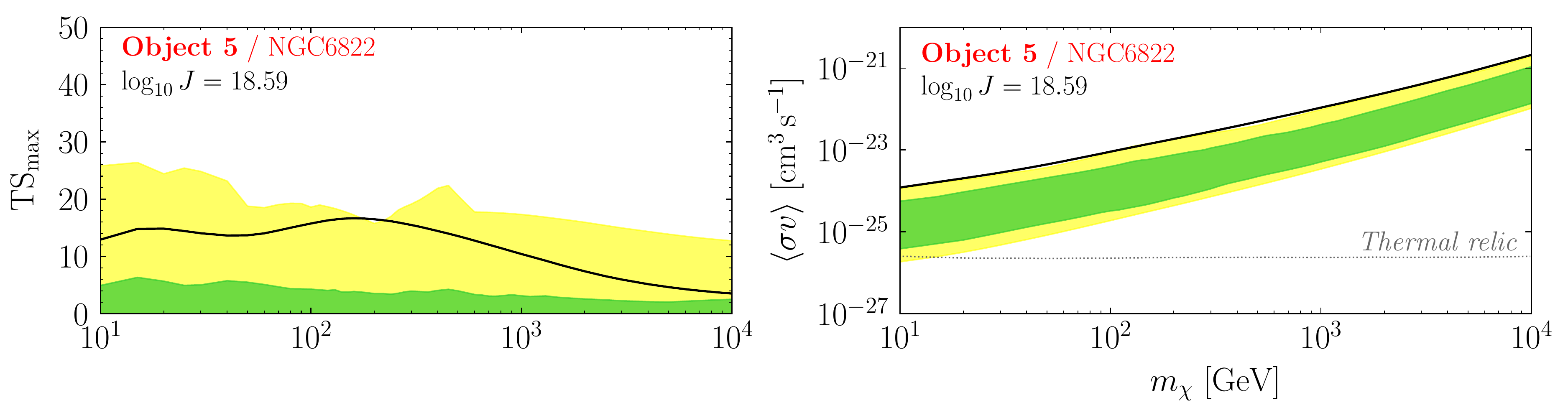}
   \caption{NGC6822 has one of the largest $J$-factors of the objects in the catalog, but it fails the selection requirements because of its proximity to the Galactic plane.  We show the analog of Fig.~\ref{fig:individual_maxts} (left) and Fig.~\ref{fig:individual_lims} (right). We see that this object  has a broad TS$_\text{max}$ excess over many masses and a weaker limit than expected from random sky locations.}
   \label{fig:maxTSoneobject}
\end{figure}

\begin{figure}[htbp]
  \centering
   \includegraphics[width=0.95\textwidth]{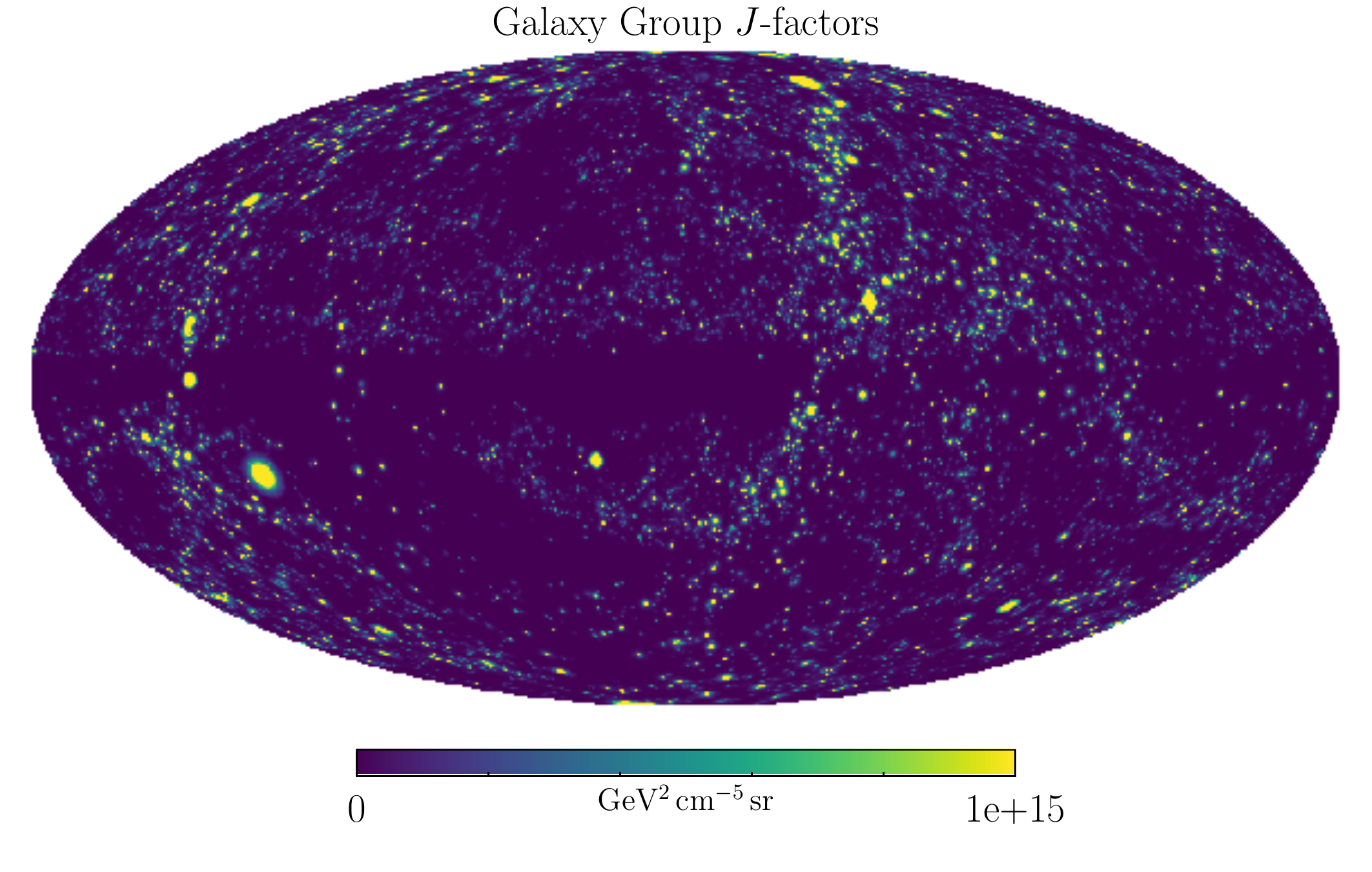}
  \caption{Mollweide projection of all the $J$-factors inferred using the T15 and T17 catalogs, smoothed at $2^\circ$ with a Gaussian kernel. If we could see beyond conventional astrophysics to an extragalactic DM signal, this is how it would appear on the sky.}
   \label{fig:jfactor_maps}
\end{figure}
\clearpage}

\afterpage{
\begin{figure}[htbp]
   \centering
   \includegraphics[width=0.95\textwidth]{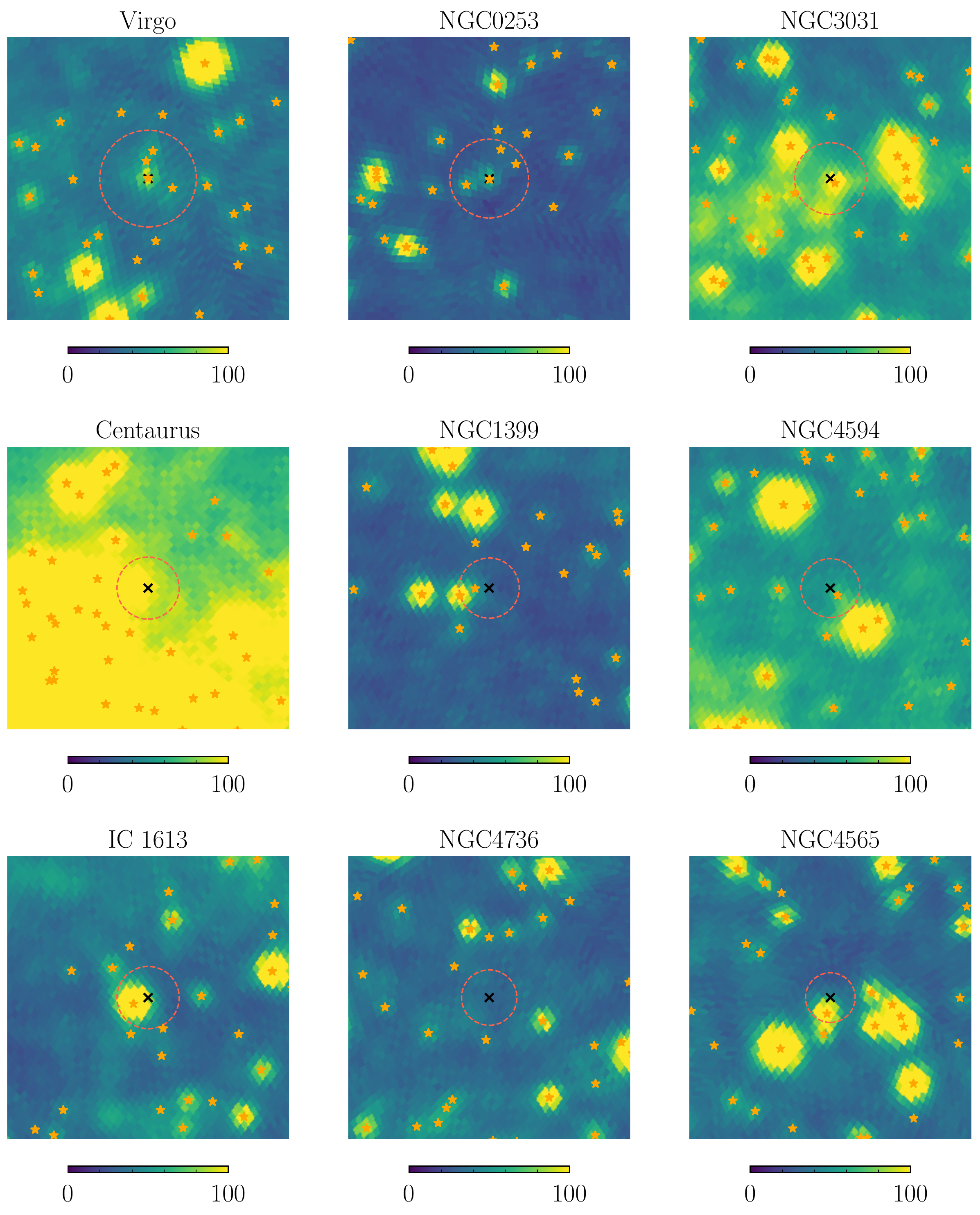}
   \caption{The \textit{Fermi}-LAT data centered on the top nine  halos that are included in the stacked sample. We show the photon counts (for the energies analyzed) within a $20^\circ$$\times$$20^\circ$ square centered on the region of interest. The dotted circle shows the scale radius $\theta_\mathrm{s}$, which is a proxy for the scale of DM annihilation, and the orange stars indicate the \emph{Fermi} 3FGL point sources.}
   \label{fig:individual_skyrois}
\end{figure}
\clearpage}

\newpage

\section{Variations on the Analysis}
\label{sec:systematics}

We have performed a variety of systematic tests to understand the robustness of the results presented in chapter~\ref{chap:fermigg}.  Several of these uncertainties are discussed in detail in our companion paper~\cite{Lisanti:2017qoz}; here, we focus specifically on how they affect the results of the data analysis.  \vspace{0.1in}

\begin{figure}[t!]
   \centering
	\includegraphics[width=.49\textwidth]{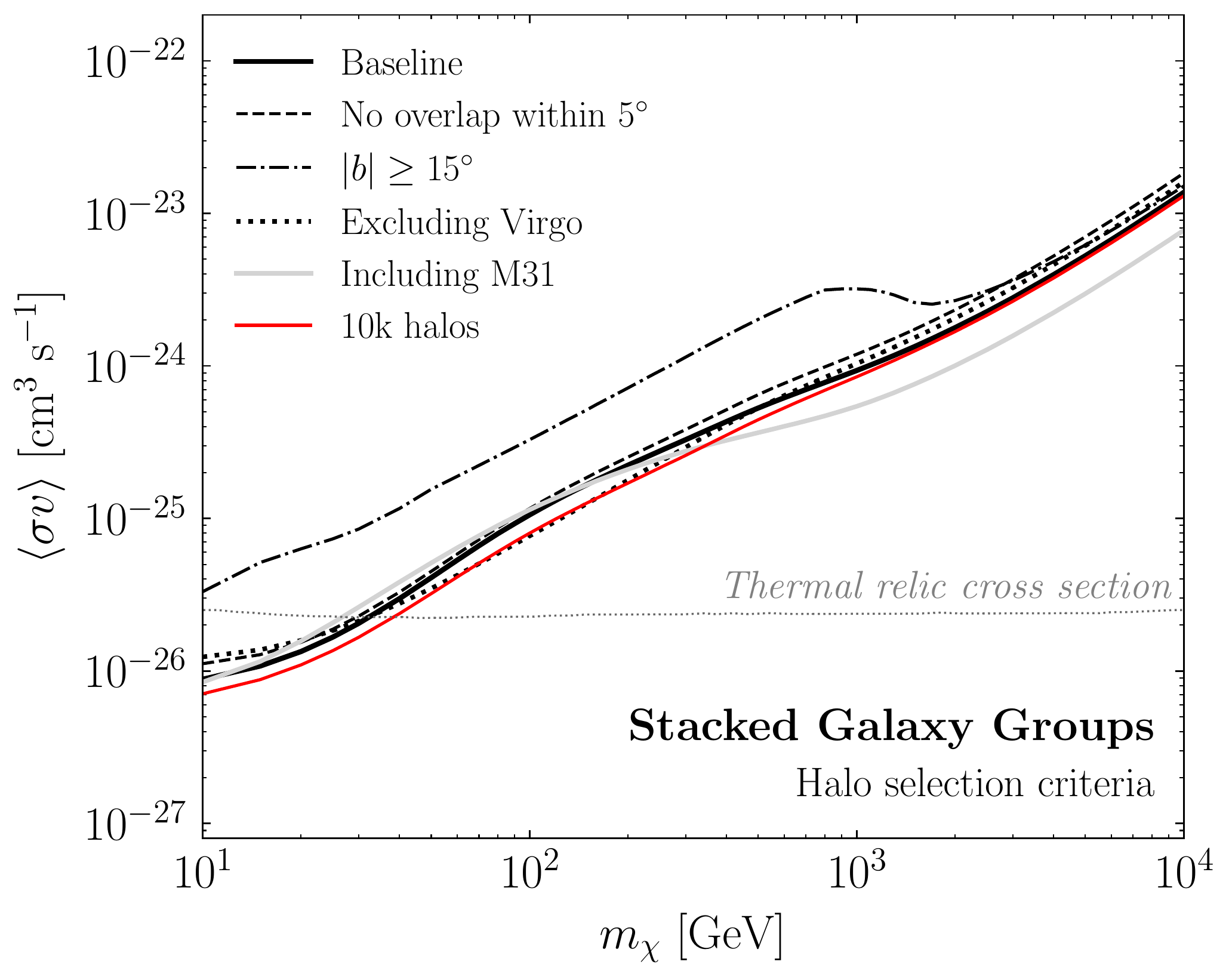} 
 	\includegraphics[width=.49\textwidth]{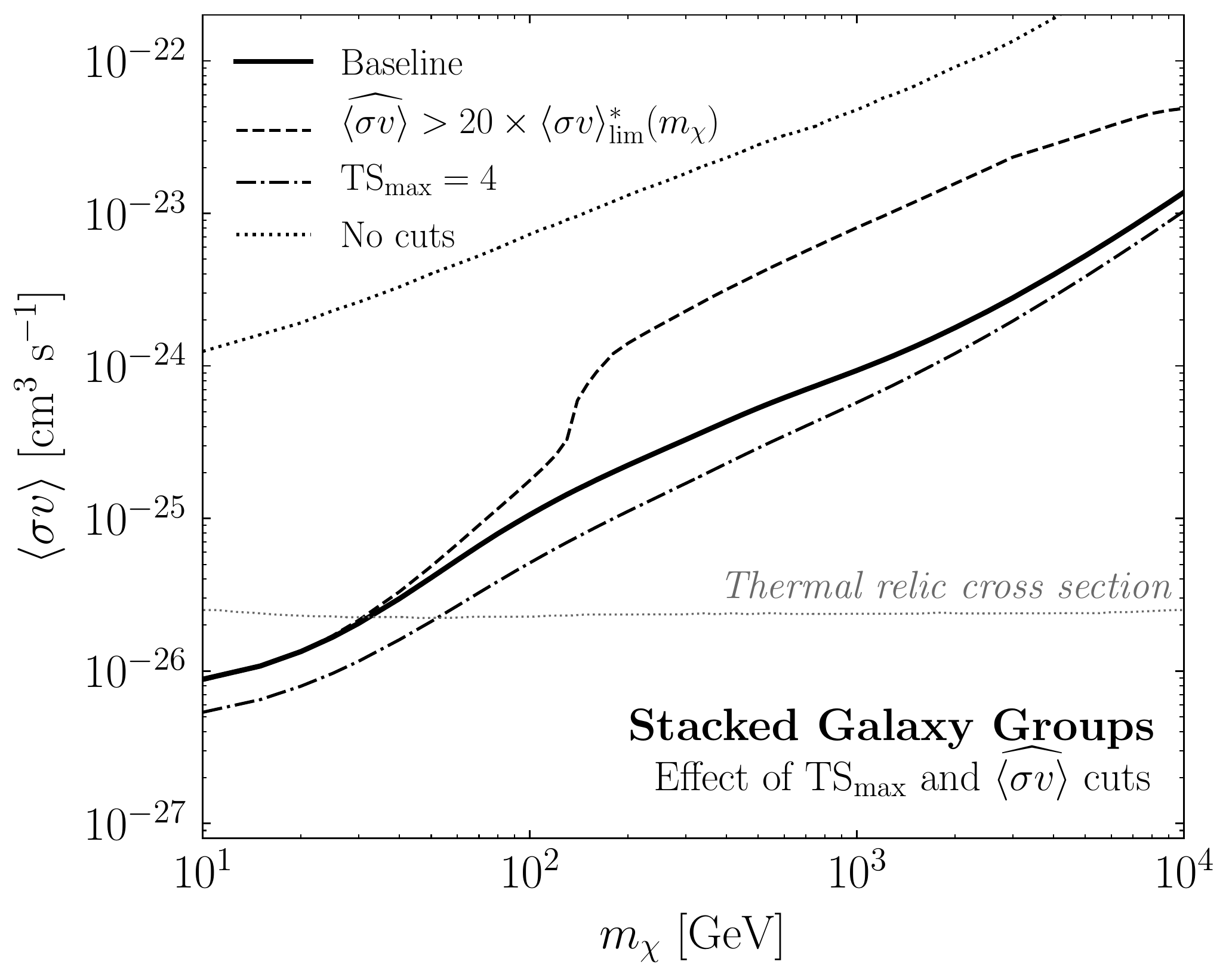} 
   \caption{The same as the baseline analysis shown in the left panel of Fig.~\ref{fig:bounds} of chapter~\ref{chap:fermigg}, except varying several assumptions made in the analysis.  (Left) We show the effect of relaxing the overlapping halo criterion to $5^\circ$ (dashed), reducing the latitude cut to $|b|\geq 15^\circ$ (dot-dashed), excluding Virgo (dotted), and including Andromeda (gray).  The limit obtained when starting from an initial 10,000 halos is shown as the red line.  (Right) We show the effect of strengthening the cross section (dashed) or weakening the TS$_\text{max}$ (dot-dashed) selection criteria, as well as completely removing the TS$_\text{max}$ and cross section cuts (dotted). }
   \label{fig:cutsandhalos}
\end{figure}

\noindent  {\bf Halo Selection Criteria.}  
Here, we demonstrate how variations on the halo selection conditions listed above affect the baseline results of Fig.~\ref{fig:bounds}.  In the left panel of Fig.~\ref{fig:cutsandhalos}, the red line shows the limit that is obtained when starting with 10,000 halos instead of 1000, but requiring the same selection conditions.  Despite the modest improvement in the limit, we choose to use 1000 halos in the baseline study because systematically testing the robustness of the analysis procedure, as done in Ref.~\cite{Lisanti:2017qoz}, becomes computationally prohibitive otherwise. In order to calibrate the analysis for higher halo numbers, it would be useful to use semi-analytic methods to project the sensitivity, such as those discussed in Ref.~\cite{Cowan:2010js,Edwards:2017mnf}, although we leave the details to future work.

Virgo is the object with the highest $J$-factor in the stacked sample. As made clear in the dedicated study of this object by the \emph{Fermi} Collaboration~\cite{Ackermann:2015fdi}, there are challenges associated with modeling the diffuse emission in Virgo's vicinity.  However, we emphasize that the baseline limit is not highly sensitive to any one halo, including the brightest in the sample.  For example, the dotted line in the left panel of Fig.~\ref{fig:cutsandhalos} shows the impact on the limit after removing Virgo from the stacking. Critically, we see that the limit is almost unchanged, highlighting that the stacked result is not solely driven by the object with the largest $J$-factor.

The effect of including Andromeda (M31) is shown as the gray solid line. We exclude Andromeda from the baseline analysis because of its large angular size, as discussed in detail above. Our analysis relies on the assumption that the DM halos are approximately point-like on the sky, which fails for Andromeda, and we therefore deem it to fall outside the scope of the systematic studies performed here.

\begin{figure}[tb]
   \centering
   \includegraphics[width=.5\textwidth]{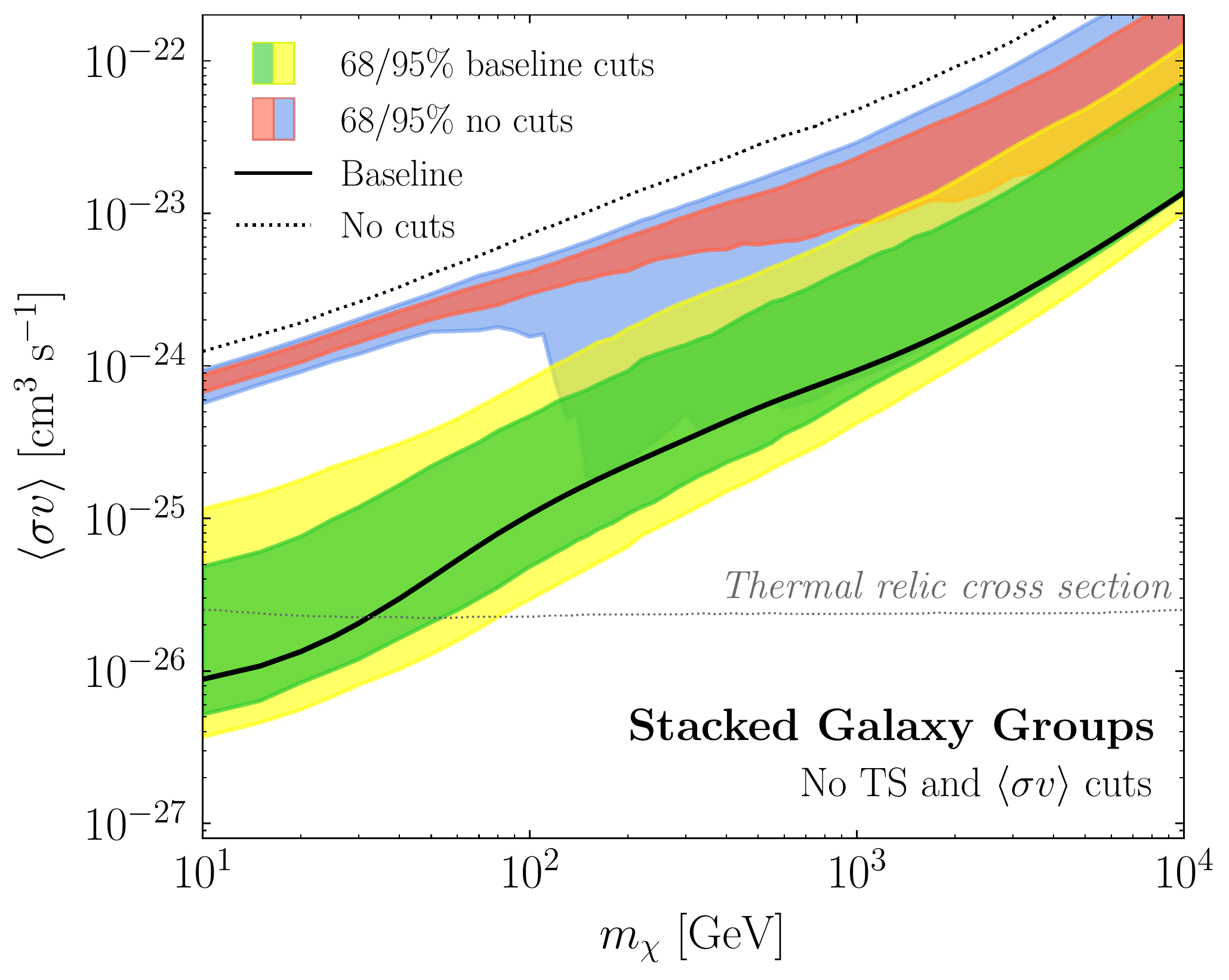}
   \caption{The results of the baseline analysis with the default cuts, as shown in the left of Fig.~\ref{fig:bounds}, compared to the corresponding result when no cuts are placed on the TS$_\text{max}$ or cross section of the halos in the catalog.  The significant offset between the limit obtained with no cuts (dotted line) and the corresponding expectation from random sky locations (red/blue band)  demonstrates that many of the objects that are removed by the TS$_\text{max}$ and cross section cuts are  legitimately associated with astrophysical emission. See text for details.}
   \label{fig:systematics_nots_cuts}
\end{figure}

\begin{figure}[t]
   \centering
   \includegraphics[width=.49\textwidth]{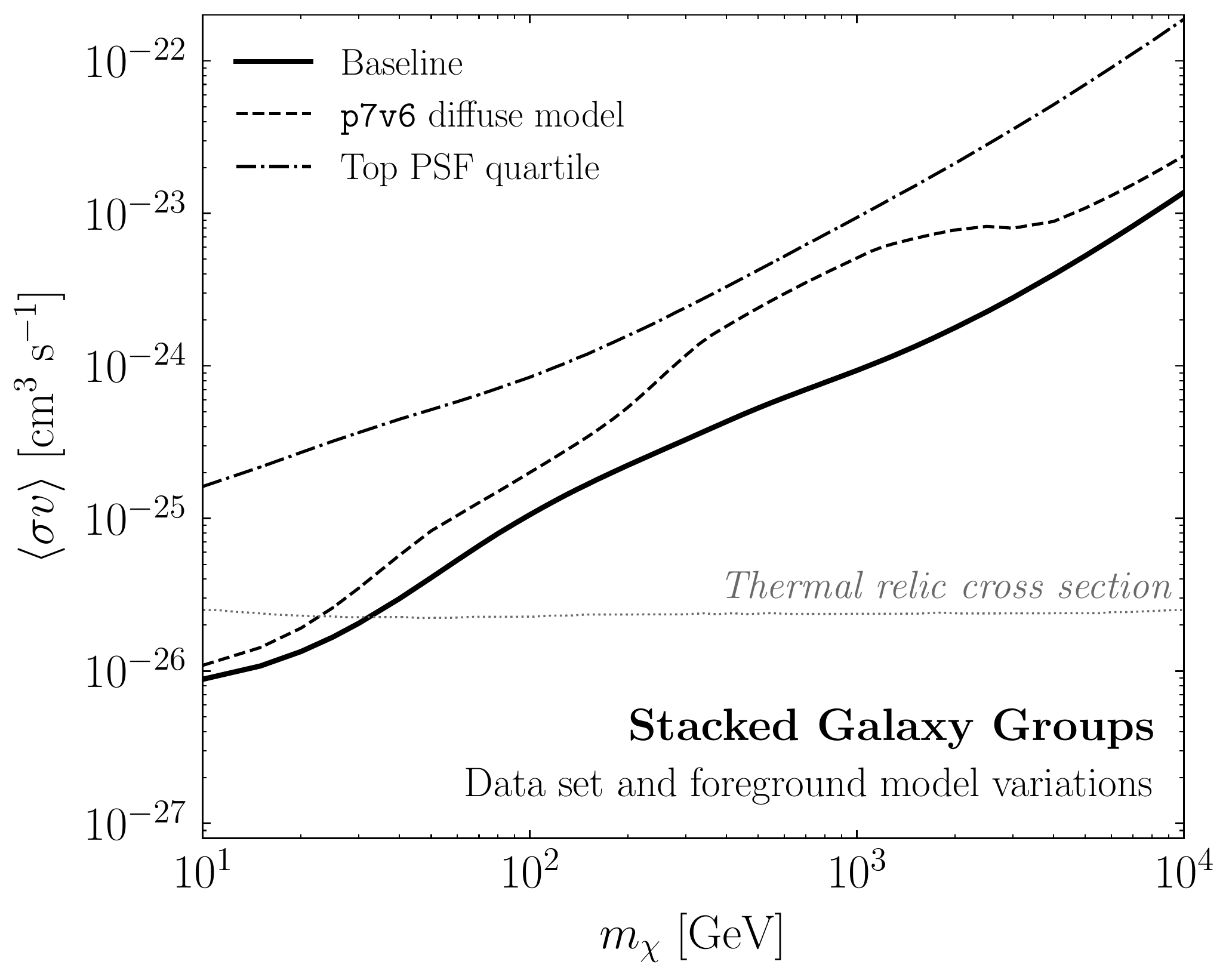}
   \includegraphics[width=.49\textwidth]{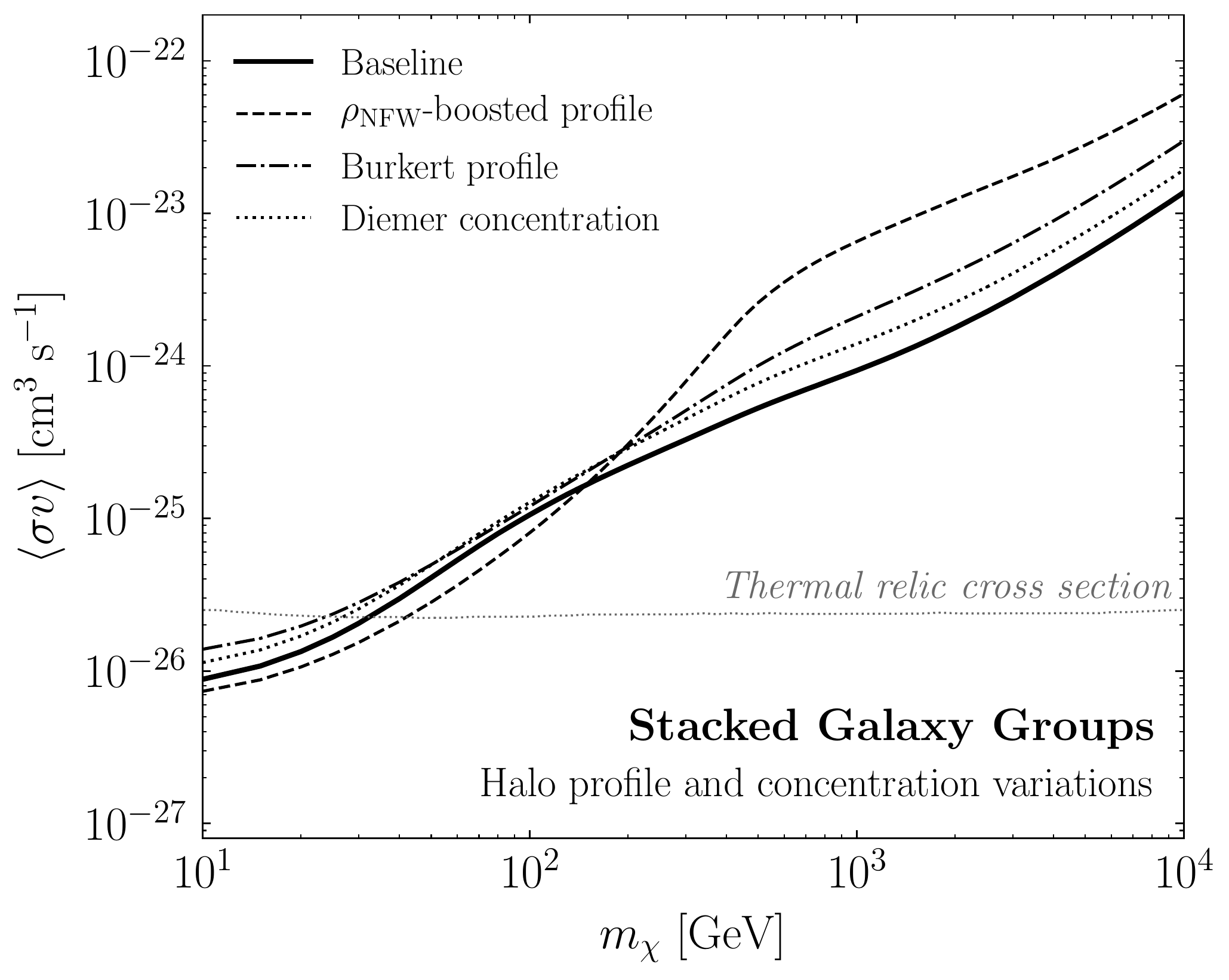}
   \caption{The same as the baseline analysis shown in the left panel of Fig.~\ref{fig:bounds} of chapter~\ref{chap:fermigg}, except varying several assumptions made in the analysis.  (Left) We show the effect of using the top PSF quartile of the UltracleanVeto data set (dot-dashed) and the \texttt{p7v6} diffuse model (dashed).  (Right) We show the effect of using the cored Burkert profile~\cite{Burkert:1995yz} (dot-dashed) and the Diemer~and~Kravtsov concentration model~\cite{Diemer:2014gba} (dotted).  The ``$\rho_\text{NFW}$-boosted profile'' (dashed) shows what happens when the annihilation flux from the subhalo boost is assumed to follow the NFW profile (as opposed to a squared-NFW profile). }
   \label{fig:systematics_data_profile}
\end{figure}

The dashed line shows the effect of tightening the condition on overlapping halos from $2^\circ$ to $5^\circ$. Predictably, the limit is slightly weakened due to the smaller pool of available targets.  We also show the effect of decreasing the latitude cut to $b\geq 15^\circ$ (dot-dashed line). In this case, the number of halos included in the stacked analysis increases, but the limit is weaker---considerably so below $m_\chi \sim 10^3$~GeV.  The weakened limits are likely due to enhanced diffuse emission along the plane as well as contributions from unresolved point sources, both of which are difficult to accurately model. In cases with such mismodeling, the addition of a DM template can generically improve the quality of the fit, which leads to excesses at low energies, in particular.  The baseline latitude cut ameliorates  precisely these concerns.

The right  panel of Fig.~\ref{fig:cutsandhalos} illustrates the effects of changing, or removing completely, the cross section and TS$_\text{max}$ cuts on the halos.  Specifically, the dashed black line shows what happens when we require that a halo's excess be even more inconsistent with the limits set by other galaxy groups; specifically, requiring that $(\sigma v)_\text{best} > 20 \times (\sigma v)^*_\text{lim}$. The dot-dashed line shows the limit when we decrease the statistical significance requirement to $\text{TS}_\text{max} > 4$.  
Note that the two changes have opposite effects on the limits.  This is expected because more halos with excesses are included in the stacking procedure with the more stringent cross section requirement, which weakens the limit, whereas fewer are included if we reduce the TS$_\text{max}$ cut, strengthening the limit.  

The dotted line in the right panel of Fig.~\ref{fig:cutsandhalos} shows  what happens when no requirement at all is placed on the TS$_\text{max}$ and cross section; in this case, the limit is dramatically weakened by several orders of magnitude.   We show the same result in Fig.~\ref{fig:systematics_nots_cuts} (dotted line), but with a comparison to the null hypothesis corresponding to no TS$_\text{max}$ and cross section cuts, which is shown as the 68\%~(95\%) red~(blue) bands.\footnote{We thank A.~Drlica-Wagner for suggesting this test.}    In the baseline case, the limit is consistent with the random sky locations---\emph{i.e.}, the solid black line falls within the green/yellow bands.  However, with no TS$_\text{max}$ and cross section cuts, this is no longer true---\emph{i.e.}, the dotted black line falls outside the red/blue bands.  Clear excesses are observed above the background expectation in this case, but they are inconsistent with a DM interpretation as they are strongly excluded by other halos in the stack.  When deciding on the TS$_\text{max}$ and cross section requirements that we used for the baseline analysis in Fig.~\ref{fig:bounds}, our goal was to maximize the sensitivity reach while simultaneously ensuring that an actual DM signal would not be excluded.  We verified the selection criteria thoroughly by performing injected signal tests on the data (discussed above) as well as on mock data (discussed in Ref.~\cite{Lisanti:2017qoz}).  Ideally, galaxy groups would be excluded from the stacking based on the specific properties of the astrophysical excesses that they exhibit, as opposed to the TS$_\text{max}$ and cross section requirements used here.  For example, one can imagine excluding groups that are known to host  AGN or galaxies with high amounts of star-formation activity.  We plan to study such possibilities in future work.        \vspace{0.1in}

\noindent  {\bf Data Set and Foreground Models.}  
In the results presented thus far, we have used all quartiles of the UltracleanVeto event class of the {\it Fermi} data.  Alternatively, we can restrict ourselves to the top quartile of events, as ranked by PSF.  Using this subset of data has the advantage of improved angular resolution, but the disadvantage of a $\sim$75\% reduction in statistics.  The  left panel of Fig.~\ref{fig:systematics_data_profile} shows the limit (dot-dashed line) obtained by repeating the analysis with the top quartile of UltracleanVeto data; the bounds are weaker than in the all-quartile case, as would be expected.  However, the amount by which the limit weakens is not completely consistent with the decrease in statistics.  Rather, it appears that when we lower the photon statistics, more halos that were previously excluded by the cross section and TS$_\text{max}$ criteria in the baseline analysis are allowed into the stacking and collectively weaken the limit.

Another choice that we made for the baseline analysis was to use the \texttt{p8r2} foreground model for gamma-ray emission from cosmic-ray processes in the Milky Way.   In this model, the bremsstrahlung and boosted pion emission are traced with gas column-density maps and the IC emission is modeled using \texttt{Galprop}~\cite{Strong:2007nh}.  After fitting the data with these three components, any `extended emission excesses' are identified and added back into the foreground model~\cite{Acero:2016qlg}.  To study the dependence of the results on the choice of foreground model, we repeat the analysis using the Pass~7 \emph{gal\_2yearp7v6\_v0.fits} (\texttt{p7v6}) model, which includes large-scale structures like Loop~1 and the \emph{Fermi} bubbles---in addition to the bremsstrahlung, pion, and IC emission---but does not account for any data-driven excesses as is done in \texttt{p8r2}.  The results of the stacked analysis using the \texttt{p7v6} model are shown in the left panel of Fig.~\ref{fig:systematics_data_profile} (dashed line).  The limit is somewhat weaker to that obtained using \texttt{p8r2}, though it is broadly similar to the latter.  This is to be expected for stacked analyses, where the dependence on mismodeling of the foreground emission is reduced because the fits are done on small, independent regions of the sky, so that offsets in the point-to-point normalizations of the diffuse model can have less impact. For more discussion of this point, see Ref.~\cite{Daylan:2014rsa,Linden:2016rcf,Narayanan:2016nzy,Cohen:2016uyg}.\vspace{0.1in}

\noindent  {\bf Halo Density Profile and Concentration.} Our baseline analysis makes two assumptions about the profiles of gamma-ray emission from the extragalactic halos.  The first assumption is that the DM profile of the smooth halo is described by an NFW profile:
\begin{equation}
\rho_{\rm NFW}(r) = \frac{\rho_s}{r/r_s\,(1+r/r_s)^2}\,,
\end{equation}
where $\rho_s$ is the normalization and $r_s$ the scale radius~\cite{Navarro:1996gj}.  The NFW profile successfully describes the shape of cluster-size DM halos in $N$-body simulations with and without baryons (see, {\it e.g.}, Ref.~\cite{Springel:2008cc,Schaller:2014uwa}).  However, some evidence exists pointing to cored density profiles on smaller scales ({\it e.g.}, dwarf galaxies), and the density profiles in these systems may be better described by the phenomenological Burkert profile~\cite{Burkert:1995yz}:
\begin{equation}
\rho_{\rm Burkert}(r) = \frac{\rho_B}{(1+r/r_B)(1+(r/r_B)^2)}\,,
\end{equation}
where $\rho_B$ and $r_B$ are the Burkert corollaries to the NFW $\rho_s$ and $r_s$, but have numerically different values. While it appears unlikely that the Burkert profile is a good description of the DM profiles of the cluster-scale halos considered here, using this profile provides a useful systematic variation because it predicts less annihilation flux than the NFW profile does.  The right panel of Fig.~\ref{fig:systematics_data_profile} shows the effect of using the Burkert profile to describe the halos in the T15 and T17 catalogs (dot-dashed line); the limit is slightly weaker, as expected.

The second assumption we made is that the shape of the gamma-ray emission from DM annihilation follows the projected integral of the DM-distribution squared.  This is likely incorrect because the contribution from the boost factor, which can be substantial, should have the spatial morphology of the distribution of DM subhalos.  Neglecting tidal effects, we expect the subhalos to follow the DM distribution (instead of the squared distribution).  Including tidal effects is complicated, as subhalos closer to the halo center are more likely to be tidally stripped, which both increases their concentration and decreases their number density.  We do not attempt to model the change in the spatial morphology of the subhalo distribution from tidal stripping and instead consider the limit where the annihilation flux from the subhalo boost follows the NFW distribution.  This gives a much wider angular profile for the annihilation flux for large clusters,  compared to the case where the boost is simply a multiplicative factor.  The dashed line in the right panel of Fig.~\ref{fig:systematics_data_profile} shows the effect on the limit of modeling the gamma-ray emission in this way (labeled ``$\rho_\text{NFW}$-boosted profile'').  The extended spatial profile leads to a minimal change in the limit over most of the mass range, which is to be expected given that most of the galaxy groups can be well-approximated as point sources.

A halo's virial concentration is an indicator of its overall density and is defined as $c_\text{vir} \equiv r_\text{vir}/r_s$, where $r_\text{vir}$ is the virial radius and $r_s$ the NFW scale radius of the halo.  A variety of models exist in the literature that map from halo mass to concentration.  Our fiducial case is the Correa \emph{et al.} model from Ref.~\cite{Correa:2015dva}.  Here we show how the limit (dotted line) changes when we use the model of Diemer and Kravtsov~\cite{Diemer:2014gba}, updated with the Planck 2015 cosmology~\cite{Ade:2015xua}.  The change to the limit is minimal, which is perhaps a reflection of the fact that the change in the mean concentrations between the concentration-mass models is small compared to the statistical spread predicted in these models, which is incorporated into the $J$-factor uncertainties.  We have also verified that increasing the dispersion on the concentration for the Correa~\emph{et al.} model to 0.24~\cite{Bullock:1999he}, which is above the 0.14--0.19 range used in the baseline study, worsens the limit by a $\mathcal{O}(1)$ factor.\vspace{0.1in}

\noindent  {\bf Substructure Boost.}  
Hierarchical structure formation implies that larger structures can host smaller substructures, the presence of which can significantly enhance signatures of DM annihilation in host halos. Although several models exist in the literature to characterize this effect, the precise enhancement sensitively depends on the methods used as well as the astrophysical and particle physics properties that are assumed.  Phenomenological extrapolation of subhalo properties (\emph{e.g.}, the concentration-mass relation) over many orders of magnitude down to very small masses $\mathcal O(10^{-6}$)~M$_{\odot}$ lead to large enhancements of $\mathcal O(10^{2})$ and $\mathcal O(10^{3})$ for galaxy- and cluster-sized halos, respectively~\cite{Gao:2011rf}. Recent numerical simulations and analytic studies~\cite{Anderhalden:2013wd,Correa:2015dva,Ludlow:2013vxa} suggest that the concentration-mass relation flattens at smaller masses, yielding boosts that are much more modest, about an order-of-magnitude below phenomenological extrapolations~\cite{Nezri:2012tu,Sanchez-Conde:2013yxa}.  In addition, the concentration-mass relation for field halos cannot simply be applied to subhalos, because the latter undergo tidal stripping as they fall into and orbit their host.  Such effects tend to make the subhalos more concentrated---and therefore more luminous---than their field-halo counterparts, though the number-density of such subhalos is also reduced~\cite{Bartels:2015uba}.    

\begin{figure}[t]
   \centering
   \includegraphics[width=0.49\textwidth]{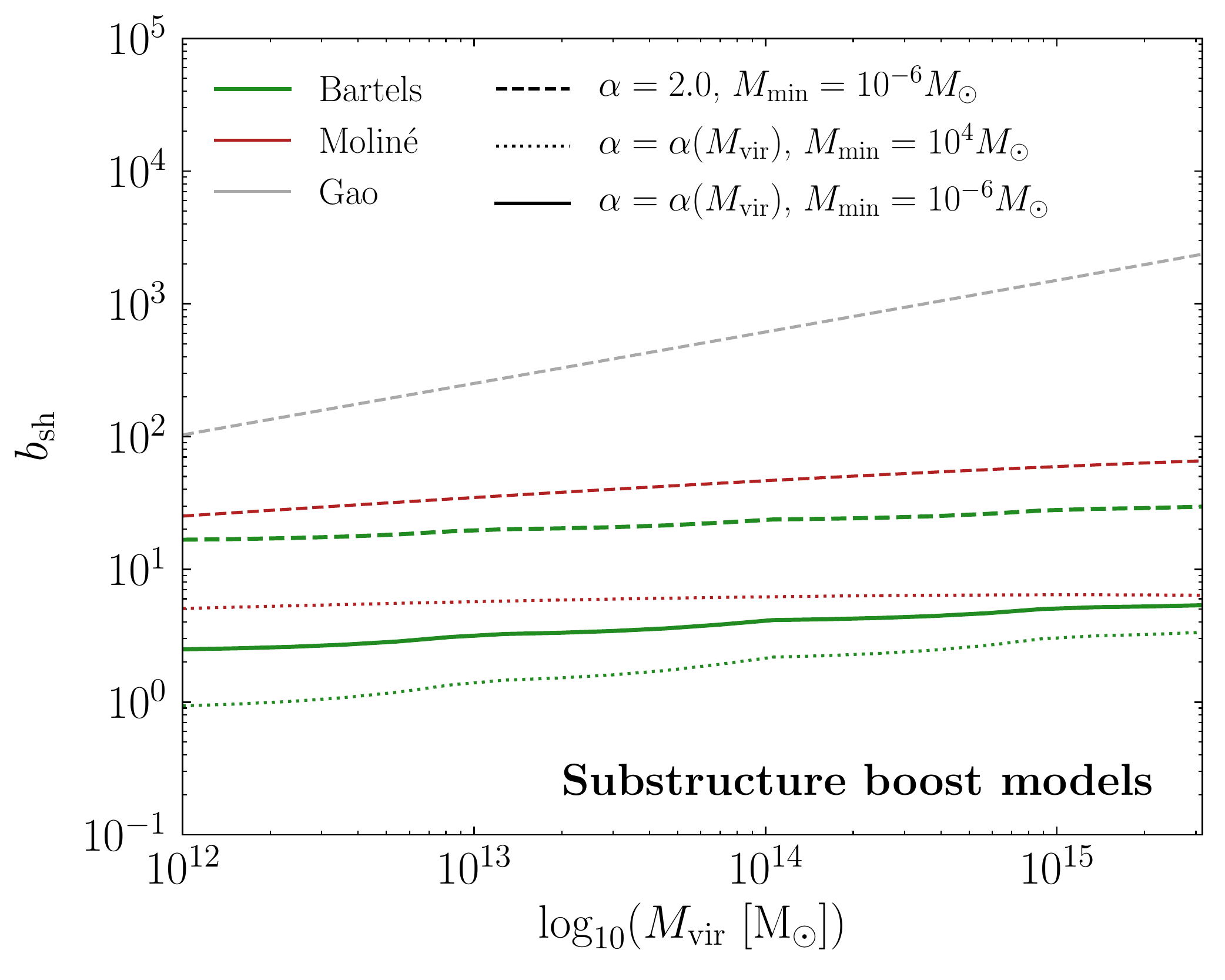}
      \includegraphics[width=.48\textwidth]{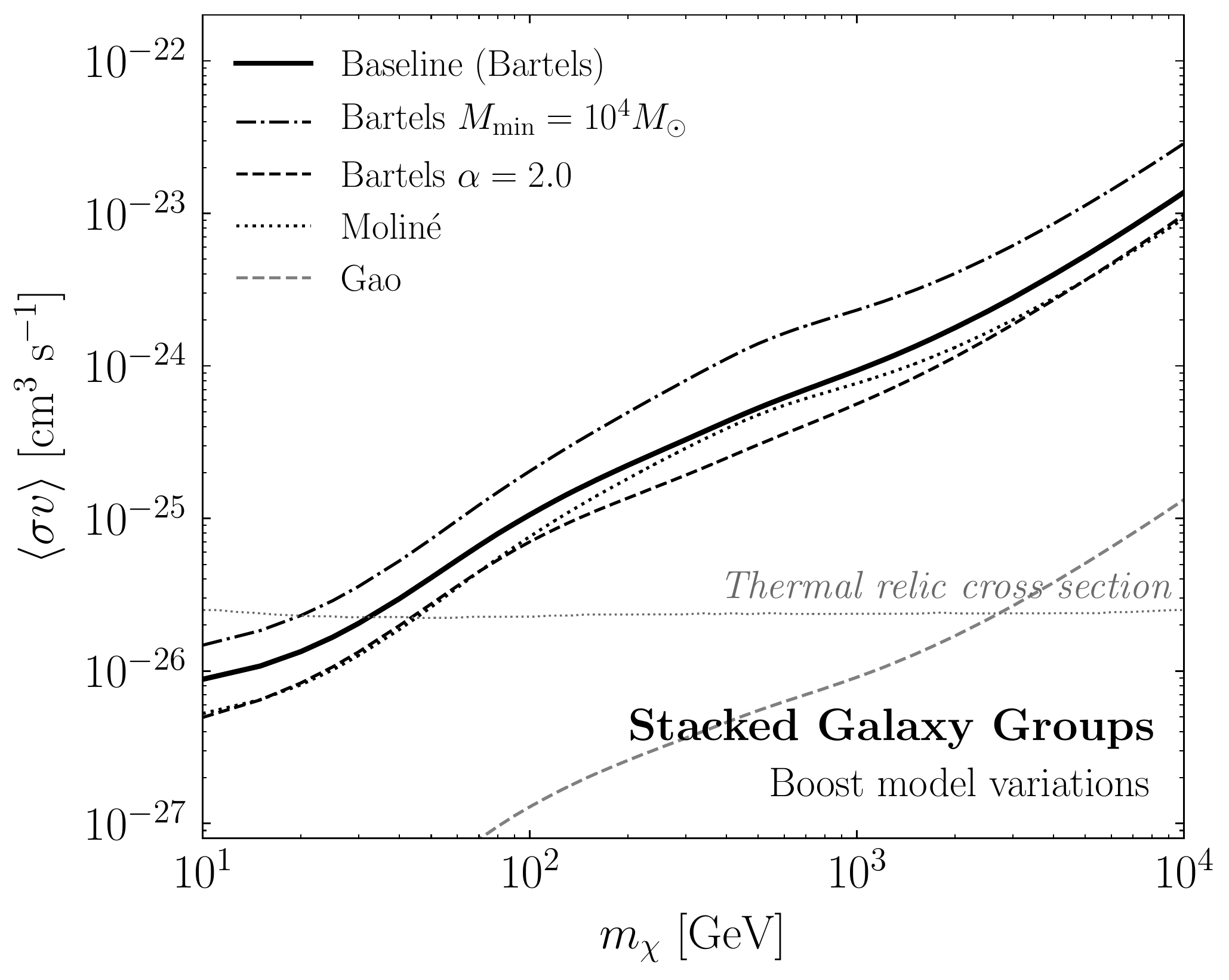}
   \caption{(Left) Examples of substructure boost models commonly used in the literature, reproduced from \cite{Lisanti:2017qoz}. Our fiducial model, based on Ref.~\cite{Bartels:2015uba} using $M_\text{min} = 10^{-6}$~M$_\odot$ and self-consistently computing $\alpha$, is shown as the thick green solid line. Variations on $M_\text{min}$ and $\alpha$ are shown with the dotted and dashed lines, respectively. Also plotted are the boost models of Molin\'e~\cite{Moline:2016pbm} (red) and Gao~\cite{Gao:2011rf} (grey).  (Right) The same as the baseline analysis shown in the left panel of Fig.~\ref{fig:bounds} of chapter~\ref{chap:fermigg}, except varying the boost model.
   }
   \label{fig:systematics_boost}
\end{figure}

When taken together, the details of the halo formation process shape the subhalo mass function $dn/dM_\text{sh}\propto M_\text{sh}^{-\alpha}$, where $\alpha  \in \left[1.9, 2.0\right]$.  The mass function does not follow a power-law to arbitrarily low masses, however, because the underlying particle physics model for the DM can place a minimum cutoff on the subhalo mass, $M_\text{min}$.  For example, DM models with longer free-streaming lengths wash out smaller-scale structures, resulting in higher cutoffs.

The left panel of Fig.~\ref{fig:systematics_boost} shows a variety of boost models commonly used in DM studies. The fiducial boost model used here~\cite{Bartels:2015uba} is shown as the thick green solid line and variations on $M_\text{min}$ and $\alpha$ are also plotted. The right panel of Fig.~\ref{fig:systematics_boost} shows that the expected limit when $M_\text{min} = 10^4$~M$_\odot$ instead of $M_\text{min} = 10^{-6}$~M$_\odot$ (dot-dashed) is weaker across all masses.  While a minimum subhalo mass of  $10^{4}$~M$_\odot$ is likely inconsistent with bounds on the kinetic decoupling temperature of thermal DM, this example illustrates the importance played by $M_\text{min}$ in the sensitivity reach.  Additionally, Fig.~\ref{fig:systematics_boost} demonstrates the case where $\alpha=2.0$ (dashed line).  Increasing the inner slope of the subhalo mass function leads to a correspondingly stronger limit, however observations tend to favor a slope closer to $\alpha = 1.9$ (which is what the most massive halos correspond to in our fiducial case).

Ref.~\cite{Sanchez-Conde:2013yxa} derived a boost factor model that accounts for the flattening of the concentration-mass relation at low masses, but does not include the effect of tidal stripping.  They assume a minimum sub-halo mass of $10^{-6}$~M$_\odot$ and a halo-mass function $dN/dM \sim M^{-2}$.  This was updated by Ref.~\cite{Moline:2016pbm} to account for the effect of tidal disruption. This updated boost factor model, which takes $\alpha = 1.9$, gives the constraint shown in Fig.~\ref{fig:systematics_boost} labeled ``Molin\'e'' (dotted).  This model is to be contrasted with the boost factor model of Ref.~\cite{Gao:2011rf}, labeled ``Gao'' in Fig.~\ref{fig:systematics_boost} (grey-dashed), which uses a phenomenological power-law extrapolation of the concentration-mass relation to low sub-halo masses.  Because the annihilation rate increases with increasing concentration parameter, the model in Ref.~\cite{Gao:2011rf} predicts substantially larger boosts than other scenarios that take into account a more realistic flattening of the concentration-mass relation at low subhalo masses.\vspace{0.1in}

\begin{figure}[t]
   \centering
   \includegraphics[width=.49\textwidth]{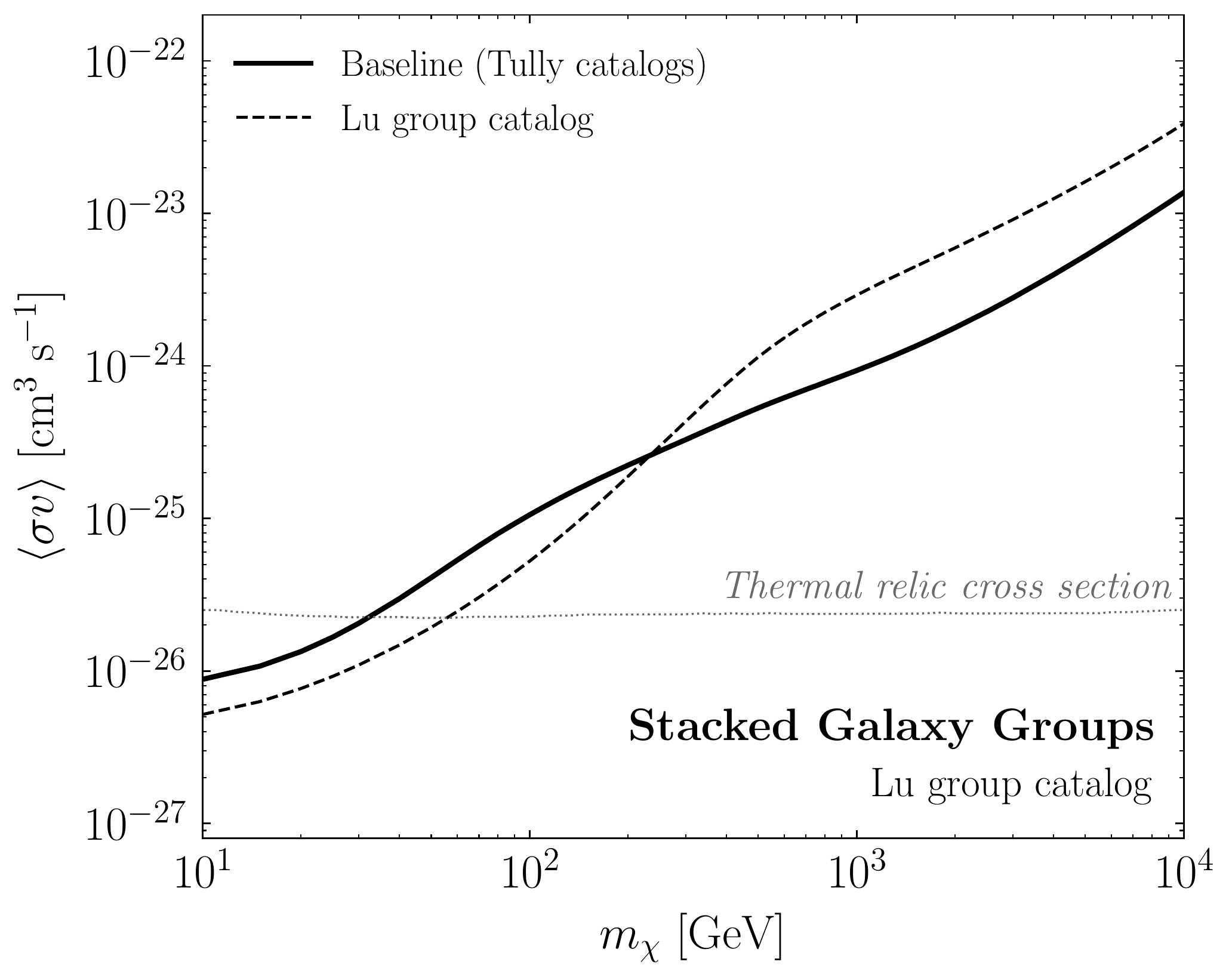}
    \includegraphics[width=.49\textwidth]{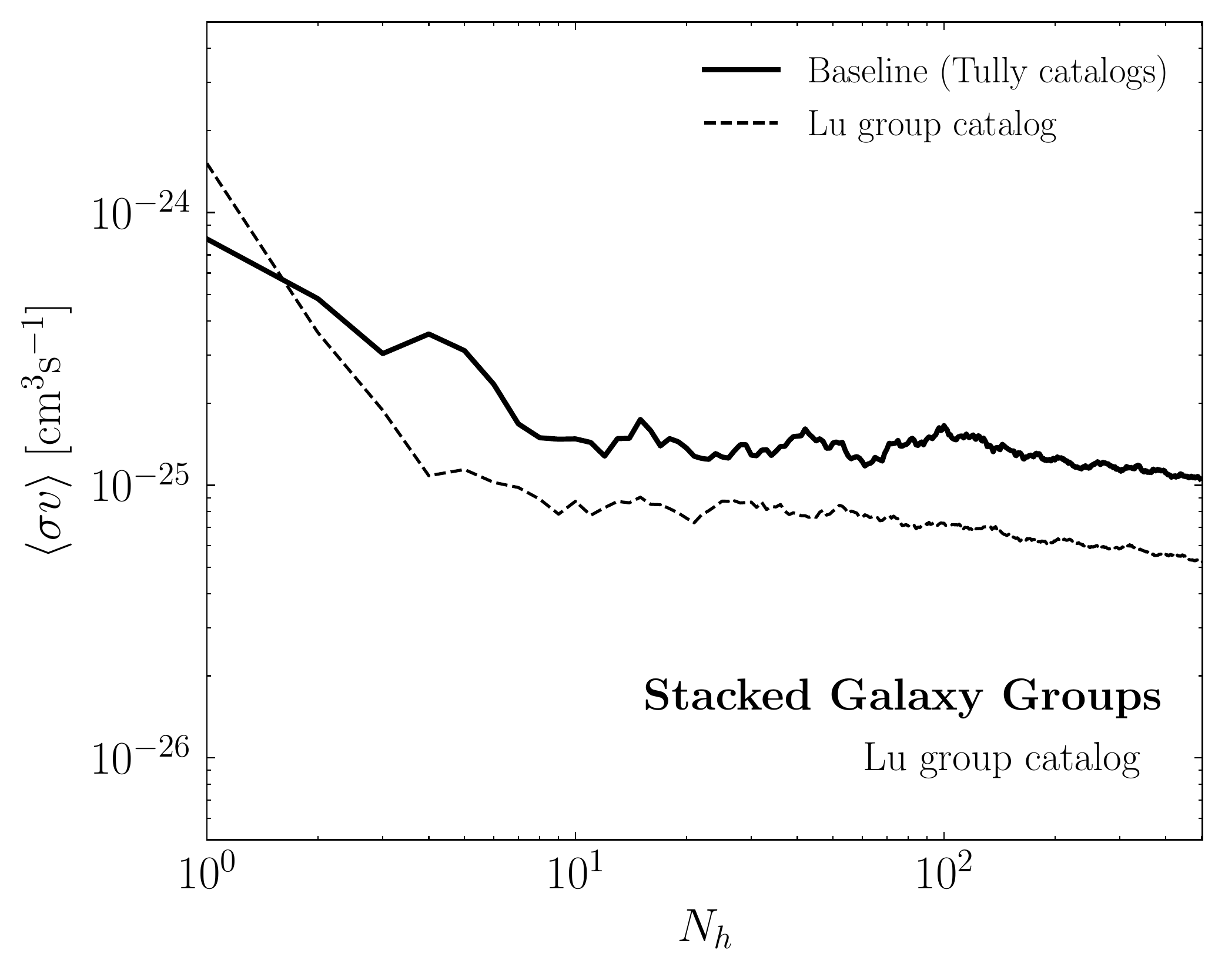} 
   \caption{The same as Fig.~\ref{fig:bounds} of chapter~\ref{chap:fermigg}, except using the Lu~\emph{et al.} galaxy group catalog~\cite{Lu:2016vmu} (dashed) instead of the T15 and T17 catalogs in the baseline analysis. }
   \label{fig:lucatalog}
\end{figure}

\noindent  {\bf Galaxy Group Catalog.}  
We now explore the dependence of the results on the group catalog that is used to select the halos.  In this way, we can better understand how the DM bounds are affected by uncertainties on galaxy clustering algorithms and the inference of the virial mass of the halos.  The baseline limits are based on the T15 and T17 catalogs, but here we repeat the analysis using the Lu~\emph{et al.} catalog~\cite{Lu:2016vmu}, which solely relies on 2MRS observations.  The group-finding algorithm used by Ref.~\cite{Lu:2016vmu} is different to that of T15 and T17 in many ways, relying on a friends-of-friends algorithm as opposed to one based on matching group properties at different scales to $N$-body simulations. Lu~\emph{et al.} also use a different halo mass determination.  For these reasons, it provides a good counterpoint to T15 and T17 for estimating systematic uncertainties associated with the identification of galaxy groups. While T17 includes measured distances for nearby groups, the Lu catalog corrects for the effect of peculiar velocities following the prescription in Ref.~\cite{1996AJ....111..794K} and the effect of Virgo infall as in Ref.~\cite{Karachentsev:2013jca}. Figure~\ref{fig:lucatalog} is a repeat of Fig.~\ref{fig:bounds} in chapter~\ref{chap:fermigg}, except using the Lu~\emph{et al.} catalog.  Despite important differences between the group catalogs used, the Lu~\emph{et al.} results are very similar to the baseline case. \\

\noindent There are a variety of sources of systematic uncertainty beyond those described here that  deserve further study.  For example, a systematic bias in the $J$-factor determination due to offsets in either the mass inference or the concentration-mass relation can be a potential source of uncertainty. A better understanding of the galaxy-halo connection and the small-scale structure of halos is required to mitigate this. Furthermore, we assumed distance uncertainties to be subdominant in our analysis. While this is certainly a good assumption over the redshift range of interest---nearby groups have measured distances, while groups further away come with spectroscopic redshift measurements with small expected peculiar velocity contamination---uncertainties on these do exist. We have also assumed that our targets consist of virialized halos and have not accounted for possible out-of-equilibrium effects in modeling these~\cite{1993AJ....105.2035D}.

%% file: dmdecay-app.tex
\chapter{Dark Matter Decay}

The appendix is organized as follows.  In Sec.~\ref{sec: methods}, we provide more detail regarding the methods used in chapter~\ref{chap:dmdecay}.  In particular, we discuss the calculations of the gamma-ray spectra and the data analysis.  In Sec.~\ref{sec: detailed results}, we give extended results beyond those given in chapter~\ref{chap:dmdecay}.  Then, in Sec.~\ref{sec: systematics}, we characterize and test sources of systematic uncertainty that could affect our results.  Lastly, in Sec.~\ref{sec: models}, we describe additional theory considerations for our analysis, including additional final states, extending our results to higher masses, and also additional models beyond those discussed in chapter~\ref{chap:dmdecay}.

\section{Methods}
\label{sec: methods}

We begin this section by detailing the calculations of the prompt and secondary spectra from DM decay.  Then, we discuss in detail the likelihood profile technique used in this paper.  

\subsection{Spectra}

This section provides a more detailed description of the gamma-ray spectra that result from heavy DM decay.  There is a natural decomposition into three components: (1)~prompt Galactic gamma-ray emission, (2)~Galactic inverse Compton~(IC) emission from high-energy electrons and positrons up-scattering background photons, and (3)~extragalactic flux from DM decay outside of our Galaxy.  As mentioned in chapter~\ref{chap:dmdecay}, when calculating the prompt spectrum (and the primary electron and positron flux) it is crucial, for certain final states, to included electroweak radiative processes, as these may be the only source of gamma-ray emission.  To illustrate this point, in Tab.~\ref{table:Particles} we show the average number of primary gamma-rays, neutrinos, and electrons coming from DM decay to $b \,\bar b$ and $\nu\, \bar \nu$ for various DM masses. 
We note that for $m_\chi=100\,$GeV there are in average 3\,(0) hadrons in the final state, while for $m_\chi=1\,$PeV there are  77\,(1) hadrons for the $b\bar{b}\,(\nu_e\bar{\nu}_e)$ decay mode. The energy fraction of these hadrons is 13\,(0)\,\% and 16\,(0.5)\,\% for $b\bar{b}\,(\nu_e\bar{\nu}_e)$ modes with a DM mass of 100\,GeV and 1\,PeV, respectively.  In addition, the energy fractions of photons, neutrinos and electrons are almost independent of the DM mass for the $b\bar{b}$ decay mode. This can be understood as the majority of these final states originate from pion decays.
{
\begin{table}[h]
\renewcommand{\arraystretch}{2}
\setlength{\arrayrulewidth}{.3mm}
\centering
\setlength{\tabcolsep}{0.8em}
\begin{tabular}{ c  | c | c | c | c || c | c | c | c |}
\cline{2-9}
& \multicolumn{4}{ c|| }{$\chi \to b \,\bar{b}$} & \multicolumn{4}{ c| }{$\chi \to \nu_e\, \bar{\nu}_e$} \\ 
\cline{1-1} \cline{2-9}
\multicolumn{1}{|c|}{$m_{\chi}$} & $\gamma$ & $\nu$ & $e^-/e^+$  & All & $\gamma$ & $\nu$ & $e^-/e^+$ & All \\ \hline \hline
\multicolumn{1}{|c|}{100 GeV} & 26 & 66 & 23 & 120 & 0 & 2 & 0 & 2 \\
\multicolumn{1}{|c|}{1 TeV} & 58 & 150 & 51 & 270 & 0.37 & 3 & 0.36 & 3.8 \\
\multicolumn{1}{|c|}{10 TeV} & 120 & 320 & 110 & 570 & 2.0 & 7.4 & 1.9 & 12 \\
\multicolumn{1}{|c|}{100 TeV} & 250 & 660 & 230 & 1200 & 5.1 & 15 & 4.8 & 26 \\
\multicolumn{1}{|c|}{1 PeV} & 490 & 1300 & 440 & 2300 & 9.2 & 27 & 8.7 & 46 \\ \hline
\end{tabular}
\caption{Average number of final state particles for DM decay to bottom quarks or electron neutrinos. For the neutrino case, the presence of electroweak corrections has a large impact on the resulting spectrum for higher masses, whereas for the hadronic final state the effect is less important.
}
\label{table:Particles}
\end{table}
}

\begin{figure}[t]
	\leavevmode
	\begin{center}$
	\begin{array}{cc}
	\scalebox{0.4}{\includegraphics{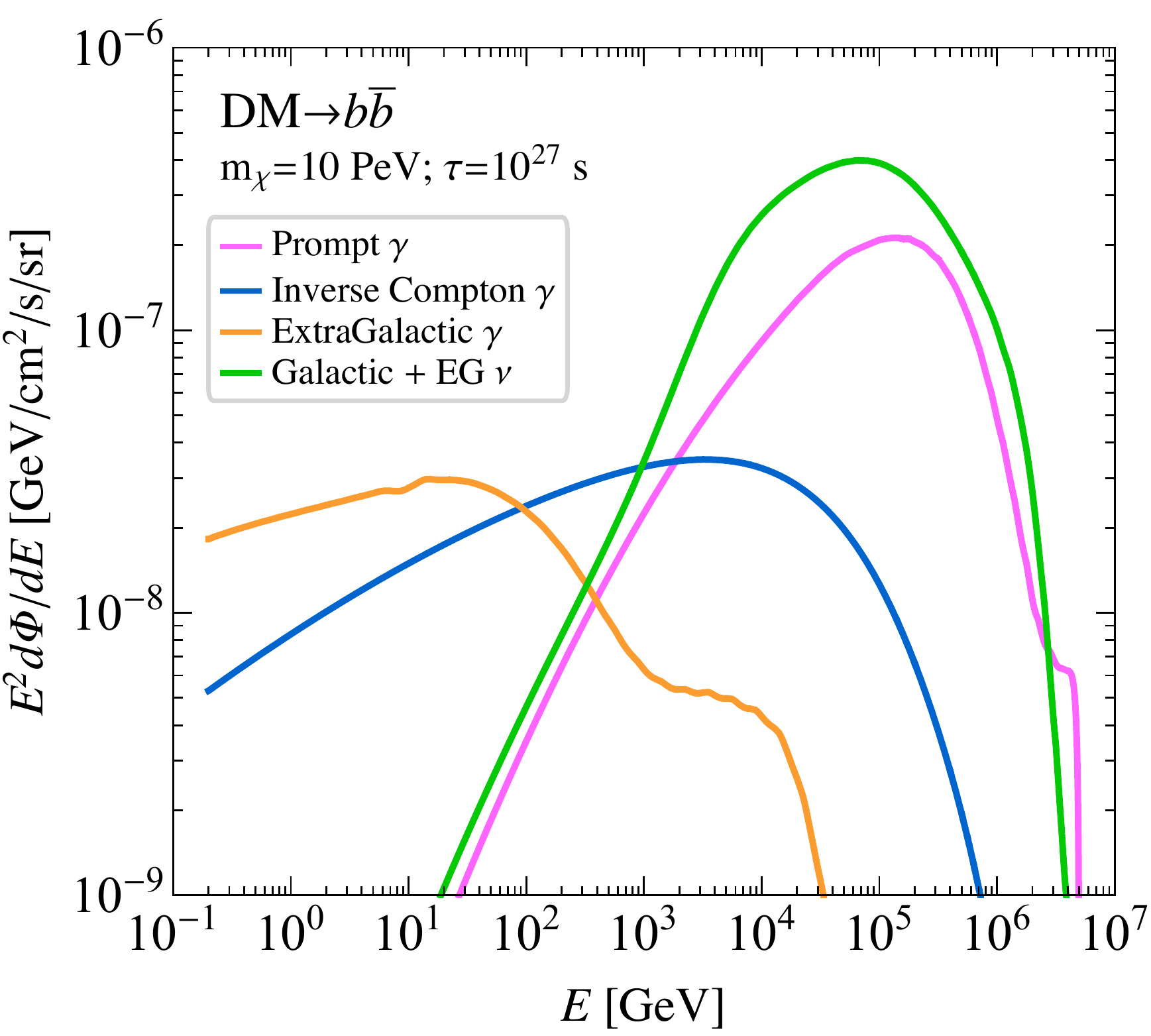}} &
	\scalebox{0.4}{\includegraphics{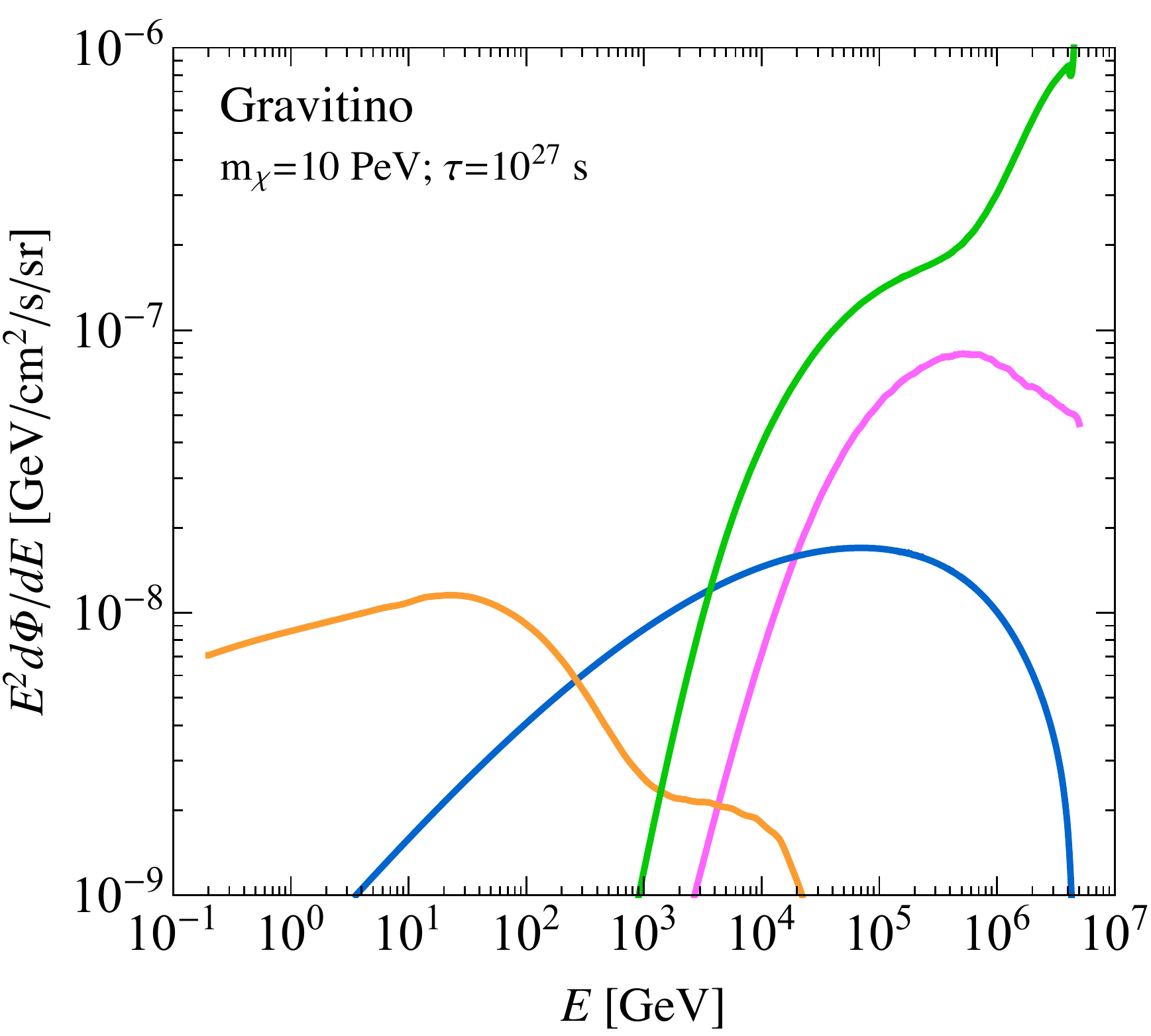}}
	\end{array}$
	\end{center}
	\vspace{-.70cm}
	\caption{Gamma-ray and neutrino spectra for DM decaying to $b\,\bar{b}$ (left) and a model of gravitino DM (right) as detailed in Sec.~\ref{sec: models} below, with $m_\chi = 10$ PeV and $\tau = 10^{27}$ s.  All fluxes are normalized within the ROI used in our main analysis. {\it Fermi} can detect photons in the range $\sim 0.2 - 2000$\,GeV. For heavy DM decays, the flux in the {\it Fermi} energy range is dominated by the IC and extragalactic contributions, rather than the prompt Galactic emission.
	}
	\vspace{-0.15in}
	\label{Fig: BasicSpec}
\end{figure}

\begin{figure}[t]
	\leavevmode
	\begin{center}$
	\begin{array}{cc}
	\scalebox{0.4}{\includegraphics{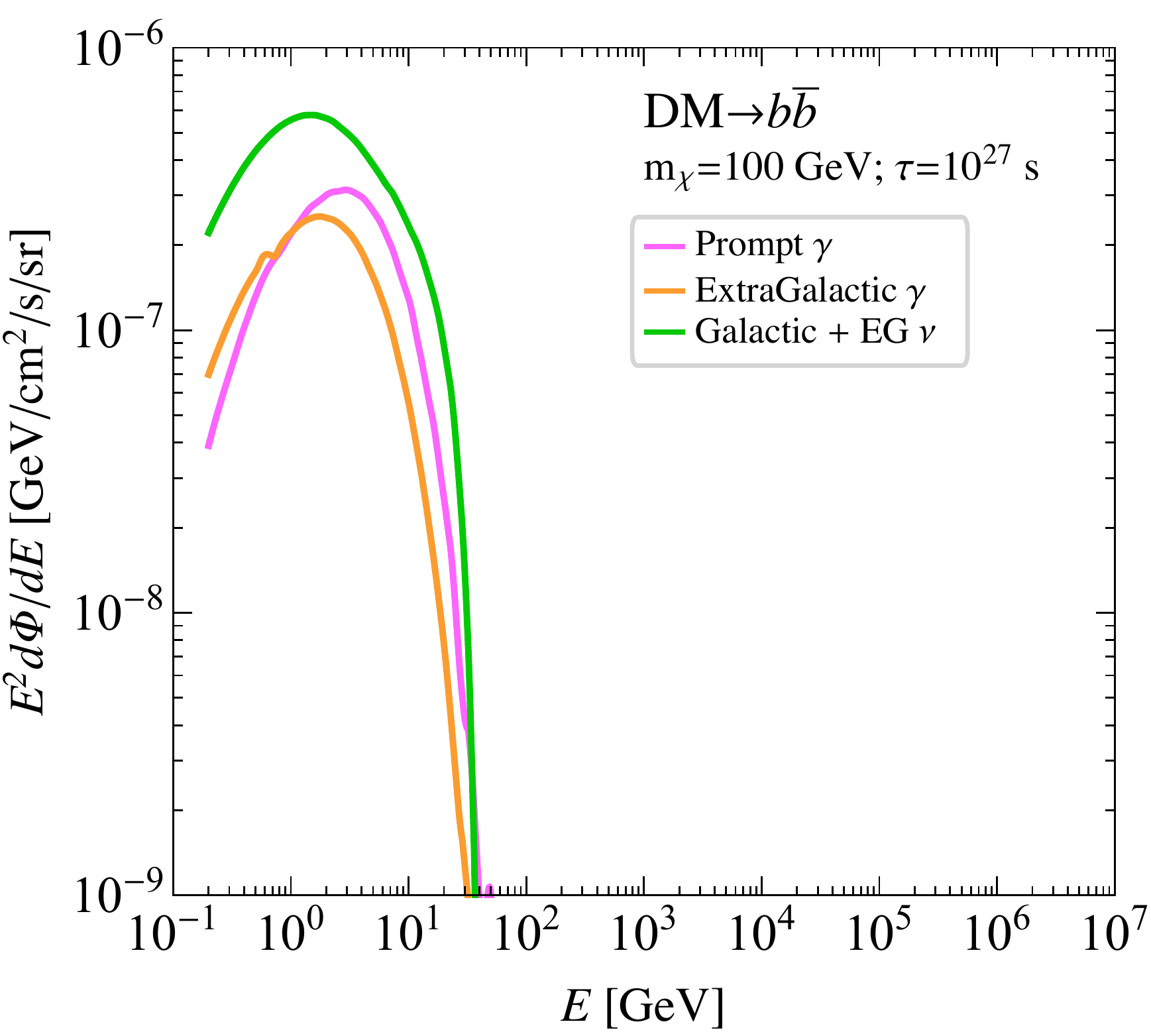}}  &
	\scalebox{0.4}{\includegraphics{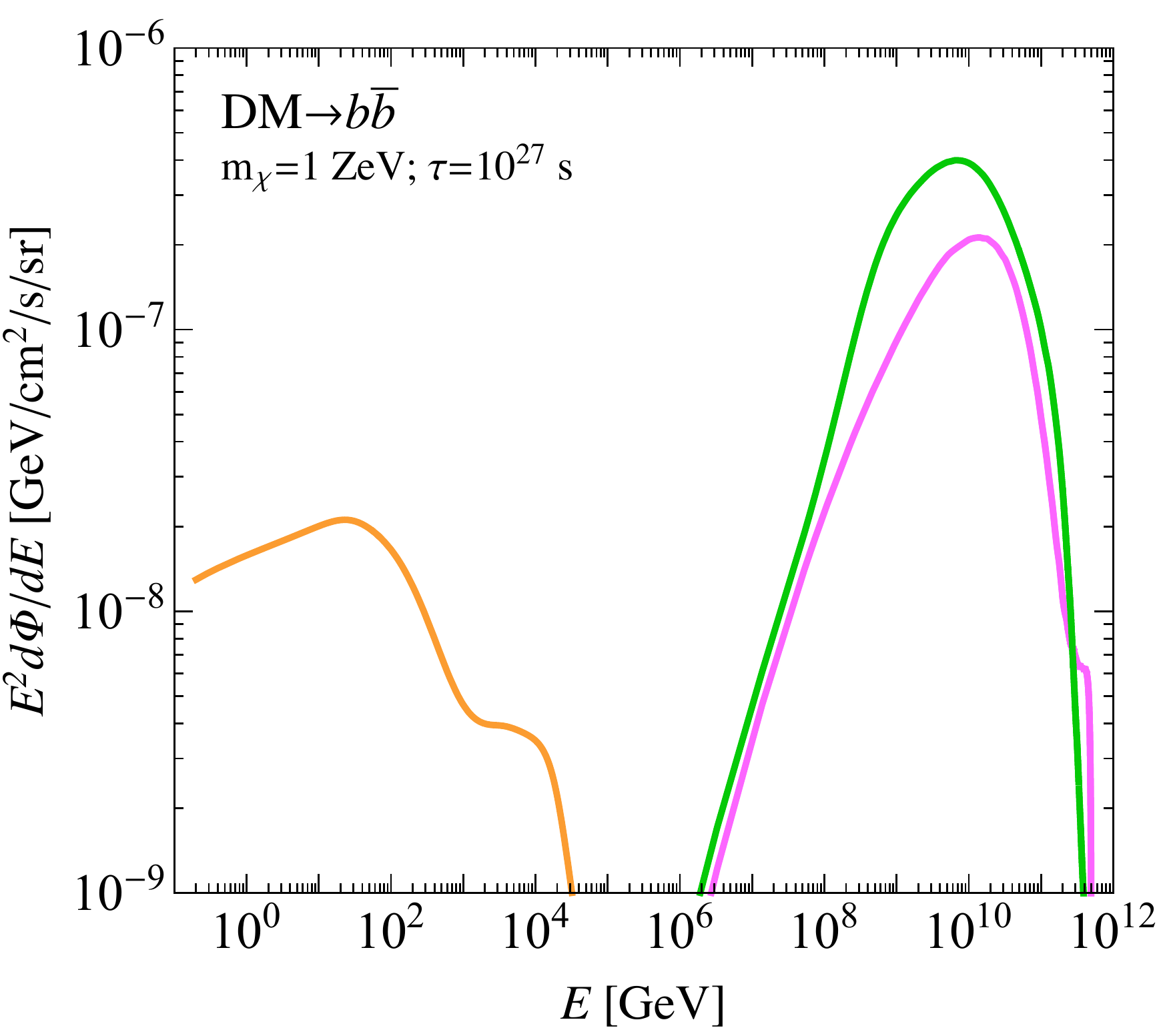}}
	\end{array}$
	\end{center}
	\vspace{-.70cm}
	\caption{Gamma-ray and neutrino spectra for DM decaying to $b\,\bar{b}$ for two different DM masses: $100$ GeV (left) and $10^{12}$ GeV (right). These should be compared to the {\it Fermi} energy range of $\sim 0.2 - 2000$ GeV. For the lighter DM case, prompt emission dominates, whilst at higher masses the dominant contribution is from the extragalactic flux. In neither of these cases is IC emission relevant, this only contributes meaningfully for intermediate $\mathcal{O}({\rm PeV-EeV})$ masses, as seen on the left of Fig.~\ref{Fig: BasicSpec}.
	}
	\vspace{-0.15in}
	\label{Fig: highlowbSpec}
\end{figure}

Additionally, we show the typical number of radiated $W$ and $Z$ bosons.  In the $b \bar b$ case, electroweak corrections are not significant even for 1\,PeV DM.  However, in the $\nu \bar \nu$ case the radiated $W$ and $Z$ bosons are responsible for the majority of the primary particles (and all of the gamma-rays and electrons) at masses above the electroweak scale. The importance of these electroweak corrections on dark matter annihilation/decay spectra have been previously noted (see~\emph{e.g.}~\cite{Kachelriess:2007aj,Regis:2008ij,Mack:2008wu,Bell:2008ey,Dent:2008qy,Borriello:2008gy,Bertone:2008xr,Bell:2008vx,Cirelli:2009vg,Kachelriess:2009zy,Ciafaloni:2010ti}). For DM masses above 10\,PeV, the large number of final states implies that generation of the spectra through showering in \textsc{Pythia} is no longer practical. We discuss in Appendix~\ref{sec:ExtendingToGUTScale} how we extend our spectra beyond these masses.\footnote{Publicly available DM spectra, such as those in~\cite{Cirelli:2010xx,Elor:2015tva,Elor:2015bho}, do not extend up to these high masses, which is why we have recalculated them. While there are certainly modeling errors associated with running \textsc{Pythia} at these very high energies, they are expected to be subdominant to the astrophysical uncertainties inherent in this analysis. We extend the spectra above 10 PeV by rescaling the appropriately normalized spectrum, as described and validated in this appendix.}

As was shown in Fig.~\ref{Fig: glue} in chapter~\ref{chap:dmdecay}, the prompt flux tends to be most important for lower DM masses near the {\it Fermi} energy range, while the IC emission may play a leading role for DM masses near the PeV scale.  The extragalactic flux is important over the whole mass range, but at very high masses -- well above the PeV scale -- the extragalactic flux may be the only source of gamma-rays in the {\it Fermi} energy range.  To illustrate these points, Fig.~\ref{Fig: BasicSpec} shows the gamma-ray and neutrino spectra at Earth, normalized to within the ROI used in our main analysis, for 10\,PeV DM decay with $\tau = 10^{27}$ s.  We consider two final states, $b \,\bar b$ (left) and the gravitino model (right), which is described in more detail later in this appendix.   

Importantly, for DM masses  $\gtrsim$1\,TeV, the gravitino decays roughly 50\% of the time into $W^{\pm}\, \ell^{\mp}$, where $\ell^{\mp}$ are SM leptons, and 50\% of the time into $Z^0\, \nu$ and $h \, \nu$.  These latter two final states are responsible for the sharp rise in the Galactic and extragalactic neutrino spectrum in the gravitino model at energies approaching the DM mass (10\,PeV in this case).  In both cases, however, the prompt gamma-ray spectra are seen to be sub-dominant within the {\it Fermi} energy range, which extends up to $\sim$2\,TeV.  At the upper end of the {\it Fermi} energy range, the IC emission is the dominant source of flux, while the extragalactic emission extends to much lower energies. 

To illustrate this point further, we show in Fig.~\ref{Fig: highlowbSpec} the $b\, \bar b$ final-state spectra for $m_\chi=100$\,GeV and $1$\,ZeV $\big(=10^{12} \text{ GeV}\big)$.  In the low-mass case, the IC emission is produced in the Thomson regime and peaks well below the {\it Fermi} energy range. Furthermore, in this case the extragalactic spectrum is generally sub-dominant to the prompt Galactic emission. In the high-mass case, the extragalactic flux is the only source of emission within the {\it Fermi} energy range. Indeed, it is well known that the extragalactic spectrum approaches a universal form, regardless of the primary spectra (\emph{e.g.} see~\cite{Murase:2012xs}; also as plotted in Fig.~\ref{Fig:UniversalSpec}). This can be seen by comparing the extragalactic spectrum on the right of Fig.~\ref{Fig: highlowbSpec} to that on the left on Fig.~\ref{Fig: BasicSpec}, and this is explored in more detail in Sec.~\ref{sec:ExtendingToGUTScale}. Finally for the ZeV DM decays, the IC emission is still largely peaked in the {\it Fermi} energy range, but has now transitioned completely to the Klein-Nishina regime, where the cross section is greatly reduced. As such its contribution is several orders of magnitude sub-dominat to the extragalactic flux. Note that in Fig.~\ref{Fig: highlowbSpec}, and in subsequent spectral plots, we have used a galactic $J$-factor that is averaged over our ROI. In detail, if we define $\rho(s,l,b)$ to be the DM density as a function of distance from Earth $s$, as well as galactic longitude $l$ and latitude $b$, then we used:
\begin{equation}
J = \int_{\rm ROI} \text{d} \Omega \int \text{d}s\, \rho(s,l,b) / \int_{\rm ROI} \text{d} \Omega \simeq 4.108 \times 10^{22}~{\rm GeV}~{\rm cm}^{-2}\,.
\end{equation}
This is larger than the all-sky averaged value by a factor of 2.6.

\begin{figure}[t]
        \leavevmode
        \begin{center}$
        \begin{array}{cc}
        \scalebox{0.4}{\includegraphics{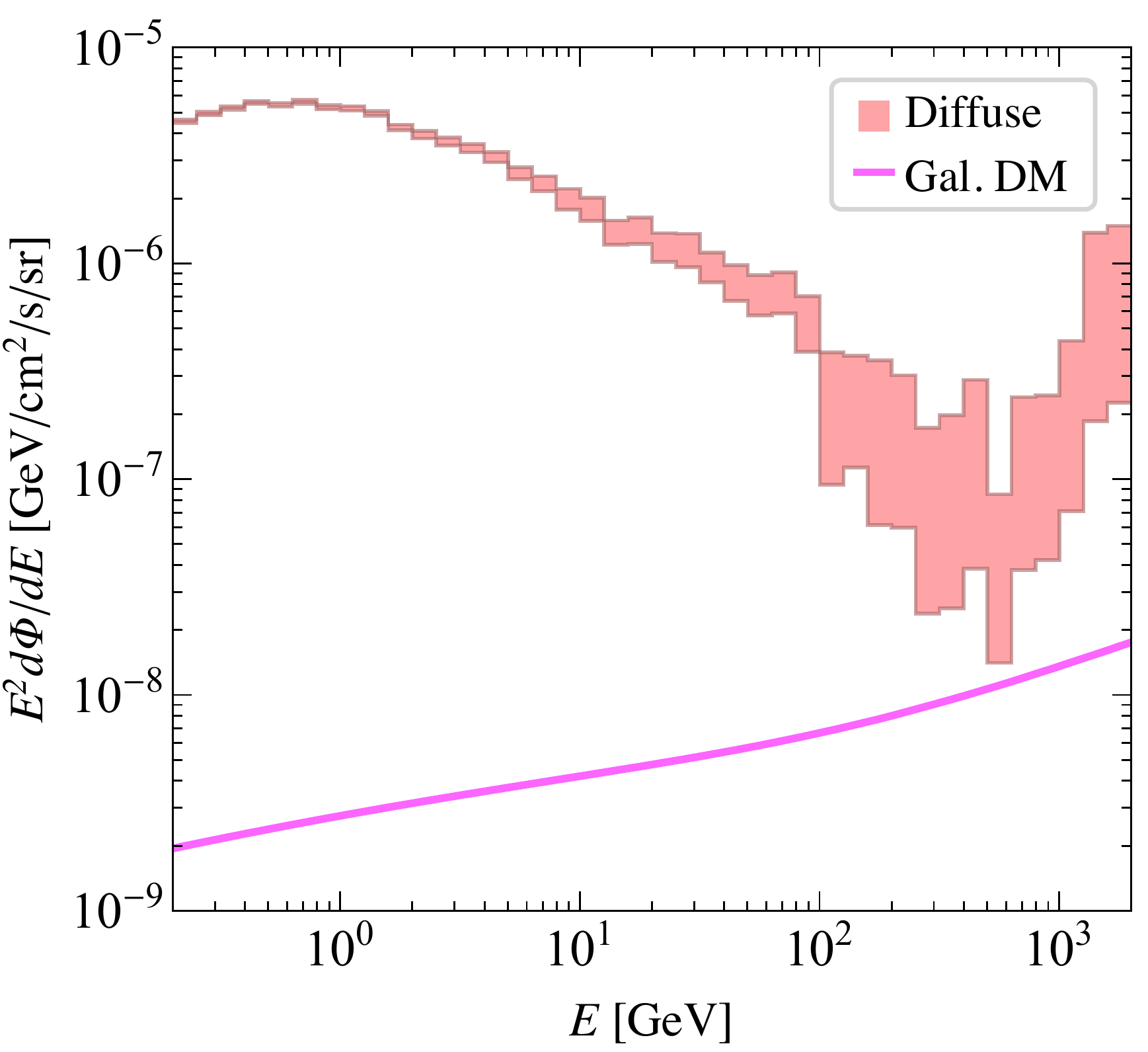}}  &
        \scalebox{0.4}{\includegraphics{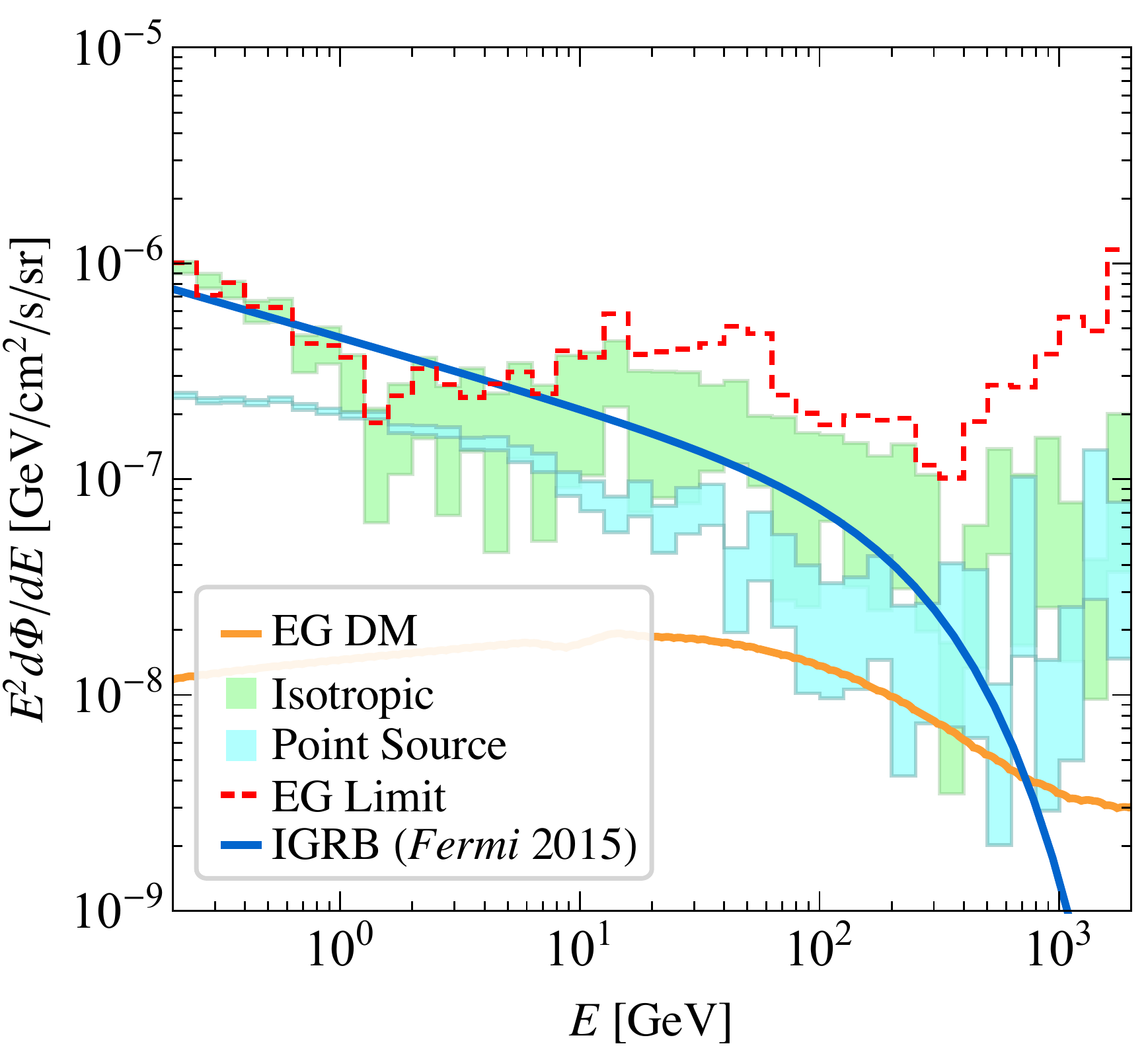}}
        \end{array}$
        \end{center}
        \vspace{-.70cm}
        \caption{A comparison between the $1$ PeV $b\, \bar{b}$ DM spectrum and that from our background models, for the largest lifetime we can constrain using only either Galactic ($\tau = 8.4 \times 10^{27}$ s) or extragalactic ($\tau = 1.7 \times 10^{27}$ s) DM flux. Spectra are averaged over the ROI used in our analysis. Left: Here we show the diffuse Galactic spectrum, compared to the smallest Galactic (prompt and IC) flux we can constrain. For the diffuse model we show the 68\% confidence interval determined from the posterior of our fit in each energy bin. Diffuse emission is responsible for the vast majority of the photons seen in our analysis, and it sits several orders of magnitude above the DM flux we can constrain in most energy bins. Right: The 68\% confidence intervals on the spectrum of our isotropic and point source models, compared to the weakest extragalactic DM flux we can constrain.  We also show in this plot the bin-by-bin 95\% limit we set on extragalactic flux, homogenious across the northern and southern sub-regions. Further, we illustrate the IGRB as measured by {\it Fermi} \cite{Ackermann:2014usa}, which is in good agreement with our isotropic spectrum across most of the energy range. See text for details. }
        \vspace{-0.15in}
        \label{Fig: DMvsBkgSpec}
\end{figure}

In chapter~\ref{chap:dmdecay}, we assumed that for the energies relevant for {\it Fermi}, the IC morphology will be effectively identical to that of the prompt DM decay flux. This justified the combination of the prompt and IC flux into a single spatial template which followed the above $J$-factor. 
In principle there are at least three places additional spatial dependence could enter, beyond the prompt $e^{\pm}$ spatial distribution injected by DM decays: 1. the distribution of the seed photon fields; 2. the distribution of the magnetic fields under which the electrons cool; and 3. the diffusion of the $e^{\pm}$. 
Referring to the first of these, there are three fields available to up-scatter off: the CMB, the integrated stellar radiation, and the infrared background due to the irradiated stellar radiation. These last two are position dependent and tend to decrease rapidly off the plane. So as long as we look off the plane, as we do, the CMB dominates and is position independent. Importantly, neglecting the other contributions is conservative, as they would only contribute additional flux. 
Regarding the second point, the regular and halo magnetic fields play an important role in the $e^{\pm}$ cooling. The former component highly depends on the Galactic latitude and decays off the plane; it is subdominant to the halo magnetic field in our ROI, so we ignore it.  
Finally, for the energies of interest, the diffusion of the $e^{\pm}$ can be neglected to a good approximation on the scales of interest, as discussed in \cite{Esmaili:2015xpa}. The halo field is expected to be strong enough for electrons and positrons to lose their energy in the halo.

Finally, Fig.~\ref{Fig: DMvsBkgSpec} shows the spectrum of the weakest Galactic and extragalactic DM fluxes we can constrain for $1$\,PeV DM decaying to bottom quarks, directly compared to the background contributions. In these figures, the three background components from our fits -- diffuse, isotropic, and point source emission -- are shown via a band between the 16 and 84 percentiles on these parameters extracted from the posterior, where the values are given directly in each of our 40 energy bins. Between these figures we see that diffuse emission dominates over essentially the entire energy range. We also see that the value of the isotropic flux is not particularly well constrained within our small ROI, especially at higher energy.  It is is important to note that our isotropic spectrum is found by averaging the spectra in the north and south, which are fit independently.  As a comparison, we also show the 95\% limit on homogenous extragalactic emission, which is by definition the same in the northern and southern hemispheres.  Reassuringly, our limit on extragalactic emission tends to be weaker than the isotropic gamma-ray background (IGRB) as measured by {\it Fermi} \cite{Ackermann:2014usa}, which is also shown in that figure.  The IGRB was determined from a dedicated analysis at high-latitudes using a data set with very low cosmic-ray contamination.  Even though our ROI and data set are far from ideal for determining the IGRB, we see that our isotropic spectrum is generally in very good agreement with the {\it Fermi} IGRB up to energies of around a few hundred GeV; at higher energies, our isotropic spectrum appears higher than the IGRB, perhaps because of cosmic ray contamination.  However, this should only make our high-energy extragalactic results conservative.

\newpage
\subsection{Data analysis}

\begin{figure}[b!]
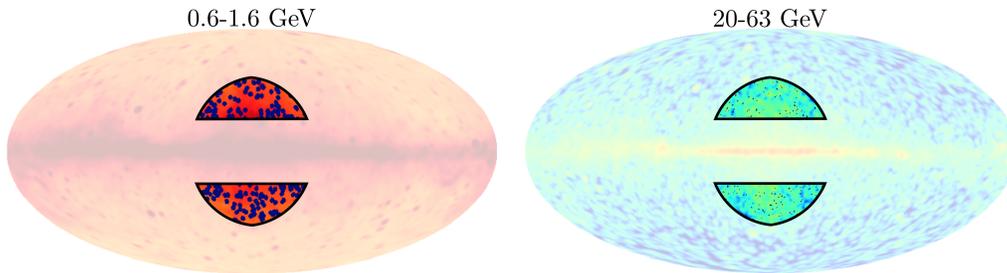

	\leavevmode
	\begin{center}$
	\begin{array}{cc}
	\scalebox{0.18}{\includegraphics{Figures/dmdecay-app/ROI_LowE.pdf}} &
	\scalebox{0.18}{\includegraphics{Figures/dmdecay-app/ROI_HighE.pdf}}
	\end{array}$
	\end{center}
	\vspace{-.40cm}
	\caption{The data within our Region of Interest (ROI), defined by $|b| > 20^{\circ}$ and $r < 45^{\circ}$, where $r$ is the angular distance from the GC. This ROI is shown in the context of the full data, shown with a lower opacity, for two different energy ranges: 0.6-1.6 GeV (left) and 20-63 GeV (right). In both cases the data has been smoothed to $2^{\circ}$, and all 3FGL point sources within our ROI have been masked at their 95\% containment radius. These are shown in blue, and are much larger on the left than the right as the {\it Fermi} PSF increases with decreasing energy. In our lowest energy bin (not shown), the point source mask covers most of our ROI.  In both figures, red shades indicate increased photon counts, while in the left (right) orange (blue) shades illustrate regions of low photon counts.}
	\vspace{-0.15in}
	\label{Fig:ROI}
\end{figure}
 
In this section, we expand upon the profile-likelihood analysis technique used in this work (see~\cite{Rolke:2004mj} for comments on this method). The starting point for this is the data itself, which we show in Fig.~\ref{Fig:ROI}. There we show our ROI in the context of the full dataset. Recall this ROI is defined by $|b| > 20^{\circ}$ and $r < 45^{\circ}$, with 3FGL PSs masked; this particular choice is discussed in detail in Sec.~\ref{sec: systematics}. The raw {\it Fermi} data is a list of photons with associated energies and positions on the sky. We bin these photons into 40 energy bins, indexed by $i$, that are equally log spaced from 200 MeV and 2 TeV. In each energy bin we then take the resulting data $d_i$, and spatially bin it using a HEALPix~\cite{Gorski:2004by} pixelation with \texttt{nside}=128. This divides our ROI into 12,474 pixels (before the application of a point source mask), which we index with $p$. The result of this energy and spatial binning reduces the raw data into a list of integers $n^p_i$ for the number of photons in pixel $p$ in the $i^\text{th}$ energy bin.

To determine the allowable DM decay contribution to this data, we need to describe it with a set of model parameters ${\bf \theta} = \{ {\bf \psi}, {\bf \lambda} \}$. As discussed in chapter~\ref{chap:dmdecay}, ${\bf \psi}$ are the parameters of interest which describe the DM flux, while ${\bf \lambda}$ are the set of nuisance parameters. In detail ${\bf \psi}$ accounts for the Galactic and separately extragalactic DM decay flux, and ${\bf \lambda}$ models the Galactic diffuse emission, {\it Fermi} bubbles, isotropic flux, and emission from PSs. Recall that each of the nuisance parameters is given a separate degree of freedom in the northern and southern Galactic hemispheres.

In terms of these model parameters, we can then build up a likelihood function in terms of the binned data. In doing so, we treat each energy bin independently, so that in the $i^\text{th}$ bin we have:
\es{LLfunction}{
p_i\big(d_i \big| {\theta}_i\big)  = \prod_p { \mu^p_i(\theta_i)^{n^p_i} \,e^{-\mu^p_i(\theta_i)} \over n^p_i!} \,,
}
where $\mu^p_i(\theta_i)$ is the mean predicted number of photon counts in that pixel as a function of the model parameters $\theta_i = \{ {\bf \psi}_i, {\bf \lambda}_i \}$.  The $\mu^p_i(\theta_i)$ are calculated from the set of templates used in the fit, which describe the spatial distribution of the various contributions described above.  More specifically, if the $j^\text{th}$ template in energy bin $i$ predicts $T^{j,p}_i$ counts in the pixel $p$, then $\mu^p_i(\theta_i) = \sum_j A^j_i(\theta_i) \,T^{j,p}_i$, where $A^j_i(\theta_i)$ is the normalization of the $j^\text{th}$ template as a function of the model parameters.  In our analysis, all of the normalization functions are linear in the model parameters, and in particular there is a model parameter that simply rescales the normalization of each template in each energy bin.

The likelihood profile in the single energy bin, as a function of the parameters of interest $\psi_i$, is then given by maximizing the log likelihood over the nuisance parameters $\lambda_i$:
\es{LLprofile}{
\log  p_i\big(d_i \big| {\psi}_i\big) = \max_{\lambda_i} \log p_i\big(d_i \big| {\theta}_i\big) \,.
}
This choice to remove the nuisance parameters by taking their maximum is what defines the profile-likelihood method. After doing so we have reduced the likelihood to a function of just the DM parameters, which are equivalent to the isotropic and LOS integrated NFW correlated flux coming from DM decay. As such, we can write
\es{LLprofile2}{
\log  p_i\big(d_i \big| {\psi}_i\big) = \log p_i\Big(d_i \Big| \Big\{I^i_\text{iso}, I^i_\text{NFW}\Big\}\Big) \,.
}
For a given DM decay model, ${\cal M}$, there will be a certain set of values for $\{I^i_\text{iso}, I^i_\text{NFW}\}$ in each energy bin. Given these, the likelihood associated with that model is given by:
\es{LLprofile3}{
\log  p\big(d \big| {\cal M},  \{\tau, m_\chi\}\big) = \sum_{i=0}^{39}  \log p_i\Big(d_i \Big| \Big\{I^i_\text{iso}, I^i_\text{NFW}\Big\}\Big) \,,
}
where we have made explicit the fact that in most models the lifetime $\tau$ and mass $m_{\chi}$ are free parameters. We then define the test statistic~(TS) used to constrain the model ${\cal M}$ by\footnote{Note that this TS stands in contrast to that used in~\cite{Ackermann:2012rg}; in that work, the TS was similarly defined, except that instead of using $\tau = \infty$ as a reference the $\tau$ of maximal likelihood was used.  The definition of TS used here is more conservative than that in~\cite{Ackermann:2012rg}, though formally, with Wilk's theorem in mind, our limits do not have the interpretation of 95\% constraints.}
\es{TS}{
\text{TS}\big( {\cal M},  \{\tau, m_\chi \} \big) = 2 \times \Big[ \log  p\big(d \big| {\cal M},  \{\tau, m_\chi\}\big) - \log  p\big(d \big| {\cal M},  \{\tau = \infty, m_\chi\}\big)  \Big] \,.
}
Note that fundamentally it is the list of values $\{I^i_\text{iso}, I^i_\text{NFW}\}$ that determine the TS. This means we can build a 2-d table of TS values in each energy bin as a function of the extragalactic and Galactic DM flux. This table only needs to be computed once; afterwards a given model can be mapped onto a set of flux values, which has an associated TS in the tables. This is the approach we have followed, and we show these DM flux versus TS functions in Sec.~\ref{sec: detailed results}.  The table of TS values is also available as Supplementary Data~\cite{supp-data}.

\section{Likelihood Profiles}
\label{sec: detailed results}
\begin{figure}[t!]
	\leavevmode
	\begin{center}$
	\begin{array}{c}
	\scalebox{0.28}{\includegraphics{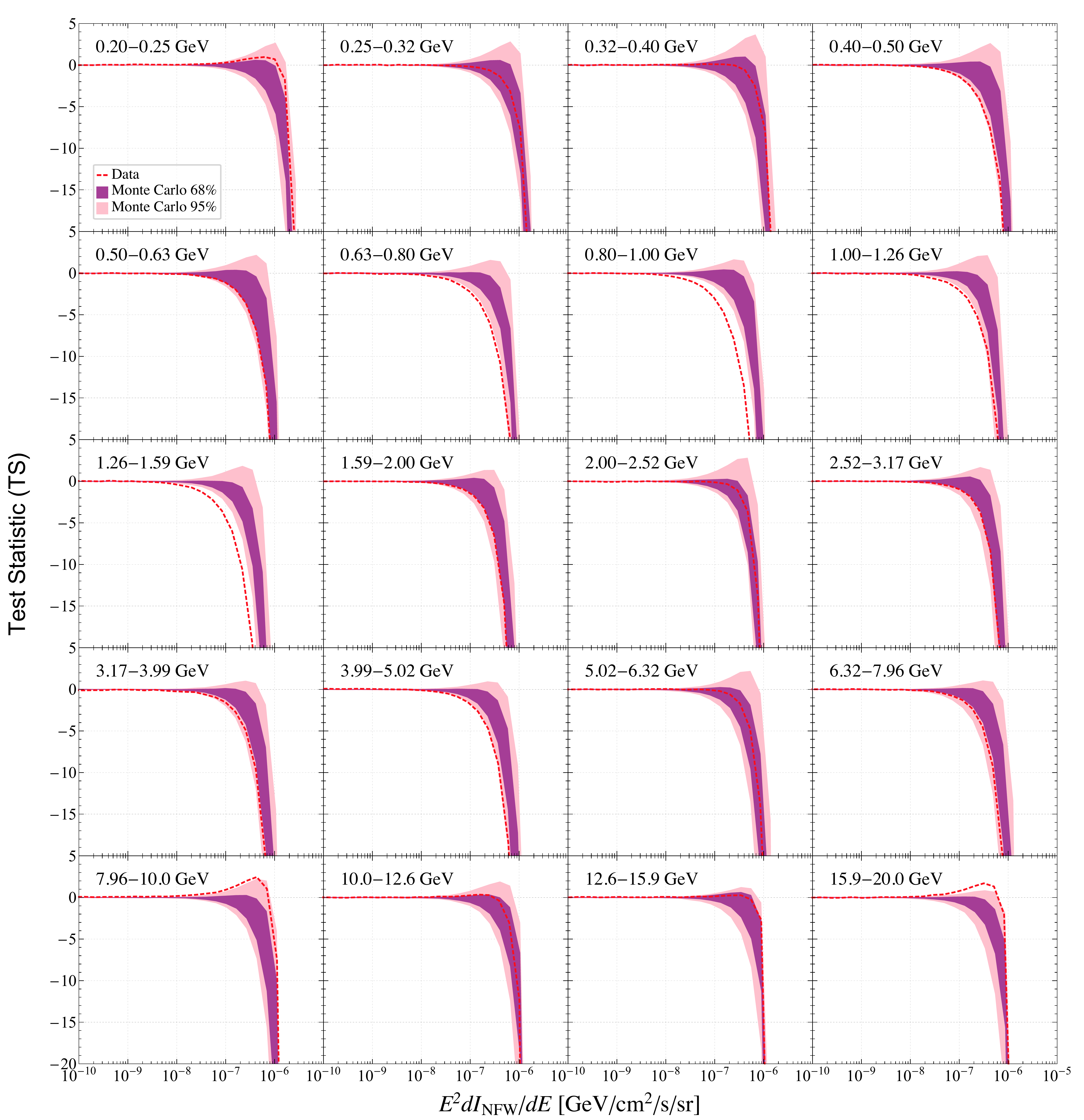}} 
	\end{array}$
	\end{center}
	\vspace{-.70cm}
	\caption{The change in log-likelihood, $\text{TS} \equiv p_i(d_i | \{ I_\text{NFW}^i \} ) -  p_i(d_i | \{ I_\text{NFW}^i = 0 \} )$, as a function of the intensity $I_\text{NFW}^i$ of NFW-correlated emission in the first 20 energy bins.  The measurement is given by the dashed red line, and the 68\% and 95\% confidence regions as derived from MC are given by the purple and pink bands respectively.  In most energy bins, the likelihood curves from the analysis of the data is seen to agree, within statistical uncertainties, with the expectation from the background templates only, as indicated by the MC bands. }
	\vspace{-0.15in}
	\label{Fig: LL_profileG0to19}
\end{figure}

\begin{figure}[t!]
        \leavevmode
        \begin{center}$
        \begin{array}{c}
        \scalebox{0.28}{\includegraphics{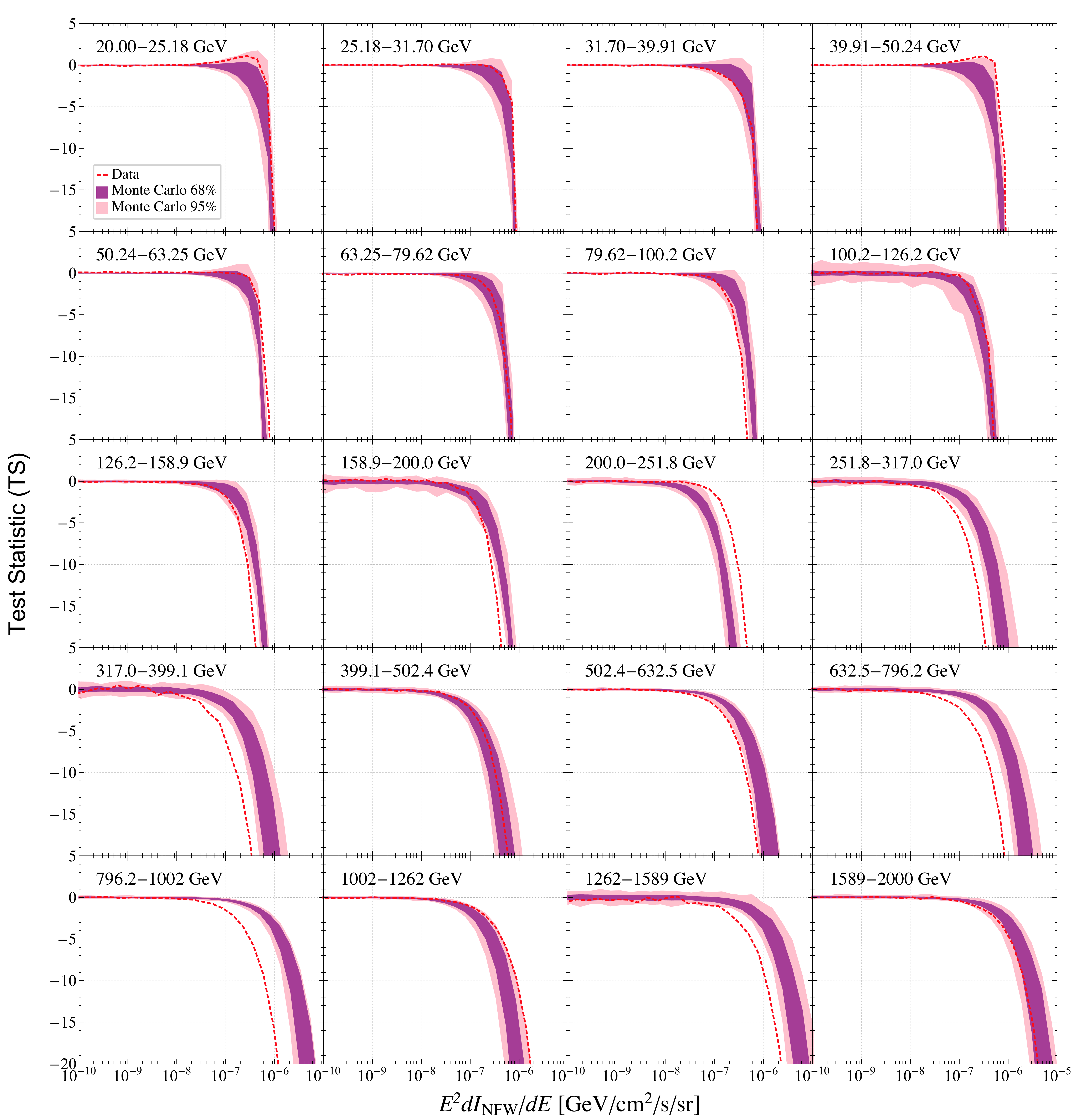}}
        \end{array}$
        \end{center}
        \vspace{-.70cm}
        \caption{As in~\ref{Fig: LL_profileG0to19}, except for the later 20 energy bins.}
        \vspace{-0.15in}
        \label{Fig: LL_profileG20to39}
\end{figure}

As described in chapter~\ref{chap:dmdecay}, our limits on specific DM final states and models are obtained from 2-d likelihood profiles, where the two dimensions encompass LOS integrated NFW correlated Galactic gamma-ray flux and extragalactic gamma-ray flux.  In Figs.~\ref{Fig: LL_profileG0to19} and~\ref{Fig: LL_profileG20to39} we show slices of these log-likelihood profiles when the extragalactic DM-induced flux is set to zero.  The bands indicate the 68\% and 95\% confidence intervals for the expected profiles obtained from background-only MC simulations.  The simulations use the set of background (``nuisance'') templates normalized to the best-fit values obtained from a template analysis of the data in the given energy bin.  In most energy bins, the results obtained on the real data are consistent with the MC expectations, showing that -- for the most part -- we are in a statistics-dominated regime.  In some energy bins, such as that from $15.9$--$20.0$ GeV, the data shows a small excess in the TS compared to the MC expectation.  While such an excess is perhaps not surprising since we are looking at multiple independent energy bins, it could also arise from a systematic discrepancy between the background templates and the real data.  More of a concern are energy bins where the limits set from the real data are more constraining than the MC expectation, such as the energy bin from $0.5$--$0.63$ GeV.  It is possible that this discrepancy, in part, arises from an over-subtraction of diffuse emission in certain regions of sky since the diffuse template is not a perfect match for the real cosmic-ray induced emission in our Galaxy.  This possibility -- and the efforts that we have taken to minimize its impact---is discussed further in Sec.~\ref{sec: systematics}.

\begin{figure}[t!]
	\leavevmode
	\begin{center}$
	\begin{array}{c}
	\scalebox{0.28}{\includegraphics{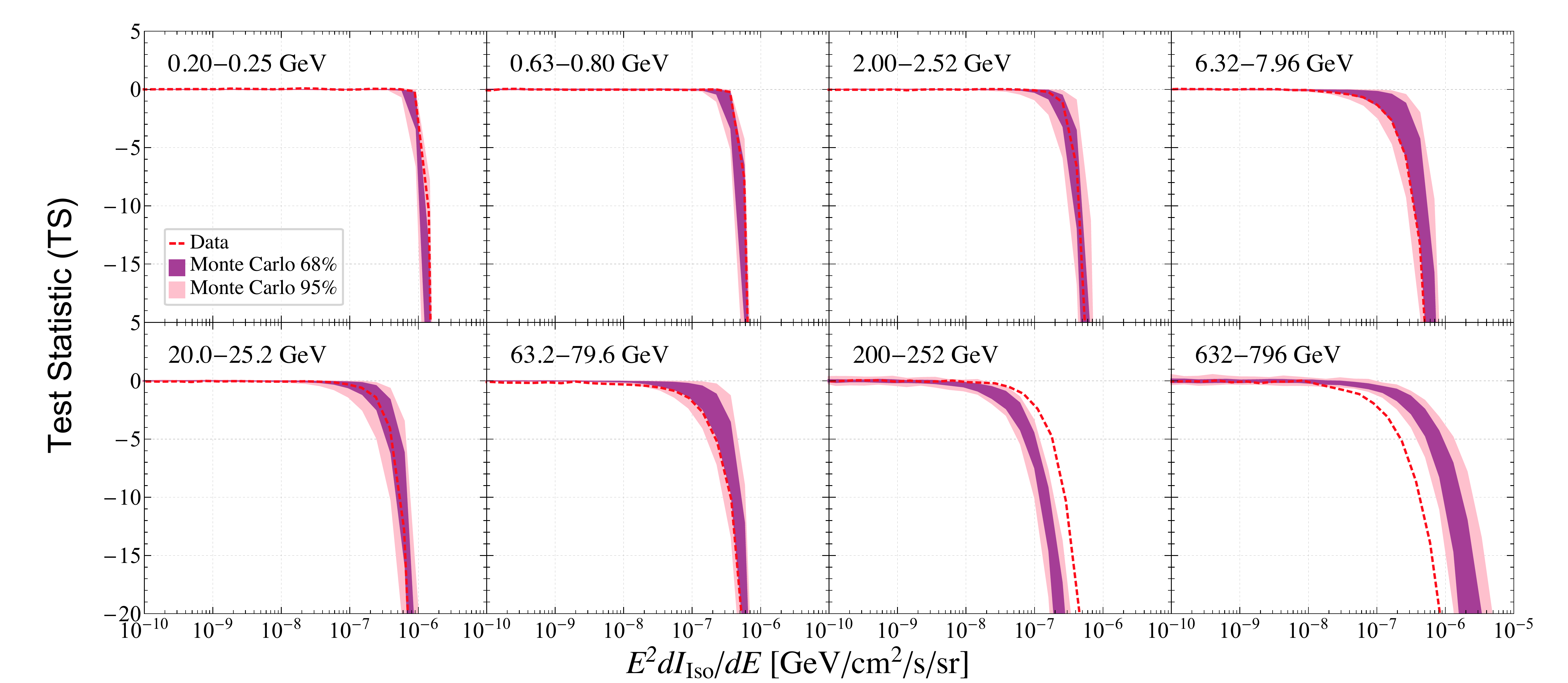}} 
	\end{array}$
	\end{center}
	\vspace{-.70cm}
	\caption{As in~\ref{Fig: LL_profileG0to19}, except for a selection of energy bins for the extragalactic only flux.}
	\vspace{-0.15in}
	\label{Fig: LL_profileEG}
\end{figure}

In Fig.~\ref{Fig: LL_profileEG}, we show a selection of the log-likelihood profiles found for vanishing Galactic DM-induced gamma-ray flux and shown instead as functions of the extragalactic DM-induced flux. It is important to remember that in the template fit we marginalize over isotropic emission.  As a result, it is impossible with our method to find a positive change in the TS as we increase the DM-induced isotropic flux $I_\text{iso}$.  In words, we remain completely agnostic towards the origin of the IGRB in our analysis.  That is, we do not assume the IGRB is due to standard astrophysical emission but we also do not assume it is due to DM decay. The 1-d likelihood profiles as functions of $I_\text{iso}$ instead show the limits obtained for the isotropic flux coming simply from the requirement that they do not overproduce the observed data.

In some energy bins, particularly at high energies (such as the energy bin from 632-796 GeV in Fig.~\ref{Fig: LL_profileEG}), the data is seen to be more constraining than the MC expectation.  However, we stress that the isotropic flux is not well determined, especially at these high energies, in our small region.  With that said, the isotropic flux determined in this small region tends to be larger than the IGRB determined from a dedicated analysis at high latitudes (see Fig.~\ref{Fig: DMvsBkgSpec}).  As a result, our limits on the extragalactic flux are likely conservative.

The full 2-d likelihood profiles are available as Supplementary Data~\cite{supp-data}. These are given as a function of the average Galactic and extragalactic DM flux in our ROI, without including any point source mask. The absence of the point source mask is chosen to simplify the use of our flux-TS tables.

\section{Systematics Tests}
\label{sec: systematics}

\begin{figure}[t!]
        \leavevmode
        \begin{center}$
        \begin{array}{cc}
        \scalebox{0.28}{\includegraphics{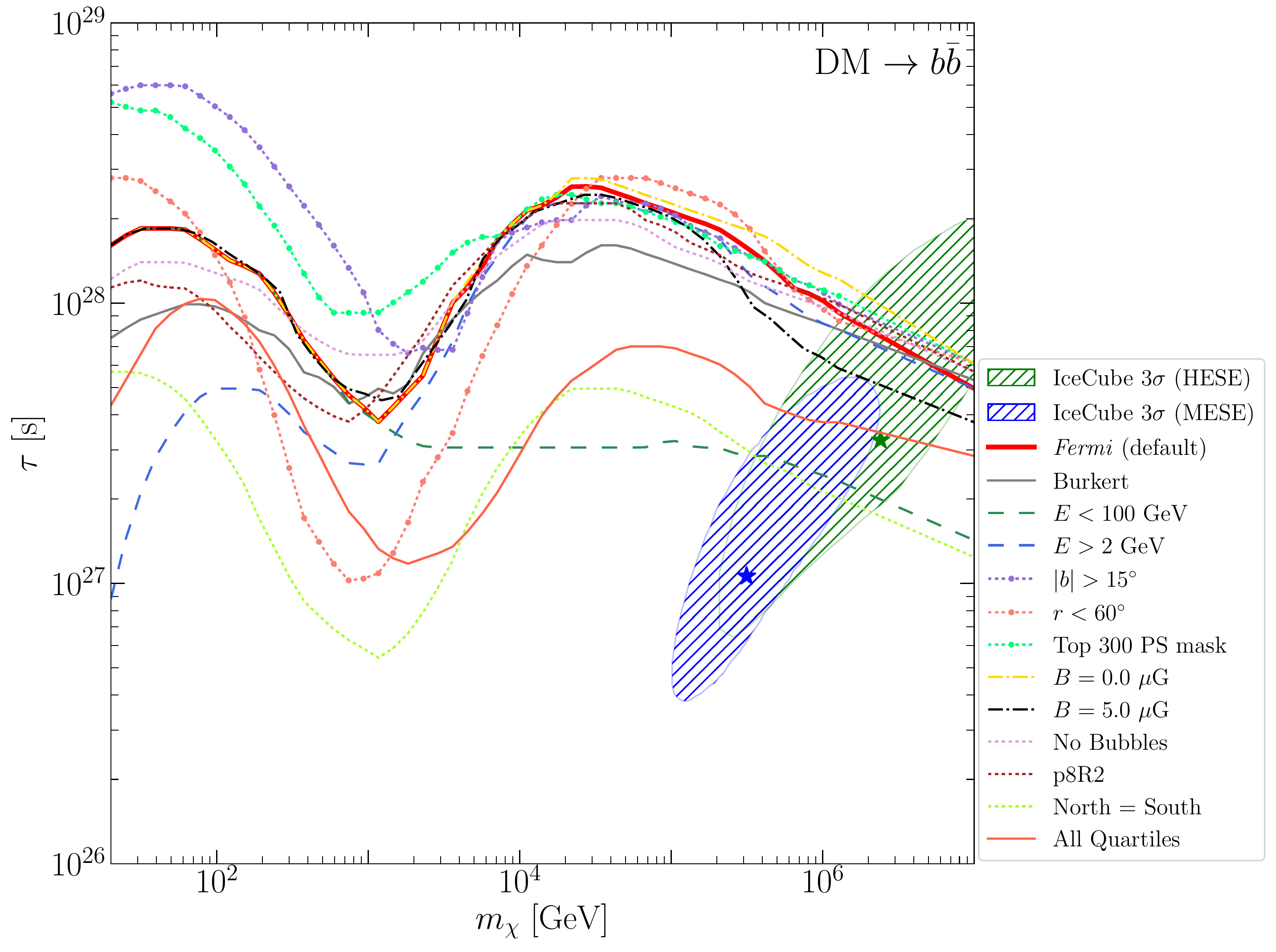}} & \scalebox{0.28}{\includegraphics{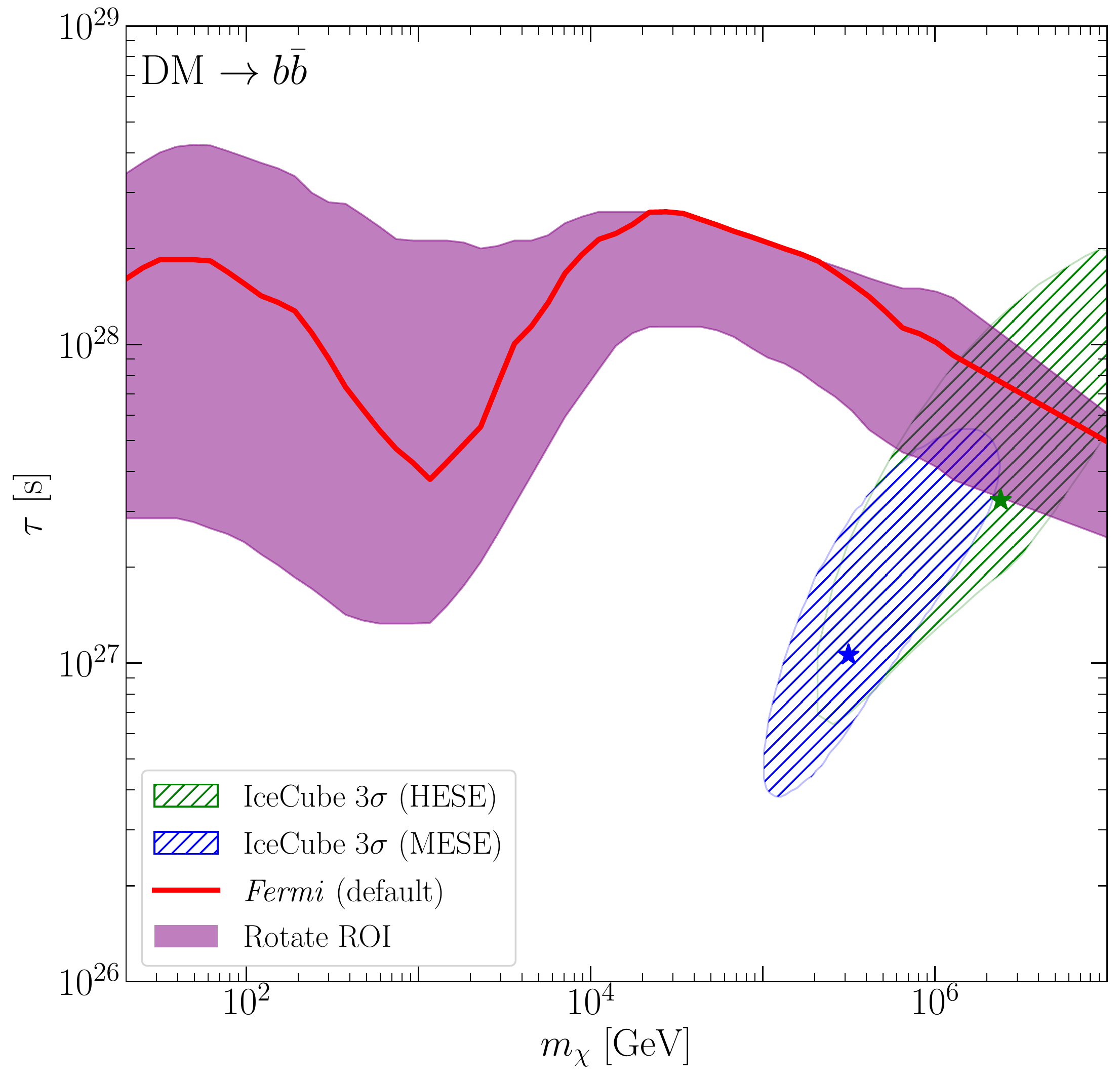}}
        \end{array}$
        \end{center}
        \vspace{-.70cm}
        \caption{Left: The limit derived for DM decay to $b \bar{b}$ for ten systematic variations on our analysis, as compared to our default analysis. Right: A purely data driven systematic cross check, where we have moved the position of our default ROI to five non-overlapping locations around the Galactic plane ($b=0$) and show the band of the limits derived from these regions is consistent with what we found for an ROI centered at the GC. See text for details.}
        \vspace{-0.15in}
        \label{Fig: DataSystematics}
\end{figure}

We have performed a variety of systematic tests to understand the robustness of our analysis.  Figure~\ref{Fig: DataSystematics} summarizes the results of some of the more important tests. 

In Fig.~\ref{Fig: DataSystematics}, we show limits on the $b\, \bar b$ final state with a variety of different variations on the analysis method.  Certain variations are shown to cause very little difference, such as not including an extra {\it Fermi} bubbles template, taking $B = 0.0$ $\mu$G when computing the IC flux, and using the more up-to-date Pass 8 model \texttt{gll\_iem\_v06} (\texttt{p8r2}) diffuse model instead of the \texttt{p7v6} model. As the \texttt{p8r2} model identifies regions of extended excess emission in the data and adds these back to the model, it is unclear if such a model would absorb a potential DM signal. Due to this concern, we used the \texttt{p7v6} model as our default in the main analyses.

Assuming $B = 5.0$ $\mu$G when computing the IC flux leads to slightly weaker constraints at higher masses due to the decrease in the IC contribution, as would be expected.  However, we emphasize that Faraday rotation measurements suggest that $B \leq 2.0$ $\mu$G across most of our ROI~\cite{Mao:2012hx}, so $5.0$ $\mu$G is likely overly conservative.

We also note that the limit $B \to 0.0$ $\mu$G must be taken with care.  Without any magnetic field, the energy loss rate of high energy electrons and positrons from IC emission alone is not sufficient to keep the leptons confined to the halo.  However, even taking $B \sim 0.1-1$ nG, which is a typical value quoted for intergalactic magnetic fields, the Larmor radius $\sim$$0.1~(E_e /100 \, \, \text{TeV})(1\, \, \text{nG}/B)$ kpc, with $E_e$ the lepton energy, is sufficiently small to confine the $e^{\pm}$ in our ROI.  Larger values of circumgalactic magnetic fields in the halo are more likely.

An additional systematic is the assumption of the DM profile, as direct observations do not sufficiently constrain the profile over our ROI and we must rely on models.  In this work, we have assumed the NFW profile.  Another well-motivated profile is the Burkert profile~\cite{Burkert:1995yz}, which is similar to the NFW at large distances but has an inner core that results in less DM towards the center of the Galaxy.  In Fig.~\ref{Fig: DataSystematics} we show the limit we obtain using the Burkert profile with scale radius $r_0 = 13.33$ kpc.  From this analysis we conclude that the systematic uncertainty from the DM profile is less significant than other sources of uncertainty associated with the data analysis.

Masking the top 300 brightest and most variable PSs across the full sky, instead of masking all PSs, and masking the Galactic plane at $|b| > 15^\circ$, instead of $20^\circ$, both lead to stronger constraints at low energies. This is not surprising considering that the PS mask at low energies significantly reduces the ROI, and so any increase to the size of the ROI helps strengthen the limit.  Going out to distances within $60^\circ$ of the GC, on the other hand, slightly strengthens the limit at low masses, gives a similar limit at high masses, but weakens the limit at intermediate masses.  This is due to the fact that the diffuse templates often provide poor fits to the data when fit over too large of regions.  As a result, it becomes more probable that the added NFW-correlated template can provide an improved fit to the data, which is the case at a few intermediate energies.  This is also the reason why the limit is found to be slightly worse when the templates are not floated separately in the North and South, but rather floated together across the entire ROI (North=South in Fig.~\ref{Fig: DataSystematics}).  As a result, we find that the addition of the NFW-correlated template often slightly improves the overall fit to the data in this case.  Since it is hard to imagine a scenario where a DM signal would show up in the North=South fit and not in the fit where the North and South are floated independently, and since the latter analysis provides a better fit to the data, we float the templates independently above and below the plane in our main analysis.
Reassuringly, most of the systematics do not have significant effects at high masses, where we are generally in the statistics dominated regime.  

\begin{figure}[t!]
        \leavevmode
        \begin{center}$
        \begin{array}{cc}
        \scalebox{0.4}{\includegraphics{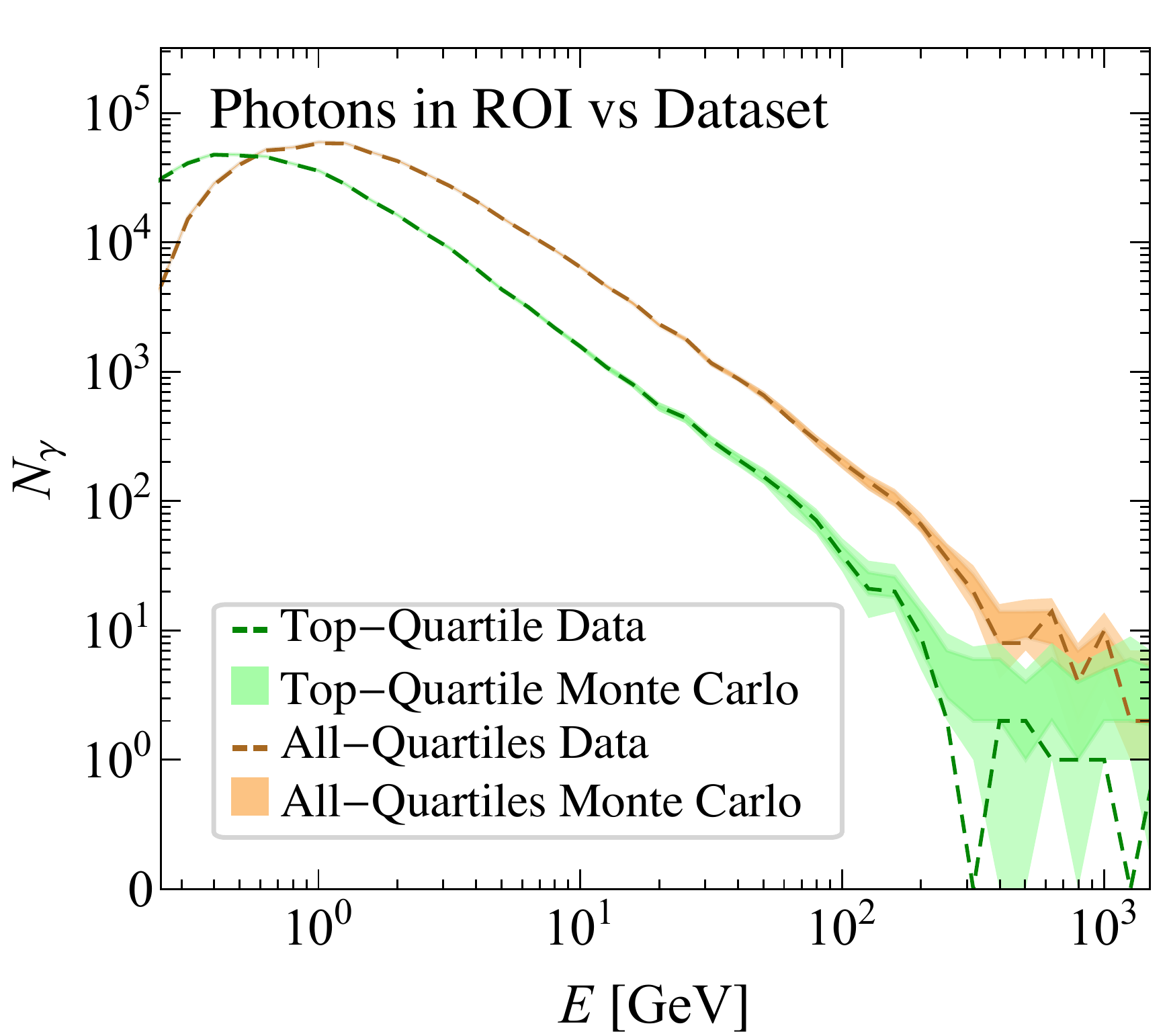}}  & \scalebox{0.4}{\includegraphics{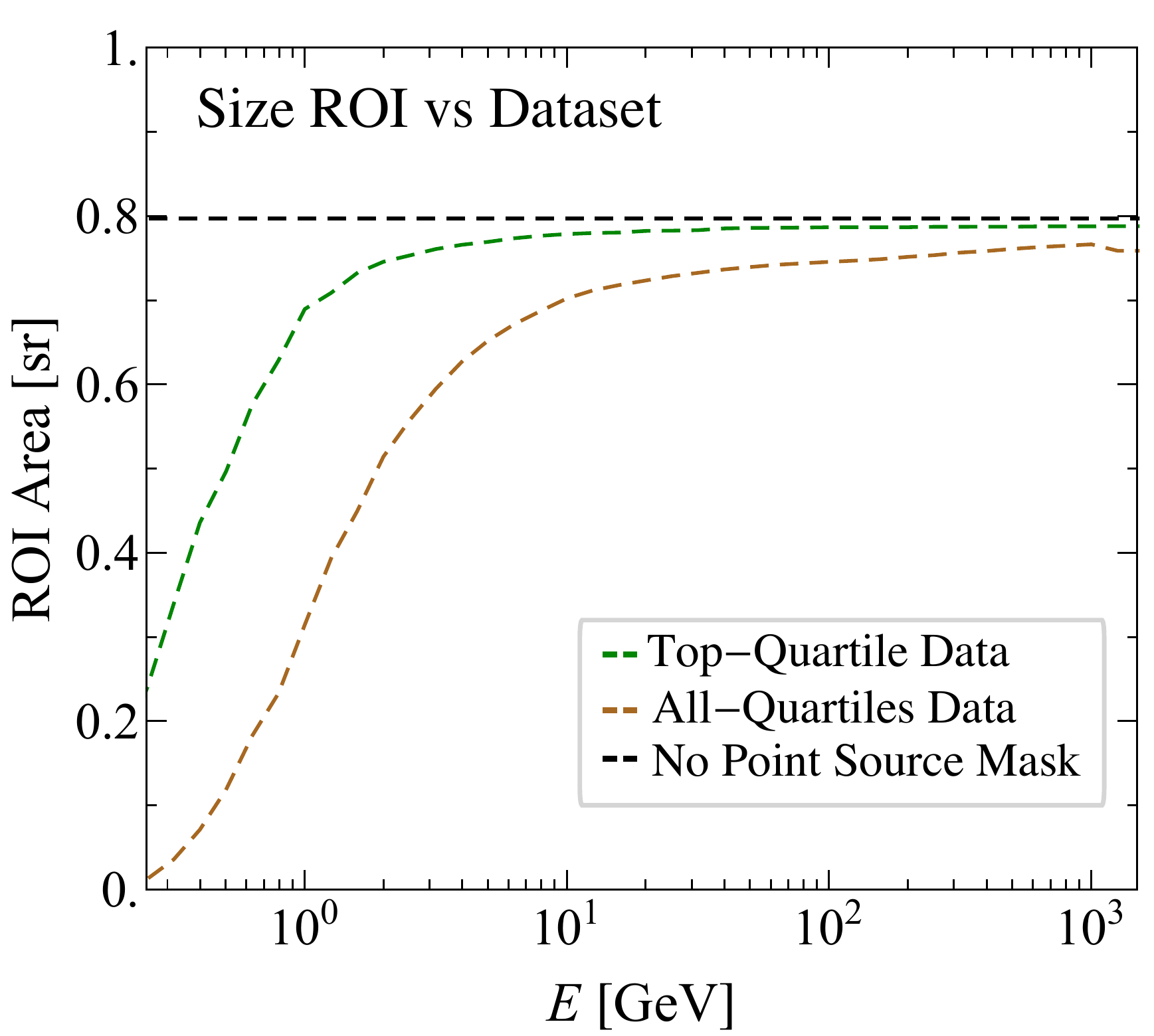}} \\
        \scalebox{0.4}{\includegraphics{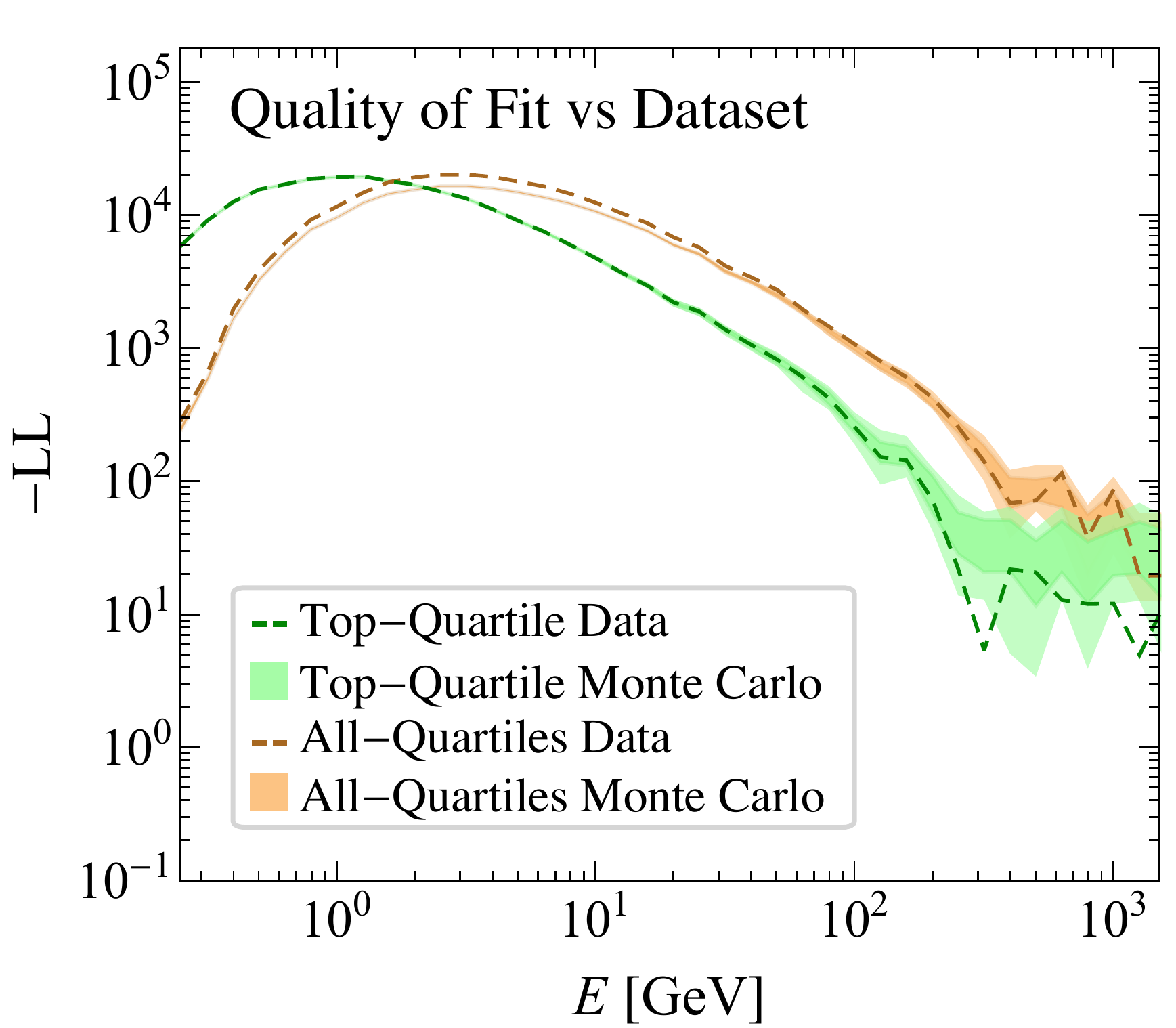}}  & \scalebox{0.41}{\includegraphics{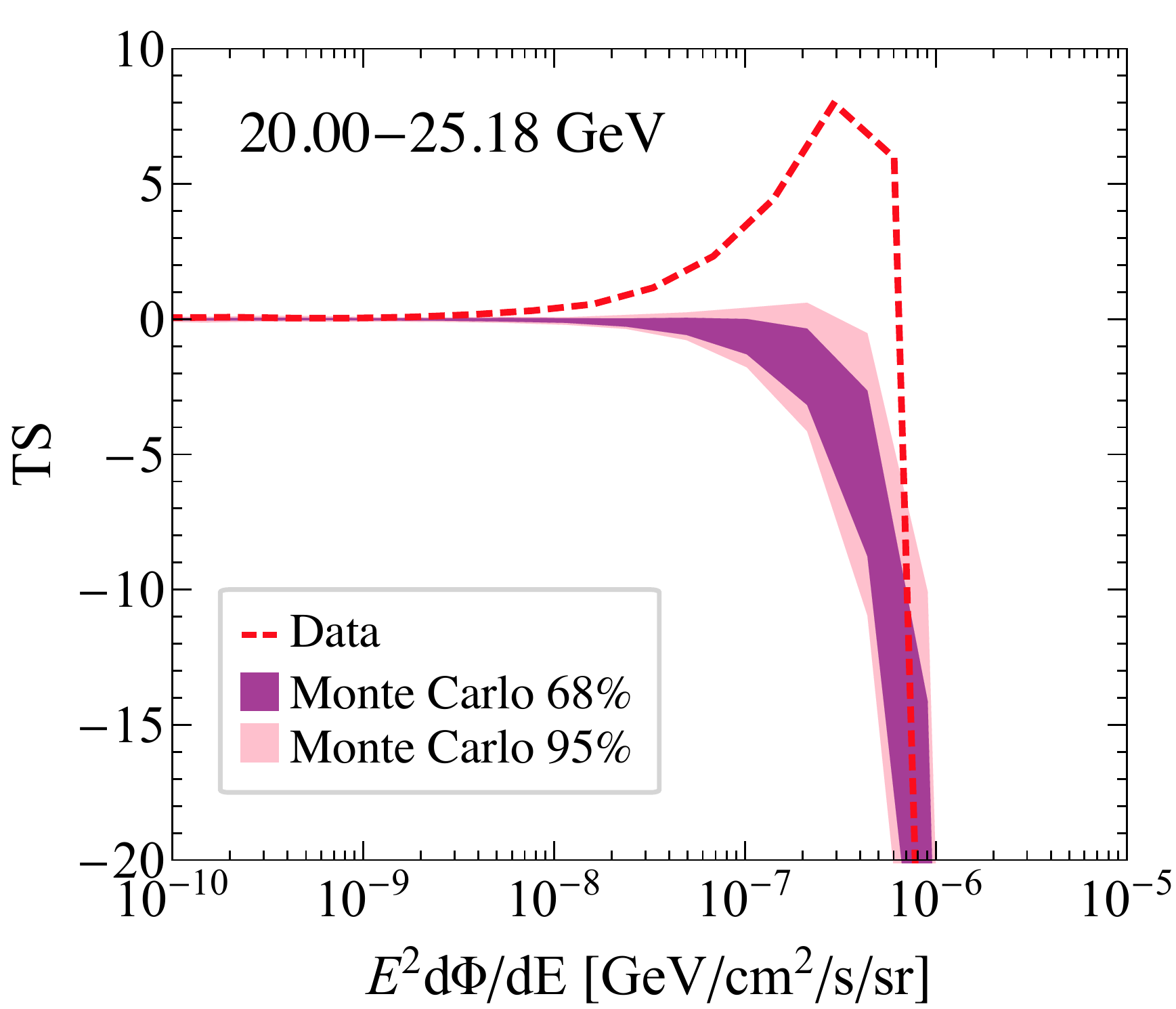}}
        \end{array}$
        \end{center}
        \vspace{-.70cm}
        \caption{Top left: the number of photons in our ROI as a function of energy for the top-quartile and all-quartiles. We show both the result in data and MC, where for the MC we indicate the 68\% and 95\% confidence intervals constructed from multiple MC realizations in each energy bin. Top right: the size of the ROI, in sr, as a function of energy. The variation with energy is due to the changing size of the PS mask. Bottom left: As in the top left plot, but here we show the quality of fit (the negative of the log-likelihood) as a function of energy. Bottom right: At intermediate energies there are residuals in the all-quartile data that can be absorbed by our Galactic DM template, leading to large excesses such as the one shown here. Such excesses play a role in the all-quartile limit being weaker than the top-quartile limit, as shown in Fig.~\ref{Fig: DataSystematics}.  However, the all-quartile limits are also weaker in part due to the reduced ROI at low energies.}
        \vspace{-0.15in}
        \label{Fig: UCVA}
\end{figure}

Many of the variations discussed above are associated with minimizing the impact of over-subtraction as discussed in chapter~\ref{chap:dmdecay}. Fundamentally, we do not possess a background model that describes the gamma-ray sky to the level of Poisson noise, and the choice of ROI can exacerbate the issues associated with having a poor background model. To determine our default ROI we considered a large number of possibilities and chose the one where we had the best agreement between data and MC, which ultimately led us to the relatively small ROI shown in Fig.~\ref{Fig:ROI} used for our default analysis. We emphasize that we did not choose the ROI where we obtained the strongest limits, as is clear from Fig.~\ref{Fig: DataSystematics}, and as such we do not need to impose a trials factor from considering many different limits.

A further important systematic is our choice of data set.  In our main analysis, we used the top quartile of events, as ranked by the PSF, from the UltracleanVeto class.  Roughly four times as much data is available, within the same event class, if we take all photons regardless of their PSF ranking.  Naively using all of the available data would strengthen our bound.  However, as we show in Fig.~\ref{Fig: DataSystematics}, this is not the case---in fact, the limit we obtain using all photons is weaker than the limit we obtain using the top quartile of events.  There are two reasons for this, both of which are illustrated in Fig.~\ref{Fig: UCVA}. The first reason is simply that since we mask PSs at 95\% containment, as determined by the PSF, there is less area in our ROI in the analysis that uses all quartiles of events relative to the analysis using the top quartile.  Indeed, in Fig.~\ref{Fig: UCVA} we show the number of counts $N_\gamma$ in the different energy bins in our ROI for the top-quartile and all-quartile analyses.  At high energies, the top-quartile analysis has fewer photons than the all-quartile analysis, as would be expected.  However, since the PSF becomes increasingly broad at low energies, we find that at energies less than around 1 GeV, the top-quartile data has a larger $N_\gamma$.  Since both the IC and extragalactic emission tends to be quite soft, the data at low energies has an important impact on the limits. We further emphasize this by showing the size of the ROI as a function of energy in Fig.~\ref{Fig: UCVA}.
 
The second difference between the two data sets is that with the top-quartile only we find that the data is generally consistent with the background models, up to statistical uncertainties, while with the full data set there are systematic differences between the data and background models across almost all energies.  This is illustrated in the bottom left panel of Fig.~\ref{Fig: UCVA}, where we compare the data result to expectations from MC (68\% and 95\% statistical confidence intervals) from the background templates only for the maximum log-likelihood. There we see that in the top quartile case the data is consistent with MC up to energies $\sim$100 GeV.  In the all-quartile case, on the other hand, the data appears to systematically have a larger log-likelihood than the MC at energies less than around 50 GeV.  This difference could again be due to the increased PSF in the all quartile case, which smears out small errors in background mis-modeling onto larger scales. The addition of a Galactic DM template can then be used to improve this mis-modeling, which can lead to a strong preference for the DM decay flux in isolated energy bins, an example of which is shown in the bottom right panel of Fig.~\ref{Fig: UCVA}. Such excesses weaken the limit that can be set and ultimately play a central role in the all-quartile limit being weaker than naively expected.
 
We note that even in the top-quartile case there does appear to be some systematic difference between the MC expectation and the data at energies greater than around 100 GeV.  In particular, the data appears to generally have fewer photons than expected from MC.  With that said, this is a low-statistics regime where some energy bins have $N_\gamma = 0$.  This difference is also not too surprising, considering that the PS model and diffuse model were calibrated at lower energies and simply extrapolated to such high energies.  Part of this difference could also be due to cosmic ray contamination.  Thus, systematic discrepancies between data and MC at energies greater than around 100 GeV should be expected.  To illustrate the importance of this high energy data on our results, we show in Fig.~\ref{Fig: DataSystematics} the limit obtained when only including data with photon energies less than $100$ GeV; at 10 PeV, the limit is around 5 times weaker without the high-energy data. We also show in that plot the impact of removing the data below $2$ GeV, which has a large impact at lower masses but a minimal impact at higher masses.

In addition to the numerous variations of our modeling discussed above, we have also performed a purely data driven systematic cross check on our analysis shown on the right of Fig.~\ref{Fig: DataSystematics}, similar to that used in \cite{Calore:2014xka}. In the absence of any DM decay flux in the {\it Fermi} data, there should be nothing particularly special about the ROI near the Inner Galaxy that we have used---shown in Fig.~\ref{Fig:ROI}---and we should be able to set similar limits in other regions of the sky. This is exactly what we confirm in Fig.~\ref{Fig: DataSystematics}, where in addition to our default limit we show the band of limits derived from moving our ROI to five non-overlapping regions around the Galactic plane ($b=0$). As shown in the figure, even allowing for this data driven variation, the best fit IceCube points always remain in tension with the limit we would derive.

As a final note, we emphasize the importance of modeling non-DM contributions to the gamma-ray data in addition to the spatial morphology of the signal.  The limits on the DM lifetime would be weaker if we used a more simplistic analysis that did not incorporate background modeling and spatial dependence into the likelihood.  For example, we may set a conservative limit on the DM lifetime by using a likelihood function 
\es{LLsimp}{
\log p_i(d | \psi) = \sum_i \max_{\lambda_i}  \left[- {\left(  \sum_p \mu_i^p(\psi, \lambda_i) - \sum_p n_i^p \right)^2 \over 2  \sum_p \mu_i^p(\psi, \lambda_i)  }  - {1 \over 2} \log \left( 2 \pi  \sum_p \mu_i^p(\psi, \lambda_i) \right) \right]\,. 
}
The likelihood function depends on $ \sum_p \mu_i^p(\psi, \lambda_i) \equiv \sum_p \mu_i^p(\psi) +  \lambda_i$, which is a function of the  DM model parameters $\psi$.  The $ \lambda_i$ are nuisance parameters that allow us to add an arbitrary (positive) amount of emission in each energy bin.  These nuisance parameters account for the fact that we are assuming no knowledge of the mechanisms that would yield the gamma-rays recorded in this data set---the data may arise from DM decay or from something else.  As a corollary to this point, we may only determine limits with this likelihood function; by construction, we cannot find evidence for decaying DM.   Using~\eqref{LLsimp} within our ROI, we estimate a limit $\tau \approx 1 \times 10^{27}$ s for DM decay to $b \, \bar{b}$ with $m_\chi=1\,$PeV.  This should be contrasted with the limit $\tau \approx 1 \times 10^{28}$ s that we obtain with the full likelihood function, as given in Eq.~\eqref{LLfunction}.  This emphasizes the importance of including spatial dependence and background modeling in the likelihood analysis, as this knowledge increases the limit by around an order of magnitude in this example.  Even more important is the inclusion of energy dependence in~\eqref{LLsimp}.  Were we to modify~\eqref{LLsimp} to only use one large energy bin from $200$ MeV to $2$ TeV, then the limit would drop to $\sim$$10^{25}$ s in this example.  However, it is important to emphasize that the DM-induced flux is orders of magnitude larger than the data at high energies for this lifetime.

\newpage
\section{Extended Theory Interpretation}
\label{sec: models}

In this section, we expand upon the decaying DM interpretation of our results in the context of additional final states and also specific simplified models.  We begin by giving limits on a variety of two-body final states.  Then, we comment on how we may use universal scaling relations to extend our result to high DM masses, beyond where it is possible to generate the spectra in \textsc{Pythia}. Finally, we illustrate this point by providing two example models. The limits on all final states and models considered in this work are provided as part of the Supplementary Data~\cite{supp-data}.

\subsection{Additional Final States}
\label{sec: additional}
\begin{figure}[t!]
	\leavevmode
	\begin{center}$
	\begin{array}{ccc}
	\scalebox{0.34}{\includegraphics{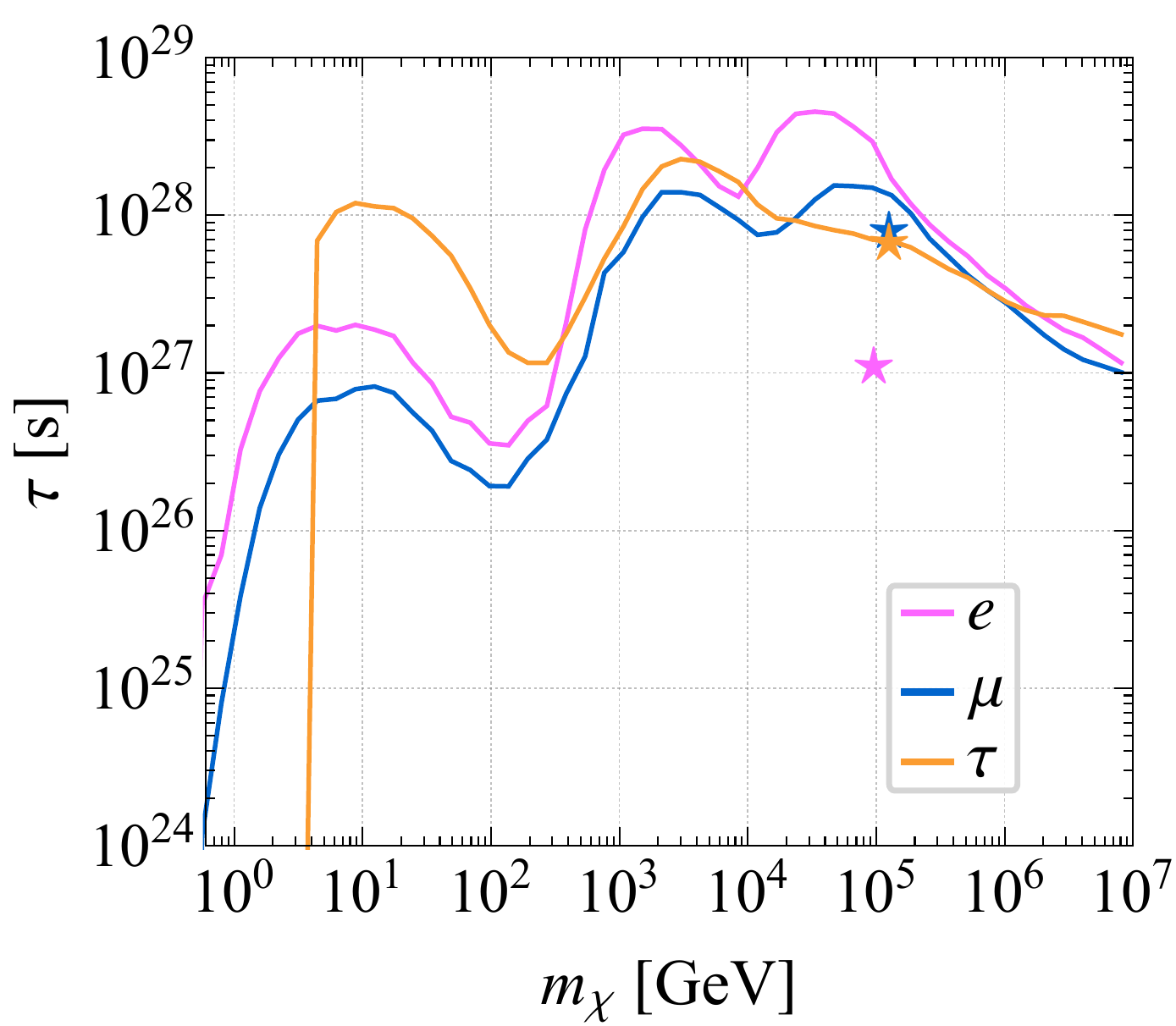}} & \scalebox{0.34}{\includegraphics{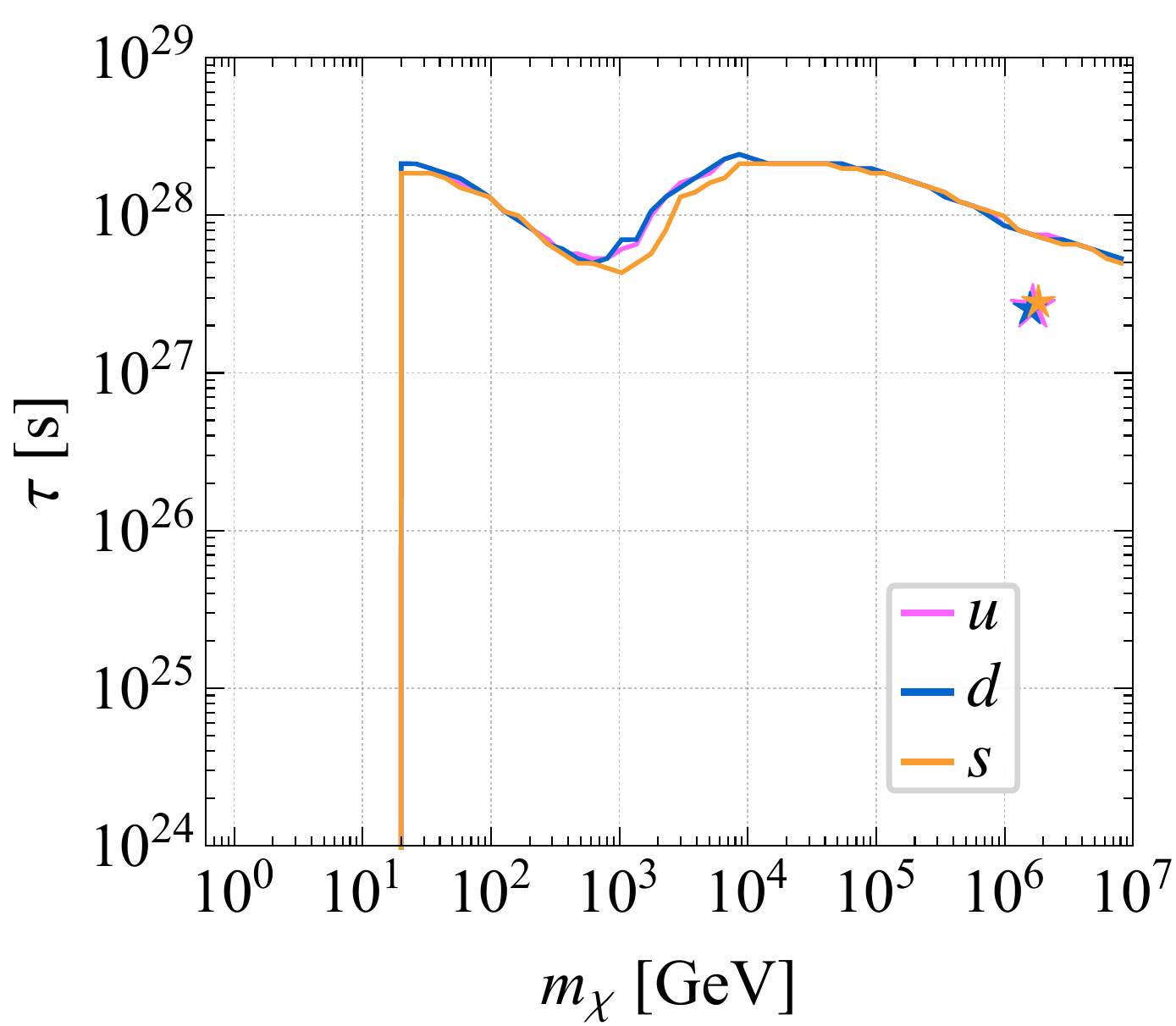}} & \scalebox{0.34}{\includegraphics{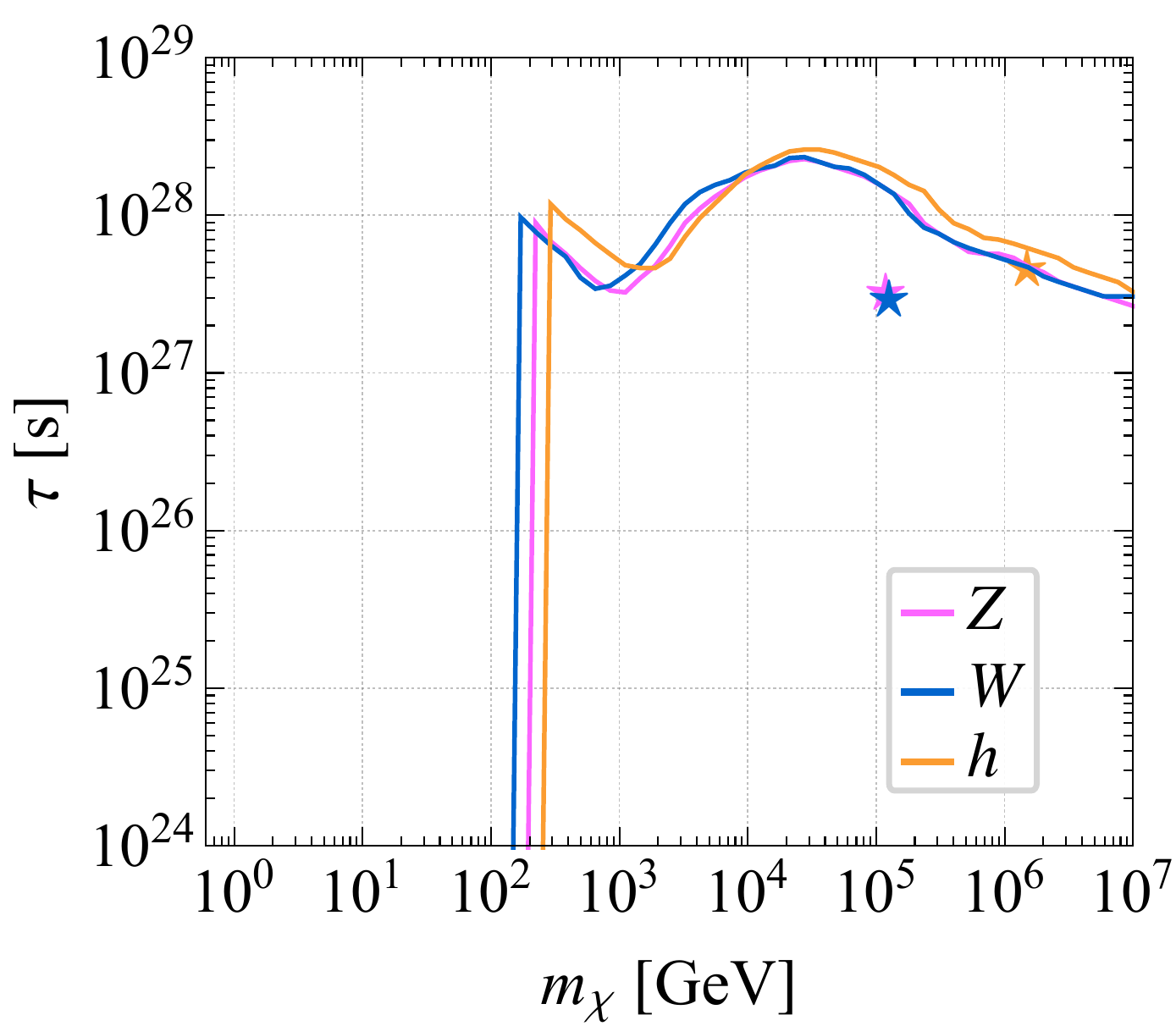}} \\
	\scalebox{0.34}{\includegraphics{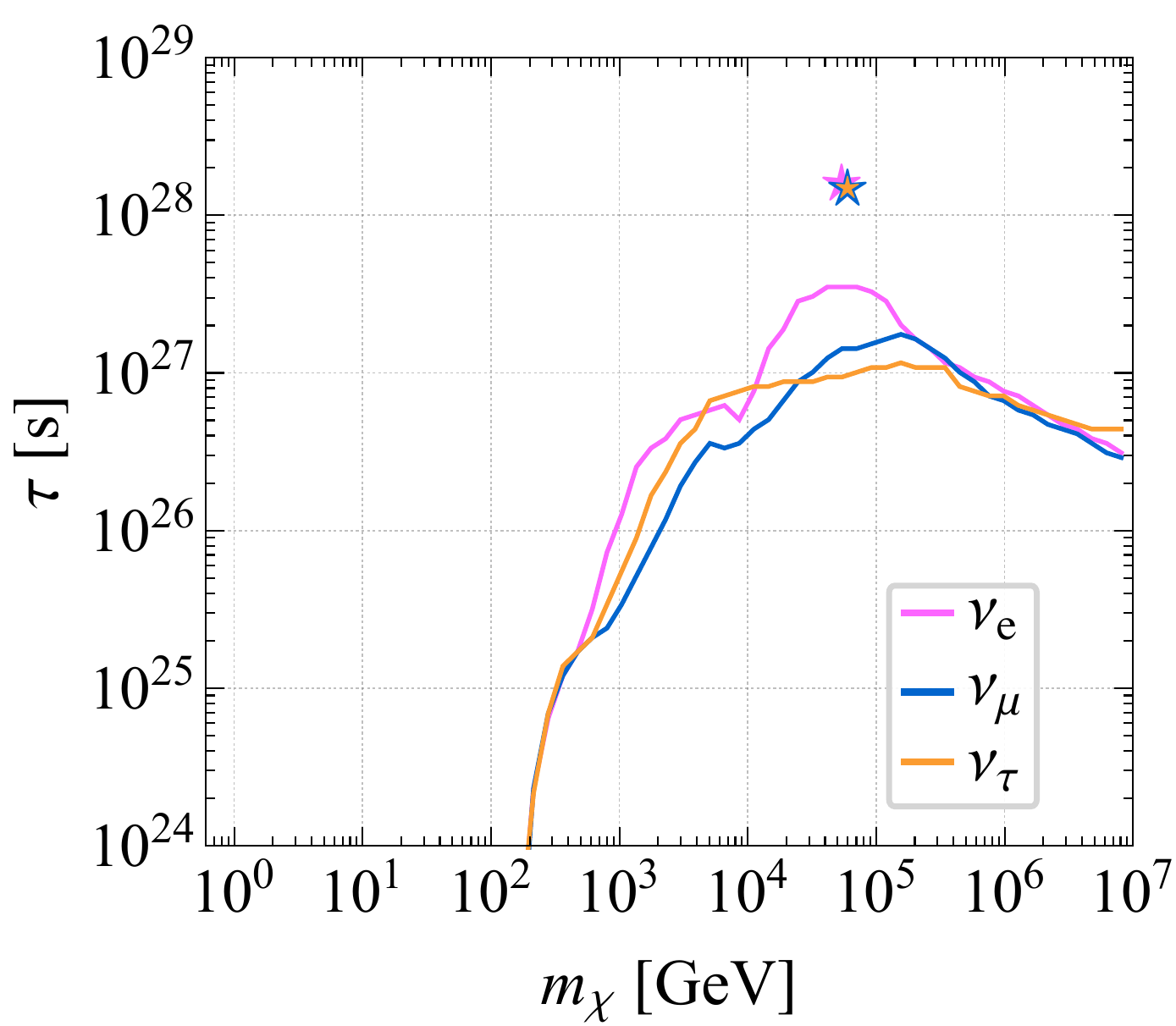}} & \scalebox{0.34}{\includegraphics{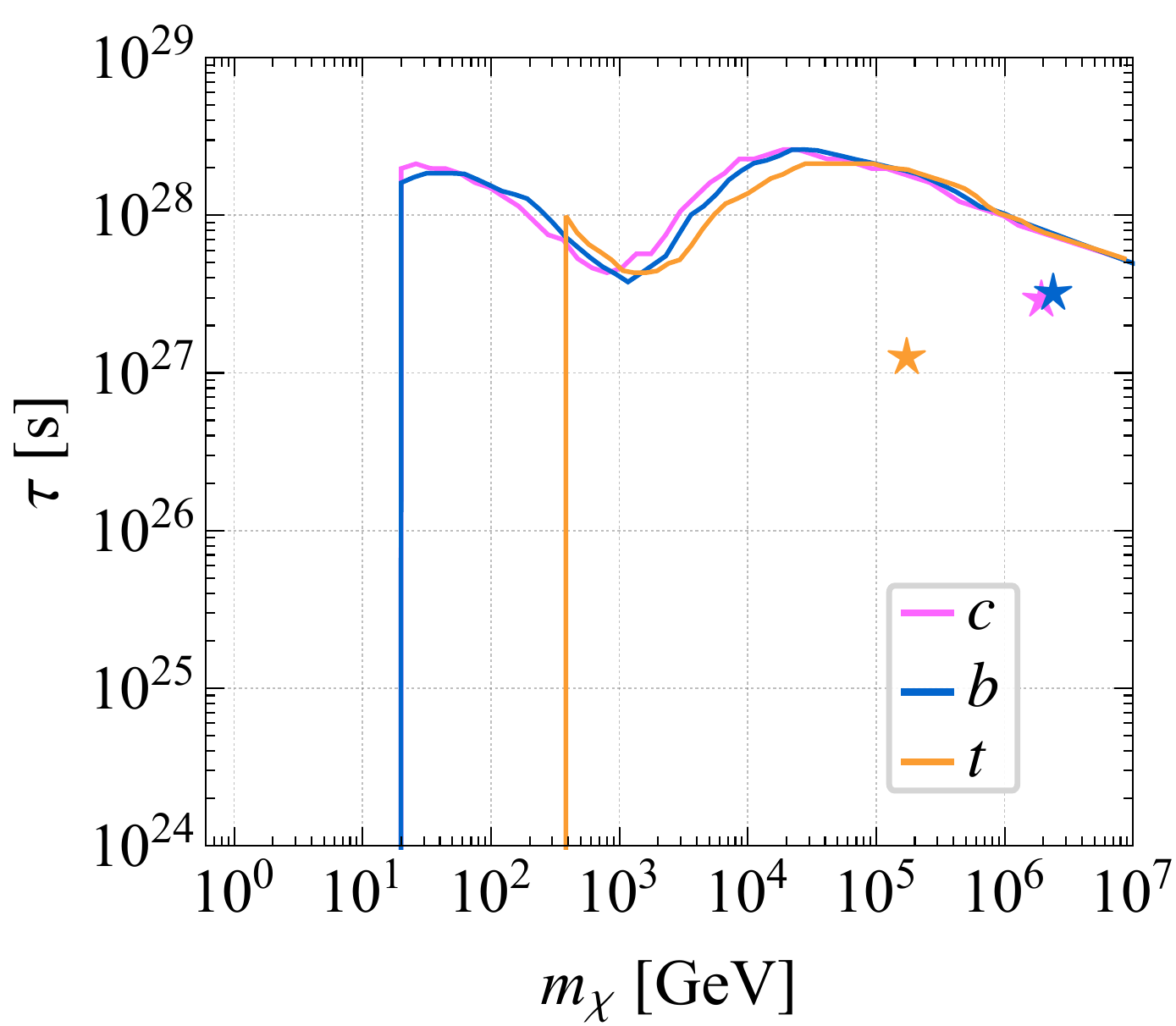}} & \scalebox{0.34}{\includegraphics{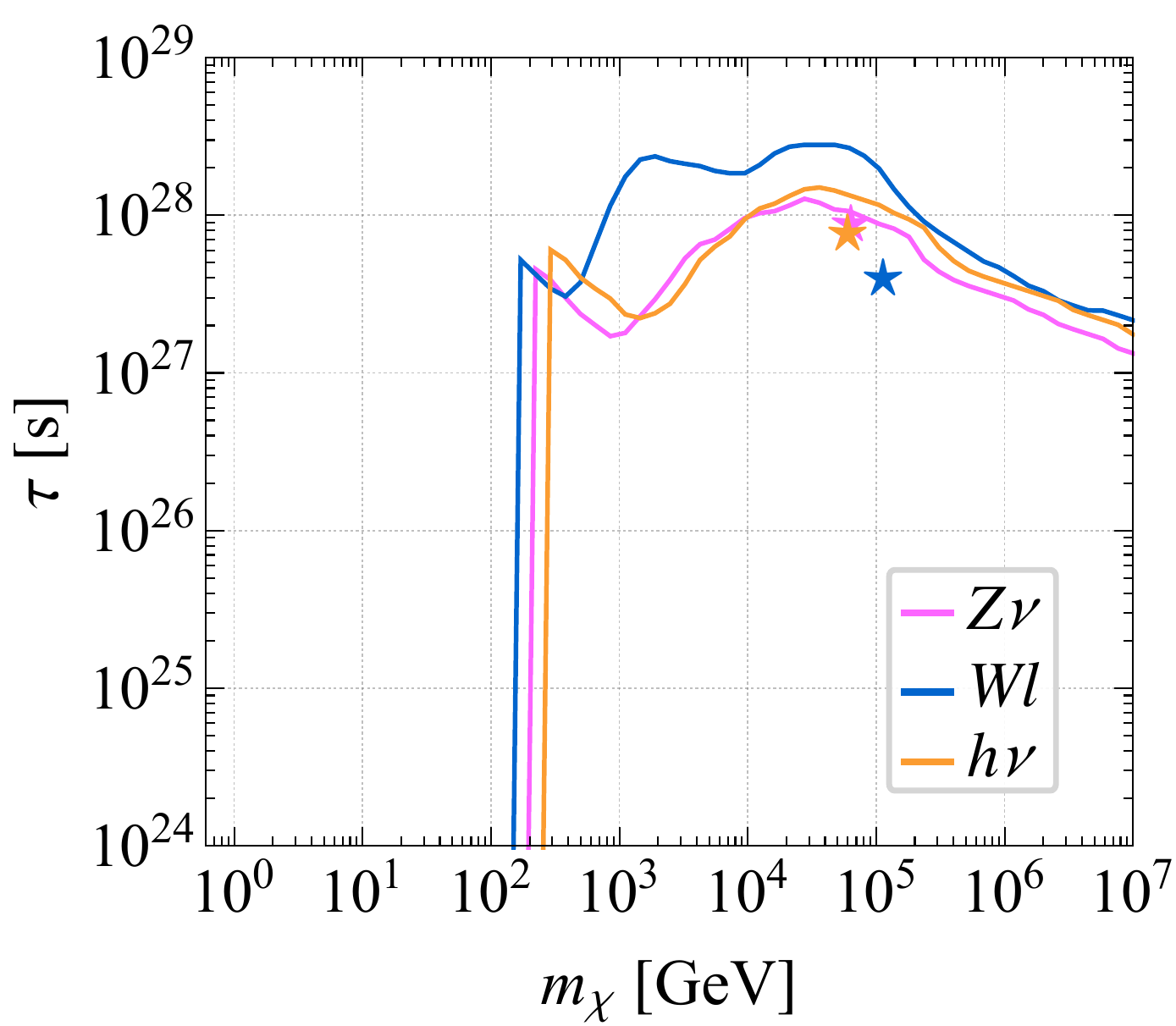}} \\ 
	\end{array}$
	\end{center}
	\vspace{-.70cm}
	\caption{Limits on all final states considered in this work. For each final state we show both the limit on the decay lifetime as a function of the DM mass, and also the best fit point for an interpretation of the IceCube flux with this channel as a star. Except for decay directly into neutrinos, for every other final state this best fit point is in tension with the limit we derive from {\it Fermi}.}
	\vspace{-0.15in}
	\label{Fig:OtherFSLims}
\end{figure}

In addition to DM decays directly to bottom quarks, the benchmark final state used extensively in this work, we also determine the {\it Fermi} limits on DM decay into a number of two-body final states. In detail we consider all flavor conserving decays to charged leptons, neutrinos, quarks, electroweak bosons, and Higgs bosons.  Due to our emphasis on modes that yield high energy neutrinos, we also include three mixed final states, $Z \nu$, $W \ell$ and $h \nu$. For these last three cases we consider an equal admixture of lepton and neutrino flavors. These limits are all shown in Fig.~\ref{Fig:OtherFSLims}. 

\begin{figure}[t!]
	\leavevmode
	\begin{center}$
	\begin{array}{c}
	\scalebox{0.3}{\includegraphics{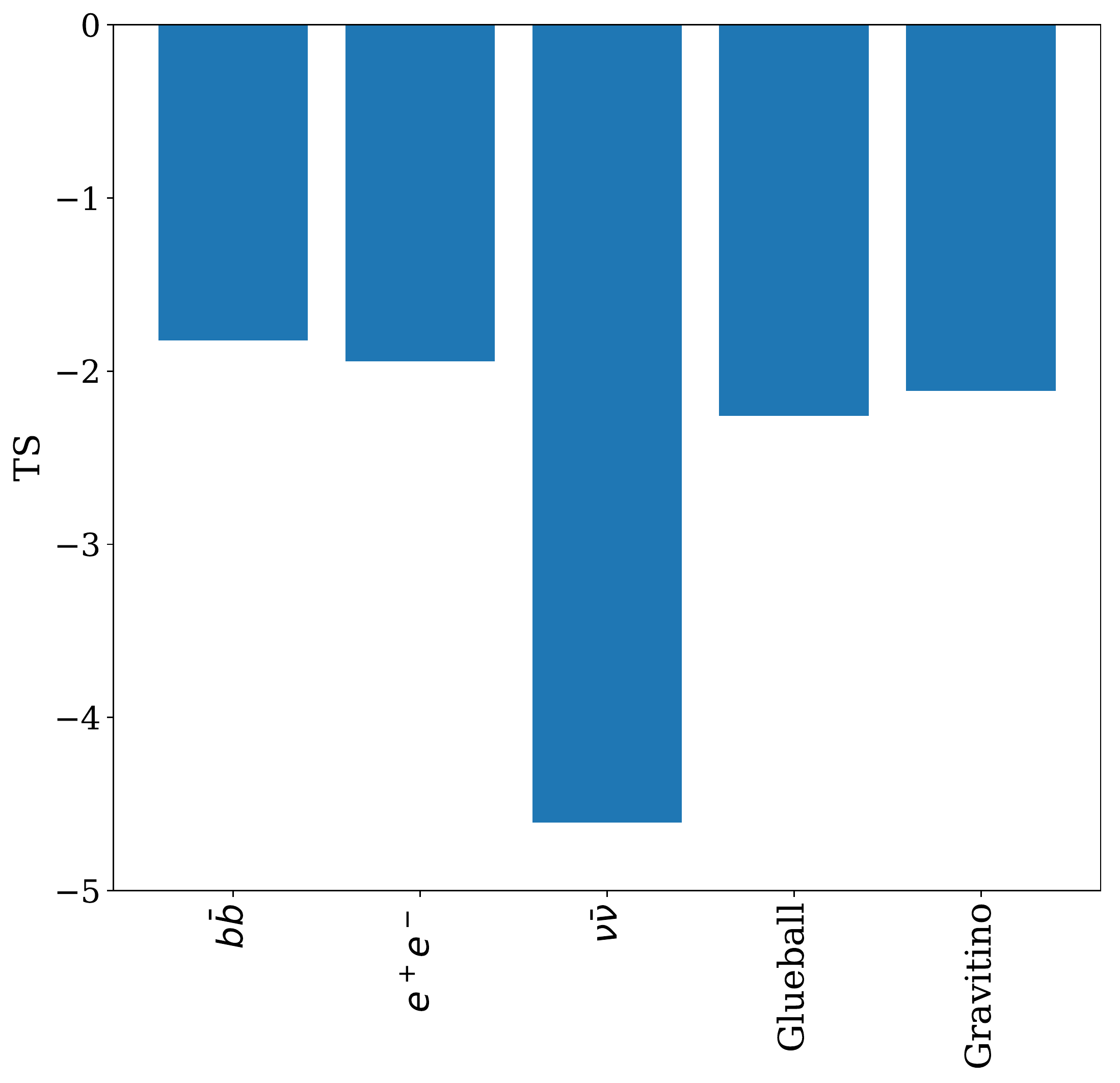}}
	\end{array}$
	\end{center}
	\vspace{-.70cm}
	\caption{Quality of the best fit to the combined IceCube data for a selection of final states and models. The quality of the fit is represented as a TS for the DM-only model defined with respect to the best fit power law with an exponential cutoff; this simple model is meant to represent an astrophysical fit to the data. Among the DM-only models, $b\, \bar b$ provides the best fit to the data, which motivated our choice to focus on it in chapter~\ref{chap:dmdecay}. Other final states and models give a comparative goodness of fit, except for the case of decay directly to neutrinos which gives a sharp spectrum and consequently a poor fit.}
	\vspace{-0.15in}
	\label{Fig:bestFit}
\end{figure}

Figure~\ref{Fig:OtherFSLims} has some interesting features. Channels which produce more electrons and positrons tend to have stronger limits at high masses due to the associated Galactic IC flux. This is clear for DM decays to $e^+ e^-$ and also to $\nu_e\, \bar{\nu}_e$. Most of the quark final states lead to nearly identical limits; these channels produce a large number of pions regardless of flavor yielding a similar final state spectrum. The only difference is for the top quark, which first decays to $b W$, thereby generating a prompt spectrum which differs from the lighter quarks. Note that for the lighter quarks, the threshold is still always set at 20 GeV; \textsc{Pythia} does not operate below this energy since they do not simulate the full spectrum of QCD resonances.  We leave the extension of our results to lower masses for colored final states to future work.

In addition to the limits, we also show the best fit point for a fit to the IceCube data as a star for each channel.  The best fit point is always in tension with the limits we derive from {\it Fermi}, except for decays directly into neutrinos.  However, as we show in the next subsection, when modeling the DM interactions in a consistent theory context, one must rely on a very restricted setup to manifest exclusive decays into neutrino pairs.

The quality of fit for the different stars represented in Fig.~\ref{Fig:OtherFSLims} are not identical. This point is highlighted in Fig.~\ref{Fig:bestFit} where we show the quality of fit (for DM only) for three two-body final states, $b$, $e$, and $\nu_e$, as well as two models, glueball and gravitino dark matter. The quality of fit is shown with respect to the best fit power law multiplied by an exponential cutoff, chosen to represent an astrophysical fit to the data. The astrophysical model always gives the best fit, with the $b\, \bar{b}$ DM-only model a worse fit to the data by a TS $\sim 1.9$. A number of the other final states and models also give a comparable quality of fit to the $b \, \bar b$ final state, as their neutrino spectra are all broad enough to fit the data in a number of energy bins. This is not the case for $\nu_e\, \bar{\nu}_e$---the only final state not in tension with our limits from \textit{Fermi}---where the sharp neutrino spectrum can at most meaningfully contribute to a single energy bin. 

\subsection{Extending \textbf{\textit{Fermi}} limits beyond 10 PeV}
\label{sec:ExtendingToGUTScale}

As discussed above, generating the prompt spectra much above 10\,PeV in \textsc{Pythia} is not feasible. The issue is already clear in Tab.~\ref{table:Particles}: as the DM mass is increased, so is the energy injected into the final state decays which leads to a large number of final states resulting from the showering and hadronization processes. At some point this process simply takes too long to generate directly. Nevertheless, this section provides the details of the spectrum generation for $b\, \bar{b}$, and then how these are utilized to extend our {\it Fermi} limits up to energies $\sim$$10^{12}$ GeV.

\begin{figure}[t]
        \leavevmode
        \begin{center}$
        \begin{array}{cc}
        \scalebox{0.4}{\includegraphics{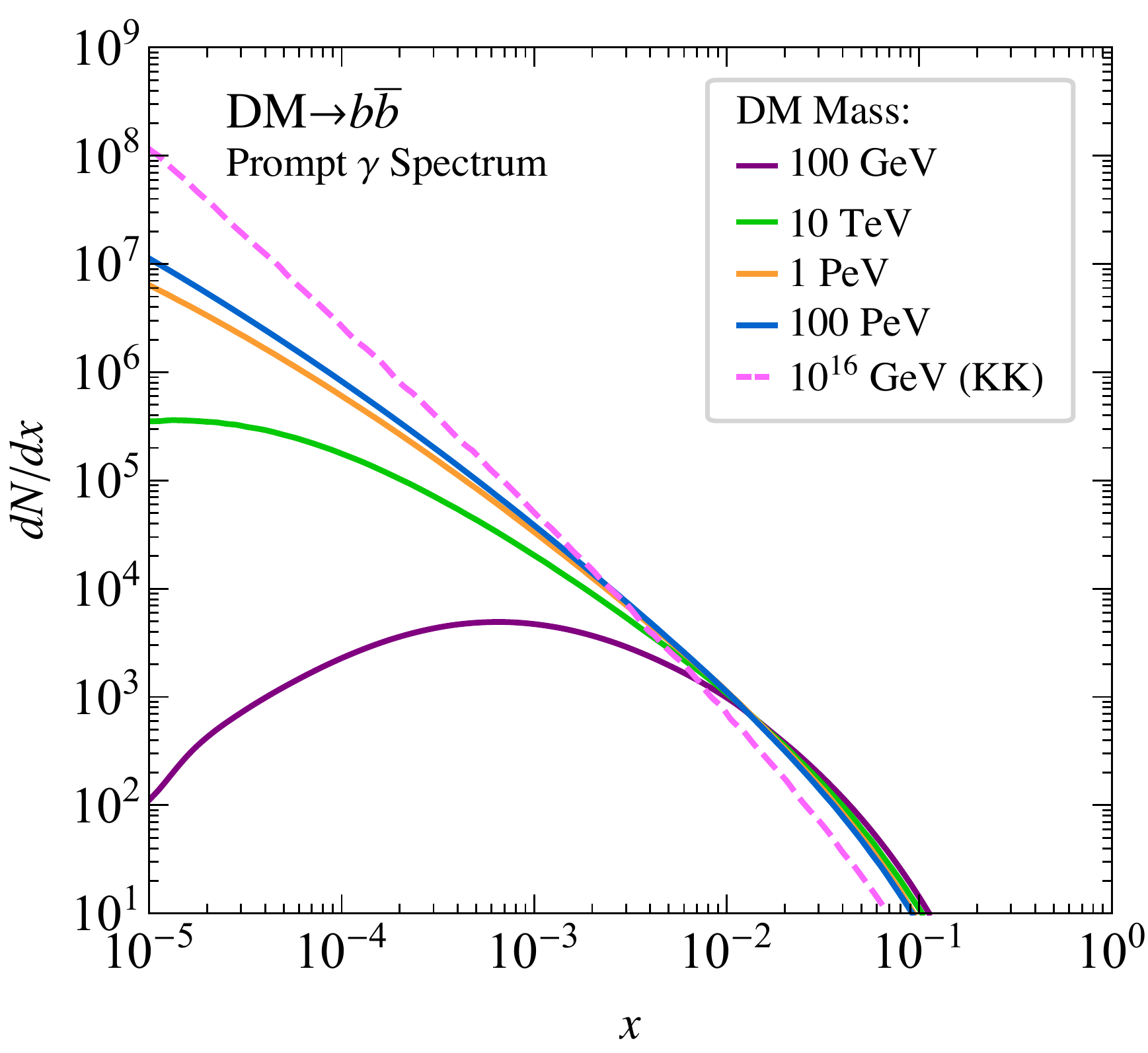}}  &
        \scalebox{0.41}{\includegraphics{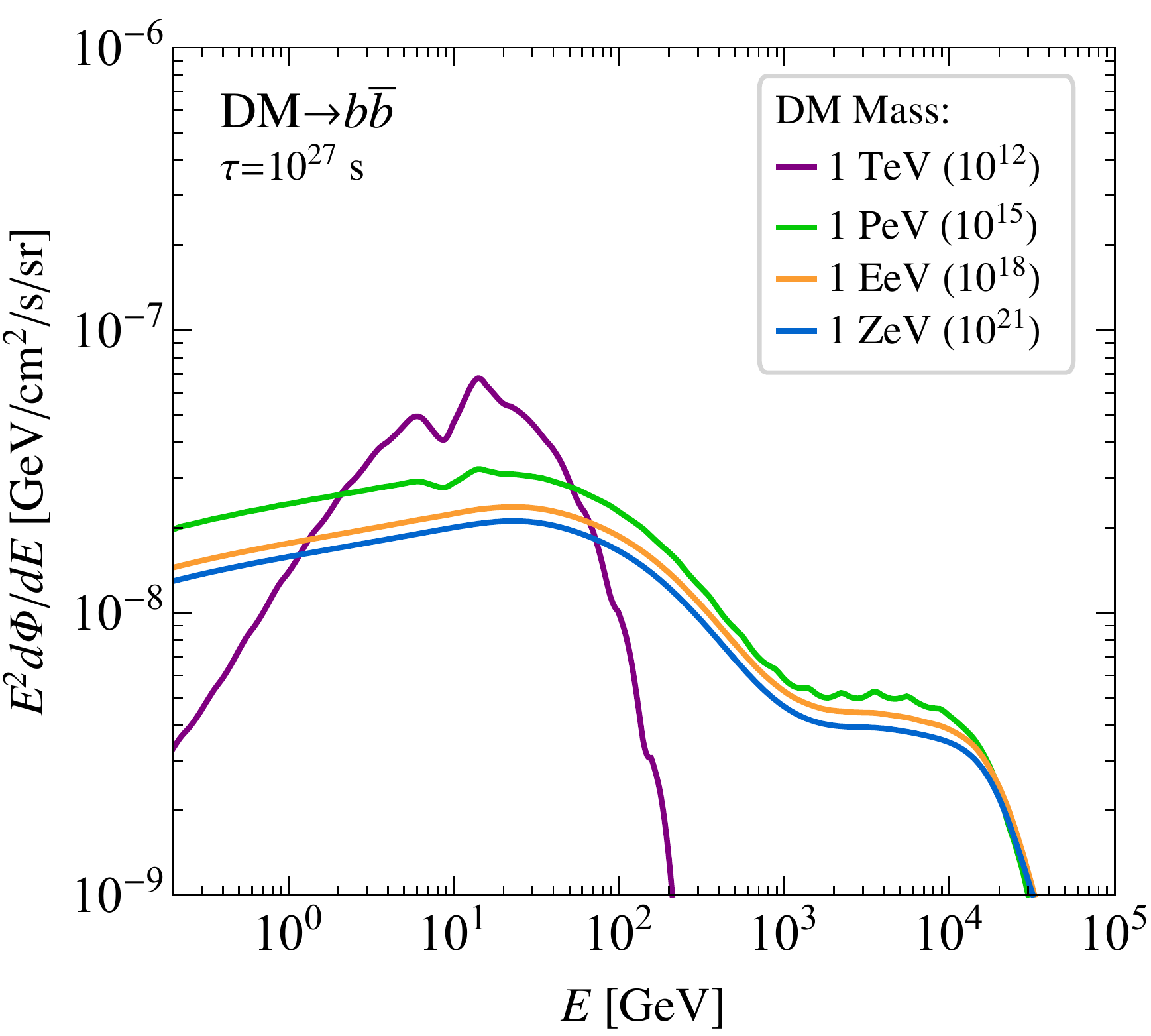}}
        \end{array}$
        \end{center}
        \vspace{-.70cm}
        \caption{Left: The prompt photon DM decay $b \,\bar{b}$ spectrum approaches a universal form in $dN/dx$, where $x = E/m$. All spectra except for the one at $10^{16}$ GeV were determined using \textsc{Pythia}; the spectrum at the GUT scale is taken from \cite{Kalashev:2016cre} and labelled as KK. The prompt $e^{\pm}$ and neutrino spectra also approach universal forms. Taken together this allows us to determine the $b \bar{b}$ spectrum up to masses $\sim$$10^{12}$ GeV. Right: At very high masses the Galactic flux from DM decay expected in the {\it Fermi} energy range is negligible. Nevertheless due to cascade processes, the extragalactic flux, shown here here for DM with $\tau = 10^{27}$ s, approaches a universal form. This implies {\it Fermi} can set an essentially mass independent limit on very heavy dark matter, as shown in Fig.~\ref{Fig:GUTLim}.}
        \vspace{-0.15in}
        \label{Fig:UniversalSpec}
\end{figure}

The key observation is that when the prompt photon, electron/positron, or neutrino spectra are considered in terms of $dN/dx$ where $x=E/m_\chi$, for $b\, \bar{b}$, and likely many other channels though we have not fully characterized this for all final states, they approach a universal form independent of mass. This is shown for the case of photons on the left of Fig.~\ref{Fig:UniversalSpec}. There we show \textsc{Pythia} generated spectra up to 100 PeV, and compare them to a spectrum at the GUT scale determined in \cite{Kalashev:2016cre}. The computation in~\cite{Kalashev:2016cre} takes the fragmentation function for bottom quarks at lower energies, and then runs them to the GUT scale using the DGLAP evolution equations. This universality allows us to determine the prompt spectra for ${\rm DM} \to b\, \bar{b}$ with $m_\chi$ well above the PeV scale.

Given these spectra, the next consideration is whether a meaningful flux from these decays populates the {\it Fermi} energy range. For prompt and IC flux from the Milky Way the answer is no, as is evident already in Fig.~\ref{Fig: highlowbSpec}. The synchrotron flux from electrons and positrons is expected in the Fermi range or even higher energies for DM mass of $\gtrsim{10}^9$~GeV, which can improve the lifetime limits by a factor of 2-3~\cite{Murase:2012xs}. However, the results depend on halo magnetic fields that are uncertain. Thus, we here consider conservative constraints without the Galactic synchrotron component.  Nevertheless the situation is different for the extragalactic flux, as shown on the right of Fig.~\ref{Fig:UniversalSpec}. There we see that the amount of flux approaches a universal form, essentially independent of the DM mass. The intuition for how this is possible is as follows. The total DM energy injected in decays is independent of mass: as we increase the mass of each DM particle, the number density decreases as $1/m_{\chi}$, but at the same time the power injected per decay increases as $m_{\chi}$, keeping the total injected power constant. Extragalactic cascades reprocess this power into the universal spectrum shown, and this implies that above a certain mass the extragalactic flux seen in the {\it Fermi} energy range becomes a constant.

\begin{figure}[t]
        \leavevmode
        \begin{center}$
        \begin{array}{c}
        \scalebox{0.4}{\includegraphics{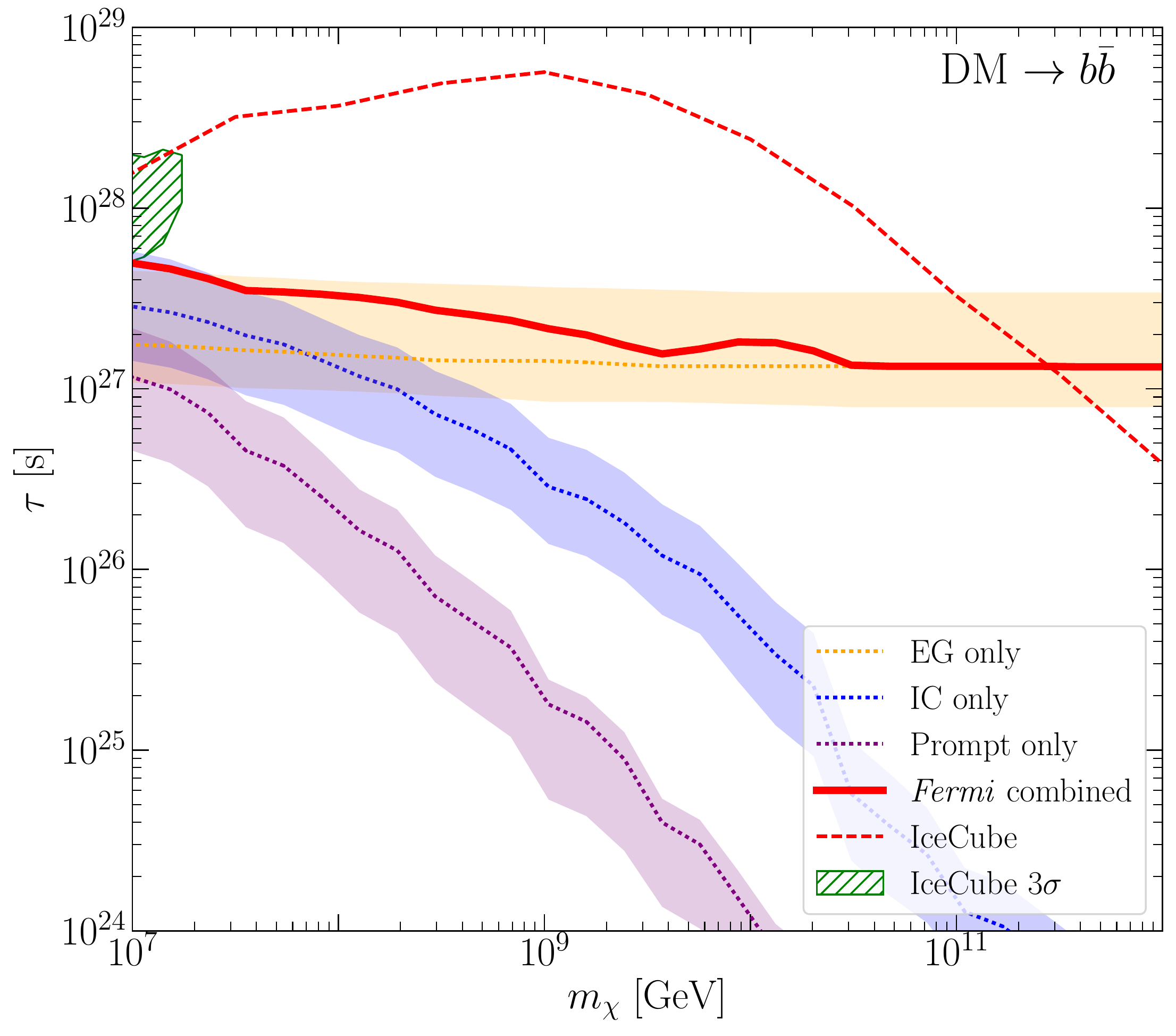}}
        \end{array}$
        \end{center}
        \vspace{-.70cm}
        \caption{Using the universal form of the spectra shown in Fig.~\ref{Fig:UniversalSpec}, {\it Fermi} can set limits on DM decays up to masses well above the PeV scale. At higher masses this limit comes only from the extragalactic contribution, such that after about $10^{10}$ GeV, the limit set becomes essentially independent of mass. Note that at these high masses, the {\it Fermi} limits are noticeably weaker than those obtained by direct searches for prompt Galactic photons from the decay of these heavy particles, as determined in \cite{Kalashev:2016cre}.  Note that the labeling is the same as in Fig.~\ref{Fig: glue} of chapter~\ref{chap:dmdecay}.}
        \vspace{-0.15in}
        \label{Fig:GUTLim}
\end{figure}

Using this, we extend our limits on the $b\, \bar{b}$ final state up to the masses $\sim$$10^{12}$ GeV in Fig.~\ref{Fig:GUTLim}. There we see that above $\sim$$10^{10}$ GeV, the limit becomes independent of mass and is coming only from the extragalactic contribution, exactly because of the universal form of the extragalactic flux. The same is not true for the neutrino spectrum---there is no significant reprocessing of the Galactic or extragalactic neutrino flux---and as such the limits IceCube would be able to set decrease with increasing mass. Despite this, limits determined from direct searches for prompt Galactic photons, which at these high energies are not significantly attenuated, set considerably stronger limits as shown in~\cite{Kalashev:2016cre}. Nevertheless, given that {\it Fermi} cannot see photons much above 2 TeV, we find it impressive that the instrument can set limits up to these masses.

We have cut Fig.~\ref{Fig:GUTLim} off at masses $\sim$$10^{12}$ GeV because at higher energies processes such as double pair-production may become important (see~\cite{Bhattacharjee:1998qc} for a review and references therein).  The neutrino limits may also be affected by scattering off the cosmic neutrino background at very high masses.  We leave such discussions to future work.

\subsection{Additional Models}
\label{sec: add models} 

In this section, we give limits on two additional DM models of interest beyond the example of a hidden sector glueball decaying via the operator $\lambda_D  \,G_{D\mu\nu}\,G_D^{\mu \nu}\, |H|^2 /\Lambda^2$ discussed in chapter~\ref{chap:dmdecay}.

Gravitino DM whose decay is due to the presence of bi-linear $R$-parity violation (via the super-potential coupling $W \supset H_u\, L$) is a well studied scenario. If the gravitino, denoted by $\psi_{3/2}$, is very heavy, it will decay via the following four channels: $\psi_{3/2}~\rightarrow~\nu\,\gamma, \nu \,Z^0, \nu \,h, \ell^\pm \,W^{\mp}$~\cite{Ishiwata:2008cu, Grefe:2008zz}.  For $m_{3/2}$ near the weak scale, the branching ratios are somewhat sensitive to the details of the SUSY breaking masses.  However, once $m_{3/2} \gg v$, the decay pattern quickly asymptotes to $1:1:2$ for the $\nu \,Z^0, \nu\, h, \ell^\pm\, W^{\mp}$ channels respectively, as expected from the Goldstone equivalence theorem.

\begin{figure}[t!]
	\leavevmode
	\begin{center}$
	\begin{array}{cc}
	\scalebox{0.3}{\includegraphics{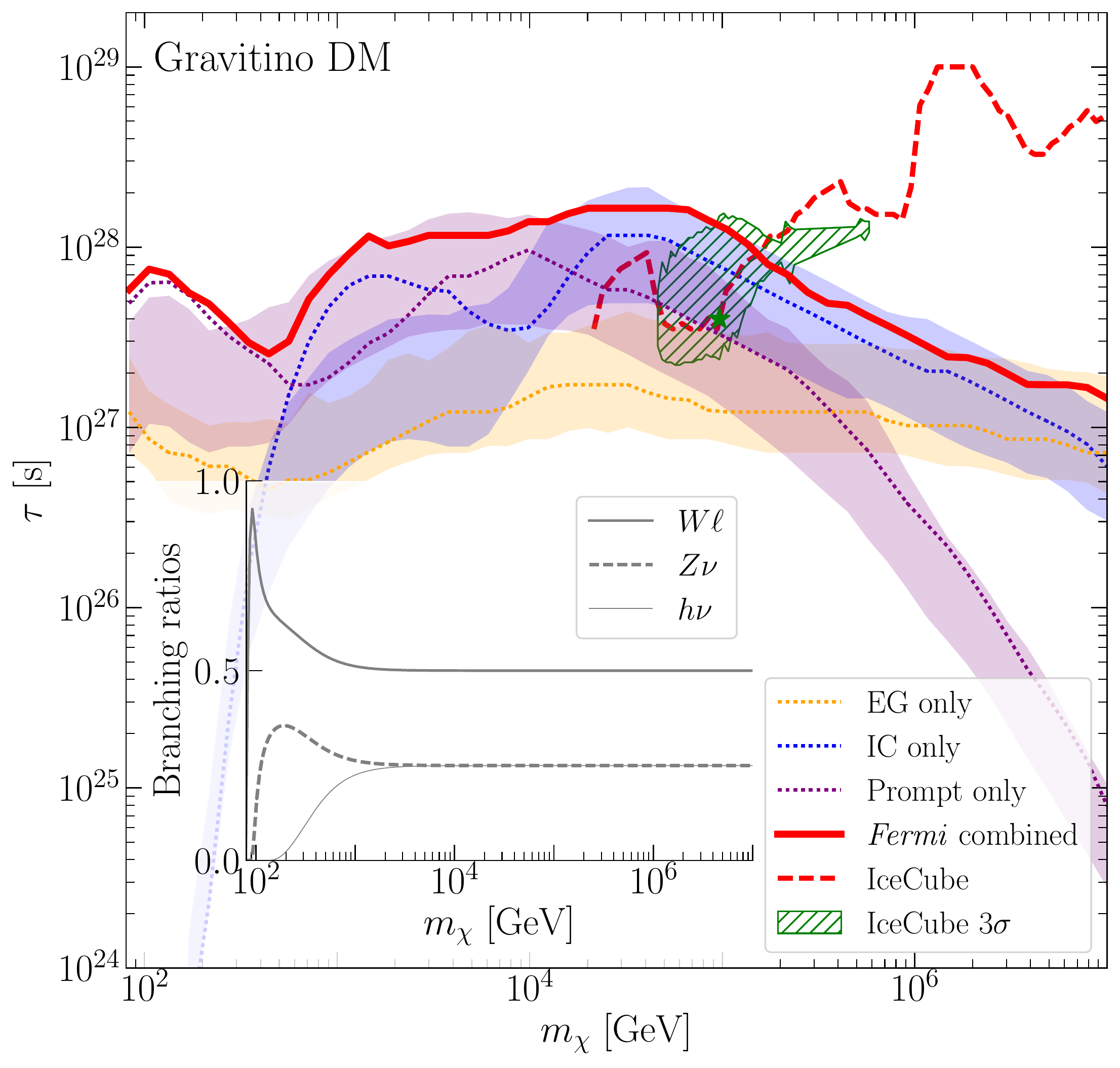}} &\scalebox{0.3}{\includegraphics{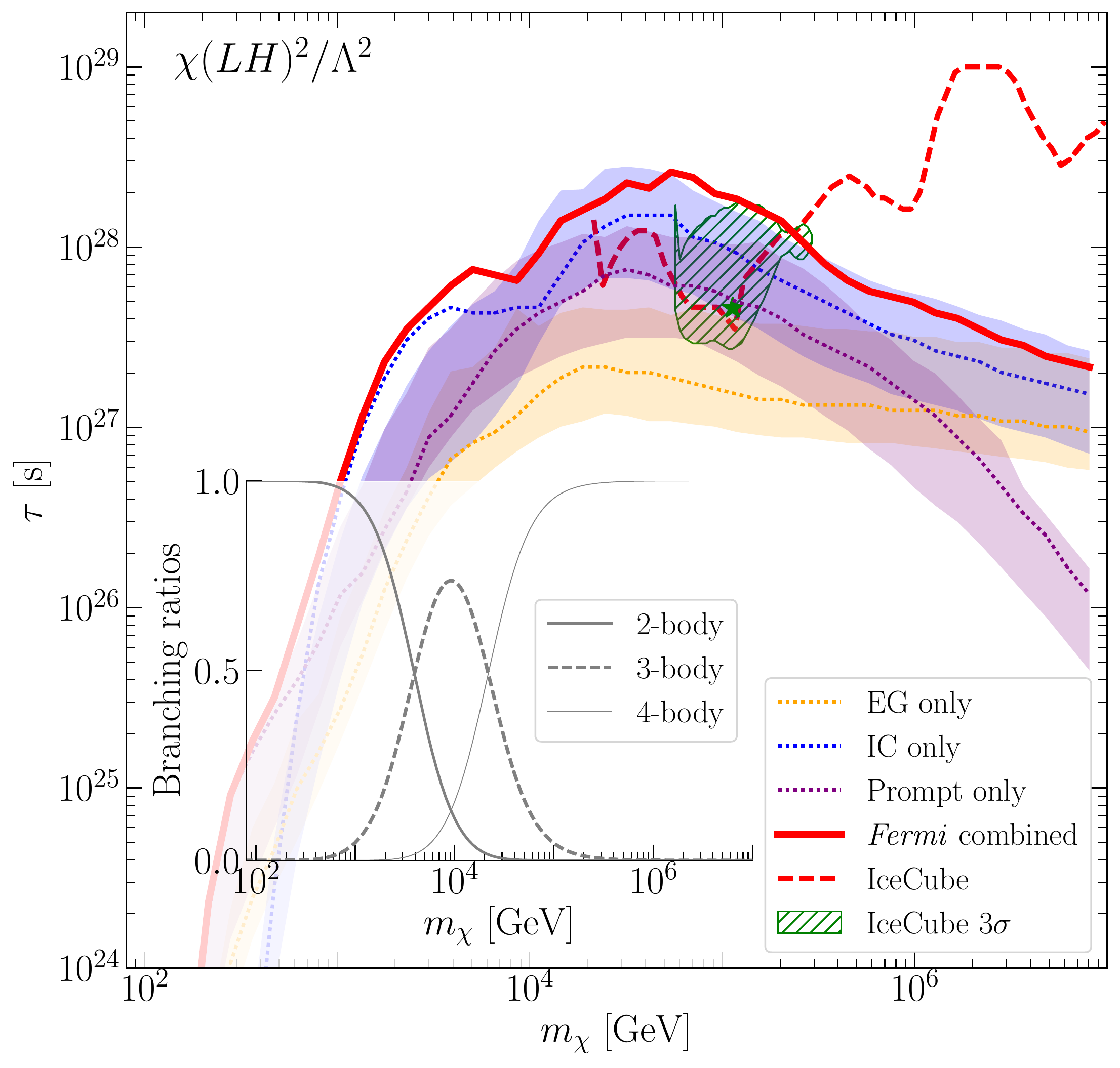}}
	\end{array}$
	\end{center}
	\vspace{-.70cm}
	\caption{Constraints on decaying gravitino DM (left) and DM decaying via the operator $\chi \,(L\,H)^2$ (right).  Notation and labeling is as in Fig.~\ref{Fig: glue} in chapter~\ref{chap:dmdecay}.}
	\vspace{-0.15in}
	\label{Fig: SM results}
\end{figure}

In Fig.~\ref{Fig: SM results} we show the constraints on decaying gravitino DM assuming the above decay modes, with branching ratios given as functions of mass in the inset, using the benchmark parameters of~\cite{Ishiwata:2008cu}.  At masses below the electroweak scale, the $\gamma \nu$ final state dominates.  This channel is best searched for using a gamma-ray line search, which is beyond the scope of this work. For this reason, we only show our constraints for masses above $m_W$.  Note that the region where decaying DM could provide a $\sim$3$\sigma$ improvement over the null hypothesis for the IceCube, the ultra-high-energy neutrino flux (green hashed region) is almost completely excluded by our gamma-ray constraints.  The IceCube constraints, determined using the same methods discussed in chapter~\ref{chap:dmdecay}, begin to dominate at scales above $\sim$$100$\,TeV.  

In Fig.~\ref{Fig: SM results}, we also show limits obtained on the lifetime of the DM $\chi$ under the assumption that $\chi$ interacts with the SM through the operator in Eq.~\eqref{sp0}---this model was discussed in detail in the previous subsection, see Tab.~\ref{tab:ModelBuilding}. The inset plot shows the branching ratios as a function of energy, and illustrates the transition from two- to three- to four-body decays dominating as the mass is increased.  In this case as well, almost the entire range of parameter space relevant for IceCube is disfavored by our gamma-ray limits.

We use \textsc{FeynRules} 2.0~\cite{Alloul:2013bka} to generate the UFO model files, which are then fed to \textsc{MadGraph5\_aMC{@}NLO}~\cite{Alwall:2011uj, Alwall:2014hca} to compute  the parton-level decay interfaced with \textsc{Pythia} for the showering/hadronization of the decays $\chi \to \nu\, \nu\, h \,h$, $\ \nu\, \nu\, Z^0 \,h$, $\ \nu\, \nu\, Z^0 \,Z^0$, $\nu\, e^-\, h\, W^+$, $\ \nu\, e^-\, Z^0\,W^+$, $e^-\,e^- \,W^+\, W^+$ and $\chi \to \nu\, \nu \,h$, $\nu\,\nu\,Z^0$, $\nu\,e^- \, W^+ $.

%% file: cascsig-app.tex
\chapter{Cascade Spectra and the GCE}\label{app:casclim-app}

\section{0-step Spectra}
\label{app:zerostep}
In order to calculate the photon spectrum, it is more straightforward to first determine the density of states according to:
\be\begin{aligned}
{\rm annihilations:}~\frac{1}{N_{\gamma}} \frac{dN_{\gamma}}{dE_{\gamma}} &= \frac{1}{\langle \sigma v \rangle} \frac{d \langle \sigma v \rangle}{d E_{\gamma}} \\
{\rm decays:}~\frac{1}{N_{\gamma}} \frac{dN_{\gamma}}{dE_{\gamma}} &= \frac{1}{\Gamma} \frac{d \Gamma}{d E_{\gamma}}
\label{eq:spectra}
\end{aligned}\ee
from which the spectrum can be easily backed out. Note that as pointed out in \cite{Fortin:2009rq}, the case of decays as a zeroth step in a cascade of $\chi \to \phi_n \phi_n$ will give an identical photon spectrum to the annihilation scenario for a DM particle with twice the mass as for annihilations. This is the sense in which our results are readily transferred to the case of decaying DM. The key difference for the decaying case is the spatial morphology of the signal will generically require a line of sight integral over the DM density, rather than density squared as appears in the $J$-factor in Eq.~\ref{eq:Jfactor}. The observed spatial morphology of the GCE appears to disfavour decaying scenarios, which is why we do not mention them further here, although see \cite{Finkbeiner:2014sja} for a novel decay scenario that is distributed like density squared.

The result of Eq.~\ref{eq:spectra} is that in some circumstances it is possible to calculate various step cascades analytically. This approach is shown for several cases in \cite{Fortin:2009rq}. Yet in many cases - most notably those involving hadronic processes in their final states - analytic calculations are not feasible. For the present analysis we used a combination of analytic and numeric results depending on the final state employed. The details for each case is outlined below.

\subsection{Annihilations to $e^+ e^-$ }
The only contribution to the photon spectrum arises from FSR via the decay $\phi_1 \to e^+ e^- \gamma$. The spectrum in this case can be calculated analytically using Eq.~\ref{eq:spectra}, which was done in \cite{Mardon:2009rc} for the generic case of $\phi_1 \to f^+ f^- \gamma$. As pointed out there, when using the simple convolution formula Eq.~\ref{eq:boosteq}, consistency requires throwing away terms $\mathcal{O}(\epsilon_f^2)$ and higher, where $\epsilon_f = 2m_f/m_1$. Doing so they obtained the following expression for the spectrum that we include for completeness:
\be
\frac{dN_{\gamma}^{\rm FSR}}{dx_0} = \frac{\alpha_{\rm EM}}{\pi} \frac{1+(1-x_0)^2}{x_0} \left[ \ln \left( \frac{4(1-x_0)}{\epsilon_{f}^2} \right) -1 \right]\,.
\label{eq:FSR}
\ee
Note the $\ln$ term will dominate for small $\epsilon_f$, and the $-1$ is simply included to ensure consistency with the large hierarchies approximation. We confirmed that this spectrum is in agreement with the output from  \texttt{Pythia8} in the case of final state electrons. From here, by repeated use of the convolution formula it is possible to obtain completely analytic formula for the $n$-step cascade, which were used in our fits. For example, the first two steps are shown in \cite{Mardon:2009rc}.

\subsection{Annihilations to $\mu^+ \mu^-$ }
For final state muons, in addition to FSR, as pointed out in \cite{Mardon:2009rc} the radiative decay of the muon $\mu \rightarrow e \bar{\nu}_e \nu_{\mu} \gamma$ will meaningfully contribute to the photon spectrum. This decay was calculated in \cite{Kuno:1999jp}, and again for completeness we include it here as it was presented in \cite{Mardon:2009rc}:
\be
\frac{dN_{\mu\to\gamma}}{dx_{-1}} = \frac{\alpha_{\rm EM}}{3\pi} \frac{1}{x_{-1}} \left( T_{-1}(x_{-1}) \ln \frac{1}{r} + U_{-1}(x_{-1}) \right)\,,
\ee
where $r=m_e^2/m_{\mu}^2$ and
\be\begin{aligned}
T_{-1}(x) =& (1-x)(3-2x+4x^2 - 2x^3) \\
U_{-1}(x) =& (1-x) \left( -\frac{17}{2} + \frac{23}{6} x - \frac{101}{12} x^2 + \frac{55}{12} x^3 \right. \\
&\left.+ (3-2x + 4x^2 - 2x^3) \ln(1-x) \right)
\end{aligned}\ee
Note the subscript $-1$ here is used to remind us this is the spectrum calculated in the rest frame of the muon. To then obtain the 0-step cascade we would have to apply Eq.~\ref{eq:boosteq} once, assuming $\epsilon_{\mu} = 2m_{\mu}/m_1 \ll 1$, and then combine this with the FSR spectrum in Eq.~\ref{eq:FSR}.

\subsection{Annihilations to $\tau^+ \tau^-$ }
For the case of final state taus, FSR will now be a subdominant contribution. Instead the spectrum will have a much larger contribution from leptonic and semi-leptonic tau decays: $\tau^- \rightarrow \nu_\tau  l^- \bar{\nu_l}$ and $\nu_\tau d \bar{u}$. The quarks will then hadronize (dominantly to pions) which will result in large contributions to the photon spectrum. We simulated this final state in  \texttt{Pythia8} to generate an initial spectrum, to which we could then apply the convolution formula.

\subsection{Annihilations to $b \bar{b}$}
Much like for taus, in the case of final state $b$-quarks FSR is a subdominant contribution, and instead the spectrum is largely determined by hadronic processes. As such we again utilize  \texttt{Pythia8} to obtain the initial spectrum.

\section{Kinematics of a Multi-step Cascade}
\label{app:boost}

As already emphasized the utility of the small $\epsilon_i=2m_i/m_{i+1}$ - or large hierarchies - approximation is threefold: 
\begin{enumerate}
\item It simplifies calculations in that we can use Eq.~\ref{eq:boosteq}, rather than the general formula we display below; 
\item More importantly it allows us to describe a cascade using just the identity of the final state $f$, the value of $\epsilon_f$, and the number of steps $n$, in contrast to the many possible parameters of the generic case; 
\item Despite the simplifications afforded, results in this framework can be used to estimate the results even for general $\epsilon_i$, as described in Sec.~\ref{sec:generalcascade}.
\end{enumerate}
 In this appendix we show how the kinematics of scalar cascade decays lead to an expression for the $n$-step spectrum in terms of the $(n-1)$-step result. In addition we outline how Eq.~\ref{eq:boosteq} emerges in the small $\epsilon$ limit, with error $\mathcal{O}(\epsilon_i^2)$, as well as how the transition to the degenerate case as $\epsilon \to 1$ occurs.

Our starting point is the 0-step spectrum $dN_\gamma / dx_0$ where $x_0 = 2 E_0 / m_1$ and $E_0$ is the photon energy in the rest frame of $\phi_1$. This results from the process $\phi_1 \to \gamma X$, where the identity of $X$ depends on the final state considered. From here we want to calculate $dN_\gamma / dx_1$ - the spectrum from a cascade that includes $\phi_2 \to \phi_1 \phi_1$ and so is one step longer - where $x_1 = 2 E_1/m_2$ and $E_1$ is the photon energy in the $\phi_2$ rest frame. If we assume isotropic scalar decays, then we can obtain this by simply integrating the 0-step result over all allowed energies and emission angles:
\be\begin{aligned}
\frac{dN_{\gamma}}{dx_1} = & 2 \int_{-1}^1 d \cos \theta \int_{0}^1 dx_{0} \frac{dN_{\gamma}}{dx_{0}} \\
&\delta \left( 2x_1 - x_0 - \cos \theta x_0 \sqrt{1-\epsilon_1^2} \right)\,,
\end{aligned}\ee
where $\theta$ is defined as the angle between the photon momentum and the $\phi_1$ boost axis as it is measured in the $\phi_1$ rest frame. The limits of integration $0 \leq x_0 \leq 1$ reflect the fact that the photon energy in the $\phi_1$ rest frame can be arbitrarily soft on the one side, and on the other it can have an energy at most half the mass of the initial particle, $m_1/2$ here. The $\delta$ function is simply enforcing how the photon energy changes when we move from the $\phi_1$ to the $\phi_2$ rest frame, i.e. from $E_0$ to $E_1$. It also sets the kinematic range for $x_1$, which is:
\be
0 \leq x_1 \leq \frac{1}{2} \left(1+ \sqrt{1-\epsilon_1^2}\right)\,.
\label{x0Range}
\ee
Now if we then use the $\delta$ function to perform the angular integral, the one step spectrum reduces to:
\be
\frac{dN_\gamma}{dx_1} = 2 \int_{t_{1,{\rm min}}}^{t_{1,{\rm max}}} \frac{dx_{0}}{x_0 \sqrt{1- \epsilon_{1}^2}} \frac{dN_\gamma}{dx_{0}}~\,,
\label{eq:fullepsilonstep1}
\ee
where we have introduced:
\be\begin{aligned}
t_{1,{\rm max}} &= \min \left[ 1,\, \frac{2x_1}{\epsilon_1^2} \left( 1 + \sqrt{1-\epsilon_1^2} \right) \right] \\
t_{1,{\rm min}} &= \frac{2 x_1}{\epsilon_1^2}\left( 1 - \sqrt{1-\epsilon_1^2} \right)
\label{eq:1stepminmax}
\end{aligned}\ee
The maximum here is either set by the maximum physical value of $x_0$, which is $1$, or alternatively by where the $\delta$ function loses support. We can then repeat this process to recursively obtain the $i$th order spectrum from the $(i-1)$th order result. Explicitly we find:
\be
\frac{dN_\gamma}{dx_i} = 2 \int_{t_{i,{\rm min}}}^{t_{i,{\rm max}}} \frac{dx_{i-1}}{x_{i-1} \sqrt{1- \epsilon_{i}^2}} \frac{dN_\gamma}{dx_{i-1}}~\,,
\label{eq:fullepsilon}
\ee
where we have defined:
\be\begin{aligned}
t_{i,{\rm max}} &=  \min \left[ \frac{1}{2^{i-1}} \prod_{k=1}^{i-1} \left( 1 + \sqrt{1- \epsilon_k^2} \right),\right. \\
&\left.\;\;\;\;\;\;\;\;\;\;\;\;\;\frac{2x_i}{\epsilon_i^2} \left( 1+\sqrt{1-\epsilon_i^2}\right) \right] \\
t_{i,{\rm min}} &= \frac{2 x_i}{\epsilon_i^2}\left( 1 - \sqrt{1-\epsilon_i^2} \right)
\label{eq:nstepximin}
\end{aligned}\ee
and now the kinematic range of $x_i$ is
\be
0 \leq x_i \leq \frac{1}{2^i} \prod_{k=1}^i \left( 1 + \sqrt{1- \epsilon_k^2} \right)\,.
\label{eq:xnRange}
\ee
With the exact result of Eq.~\ref{eq:fullepsilon}, we can now see that in the small $\epsilon$ limit the result reduces to Eq.~\ref{eq:boosteq} with corrections at most of order $\epsilon^2$, as claimed. The exact result also captures an additional feature that the large hierarchies result does not: the emergence of a degenerate step in the cascade as $\epsilon_i\to 1$ for some $i$. As discussed in Sec.~\ref{sec:generalcascade}, when this occurs, just from the kinematics we can see that the $(i+1)$-step result will reduce to the $i$-step spectrum, but shifted in energy and normalisation. Starting with Eq.~\ref{eq:fullepsilon}, setting $1-\epsilon_i^2 \equiv z$ and then taking $z \to 0$ it is straightforward to confirm that the exact result also reproduces this behaviour.

As discussed in Sec.~\ref{sec:generalcascade}, there should be a smooth interpolation between the two extreme cases of $\epsilon_i=0$ and $\epsilon_i=1$, and using Eq.~\ref{eq:fullepsilon} we can demonstrate that indeed there is. This is shown in Fig.~\ref{fig:fullEpsilon0p1}, where we take the case of a 1-step cascade for final state taus with $\epsilon_{\tau}=0.1$. We plot the two extreme cases and show how intermediate $\epsilon$ transition between these by plotting five values: $0.3$, $0.5$, $0.7$, $0.9$ and $0.99$. Note that as claimed earlier, the transition is roughly quadratic in $\epsilon$; for small and intermediate values of $\epsilon$, the result is well approximated by the $\epsilon=0$ result, again highlighting the utility of the large hierarchies approximation.

\section{Model-Building Considerations}
\label{app:models}

\subsection{A Simple Model}

Let us extend the usual Higgs Portal \cite{Patt:2006fw, MarchRussell:2008yu} model to include a rich dark sector with $n$ scalar mediators and a set of $n$ $\mathbb{Z}_2$ symmetries.\footnote{A more complex symmetry structure could allow off-diagonal couplings between the scalars and the Higgs, with potentially rich observational signatures. We thank Jessie Shelton for this observation.} This will serve as an illustrative example of how different observable signatures depend on different model parameters, as discussed in the main text.

Consider the potential:
\be\begin{aligned}
&V\left(\chi, \phi_1, H\right )  = V_\chi + V_H + c_k \phi_1^2 |H|^2 \\
&+ \sum_{i = 1}^{n}\left(\frac{\lambda_{4,i}}{2} \chi^2 \phi_{i}^2 -\frac{1}{2}m_{i}^2 \phi_i^2\right) + \sum_{i,j = 1}^{n}  \frac{\lambda_{ij}}{4 !}\phi_i^2 \phi_j^2\,, 
\label{eq:HiggsPortal}
\end{aligned}\ee
Here $V_\chi$ and $V_H$ contain the usual mass and quartic terms for the DM and Higgs fields. As discussed previously it is reasonable that the dark sector is secluded such that the dominant portal coupling is $c_k \phi_1^2 |H|^2$. Upon electroweak and $\mathbb{Z}_2$ symmetry breaking the $\lambda_{4,i}$ couplings allow annihilations $\chi \chi \rightarrow \phi_i \phi_i$. We assume that DM annihilates preferentially to the heaviest mediator through $\lambda_{4,n} \chi^2 \phi_n^2$. So it is $\lambda_{4,n}$ that dominantly controls the thermal annihilation cross-section and therefore the DM relic abundance $\Omega_\chi h^2 \sim 0.11$. The dark sector quartic term will generate interactions of the form $\lambda_{ij}  \langle \phi_i \rangle \phi_i \phi_{j}^2 $, allowing the mediators to cascade decay in the dark sector. Additionally the Higgs Portal interaction will generate a mixing between $\phi_1$ and the Higgs. The end result will be a dark cascade ending in the $c_k$ suppressed decay $\phi_1 \rightarrow f \bar{f}$, with a subsequent photon spectrum that can be fit to the GCE. 

While the thermal relic cross-section depends on $\lambda_{4,n}$, the direct detection cross-section will also depend on the portal coupling $c_k$. This additional small parameter gives us the needed freedom to explain the GCE while alleviating constraints from direct detection. Additionally we point out that the size of the couplings $\lambda_{ij}$ will need to be large enough such that  decays of the new light states occur before BBN. Given the number of new free parameters, this setup should not be difficult to construct. Finally we point out that the Higgs Portal interaction also contains a coupling which leads to the decay $h \rightarrow \phi_1 \phi_1$. Invisible Higgs decay is constrained by collider searches which impose an upper bound of about $c_k \lesssim 10^{-2}$ \cite{Martin:2014sxa}.

\subsection{The Sommerfeld Enhancement}

We have seen that the preferred cross-section steadily increases with the number of steps in the cascade, moving away from the thermal relic value that is favored for the direct case. This increased cross-section is also accompanied by an increase in the preferred mass scale for the DM (indeed, the requirement for a larger cross-section is largely driven by the reduced number density of heavier DM). In the presence of a mediator much lighter than the DM, exchange of such a mediator could enhance the present-day annihilation cross-section via the Sommerfeld enhancement (e.g. \cite{Sommerfeld:1938,Hisano:2003ec, Hisano:2004ds,ArkaniHamed:2008qn, Pospelov:2008jd}), naturally leading to an apparently larger-than-thermal annihilation signal.

However, there are some obstacles to such an interpretation, at least in the simple case we have studied where the particles involved in the cascade are all scalars. For the case of fermionic DM coupled to a light scalar or vector of mass $m_\phi$ with coupling $\alpha_D$, the Sommerfeld enhancement at low velocity is parametrically given by $m_\phi/\alpha_D m_{\chi}$. A large enhancement thus requires $\alpha_D \gtrsim m_\phi / m_\chi$. In order to obtain the correct relic density, we typically require $\alpha_D$ to be $\mathcal{O}(0.01)$, and so a significant Sommerfeld enhancement would require the \emph{first} step in the cascade to involve a mass gap of two orders of magnitude. This may be plausible for the electron and even muon channels, but is challenging for final states involving heavier particles such as taus and $b$-quarks; if the mediator is heavy enough to decay to these particles, the required DM mass becomes much too large to fit the GCE even for a one-step cascade, and adding more hierarchical steps only exacerbates the self-consistency issue (as discussed in Secs.~\ref{sec:methods}-\ref{sec:results}).

Furthermore, if the DM is a fermion, its annihilation into scalars is generically $p$-wave suppressed, making it difficult to obtain a large enough cross-section to obtain the GCE. If instead the DM is a heavy (singlet) scalar, the simplest way to couple it to the light scalar to which it annihilates is an interaction of the form $\mathcal{L}_\mathrm{quartic} = \frac{\lambda_4}{2} \chi^2 \phi_{n}^2$. When the light scalar obtains a vacuum expectation value, this gives rise to an interaction of the form $\lambda_4 \langle \phi_n \rangle \phi_n \chi^2$, and repeated exchanges of the light scalar $\phi_n$ can give rise to enhanced annihilation. However, assuming $\langle \phi_n \rangle \sim m_n$, the size of the coupling is suppressed by the small mass of the light scalar, even as its range is enhanced. Accordingly, a large enhancement to annihilation is not expected, at least in this simple scenario.

As discussed in Sec.~\ref{sec:generalcascade}, our results can be extended to cascades including particles other than scalars, in which these later issues do not arise; for example, in the axion portal \cite{Nomura:2008ru}, two-step cascades occur through $\chi \chi \rightarrow s a$, $s \rightarrow a a$, $a \rightarrow f \bar{f}$, where $s$ is a dark scalar and $a$ a dark pseudoscalar. This annihilation channel is $s$-wave and can be Sommerfeld-enhanced by exchange of the $s$. However, the first difficulty described above may still apply, with the large hierarchy between the $\chi$ and $s$ potentially implying a DM mass too large to easily fit the GCE.

%% file: casclim-app.tex
\chapter{Limits on Cascade Spectra}

\section{Details of $n$-body Cascades}
\label{app:multibody}

In this appendix we will derive Eq.~\ref{eq:nbodyboosteq} and provide some additional intuition for this case as well as pointing out that for a small number of steps, the cascade setup can provide an excellent approximation (albeit with some dependence on the channel). To set up this problem, firstly recall that the key physics encapsulated in Eq.~\ref{eq:boosteq} is that when we add in a cascade step we need to boost the spectrum to the new rest frame. In the case of 2-body decays this is particularly simple, because we know exactly how much to boost by. Explicitly, if we have added in a step of the form $\phi_i \to \phi_{i-1} \phi_{i-1}$, then in the $\phi_i$ rest frame we know the $\phi_{i-1}$ particles must be emitted back to back, meaning we know their energy and hence their boost. If instead we introduce a step via $\phi_i \to \phi_{i-1} \phi_{i-1} \phi_{i-1}$, we no longer know the boost exactly, instead we can only associate a probability with any boost which we can determine from the energy distribution for a given $\phi_{i-1}$. Accordingly what we need to calculate is the energy spectrum of a particular $\phi$ in the decay $\chi \chi \to n \times \phi$, and then combine this with a version of Eq.~\ref{eq:boosteq} suitable for a general boost. Below we will firstly do this exactly for the case of a 3-body decay, show what this becomes after applying the large hierarchies approximation, and then we will show the general $n$-body result assuming hierarchical decays.

As discussed, our starting point is the energy spectrum of a particular $\phi$ in the decay $\chi \chi \to 3 \times \phi$, which can be determined from the three body phase space. For this purpose we make use of the analytic formula for the $n$-body phase space outlined in \cite{Byckling:1969sx,Kersevan:2004yh}. In the case where our three final state scalars have mass $m$, we can write the 3-body phase space as:
\begin{equation}\begin{aligned}
\Phi_3 = (4\pi)^2 \int_{4m^2}^{(M_3-m)^2} &dM_2^2 \frac{\sqrt{\lambda(M_3^2,M_2^2,m^2)}}{8M_3^2} \\
\times &\frac{\sqrt{\lambda(M_2^2,m^2,m^2)}}{8M_2^2}\,,
\label{eq:3bodyphasespace}
\end{aligned}\end{equation}
where $\lambda(x,y,z)=x^2 + y^2 + z^2 - 2xy - 2yz - 2zx$ and if we say the mass of the dark matter is $m_{\chi}$ and the energy of one $\phi$ particle is $E$, then $M_3^2=4m_{\chi}^2$ and $M_2^2 = 4m_{\chi}^2 + m^2 - 4 m_{\chi} E$. Using this, the energy spectrum of the scalars is simply:
\begin{equation}
\frac{dN_{\phi}}{dE} \propto \frac{d\Phi_3}{dE}\,,
\label{eq:3bodytospec}
\end{equation}
where the constant of proportionality can be determined by normalising the spectrum. Before proceeding, it is useful to introduce a set of dimensionless variables to work with as we did in the 2-body case. As there, we firstly define $\epsilon_1= m/m_{\chi}$, but note here that $\epsilon_1 \in [0,2/3]$, rather than $[0,1]$ as in the 2-body case. To play a similar role to $x$, we also introduce $\xi = E/m_{\chi} \in [\epsilon_1,1-3\epsilon_1^2/4]$, where the limits here are fixed by Eq.~\ref{eq:3bodyphasespace} and can also be seen from the kinematics. In terms of these variables, we can use Eq.~\ref{eq:3bodytospec} and Eq.~\ref{eq:3bodyphasespace} to arrive at:
\begin{equation}
\frac{dN_{\phi}}{d\xi} = C \sqrt{\frac{(\xi^2-\epsilon_1^2)(4-3\epsilon_1^2-4\xi)}{4+\epsilon_1^2-4\xi}}\,,
\label{eq:3bodyspec}
\end{equation}
where $C$ is a constant that normalises the spectrum and can be determined numerically. Note that when $\epsilon_1 \to 2/3$, this distribution approaches a $\delta$ function, as expected when the particles are all produced at rest. We will return to the limit of small $\epsilon_1$ shortly.

Using this result, we can then revisit the derivation of the boost formula given in \cite{Elor:2015tva}, a hierarchical version of which is given in Eq.~\ref{eq:boosteq}, and derive the analogue for an arbitrary boost. Doing so, if we label the spectrum of the decay of $\phi \to 2 \times \text{(SM final state)}$ as $dN/dx_0$, we can write the spectrum of the same particle from the decay $\chi \chi \to 3 \times \phi \to 6 \times \text{(SM final state)}$ as:
\begin{equation}\begin{aligned}
\frac{dN}{dx_1} = 3 &\int_{\epsilon_1}^{1-(3/4)\epsilon_1^2} d \xi C \sqrt{\frac{(\xi^2-\epsilon_1^2)(4-3\epsilon_1^2-4\xi)}{4+\epsilon_1^2-4\xi}} \\
\times &\int_{t_{\rm min}}^{t_{\rm max}} \frac{dx_0}{x_0 \sqrt{\xi^2-\epsilon_1^2}} \frac{dN}{dx_0}\,,
\label{eq:3bodyspecfull}
\end{aligned}\end{equation}
where we have defined:
\begin{equation}\begin{aligned}
t_{\rm max} &\equiv {\rm min} \left[ 1, \frac{2x_1}{\epsilon_1^2} \left( \xi + \sqrt{\xi^2-\epsilon_1^2} \right) \right] \\
t_{\rm min} &\equiv \frac{2x_1}{\epsilon_1^2} \left( \xi - \sqrt{\xi^2 - \epsilon_1^2} \right)
\end{aligned}\end{equation}
There are two directions the above result can be generalised. For one, we could extend this to a longer cascade of 3-body decays, although the logic here is identical to the general 2-body case discussed in \cite{Elor:2015tva}, so we will not repeat that here. Secondly we can look to extend this to 4-body decays and higher. The difficulty with this is that the $n$-body phase space quickly becomes analytically intractable. Nevertheless as observed in \cite{Liu:2014cma}, in the large hierarchies regime ($\epsilon_1 \ll 1$) we regain analytic control as we will now outline.

\begin{figure}[t!]
\centering
\begin{minipage}{.45\textwidth}
  \centering
  \includegraphics[scale=0.6]{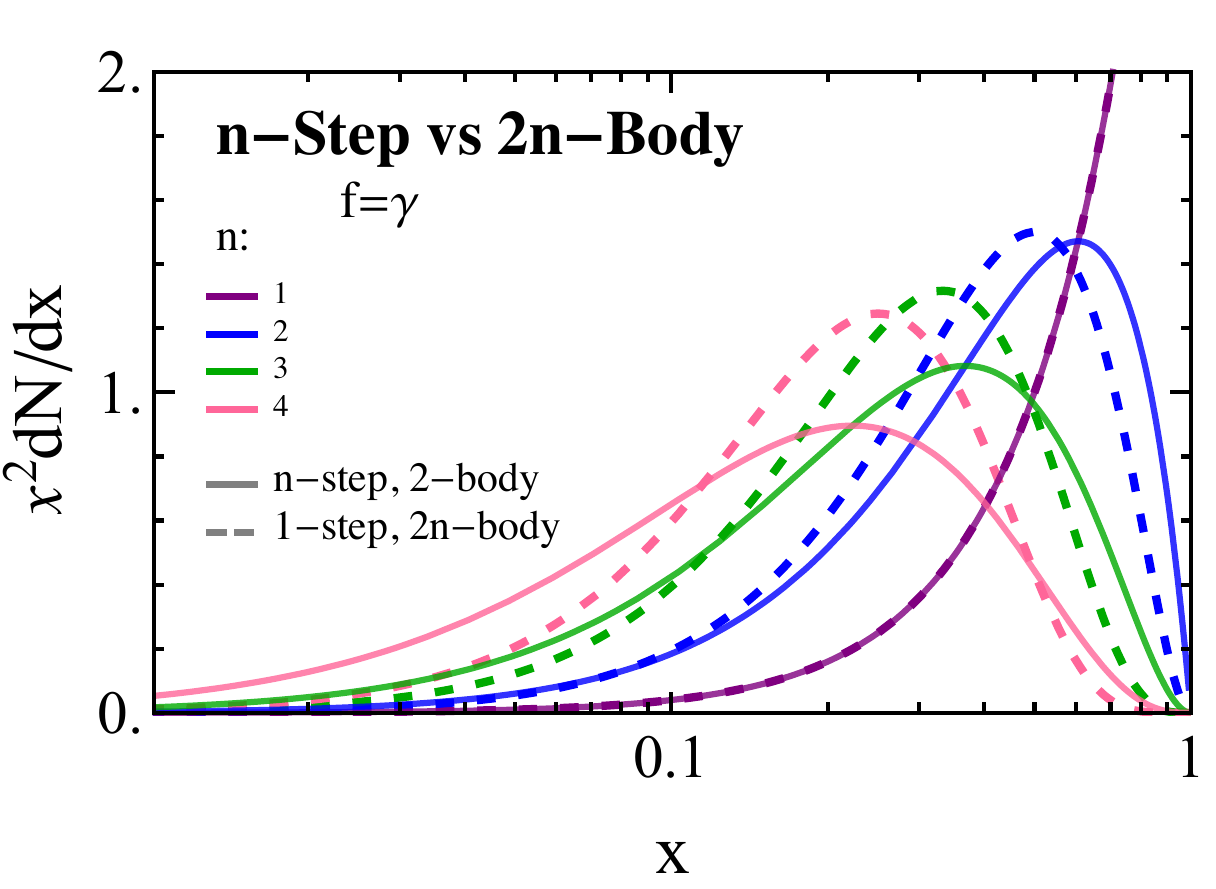}
  \captionof{figure}{\footnotesize{Spectra for a cascades containing $n$-step 2-body decay and a 1-step $2n$-body decay, both to $\gamma \gamma$, are shown as the solid and dashed curves respectively, for the case of $n=$ (1,2,3,4) in (purple, blue, green, pink). We see that for $n=$ 1 and 2 we can approximate one of these types of spectra by the other (with the $n=$ 1 case being exact by definition).}}
  \label{fig:nbodyvsnstepgamma}
\end{minipage}
\hspace{0.4in}
\begin{minipage}{.45\textwidth}
\vspace{-0.8in}
  \centering
  \includegraphics[scale=0.6]{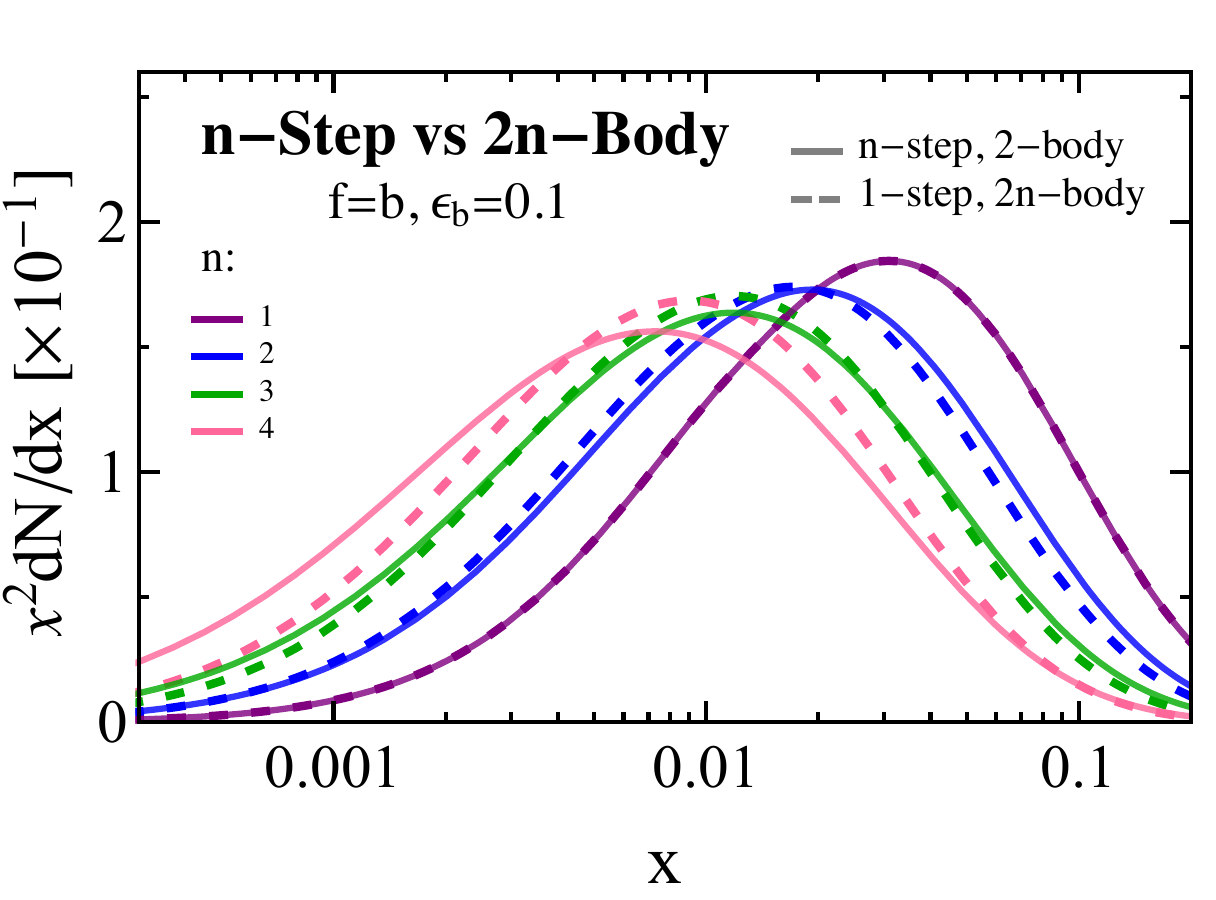}
  \captionof{figure}{\footnotesize{The same as Fig.~\ref{fig:nbodyvsnstepgamma}, but for a final state $b\bar{b}$ with $\epsilon_b=0.1$. Note that again we get close agreement in the $n=$ 1 and 2 case.}}
  \label{fig:nbodyvsnstep}
\end{minipage}
\end{figure}

Returning to Eq.~\ref{eq:3bodyspec}, taking the $\epsilon_1 \to 0$ limit we find that:
\begin{equation}
\frac{dN_{\phi}}{d\xi} = 2 \xi + \mathcal{O}(\epsilon_1^2)\,,
\label{eq:3bodysmalleps}
\end{equation}
where now $\xi \in [0,1]$. Following \cite{Liu:2014cma}, this can then be generalised to the $n$-body case, where we find:
\begin{equation}
\frac{dN_{\phi}}{d\xi} = (n-1)(n-2)(1-\xi)^{n-3} \xi + \mathcal{O}(\epsilon_1^2)\,,
\label{eq:nbodysmalleps}
\end{equation}
where again $\xi \in [0,1]$. Using this we can finally give the equivalent expression of Eq.~\ref{eq:boosteq} for the $n$-body case:
\begin{equation}
\frac{dN}{dx_1} = n(n-1)(n-2) \int_0^1 d \xi (1-\xi)^{n-3} \int_{x_1/\xi}^1 \frac{dx_0}{x_0} \frac{dN}{dx_0} + \mathcal{O}(\epsilon_1^2)
\label{eq:nbodyboosteqapp}
\end{equation}
thereby demonstrating Eq.~\ref{eq:nbodyboosteq}.

As a simple example of how this can be used, consider the decay $\phi \to \gamma \gamma$ which has the spectrum $dN_{\gamma}/dx_0 = 2 \delta(x_0-1)$. If we substitute this in, we find the spectrum for $\chi \chi \to n \times \phi \to 2n \times \gamma$ is just
\begin{equation}
\frac{dN_{\gamma}}{dx_1} = 2n(n-1)(1-x_1)^{n-2}\,.
\label{eq:nbodydelta}
\end{equation}
Integrating this over $x_1 \in [0,1]$, we find $N_{\gamma}=2n$, as expected.

\begin{figure}[t!]
\centering
  \includegraphics[scale=0.6]{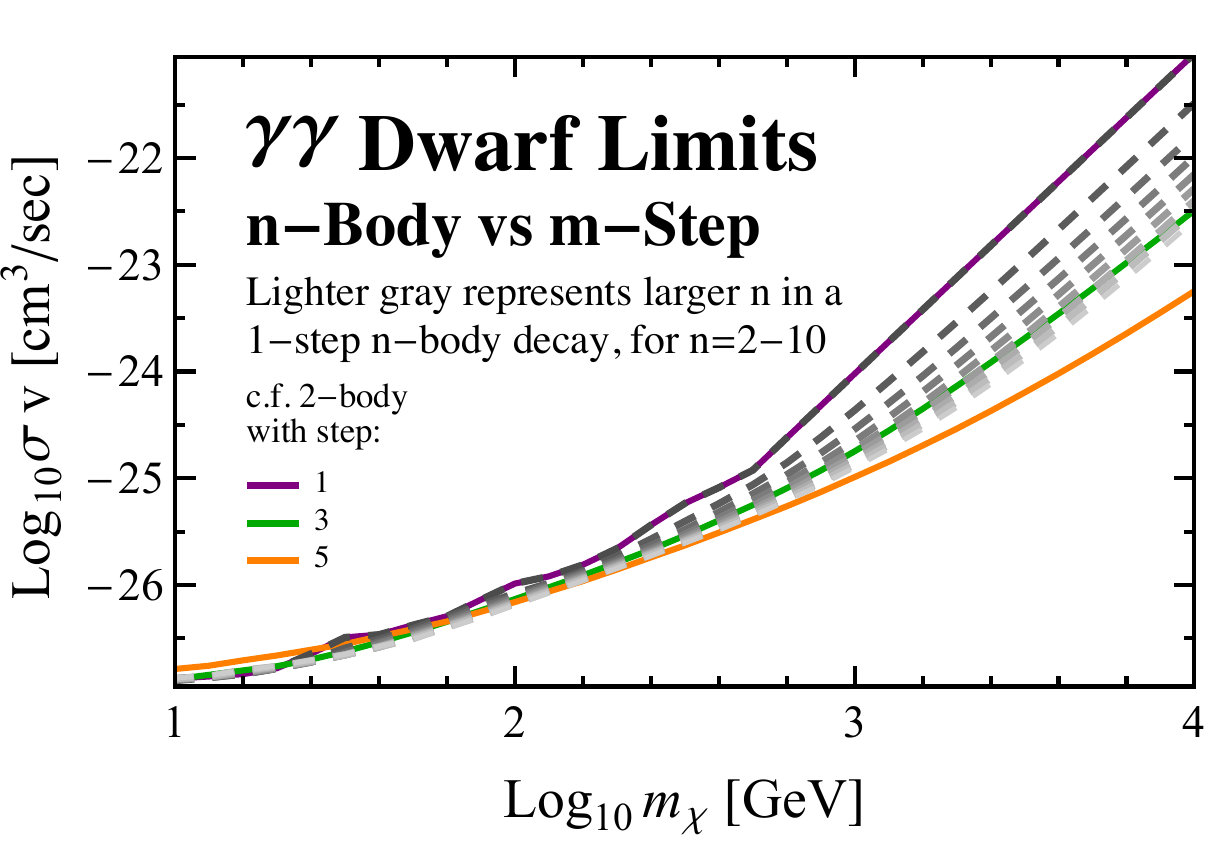}
  \caption{\footnotesize{Dwarf limits for n-body vs m-step cascades for the $\gamma \gamma$ final state. We show the multi-body case for $n=$ 2-10 in gray, with lighter gray corresponding to larger $n$. In orange, green and purple we also show the 1, 3 and 5-step 2-body cascade for the same final state. As discussed in the text, for the multi-body case the spectrum sits in between the cascade spectra, and thus we expect the limits to do the same. The figure makes this clear and emphasizes how the multi-body framework is captured within the cascade setup.}}
  \label{fig:DwarfdeltaNBody}
\end{figure}

To follow on from this, consider the spectrum derived by repeated application of the boost formula in Eq.~\ref{eq:boosteq} to the same $\phi \to \gamma \gamma$ spectrum, $dN_{\gamma}/dx_0 = 2 \delta(x_0-1)$. Doing so we obtain:
\begin{equation}
\frac{dN_{\gamma}}{dx_n} = \frac{(-2)^{n+1}}{(n-1)!} \ln^{n-1} x_n\,.
\label{eq:nstepdelta}
\end{equation}
Note that if we integrate this over $x_n \in [0,1]$, we find $N_{\gamma}=2^{n+1}$. Now by definition Eq.~\ref{eq:nbodydelta} with $n=$ 2 is identical Eq.~\ref{eq:nstepdelta} with $n=$ 1, as in this case they both represent a 2-body 1-step cascade. Note also though that if we take Eq.~\ref{eq:nbodydelta} with $n=$ 4 and Eq.~\ref{eq:nstepdelta} with $n=$ 2, then both situations have the same number of final state photons from different kinematic setups. In Fig.~\ref{fig:nbodyvsnstepgamma} we compare an $n$-step 2-body decay and a 1-step $2n$-body decay for final state photons, for $n=$ (1,2,3,4). We see that whilst they agree for $n=$ 1 (by definition), and are quite similar to each other for $n=$ 2, this similarity breaks down rapidly. This is not entirely surprising as a 6-body 1-step cascade has a different number of final state photons to a 2-body 3-step cascade, but one can also check that this latter spectrum also does not agree well with an 8-body 1-step result which would have the same number of photons. In Fig.~\ref{fig:nbodyvsnstep} we show the same comparison for final state $b\bar{b}$ with $\epsilon_b=0.1$. Here we see the agreement is better although still beginning to break down for larger $n$. We also tested this for several other final states, with the common theme that the spectrum of a 4-body 1-step decay is often well approximated by that of a 2-step 2-body decay.

\begin{figure}[t!]
\centering
\begin{minipage}{.45\textwidth}
  \centering
  \includegraphics[scale=0.54]{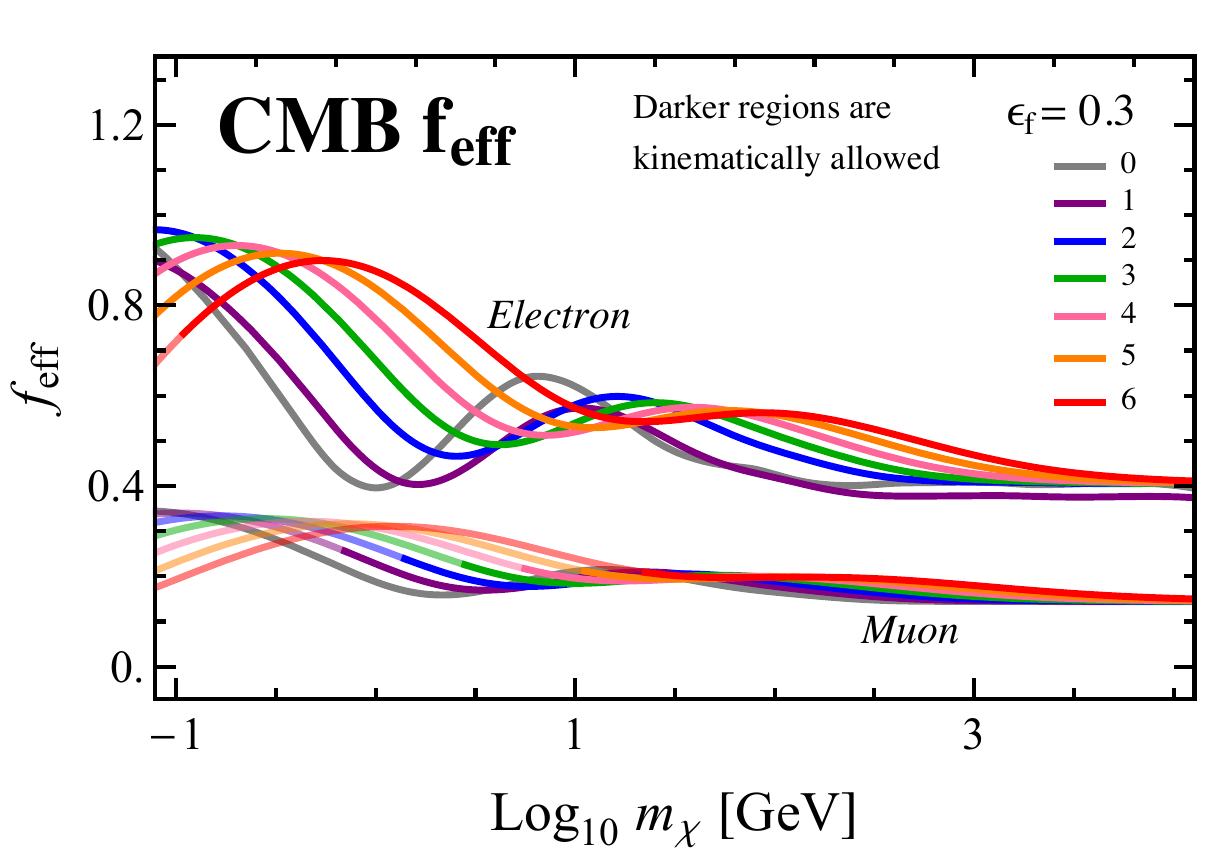}
  \captionof{figure}{\footnotesize{$f_\mathrm{eff}$ for $n =$ 0-6 step cascades to final state electrons and muons, with $\epsilon_f = 0.3$. Note that the difference in pattern between $f_\mathrm{eff}$ for direct and single step electrons and higher step cascades can be understood by recalling that the direct electron FSR spectrum is sharply peaked. Each subsequent cascade step smooths out this spectrum thus changing the shape significantly.}}
  \label{fig:feffElectronMuon}
\end{minipage}
\hspace{0.4in}
\begin{minipage}{.45\textwidth}
\vspace{-0.98in}
  \centering
  \includegraphics[scale=0.555]{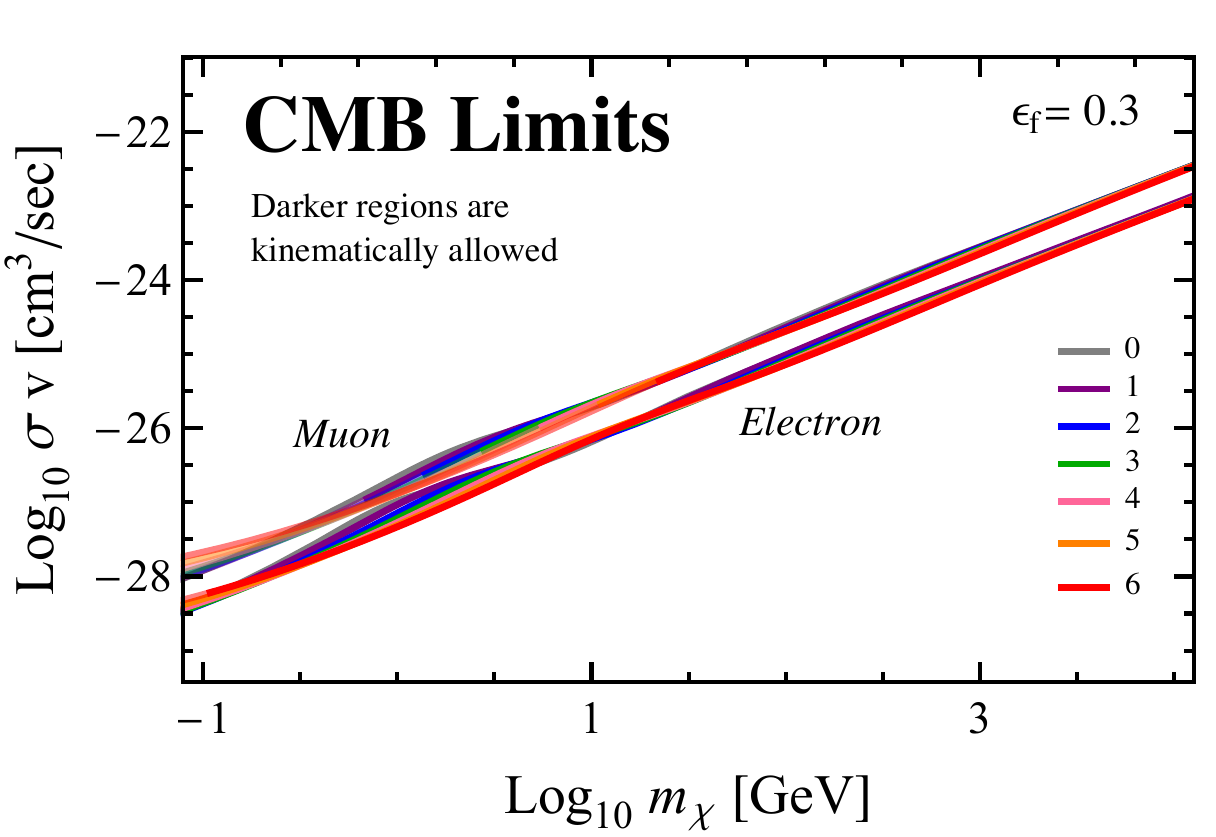}
  \captionof{figure}{\footnotesize{The bound on $\langle \sigma v \rangle$ for $n=$ 0-6 step cascades to final state electrons and muons, with $\epsilon_f = 0.3$.}}
  \label{fig:BoundElectronMuon}
\end{minipage}
\end{figure}

Lastly let us confirm the claim from the main text that the results for multi-body decays sit in between our multi-step cascade results. For this purpose consider again the photon spectrum obtained from $\phi$ decaying into $b\bar{b}$ with $\epsilon_b=0.1$. In this case we plot the $n$-body spectrum for $n=$ 2 to 10 in Fig.~\ref{fig:nbody}, which we presented in the main text. In the figure we also plot the case of a 1-step, 3-step and 5-step 2-body cascade. Clearly the $n$-body results sit in between these multi step cascade cases, which indicates that the constraints on an $n$-body 1-step cascade will be largely contained within the limits on a 2-body $n$-step cascade. To demonstrate this, in Fig.~\ref{fig:DwarfdeltaNBody} we show that for the case of the $\gamma \gamma$ limits extracted from the dwarfs, the $n=$ 2-10 multi-body decays sit exactly in between the 1, 3 and 5-step cascades as claimed.

\section{Details of the CMB Results}
\label{app:CMBdetails}

In this appendix we present additional results from the CMB analysis. For many of the final states considered in the main text, the kinematic threshold on their production means that it is not sensible to go to lower masses than we presented. This is not the case, however, for electrons and muons, where we can take our results to much lower masses. 

\begin{figure}[htbp]
\begin{center}
\includegraphics[scale=0.64]{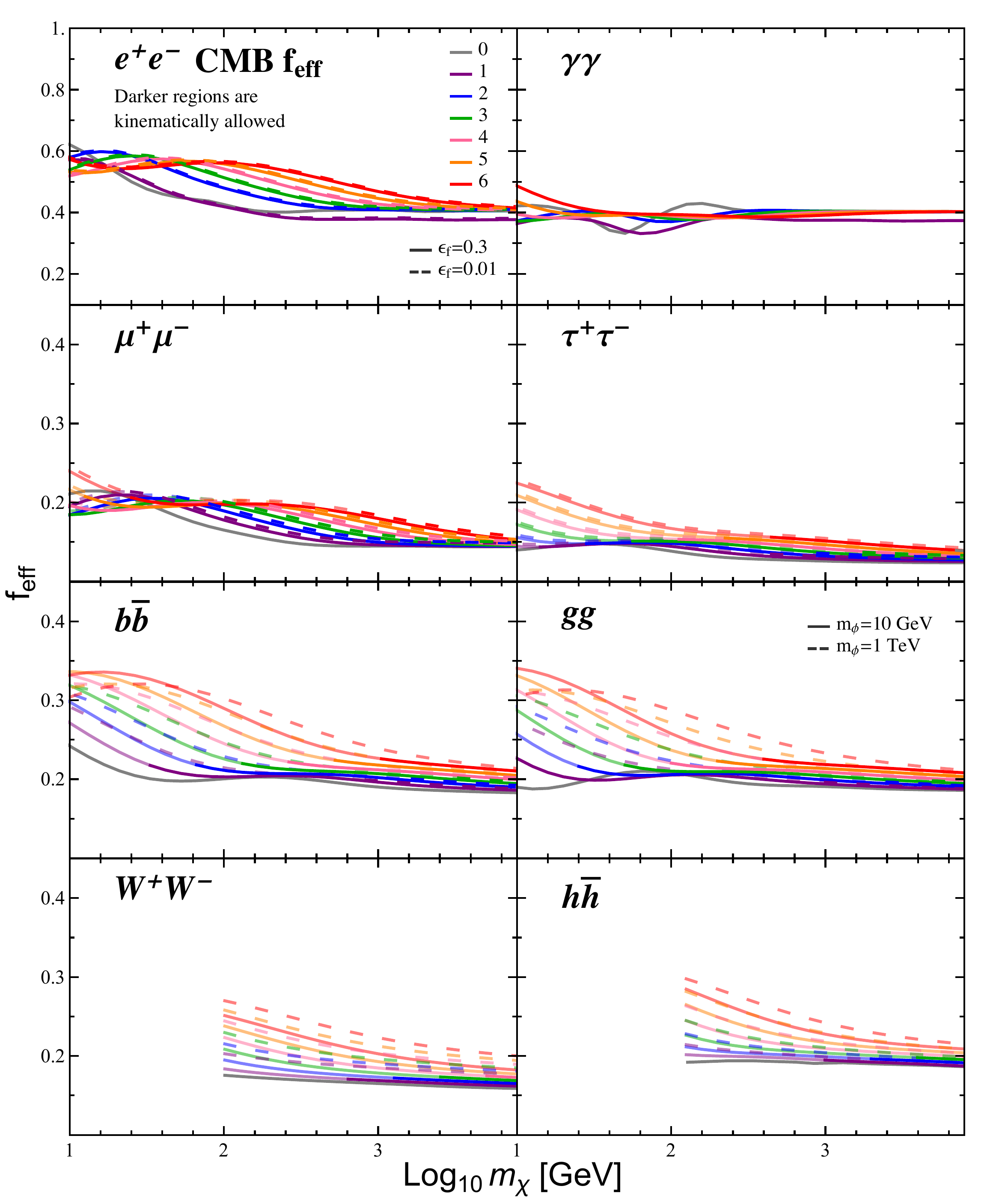}
\end{center}
\caption{\footnotesize{$f_\mathrm{eff}$ for $n=$ 1-6 step cascade for various final states, with $\epsilon_f = 0.3$ (solid) and  $\epsilon_f = 0.01$ (dashed). The shaded out portions of the plot correspond to values of $m_\chi$ that are kinematically forbidden. For the case of direct annihilation (gray line) only the spectrum for $m_\chi >10$GeV is displayed, since for lower values of $m_\chi$ the PPPC is unreliable. For direct annihilations to photons the spectrum is simply a delta function so in this case we plot $f_\mathrm{eff}$ down to lower masses as well.}}
\label{fig:feffCMB0p3}
\end{figure}

In Fig.~\ref{fig:feffElectronMuon} we present the value of $f_{\textrm{eff}}$ for cascades ending in final state electrons and muons with $\epsilon_f = 0.3$. Here we consider DM with mass as low $\mathcal{O}({\rm keV})$ which is relevant to various CMB studies, which should be compared with our general results for $f_\textrm{eff}$ in Fig.~\ref{fig:feffCMB0p3} for DM annihilations into the eight final states considered in the main text. As expected $f_{\textrm{eff}}$ is largest for annihilations to final state electrons.

\begin{figure}[htbp]
\begin{center}
\includegraphics[scale=0.65]{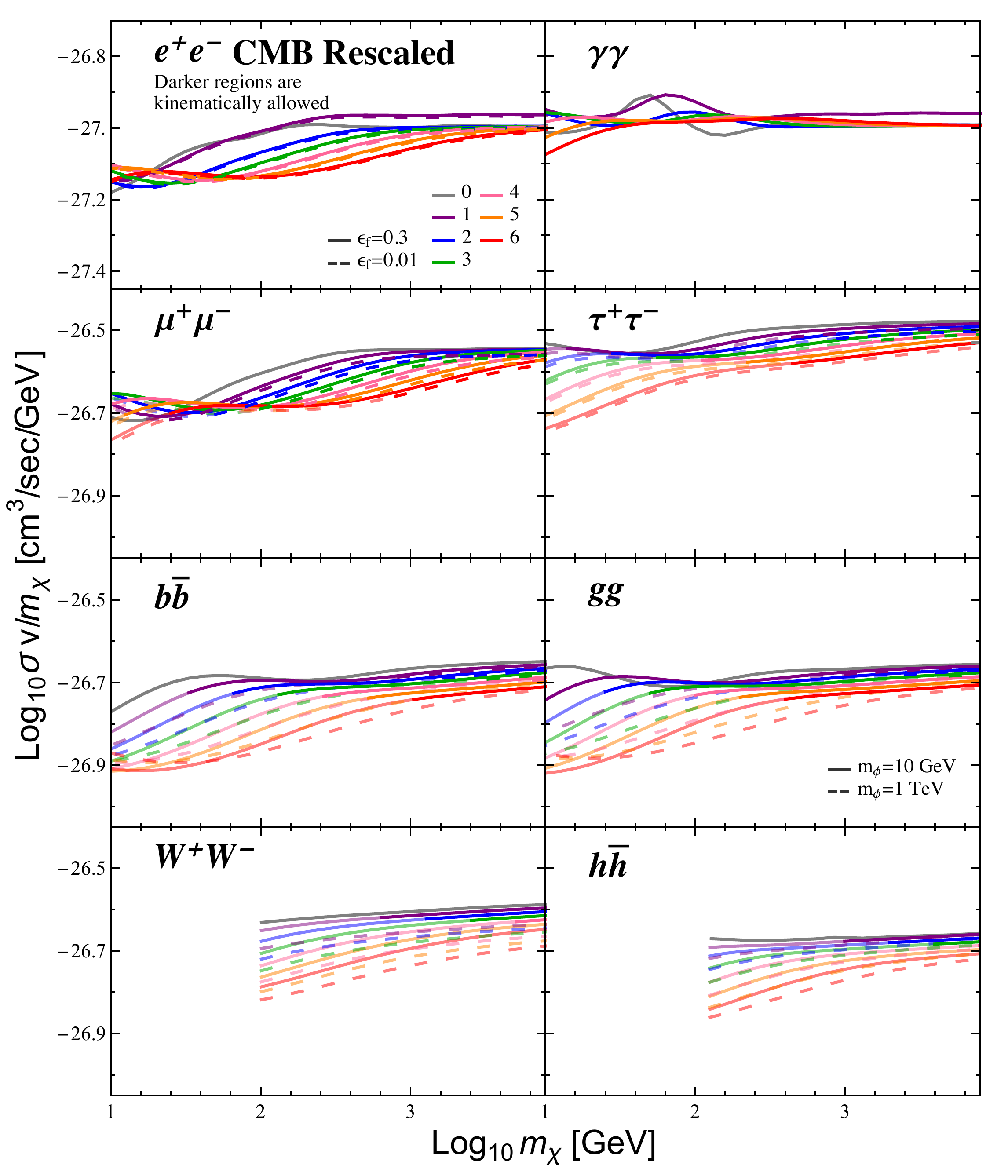}
\end{center}
\caption{\footnotesize{Values of the bound on $\langle \sigma v \rangle / m_\chi$ for various final states. The bound in very robust and we find of order roughly $10^{-27.3} - 10^{-26.6}~\textrm{cm}^3/\textrm{s}/\textrm{GeV}$, independent of final state (although the bound is slightly higher for electrons and photons), number of steps, or $\epsilon_f$.
}}
\label{fig:ScaledCMB0p3}
\end{figure}

The corresponding bound $\langle \sigma v \rangle$ for a given $m_{\chi}$ for light DM annihilating to final state electrons and muons is displayed in Fig ~\ref{fig:BoundElectronMuon}, and more generally in Fig ~ \ref{fig:CMB0p3}. As this bound is fairly insensitive to the final state and number of steps it is interesting to examine the rescaled bound on $\langle \sigma v \rangle / m_\chi$ which we display in Fig.~\ref{fig:ScaledCMB0p3} for $n =$ 0-6 step cascades to various final states. We find that generically the re-scaled bound for all SM final states, $\epsilon_f$ values, and cascade step falls within the narrow range $\langle \sigma v \rangle / m_\chi = 10^{-27.3} -  10^{-26.6}~\textrm{cm}^3/\textrm{s}/\textrm{GeV}$.

\section{Pass 7 versus Pass 8 for the Dwarfs}
\label{app:p7p8}

\begin{figure}[t!]
\centering
\begin{tabular}{c}
\includegraphics[scale=0.6]{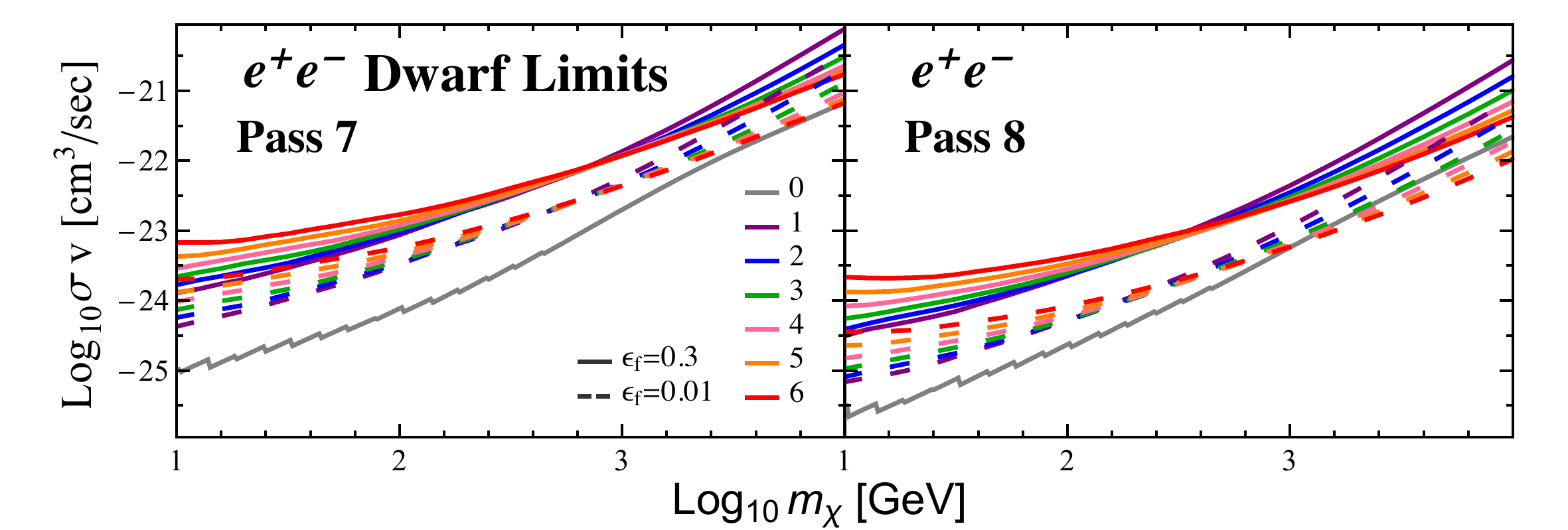}
\end{tabular}
\caption{\footnotesize{Here we recreate the results shown in Fig.~\ref{fig:DwarfLimits} for the case of final state $e^+ e^-$, for the case of the 4 years of Pass 7 data analysed in \cite{Ackermann:2013yva} (left) and for the 6 years of Pass 8 data considered in \cite{Ackermann:2015zua} (right). We see that the updated dataset essentially just strengthens the limits by roughly half an order of magnitude, without noticeably changing other basic features.}}
\label{fig:P7vsP8Dwarfs}
\end{figure}

As discussed in the main body, the limits displayed in Fig.~\ref{fig:DwarfLimits} were derived using 6 years of Pass 8 data collected using the {\it Fermi} Gamma-Ray Space Telescope; more specifically using the publicly available results of \cite{Ackermann:2015zua} made from analysing this data. This work was an updated version of the analysis that appeared in \cite{Ackermann:2013yva}, which set limits using the same 15 dwarf spheroidal galaxies, but only with 4 years of Pass 7 data. These results are also publicly available,\footnote{Both can be obtained from \\ http://www-glast.stanford.edu/pub\_data/} meaning we can cross check how much our results change when between datasets. We did this for each of the final states considered in Fig.~\ref{fig:DwarfLimits} and found generically the shape of the limit curves were unchanged, but that the limits themselves improved by roughly half an order of magnitude when using the updated analysis. We show an example of this for the case of electrons in Fig.~\ref{fig:P7vsP8Dwarfs}, and we see that the generic features of the limits are unchanged but the results strengthen as we move from the Pass 7 to the Pass 8 dataset.

\section{Description of Cascade Spectra Files}
\label{app:fileoutline}

All of the spectra used in this work are publicly available in .dat format at: \\ http://web.mit.edu/lns/research/CascadeSpectra.html. \\ The details of how these spectra were generated from the direct spectra mentioned in Sec.~\ref{sec:Spec} is discussed in Sec.~\ref{sec:Review} and more comprehensively in \cite{Elor:2015tva}. The format of the spectra files has been modeled after those made available by \cite{Cirelli:2010xx}, in the hope that anyone who has used the results of that paper should have no difficulty using ours. In addition to the files themselves we have also included two example files showing how to load the spectra in Mathematica and Python.

There are four basic file types included, which we describe briefly in turn.
\begin{itemize}
\item AtProduction\_\{gammas,positrons,antiprotons\}.dat: these are the files provided by \cite{Cirelli:2010xx} and contain the 0-step or direct annihilation spectrum of \{photons, positrons, antiprotons\} for various final states;
\item Cascade\_\{Gam,E,Mu,Tau,B,W,H,G\}\_gammas.dat: photon spectrum from final state \{photons, electrons, muons, taus, $b$-quarks, Ws, Higgs, gluons\};
\item Cascade\_\{Gam,E,Mu,Tau,B,W,H,G\}\_positrons.dat - positron spectrum from final state \{photons, electrons, muons, taus, $b$-quarks, $W$s, Higgs, gluons\}; and
\item Cascade\_\{B,W,H,G\}\_antiprotons.dat - antiproton spectrum from final state \{$b$-quarks, $W$s, Higgs, gluons\}.
\end{itemize}
Again we emphasize that the AtProduction files were created by the authors of \cite{Cirelli:2010xx}, we only include them in our results as it is convenient to store the 0-step spectra in separate files from the cascade results, yet having them in the same place is useful.

The contents of the three AtProduction\_\{gammas,positrons,antiprotons\}.dat have the following format:
\begin{itemize}
\item Each file has 30 columns and 11099 rows, where the first row contains column labels and all others contain numerical values.
\item The first column contains the dark matter mass in GeV, running from 5 GeV up to 100 TeV. Note using these direct spectra below 5 GeV is not advised as the extrapolation is often unreliable.
\item The second column contains $\log_{10}(x)$ values, where $x=E/m_{\chi}$. This ranges from -8.9 to 0 in steps of 0.05.
\item Finally the columns 3-30 contain the value of the spectrum in $dN/d\log_{10}(x) = \ln(10) x dN/dx$ of the spectrum at that value of $m_{\chi}$ and $x$. The columns of relevance for us are 5 (electrons), 8 (muons), 11 (taus), 14 ($b$-quarks), 18 ($W$-bosons), 22 (gluons), 23 (photons) and 24 (Higgs).
\end{itemize}

The contents of the 19 Cascade\_\{Final State\}\_\{Spectrum Type\}.dat has been modeled on these files. To be explicit we have:
\begin{itemize}
\item Each file has 8 columns and 1612 rows, where the first row contains column labels and all others contain numerical values.
\item The first column contains the value of $\epsilon_f$. We include the spectra for the values 0.01, 0.03, 0.05, 0.07, 0.1, 0.2, 0.3, 0.4 and 0.5. The only exception to this is for gluons or the positron spectrum from photons, where the first column contains $m_{\phi}$ values instead, and we include values of 10, 20, 40, 50, 80, 100, 500, 1000 and 2000 GeV. Within these parameter ranges the interpolation is quite reliable, but outside these ranges linear interpolation is recommended. Note that several spectra, such as the $\gamma \gamma$ photons spectrum or the electron positron spectrum have no dependence on $\epsilon_f$ or $m_{\phi}$. Nevertheless we still include an $\epsilon_f$ column in those files for consistency, and note picking any value of this parameter will result in an identical spectrum.
\item The second column contains $\log_{10}(x)$ values, where $x=E/m_{\chi}$. This ranges from -8.9 to 0 in steps of 0.05.
\item Finally the columns 3-8 contain the value of the spectrum in $dN/d\log_{10}(x) = \ln(10) x dN/dx$ of the spectrum at that value of $m_{\chi}$ and $\epsilon_f$ or $m_{\phi}$. The columns represent an $n=$ 1 cascade (column 3) up to an $n=$ 6 one (column 8).
\end{itemize}

%% file: dm1loop-app.tex
\chapter{Dark Matter Annihilation at One-Loop}

\section{One-loop Calculation of $\chi^a \chi^b \to W^c W^d$ in the Full Theory}
\label{app:oneloopfull}

In this appendix we outline the details of the high-scale matching calculation, which gives rise to the Wilson coefficients stated in Eq.~\eqref{eq:WilsonCoeff}. These coefficients are determined solely by the ultraviolet (UV) physics, allowing us to simplify the calculation by working in the unbroken theory with $m_W=m_Z=\delta m=0$. Combining this with the heavy Majorana fermion DM being non-relativistic, there are only two possible Dirac structures that can appear in the result:
\begin{eqnarray}
\mathcal{M}_A &=& \epsilon_{\mu}^*(p_3) \epsilon_{\nu}^*(p_4) \epsilon^{\sigma \mu \nu \alpha} p_{3\alpha} i \bar{v}(p_2) \gamma_{\sigma} \gamma_5 u(p_1)\,, \nn
\mathcal{M}_B &=& \epsilon_{\mu}^*(p_3) \epsilon_{\nu}^*(p_4) g^{\mu \nu} \bar{v}(p_2) \slashed{p}_3 u(p_1)\,,
\label{eq:FullTheoryOps}
\end{eqnarray}
where $p_1$ and $p_2$ are the momenta of the incoming fermions, whilst $p_3$ and $p_4$ correspond to the outgoing bosons. The symmetry properties of these structures under the interchange of initial and final state particles allow us to write our full amplitude as:
\begin{eqnarray}
\mathcal{M}_{abcd} &=& \frac{4\pi\alpha_2}{m_{\chi}^2} \left\{ \left[ B_1 \delta_{ab} \delta_{cd} + B_2 \left( \delta_{ac} \delta_{bd} + \delta_{ad} \delta_{bc} \right) \right] \mathcal{M}_A \right.\nn
&&\left.+ B_3 \left( \delta_{ac} \delta_{bd} - \delta_{ad} \delta_{bc} \right) \mathcal{M}_B \right\}\,.
\label{eq:FullTheoryStructure}
\end{eqnarray}
The above equation serves to define the Wilson coefficients $B_r$ in a convenient form. These coefficients are related to the EFT coefficients of the operators defined in Eq.~\eqref{eq:Ops} and \eqref{eq:GaugeIndex} via:
\begin{equation}\begin{aligned}
C_1 = (- \pi \alpha_2/m_{\chi}) B_1\,,\;\;\;\;C_2 = (-2\pi \alpha_2/m_{\chi}) B_2\,.
\label{eq:CoeffMapping}
\end{aligned}\end{equation}
For NLL accuracy we only need the tree-level value of these coefficients, which receive a contribution from $s$, $t$ and $u$-channel type graphs and were calculated in \cite{Ovanesyan:2014fwa}. For completeness we state their values here:
\begin{equation}\begin{aligned}
B_1^{(0)} = 1\,,\;\;\; B_2^{(0)} = - \frac{1}{2}\,,\;\;\; B_3^{(0)} = 0\,.
\label{eq:TreeLevelCoeff}
\end{aligned}\end{equation}
Combining these with Eq.~\eqref{eq:CoeffMapping}, we see that the first terms in Eq.~\eqref{eq:WilsonCoeff} are indeed the tree-level contributions as claimed.

The operator associated with $B_3$ was not discussed in the earlier work of \cite{Ovanesyan:2014fwa} as it cannot contribute to the high-scale matching calculation at any order, as we will now argue. Firstly note that the $B_3$ operator is skew under the interchange $a \leftrightarrow b$. Due to the mass splitting between the neutral and charged states, present day annihilation is initiated purely by $\chi^0 \chi^0 = \chi^3 \chi^3$, a symmetric state that cannot overlap with $B_3$. One may worry that exchange of one or more weak bosons between the initial states -- the hallmark of the Sommerfeld enhancement -- may nullify this argument. But it can be checked that if the initial states to such an exchange have identical gauge indices, then so will the final states. As such $B_3$ is not relevant for calculating high-scale matching.\footnote{Diagrams where a soft gauge boson is exchanged between an initial and final state particle would in principle allow $B_3$ to contribute. Such a contributions would however be to the low-scale matching, which we discuss in App.~\ref{app:lowscalematching}. As discussed there, $B_3$ contributions to present day DM annihilation are power suppressed, and therefore do not contribute at any order in the leading power effective theory.}

In spite of this there are several reasons to calculate $B_3$ here. From a practical point of view $B_3$ gives us an additional handle on the consistency of our result, which we check in App.~\ref{app:consistency}. Given that many graphs that generate $B_1$ and $B_2$ also contribute to $B_3$, the consistency of $B_3$ provides greater confidence in the results for the operators we are interested in. Further, from a physics point of view, although $B_3$ is not relevant for high-scale matching when considering present day indirect detection experiments, it could be relevant for calculating the annihilation rate in the early universe, where all states in the DM triplet were present, to the extent that the non-relativistic approximation is still relevant. For this reasons we state it in case it is of interest for future work, such as expanding on calculations of the relic density at one loop (see for example \cite{Boudjema:2005hb,Baro:2007em,Baro:2009na}).

\subsection*{Determining Matching Coefficients}

Let us briefly review how matching coefficients are calculated at one loop. To begin with we can write the general structure of the UV and infrared (IR) divergences of the one-loop result in the full theory as:
\begin{equation}
\mathcal{M}_{\rm bare}^{\rm full} = \frac{K}{\epsilon_{\rm IR}^2} + \frac{L}{\epsilon_{\rm IR}} + \frac{M}{\epsilon_{\rm UV}} + N \left( \frac{1}{\epsilon_{\rm UV}} - \frac{1}{\epsilon_{\rm IR}} \right) + C\,,
\label{eq:barefull}
\end{equation}
where $N$ is the coefficient associated with the various scaleless integrals, and $C$ is the finite contribution. Now the full theory is a renormalizable gauge theory, so we know the counter-term must be of the form:
\begin{equation}
\delta^{\rm full} = - \frac{M+N}{\epsilon_{\rm UV}} + D + \frac{E}{\epsilon_{\rm IR}^2} + \frac{F}{\epsilon_{\rm IR}}\,,
\label{eq:fullct}
\end{equation}
where the values of $D$, $E$ and $F$ are scheme dependent. Nonetheless when calculating matching coefficients it is easiest to work in the on-shell scheme for the wave-function renormalization factors, so below to denote this we add an ``os'' subscript to $D$, $E$ and $F$. The reason this scheme is the most straightforward, is that in any other scheme when we map our Feynman amplitude calculation for $\mathcal{M}^{\rm full}$ onto the $S$-matrix elements we want via the LSZ reduction, there will be non-trivial residues corresponding to the external particles. When using the on-shell scheme for the wave-function renormalization factors, however, these residues are just unity, which simplifies the calculation as we can then ignore them. We emphasize that whatever scheme one uses, the final result for the Wilson coefficients in $\overline{\rm MS}$ will be the same.

With this in mind, if we then combine $\delta^{\rm full}$ with the bare results we obtain a UV finite answer:
\begin{equation}
\mathcal{M}_{\rm ren.}^{\rm full} = \frac{K+E_{\rm os}}{\epsilon_{\rm IR}^2} + \frac{L-N+F_{\rm os}}{\epsilon_{\rm IR}} + C + D_{\rm os}\,.
\label{eq:renfull}
\end{equation}
In our calculation we will use dimensional regularization to regulate both UV and IR divergences, which effectively sets $\epsilon_{\rm UV} = \epsilon_{\rm IR}$, causing all scaleless integrals to vanish. Naively this seems to change the above argument, but as long as we still use the correct counter-term in Eq.~\eqref{eq:fullct} we find:
\begin{equation}\begin{aligned}
\mathcal{M}_{\rm ren.}^{\rm full} =& \frac{K}{\epsilon^2} + \frac{L}{\epsilon} + \frac{M}{\epsilon} + C - \frac{M+N}{\epsilon} \\
&+ D_{\rm os} + \frac{E_{\rm os}}{\epsilon^2} + \frac{F_{\rm os}}{\epsilon} \\
=& \frac{K+E_{\rm os}}{\epsilon^2} + \frac{L-N+F_{\rm os}}{\epsilon} + C + D_{\rm os}\,.
\end{aligned}\end{equation}
Comparing this with Eq.~\eqref{eq:renfull}, we see that if we interpret all of the divergences in the final result as IR, then this method is equivalent to carefully distinguishing $\epsilon_{\rm UV}$ and $\epsilon_{\rm IR}$ throughout.

In the EFT, with the above choice of zero masses and working on-shell with dimensional regularization, all graphs are scaleless. At one loop they have the general form:\footnote{One may worry there could also be scaleless integrals of the form $\left( \epsilon_{\rm UV}^{-1} - \epsilon_{\rm IR}^{-1} \right)^2$, but the use of the zero-bin subtraction \cite{Manohar:2006nz} ensures such contributions cannot appear.}
\begin{equation}
\mathcal{M}_{\rm bare}^{\rm EFT} = O \left( \frac{1}{\epsilon_{\rm UV}^2} - \frac{1}{\epsilon_{\rm IR}^2} \right) + P \left( \frac{1}{\epsilon_{\rm UV}} - \frac{1}{\epsilon_{\rm IR}} \right)\,.
\label{eq:bareeft}
\end{equation}
Importantly if we have the correct EFT description of the full theory, then the two theories must have the same IR divergences. Comparing Eq.~\eqref{eq:bareeft} to Eq.~\eqref{eq:renfull}, we see this requires $O = - K-E_{\rm os}$ and $P=N-L-F_{\rm os}$. The EFT is again a renormalizable theory, so we can cancel the UV divergences using $\delta^{\rm EFT} = (K+E_{\rm os}) \epsilon_{\rm UV}^{-2} +(L+F_{\rm os}-N) \epsilon_{\rm UV}^{-1}$. Note as all EFT graphs are scaleless there are no finite contributions that could be absorbed into the counter-term, so in any scheme there is no finite correction to $\delta^{\rm EFT}$. Using this counter-term, we conclude:
\begin{equation}
\mathcal{M}_{\rm ren.}^{\rm EFT} = \frac{K+E_{\rm os}}{\epsilon_{\rm IR}^2} + \frac{L-N+F_{\rm os}}{\epsilon_{\rm IR}} \,.
\label{eq:reneft}
\end{equation}
Again note that for a similar argument to that in the full theory, if we had set $\epsilon_{\rm UV} = \epsilon_{\rm IR}$ at the outset, then as long as we still used the correct counter-term we would arrive at the same result.

The matching coefficient is then obtained from subtracting the renormalized EFT from the renormalized full theory result, so taking the appropriate results above we conclude:
\begin{equation}
\mathcal{M}_{\rm ren.}^{\rm full} - \mathcal{M}_{\rm ren.}^{\rm EFT} = C+D_{\rm os}\,.
\label{eq:sketchmatching}
\end{equation}
Comparing this with Eq.~\eqref{eq:renfull}, we see that provided we have the correct EFT, then the matching coefficient is just the finite contribution to the renormalized full theory amplitude in the on-shell scheme. Even though this result makes explicit reference to a scheme in $D_{\rm on-shell}$, it is in fact scheme independent. The reason for this is that if we worked in a different scheme, although $D$ would change, we would also have to account for the now non-trivial external particle residues that enter via LSZ. Their contribution is what ensures Eq.~\eqref{eq:sketchmatching} is scheme independent.

\subsection*{Results of the Calculation}

As outlined above, in order to obtain the matching coefficients we need the finite contribution to the renormalized full theory amplitude. Now to compute this in the particular theory we consider in this work, we need to calculate the 25 diagrams that contribute to the one-loop correction to $\chi^a \chi^b \to W^c W^d$. The diagrams are identical to those considered in \cite{Fuhrer:2010eu}, where they defined a numbering scheme for the diagrams, grouping them by topology and labelling them as $T_i$ for various $i$. We follow that numbering scheme here, but cannot use their results as they considered massless initial state fermions whilst ours are massive and non-relativistic. In general we calculate the diagrams using dimensional regularization with $d=4-2\epsilon$ to regulate the UV and IR, and work in `t Hooft-Feynman gauge. Loop integrals are determined using Passarino-Veltman reduction \cite{Passarino:1978jh}, and we further make use of the results in \cite{Denner:1992vza,Ellis:2007qk,'tHooft:1978xw,Ellis:2011cr} as well as FeynCalc \cite{Mertig:1990an,Shtabovenko:2016sxi} and Package-X \cite{Patel:2015tea}.

In the EFT description of the full theory outlined in Sec.~\ref{sec:NLL}, the factorization of the matrix elements ensured a separation between the Sommerfeld and Sudakov contributions. Yet for the full theory no clear separation exists and there will be graphs that contribute to both effects -- in particular the graph $T_{1c}$ considered below. The purpose of the Wilson coefficients we are calculating here is to provide corrections to the Sudakov contribution -- we do not want to spoil the EFT distinction by including Sommerfeld effects in these coefficients. In order to cleanly separate the contributions we take the relative velocity of our non-relativistic initial states to be zero. This ensures that any contribution of the form $1/v$, characteristic of Sommerfeld enhancement, become power divergences and therefore vanish in dimensional regularization. This is different to the treatment in HI, where they calculated the diagram without sending $v \to 0$ and subtracted the Sommerfeld contribution by hand.

In our calculation the DM is a Majorana fermion. It turns out that for almost all the graphs below the result is identical regardless of whether we think of the fermion as Majorana or Dirac -- a result that is also true at tree-level. The additional symmetry factors in the Majorana case are exactly cancelled by the factors of $1/2$ entering from the Majorana Lagrangian. The exceptions to this are for graphs containing a closed loop of fermions, specifically $T_{2d}$ and $T_{6d}$ below, as well as closed fermion loop contributions to the counter-terms.

Using the approach outlined above we now state the contribution to $B_r$ as defined in Eq.~\eqref{eq:FullTheoryStructure} graph by graph. Throughout we define $L\equiv \ln \mu/2m_{\chi}$.

\subsection*{\large $T_{1a}$}
\begin{center}
\includegraphics[height=0.15\columnwidth]{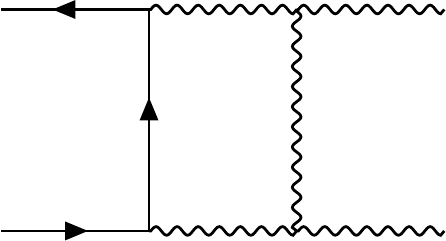}
\end{center}
The result for this graph and its cross term is:
\begin{equation}\begin{aligned}
B_1^{[1a]} &= \frac{\alpha_2}{4\pi} \left[ - \frac{2}{\epsilon^2} - \frac{1}{\epsilon} \left( 4L + 2 i \pi + 2 \right) - 4 L^2 \right. \\
&\hspace{0.4in} \left.- 4 L - 4 i \pi L - 4 + \frac{7\pi^2}{6} + 4 \ln 2  \right]\,, \\
B_2^{[1a]} &= \frac{1}{2} B_1^{[1a]}\,, \\
B_3^{[1a]} &= \frac{\alpha_2}{4\pi} \left[ \frac{1}{4 \epsilon^2} + \frac{1}{4 \epsilon} \left( 2L - 3 i \pi - 2 \right) + \frac{1}{2} L^2 \right. \\
&\hspace{1.04in}- L - \frac{3}{2} i \pi L + \frac{17\pi^2}{48}\\
&\hspace{1.04in}\left. - \frac{1}{6} (2 + 7 i \pi - 8 \ln 2 ) \right]\,.
\end{aligned}\end{equation}
In calculating this graph in the non-relativistic limit via Passarino-Veltman reduction there are additional spurious divergences that must be regulated. The origin of these divergences is that Passarino-Veltman assumes the momenta appearing in the integrals to be linearly independent. But in the center of momentum frame if we take $v=0$, then $p_1$ and $p_2$ are identical and this assumption breaks down, leading to the divergences of the form $(s-4m_{\chi}^2)^{-1}$, where $s=(p_1+p_2)^2$. A simple way to regulate them is to give the initial states a small relative velocity. This does not lead to a violation of our separation of Sommerfeld and Sudakov effects as this graph does not contribute to the Sommerfeld enhancement. As such this procedure introduces no $1/v$ contributions to the final result and the regulator can be safely removed at the end. This is the only diagram where this issue appears -- if it occurred in a graph that did contribute to the Sommerfeld effect we would need to use a different regulator.

\subsection*{\large $T_{1b}$}
\begin{center}
\includegraphics[height=0.15\columnwidth]{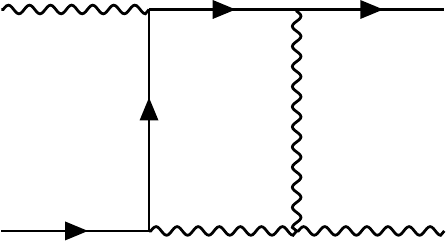}
\end{center}
This graph has a single crossed term and combining the two yields:
\begin{equation}\begin{aligned}
B_1^{[1b]} &= B_3^{[1b]} = 0\,, \\
B_2^{[1b]} &= \frac{\alpha_2}{4\pi} \left[ \frac{2}{\epsilon^2} + \frac{4L+2}{\epsilon} + 4L(L+1) \right. \\
&\hspace{0.85in} \left. - \frac{2\pi^2}{3} + 4 - 8 \ln 2 \right]\,.
\end{aligned}\end{equation}

\subsection*{\large $T_{1c}$}
\begin{center}
\includegraphics[height=0.15\columnwidth]{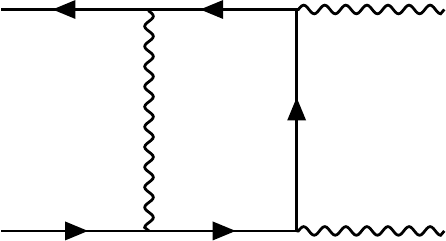}
\end{center}
The combination of this graph and its crossed term is:
\begin{equation}\begin{aligned}
B_1^{[1c]} &= \frac{\alpha_2}{4\pi} \left[ \frac{2}{\epsilon} - 4 + 4L + 4 \ln 2 \right]\,, \\
B_2^{[1c]} &= \frac{1}{2} B_1^{[1c]}\,, \\
B_3^{[1c]} &= \frac{\alpha_2}{4\pi} \left[ \frac{1}{\epsilon} - 2 + 2L + \frac{\pi^2}{4} - 2 \ln 2 \right]\,.
\end{aligned}\end{equation}
Formally this graph also gives a contribution to the Sommerfeld enhancement in the full theory. Nevertheless as we take $v=0$ at the outset, the contribution here is purely to the Sudakov terms.

\subsection*{\large $T_{1d}$}
\begin{center}
\includegraphics[height=0.15\columnwidth]{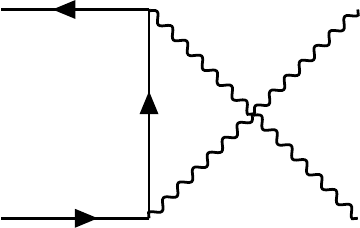}
\end{center}
The contribution from this diagram vanishes in the non-relativistic limit, i.e.
\begin{equation}\begin{aligned}
B_1^{[1d]} = B_2^{[1d]} = B_3^{[1d]} = 0\,.
\end{aligned}\end{equation}

\subsection*{\large $T_{2a}$}
\begin{center}
\includegraphics[height=0.15\columnwidth]{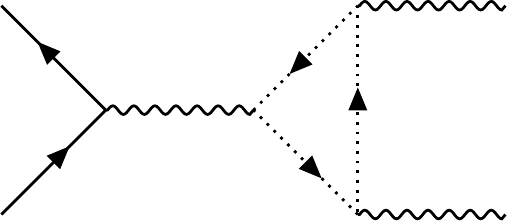}
\end{center}
For the case of ghosts running in the loop of the above graph we have its contribution and the crossed term giving
\begin{equation}\begin{aligned}
B_1^{[2a]} &= B_2^{[2a]} = 0\,, \\
B_3^{[2a]} &= \frac{\alpha_2}{4\pi} \left[ \frac{1}{24\epsilon} + \frac{2L + i \pi}{24} + \frac{11}{72} \right]\,. 
\end{aligned}\end{equation}

\subsection*{\large $T_{2b}$}
\begin{center}
\includegraphics[height=0.15\columnwidth]{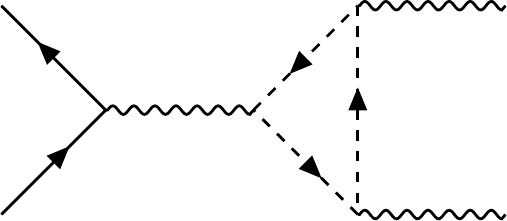}
\end{center}
For a scalar Higgs in the loop, the graph and its cross term contribute:
\begin{equation}\begin{aligned}
B_1^{[2b]} &= B_2^{[2b]} = 0\,, \\
B_3^{[2b]} &= \frac{\alpha_2}{4\pi} \left[ \frac{1}{12\epsilon} + \frac{2L + i \pi}{12} + \frac{11}{36} \right]\,.
\end{aligned}\end{equation}

\subsection*{\large $T_{2c}$}
\begin{center}
\includegraphics[height=0.15\columnwidth]{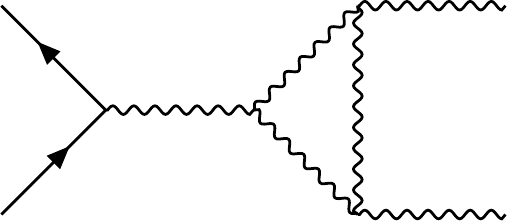}
\end{center}
There is no crossed graph associated with the graph above as the gauge bosons running in the loop are real fields. As such taking just this graph gives
\begin{equation}\begin{aligned}
B_1^{[2c]} &= B_2^{[2c]} = 0\,, \\
B_3^{[2c]} &= \frac{\alpha_2}{4\pi} \left[ \frac{3}{4\epsilon^2} + \frac{1}{\epsilon} \left( \frac{3}{4} (2L + i \pi) + \frac{17}{8} \right) \right.\\
&\hspace{0.5in}+ \frac{3}{8} \left( 2L + i \pi \right)^2 \\
&\hspace{0.5in}\left. + \frac{17}{8} \left( 2L + i \pi \right) + \frac{95}{24} - \frac{\pi^2}{16} \right]\,.
\end{aligned}\end{equation}

\subsection*{\large $T_{2d}$}
\begin{center}
\includegraphics[height=0.15\columnwidth]{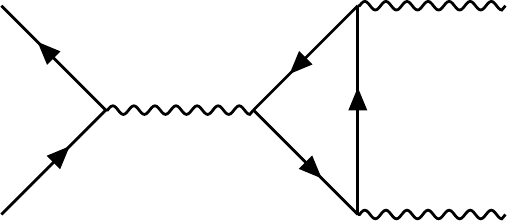}
\end{center}
There are two types of fermions that can run in the loop: the Majorana triplet fermion that make up our DM or left-handed SM doublets. As with the gauge bosons these SM fermions are taken to be massless and for generality we say there are $n_D$ of them.\footnote{For the SM well above the electroweak scale $n_D=12$. In detail, for each generation there are four doublets: the lepton doublet and due to color, three quark doublets. As such for three generations we have twelve left-handed SM doublets.} For the SM doublets there is a crossed graph, whilst for the Majorana DM field there is not, so that:
\begin{equation}\begin{aligned}
B_1^{[2d]} &= B_2^{[2d]} = 0\,, \\
B_3^{[2d]} &= \frac{\alpha_2}{4\pi} \left[ - \left( \frac{2}{3\epsilon} + \frac{4}{3} L + \frac{4}{3} \ln 2 - \frac{5}{9} + \frac{\pi^2}{4} \right) \right. \\
&\hspace{0.5in} \left.- n_D \left( \frac{1}{6\epsilon} + \frac{1}{6} \left( 2L + i \pi \right) + \frac{7}{36} \right) \right]\,.
\end{aligned}\end{equation}
If the DM had been a Dirac field instead, there would have been a crossed graph and the result would be modified such that the first line of $B_3^{[2d]}$ gets multiplied by 2.

The factor of $7/36$ we find in the last line of $B_3^{[2d]}$ is consistent with the expression found for this graph, but with different kinematics, in \cite{Fuhrer:2010eu}, but disagrees with \cite{Bardin:1999ak}.

\subsection*{\large $T_{2e-h}$}
\begin{center}
\includegraphics[height=0.15\columnwidth]{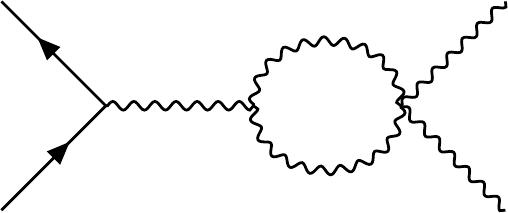} \hspace{0.1in}
\includegraphics[height=0.15\columnwidth]{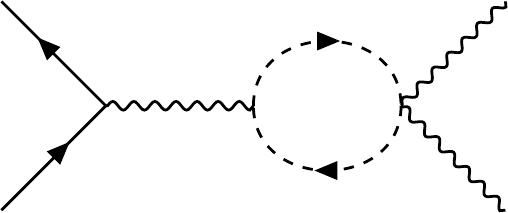} \\
\vspace{0.1in}
\includegraphics[height=0.15\columnwidth]{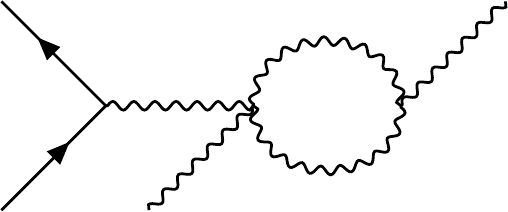} \hspace{0.1in}
\includegraphics[height=0.15\columnwidth]{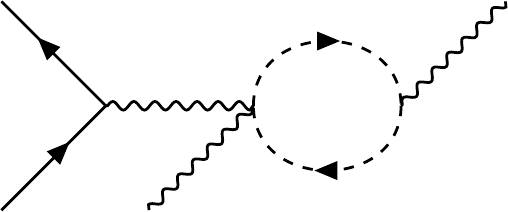}
\end{center}
The four graphs shown above do not contribute to our one-loop result; the graphs on the top row do not generate either $\mathcal{M}_A$ or $\mathcal{M}_B$, whilst the loops on the second line are both scaleless and so vanish in dimensional regularization. As such we have:
\begin{equation}\begin{aligned}
B_1^{[2e-f]} = B_2^{[2e-f]} = B_3^{[2e-f]} = 0\,.
\end{aligned}\end{equation}

\subsection*{\large $T_{3a}$ and $T_{4a}$}
\begin{center}
\includegraphics[height=0.15\columnwidth]{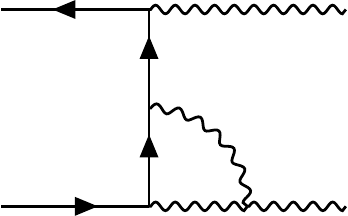} \hspace{0.1in}
\includegraphics[height=0.15\columnwidth]{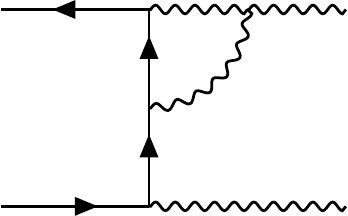}
\end{center}
The two graphs shown above have identical amplitudes. For each graph independently, the sum of it and its crossed graph is:
\begin{equation}\begin{aligned}
B_1^{[3a/4a]} &= \frac{\alpha_2}{4\pi} \left[ - \frac{1}{\epsilon^2} + \frac{2-2L}{\epsilon} - 2L^2 \right. \\
&\hspace{0.3in} \left. + 4L - 2\ln 2 + 4 + \frac{\pi^2}{12} \right]\,, \\
B_2^{[3a/4a]} &= - \frac{1}{2} B_1^{[3a/4a]}\,, \\
B_3^{[3a/4a]} &= \frac{1}{2} B_1^{[3a/4a]}\,.
\end{aligned}\end{equation}

\subsection*{\large $T_{3b}$ and $T_{4b}$}
\begin{center}
\includegraphics[height=0.15\columnwidth]{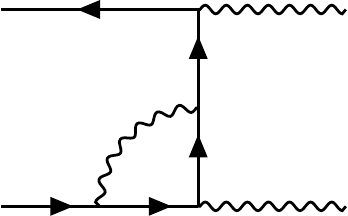} \hspace{0.1in}
\includegraphics[height=0.15\columnwidth]{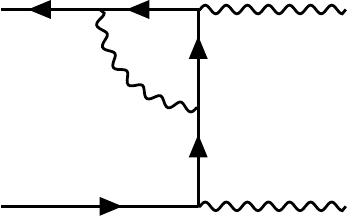}
\end{center}
As for $T_{3a}$ and $T_{4a}$, these two graphs also have equal amplitudes. Again we provide the combination of each with its crossed graph:
\begin{equation}\begin{aligned}
B_1^{[3b/4b]} &= \frac{\alpha_2}{4\pi} \left[ \frac{1}{\epsilon} + 2L - 2\ln 2 + \frac{\pi^2}{4} \right]\,, \\
B_2^{[3b/4b]} &= -\frac{1}{2} B_1^{[3b/4b]}\,, \\
B_3^{[3b/4b]} &= \frac{1}{2} B_1^{[3b/4b]}\,.
\end{aligned}\end{equation}

\subsection*{\large $T_{5a}$}
\begin{center}
\includegraphics[height=0.15\columnwidth]{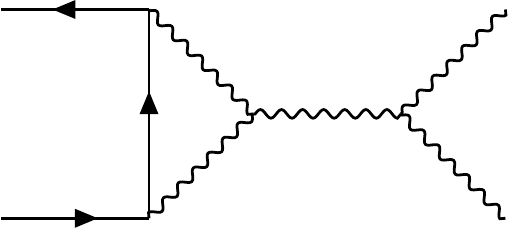}
\end{center}
Whether the above graph has a crossed graph associated with interchanging the initial states depends on the identity of the initial state fermions. For Majorana fermions there is such a crossing, whilst for Dirac there is not. Despite this, in either case the combination of the graph and its crossing (where it exists) is the same in both cases and is simply:
\begin{equation}\begin{aligned}
B_1^{[5a]} &= B_2^{[5a]} = 0\,, \\ 
B_3^{[5a]} &= \frac{\alpha_2}{4\pi} \left[ - \frac{3}{2\epsilon} - 3 L - \frac{13}{3} \ln 2 - \frac{8}{3} + \frac{2}{3} i \pi \right]\,.
\end{aligned}\end{equation}

\subsection*{\large $T_{5b}$}
\begin{center}
\includegraphics[height=0.15\columnwidth]{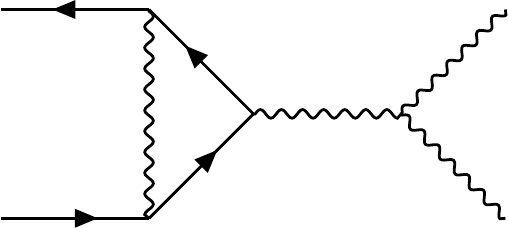}
\end{center}
As for $T_{5a}$ the existence of a crossed graph depends on the nature of the DM. Regardless again the result is the same if we take it to be Dirac or Majorana, which is:
\begin{equation}\begin{aligned}
B_1^{[5b]} &= B_2^{[5b]} = 0\,, \\ 
B_3^{[5b]} &= \frac{\alpha_2}{4\pi} \left[ \frac{3}{2\epsilon} + 3L + 3 \ln 2 - 2 \right]\,.
\end{aligned}\end{equation}

\subsection*{\large $T_{6a}$}
\begin{center}
\includegraphics[height=0.15\columnwidth]{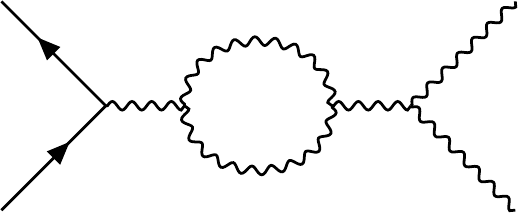}
\end{center}
For a gauge boson in the loop we have:
\begin{equation}\begin{aligned}
B_1^{[6a]} &= B_2^{[6a]} = 0\,, \\ 
B_3^{[6a]} &= \frac{\alpha_2}{4\pi} \left[ - \frac{19}{12\epsilon} - \frac{19}{6} L - \frac{29}{9} - \frac{19}{12} i \pi \right]\,.
\end{aligned}\end{equation}
Note this graph and the remaining $T_6$ type topologies have no crossed graphs.

\subsection*{\large $T_{6b}$}
\begin{center}
\includegraphics[height=0.15\columnwidth]{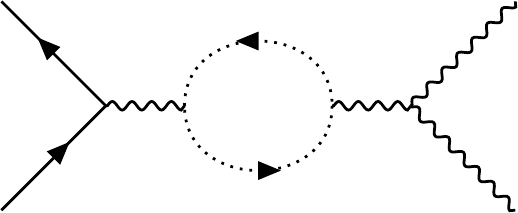}
\end{center}
In the case of a ghost loop we have:
\begin{equation}\begin{aligned}
B_1^{[6b]} &= B_2^{[6b]} = 0\,, \\ 
B_3^{[6b]} &= \frac{\alpha_2}{4\pi} \left[ - \frac{1}{12\epsilon} - \frac{1}{6} L - \frac{2}{9} - \frac{1}{12} i \pi \right]\,.
\end{aligned}\end{equation}

\subsection*{\large $T_{6c}$}
\begin{center}
\includegraphics[height=0.15\columnwidth]{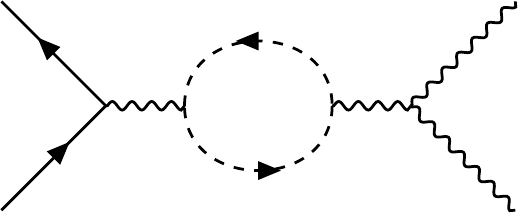}
\end{center}
For a scalar Higgs we have an identical contribution to $T_{6b}$:
\begin{equation}\begin{aligned}
B_1^{[6c]} &= B_2^{[6c]} = 0\,, \\ 
B_3^{[6c]} &= \frac{\alpha_2}{4\pi} \left[ - \frac{1}{12\epsilon} - \frac{1}{6} L - \frac{2}{9} - \frac{1}{12} i \pi \right]\,.
\end{aligned}\end{equation}

\subsection*{\large $T_{6d}$}
\begin{center}
\includegraphics[height=0.15\columnwidth]{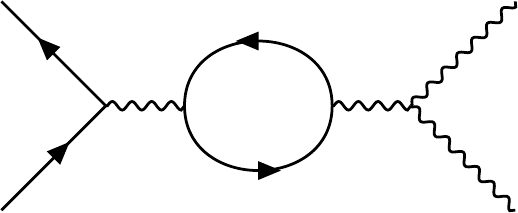}
\end{center}
As for $T_{2d}$ the fermion in the loop could again be either DM or SM. Allowing there to be $n_D$ left-handed SM doublets we have
\begin{equation}\begin{aligned}
B_1^{[6d]} &= B_2^{[6d]} = 0\,, \\ 
B_3^{[6d]} &= \frac{\alpha_2}{4\pi} \left[ \left( \frac{2}{3\epsilon} + \frac{4}{3} L + \frac{4}{3} \ln 2 + \frac{16}{9} \right) \right. \\
&\hspace{0.5in} \left.+ n_D \left( \frac{1}{6\epsilon} + \frac{1}{3} L + \frac{5}{18} + \frac{1}{6} i \pi \right) \right]\,.
\end{aligned}\end{equation}
Here there is a symmetry factor of $1/2$ for the loop in the case of the Majorana DM field. If the DM was a Dirac fermion instead, the first line of $B_3^{[6d]}$ would get multiplied by $2$ as this symmetry factor would not be present.

\subsection*{\large $T_{6e}$ and $T_{6f}$}
\begin{center}
\includegraphics[height=0.15\columnwidth]{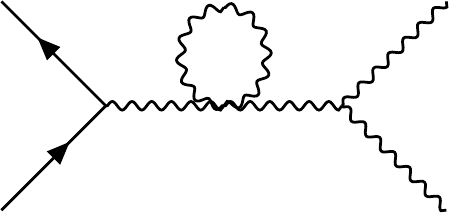} \hspace{0.1in}
\includegraphics[height=0.15\columnwidth]{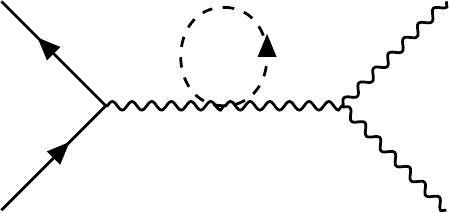}
\end{center}
Both of these integrals are scaleless and vanish in dimensional regularization, so:
\begin{equation}\begin{aligned}
B_1^{[6e-f]} = B_2^{[6e-f]} = B_3^{[6e-f]} = 0\,.
\end{aligned}\end{equation}

\subsection*{\large $T_{7}$}
\begin{center}
\includegraphics[height=0.18\columnwidth]{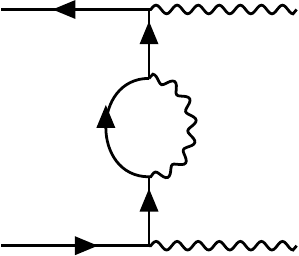}
\end{center}
For the final graph we again have a crossed contribution, and combining the two gives:
\begin{equation}\begin{aligned}
B_1^{[7]} &= \frac{\alpha_2}{4\pi} \left[ -\frac{8}{\epsilon} - 16L - 12 \right]\,, \\
B_2^{[7]} &= - \frac{1}{2} B_1^{[7]}\,, \\
B_3^{[7]} &= \frac{1}{2} B_1^{[7]}\,.
\end{aligned}\end{equation}

\newpage
\subsection*{Counter-terms}
To begin with, as $B_3$ vanishes at tree level there are no counter-term corrections to its value at one loop. Instead we only need to consider graphs that would contribute to $B_1$ and $B_2$, of which there are three:
\begin{center}
\includegraphics[height=0.15\columnwidth]{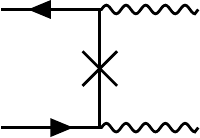} \hspace{0.1in}
\includegraphics[height=0.16\columnwidth]{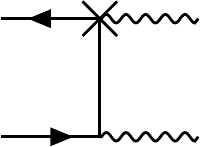} \hspace{0.1in}
\includegraphics[height=0.16\columnwidth]{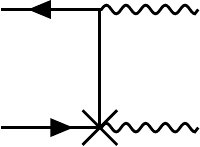} \hspace{0.1in}
\end{center}
The graph on the left corresponds to the wave-function and mass renormalization of the DM -- denoted as $Z_{\chi}$ and $Z_m$ -- whilst the remaining two graphs account for the renormalization of the DM and electroweak gauge boson interaction vertex $g_2 \bar{\chi} \slashed{W} \chi$ -- here $Z_1$. Now if we calculate the above three graphs, we find a contribution proportional to the tree-level amplitude $\mathcal{M}_{\rm tree}$, as well as a term that would contribute to $B_3$. As we know this latter term must be cancelled by other graphs given $B_3^{(0)}=0$, we keep only the former term which gives:
\begin{equation}
\left(2 \delta_1 - \delta_{\chi} - \delta_m \right) \mathcal{M}_{\rm tree}\,,
\label{eq:CTgraph}
\end{equation}
where we have used $Z_i = 1 + \delta_i$.

Next, when determining the $\delta_i$ we need to pick a scheme. As explained above, when calculating matching coefficients it is easiest to work in the on-shell scheme to ensure we do not have to worry about residues from the LSZ reduction. The meaning of the on-shell values of $\delta_{\chi}$ and $\delta_m$ is clear, whereas the interpretation of the on-shell $\delta_1$ is ambiguous in a non-abelian theory. Here we treat this counter-term as follows. By definition we know $\delta_1 = \delta_{g_2} + \frac{1}{2} \delta_W + \delta_{\chi}$, where $\delta_{g_2}$ and $\delta_W$ are the counter-terms for the coupling and gauge boson wave-functions respectively. For the gauge-boson wave-function we use the on-shell scheme as usual. For the coupling counter-term, however, we define it to be purely UV, in the $\overline{\rm MS}$ scheme and as our full theory is defined with the DM a propagating degree of freedom, this coupling is defined above the $m_{\chi}$. In the EFT the DM is integrated out, so the appropriate coupling for the matching is one defined below $m_{\chi}$. We put this issue aside for now and return to it in the next section.

The above choices then define our scheme for $\delta_1$ in a manner that ensures all residues are still 1. With this scheme, we can then calculate the relevant counter-terms and find:
\begin{eqnarray}
\delta_{\chi} &=& - \frac{\alpha_2}{4\pi} \left[ \frac{2}{\epsilon_{\rm UV}} +  4L + 4 \ln 2 + 4 \right]\,, \label{eq:CTvalues} \\
\delta_m &=& - \frac{\alpha_2}{4\pi} \left[ \frac{6}{\epsilon_{\rm UV}} + 12L + 12 \ln 2 + 8 \right]\,, \nn
\delta_W &=& - \frac{\alpha_2}{4\pi} \left[ \frac{2n_D-3}{6\epsilon_{\rm UV}} + \frac{19-2n_f}{6\epsilon_{\rm IR}} + \frac{16}{3} L + \frac{16}{3} \ln 2 \right]\,, \nn
\delta_g &=& - \frac{\alpha}{4\pi} \left[ \frac{27-2n_D}{12\epsilon_{\rm UV}} \right]\,, \nn
\delta_1 &=& - \frac{\alpha_2}{4\pi} \left[ \frac{4}{\epsilon_{\rm UV}} + \frac{19-2n_D}{12\epsilon_{\rm IR}} + \frac{20}{3} L + \frac{20}{3} \ln 2 + 4 \right]\,, \nonumber
\end{eqnarray}
where $n_D$ is again the number of left-handed SM doublets. Recall that in determining the counter-terms we cannot neglect scaleless integrals as we did for the main calculation, so their contribution has been included here and we explicitly distinguish $\epsilon_{\rm UV}$ from $\epsilon_{\rm IR}$. Subbing these results into Eq.~\eqref{eq:CTgraph}, we find the crossed contribution is:
\begin{equation}\begin{aligned}
B_1^{[{\rm CT}]} &= \frac{\alpha_2}{4\pi} \left[ \frac{2n_D-19}{6\epsilon_{\rm IR}} + \frac{8}{3} L + \frac{8}{3} \ln 2 + 4 \right]\,, \\
B_2^{[{\rm CT}]} &= \frac{\alpha_2}{4\pi} \left[ \frac{19-2n_D}{12\epsilon_{\rm IR}} - \frac{8}{6} L - \frac{8}{6} \ln 2 - 2 \right]\,, \\
B_3^{[{\rm CT}]} &= 0\,.
\label{eq:CTcontribution}
\end{aligned}\end{equation}
Interestingly the counter-term contribution is UV finite. This implies that the sum of all one-loop graphs before adding in counter-terms must be UV finite. Given that we used dimensional regularization to regulate both UV and IR divergences this cannot be immediately read off from our results, but going back to the integrals and keeping track of the UV divergences we confirmed that the sum is indeed UV finite. This cancellation appears to be purely accidental.

Note if our DM field had instead been a Dirac fermion, there would be several modifications to the above. Firstly the $L$ and $\ln 2$ dependence in $\delta_W$ and $\delta_1$ would be modified, whilst the $\epsilon_{\rm UV}$ dependence in $\delta_W$ and $\delta_g$ would also change. In the combination stated in Eq.~\eqref{eq:CTcontribution} this only changes the $L$ and $\ln 2$ dependence, but in a way that is exactly cancelled when we account for the scale of the coupling in the next section.

\subsection*{Scale of the Coupling}

Throughout the above calculation we have treated the DM as a propagating degree of freedom and included its effects in loop diagrams. This implies that the coupling used so far throughout this appendix implicitly depends on $n_D+1$ flavors -- $n_D$ left-handed SM doublets and one Majorana DM fermion -- i.e. we have used $\alpha_2 = \alpha_2^{(n_D+1)}(\mu)$. In the EFT however, the DM is no longer a propagating field and so the appropriate coupling is $\alpha_2^{(n_D)}(\mu)$. At order $\alpha_2^2$, which we are working to at one loop, the distinction will lead to a finite contribution because of the matching at the scale $\mu=m_{\chi}$, which we calculate in this section.

Let us start by reviewing the treatment of the running coupling in general. This running is captured by the $\beta$-function, which is defined by $\beta(\alpha_2) = \mu d\alpha_2/d\mu$, where here $\alpha_2$ is the renormalized coupling; the bare coupling is independent of $\mu$. In general the $\beta$-function can be written as:
\begin{equation}\begin{aligned}
\beta(\alpha_2) &= - 2 \alpha_2 \left[ \epsilon + \sum_{n=1}^{\infty} \left( \frac{\alpha_2}{4\pi} \right)^n b_{n-1} \right] \\
&= -2 \epsilon \alpha_2 - \frac{b_0}{2\pi} \alpha_2^2 + \ldots\,,
\end{aligned}\end{equation}
where we have expanded it to the order we work to in the second equality. At this order we can solve for the running of the coupling as:
\begin{equation}
\alpha_2(\mu) = \frac{\alpha_2(\mu_0)}{1+\alpha_2(\mu_0) \frac{b_0}{2\pi} \ln \frac{\mu}{\mu_0} + \ldots}\,.
\label{eq:coupling}
\end{equation}
Now the above is completely general, so let us focus in to the specific problem we have. At this order it suffices to demand that the couplings match at the scale $m_{\chi}$, and at one-loop this is captured by a difference in $b_0$. For our problem we define $b_0^{(n_D+1)}$ to be the value above $m_{\chi}$ and $b_0^{(n_D)}$ the value below. Then using Eq.~\eqref{eq:coupling} to define $\alpha_2^{(n_D+1)}(\mu)$ and $\alpha_2^{(n_D)}(\mu)$, we demand they match at a scale $m_{\chi}$, which gives:
\begin{equation}\begin{aligned}
&\alpha_2^{(n_D+1)}(\mu) = \alpha_2^{(n_D)}(\mu) \left[ 1 \vphantom{\frac{\mu}{m_{\chi}}} \right. \\
&\left. + \frac{\alpha_2^{(n_D)}(\mu)}{2\pi} \left( b_0^{(n_D+1)} - b_0^{(n_D)} \right) \ln \frac{\mu}{m_{\chi}} + \ldots \right]\,.
\label{eq:genalphamatch}
\end{aligned}\end{equation}
So now we just need to determine $b_0^{(n_D+1)} - b_0^{(n_D)}$. In general for a theory containing just gauge bosons, Weyl fermions (WF), Majorana fermions (MF) and charged scalars (CS), we can write:
\begin{equation}\begin{aligned}
b_0 &= \frac{11}{3} C_A - \frac{2}{3} \sum_{i\in {\rm WF}} C(R_i) \\
&- \frac{2}{3} \sum_{i\in {\rm MF}} C(R_i) - \frac{1}{3} \sum_{i\in {\rm CS}} C(R_i)\,.
\end{aligned}\end{equation}
Our calculation has all four of these ingredients: electroweak gauge bosons, the left-handed SM fermions (which are Weyl because only one chirality couples to the gauge bosons), the Majorana DM fermion and the Higgs. Then using $C_A = 2$, $C(R) = 1/2$ for the SM left-handed fermions and the Higgs, and $C(R)=2$ for the adjoint Wino, we conclude:
\begin{equation}\begin{aligned}
b_0^{(n_D)} &= \frac{43-2n_D}{6}\,, \\
b_0^{(n_D+1)} &= \frac{35-2n_D}{6}\,.
\end{aligned}\end{equation}
From this Eq.~\eqref{eq:genalphamatch} tells us that to the order we are working:
\begin{equation}\begin{aligned}
\alpha_2^{(n_D+1)}(\mu) = \alpha_2^{(n_D)}(\mu) \left[ 1 - \frac{\alpha_2^{(n_D)}(\mu)}{4\pi} \left( \frac{8}{3} L + \frac{8}{3} \ln 2 \right) \right]\,.
\end{aligned}\end{equation}
Now as there is only a difference between the couplings at next to leading order, this only corrects the tree level result stated in Eq.~\eqref{eq:TreeLevelCoeff}. As such the impact of changing to the coupling defined below $m_{\chi}$, which is relevant for the matching, is to add the following contribution:
\begin{equation}\begin{aligned}
B_1^{[{\rm Matching}]} &= \frac{\alpha_2}{4\pi} \left[ - \frac{8}{3} L - \frac{8}{3} \ln 2 \right]\,, \\
B_2^{[{\rm Matching}]} &= - \frac{1}{2} B_1^{[{\rm Matching}]}\,, \\
B_3^{[{\rm Matching}]} &= 0\,,
\label{eq:matchcontribution}
\end{aligned}\end{equation}
where now here and in all earlier one-loop results we can take $\alpha_2 = \alpha_2^{(n_D)}$. As alluded to above, this result is modified for a Dirac DM fermion, but in a way exactly compensated by a change in the counter-term contribution.

\subsection*{Combination}
Combining the 25 graphs above with the counter-terms and the matching contributions, we arrive at the following result:
\begin{align}
B_1^{(1)} &= \frac{\alpha_2}{4\pi} \left[ - \frac{4}{\epsilon^2} - \frac{48L+12i\pi+31-2n_D}{6\epsilon} - 8 L^2 - 4L - 4 i \pi L - 8 + \frac{11\pi^2}{6} \right]\,, \nonumber \\
B_2^{(1)} &= \frac{\alpha_2}{4\pi} \left[ \frac{2}{\epsilon^2} + \frac{48L-12i\pi+55-2n_D}{12\epsilon} + 4L^2 + 6L - 2 i \pi L - \frac{5\pi^2}{12} \right]\,, \nonumber \\
B_3^{(1)} &= \frac{\alpha_2}{4\pi} \left[ \frac{n_D - 72 \ln 2 - 71 + 3\pi^2}{12} \right]\,,
\label{eq:FullTheoryOneLoop}
\end{align}
where recall $L=\ln \mu/2m_{\chi}$, $n_D$ is the number of SM left-handed doublets and now all $\epsilon = \epsilon_{\rm IR}$.

As explained in detail at the outset of the calculation, the one-loop contribution to the matching coefficient is just the finite part of this result. Combining this with the tree-level term in Eq.~\eqref{eq:TreeLevelCoeff} and mapping back to $C_r$ using Eq.~\eqref{eq:CoeffMapping} then gives us the Wilson coefficients in Eq.~\eqref{eq:WilsonCoeff}, which we set out to justify.

If instead we had a Dirac DM triplet rather than a Majorana, then the only impact on the above would be for $B_3^{(1)}$, and we would instead have
\begin{equation}
B_3^{(1)} = \frac{\alpha_2}{4\pi} \left[ \frac{n_D - 72 \ln 2 - 43}{12} \right]\,.
\end{equation}

\section{Consistency Check on the High-Scale Matching}
\label{app:consistency}

As a non-trivial check on our high-scale calculation, we can calculate the $\ln \mu$, or $L$ in our case, pieces of Eq.~\eqref{eq:FullTheoryOneLoop} independently using the NLL results. To begin with, if we define $C \equiv (C_1~C_2~C_3)^{T}$, then from the definition of the anomalous dimension we have:
\begin{equation}
\mu \frac{d}{d\mu} C(\mu) = \hat{\gamma}(\mu) C(\mu)\,.
\label{eq:DerivRunning}
\end{equation}
Next we expand the coefficients as a series in $\alpha_2$: $C(\mu) = C^{(0)}(\mu) + C^{(1)}(\mu) + ...$, where $C^{(0)}(\mu)$ is the tree-level contribution and $C^{(1)}(\mu)$ the one-loop result. Now we want a cross check on the one-loop contribution, so we evaluate Eq.~\eqref{eq:DerivRunning} at $\mathcal{O}(\alpha_2)$, giving
\begin{equation}
\mu \frac{d\alpha_2}{d\mu} \frac{\partial C^{(0)}}{\partial \alpha_2} + \mu \frac{\partial C^{(1)}(\mu)}{\partial \mu} = \hat{\gamma}_{\rm 1-loop}(\mu) C^{(0)}(\mu)\,,
\label{eq:DerivRunningorder2}
\end{equation}
and rearranging we arrive at:
\begin{equation}
\mu \frac{\partial C^{(1)}(\mu)}{\partial \mu} = \hat{\gamma}_{\rm 1-loop}(\mu) C^{(0)}(\mu) - \mu \frac{d\alpha_2}{d\mu} \frac{\partial C^{(0)}}{\partial \alpha_2}\,.
\label{eq:Consistency}
\end{equation}
This equation shows that we can derive the $\mu$ and hence $L$ dependence of the one-loop Wilson coefficient from the one-loop anomalous dimension and tree-level Wilson coefficient, both of which are known from the NLL result. To be more explicit, we can write the bare Wilson coefficient as
\begin{equation}\begin{aligned}
C_{\rm bare} = &\mu^{2\epsilon} \left( \frac{a}{\epsilon^2} + \frac{b}{\epsilon} + \mu {\rm -independent} \right) \\
= &\frac{a}{\epsilon^2} + \frac{b+2a L}{\epsilon} + 2 a L^2 + 2 b L \\
&+ \mu {\rm -independent}\,,
\end{aligned}\end{equation}
where in the second equality we swapped form $\ln \mu$ to $L$ and absorbed the additional $\ln 2$ factors into the $\mu$-independent term. From here we can write the renormalized Wilson coefficient as
\begin{equation}
C_{\rm ren.} = 2 a L^2 + 2 b L + \mu {\rm -independent}\,,
\end{equation}
which we can then substitute into the left-hand side of Eq.~\eqref{eq:Consistency} to derive $a$ and $b$ for each Wilson coefficient. Doing this and then mapping back to $B_r$ using Eq.~\eqref{eq:CoeffMapping}, we find
\begin{eqnarray}
B_1^{(1)} &=& \frac{\alpha_2}{4\pi} \left[ - \frac{8L}{\epsilon} - 8L^2 -4L - 4i \pi L + \mu {\rm -ind.} \right]\,, \nn
B_2^{(1)} &=& \frac{\alpha_2}{4\pi} \left[ \frac{4L}{\epsilon} + 4L^2 + 6L - 2i \pi L + \mu {\rm -ind.} \right]\,, \nn
B_3^{(1)} &=& \frac{\alpha_2}{4\pi} \left[ 0 + \mu {\rm -ind.} \right]\,,
\end{eqnarray}
in exact agreement with Eq.~\eqref{eq:FullTheoryOneLoop}. In particular, as $B_3^{(0)}=0$, we needed $B_3^{(1)}$ to be independent of $L$, as we found.

\section{Low-Scale Matching Calculation}
\label{app:lowscalematching}

The focus of this appendix is to derive the low-scale matching conditions stated in Eqs.~\eqref{eq:lowbreakdown}, \eqref{eq:LowSoft}, \eqref{eq:LowColinear}, \eqref{eq:LowColinearConsts1}, and \eqref{eq:LowColinearConsts2}. At this scale, the matching is from an effective theory where the $W$, $Z$, top and Higgs are dynamical degrees of freedom -- NRDM-SCET$_{\rm EW}$ -- onto a theory where these electroweak modes have been integrated out -- NRDM-SCET$_{\gamma}$.

In order to perform the calculation we will make use of the formalism of electroweak SCET developed in \cite{Chiu:2007yn,Chiu:2007dg,Chiu:2008vv,Chiu:2009mg,Chiu:2009ft}. As we are working in SCET, there are both collinear and soft gauge boson diagrams that will appear in the one-loop matching. In \cite{Chiu:2009mg} it was proven that at one-loop the total low-scale matching contribution from these soft and collinear SCET modes can always be decomposed into a contribution that is diagonal, in that it leads to no operator mixing, and another that is non-diagonal, as it does induce mixing. In their works, they then refer to the diagonal parts as \textit{collinear} and non-diagonal ones as \textit{soft}, a notation we follow.\footnote{This explains the notation used in the main text, and caution is required to distinguish the two. As an example, our initial heavy DM states can have a soft wave-function type one-loop correction. But as this cannot mix operators, we call it a collinear contribution and label it as such in Eq.~\eqref{eq:lowbreakdown}.}

At one loop the matching amounts to evaluating the diagrams that appear in NRDM-SCET$_{\rm EW}$ but not NRDM-SCET$_{\gamma}$. These diagrams can be broken into three classes:
\begin{enumerate}
\item Wave-function diagrams correcting our initial non-relativistic states;
\item Diagrams where a soft gauge boson is exchanged between two different external states; and
\item Final state wave-function diagrams, which are now corrections to collinear states.
\end{enumerate}
Each class will be discussed separately below. Before doing so, however, we first define our operators and outline how the low-scale matching proceeds at tree level.

Unlike for the high-scale matching, here we only consider the two operators that match onto $\mathcal{M}_A$ in Eq.~\eqref{eq:FullTheoryOps}, as opposed to the third operator coming from $\mathcal{M}_B$. The reason for this is the additional operator does not contribute to the low-scale matching calculation for present day DM annihilation at any order in leading power NRDM-SCET. To understand this note that the operators coming from $\mathcal{M}_A$ and $\mathcal{M}_B$ have different spin structures. In order to mix these structures we need to transfer angular momentum between the states. The only low-scale graphs we can write down to do this are soft gauge boson exchanges. The spin structure of the coupling of a soft exchange to an $n$-collinear gauge boson is $\slashed{n}$ and the corresponding coupling to our non-relativistic DM field is $\slashed{v}$. Neither coupling allows for a transfer of angular momentum, demonstrating that these operators cannot mix. Unlike for the high-scale matching, we will not make use of the operator corresponding to $\mathcal{M}_B$ for our low-scale consistency check, so we drop it from consideration at the outset.

\newpage
\subsection*{Operator Definition and Tree-level Matching}

Prior to electroweak symmetry breaking, the two relevant operators in NRDM-SCET$_{\rm EW}$ can be written schematically as:
\begin{equation}\begin{aligned}
\mathcal{O}_1 &= \frac{1}{2} \delta_{ab} \delta_{cd} \chi^a \chi^b W_3^c W_4^d\,, \\
\mathcal{O}_2 &= \frac{1}{4} \left( \delta_{ac} \delta_{bd} + \delta_{ad} \delta_{bc} \right) \chi^a \chi^b W_3^c W_4^d\,.
\label{eq:UnbrokenOps}
\end{aligned}\end{equation}
Our notation here is schematic in the sense that we have suppressed the Lorentz structure and soft Wilson lines. The form of these is written out explicitly in Eq.~\eqref{eq:Ops} and is left out for convenience as it appears in every operator written down in this appendix. Further, in this equation the factor of $1/2$ is introduced for convenience; as $\chi$ is a Majorana field this factor ensures the Feynman rule associated with these operators has no additional numerical factor. Note also that the gauge bosons are labelled as they are associated with a collinear direction. At tree-level the low-scale matching is effected simply by mapping the fields in these operators onto their broken form. Explicitly we have:
\begin{equation}\begin{aligned}
\chi^1 &= \frac{1}{\sqrt{2}} \left( \chi^+ + \chi^- \right)\,,\;\;\;\;\;\;
\chi^2 = \frac{i}{\sqrt{2}} \left( \chi^+ - \chi^- \right)\,,\;\;\;\;\;\;
\chi^3 = \chi^0\,, \\
W^1 &= \frac{1}{\sqrt{2}} \left( W^+ + W^- \right)\,,\;\;
W^2 = \frac{i}{\sqrt{2}} \left( W^+ - W^- \right)\,,\;\;
W^3 = s_W A + c_W Z\,.
\label{eq:BrokenFields}
\end{aligned}\end{equation}
Substituting these into Eq.~\eqref{eq:UnbrokenOps} yields 22 operators in the broken theory. Of these, 14 involve a $W^{\pm}$ in the final state, so we will not consider them further. We define the remaining 8 as:
\begin{equation}\begin{aligned}
\hat{\mathcal{O}}_1 &= \frac{1}{2} \chi^0 \chi^0 A_3 A_4\,,\hspace{0.2in}\hat{\mathcal{O}}_2 = \frac{1}{2} \chi^0 \chi^0 Z_3 A_4\,,\\
\hat{\mathcal{O}}_3 &= \frac{1}{2} \chi^0 \chi^0 A_3 Z_4\,,\hspace{0.21in}\hat{\mathcal{O}}_4 = \frac{1}{2} \chi^0 \chi^0 Z_3 Z_4\,,\\
\hat{\mathcal{O}}_5 &= \chi^+ \chi^- A_3 A_4\,,\hspace{0.245in}\hat{\mathcal{O}}_6 = \chi^+ \chi^- Z_3 A_4\,,\\
\hat{\mathcal{O}}_7 &= \chi^+ \chi^- A_3 Z_4\,,\hspace{0.255in}\hat{\mathcal{O}}_8 = \chi^+ \chi^- Z_3 Z_4\,,
\label{eq:BrokenOps}
\end{aligned}\end{equation}
where again we have used the schematic notation of Eq.~\eqref{eq:UnbrokenOps}, as we will for all operators in this appendix. At tree level, the operators in Eq.~\eqref{eq:UnbrokenOps} and \eqref{eq:BrokenOps} are related simply by the change of variables in Eq.~\eqref{eq:BrokenFields}. This mapping is performed by a $22 \times 2$ matrix, but again we only state the part of this matrix we are interested in:
\begin{equation}
\hat{D}_{s,1-8}^{(0)} = \begin{bmatrix} 
s_W^2 & s_W^2 \\
s_W c_W & s_W c_W \\
s_W c_W & s_W c_W \\
c_W^2 & c_W^2 \\
s_W^2 & 0 \\
s_W c_W & 0 \\
s_W c_W & 0 \\
c_W^2 & 0 
\end{bmatrix}\,.
\label{eq:TreeLevelMapping}
\end{equation}
In terms of the calculation presented in the main text, what we actually want is the mapping onto the Sudakov factors $\Sigma$, defined in Eq.~\eqref{eq:Factorized}, not the broken operators in Eq.~\eqref{eq:BrokenOps}. As given there, the $s_W$ and $c_W$ factors are absorbed into $P_X$, and so will not contribute to the $\Sigma$ factors. Then $\hat{\mathcal{O}}_{1-4}$ represent the contributions to neutral annihilation $\chi^0 \chi^0 \to X$, represented by $\Sigma_1 - \Sigma_2$, and $\hat{\mathcal{O}}_{5-8}$ the contributions to charged annihilation $\chi^+ \chi^- \to X$, represented by $\Sigma_1$. Accordingly we have:
\begin{equation}
\hat{D}^{(0)} = \ln \begin{bmatrix}
1 & 0 \\
1 & 1
\end{bmatrix}\,.
\label{eq:TreeLevelLowScale}
\end{equation}
This provides the tree-level result we could use in Eq.~\eqref{eq:Running}, where the $\ln$ is used to remove the exponential in that equation. Next we turn to calculating this one-loop low-scale matching in full, considering the three classes of diagrams that can contribute in turn.

\subsection*{Initial State Wave-function Graphs}

There are two graphs that fall under the category of initial state wave-function corrections, and these are shown below.
\begin{center}
\includegraphics[height=0.18\columnwidth]{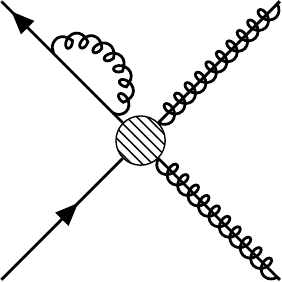} \hspace{0.1in}
\includegraphics[height=0.18\columnwidth]{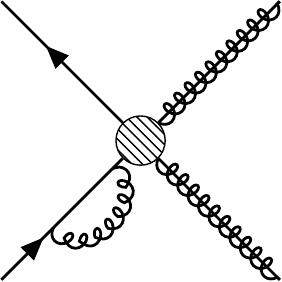}
\end{center}
Note here we follow the standard SCET conventions of drawing collinear fields as gluons with a solid line through them, whereas soft fields are represented simply by gluon lines. In these graphs, the soft gauge field can be either a $W$ or $Z$ boson. In either case the integral to be calculated is:
\begin{equation}
- g^2 \int \dbar^d k \frac{\mu^{2\epsilon}}{[k^2-m^2]v \cdot (k + p)}\,,
\end{equation}
where $g$ is the coupling -- $g_2$ for a $W$ boson, $c_W g_2$ for a $Z$ boson, $p$ is the external momentum, $k$ is the loop momentum, $m$ the gauge boson mass, and $v$ is the velocity associated with the non-relativistic $\chi$ field. Given our initial state is heavy, this is unsurprisingly exactly the heavy quark effective theory wave-function renormalization graph. The analytic solution can be found in e.g. \cite{Stewart:1998ke,Manohar:2000dt}, and using this we find:
\begin{equation}
= - i v \cdot p \frac{\alpha}{2\pi} \left[ \frac{1}{\epsilon} + \ln \frac{\mu^2}{m^2} \right]\,,
\end{equation}
where $\alpha = g^2/4\pi$. Now in addition to the one-loop graphs we drew above, at this order there will also be a counter-term of the form $i v \cdot p (Z_{\chi} - 1)$. Again working in the on-shell scheme so that we do not need to consider the residues, we conclude:
\begin{equation}
Z_{\chi} = 1 + \frac{\alpha_2(\mu)}{2\pi} \left[ \frac{1}{\epsilon} - \ln \frac{m_W^2}{\mu^2} - c_W^2 \ln \frac{m_Z^2}{\mu^2} \right]\,.
\end{equation}
Now each of our initial states will contribute $Z_{\chi}^{1/2}$, implying that the contribution to $\hat{D}(\mu)$ given in Eq.~\eqref{eq:lowbreakdown} is $\ln D_c^{\chi}(\mu) \mathbb{I}$, where
\begin{equation}
D_c^{\chi}(\mu) = 1 - \frac{\alpha_2(\mu)}{2\pi} \left[\ln \frac{m_W^2}{\mu^2} + c_W^2 \ln \frac{m_Z^2}{\mu^2} \right]\,,
\label{eq:iswfgderivation}
\end{equation}
and the subscript $c$ indicates this is a collinear contribution in the sense that it leads to no operator mixing. This is exactly as in Eq.~\eqref{eq:LowColinear} and justifies this part of the low-scale matching.

\subsection*{Soft Gauge Boson Exchange Graphs}

In this section we calculate the contribution from the exchange of a soft $W$ or $Z$ gauge boson between different external final states. As these gauge bosons carry SU(2)$_{\rm L}$ gauge indices, unsurprisingly these graphs will lead to operator mixing. Consequently, in terms of the notation introduced above these graphs will lead to non-diagonal or soft contributions. Nonetheless they will also induce diagonal or collinear terms, and we will carefully separate the two below.

Once separated, we will group the collinear contribution with those we get from the final state wave-function graphs we consider in the next subsection. The reason for this is that these collinear contributions for photon and $Z$ final states, as we have, were already evaluated in \cite{Chiu:2009ft}, and we will not fully recompute them here. In that work, however, the collinear contribution was only stated in full. The breakdown into the soft boson exchange and final state wave-function graphs was not provided. This raises a potential issue because in that work all external states were taken to be collinear, not non-relativistic. As such, in this section we will explicitly calculate the soft gauge boson exchange graphs for both kinematics and demonstrate that the diagonal contribution is identical in the two cases.

Before calculating the graphs, we first introduce some useful notation. At one loop the gauge bosons will have two couplings to the four external states. Each of these couplings will have an associated gauge index structure, and in order to deal with this it is convenient to introduce gauge index or color operators $\mathbf{T}$. This notation was first introduced in \cite{Catani:1996jh,Catani:1996vz}, and it allows the gauge index structure to be organized generally rather than case by case. Examples can be found in the original papers and also in e.g. \cite{Chiu:2009mg,Chiu:2009ft,Moult:2015aoa}. An example relevant for our purposes is the action of $\mathbf{T}$ on an SU(2)$_{\rm L}$ adjoint, which is the representation of both our initial and final states:
\begin{equation}\begin{aligned}
\mathbf{T} \chi^a &= (T_A^c)_{a a^{\prime}} \chi^{a^{\prime}} = - i\epsilon_{ca a^{\prime}} \chi^{a^{\prime}}\,, \\
\mathbf{T} W^a &= (T_A^c)_{a a^{\prime}} W^{a^{\prime}} = - i\epsilon_{ca a^{\prime}} W^{a^{\prime}}\,.
\label{eq:gaugeindexopaction}
\end{aligned}\end{equation}
In terms of this notation then, we can write the gauge index structure of all relevant one-loop low-scale matching graphs as $\mathbf{T}_i \cdot \mathbf{T}_j$, where $i,j$ label any of the four external legs. Because of this we label the result from these soft exchange diagrams as $S_{ij}$ for the case of our kinematics -- non-relativistic initial states and collinear final states -- and we use $S^{\prime}_{ij}$ to denote the kinematics of \cite{Chiu:2009ft} -- all external states collinear. Following \cite{Chiu:2009mg,Chiu:2009ft}, we take all external momenta to be incoming and further rapidity divergences will be regulated with the $\Delta$-regulator \cite{Chiu:2009yx}. Now let us turn to the graphs one by one.

\subsection*{\large $S^{(\prime)}_{12}$}

\begin{center}
\includegraphics[height=0.18\columnwidth]{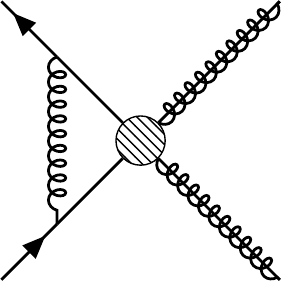}
\end{center}
In this graph the soft gauge boson exchanged between the initial state can be a $W$ or $Z$ boson. In either case, the value of this graph is:
\begin{eqnarray}
S_{12} &=& \frac{\alpha}{2\pi} \mathbf{T}_1 \cdot \mathbf{T}_2 \left[ \frac{1}{\epsilon} - \ln \frac{m^2}{\mu^2} \right]\,, \label{eq:S12} \\
S^{\prime}_{12} &=& \frac{\alpha}{2\pi} \mathbf{T}_1 \cdot \mathbf{T}_2 \left[ \frac{1}{\epsilon^2} - \frac{1}{\epsilon} \left( \ln \frac{\delta_1 \delta_2}{\mu^2} + i \pi \right) - \frac{1}{2} \ln^2 \frac{m^2}{\mu^2}  \right. \nn
&&\hspace{0.83in} \left.+ i \pi \ln \frac{m^2}{\mu^2} + \ln \frac{m^2}{\mu^2} \ln \frac{\delta_1 \delta_2}{\mu^2} - \frac{\pi^2}{12} \right]\,, \nonumber
\end{eqnarray}
where as above $\alpha = g^2/4\pi$ and the identity $g$ and $m$ depend on whether this is for a $W$ or $Z$. In $S^{\prime}_{12}$, $\delta_{1/2}$ are the $\Delta$-regulators and unsurprisingly these only appear for the collinear kinematics.

\subsection*{\large $S^{(\prime)}_{13}$, $S^{(\prime)}_{14}$, $S^{(\prime)}_{23}$, and $S^{(\prime)}_{24}$}

\begin{center}
\includegraphics[height=0.18\columnwidth]{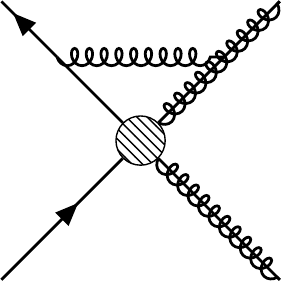} \hspace{0.1in}
\includegraphics[height=0.18\columnwidth]{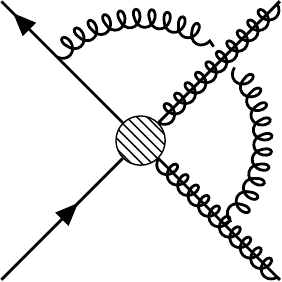} \\
\vspace{0.1in}
\includegraphics[height=0.18\columnwidth]{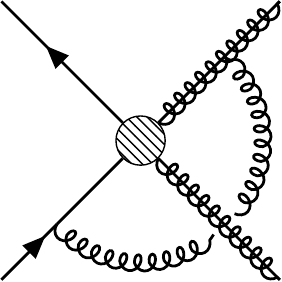} \hspace{0.1in}
\includegraphics[height=0.18\columnwidth]{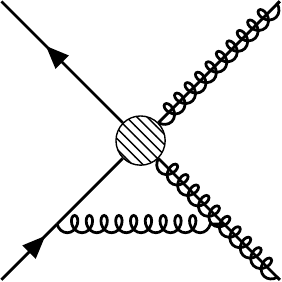}
\end{center}
Again the exchanged soft boson can be a $W$ or $Z$. These four graphs are grouped together as they have a common form, for example:
\begin{eqnarray}
S_{13} &=& \frac{\alpha}{2\pi} \mathbf{T}_1 \cdot \mathbf{T}_3 \left[ \frac{1}{2\epsilon^2} - \frac{1}{2\epsilon} \ln \frac{\delta_3^2}{\mu^2} - \frac{1}{4} \ln^2 \frac{m^2}{\mu^2} \right. \label{eq:S13} \\
&&\hspace{1.00in}\left.+ \frac{1}{2} \ln \frac{\delta_3^2}{\mu^2} \ln \frac{m^2}{\mu^2} - \frac{\pi^2}{24} \right]\,, \nn
S_{13}^{\prime} &=& \frac{\alpha}{2\pi} \mathbf{T}_1 \cdot \mathbf{T}_3 \left[ \frac{1}{\epsilon^2} - \frac{1}{\epsilon} \ln \left( - \frac{\delta_1 \delta_3}{\mu^2 w_{13}} \right) - \frac{1}{2} \ln^2 \frac{m^2}{\mu^2} \right. \nn
&&\hspace{0.97in}\left.+ \ln \frac{m^2}{\mu^2} \ln \left( - \frac{\delta_1 \delta_3}{\mu^2 w_{13}} \right) - \frac{\pi^2}{12} \right]\,. \nonumber
\end{eqnarray}
Then $S^{(\prime)}_{14}$ is given by the same expressions but with $3 \to 4$, whilst $S^{(\prime)}_{23}$ and $S^{(\prime)}_{24}$ are given by similar replacements. For the all collinear case we have defined the following functions of the kinematics:
\begin{equation}\begin{aligned}
w_{13} &= w_{24} \equiv \frac{1}{2} n_1 \cdot n_3 = \frac{1}{2} n_2 \cdot n_4 = \frac{t}{s}\,, \\
w_{14} &= w_{23} \equiv \frac{1}{2} n_1 \cdot n_4 = \frac{1}{2} n_2 \cdot n_3 = \frac{u}{s}\,,
\label{eq:wkin}
\end{aligned}\end{equation}
where $s$, $t$, and $u$ are the Mandelstam variables relevant for all incoming momenta. The signs inside the logs in Eq.~\eqref{eq:S13} can be understood by noting that as $t < 0$, $u < 0$, whilst $s > 0$, we have $w_{ij} < 0$.

\subsection*{\large $S^{(\prime)}_{34}$}

\begin{center}
\includegraphics[height=0.18\columnwidth]{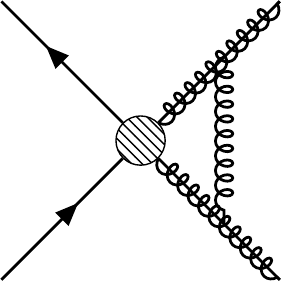}
\end{center}
Finally we have the graph above, which yields:
\begin{eqnarray}
S_{34} &=& \frac{\alpha}{2\pi} \mathbf{T}_3 \cdot \mathbf{T}_4 \left[ \frac{1}{\epsilon^2} - \frac{1}{\epsilon} \left( \ln \frac{\delta_3 \delta_4}{\mu^2} + i \pi \right) - \frac{1}{2} \ln^2 \frac{m^2}{\mu^2} \right. \nn
&&\hspace{0.83in} \left.+ i \pi \ln \frac{m^2}{\mu^2} + \ln \frac{m^2}{\mu^2} \ln \frac{\delta_3 \delta_4}{\mu^2} - \frac{\pi^2}{12} \right]\,, \nn
S_{34}^{\prime} &=& S_{34}\,.
\label{eq:S34}
\end{eqnarray}

This completes the list of graphs to evaluate. As written it appears that all graphs are non-diagonal from their gauge index structure. However as we will now show, the combinations of all graphs can be reduced to a diagonal and non-diagonal piece. Firstly for the case of all collinear external states we have:
\begin{equation}
S^{\prime}_{12} + S^{\prime}_{13} + S^{\prime}_{14} + S^{\prime}_{23} + S^{\prime}_{24} + S^{\prime}_{34} \equiv \sum_{\langle i j \rangle} S^{\prime}_{ij}\,,
\label{eq:angsum}
\end{equation}
which serves to define $\langle i j \rangle$. The part of this sum that involving the rapidity regulators can be written as
\begin{equation}
\frac{\alpha}{2\pi} \ln \frac{m^2}{\mu^2} \sum_{\langle i j \rangle} \mathbf{T}_i \cdot \mathbf{T}_j \left( \ln \frac{\delta_i}{\mu} + \ln \frac{\delta_j}{\mu} \right)\,.
\label{eq:acdeltareg}
\end{equation}
This can be simplified using the following identity:\footnote{This and the gauge index identity stated below in Eq.~\eqref{eq:gaugeindexid2} follow simply from the fact $\sum_i \mathbf{T}_i = 0$ when it acts on gauge index singlet operators, see for example \cite{Chiu:2009mg}.}
\begin{equation}
\sum_{\langle i j \rangle } \left( f_i + f_j \right) \mathbf{T}_i \cdot \mathbf{T}_j = - \sum_i f_i \mathbf{T}_i \cdot \mathbf{T}_i\,.
\label{eq:gaugeindexid1}
\end{equation}
If we identify $f_i = \ln \delta_i / \mu$, then Eq.~\eqref{eq:acdeltareg} becomes:
\begin{equation}
= - \frac{\alpha}{2\pi} \ln \frac{m^2}{\mu^2} \sum_{\langle i j \rangle} \mathbf{T}_i \cdot \mathbf{T}_i \ln \frac{\delta_i}{\mu} \,,
\label{eq:acdeltaregsimp}
\end{equation}
which is now diagonal in the gauge indices. For the remaining terms that are independent of $\delta$, we organize them as follows:
\begin{equation}\begin{aligned}
\sum_{\langle i j \rangle} S^{\prime}_{ij}
&= \frac{1}{2} \left[ S^{\prime}_{12} + S^{\prime}_{13} + S^{\prime}_{14} \right] \\
&+ \frac{1}{2} \left[ S^{\prime}_{21} + S^{\prime}_{23} + S^{\prime}_{24} \right] \\
&+ \frac{1}{2} \left[ S^{\prime}_{31} + S^{\prime}_{32} + S^{\prime}_{34} \right] \\
&+ \frac{1}{2} \left[ S^{\prime}_{41} + S^{\prime}_{42} + S^{\prime}_{43} \right] \,,
\label{eq:gaugeindexgboost}
\end{aligned}\end{equation}
where we used the fact $S^{\prime}_{ij} = S^{\prime}_{ji}$. Each of these groups can now be simplified. For example, the first group can be written as:
\begin{eqnarray}
S^{\prime}_{12} + S^{\prime}_{13} + S^{\prime}_{14} &= &\frac{\alpha}{2\pi} \left( \mathbf{T}_1 \cdot \mathbf{T}_2 + \mathbf{T}_1 \cdot \mathbf{T}_3 + \mathbf{T}_1 \cdot \mathbf{T}_4 \right) \nn
&&\times \left[ - \frac{1}{2} \ln^2 \frac{m^2}{\mu^2} - \frac{\pi^2}{12} \right] \nn
&&+ \frac{\alpha}{2\pi} \mathbf{T}_1 \cdot \mathbf{T}_2 \left[ i \pi \ln \frac{m^2}{\mu^2} \right] \label{eq:boostorg} \\
&&- \frac{\alpha}{2\pi} \mathbf{T}_1 \cdot \mathbf{T}_3 \left[ \ln \left( - \frac{t}{s} \right) \ln \frac{m^2}{\mu^2} \right] \nn
&&- \frac{\alpha}{2\pi} \mathbf{T}_1 \cdot \mathbf{T}_4 \left[ \ln \left( - \frac{u}{s} \right) \ln \frac{m^2}{\mu^2} \right]\,, \nonumber
\end{eqnarray}
If we then use
\begin{equation}
\sum_{j,j\neq i} \mathbf{T}_i \cdot \mathbf{T}_j = - \mathbf{T}_i \cdot \mathbf{T}_i\,,
\label{eq:gaugeindexid2}
\end{equation}
Eq.~\eqref{eq:boostorg} can be rewritten as:
\begin{eqnarray}
&=&\frac{\alpha}{2\pi} \mathbf{T}_1 \cdot \mathbf{T}_1 \left[ \frac{1}{2} \ln^2 \frac{m^2}{\mu^2} + \frac{\pi^2}{12} \right] \nn
&&+ \frac{\alpha}{2\pi} \mathbf{T}_1 \cdot \mathbf{T}_2 \left[ i \pi \ln \frac{m^2}{\mu^2} \right] \label{eq:boostorg2} \\
&&- \frac{\alpha}{2\pi} \mathbf{T}_1 \cdot \mathbf{T}_3 \left[ \ln \left( - \frac{t}{s} \right) \ln \frac{m^2}{\mu^2} \right] \nn
&&- \frac{\alpha}{2\pi} \mathbf{T}_1 \cdot \mathbf{T}_4 \left[ \ln \left( - \frac{u}{s} \right) \ln \frac{m^2}{\mu^2} \right]\,. \nonumber
\end{eqnarray}
Repeating this for the remaining three terms in Eq.~\eqref{eq:gaugeindexgboost} and reinserting the $\delta$ contributions, we can rewrite the combination of all terms as:
\begin{equation}
\sum_{\langle i j \rangle} S^{\prime}_{ij} \equiv \sum_{\langle i j \rangle} \hat{S}^{\prime}_{ij} + \sum_i C_i\,,
\label{eq:boostreduce}
\end{equation}
where we have defined:
\begin{eqnarray}
\hat{S}^{\prime}_{ij} &\equiv& - \frac{\alpha}{2\pi} \ln \frac{m^2}{\mu^2} \mathbf{T}_i \cdot \mathbf{T}_j U_{ij}^{\prime}\,, \label{eq:boostsoftcollinear} \\
C_i &\equiv& \frac{\alpha}{2\pi} \mathbf{T}_i \cdot \mathbf{T}_i \left[ \frac{1}{4} \ln^2 \frac{m^2}{\mu^2} + \frac{\pi^2}{24} - \frac{1}{2} \ln \frac{m^2}{\mu^2} \ln \frac{\delta_i^2}{\mu^2} \right]\,, \nonumber
\end{eqnarray}
and from the above we can see that:
\begin{equation}\begin{aligned}
U_{12}^{\prime} &= U_{34}^{\prime} = - i \pi\,, \\
U_{13}^{\prime} &= U_{24}^{\prime} = \ln \left( - \frac{t}{s} \right)\,, \\
U_{14}^{\prime} &= U_{23}^{\prime} = \ln \left( - \frac{u}{s} \right)\,.
\label{eq:boostUij}
\end{aligned}\end{equation}
Thus as claimed, we have reduced $\sum_{\langle i j \rangle} S^{\prime}_{ij}$ in Eq.~\eqref{eq:boostreduce} into a diagonal and non-diagonal piece. Importantly we have explicitly isolated the collinear contribution $C_i$, and as we will now show we get exactly the same diagonal contribution for the kinematics of interest in this work.

Before doing so, however, note that the irreducibly non-diagonal contribution given in Eq.~\eqref{eq:boostsoftcollinear} and Eq.~\eqref{eq:boostUij} agrees with Eq.~(150) in \cite{Chiu:2009mg}, where they gave the general form of $U_{ij}^{\prime}$ for the case of all external collinear particles:
\begin{equation}
U_{ij}^{\prime} = \ln \frac{-n_i \cdot n_j - i0^+}{2}\,.
\end{equation}

Next we repeat this procedure for $\sum_{\langle i j \rangle} S_{ij}$, where we have non-relativistic fields in the initial state. As before we consider the contribution from the rapidity regulators at the outset, which for $\delta_3$ yield:
\begin{eqnarray}
&&\frac{\alpha}{2\pi} \left( \mathbf{T}_1 \cdot \mathbf{T}_2 + \mathbf{T}_1 \cdot \mathbf{T}_3 + \mathbf{T}_1 \cdot \mathbf{T}_4 \right) \left[ \frac{1}{2} \ln \frac{m^2}{\mu^2} \ln \frac{\delta_3^2}{\mu^2} \right] \nn
&=&- \frac{\alpha}{2\pi} \mathbf{T}_3 \cdot \mathbf{T}_3 \left[ \frac{1}{2} \ln \frac{m^2}{\mu^2} \ln \frac{\delta_3^2}{\mu^2} \right]\,,
\label{eq:nrdeltaorg}
\end{eqnarray}
where we again used Eq.~\eqref{eq:gaugeindexid2}. An identical relation will hold for $\delta_4$, and this time there is no $\delta_{1/2}$ as the non-relativistic fields do not lead to rapidity divergences. For the remaining terms, we now organize them as follows:
\begin{equation}\begin{aligned}
\sum_{\langle i j \rangle} S_{ij}
= S_{12} 
+ &\left[ S_{31} + S_{32} + \frac{1}{2} S_{34} \right] \\
+ &\left[ S_{41} + S_{42} + \frac{1}{2} S_{43} \right]\,.
\label{eq:gaugeindexgnr}
\end{aligned}\end{equation}
Evaluating each of the terms in square brackets and simplifying the gauge index structure as before, we arrive at the following:
\begin{equation}
\sum_{\langle i j \rangle} S_{ij} \equiv \sum_{\langle i j \rangle} \hat{S}_{ij} + C_3 + C_4\,,
\label{eq:nrreduce}
\end{equation}
where we again have:
\begin{eqnarray}
\hat{S}_{ij} &\equiv& - \frac{\alpha}{2\pi} \ln \frac{m^2}{\mu^2} \mathbf{T}_i \cdot \mathbf{T}_j U_{ij}\,, \label{eq:nrsoftcollinear} \\
C_i &\equiv& \frac{\alpha}{2\pi} \mathbf{T}_i \cdot \mathbf{T}_i \left[ \frac{1}{4} \ln^2 \frac{m^2}{\mu^2} + \frac{\pi^2}{24} - \frac{1}{2} \ln \frac{m^2}{\mu^2} \ln \frac{\delta_i^2}{\mu^2} \right]\,, \nonumber
\end{eqnarray}
and now
\begin{equation}\begin{aligned}
U_{12} &= 1\,, \\
U_{34} &= - i \pi\,, \\
U_{13} &= U_{24} = U_{14} = U_{23} = 0 \,.
\label{eq:noboostUij}
\end{aligned}\end{equation}
Critically, although the non-diagonal contribution is different to the case of all collinear kinematics, we see that the collinear function defined in Eq.~\eqref{eq:nrsoftcollinear} is identical to that in Eq.~\eqref{eq:boostsoftcollinear}. This justifies the claim made earlier that the diagonal part of this calculation is the same for both kinematics. As such we put the $C_i$ terms aside for the moment, and return to them when we consider the final state wave-function graphs.

What remains here then is to evaluate the irreducibly non-diagonal contribution: $\sum_{\langle i j \rangle} \hat{S}_{ij}$. This essentially amounts to calculating the gauge index structure, which the use of gauge index operators has allowed us to put off until now. In addition we need to recall that we have a contribution to each graph from a $W$ and $Z$ boson exchange. As above we closely follow the approach in \cite{Chiu:2009mg,Chiu:2009ft}, except accounting for the differences in our kinematics. To this end, we begin by observing that after electroweak symmetry breaking the unbroken SU(2)$_{\rm L}$ and U(1)$_Y$ generators, $\mathbf{t}$ and $Y$, become
\begin{equation}\begin{aligned}
\alpha_2 \mathbf{t} \cdot \mathbf{t} + \alpha_1 Y \cdot Y \to &\frac{1}{2} \alpha_W (t_+ t_- + t_- t_+) \\
&+ \alpha_Z t_Z \cdot t_Z + \alpha_{\rm em} Q \cdot Q\,,
\end{aligned}\end{equation}
where $\alpha_2 = \alpha_{\rm em}/s_W^2$, $\alpha_1 = \alpha_{\rm em}/c_W^2$, $\alpha_W = \alpha_2$, $\alpha_Z = \alpha_2/c_W^2$, and $t_Z = t_3 - s_W^2 Q$. This implies that we can write the full contribution as:
\begin{eqnarray}
\hat{D}_s^{(1)} &= &\frac{\alpha_W(\mu)}{2\pi} \ln \frac{m_W^2}{\mu^2} \left[ - \sum_{\langle i j \rangle} \frac{1}{2} (t_+ t_- + t_- t_+) U_{ij} \right] \nn
&&+ \frac{\alpha_Z(\mu)}{2\pi} \ln \frac{m_Z^2}{\mu^2} \left[ - \sum_{\langle i j \rangle} t_{Zi} t_{Zj} U_{ij} \right]\,.
\end{eqnarray}
Now the contribution on the first line is more complicated, because $(t_+ t_- + t_- t_+)U_{ij}$ is a non-diagonal $22 \times 22$ matrix, whereas as we will see $t_{Zi} t_{Zj} U_{ij}$ is diagonal. Nevertheless we can simplify the non-diagonal part by using the following relation:
\begin{equation}\begin{aligned}
\frac{1}{2} (t_+ t_- + t_- t_+) = \mathbf{t} \cdot \mathbf{t} - t_3 \cdot t_3\,.
\end{aligned}\end{equation}
Here $t_3 \cdot t_3$ is again diagonal, and whilst $\mathbf{t} \cdot \mathbf{t}$ is non-diagonal, it is written in terms of the unbroken operators so that we can calculate it in the unbroken theory where we only have 2 operators not 22. Thus it is now a $2 \times 2$ matrix. In terms of this we can now write the non-diagonal contribution to the low-scale matching as:
\begin{equation}\begin{aligned}
\hat{D}_s &= \hat{D}_s^{(0)} + \hat{D}_{s,W}^{(1)} + \hat{D}_{s,Z}^{(1)}\,, \\
\hat{D}_{s,W}^{(1)} &= \frac{\alpha_W(\mu)}{2\pi} \ln \frac{m_W^2}{\mu^2} \left[ \hat{D}_s^{(0)} \mathfrak{S} + \mathfrak{D}_W \hat{D}_s^{(0)} \right]\,, \\
\hat{D}_{s,Z}^{(1)} &= \frac{\alpha_Z(\mu)}{2\pi} \ln \frac{m_Z^2}{\mu^2} \left[ \mathfrak{D}_Z \hat{D}_s^{(0)} \right]\,,
\label{eq:ndlsfulldef}
\end{aligned}\end{equation}
where $\hat{D}_s^{(0)}$ is given in Eq.~\eqref{eq:TreeLevelMapping} and as we will now demonstrate $\hat{D}_s$ is effectively the matrix given in Eq.~\eqref{eq:LowSoft} that we set out to justify. In order to do this we have to evaluate the remaining terms:
\begin{equation}\begin{aligned}
\mathfrak{S} &\equiv - \sum_{\langle i j \rangle} \mathbf{t}_i \cdot \mathbf{t}_j U_{ij}\,, \\
\mathfrak{D}_W &\equiv \sum_{\langle i j \rangle} \mathbf{t}_{3i} \cdot \mathbf{t}_{3j} U_{ij}\,, \\
\mathfrak{D}_Z &\equiv - \sum_{\langle i j \rangle} \mathbf{t}_{Zi} \cdot \mathbf{t}_{Zj} U_{ij}\,.
\label{eq:ndlsfulldef2}
\end{aligned}\end{equation}
The form of each of these matrices can be evaluated by acting with them on the operators -- the unbroken operators in Eq.~\eqref{eq:UnbrokenOps} for $\mathfrak{S}$ and the broken operators in Eq.~\eqref{eq:BrokenOps} for $\mathfrak{D}_{W/Z}$ -- where the action of the gauge index operators is given by Eq.~\eqref{eq:gaugeindexopaction}. Doing this, we find:
\begin{equation}
\mathfrak{S} = \begin{bmatrix}
2-2i\pi & 1-i\pi \\
0 & i\pi - 1
\end{bmatrix}\,,
\label{eq:frakS}
\end{equation}
whilst
\begin{equation}\begin{aligned}
\mathfrak{D}_{W,1-8} &= {\rm diag} \left( 0, 0, 0, 0, -1, -1, -1, -1 \right)\,, \\
\mathfrak{D}_Z &= - c_W^4 \mathfrak{D}_W\,.
\label{eq:frakD}
\end{aligned}\end{equation}
Substituting these results into Eq.~\eqref{eq:ndlsfulldef}, we find:
\begin{equation}
\hat{D}_{s,1-8} = \begin{bmatrix}
s_W^2 \left[ 1 + G(\mu) \right] & s_W^2 \\
s_W c_W \left[ 1 + G(\mu) \right] & s_W c_W \\
s_W c_W \left[ 1 + G(\mu) \right] & s_W c_W \\
c_W^2 \left[ 1 + G(\mu) \right] & c_W^2 \\
s_W^2 \left[ 1 + H(\mu) \right] & s_W^2 I(\mu) \\
s_W c_W \left[ 1 + H(\mu) \right] & s_W c_W I(\mu) \\
s_W c_W \left[ 1 + H(\mu) \right] & s_W c_W I(\mu) \\
c_W^2 \left[ 1 + H(\mu) \right] & c_W^2 I(\mu)
\end{bmatrix}\,,
\label{eq:softmatrixfull}
\end{equation}
where we have defined:
\begin{equation}\begin{aligned}
G(\mu) \equiv& \frac{\alpha_W(\mu)}{2\pi} \ln \frac{m_W^2}{\mu^2} \left( 2-2i\pi \right)\,, \\
H(\mu) \equiv& \frac{\alpha_W(\mu)}{2\pi} \ln \frac{m_W^2}{\mu^2} \left( 1-2i\pi \right) \\
&+ c_W^4 \frac{\alpha_Z(\mu)}{2\pi} \ln \frac{m_Z^2}{\mu^2}\,, \\
I(\mu) \equiv& \frac{\alpha_W}{2\pi} \ln \frac{m_W^2}{\mu^2} (1-i\pi)\,.
\label{eq:softmatrixfullentries}
\end{aligned}\end{equation}
From the form of $\hat{D}_s$ given in Eq.~\eqref{eq:softmatrixfull}, we can again reduce this to a $2 \times 2$ matrix which maps onto $\Sigma_1$ and $\Sigma_1 - \Sigma_2$, exactly as we did for the tree-level low-scale matching. Doing this, the $2 \times 2$ matrix we obtain is exactly Eq.~\eqref{eq:LowSoft}, which we set out to justify.

\subsection*{Final State Wave-function Graphs}

Finally we have the last contribution, which is the combination of final state wave-function graphs as well as $C_3 + C_4$, as defined in Eq.~\eqref{eq:nrsoftcollinear}. As mentioned in the previous subsection, this calculation has already been performed in \cite{Chiu:2009ft}, and given that the form of $C_i$ is the same for our kinematics as it is for theirs, we take the result from their work. In that paper they calculated this collinear contribution for all possible weak bosons. For our calculation we are only interested in a final state photon or $Z$, for which they give:
\begin{equation}\begin{aligned}
D_c^Z = &\frac{\alpha_2}{2\pi} \left[ F_W + f_S \left( \frac{m_Z^2}{m_W^2}, 1 \right) \right] \\
&+ \frac{1}{2} \delta \mathfrak{R}_Z + \tan \bar{\theta}_W \mathfrak{R}_{\gamma \to Z}\,, \\
D_c^{\gamma} = &\frac{\alpha_2}{2\pi} \left[ F_W + f_S \left( 0, 1 \right) \right] \\
&+ \frac{1}{2} \delta \mathfrak{R}_{\gamma} + \cot \bar{\theta}_W \mathfrak{R}_{Z \to \gamma}\,.
\label{eq:CFKMcollinear}
\end{aligned}\end{equation}
The various terms in these equations are outlined below. Nonetheless, once the full expressions are written out the analytic result for the terms in Eq.~\eqref{eq:LowColinearConsts2} can be extracted as the terms independent of $\ln \mu^2$. 

To begin with we have:
\begin{equation}\begin{aligned}
F_W \equiv& \ln \frac{m_W^2}{\mu^2} \ln \frac{s}{\mu^2} - \frac{1}{2} \ln^2 \frac{m_W^2}{\mu^2} \\
&- \ln \frac{m_W^2}{\mu^2} - \frac{5\pi^2}{12} + 1\,,
\end{aligned}\end{equation}
where note for our calculation $s = 4 m_{\chi}^2$. Next $f_S(w, z)$ is defined as:
\begin{equation}
f_S(w,z) \equiv \int_0^1 dx \frac{(2-x)}{x} \ln \frac{1-x+zx-wx(1-x)}{1-x}\,,
\end{equation}
such that an explicit calculation gives us
\begin{equation}\begin{aligned}
f_S \left( \frac{m_Z^2}{m_W^2}, 1 \right) &= 1.08355\,, \\
f_S \left( 0, 1 \right) &= \frac{\pi^2}{3} - 1\,.
\end{aligned}\end{equation}
Finally the $\mathfrak{R}$ contributions are defined by:\footnote{Note there is a typo in Eq.~B2 of \cite{Chiu:2009ft}, where $\mathfrak{R}_{\gamma \to Z}$ and $\mathfrak{R}_{Z \to \gamma}$ involved $\Pi^{\prime}$ rather than $\Pi$. The expressions stated here are the correct ones, and we thank Aneesh Manohar for confirming this and for  providing a numerical cross check on our results for these terms.}
\begin{equation}\begin{aligned}
\delta \mathfrak{R}_Z &\equiv \Pi^{\prime}_{ZZ}(m_Z^2)\,, \\
\delta \mathfrak{R}_\gamma &\equiv \Pi^{\prime}_{\gamma \gamma}(0)\,, \\
\mathfrak{R}_{\gamma \to Z} &\equiv \frac{1}{m_Z^2} \Pi_{Z\gamma} (m_Z^2)\,, \\
\mathfrak{R}_{Z \to \gamma} &\equiv -\frac{1}{m_Z^2} \Pi_{\gamma Z} (0)\,,
\end{aligned}\end{equation}
where $\Pi^{\prime} \equiv \partial \Pi(k^2)/\partial k^2$ and the various $\Pi$ functions are defined via the inverse of the transverse gauge boson propagator
\begin{equation}
-i \left( g_{\mu \nu} - \frac{k_{\mu} k_{\nu}}{k^2} \right) \begin{bmatrix} k^2 - m_Z^2 - \Pi_{ZZ}(k^2) & - \Pi_{Z\gamma}(k^2) \\ - \Pi_{\gamma Z}(k^2) & k^2 - \Pi_{\gamma \gamma}(k^2) \end{bmatrix}\,.
\end{equation}
The form of the $\Pi$ functions is not given explicitly in \cite{Chiu:2009ft}, but can be determined from the results of e.g. \cite{Denner:1991kt,Bardin:1999ak}. When doing so, there are two factors that must be accounted for. Firstly the $\Pi$ functions must be calculated in $\overline{\rm MS}$. This is because \cite{Chiu:2009ft} accounts for the residues explicitly in \eqref{eq:CFKMcollinear}. If we used the on-shell scheme for external particles, as we did for the high-scale matching, we would double count the contribution from the residues. Secondly we need to respect that the low-scale matching is performed above and below the electroweak scale, which means the $\Pi$ functions for the photon and $Z$ must be treated differently. Above the matching scale the $W$, $Z$, top and Higgs are dynamical degrees of freedom, but below it they are not. Light degrees of freedom like the photon, bottom quark or electron are dynamical above and below. This means for the $Z$ contributions, we need to include all degrees of freedom -- heavy and light -- in the loops, as the $Z$ itself does not propagate below the matching. For the photon contributions, however, only the heavy degrees of freedom should be included. Accounting for these factors, we arrive at the following:
\begin{eqnarray}
\delta \mathfrak{R}_Z &=& \frac{\alpha_2}{4\pi} \left[ \frac{5-10s_W^2+46s_W^4}{6c_W^2} \ln \frac{m_Z^2}{\mu_Z^2} \right. \nn
&&\hspace{0.67in}\left.+ 1.5077 - 9.92036 i \vphantom{\frac{m_Z^2}{\mu_Z^2}}\right]\,,\nn
\delta \mathfrak{R}_\gamma &=& \frac{\alpha_2}{4\pi} \left[ - \frac{11}{9} s_W^2 \ln \frac{m_Z^2}{\mu_Z^2} + 0.8257 \right]\,, \nn
\mathfrak{R}_{\gamma \to Z} &=& \frac{\alpha_2}{4\pi}\left[ - \frac{7s_W^2+34s_W^4}{6c_W^2 \tan \bar{\theta}_W} \ln \frac{m_Z^2}{\mu_Z^2} \right. \nn 
&&\hspace{0.47in}\left.+ 0.3678 - 2.2748 i \vphantom{\frac{m_Z^2}{\mu_Z^2}}\right]\,, \nn
\mathfrak{R}_{Z \to \gamma} &=& \frac{\alpha_2}{4\pi} \left[ 2 s_W c_W \ln \frac{m_Z^2}{\mu_Z^2} - 0.2099 \right]\,.
\end{eqnarray}
Analytic forms for the $\Pi$ functions are provided in App.~\ref{app:PiFunctions}, we do not provide the full expressions here as they are lengthy. In order to determine the numerical values above we have used the following:
\begin{equation}\begin{aligned}
m_Z &= 91.1876~{\rm GeV}\,, \\
m_W &= 80.385~{\rm GeV}\,, \\
m_t &= 173.21~{\rm GeV}\,, \\
m_H &= 125~{\rm GeV}\,, \\
m_b &= 4.18~{\rm GeV}\,, \\
m_c &= 1.275~{\rm GeV}\,, \\
m_{\tau} &= 1.77682~{\rm GeV}\,, \\
m_s &= m_d = m_u = m_{\mu} = m_e = 0~{\rm GeV}\,, \\
c_W &= m_W/m_Z\,.
\end{aligned}\end{equation}
This completes the list of ingredients for Eq.~\eqref{eq:CFKMcollinear}. Substituting them into that equation gives exactly the relevant terms in Eqs.~\eqref{eq:LowColinear}, \eqref{eq:LowColinearConsts1}, and \eqref{eq:LowColinearConsts2}, justifying the collinear part of the low-scale matching.

We have now justified each of the pieces making up the low-scale one-loop matching. All that remains is to cross check this result, which we turn to in the next appendix.

\section{Consistency Check on the Low-Scale Matching}
\label{app:consistencylow}

In this appendix we provide a cross check on the low-scale one-loop matching calculation, much as we did for the high-scale result in App.~\ref{app:consistency}. Given that we already cross checked the high-scale result, we here make use of that to determine whether the $\ln \mu$ contributions at the low scale are correct. In order to do this, we take Eq.~\eqref{eq:Running} and turn off the running, which amounts to setting $\mu_{m_{\chi}} = \mu_Z \equiv \mu$. In detail we obtain:
\begin{equation}
\begin{bmatrix} C_{\pm}^X \vspace{0.1cm}\\ C^X_{0} \end{bmatrix} = e^{\hat{D}^X(\mu)} \begin{bmatrix} C_1(\mu) \\ C_2(\mu) \end{bmatrix}\,.
\label{eq:norunning}
\end{equation}
Now as we have the full one-loop result, the $\ln \mu$ dependence between these two terms must cancel at $\mathcal{O}(\alpha_2)$ for any $X$, which we will now demonstrate.

Before doing this in general, we first consider the simpler case where electroweak symmetry remains unbroken and we just have a $W^3 W^3$ final state. In this case, as in general, to capture all $\mu$ dependence at $\mathcal{O}(\alpha_2)$ we also need to account for the $\beta$-function. If SU(2)$_{\rm L}$ remains unbroken, however, this is just simply captured in:
\begin{equation}
\alpha_2(\mu) = \alpha_2(m_Z) + \alpha_2^2(m_Z)^2 \frac{b_0}{4\pi} \ln \frac{m_Z^2}{\mu^2}\,,
\label{eq:coup1loop}
\end{equation}
where $b_0 = (43-2 n_D)/6$, with $n_D$ the number of SM doublets. This follows directly from Eq.~\eqref{eq:coupling}. In the unbroken theory we can simply set $c_W = 1$ and $s_W = 0$, so if we do this and substitute our results from Eqs.~\eqref{eq:WilsonCoeff}, \eqref{eq:lowbreakdown}, \eqref{eq:LowSoft}, \eqref{eq:LowColinear}, \eqref{eq:LowColinearConsts1} into Eq.~\eqref{eq:norunning}, then we find:
\begin{equation}\begin{aligned}
C_{\pm}^{W^3} &= \frac{1}{m_{\chi}} \left( \frac{b_0}{4} + c_1^{W^3} - 1 \right) \ln \mu^2 + \mu {\rm -ind.}\,, \\
C_{0}^{W^3} &= \mu {\rm -ind.}\,, \\
\end{aligned}\end{equation}
Now we can calculate that $c_1^{W^3} = (2 n_D - 19)/24$, which taking $n_D=12$ exactly agrees with $c_1^Z$ in Eq.~\eqref{eq:LowColinearConsts1} when $c_W=1$ and $s_W=0$ as it must. Then recalling $b_0$ from above we see that both coefficients are then $\mu$ independent at this order, demonstrating the required consistency.

We now consider the same cross check in the full broken theory. The added complication here is that for our different final states, $\gamma \gamma$, $\gamma Z$, and $ZZ$, the coupling is actually $s_W^2 \alpha_2$, $s_W c_W \alpha_2$, and $c_W^2 \alpha_2$ respectively. As we work in $\overline{\rm MS}$, we need to account for the fact that $s_W$ and $c_W$ are functions $\mu$. Explicit calculation demonstrates that the running is only relevant for the consistency of $C_{\pm}^X$ -- the cancellation in $C_{0}^X$ is independent of the $\beta$-function at this order -- and in fact we find:
\begin{equation}\begin{aligned}
C_{\pm}^X &= \frac{1}{m_{\chi}} \left( \frac{b_0^{(X)}}{4} + \frac{1}{2} \sum_{i \in X} c_1^i - 1 \right) \ln \mu^2 + \mu {\rm -ind.}\,.
\label{eq:genlowcc}
\end{aligned}\end{equation}
To derive this we simply used Eq.~\eqref{eq:coup1loop}, with $b_0 \to b_0^{(X)}$, leaving us to derive the appropriate for of $b_0^{(X)}$ for $X = \gamma \gamma$, $\gamma Z$, $ZZ$. Firstly note that
\begin{equation}\begin{aligned}
s_W^2(\mu) &= \frac{\alpha_1(\mu)}{\alpha_1(\mu)+\alpha_2(\mu)}\,, \\
c_W^2(\mu) &= \frac{\alpha_2(\mu)}{\alpha_1(\mu)+\alpha_2(\mu)}\,,
\label{eq:swcwrun}
\end{aligned}\end{equation}
where $\alpha_1$ is the U(1)$_Y$ coupling. We can write a similar expression to Eq.~\eqref{eq:coup1loop} for $\alpha_1$, but this time we have $b_0^{(1)} = -41/6$. To avoid confusion we also now refer to the SU(2)$_{\rm L}$ $b_0$ as $b_0^{(2)}=19/6$.

Now for the case of two $Z$ bosons in the final state, the appropriate $\beta$-function is:
\begin{equation}
\beta_{ZZ} = \mu \frac{d}{d\mu} \left[ c_W^2 \alpha_2 \right]\,.
\end{equation}
Combining this with Eq.~\eqref{eq:swcwrun}, we conclude that:
\begin{equation}\begin{aligned}
b^{(ZZ)}_0 =& \left( s_W^2 + 1 \right) b_0^{(2)} - \frac{s_W^4}{c_W^2} b_0^{(1)} \\
=& \frac{19+22s_W^4}{6c_W^2}\,.
\end{aligned}\end{equation}
There is an additional factor of $c_W^2$ in this expression than if we were just calculating the $\beta$-function for $\alpha_Z$. The reason for this is that $b^{(ZZ)}_0$ is the appropriate replacement for $b_0$ in Eq.~\eqref{eq:coup1loop}, which represents the correction to $\alpha_2 = c_W^2 \alpha_Z$ not $\alpha_Z$. Substituting this into Eq.~\eqref{eq:genlowcc} along with the definition of $c_1^Z$ from Eq.~\eqref{eq:LowColinearConsts1} demonstrates consistency for the $ZZ$ case.

The case of two final state photons has to be treated differently, because of the fact our low-scale matching integrated out the electroweak degrees of freedom, which did not include the photon. This means we need to use a modified version of the SU(2)$_{\rm L}$ and U(1)$_Y$ couplings that only include the running due to the modes being removed. This amounts to accounting for the running from the Higgs, $W$ and $Z$ bosons, and the top quark, which we treat as an SU(2)$_{\rm L}$ singlet Dirac fermion to ensure it is entirely removed through the matching. Doing so, the SM $\beta$-functions now evaluate to $b_0^{(2) \prime} = 43/6$ and $b_0^{(1) \prime} = -35/18$. Repeating the same calculation as we used to determine $b^{(ZZ)}_0$, we find that:
\begin{equation}
b^{(\gamma \gamma)}_0 = \left(b_0^{(1) \prime} + b_0^{(2) \prime} \right) s_W^2 = \frac{47}{9} s_W^2\,.
\end{equation}
Again, substituting this into Eq.~\eqref{eq:genlowcc} shows that the two photon case is also consistent. The final case $\gamma Z$, but it is straightforward to see that in this case Eq.~\eqref{eq:genlowcc} breaks into two conditions that are satisfied if the $ZZ$ and $\gamma \gamma$ cases are, so this is not an independent cross check. 

As such, in the absence of running, all the $\mu$ dependence in our calculation vanishes at $\mathcal{O}(\alpha_2)$, as it must. But we emphasize that this is a non-trivial cross check, that involves all aspects of the calculation in the full broken theory.

\section{Analytic Form of $\Pi$}
\label{app:PiFunctions}

Here we state the analytic expressions for the $\overline{\rm MS}$ electroweak $\Pi$ functions for photon and $Z$ boson, appropriate for the matching from SCET$_{\rm EW}$ to SCET$_{\gamma}$. These results can be determined using standard references, such as \cite{Denner:1991kt,Bardin:1999ak}. As the photon is a dynamical degree of freedom above and below the matching, we only need to consider loop diagrams involving electroweak modes that are integrated out through the matching. This simplifies the evaluation, and we have the following two functions:
\begin{equation}\begin{aligned}
\Pi^{\prime}_{\gamma \gamma}(0) = \frac{\alpha_2s_W^2}{4\pi} &\left\{ - \frac{16}{9} \ln \frac{\mu^2}{m_t^2} + 3 \ln \frac{\mu^2}{m_W^2} + \frac{2}{3} \right\}\,, \\
\Pi_{\gamma Z}(0) = \frac{\alpha_2s_W^2}{4\pi} &\left\{ \frac{2m_W^2}{s_W c_W} \ln \frac{\mu^2}{m_W^2} \right\}\,.
\end{aligned}\end{equation}
As the $Z$ itself is being integrated out, we need to include all relevant loops when calculating $\Pi_{Z \gamma}$ and $\Pi^{\prime}_{ZZ}$. In order to simplify the expressions, we firstly introduce the following expressions:
\begin{eqnarray}
\beta &\equiv &\sqrt{\frac{4m^2}{s}-1}\,,\;\;\; \xi \equiv \sqrt{1-\frac{4m^2}{s}}\,, \\
\lambda_{\pm} &\equiv &\frac{1}{2s} \left( s - m_2^2 + m_1^2 \pm \sqrt{(s-m_2^2+m_1^2)^2-4s (m_1^2-i\epsilon)} \right)\,. \nonumber
\end{eqnarray}
In terms of these we then define:
\begin{align}
a(m_1,m_2) \equiv &1 + \frac{m_1^2}{m_2^2-m_1^2} \ln \frac{m_1^2}{m_2^2}\,, \\
b(s,m) \equiv &2 + i \beta \ln \left( \frac{\beta+i}{\beta-i} \right)\,,\;
b_2(s,m) \equiv 2 - \xi \ln \frac{1+\xi}{1-\xi} + i \pi \xi\,, \nonumber \\
c(s,m) \equiv &- \frac{2m^2}{s^2\beta} \left( \frac{2\beta}{1+\beta^2} + i \ln \frac{\beta+i}{\beta-i} \right)\,,\;
c_2(s,m) \equiv \frac{2m^2}{s^2 \xi} \left( \frac{2 \xi}{\xi^2-1} - \ln \frac{1+\xi}{1-\xi} \right)\,, \nonumber \\
d(s,m_1,m_2) \equiv &2  + \lambda_+ \ln \left( \frac{\lambda_+-1}{\lambda_+} \right) - \ln \left( \lambda_+ -1 \right) 
+ \lambda_- \ln \left( \frac{\lambda_--1}{\lambda_-} \right) - \ln \left( \lambda_- -1 \right)\,, \nonumber\\
e(s,m_1,m_2) \equiv &- \frac{1}{s} + \ln \left( \frac{\lambda_+-1}{\lambda_+} \right) \frac{\partial \lambda_+}{\partial s}
+ \ln \left( \frac{\lambda_--1}{\lambda_-} \right) \frac{\partial \lambda_-}{\partial s}\,. \nonumber
\end{align}

We can now write out the full expressions:
\begin{equation}\begin{aligned}
\Pi_{Z \gamma}(m_Z^2) = \frac{\alpha_2s_W^2}{4\pi} &\left\{ \frac{6-16s_W^2}{9c_W s_W} \left[ \frac{1}{3} m_Z^2 - m_Z^2 \ln \frac{\mu^2}{m_t^2} - (m_Z^2 + 2m_t^2) b(m_Z^2, m_t) \right] \right. \\
&+ \frac{3-4s_W^2}{9c_W s_W} \left[ \frac{1}{3} m_Z^2 - m_Z^2 \ln \frac{\mu^2}{m_b^2} - (m_Z^2 + 2m_b^2) b_2(m_Z^2, m_b)\right] \\
\vphantom{\Pi_{Z \gamma}(m_Z^2) = \frac{\alpha_2s_W^2}{4\pi}} \hspace{3cm} &+ \frac{6-16s_W^2}{9c_W s_W} \left[ \frac{1}{3} m_Z^2 - m_Z^2 \ln \frac{\mu^2}{m_c^2} - (m_Z^2 + 2m_c^2) b_2(m_Z^2, m_c) \right] \\
&+ \frac{1-4s_W^2}{3 c_W s_W} \left[ \frac{1}{3} m_Z^2 - m_Z^2 \ln \frac{\mu^2}{m_{\tau}^2} - (m_Z^2 + 2m_{\tau}^2) b_2(m_Z^2, m_{\tau}) \right] \\
&+ m_Z^2 \frac{16s_W^2-6}{3c_W s_W} \left[ \frac{5}{3} + i \pi + \ln \frac{\mu^2}{m_Z^2} \right] \\
&+ \frac{1}{3s_W c_W} \left\{ \left[ \left(9 c_W^2 + \frac{1}{2} \right) m_Z^2 + \left( 12 c_W^2 + 4 \right) m_W^2 \right] \right. \\
&\hspace{1.9cm} \times \left(\ln \frac{\mu^2}{m_W^2} + b(m_Z^2,m_W) \right) \\
&\hspace{1.9cm}\left. \left.- (12 c_W^2 - 2) m_W^2 \ln \frac{\mu^2}{m_W^2} + \frac{1}{3} m_Z^2 \right\} \right\}\,,
\end{aligned}\end{equation}
and finally
\begin{equation}\begin{aligned}
\Pi^{\prime}_{ZZ}(m_Z^2) = \frac{\alpha_2s_W^2}{4\pi} &\left\{ 2 \left\{ \frac{9 - 24 s_W^2 + 32 s_W^4}{36 c_W^2 s_W^2} \left[ - 
\ln \frac{\mu^2}{m_t^2} - b(m_Z^2,m_t) \hspace{6.2cm} \right.\right.\right. \\
&\left.\left.- (m_Z^2+2m_t^2) c(m_Z^2,m_t) + \frac{1}{3} \right] + \frac{3}{4s_W^2c_W^2} m_t^2 c(m_Z^2,m_t) \right\} \\
&+ 2 \left\{ \frac{9 - 12 s_W^2 + 8 s_W^4}{36 c_W^2 s_W^2} \left[ - \ln \frac{\mu^2}{m_b^2} - b_2(m_Z^2,m_b)\right.\right. \\
&\left.\left.- (m_Z^2+2m_b^2) c_2(m_Z^2,m_b) + \frac{1}{3} \right] + \frac{3}{4s_W^2c_W^2} m_b^2 c_2(m_Z^2,m_b) \right\} \\
&+ 2 \left\{ \frac{9 - 24 s_W^2 + 32 s_W^4}{36 c_W^2 s_W^2} \left[ - \ln \frac{\mu^2}{m_c^2} - b_2(m_Z^2,m_c)\right.\right. \\
&\hspace{0.8cm}\left.\left.- (m_Z^2+2m_c^2) c_2(m_Z^2,m_c) + \frac{1}{3} \right] + \frac{3}{4s_W^2c_W^2} m_c^2 c_2(m_Z^2,m_c) \right\} \\
&+ \frac{2}{3} \left\{ \frac{1 - 4 s_W^2 + 8 s_W^4}{4 c_W^2 s_W^2} \left[ - \ln \frac{\mu^2}{m_{\tau}^2} - b_2(m_Z^2,m_{\tau})\right.\right. \\
&\hspace{0.8cm}\left.\left.- (m_Z^2+2m_{\tau}^2) c_2(m_Z^2,m_{\tau}) + \frac{1}{3} \right] + \frac{3}{4s_W^2c_W^2} m_{\tau}^2 c_2(m_Z^2,m_{\tau}) \right\} \\
&+  \frac{7-12s_W^2+16s_W^4}{3 s_W^2 c_W^2} \left[ - \frac{2}{3} - \ln \frac{\mu^2}{m_Z^2} - i \pi \right] \\
&+ \frac{1}{6s_W^2c_W^2} \left\{ \left( 18 c_W^4 + 2 c_W^2 - \frac{1}{2} \right) \left( \ln \frac{\mu^2}{m_W^2} + b(m_Z^2,m_W) \right) \right. \\
&\hspace{1.9cm}+ \frac{1}{3} \left( 4 c_W^2-1 \right) \\
&\hspace{1.9cm}+ \left[ \left( 18 c_W^4 + 2 c_W^2 - \frac{1}{2} \right) m_Z^2 \right. \\
&\hspace{2.3cm} \left. \left. + \left( 24 c_W^4 + 16 c_W^2 - 10 \right) m_W^2 \vphantom{\frac{1}{6s_W^2c_W^2}}\right] c(m_Z^2,m_W) \right\}\\
&+ \frac{1}{12s_W^2c_W^2} \left\{ - \left( \ln \frac{\mu^2}{m_Z^2} + d(m_Z^2,m_Z,m_H) \right) \right. \\
&\hspace{2.07cm} + \left( 2 m_H^2 - 11 m_Z^2 \right) e(m_Z^2,m_Z,m_H) \\
&\hspace{2.07cm} - \frac{(m_Z^2-m_H^2)^2}{m_Z^2} e(m_Z^2,m_Z,m_H) - \frac{2}{3} \\
&\hspace{2.02cm}+ \frac{(m_Z^2-m_H^2)^2}{m_Z^4} \left( \ln \frac{m_H^2}{m_Z^2} + d(m_Z^2,m_Z,m_H) \right. \\
&\hspace{4.8cm} \left. \left. \left.\vphantom{\frac{(m_Z^2-m_H^2)^2}{m_Z^4}} - a(m_Z,m_H) \right) \right\} \right\}\,.
\end{aligned}\end{equation}

%% file: main.bbl
\providecommand{\href}[2]{#2}\begingroup\raggedright\begin{thebibliography}{100}

\bibitem{Bertone:2016nfn}
G.~Bertone and D.~Hooper, {\it {A History of Dark Matter}},  {\em Submitted to:
  Rev. Mod. Phys.} (2016) [\href{http://arxiv.org/abs/1605.04909}{{\tt
  arXiv:1605.04909}}].

\bibitem{Tremaine:1979we}
S.~Tremaine and J.~E. Gunn, {\it {Dynamical Role of Light Neutral Leptons in
  Cosmology}},  {\em Phys. Rev. Lett.} {\bf 42} (1979) 407--410.

\bibitem{Buckley:2017ijx}
M.~R. Buckley and A.~H.~G. Peter, {\it {Gravitational probes of dark matter
  physics}},  \href{http://arxiv.org/abs/1712.06615}{{\tt arXiv:1712.06615}}.

\bibitem{Kahlhoefer:2017dnp}
F.~Kahlhoefer, {\it {Review of LHC Dark Matter Searches}},  {\em Int. J. Mod.
  Phys.} {\bf A32} (2017), no.~13 1730006,
  [\href{http://arxiv.org/abs/1702.02430}{{\tt arXiv:1702.02430}}].

\bibitem{Undagoitia:2015gya}
T.~Marrodán~Undagoitia and L.~Rauch, {\it {Dark matter direct-detection
  experiments}},  {\em J. Phys.} {\bf G43} (2016), no.~1 013001,
  [\href{http://arxiv.org/abs/1509.08767}{{\tt arXiv:1509.08767}}].

\bibitem{Giudice:1998xp}
G.~F. Giudice, M.~A. Luty, H.~Murayama, and R.~Rattazzi, {\it {Gaugino mass
  without singlets}},  {\em JHEP} {\bf 12} (1998) 027,
  [\href{http://arxiv.org/abs/hep-ph/9810442}{{\tt hep-ph/9810442}}].

\bibitem{Randall:1998uk}
L.~Randall and R.~Sundrum, {\it {Out of this world supersymmetry breaking}},
  {\em Nucl. Phys.} {\bf B557} (1999) 79--118,
  [\href{http://arxiv.org/abs/hep-th/9810155}{{\tt hep-th/9810155}}].

\bibitem{Hui:2016ltb}
L.~Hui, J.~P. Ostriker, S.~Tremaine, and E.~Witten, {\it {Ultralight scalars as
  cosmological dark matter}},  {\em Phys. Rev.} {\bf D95} (2017), no.~4 043541,
  [\href{http://arxiv.org/abs/1610.08297}{{\tt arXiv:1610.08297}}].

\bibitem{Peccei:1977hh}
R.~D. Peccei and H.~R. Quinn, {\it {CP Conservation in the Presence of
  Instantons}},  {\em Phys. Rev. Lett.} {\bf 38} (1977) 1440--1443.

\bibitem{Peccei:1977ur}
R.~D. Peccei and H.~R. Quinn, {\it {Constraints Imposed by CP Conservation in
  the Presence of Instantons}},  {\em Phys. Rev.} {\bf D16} (1977) 1791--1797.

\bibitem{Weinberg:1977ma}
S.~Weinberg, {\it {A New Light Boson?}},  {\em Phys. Rev. Lett.} {\bf 40}
  (1978) 223--226.

\bibitem{Wilczek:1977pj}
F.~Wilczek, {\it {Problem of Strong p and t Invariance in the Presence of
  Instantons}},  {\em Phys. Rev. Lett.} {\bf 40} (1978) 279--282.

\bibitem{Preskill:1982cy}
J.~Preskill, M.~B. Wise, and F.~Wilczek, {\it {Cosmology of the Invisible
  Axion}},  {\em Phys. Lett.} {\bf 120B} (1983) 127--132.

\bibitem{Abbott:1982af}
L.~F. Abbott and P.~Sikivie, {\it {A Cosmological Bound on the Invisible
  Axion}},  {\em Phys. Lett.} {\bf 120B} (1983) 133--136.

\bibitem{Dine:1982ah}
M.~Dine and W.~Fischler, {\it {The Not So Harmless Axion}},  {\em Phys. Lett.}
  {\bf 120B} (1983) 137--141.

\bibitem{Foster:2017hbq}
J.~W. Foster, N.~L. Rodd, and B.~R. Safdi, {\it {Revealing the Dark Matter Halo
  with Axion Direct Detection}},  \href{http://arxiv.org/abs/1711.10489}{{\tt
  arXiv:1711.10489}}.

\bibitem{Gunn:1978gr}
J.~E. Gunn, B.~W. Lee, I.~Lerche, D.~N. Schramm, and G.~Steigman, {\it {Some
  Astrophysical Consequences of the Existence of a Heavy Stable Neutral
  Lepton}},  {\em Astrophys. J.} {\bf 223} (1978) 1015--1031.

\bibitem{Stecker:1978du}
F.~W. Stecker, {\it {The Cosmic Gamma-Ray Background from the Annihilation of
  Primordial Stable Neutral Heavy Leptons}},  {\em Astrophys. J.} {\bf 223}
  (1978) 1032--1036.

\bibitem{Slatyer:2017sev}
T.~R. Slatyer, {\it {TASI Lectures on Indirect Detection of Dark Matter}},  in
  {\em {Theoretical Advanced Study Institute in Elementary Particle Physics:
  Anticipating the Next Discoveries in Particle Physics (TASI 2016) Boulder,
  CO, USA, June 6-July 1, 2016}}, 2017.
\newblock \href{http://arxiv.org/abs/1710.05137}{{\tt arXiv:1710.05137}}.

\bibitem{AdamEvans}
A.~Evans, ``M31, the andromeda galaxy (now with h-alpha).''
  \url{https://www.flickr.com/photos/astroporn/4999978603/}.
\newblock Accessed: 2018-03-14.

\bibitem{Slatyer:2015jla}
T.~R. Slatyer, {\it {Indirect dark matter signatures in the cosmic dark ages.
  I. Generalizing the bound on s-wave dark matter annihilation from Planck
  results}},  {\em Phys. Rev.} {\bf D93} (2016), no.~2 023527,
  [\href{http://arxiv.org/abs/1506.03811}{{\tt arXiv:1506.03811}}].

\bibitem{Slatyer:2015kla}
T.~R. Slatyer, {\it {Indirect Dark Matter Signatures in the Cosmic Dark Ages
  II. Ionization, Heating and Photon Production from Arbitrary Energy
  Injections}},  {\em Phys. Rev.} {\bf D93} (2016), no.~2 023521,
  [\href{http://arxiv.org/abs/1506.03812}{{\tt arXiv:1506.03812}}].

\bibitem{Slatyer:2016qyl}
T.~R. Slatyer and C.-L. Wu, {\it {General Constraints on Dark Matter Decay from
  the Cosmic Microwave Background}},  {\em Phys. Rev.} {\bf D95} (2017), no.~2
  023010, [\href{http://arxiv.org/abs/1610.06933}{{\tt arXiv:1610.06933}}].

\bibitem{Liu:2016cnk}
H.~Liu, T.~R. Slatyer, and J.~Zavala, {\it {Contributions to cosmic
  reionization from dark matter annihilation and decay}},  {\em Phys. Rev.}
  {\bf D94} (2016), no.~6 063507, [\href{http://arxiv.org/abs/1604.02457}{{\tt
  arXiv:1604.02457}}].

\bibitem{Lisanti:2017qoz}
M.~Lisanti, S.~Mishra-Sharma, N.~L. Rodd, B.~R. Safdi, and R.~H. Wechsler, {\it
  {Mapping Extragalactic Dark Matter Annihilation with Galaxy Surveys: A
  Systematic Study of Stacked Group Searches}},
  \href{http://arxiv.org/abs/1709.00416}{{\tt arXiv:1709.00416}}.

\bibitem{Hisano:2003ec}
J.~Hisano, S.~Matsumoto, and M.~M. Nojiri, {\it {Explosive dark matter
  annihilation}},  {\em Phys. Rev. Lett.} {\bf 92} (2004) 031303,
  [\href{http://arxiv.org/abs/hep-ph/0307216}{{\tt hep-ph/0307216}}].

\bibitem{Hisano:2004ds}
J.~Hisano, S.~Matsumoto, M.~M. Nojiri, and O.~Saito, {\it {Non-perturbative
  effect on dark matter annihilation and gamma ray signature from galactic
  center}},  {\em Phys. Rev.} {\bf D71} (2005) 063528,
  [\href{http://arxiv.org/abs/hep-ph/0412403}{{\tt hep-ph/0412403}}].

\bibitem{Cirelli:2007xd}
M.~Cirelli, A.~Strumia, and M.~Tamburini, {\it {Cosmology and Astrophysics of
  Minimal Dark Matter}},  {\em Nucl. Phys.} {\bf B787} (2007) 152--175,
  [\href{http://arxiv.org/abs/0706.4071}{{\tt arXiv:0706.4071}}].

\bibitem{ArkaniHamed:2008qn}
N.~Arkani-Hamed, D.~P. Finkbeiner, T.~R. Slatyer, and N.~Weiner, {\it {A Theory
  of Dark Matter}},  {\em Phys. Rev.} {\bf D79} (2009) 015014,
  [\href{http://arxiv.org/abs/0810.0713}{{\tt arXiv:0810.0713}}].

\bibitem{Blum:2016nrz}
K.~Blum, R.~Sato, and T.~R. Slatyer, {\it {Self-consistent Calculation of the
  Sommerfeld Enhancement}},  {\em JCAP} {\bf 1606} (2016), no.~06 021,
  [\href{http://arxiv.org/abs/1603.01383}{{\tt arXiv:1603.01383}}].

\bibitem{Boddy:2017vpe}
K.~K. Boddy, J.~Kumar, L.~E. Strigari, and M.-Y. Wang, {\it
  {Sommerfeld-Enhanced $J$-Factors For Dwarf Spheroidal Galaxies}},  {\em Phys.
  Rev.} {\bf D95} (2017), no.~12 123008,
  [\href{http://arxiv.org/abs/1702.00408}{{\tt arXiv:1702.00408}}].

\bibitem{Boddy:2018qur}
K.~Boddy, J.~Kumar, D.~Marfatia, and P.~Sandick, {\it {Model-independent
  constraints on dark matter annihilation in dwarf spheroidal galaxies}},
  \href{http://arxiv.org/abs/1802.03826}{{\tt arXiv:1802.03826}}.

\bibitem{Lisanti:2017qlb}
M.~Lisanti, S.~Mishra-Sharma, N.~L. Rodd, and B.~R. Safdi, {\it {A Search for
  Dark Matter Annihilation in Galaxy Groups}},
  \href{http://arxiv.org/abs/1708.09385}{{\tt arXiv:1708.09385}}.

\bibitem{Fermi:1949ee}
E.~Fermi, {\it {On the Origin of the Cosmic Radiation}},  {\em Phys. Rev.} {\bf
  75} (1949) 1169--1174.

\bibitem{Ackermann:2014usa}
{\bf Fermi-LAT} Collaboration, M.~Ackermann et~al., {\it {The spectrum of
  isotropic diffuse gamma-ray emission between 100 MeV and 820 GeV}},  {\em
  Astrophys. J.} {\bf 799} (2015) 86,
  [\href{http://arxiv.org/abs/1410.3696}{{\tt arXiv:1410.3696}}].

\bibitem{Su:2010qj}
M.~Su, T.~R. Slatyer, and D.~P. Finkbeiner, {\it {Giant Gamma-ray Bubbles from
  Fermi-LAT: AGN Activity or Bipolar Galactic Wind?}},  {\em Astrophys. J.}
  {\bf 724} (2010) 1044--1082, [\href{http://arxiv.org/abs/1005.5480}{{\tt
  arXiv:1005.5480}}].

\bibitem{Cowan:2010js}
G.~Cowan, K.~Cranmer, E.~Gross, and O.~Vitells, {\it {Asymptotic formulae for
  likelihood-based tests of new physics}},  {\em Eur. Phys. J.} {\bf C71}
  (2011) 1554, [\href{http://arxiv.org/abs/1007.1727}{{\tt arXiv:1007.1727}}].
  [Erratum: Eur. Phys. J.C73,2501(2013)].

\bibitem{Goodenough:2009gk}
L.~Goodenough and D.~Hooper, {\it {Possible Evidence For Dark Matter
  Annihilation In The Inner Milky Way From The Fermi Gamma Ray Space
  Telescope}},  \href{http://arxiv.org/abs/0910.2998}{{\tt arXiv:0910.2998}}.

\bibitem{Hooper:2010mq}
D.~Hooper and L.~Goodenough, {\it {Dark Matter Annihilation in The Galactic
  Center As Seen by the Fermi Gamma Ray Space Telescope}},  {\em Phys. Lett.}
  {\bf B697} (2011) 412--428, [\href{http://arxiv.org/abs/1010.2752}{{\tt
  arXiv:1010.2752}}].

\bibitem{Boyarsky:2010dr}
A.~Boyarsky, D.~Malyshev, and O.~Ruchayskiy, {\it {A comment on the emission
  from the Galactic Center as seen by the Fermi telescope}},  {\em Phys. Lett.}
  {\bf B705} (2011) 165--169, [\href{http://arxiv.org/abs/1012.5839}{{\tt
  arXiv:1012.5839}}].

\bibitem{Hooper:2011ti}
D.~Hooper and T.~Linden, {\it {On The Origin Of The Gamma Rays From The
  Galactic Center}},  {\em Phys. Rev.} {\bf D84} (2011) 123005,
  [\href{http://arxiv.org/abs/1110.0006}{{\tt arXiv:1110.0006}}].

\bibitem{Abazajian:2012pn}
K.~N. Abazajian and M.~Kaplinghat, {\it {Detection of a Gamma-Ray Source in the
  Galactic Center Consistent with Extended Emission from Dark Matter
  Annihilation and Concentrated Astrophysical Emission}},  {\em Phys. Rev.}
  {\bf D86} (2012) 083511, [\href{http://arxiv.org/abs/1207.6047}{{\tt
  arXiv:1207.6047}}]. [Erratum: Phys. Rev.D87,129902(2013)].

\bibitem{Gordon:2013vta}
C.~Gordon and O.~Macias, {\it {Dark Matter and Pulsar Model Constraints from
  Galactic Center Fermi-LAT Gamma Ray Observations}},  {\em Phys. Rev.} {\bf
  D88} (2013), no.~8 083521, [\href{http://arxiv.org/abs/1306.5725}{{\tt
  arXiv:1306.5725}}]. [Erratum: Phys. Rev.D89,no.4,049901(2014)].

\bibitem{Abazajian:2014fta}
K.~N. Abazajian, N.~Canac, S.~Horiuchi, and M.~Kaplinghat, {\it {Astrophysical
  and Dark Matter Interpretations of Extended Gamma-Ray Emission from the
  Galactic Center}},  {\em Phys. Rev.} {\bf D90} (2014), no.~2 023526,
  [\href{http://arxiv.org/abs/1402.4090}{{\tt arXiv:1402.4090}}].

\bibitem{Daylan:2014rsa}
T.~Daylan, D.~P. Finkbeiner, D.~Hooper, T.~Linden, S.~K.~N. Portillo, N.~L.
  Rodd, and T.~R. Slatyer, {\it {The characterization of the gamma-ray signal
  from the central Milky Way: A case for annihilating dark matter}},  {\em
  Phys. Dark Univ.} {\bf 12} (2016) 1--23,
  [\href{http://arxiv.org/abs/1402.6703}{{\tt arXiv:1402.6703}}].

\bibitem{Lee:2015fea}
S.~K. Lee, M.~Lisanti, B.~R. Safdi, T.~R. Slatyer, and W.~Xue, {\it {Evidence
  for Unresolved $\gamma$-Ray Point Sources in the Inner Galaxy}},  {\em Phys.
  Rev. Lett.} {\bf 116} (2016), no.~5 051103,
  [\href{http://arxiv.org/abs/1506.05124}{{\tt arXiv:1506.05124}}].

\bibitem{Bartels:2015aea}
R.~Bartels, S.~Krishnamurthy, and C.~Weniger, {\it {Strong support for the
  millisecond pulsar origin of the Galactic center GeV excess}},  {\em Phys.
  Rev. Lett.} {\bf 116} (2016), no.~5 051102,
  [\href{http://arxiv.org/abs/1506.05104}{{\tt arXiv:1506.05104}}].

\bibitem{Linden:2016rcf}
T.~Linden, N.~L. Rodd, B.~R. Safdi, and T.~R. Slatyer, {\it {High-energy tail
  of the Galactic Center gamma-ray excess}},  {\em Phys. Rev.} {\bf D94}
  (2016), no.~10 103013, [\href{http://arxiv.org/abs/1604.01026}{{\tt
  arXiv:1604.01026}}].

\bibitem{Mishra-Sharma:2016gis}
S.~Mishra-Sharma, N.~L. Rodd, and B.~R. Safdi, {\it {NPTFit: A code package for
  Non-Poissonian Template Fitting}},  {\em Astron. J.} {\bf 153} (2017), no.~6
  253, [\href{http://arxiv.org/abs/1612.03173}{{\tt arXiv:1612.03173}}].

\bibitem{Brandt:2015ula}
T.~D. Brandt and B.~Kocsis, {\it {Disrupted Globular Clusters Can Explain the
  Galactic Center Gamma Ray Excess}},  {\em Astrophys. J.} {\bf 812} (2015),
  no.~1 15, [\href{http://arxiv.org/abs/1507.05616}{{\tt arXiv:1507.05616}}].

\bibitem{Macias:2016nev}
O.~Macias, C.~Gordon, R.~M. Crocker, B.~Coleman, D.~Paterson, S.~Horiuchi, and
  M.~Pohl, {\it {X-Shaped Bulge Preferred Over Dark Matter for the Galactic
  Center Gamma-Ray Excess}},  \href{http://arxiv.org/abs/1611.06644}{{\tt
  arXiv:1611.06644}}.

\bibitem{Bartels:2017vsx}
R.~Bartels, E.~Storm, C.~Weniger, and F.~Calore, {\it {The Fermi-LAT GeV Excess
  Traces Stellar Mass in the Galactic Bulge}},
  \href{http://arxiv.org/abs/1711.04778}{{\tt arXiv:1711.04778}}.

\bibitem{Balaji:2018rwz}
B.~Balaji, I.~Cholis, P.~J. Fox, and S.~D. McDermott, {\it {Analyzing the
  Gamma-ray Sky with Wavelets}},  \href{http://arxiv.org/abs/1803.01952}{{\tt
  arXiv:1803.01952}}.

\bibitem{Bartels:2018eyb}
R.~Bartels, F.~Calore, E.~Storm, and C.~Weniger, {\it {Galactic Binaries Can
  Explain the Fermi Galactic Center Excess and 511 keV Emission}},
  \href{http://arxiv.org/abs/1803.04370}{{\tt arXiv:1803.04370}}.

\bibitem{Fermi-LAT:2017yoi}
{\bf Fermi-LAT} Collaboration, M.~Ajello et~al., {\it {Characterizing the
  population of pulsars in the inner Galaxy with the Fermi Large Area
  Telescope}},  {\em Submitted to: Astrophys. J.} (2017)
  [\href{http://arxiv.org/abs/1705.00009}{{\tt arXiv:1705.00009}}].

\bibitem{Bartels:2017xba}
R.~Bartels, D.~Hooper, T.~Linden, S.~Mishra-Sharma, N.~L. Rodd, B.~R. Safdi,
  and T.~R. Slatyer, {\it {Comment on "Characterizing the population of pulsars
  in the Galactic bulge with the $\textit{Fermi}$ Large Area Telescope"
  [arXiv:1705.00009v1]}},  \href{http://arxiv.org/abs/1710.10266}{{\tt
  arXiv:1710.10266}}.

\bibitem{Ackermann:2015zua}
{\bf Fermi-LAT} Collaboration, M.~Ackermann et~al., {\it {Searching for Dark
  Matter Annihilation from Milky Way Dwarf Spheroidal Galaxies with Six Years
  of Fermi Large Area Telescope Data}},  {\em Phys. Rev. Lett.} {\bf 115}
  (2015), no.~23 231301, [\href{http://arxiv.org/abs/1503.02641}{{\tt
  arXiv:1503.02641}}].

\bibitem{Fermi-LAT:2016uux}
{\bf DES, Fermi-LAT} Collaboration, A.~Albert et~al., {\it {Searching for Dark
  Matter Annihilation in Recently Discovered Milky Way Satellites with
  Fermi-LAT}},  {\em Astrophys. J.} {\bf 834} (2017), no.~2 110,
  [\href{http://arxiv.org/abs/1611.03184}{{\tt arXiv:1611.03184}}].

\bibitem{Keeley:2017fbz}
R.~Keeley, K.~Abazajian, A.~Kwa, N.~Rodd, and B.~Safdi, {\it {What the Milky
  Way's Dwarfs tell us about the Galactic Center extended excess}},
  \href{http://arxiv.org/abs/1710.03215}{{\tt arXiv:1710.03215}}.

\bibitem{Cohen:2016uyg}
T.~Cohen, K.~Murase, N.~L. Rodd, B.~R. Safdi, and Y.~Soreq, {\it {Gamma-ray
  Constraints on Decaying Dark Matter and Implications for IceCube}},  {\em
  Phys. Rev. Lett.} {\bf 119} (2017), no.~2 021102,
  [\href{http://arxiv.org/abs/1612.05638}{{\tt arXiv:1612.05638}}].

\bibitem{Abeysekara:2017jxs}
{\bf HAWC} Collaboration, A.~U. Abeysekara et~al., {\it {A Search for Dark
  Matter in the Galactic Halo with HAWC}},  {\em JCAP} {\bf 1802} (2018),
  no.~02 049, [\href{http://arxiv.org/abs/1710.10288}{{\tt arXiv:1710.10288}}].

\bibitem{Elor:2015tva}
G.~Elor, N.~L. Rodd, and T.~R. Slatyer, {\it {Multistep cascade annihilations
  of dark matter and the Galactic Center excess}},  {\em Phys. Rev.} {\bf D91}
  (2015) 103531, [\href{http://arxiv.org/abs/1503.01773}{{\tt
  arXiv:1503.01773}}].

\bibitem{Elor:2015bho}
G.~Elor, N.~L. Rodd, T.~R. Slatyer, and W.~Xue, {\it {Model-Independent
  Indirect Detection Constraints on Hidden Sector Dark Matter}},  {\em JCAP}
  {\bf 1606} (2016), no.~06 024, [\href{http://arxiv.org/abs/1511.08787}{{\tt
  arXiv:1511.08787}}].

\bibitem{Ovanesyan:2016vkk}
G.~Ovanesyan, N.~L. Rodd, T.~R. Slatyer, and I.~W. Stewart, {\it {One-loop
  correction to heavy dark matter annihilation}},  {\em Phys. Rev.} {\bf D95}
  (2017), no.~5 055001, [\href{http://arxiv.org/abs/1612.04814}{{\tt
  arXiv:1612.04814}}].

\bibitem{Baumgart:2017nsr}
M.~Baumgart, T.~Cohen, I.~Moult, N.~L. Rodd, T.~R. Slatyer, M.~P. Solon, I.~W.
  Stewart, and V.~Vaidya, {\it {Resummed Photon Spectra for WIMP
  Annihilation}},  \href{http://arxiv.org/abs/1712.07656}{{\tt
  arXiv:1712.07656}}.

\bibitem{Ilten:2017rbd}
P.~Ilten, N.~L. Rodd, J.~Thaler, and M.~Williams, {\it {Disentangling Heavy
  Flavor at Colliders}},  {\em Phys. Rev.} {\bf D96} (2017), no.~5 054019,
  [\href{http://arxiv.org/abs/1702.02947}{{\tt arXiv:1702.02947}}].

\bibitem{Hooper:2013rwa}
D.~Hooper and T.~R. Slatyer, {\it {Two Emission Mechanisms in the Fermi
  Bubbles: A Possible Signal of Annihilating Dark Matter}},  {\em Phys. Dark
  Univ.} {\bf 2} (2013) 118--138, [\href{http://arxiv.org/abs/1302.6589}{{\tt
  arXiv:1302.6589}}].

\bibitem{Huang:2013pda}
W.-C. Huang, A.~Urbano, and W.~Xue, {\it {Fermi Bubbles under Dark Matter
  Scrutiny. Part I: Astrophysical Analysis}},
  \href{http://arxiv.org/abs/1307.6862}{{\tt arXiv:1307.6862}}.

\bibitem{Abazajian:2010zy}
K.~N. Abazajian, {\it {The Consistency of Fermi-LAT Observations of the
  Galactic Center with a Millisecond Pulsar Population in the Central Stellar
  Cluster}},  {\em JCAP} {\bf 1103} (2011) 010,
  [\href{http://arxiv.org/abs/1011.4275}{{\tt arXiv:1011.4275}}].

\bibitem{Hooper:2013nhl}
D.~Hooper, I.~Cholis, T.~Linden, J.~Siegal-Gaskins, and T.~Slatyer, {\it
  {Pulsars Cannot Account for the Inner Galaxy's GeV Excess}},  {\em Phys.
  Rev.} {\bf D88} (2013) 083009, [\href{http://arxiv.org/abs/1305.0830}{{\tt
  arXiv:1305.0830}}].

\bibitem{Linden:2012iv}
T.~Linden, E.~Lovegrove, and S.~Profumo, {\it {The Morphology of Hadronic
  Emission Models for the Gamma-Ray Source at the Galactic Center}},  {\em
  Astrophys. J.} {\bf 753} (2012) 41,
  [\href{http://arxiv.org/abs/1203.3539}{{\tt arXiv:1203.3539}}].

\bibitem{Macias:2013vya}
O.~Macias and C.~Gordon, {\it {Contribution of cosmic rays interacting with
  molecular clouds to the Galactic Center gamma-ray excess}},  {\em Phys. Rev.}
  {\bf D89} (2014), no.~6 063515, [\href{http://arxiv.org/abs/1312.6671}{{\tt
  arXiv:1312.6671}}].

\bibitem{Kuhlen:2007ku}
M.~Kuhlen, J.~Diemand, and P.~Madau, {\it {The shapes, orientation, and
  alignment of Galactic dark matter subhalos}},  {\em Astrophys. J.} {\bf 671}
  (2007) 1135, [\href{http://arxiv.org/abs/0705.2037}{{\tt arXiv:0705.2037}}].

\bibitem{Navarro:1995iw}
J.~F. Navarro, C.~S. Frenk, and S.~D.~M. White, {\it {The Structure of cold
  dark matter halos}},  {\em Astrophys. J.} {\bf 462} (1996) 563--575,
  [\href{http://arxiv.org/abs/astro-ph/9508025}{{\tt astro-ph/9508025}}].

\bibitem{Navarro:1996gj}
J.~F. Navarro, C.~S. Frenk, and S.~D.~M. White, {\it {A Universal density
  profile from hierarchical clustering}},  {\em Astrophys. J.} {\bf 490} (1997)
  493--508, [\href{http://arxiv.org/abs/astro-ph/9611107}{{\tt
  astro-ph/9611107}}].

\bibitem{Iocco:2011jz}
F.~Iocco, M.~Pato, G.~Bertone, and P.~Jetzer, {\it {Dark Matter distribution in
  the Milky Way: microlensing and dynamical constraints}},  {\em JCAP} {\bf
  1111} (2011) 029, [\href{http://arxiv.org/abs/1107.5810}{{\tt
  arXiv:1107.5810}}].

\bibitem{Catena:2009mf}
R.~Catena and P.~Ullio, {\it {A novel determination of the local dark matter
  density}},  {\em JCAP} {\bf 1008} (2010) 004,
  [\href{http://arxiv.org/abs/0907.0018}{{\tt arXiv:0907.0018}}].

\bibitem{Navarro:2008kc}
J.~F. Navarro, A.~Ludlow, V.~Springel, J.~Wang, M.~Vogelsberger, S.~D.~M.
  White, A.~Jenkins, C.~S. Frenk, and A.~Helmi, {\it {The Diversity and
  Similarity of Cold Dark Matter Halos}},  {\em Mon. Not. Roy. Astron. Soc.}
  {\bf 402} (2010) 21, [\href{http://arxiv.org/abs/0810.1522}{{\tt
  arXiv:0810.1522}}].

\bibitem{Diemand:2008in}
J.~Diemand, M.~Kuhlen, P.~Madau, M.~Zemp, B.~Moore, D.~Potter, and J.~Stadel,
  {\it {Clumps and streams in the local dark matter distribution}},  {\em
  Nature} {\bf 454} (2008) 735--738,
  [\href{http://arxiv.org/abs/0805.1244}{{\tt arXiv:0805.1244}}].

\bibitem{Fry:1985tp}
J.~N. Fry, {\it {Statistics of Voids in Hierarchical Universes}},  {\em Phys.
  Lett.} {\bf 163B} (1985) 331--335.

\bibitem{Ryden:1987ska}
B.~S. Ryden and J.~E. Gunn, {\it {Galaxy formation by gravitational collapse}},
   {\em Astrophys. J.} {\bf 318} (1987) 15.

\bibitem{Gnedin:2011uj}
O.~Y. Gnedin, D.~Ceverino, N.~Y. Gnedin, A.~A. Klypin, A.~V. Kravtsov,
  R.~Levine, D.~Nagai, and G.~Yepes, {\it {Halo Contraction Effect in
  Hydrodynamic Simulations of Galaxy Formation}},
  \href{http://arxiv.org/abs/1108.5736}{{\tt arXiv:1108.5736}}.

\bibitem{Gnedin:2004cx}
O.~Y. Gnedin, A.~V. Kravtsov, A.~A. Klypin, and D.~Nagai, {\it {Response of
  dark matter halos to condensation of baryons: Cosmological simulations and
  improved adiabatic contraction model}},  {\em Astrophys. J.} {\bf 616} (2004)
  16--26, [\href{http://arxiv.org/abs/astro-ph/0406247}{{\tt
  astro-ph/0406247}}].

\bibitem{Governato:2012fa}
F.~Governato, A.~Zolotov, A.~Pontzen, C.~Christensen, S.~H. Oh, A.~M. Brooks,
  T.~Quinn, S.~Shen, and J.~Wadsley, {\it {Cuspy No More: How Outflows Affect
  the Central Dark Matter and Baryon Distribution in Lambda CDM Galaxies}},
  {\em Mon. Not. Roy. Astron. Soc.} {\bf 422} (2012) 1231--1240,
  [\href{http://arxiv.org/abs/1202.0554}{{\tt arXiv:1202.0554}}].

\bibitem{Kuhlen:2012qw}
M.~Kuhlen, J.~Guedes, A.~Pillepich, P.~Madau, and L.~Mayer, {\it {An Off-center
  Density Peak in the Milky Way's Dark Matter Halo?}},  {\em Astrophys. J.}
  {\bf 765} (2013) 10, [\href{http://arxiv.org/abs/1208.4844}{{\tt
  arXiv:1208.4844}}].

\bibitem{Weinberg:2001gm}
M.~D. Weinberg and N.~Katz, {\it {Bar-driven dark halo evolution: a resolution
  of the cusp-core controversy}},  {\em Astrophys. J.} {\bf 580} (2002)
  627--633, [\href{http://arxiv.org/abs/astro-ph/0110632}{{\tt
  astro-ph/0110632}}].

\bibitem{Weinberg:2006ps}
M.~D. Weinberg and N.~Katz, {\it {The bar-halo interaction. 2. secular
  evolution and the religion of n-body simulations}},  {\em Mon. Not. Roy.
  Astron. Soc.} {\bf 375} (2007) 460--476,
  [\href{http://arxiv.org/abs/astro-ph/0601138}{{\tt astro-ph/0601138}}].

\bibitem{Sellwood:2002vb}
J.~A. Sellwood, {\it {Bars and dark matter halo cores}},  {\em Astrophys. J.}
  {\bf 587} (2003) 638--648, [\href{http://arxiv.org/abs/astro-ph/0210079}{{\tt
  astro-ph/0210079}}].

\bibitem{Valenzuela:2002np}
O.~Valenzuela and A.~Klypin, {\it {Secular bar formation in galaxies with
  significant amount of dark matter}},  {\em Mon. Not. Roy. Astron. Soc.} {\bf
  345} (2003) 406, [\href{http://arxiv.org/abs/astro-ph/0204028}{{\tt
  astro-ph/0204028}}].

\bibitem{Colin:2005rr}
P.~Colin, O.~Valenzuela, and A.~Klypin, {\it {Bars and cold dark matter
  halos}},  {\em Astrophys. J.} {\bf 644} (2006) 687--700,
  [\href{http://arxiv.org/abs/astro-ph/0506627}{{\tt astro-ph/0506627}}].

\bibitem{Springel:2008cc}
V.~Springel, J.~Wang, M.~Vogelsberger, A.~Ludlow, A.~Jenkins, A.~Helmi, J.~F.
  Navarro, C.~S. Frenk, and S.~D.~M. White, {\it {The Aquarius Project: the
  subhalos of galactic halos}},  {\em Mon. Not. Roy. Astron. Soc.} {\bf 391}
  (2008) 1685--1711, [\href{http://arxiv.org/abs/0809.0898}{{\tt
  arXiv:0809.0898}}].

\bibitem{Ackermann:2013yva}
{\bf Fermi-LAT} Collaboration, M.~Ackermann et~al., {\it {Dark matter
  constraints from observations of 25 Milky Way satellite galaxies with the
  Fermi Large Area Telescope}},  {\em Phys. Rev.} {\bf D89} (2014) 042001,
  [\href{http://arxiv.org/abs/1310.0828}{{\tt arXiv:1310.0828}}].

\bibitem{Sjostrand:2006za}
T.~Sjostrand, S.~Mrenna, and P.~Z. Skands, {\it {PYTHIA 6.4 Physics and
  Manual}},  {\em JHEP} {\bf 05} (2006) 026,
  [\href{http://arxiv.org/abs/hep-ph/0603175}{{\tt hep-ph/0603175}}].

\bibitem{Bergstrom:2004cy}
L.~Bergstrom, T.~Bringmann, M.~Eriksson, and M.~Gustafsson, {\it {Gamma rays
  from Kaluza-Klein dark matter}},  {\em Phys. Rev. Lett.} {\bf 94} (2005)
  131301, [\href{http://arxiv.org/abs/astro-ph/0410359}{{\tt
  astro-ph/0410359}}].

\bibitem{Birkedal:2005ep}
A.~Birkedal, K.~T. Matchev, M.~Perelstein, and A.~Spray, {\it {Robust gamma ray
  signature of WIMP dark matter}},
  \href{http://arxiv.org/abs/hep-ph/0507194}{{\tt hep-ph/0507194}}.

\bibitem{Cirelli:2013mqa}
M.~Cirelli, P.~D. Serpico, and G.~Zaharijas, {\it {Bremsstrahlung gamma rays
  from light Dark Matter}},  {\em JCAP} {\bf 1311} (2013) 035,
  [\href{http://arxiv.org/abs/1307.7152}{{\tt arXiv:1307.7152}}].

\bibitem{Atwood:2009ez}
{\bf Fermi-LAT} Collaboration, W.~B. Atwood et~al., {\it {The Large Area
  Telescope on the Fermi Gamma-ray Space Telescope Mission}},  {\em Astrophys.
  J.} {\bf 697} (2009) 1071--1102, [\href{http://arxiv.org/abs/0902.1089}{{\tt
  arXiv:0902.1089}}].

\bibitem{Ackermann:2012kna}
{\bf Fermi-LAT} Collaboration, M.~Ackermann et~al., {\it {The Fermi Large Area
  Telescope On Orbit: Event Classification, Instrument Response Functions, and
  Calibration}},  {\em Astrophys. J. Suppl.} {\bf 203} (2012) 4,
  [\href{http://arxiv.org/abs/1206.1896}{{\tt arXiv:1206.1896}}].

\bibitem{Ackermann:2013yma}
{\bf Fermi-LAT} Collaboration, M.~Ackermann et~al., {\it {Determination of the
  Point-Spread Function for the Fermi Large Area Telescope from On-orbit Data
  and Limits on Pair Halos of Active Galactic Nuclei}},  {\em Astrophys. J.}
  {\bf 765} (2013), no.~1 54, [\href{http://arxiv.org/abs/1309.5416}{{\tt
  arXiv:1309.5416}}].

\bibitem{Portillo:2014ena}
S.~K.~N. Portillo and D.~P. Finkbeiner, {\it {Sharper Fermi LAT Images:
  instrument response functions for an improved event selection}},  {\em
  Astrophys. J.} {\bf 796} (2014), no.~1 54,
  [\href{http://arxiv.org/abs/1406.0507}{{\tt arXiv:1406.0507}}].

\bibitem{Dobler:2009xz}
G.~Dobler, D.~P. Finkbeiner, I.~Cholis, T.~R. Slatyer, and N.~Weiner, {\it {The
  Fermi Haze: A Gamma-Ray Counterpart to the Microwave Haze}},  {\em Astrophys.
  J.} {\bf 717} (2010) 825--842, [\href{http://arxiv.org/abs/0910.4583}{{\tt
  arXiv:0910.4583}}].

\bibitem{Calore:2014xka}
F.~Calore, I.~Cholis, and C.~Weniger, {\it {Background Model Systematics for
  the Fermi GeV Excess}},  {\em JCAP} {\bf 1503} (2015) 038,
  [\href{http://arxiv.org/abs/1409.0042}{{\tt arXiv:1409.0042}}].

\bibitem{Strong:1998pw}
A.~W. Strong and I.~V. Moskalenko, {\it {Propagation of cosmic-ray nucleons in
  the galaxy}},  {\em Astrophys. J.} {\bf 509} (1998) 212--228,
  [\href{http://arxiv.org/abs/astro-ph/9807150}{{\tt astro-ph/9807150}}].

\bibitem{Strong:1999sv}
A.~W. Strong and I.~V. Moskalenko, {\it {The galprop program for cosmic ray
  propagation: new developments}},  in {\em {Proceedings, 26th International
  Cosmic Ray Conference (ICRC), August 17-25, 1999, Salt Lake City: Invited,
  Rapporteur, and Highlight Papers}}, p.~255, 1999.
\newblock \href{http://arxiv.org/abs/astro-ph/9906228}{{\tt astro-ph/9906228}}.
\newblock [4,255(1999)].

\bibitem{Strong:2007nh}
A.~W. Strong, I.~V. Moskalenko, and V.~S. Ptuskin, {\it {Cosmic-ray propagation
  and interactions in the Galaxy}},  {\em Ann. Rev. Nucl. Part. Sci.} {\bf 57}
  (2007) 285--327, [\href{http://arxiv.org/abs/astro-ph/0701517}{{\tt
  astro-ph/0701517}}].

\bibitem{Law:2008uk}
C.~J. Law, F.~Yusef-Zadeh, W.~D. Cotton, and R.~J. Maddalena, {\it {GBT
  Multiwavelength Survey of the Galactic Center Region}},  {\em Astrophys. J.
  Suppl.} {\bf 177} (2008) 255, [\href{http://arxiv.org/abs/0801.4294}{{\tt
  arXiv:0801.4294}}].

\bibitem{Fermi-LAT:2011yjw}
{\bf Fermi-LAT} Collaboration, P.~Nolan et~al., {\it {Fermi Large Area
  Telescope Second Source Catalog}},  {\em Astrophys. J. Suppl.} {\bf 199}
  (2012) 31, [\href{http://arxiv.org/abs/1108.1435}{{\tt arXiv:1108.1435}}].

\bibitem{YusefZadeh:2012nh}
F.~Yusef-Zadeh et~al., {\it {Interacting Cosmic Rays with Molecular Clouds: A
  Bremsstrahlung Origin of Diffuse High Energy Emission from the Inner 2deg by
  1deg of the Galactic Center}},  {\em Astrophys. J.} {\bf 762} (2013) 33,
  [\href{http://arxiv.org/abs/1206.6882}{{\tt arXiv:1206.6882}}].

\bibitem{Abdo:2010nz}
{\bf Fermi-LAT} Collaboration, A.~A. Abdo et~al., {\it {The Spectrum of the
  Isotropic Diffuse Gamma-Ray Emission Derived From First-Year Fermi Large Area
  Telescope Data}},  {\em Phys. Rev. Lett.} {\bf 104} (2010) 101101,
  [\href{http://arxiv.org/abs/1002.3603}{{\tt arXiv:1002.3603}}].

\bibitem{YusefZadeh:1999bh}
F.~Yusef-Zadeh, D.~Choate, and W.~Cotton, {\it {The position of sgr $a^*$ at
  the galactic center}},  {\em Astrophys. J.} {\bf 518} (1999) L33,
  [\href{http://arxiv.org/abs/astro-ph/9904142}{{\tt astro-ph/9904142}}].

\bibitem{Aalseth:2010vx}
{\bf CoGeNT} Collaboration, C.~E. Aalseth et~al., {\it {Results from a Search
  for Light-Mass Dark Matter with a P-type Point Contact Germanium Detector}},
  {\em Phys. Rev. Lett.} {\bf 106} (2011) 131301,
  [\href{http://arxiv.org/abs/1002.4703}{{\tt arXiv:1002.4703}}].

\bibitem{Aalseth:2011wp}
C.~E. Aalseth et~al., {\it {Search for an Annual Modulation in a P-type Point
  Contact Germanium Dark Matter Detector}},  {\em Phys. Rev. Lett.} {\bf 107}
  (2011) 141301, [\href{http://arxiv.org/abs/1106.0650}{{\tt
  arXiv:1106.0650}}].

\bibitem{Agnese:2013rvf}
{\bf CDMS} Collaboration, R.~Agnese et~al., {\it {Silicon Detector Dark Matter
  Results from the Final Exposure of CDMS II}},  {\em Phys. Rev. Lett.} {\bf
  111} (2013), no.~25 251301, [\href{http://arxiv.org/abs/1304.4279}{{\tt
  arXiv:1304.4279}}].

\bibitem{Angloher:2011uu}
G.~Angloher et~al., {\it {Results from 730 kg days of the CRESST-II Dark Matter
  Search}},  {\em Eur. Phys. J.} {\bf C72} (2012) 1971,
  [\href{http://arxiv.org/abs/1109.0702}{{\tt arXiv:1109.0702}}].

\bibitem{Bernabei:2008yi}
{\bf DAMA} Collaboration, R.~Bernabei et~al., {\it {First results from
  DAMA/LIBRA and the combined results with DAMA/NaI}},  {\em Eur. Phys. J.}
  {\bf C56} (2008) 333--355, [\href{http://arxiv.org/abs/0804.2741}{{\tt
  arXiv:0804.2741}}].

\bibitem{Bernabei:2010mq}
{\bf DAMA, LIBRA} Collaboration, R.~Bernabei et~al., {\it {New results from
  DAMA/LIBRA}},  {\em Eur. Phys. J.} {\bf C67} (2010) 39--49,
  [\href{http://arxiv.org/abs/1002.1028}{{\tt arXiv:1002.1028}}].

\bibitem{Steigman:2012nb}
G.~Steigman, B.~Dasgupta, and J.~F. Beacom, {\it {Precise Relic WIMP Abundance
  and its Impact on Searches for Dark Matter Annihilation}},  {\em Phys. Rev.}
  {\bf D86} (2012) 023506, [\href{http://arxiv.org/abs/1204.3622}{{\tt
  arXiv:1204.3622}}].

\bibitem{Boehm:2003bt}
C.~Boehm, D.~Hooper, J.~Silk, M.~Casse, and J.~Paul, {\it {MeV dark matter: Has
  it been detected?}},  {\em Phys. Rev. Lett.} {\bf 92} (2004) 101301,
  [\href{http://arxiv.org/abs/astro-ph/0309686}{{\tt astro-ph/0309686}}].

\bibitem{Adriani:2008zr}
{\bf PAMELA} Collaboration, O.~Adriani et~al., {\it {An anomalous positron
  abundance in cosmic rays with energies 1.5-100 GeV}},  {\em Nature} {\bf 458}
  (2009) 607--609, [\href{http://arxiv.org/abs/0810.4995}{{\tt
  arXiv:0810.4995}}].

\bibitem{Chang:2008aa}
J.~Chang et~al., {\it {An excess of cosmic ray electrons at energies of 300-800
  GeV}},  {\em Nature} {\bf 456} (2008) 362--365.

\bibitem{Weniger:2012tx}
C.~Weniger, {\it {A Tentative Gamma-Ray Line from Dark Matter Annihilation at
  the Fermi Large Area Telescope}},  {\em JCAP} {\bf 1208} (2012) 007,
  [\href{http://arxiv.org/abs/1204.2797}{{\tt arXiv:1204.2797}}].

\bibitem{Su:2012ft}
M.~Su and D.~P. Finkbeiner, {\it {Strong Evidence for Gamma-ray Line Emission
  from the Inner Galaxy}},  \href{http://arxiv.org/abs/1206.1616}{{\tt
  arXiv:1206.1616}}.

\bibitem{Finkbeiner:2004us}
D.~P. Finkbeiner, {\it {WMAP microwave emission interpreted as dark matter
  annihilation in the inner galaxy}},
  \href{http://arxiv.org/abs/astro-ph/0409027}{{\tt astro-ph/0409027}}.

\bibitem{Hooper:2007kb}
D.~Hooper, D.~P. Finkbeiner, and G.~Dobler, {\it {Possible evidence for dark
  matter annihilations from the excess microwave emission around the center of
  the Galaxy seen by the Wilkinson Microwave Anisotropy Probe}},  {\em Phys.
  Rev.} {\bf D76} (2007) 083012, [\href{http://arxiv.org/abs/0705.3655}{{\tt
  arXiv:0705.3655}}].

\bibitem{Hooper:2008kg}
D.~Hooper, P.~Blasi, and P.~D. Serpico, {\it {Pulsars as the Sources of High
  Energy Cosmic Ray Positrons}},  {\em JCAP} {\bf 0901} (2009) 025,
  [\href{http://arxiv.org/abs/0810.1527}{{\tt arXiv:0810.1527}}].

\bibitem{Profumo:2008ms}
S.~Profumo, {\it {Dissecting cosmic-ray electron-positron data with Occam's
  Razor: the role of known Pulsars}},  {\em Central Eur. J. Phys.} {\bf 10}
  (2011) 1--31, [\href{http://arxiv.org/abs/0812.4457}{{\tt arXiv:0812.4457}}].

\bibitem{Dobler:2011rd}
G.~Dobler, {\it {A Last Look at the Microwave Haze/Bubbles with WMAP}},  {\em
  Astrophys. J.} {\bf 750} (2012) 17,
  [\href{http://arxiv.org/abs/1109.4418}{{\tt arXiv:1109.4418}}].

\bibitem{Dobler:2012ef}
G.~Dobler, {\it {Identifying the Radio Bubble Nature of the Microwave Haze}},
  {\em Astrophys. J.} {\bf 760} (2012) L8,
  [\href{http://arxiv.org/abs/1208.2690}{{\tt arXiv:1208.2690}}].

\bibitem{Ackermann:2013uma}
{\bf Fermi-LAT} Collaboration, M.~Ackermann et~al., {\it {Search for Gamma-ray
  Spectral Lines with the Fermi Large Area Telescope and Dark Matter
  Implications}},  {\em Phys. Rev.} {\bf D88} (2013) 082002,
  [\href{http://arxiv.org/abs/1305.5597}{{\tt arXiv:1305.5597}}].

\bibitem{Aalseth:2012if}
{\bf CoGeNT} Collaboration, C.~E. Aalseth et~al., {\it {CoGeNT: A Search for
  Low-Mass Dark Matter using p-type Point Contact Germanium Detectors}},  {\em
  Phys. Rev.} {\bf D88} (2013) 012002,
  [\href{http://arxiv.org/abs/1208.5737}{{\tt arXiv:1208.5737}}].

\bibitem{Aprile:2012nq}
{\bf XENON100} Collaboration, E.~Aprile et~al., {\it {Dark Matter Results from
  225 Live Days of XENON100 Data}},  {\em Phys. Rev. Lett.} {\bf 109} (2012)
  181301, [\href{http://arxiv.org/abs/1207.5988}{{\tt arXiv:1207.5988}}].

\bibitem{Akerib:2013tjd}
{\bf LUX} Collaboration, D.~S. Akerib et~al., {\it {First results from the LUX
  dark matter experiment at the Sanford Underground Research Facility}},  {\em
  Phys. Rev. Lett.} {\bf 112} (2014) 091303,
  [\href{http://arxiv.org/abs/1310.8214}{{\tt arXiv:1310.8214}}].

\bibitem{Boehm:2014hva}
C.~Boehm, M.~J. Dolan, C.~McCabe, M.~Spannowsky, and C.~J. Wallace, {\it
  {Extended gamma-ray emission from Coy Dark Matter}},  {\em JCAP} {\bf 1405}
  (2014) 009, [\href{http://arxiv.org/abs/1401.6458}{{\tt arXiv:1401.6458}}].

\bibitem{Hardy:2014dea}
E.~Hardy, R.~Lasenby, and J.~Unwin, {\it {Annihilation Signals from Asymmetric
  Dark Matter}},  {\em JHEP} {\bf 07} (2014) 049,
  [\href{http://arxiv.org/abs/1402.4500}{{\tt arXiv:1402.4500}}].

\bibitem{Modak:2013jya}
K.~P. Modak, D.~Majumdar, and S.~Rakshit, {\it {A Possible Explanation of Low
  Energy $\gamma$-ray Excess from Galactic Centre and Fermi Bubble by a Dark
  Matter Model with Two Real Scalars}},  {\em JCAP} {\bf 1503} (2015) 011,
  [\href{http://arxiv.org/abs/1312.7488}{{\tt arXiv:1312.7488}}].

\bibitem{Huang:2013apa}
W.-C. Huang, A.~Urbano, and W.~Xue, {\it {Fermi Bubbles under Dark Matter
  Scrutiny Part II: Particle Physics Analysis}},  {\em JCAP} {\bf 1404} (2014)
  020, [\href{http://arxiv.org/abs/1310.7609}{{\tt arXiv:1310.7609}}].

\bibitem{Okada:2013bna}
N.~Okada and O.~Seto, {\it {Gamma ray emission in Fermi bubbles and Higgs
  portal dark matter}},  {\em Phys. Rev.} {\bf D89} (2014), no.~4 043525,
  [\href{http://arxiv.org/abs/1310.5991}{{\tt arXiv:1310.5991}}].

\bibitem{Hagiwara:2013qya}
K.~Hagiwara, S.~Mukhopadhyay, and J.~Nakamura, {\it {10 GeV neutralino dark
  matter and light stau in the MSSM}},  {\em Phys. Rev.} {\bf D89} (2014),
  no.~1 015023, [\href{http://arxiv.org/abs/1308.6738}{{\tt arXiv:1308.6738}}].

\bibitem{Buckley:2013sca}
M.~R. Buckley, D.~Hooper, and J.~Kumar, {\it {Phenomenology of Dirac Neutralino
  Dark Matter}},  {\em Phys. Rev.} {\bf D88} (2013) 063532,
  [\href{http://arxiv.org/abs/1307.3561}{{\tt arXiv:1307.3561}}].

\bibitem{Anchordoqui:2013pta}
L.~A. Anchordoqui and B.~J. Vlcek, {\it {W-WIMP Annihilation as a Source of the
  Fermi Bubbles}},  {\em Phys. Rev.} {\bf D88} (2013) 043513,
  [\href{http://arxiv.org/abs/1305.4625}{{\tt arXiv:1305.4625}}].

\bibitem{Buckley:2011mm}
M.~R. Buckley, D.~Hooper, and J.~L. Rosner, {\it {A Leptophobic Z' And Dark
  Matter From Grand Unification}},  {\em Phys. Lett.} {\bf B703} (2011)
  343--347, [\href{http://arxiv.org/abs/1106.3583}{{\tt arXiv:1106.3583}}].

\bibitem{Boucenna:2011hy}
M.~S. Boucenna and S.~Profumo, {\it {Direct and Indirect Singlet Scalar Dark
  Matter Detection in the Lepton-Specific two-Higgs-doublet Model}},  {\em
  Phys. Rev.} {\bf D84} (2011) 055011,
  [\href{http://arxiv.org/abs/1106.3368}{{\tt arXiv:1106.3368}}].

\bibitem{Marshall:2011mm}
G.~Marshall and R.~Primulando, {\it {The Galactic Center Region Gamma Ray
  Excess from A Supersymmetric Leptophilic Higgs Model}},  {\em JHEP} {\bf 05}
  (2011) 026, [\href{http://arxiv.org/abs/1102.0492}{{\tt arXiv:1102.0492}}].

\bibitem{Zhu:2011dz}
G.~Zhu, {\it {WIMPless dark matter and the excess gamma rays from the Galactic
  center}},  {\em Phys. Rev.} {\bf D83} (2011) 076011,
  [\href{http://arxiv.org/abs/1101.4387}{{\tt arXiv:1101.4387}}].

\bibitem{Buckley:2010ve}
M.~R. Buckley, D.~Hooper, and T.~M.~P. Tait, {\it {Particle Physics
  Implications for CoGeNT, DAMA, and Fermi}},  {\em Phys. Lett.} {\bf B702}
  (2011) 216--219, [\href{http://arxiv.org/abs/1011.1499}{{\tt
  arXiv:1011.1499}}].

\bibitem{Logan:2010nw}
H.~E. Logan, {\it {Dark matter annihilation through a lepton-specific Higgs
  boson}},  {\em Phys. Rev.} {\bf D83} (2011) 035022,
  [\href{http://arxiv.org/abs/1010.4214}{{\tt arXiv:1010.4214}}].

\bibitem{Allgood:2005eu}
B.~Allgood, R.~A. Flores, J.~R. Primack, A.~V. Kravtsov, R.~H. Wechsler,
  A.~Faltenbacher, and J.~S. Bullock, {\it {The shape of dark matter halos:
  dependence on mass, redshift, radius, and formation}},  {\em Mon. Not. Roy.
  Astron. Soc.} {\bf 367} (2006) 1781--1796,
  [\href{http://arxiv.org/abs/astro-ph/0508497}{{\tt astro-ph/0508497}}].

\bibitem{Bernal:2014mmt}
N.~Bernal, J.~E. Forero-Romero, R.~Garani, and S.~Palomares-Ruiz, {\it
  {Systematic uncertainties from halo asphericity in dark matter searches}},
  {\em JCAP} {\bf 1409} (2014) 004, [\href{http://arxiv.org/abs/1405.6240}{{\tt
  arXiv:1405.6240}}].

\bibitem{Law:2009yq}
D.~R. Law, S.~R. Majewski, and K.~V. Johnston, {\it {Evidence for a Triaxial
  Milky Way Dark Matter Halo from the Sagittarius Stellar Tidal Stream}},  {\em
  Astrophys. J.} {\bf 703} (2009) L67--L71,
  [\href{http://arxiv.org/abs/0908.3187}{{\tt arXiv:0908.3187}}].

\bibitem{Law:2010pe}
D.~R. Law and S.~R. Majewski, {\it {The Sagittarius Dwarf Galaxy: a Model for
  Evolution in a Triaxial Milky Way Halo}},  {\em Astrophys. J.} {\bf 714}
  (2010) 229--254, [\href{http://arxiv.org/abs/1003.1132}{{\tt
  arXiv:1003.1132}}].

\bibitem{Gregoire:2013yta}
T.~Grégoire and J.~Knödlseder, {\it {Constraining the Galactic millisecond
  pulsar population using Fermi Large Area Telescope}},  {\em Astron.
  Astrophys.} {\bf 554} (2013) A62, [\href{http://arxiv.org/abs/1305.1584}{{\tt
  arXiv:1305.1584}}].

\bibitem{collaboration:2010bb}
{\bf Fermi-LAT} Collaboration, A.~Abdo et~al., {\it {A population of gamma-ray
  emitting globular clusters seen with the Fermi Large Area Telescope}},  {\em
  Astron. Astrophys.} {\bf 524} (2010) A75,
  [\href{http://arxiv.org/abs/1003.3588}{{\tt arXiv:1003.3588}}].

\bibitem{Han:2012uw}
J.~Han, C.~S. Frenk, V.~R. Eke, L.~Gao, S.~D.~M. White, A.~Boyarsky,
  D.~Malyshev, and O.~Ruchayskiy, {\it {Constraining Extended Gamma-ray
  Emission from Galaxy Clusters}},  {\em Mon. Not. Roy. Astron. Soc.} {\bf 427}
  (2012) 1651--1665, [\href{http://arxiv.org/abs/1207.6749}{{\tt
  arXiv:1207.6749}}].

\bibitem{MaciasRamirez:2012mk}
O.~Macias-Ramirez, C.~Gordon, A.~M. Brown, and J.~Adams, {\it {Evaluating the
  Gamma-Ray Evidence for Self-Annihilating Dark Matter from the Virgo
  Cluster}},  {\em Phys. Rev.} {\bf D86} (2012) 076004,
  [\href{http://arxiv.org/abs/1207.6257}{{\tt arXiv:1207.6257}}].

\bibitem{Berlin:2013dva}
A.~Berlin and D.~Hooper, {\it {Stringent Constraints on the Dark Matter
  Annihilation Cross Section From Subhalo Searches with the Fermi Gamma-Ray
  Space Telescope}},  {\em Phys. Rev.} {\bf D89} (2014), no.~1 016014,
  [\href{http://arxiv.org/abs/1309.0525}{{\tt arXiv:1309.0525}}].

\bibitem{Cirelli:2013hv}
M.~Cirelli and G.~Giesen, {\it {Antiprotons from Dark Matter: Current
  constraints and future sensitivities}},  {\em JCAP} {\bf 1304} (2013) 015,
  [\href{http://arxiv.org/abs/1301.7079}{{\tt arXiv:1301.7079}}].

\bibitem{Fornengo:2013xda}
N.~Fornengo, L.~Maccione, and A.~Vittino, {\it {Constraints on particle dark
  matter from cosmic-ray antiprotons}},  {\em JCAP} {\bf 1404} (2014), no.~04
  003, [\href{http://arxiv.org/abs/1312.3579}{{\tt arXiv:1312.3579}}].

\bibitem{Ackermann:2015tah}
{\bf Fermi-LAT} Collaboration, M.~Ackermann et~al., {\it {Limits on Dark Matter
  Annihilation Signals from the Fermi LAT 4-year Measurement of the Isotropic
  Gamma-Ray Background}},  {\em JCAP} {\bf 1509} (2015), no.~09 008,
  [\href{http://arxiv.org/abs/1501.05464}{{\tt arXiv:1501.05464}}].

\bibitem{Zechlin:2016pme}
H.-S. Zechlin, A.~Cuoco, F.~Donato, N.~Fornengo, and M.~Regis, {\it
  {Statistical Measurement of the Gamma-ray Source-count Distribution as a
  Function of Energy}},  {\em Astrophys. J.} {\bf 826} (2016), no.~2 L31,
  [\href{http://arxiv.org/abs/1605.04256}{{\tt arXiv:1605.04256}}].

\bibitem{Lisanti:2016jub}
M.~Lisanti, S.~Mishra-Sharma, L.~Necib, and B.~R. Safdi, {\it {Deciphering
  Contributions to the Extragalactic Gamma-Ray Background from 2 GeV to 2
  TeV}},  {\em Astrophys. J.} {\bf 832} (2016), no.~2 117,
  [\href{http://arxiv.org/abs/1606.04101}{{\tt arXiv:1606.04101}}].

\bibitem{Ackermann:2012uf}
{\bf Fermi-LAT} Collaboration, M.~Ackermann et~al., {\it {Anisotropies in the
  diffuse gamma-ray background measured by the Fermi LAT}},  {\em Phys. Rev.}
  {\bf D85} (2012) 083007, [\href{http://arxiv.org/abs/1202.2856}{{\tt
  arXiv:1202.2856}}].

\bibitem{Ando:2006cr}
S.~Ando, E.~Komatsu, T.~Narumoto, and T.~Totani, {\it {Dark matter annihilation
  or unresolved astrophysical sources? Anisotropy probe of the origin of cosmic
  gamma-ray background}},  {\em Phys. Rev.} {\bf D75} (2007) 063519,
  [\href{http://arxiv.org/abs/astro-ph/0612467}{{\tt astro-ph/0612467}}].

\bibitem{Ando:2013ff}
S.~Ando and E.~Komatsu, {\it {Constraints on the annihilation cross section of
  dark matter particles from anisotropies in the diffuse gamma-ray background
  measured with Fermi-LAT}},  {\em Phys. Rev.} {\bf D87} (2013), no.~12 123539,
  [\href{http://arxiv.org/abs/1301.5901}{{\tt arXiv:1301.5901}}].

\bibitem{Xia:2011ax}
J.-Q. Xia, A.~Cuoco, E.~Branchini, M.~Fornasa, and M.~Viel, {\it {A
  cross-correlation study of the Fermi-LAT $\gamma$-ray diffuse extragalactic
  signal}},  {\em Mon. Not. Roy. Astron. Soc.} {\bf 416} (2011) 2247--2264,
  [\href{http://arxiv.org/abs/1103.4861}{{\tt arXiv:1103.4861}}].

\bibitem{Ando:2014aoa}
S.~Ando, {\it {Power spectrum tomography of dark matter annihilation with local
  galaxy distribution}},  {\em JCAP} {\bf 1410} (2014), no.~10 061,
  [\href{http://arxiv.org/abs/1407.8502}{{\tt arXiv:1407.8502}}].

\bibitem{Ando:2013xwa}
S.~Ando, A.~Benoit-Lévy, and E.~Komatsu, {\it {Mapping dark matter in the
  gamma-ray sky with galaxy catalogs}},  {\em Phys. Rev.} {\bf D90} (2014),
  no.~2 023514, [\href{http://arxiv.org/abs/1312.4403}{{\tt arXiv:1312.4403}}].

\bibitem{Xia:2015wka}
J.-Q. Xia, A.~Cuoco, E.~Branchini, and M.~Viel, {\it {Tomography of the
  Fermi-lat $\gamma$-ray Diffuse Extragalactic Signal via Cross Correlations
  With Galaxy Catalogs}},  {\em Astrophys. J. Suppl.} {\bf 217} (2015), no.~1
  15, [\href{http://arxiv.org/abs/1503.05918}{{\tt arXiv:1503.05918}}].

\bibitem{Regis:2015zka}
M.~Regis, J.-Q. Xia, A.~Cuoco, E.~Branchini, N.~Fornengo, and M.~Viel, {\it
  {Particle dark matter searches outside the Local Group}},  {\em Phys. Rev.
  Lett.} {\bf 114} (2015), no.~24 241301,
  [\href{http://arxiv.org/abs/1503.05922}{{\tt arXiv:1503.05922}}].

\bibitem{Cuoco:2015rfa}
A.~Cuoco, J.-Q. Xia, M.~Regis, E.~Branchini, N.~Fornengo, and M.~Viel, {\it
  {Dark Matter Searches in the Gamma-ray Extragalactic Background via
  Cross-correlations With Galaxy Catalogs}},  {\em Astrophys. J. Suppl.} {\bf
  221} (2015), no.~2 29, [\href{http://arxiv.org/abs/1506.01030}{{\tt
  arXiv:1506.01030}}].

\bibitem{Ando:2016ang}
S.~Ando and K.~Ishiwata, {\it {Constraining particle dark matter using local
  galaxy distribution}},  {\em JCAP} {\bf 1606} (2016), no.~06 045,
  [\href{http://arxiv.org/abs/1604.02263}{{\tt arXiv:1604.02263}}].

\bibitem{Jarrett:2000me}
T.~H. Jarrett, T.~Chester, R.~Cutri, S.~Schneider, M.~Skrutskie, and J.~P.
  Huchra, {\it {2mass extended source catalog: overview and algorithms}},  {\em
  Astron. J.} {\bf 119} (2000) 2498--2531,
  [\href{http://arxiv.org/abs/astro-ph/0004318}{{\tt astro-ph/0004318}}].

\bibitem{Bilicki:2013sza}
M.~Bilicki, T.~H. Jarrett, J.~A. Peacock, M.~E. Cluver, and L.~Steward, {\it
  {2MASS Photometric Redshift catalog: a comprehensive three-dimensional census
  of the whole sky}},  \href{http://arxiv.org/abs/1311.5246}{{\tt
  arXiv:1311.5246}}. [Astrophys. J. Suppl.210,9(2014)].

\bibitem{Ackermann:2010rg}
M.~Ackermann et~al., {\it {Constraints on Dark Matter Annihilation in Clusters
  of Galaxies with the Fermi Large Area Telescope}},  {\em JCAP} {\bf 1005}
  (2010) 025, [\href{http://arxiv.org/abs/1002.2239}{{\tt arXiv:1002.2239}}].

\bibitem{Ando:2012vu}
S.~Ando and D.~Nagai, {\it {Fermi-LAT constraints on dark matter annihilation
  cross section from observations of the Fornax cluster}},  {\em JCAP} {\bf
  1207} (2012) 017, [\href{http://arxiv.org/abs/1201.0753}{{\tt
  arXiv:1201.0753}}].

\bibitem{Ackermann:2013iaq}
{\bf Fermi-LAT} Collaboration, M.~Ackermann et~al., {\it {Search for cosmic-ray
  induced gamma-ray emission in Galaxy Clusters}},  {\em Astrophys. J.} {\bf
  787} (2014) 18, [\href{http://arxiv.org/abs/1308.5654}{{\tt
  arXiv:1308.5654}}].

\bibitem{Ackermann:2015fdi}
{\bf Fermi-LAT} Collaboration, M.~Ackermann et~al., {\it {Search for extended
  gamma-ray emission from the Virgo galaxy cluster with Fermi-LAT}},  {\em
  Astrophys. J.} {\bf 812} (2015), no.~2 159,
  [\href{http://arxiv.org/abs/1510.00004}{{\tt arXiv:1510.00004}}].

\bibitem{Anderson:2015dpc}
B.~Anderson, S.~Zimmer, J.~Conrad, M.~Gustafsson, M.~Sánchez-Conde, and
  R.~Caputo, {\it {Search for Gamma-Ray Lines towards Galaxy Clusters with the
  Fermi-LAT}},  {\em JCAP} {\bf 1602} (2016), no.~02 026,
  [\href{http://arxiv.org/abs/1511.00014}{{\tt arXiv:1511.00014}}].

\bibitem{Rephaeli:2015nca}
{\bf Fermi-LAT} Collaboration, M.~Ackermann et~al., {\it {Search for gamma-ray
  emission from the Coma Cluster with six years of Fermi-LAT data}},  {\em
  Astrophys. J.} {\bf 819} (2016), no.~2 149,
  [\href{http://arxiv.org/abs/1507.08995}{{\tt arXiv:1507.08995}}].

\bibitem{Ahnen:2016qkt}
{\bf MAGIC} Collaboration, M.~L. Ahnen et~al., {\it {Deep onservation of the
  NGC 1275 region with MAGIC: search of diffuse $\gamma$-ray emission from
  cosmic rays in the Perseus cluster}},  {\em Astron. Astrophys.} {\bf 589}
  (2016) A33, [\href{http://arxiv.org/abs/1602.03099}{{\tt arXiv:1602.03099}}].

\bibitem{Liang:2016pvm}
Y.-F. Liang, Z.-Q. Shen, X.~Li, Y.-Z. Fan, X.~Huang, S.-J. Lei, L.~Feng, E.-W.
  Liang, and J.~Chang, {\it {Search for a gamma-ray line feature from a group
  of nearby galaxy clusters with Fermi LAT Pass 8 data}},  {\em Phys. Rev.}
  {\bf D93} (2016), no.~10 103525, [\href{http://arxiv.org/abs/1602.06527}{{\tt
  arXiv:1602.06527}}].

\bibitem{Adams:2016alz}
D.~Q. Adams, L.~Bergstrom, and D.~Spolyar, {\it {Improved Constraints on Dark
  Matter Annihilation to a Line using Fermi-LAT observations of Galaxy
  Clusters}},  \href{http://arxiv.org/abs/1606.09642}{{\tt arXiv:1606.09642}}.

\bibitem{Huang:2011xr}
X.~Huang, G.~Vertongen, and C.~Weniger, {\it {Probing Dark Matter Decay and
  Annihilation with Fermi LAT Observations of Nearby Galaxy Clusters}},  {\em
  JCAP} {\bf 1201} (2012) 042, [\href{http://arxiv.org/abs/1110.1529}{{\tt
  arXiv:1110.1529}}].

\bibitem{Tully:2015opa}
R.~B. Tully, {\it {Galaxy Groups: A 2MASS Catalog}},  {\em Astron. J.} {\bf
  149} (2015) 171, [\href{http://arxiv.org/abs/1503.03134}{{\tt
  arXiv:1503.03134}}].

\bibitem{2017ApJ...843...16K}
E.~Kourkchi and R.~B. Tully, {\it {Galaxy Groups Within 3500 km s$^{-1}$}},
  {\em Ap. J.} {\bf 843} (2017) 16,
  [\href{http://arxiv.org/abs/1705.08068}{{\tt arXiv:1705.08068}}].

\bibitem{Tully:2016ppz}
R.~B. Tully, H.~M. Courtois, and J.~G. Sorce, {\it {Cosmicflows-3}},  {\em
  Astron. J.} {\bf 152} (2016) 50, [\href{http://arxiv.org/abs/1605.01765}{{\tt
  arXiv:1605.01765}}].

\bibitem{Cirelli:2010xx}
M.~Cirelli, G.~Corcella, A.~Hektor, G.~Hutsi, M.~Kadastik, P.~Panci, M.~Raidal,
  F.~Sala, and A.~Strumia, {\it {PPPC 4 DM ID: A Poor Particle Physicist
  Cookbook for Dark Matter Indirect Detection}},  {\em JCAP} {\bf 1103} (2011)
  051, [\href{http://arxiv.org/abs/1012.4515}{{\tt arXiv:1012.4515}}].
  [Erratum: JCAP1210,E01(2012)].

\bibitem{Correa:2015dva}
C.~A. Correa, J.~S.~B. Wyithe, J.~Schaye, and A.~R. Duffy, {\it {The accretion
  history of dark matter haloes – III. A physical model for the
  concentration–mass relation}},  {\em Mon. Not. Roy. Astron. Soc.} {\bf 452}
  (2015), no.~2 1217--1232, [\href{http://arxiv.org/abs/1502.00391}{{\tt
  arXiv:1502.00391}}].

\bibitem{Ade:2015xua}
{\bf Planck} Collaboration, P.~A.~R. Ade et~al., {\it {Planck 2015 results.
  XIII. Cosmological parameters}},  {\em Astron. Astrophys.} {\bf 594} (2016)
  A13, [\href{http://arxiv.org/abs/1502.01589}{{\tt arXiv:1502.01589}}].

\bibitem{Lehmann:2015ioa}
B.~V. Lehmann, Y.-Y. Mao, M.~R. Becker, S.~W. Skillman, and R.~H. Wechsler,
  {\it {The Concentration Dependence of the Galaxy-Halo Connection: Modeling
  Assembly Bias with Abundance Matching}},  {\em Astrophys. J.} {\bf 834}
  (2017), no.~1 37, [\href{http://arxiv.org/abs/1510.05651}{{\tt
  arXiv:1510.05651}}].

\bibitem{Gao:2011rf}
L.~Gao, C.~S. Frenk, A.~Jenkins, V.~Springel, and S.~D.~M. White, {\it {Where
  will supersymmetric dark matter first be seen?}},  {\em Mon. Not. Roy.
  Astron. Soc.} {\bf 419} (2012) 1721,
  [\href{http://arxiv.org/abs/1107.1916}{{\tt arXiv:1107.1916}}].

\bibitem{Anderhalden:2013wd}
D.~Anderhalden and J.~Diemand, {\it {Density Profiles of CDM Microhalos and
  their Implications for Annihilation Boost Factors}},  {\em JCAP} {\bf 1304}
  (2013) 009, [\href{http://arxiv.org/abs/1302.0003}{{\tt arXiv:1302.0003}}].
  [Erratum: JCAP1308,E02(2013)].

\bibitem{Ludlow:2013vxa}
A.~D. Ludlow, J.~F. Navarro, R.~E. Angulo, M.~Boylan-Kolchin, V.~Springel,
  C.~Frenk, and S.~D.~M. White, {\it {The mass–concentration–redshift
  relation of cold dark matter haloes}},  {\em Mon. Not. Roy. Astron. Soc.}
  {\bf 441} (2014), no.~1 378--388, [\href{http://arxiv.org/abs/1312.0945}{{\tt
  arXiv:1312.0945}}].

\bibitem{Bartels:2015uba}
R.~Bartels and S.~Ando, {\it {Boosting the annihilation boost: Tidal effects on
  dark matter subhalos and consistent luminosity modeling}},  {\em Phys. Rev.}
  {\bf D92} (2015), no.~12 123508, [\href{http://arxiv.org/abs/1507.08656}{{\tt
  arXiv:1507.08656}}].

\bibitem{Crook:2006sw}
A.~C. Crook, J.~P. Huchra, N.~Martimbeau, K.~L. Masters, T.~Jarrett, and L.~M.
  Macri, {\it {Groups of Galaxies in the Two Micron All-Sky Redshift Survey}},
  {\em Astrophys. J.} {\bf 655} (2007) 790--813,
  [\href{http://arxiv.org/abs/astro-ph/0610732}{{\tt astro-ph/0610732}}].

\bibitem{Ackermann:2017nya}
{\bf Fermi-LAT} Collaboration, M.~Ackermann et~al., {\it {Observations of M31
  and M33 with the Fermi Large Area Telescope: A Galactic Center Excess in
  Andromeda?}},  {\em Astrophys. J.} {\bf 836} (2017), no.~2 208,
  [\href{http://arxiv.org/abs/1702.08602}{{\tt arXiv:1702.08602}}].

\bibitem{Gorski:2004by}
K.~M. Gorski, E.~Hivon, A.~J. Banday, B.~D. Wandelt, F.~K. Hansen, M.~Reinecke,
  and M.~Bartelman, {\it {HEALPix - A Framework for high resolution
  discretization, and fast analysis of data distributed on the sphere}},  {\em
  Astrophys. J.} {\bf 622} (2005) 759--771,
  [\href{http://arxiv.org/abs/astro-ph/0409513}{{\tt astro-ph/0409513}}].

\bibitem{Acero:2015hja}
{\bf Fermi-LAT} Collaboration, F.~Acero et~al., {\it {Fermi Large Area
  Telescope Third Source Catalog}},  {\em Astrophys. J. Suppl.} {\bf 218}
  (2015), no.~2 23, [\href{http://arxiv.org/abs/1501.02003}{{\tt
  arXiv:1501.02003}}].

\bibitem{Jeltema:2008vu}
T.~E. Jeltema, J.~Kehayias, and S.~Profumo, {\it {Gamma Rays from Clusters and
  Groups of Galaxies: Cosmic Rays versus Dark Matter}},  {\em Phys. Rev.} {\bf
  D80} (2009) 023005, [\href{http://arxiv.org/abs/0812.0597}{{\tt
  arXiv:0812.0597}}].

\bibitem{Huber:2013cia}
B.~Huber, C.~Tchernin, D.~Eckert, C.~Farnier, A.~Manalaysay, U.~Straumann, and
  R.~Walter, {\it {Probing the cosmic-ray content of galaxy clusters by
  stacking Fermi-LAT count maps}},  {\em Astron. Astrophys.} {\bf 560} (2013)
  A64, [\href{http://arxiv.org/abs/1308.6278}{{\tt arXiv:1308.6278}}].

\bibitem{Geringer-Sameth:2014qqa}
A.~Geringer-Sameth, S.~M. Koushiappas, and M.~G. Walker, {\it {Comprehensive
  search for dark matter annihilation in dwarf galaxies}},  {\em Phys. Rev.}
  {\bf D91} (2015), no.~8 083535, [\href{http://arxiv.org/abs/1410.2242}{{\tt
  arXiv:1410.2242}}].

\bibitem{TheFermi-LAT:2015kwa}
{\bf Fermi-LAT} Collaboration, M.~Ajello et~al., {\it {Fermi-LAT Observations
  of High-Energy $\gamma$-Ray Emission Toward the Galactic Center}},  {\em
  Astrophys. J.} {\bf 819} (2016), no.~1 44,
  [\href{http://arxiv.org/abs/1511.02938}{{\tt arXiv:1511.02938}}].

\bibitem{Karwin:2016tsw}
C.~Karwin, S.~Murgia, T.~M.~P. Tait, T.~A. Porter, and P.~Tanedo, {\it {Dark
  Matter Interpretation of the Fermi-LAT Observation Toward the Galactic
  Center}},  {\em Phys. Rev.} {\bf D95} (2017), no.~10 103005,
  [\href{http://arxiv.org/abs/1612.05687}{{\tt arXiv:1612.05687}}].

\bibitem{Aartsen:2013bka}
{\bf IceCube} Collaboration, M.~G. Aartsen et~al., {\it {First observation of
  PeV-energy neutrinos with IceCube}},  {\em Phys. Rev. Lett.} {\bf 111} (2013)
  021103, [\href{http://arxiv.org/abs/1304.5356}{{\tt arXiv:1304.5356}}].

\bibitem{Aartsen:2013jdh}
{\bf IceCube} Collaboration, M.~G. Aartsen et~al., {\it {Evidence for
  High-Energy Extraterrestrial Neutrinos at the IceCube Detector}},  {\em
  Science} {\bf 342} (2013) 1242856,
  [\href{http://arxiv.org/abs/1311.5238}{{\tt arXiv:1311.5238}}].

\bibitem{Aartsen:2015knd}
{\bf IceCube} Collaboration, M.~G. Aartsen et~al., {\it {A combined
  maximum-likelihood analysis of the high-energy astrophysical neutrino flux
  measured with IceCube}},  {\em Astrophys. J.} {\bf 809} (2015), no.~1 98,
  [\href{http://arxiv.org/abs/1507.03991}{{\tt arXiv:1507.03991}}].

\bibitem{Aartsen:2015rwa}
{\bf IceCube} Collaboration, M.~G. Aartsen et~al., {\it {Evidence for
  Astrophysical Muon Neutrinos from the Northern Sky with IceCube}},  {\em
  Phys. Rev. Lett.} {\bf 115} (2015), no.~8 081102,
  [\href{http://arxiv.org/abs/1507.04005}{{\tt arXiv:1507.04005}}].

\bibitem{Kalashev:2016cre}
O.~K. Kalashev and M.~{\relax Yu}. Kuznetsov, {\it {Constraining heavy decaying
  dark matter with the high energy gamma-ray limits}},  {\em Phys. Rev.} {\bf
  D94} (2016), no.~6 063535, [\href{http://arxiv.org/abs/1606.07354}{{\tt
  arXiv:1606.07354}}].

\bibitem{Aab:2015bza}
{\em {The Pierre Auger Observatory: Contributions to the 34th International
  Cosmic Ray Conference (ICRC 2015)}}, 2015.

\bibitem{Kang:2015gpa}
D.~Kang et~al., {\it {A limit on the diffuse gamma-rays measured with
  KASCADE-Grande}},  {\em J. Phys. Conf. Ser.} {\bf 632} (2015), no.~1 012013.

\bibitem{Chantell:1997gs}
{\bf CASA-MIA} Collaboration, M.~C. Chantell et~al., {\it {Limits on the
  isotropic diffuse flux of ultrahigh-energy gamma radiation}},  {\em Phys.
  Rev. Lett.} {\bf 79} (1997) 1805--1808,
  [\href{http://arxiv.org/abs/astro-ph/9705246}{{\tt astro-ph/9705246}}].

\bibitem{Ando:2015qda}
S.~Ando and K.~Ishiwata, {\it {Constraints on decaying dark matter from the
  extragalactic gamma-ray background}},  {\em JCAP} {\bf 1505} (2015), no.~05
  024, [\href{http://arxiv.org/abs/1502.02007}{{\tt arXiv:1502.02007}}].

\bibitem{Murase:2012xs}
K.~Murase and J.~F. Beacom, {\it {Constraining Very Heavy Dark Matter Using
  Diffuse Backgrounds of Neutrinos and Cascaded Gamma Rays}},  {\em JCAP} {\bf
  1210} (2012) 043, [\href{http://arxiv.org/abs/1206.2595}{{\tt
  arXiv:1206.2595}}].

\bibitem{Ackermann:2012rg}
{\bf Fermi-LAT} Collaboration, M.~Ackermann et~al., {\it {Constraints on the
  Galactic Halo Dark Matter from Fermi-LAT Diffuse Measurements}},  {\em
  Astrophys. J.} {\bf 761} (2012) 91,
  [\href{http://arxiv.org/abs/1205.6474}{{\tt arXiv:1205.6474}}].

\bibitem{Hutsi:2010ai}
G.~Hutsi, A.~Hektor, and M.~Raidal, {\it {Implications of the Fermi-LAT diffuse
  gamma-ray measurements on annihilating or decaying Dark Matter}},  {\em JCAP}
  {\bf 1007} (2010) 008, [\href{http://arxiv.org/abs/1004.2036}{{\tt
  arXiv:1004.2036}}].

\bibitem{Cirelli:2012ut}
M.~Cirelli, E.~Moulin, P.~Panci, P.~D. Serpico, and A.~Viana, {\it {Gamma ray
  constraints on Decaying Dark Matter}},  {\em Phys. Rev.} {\bf D86} (2012)
  083506, [\href{http://arxiv.org/abs/1205.5283}{{\tt arXiv:1205.5283}}].

\bibitem{Kalashev:2016xmy}
O.~Kalashev, {\it {Constraining Dark Matter and Ultra-High Energy Cosmic Ray
  Sources with Fermi-LAT Diffuse Gamma Ray Background}},  {\em EPJ Web Conf.}
  {\bf 125} (2016) 02012, [\href{http://arxiv.org/abs/1608.07530}{{\tt
  arXiv:1608.07530}}].

\bibitem{ams02pos}
 Ams-02 Collaboration, talks at the `AMS Days at CERN', 2015, 15-17 april.

\bibitem{Giesen:2015ufa}
G.~Giesen, M.~Boudaud, Y.~Génolini, V.~Poulin, M.~Cirelli, P.~Salati, and
  P.~D. Serpico, {\it {AMS-02 antiprotons, at last! Secondary astrophysical
  component and immediate implications for Dark Matter}},  {\em JCAP} {\bf
  1509} (2015), no.~09 023, [\href{http://arxiv.org/abs/1504.04276}{{\tt
  arXiv:1504.04276}}].

\bibitem{Aguilar:2013qda}
{\bf AMS} Collaboration, M.~Aguilar et~al., {\it {First Result from the Alpha
  Magnetic Spectrometer on the International Space Station: Precision
  Measurement of the Positron Fraction in Primary Cosmic Rays of 0.5–350
  GeV}},  {\em Phys. Rev. Lett.} {\bf 110} (2013) 141102.

\bibitem{Ibarra:2013zia}
A.~Ibarra, A.~S. Lamperstorfer, and J.~Silk, {\it {Dark matter annihilations
  and decays after the AMS-02 positron measurements}},  {\em Phys. Rev.} {\bf
  D89} (2014), no.~6 063539, [\href{http://arxiv.org/abs/1309.2570}{{\tt
  arXiv:1309.2570}}].

\bibitem{Chianese:2016kpu}
M.~Chianese, G.~Miele, and S.~Morisi, {\it {Dark Matter interpretation of low
  energy IceCube MESE excess}},  {\em JCAP} {\bf 1701} (2017), no.~01 007,
  [\href{http://arxiv.org/abs/1610.04612}{{\tt arXiv:1610.04612}}].

\bibitem{Sjostrand:2007gs}
T.~Sjostrand, S.~Mrenna, and P.~Z. Skands, {\it {A Brief Introduction to PYTHIA
  8.1}},  {\em Comput. Phys. Commun.} {\bf 178} (2008) 852--867,
  [\href{http://arxiv.org/abs/0710.3820}{{\tt arXiv:0710.3820}}].

\bibitem{Sjostrand:2014zea}
T.~Sjöstrand, S.~Ask, J.~R. Christiansen, R.~Corke, N.~Desai, P.~Ilten,
  S.~Mrenna, S.~Prestel, C.~O. Rasmussen, and P.~Z. Skands, {\it {An
  Introduction to PYTHIA 8.2}},  {\em Comput. Phys. Commun.} {\bf 191} (2015)
  159--177, [\href{http://arxiv.org/abs/1410.3012}{{\tt arXiv:1410.3012}}].

\bibitem{Christiansen:2014kba}
J.~R. Christiansen and T.~Sjöstrand, {\it {Weak Gauge Boson Radiation in
  Parton Showers}},  {\em JHEP} {\bf 04} (2014) 115,
  [\href{http://arxiv.org/abs/1401.5238}{{\tt arXiv:1401.5238}}].

\bibitem{Kachelriess:2007aj}
M.~Kachelriess and P.~D. Serpico, {\it {Model-independent dark matter
  annihilation bound from the diffuse $\gamma$ ray flux}},  {\em Phys. Rev.}
  {\bf D76} (2007) 063516, [\href{http://arxiv.org/abs/0707.0209}{{\tt
  arXiv:0707.0209}}].

\bibitem{Regis:2008ij}
M.~Regis and P.~Ullio, {\it {Multi-wavelength signals of dark matter
  annihilations at the Galactic center}},  {\em Phys. Rev.} {\bf D78} (2008)
  043505, [\href{http://arxiv.org/abs/0802.0234}{{\tt arXiv:0802.0234}}].

\bibitem{Mack:2008wu}
G.~D. Mack, T.~D. Jacques, J.~F. Beacom, N.~F. Bell, and H.~Yuksel, {\it
  {Conservative Constraints on Dark Matter Annihilation into Gamma Rays}},
  {\em Phys. Rev.} {\bf D78} (2008) 063542,
  [\href{http://arxiv.org/abs/0803.0157}{{\tt arXiv:0803.0157}}].

\bibitem{Bell:2008ey}
N.~F. Bell, J.~B. Dent, T.~D. Jacques, and T.~J. Weiler, {\it {Electroweak
  Bremsstrahlung in Dark Matter Annihilation}},  {\em Phys. Rev.} {\bf D78}
  (2008) 083540, [\href{http://arxiv.org/abs/0805.3423}{{\tt
  arXiv:0805.3423}}].

\bibitem{Dent:2008qy}
J.~B. Dent, R.~J. Scherrer, and T.~J. Weiler, {\it {Toward a Minimum Branching
  Fraction for Dark Matter Annihilation into Electromagnetic Final States}},
  {\em Phys. Rev.} {\bf D78} (2008) 063509,
  [\href{http://arxiv.org/abs/0806.0370}{{\tt arXiv:0806.0370}}].

\bibitem{Borriello:2008gy}
E.~Borriello, A.~Cuoco, and G.~Miele, {\it {Radio constraints on dark matter
  annihilation in the galactic halo and its substructures}},  {\em Phys. Rev.}
  {\bf D79} (2009) 023518, [\href{http://arxiv.org/abs/0809.2990}{{\tt
  arXiv:0809.2990}}].

\bibitem{Bertone:2008xr}
G.~Bertone, M.~Cirelli, A.~Strumia, and M.~Taoso, {\it {Gamma-ray and radio
  tests of the e+e- excess from DM annihilations}},  {\em JCAP} {\bf 0903}
  (2009) 009, [\href{http://arxiv.org/abs/0811.3744}{{\tt arXiv:0811.3744}}].

\bibitem{Bell:2008vx}
N.~F. Bell and T.~D. Jacques, {\it {Gamma-ray Constraints on Dark Matter
  Annihilation into Charged Particles}},  {\em Phys. Rev.} {\bf D79} (2009)
  043507, [\href{http://arxiv.org/abs/0811.0821}{{\tt arXiv:0811.0821}}].

\bibitem{Cirelli:2009vg}
M.~Cirelli and P.~Panci, {\it {Inverse Compton constraints on the Dark Matter
  e+e- excesses}},  {\em Nucl. Phys.} {\bf B821} (2009) 399--416,
  [\href{http://arxiv.org/abs/0904.3830}{{\tt arXiv:0904.3830}}].

\bibitem{Kachelriess:2009zy}
M.~Kachelriess, P.~D. Serpico, and M.~A. Solberg, {\it {On the role of
  electroweak bremsstrahlung for indirect dark matter signatures}},  {\em Phys.
  Rev.} {\bf D80} (2009) 123533, [\href{http://arxiv.org/abs/0911.0001}{{\tt
  arXiv:0911.0001}}].

\bibitem{Ciafaloni:2010ti}
P.~Ciafaloni, D.~Comelli, A.~Riotto, F.~Sala, A.~Strumia, and A.~Urbano, {\it
  {Weak Corrections are Relevant for Dark Matter Indirect Detection}},  {\em
  JCAP} {\bf 1103} (2011) 019, [\href{http://arxiv.org/abs/1009.0224}{{\tt
  arXiv:1009.0224}}].

\bibitem{Murase:2015gea}
K.~Murase, R.~Laha, S.~Ando, and M.~Ahlers, {\it {Testing the Dark Matter
  Scenario for PeV Neutrinos Observed in IceCube}},  {\em Phys. Rev. Lett.}
  {\bf 115} (2015), no.~7 071301, [\href{http://arxiv.org/abs/1503.04663}{{\tt
  arXiv:1503.04663}}].

\bibitem{Esmaili:2015xpa}
A.~Esmaili and P.~D. Serpico, {\it {Gamma-ray bounds from EAS detectors and
  heavy decaying dark matter constraints}},  {\em JCAP} {\bf 1510} (2015),
  no.~10 014, [\href{http://arxiv.org/abs/1505.06486}{{\tt arXiv:1505.06486}}].

\bibitem{Mao:2012hx}
S.~A. Mao, N.~M. McClure-Griffiths, B.~M. Gaensler, J.~C. Brown, C.~L. van Eck,
  M.~Haverkorn, P.~P. Kronberg, J.~M. Stil, A.~Shukurov, and A.~R. Taylor, {\it
  {New Constraints on the Galactic Halo Magnetic Field using Rotation Measures
  of Extragalactic Sources Towards the Outer Galaxy}},  {\em Astrophys. J.}
  {\bf 755} (2012) 21, [\href{http://arxiv.org/abs/1206.3314}{{\tt
  arXiv:1206.3314}}].

\bibitem{Haverkorn:2014jka}
M.~Haverkorn, {\it {Magnetic Fields in the Milky Way}},
  \href{http://arxiv.org/abs/1406.0283}{{\tt arXiv:1406.0283}}.

\bibitem{Beck:2014pma}
M.~C. Beck, A.~M. Beck, R.~Beck, K.~Dolag, A.~W. Strong, and P.~Nielaba, {\it
  {New constraints on modelling the random magnetic field of the MW}},  {\em
  JCAP} {\bf 1605} (2016), no.~05 056,
  [\href{http://arxiv.org/abs/1409.5120}{{\tt arXiv:1409.5120}}].

\bibitem{Kribs:1996ac}
G.~D. Kribs and I.~Z. Rothstein, {\it {Bounds on longlived relics from diffuse
  gamma-ray observations}},  {\em Phys. Rev.} {\bf D55} (1997) 4435--4449,
  [\href{http://arxiv.org/abs/hep-ph/9610468}{{\tt hep-ph/9610468}}]. [Erratum:
  Phys. Rev.D56,1822(1997)].

\bibitem{Rolke:2004mj}
W.~A. Rolke, A.~M. Lopez, and J.~Conrad, {\it {Limits and confidence intervals
  in the presence of nuisance parameters}},  {\em Nucl. Instrum. Meth.} {\bf
  A551} (2005) 493--503, [\href{http://arxiv.org/abs/physics/0403059}{{\tt
  physics/0403059}}].

\bibitem{Narayanan:2016nzy}
S.~A. Narayanan and T.~R. Slatyer, {\it {A Latitude-Dependent Analysis of the
  Leptonic Hypothesis for the Fermi Bubbles}},  {\em Mon. Not. Roy. Astron.
  Soc.} {\bf 468} (2017), no.~3 3051--3070,
  [\href{http://arxiv.org/abs/1603.06582}{{\tt arXiv:1603.06582}}].

\bibitem{Casandjian:2009wq}
{\bf Fermi-LAT} Collaboration, J.-M. Casandjian and I.~Grenier, {\it {High
  Energy Gamma-Ray Emission from the Loop I region}},
  \href{http://arxiv.org/abs/0912.3478}{{\tt arXiv:0912.3478}}.

\bibitem{supp-data}
Supplementary Data for ``Gamma-ray Constraints on Decaying Dark Matter and
  Implications for IceCube".

\bibitem{Aartsen:2016ngq}
{\bf IceCube} Collaboration, M.~G. Aartsen et~al., {\it {Constraints on
  Ultrahigh-Energy Cosmic-Ray Sources from a Search for Neutrinos above 10 PeV
  with IceCube}},  {\em Phys. Rev. Lett.} {\bf 117} (2016), no.~24 241101,
  [\href{http://arxiv.org/abs/1607.05886}{{\tt arXiv:1607.05886}}]. [Erratum:
  Phys. Rev. Lett.119,no.25,259902(2017)].

\bibitem{Aartsen:2014muf}
{\bf IceCube} Collaboration, M.~G. Aartsen et~al., {\it {Atmospheric and
  astrophysical neutrinos above 1 TeV interacting in IceCube}},  {\em Phys.
  Rev.} {\bf D91} (2015), no.~2 022001,
  [\href{http://arxiv.org/abs/1410.1749}{{\tt arXiv:1410.1749}}].

\bibitem{Esmaili:2013gha}
A.~Esmaili and P.~D. Serpico, {\it {Are IceCube neutrinos unveiling PeV-scale
  decaying dark matter?}},  {\em JCAP} {\bf 1311} (2013) 054,
  [\href{http://arxiv.org/abs/1308.1105}{{\tt arXiv:1308.1105}}].

\bibitem{Feldstein:2013kka}
B.~Feldstein, A.~Kusenko, S.~Matsumoto, and T.~T. Yanagida, {\it {Neutrinos at
  IceCube from Heavy Decaying Dark Matter}},  {\em Phys. Rev.} {\bf D88}
  (2013), no.~1 015004, [\href{http://arxiv.org/abs/1303.7320}{{\tt
  arXiv:1303.7320}}].

\bibitem{Ema:2013nda}
Y.~Ema, R.~Jinno, and T.~Moroi, {\it {Cosmic-Ray Neutrinos from the Decay of
  Long-Lived Particle and the Recent IceCube Result}},  {\em Phys. Lett.} {\bf
  B733} (2014) 120--125, [\href{http://arxiv.org/abs/1312.3501}{{\tt
  arXiv:1312.3501}}].

\bibitem{Zavala:2014dla}
J.~Zavala, {\it {Galactic PeV neutrinos from dark matter annihilation}},  {\em
  Phys. Rev.} {\bf D89} (2014), no.~12 123516,
  [\href{http://arxiv.org/abs/1404.2932}{{\tt arXiv:1404.2932}}].

\bibitem{Bhattacharya:2014vwa}
A.~Bhattacharya, M.~H. Reno, and I.~Sarcevic, {\it {Reconciling neutrino flux
  from heavy dark matter decay and recent events at IceCube}},  {\em JHEP} {\bf
  06} (2014) 110, [\href{http://arxiv.org/abs/1403.1862}{{\tt
  arXiv:1403.1862}}].

\bibitem{Higaki:2014dwa}
T.~Higaki, R.~Kitano, and R.~Sato, {\it {Neutrinoful Universe}},  {\em JHEP}
  {\bf 07} (2014) 044, [\href{http://arxiv.org/abs/1405.0013}{{\tt
  arXiv:1405.0013}}].

\bibitem{Rott:2014kfa}
C.~Rott, K.~Kohri, and S.~C. Park, {\it {Superheavy dark matter and IceCube
  neutrino signals: Bounds on decaying dark matter}},  {\em Phys. Rev.} {\bf
  D92} (2015), no.~2 023529, [\href{http://arxiv.org/abs/1408.4575}{{\tt
  arXiv:1408.4575}}].

\bibitem{Fong:2014bsa}
C.~S. Fong, H.~Minakata, B.~Panes, and R.~Zukanovich~Funchal, {\it {Possible
  Interpretations of IceCube High-Energy Neutrino Events}},  {\em JHEP} {\bf
  02} (2015) 189, [\href{http://arxiv.org/abs/1411.5318}{{\tt
  arXiv:1411.5318}}].

\bibitem{Dudas:2014bca}
E.~Dudas, Y.~Mambrini, and K.~A. Olive, {\it {Monochromatic neutrinos generated
  by dark matter and the seesaw mechanism}},  {\em Phys. Rev.} {\bf D91} (2015)
  075001, [\href{http://arxiv.org/abs/1412.3459}{{\tt arXiv:1412.3459}}].

\bibitem{Ema:2014ufa}
Y.~Ema, R.~Jinno, and T.~Moroi, {\it {Cosmological Implications of High-Energy
  Neutrino Emission from the Decay of Long-Lived Particle}},  {\em JHEP} {\bf
  10} (2014) 150, [\href{http://arxiv.org/abs/1408.1745}{{\tt
  arXiv:1408.1745}}].

\bibitem{Esmaili:2014rma}
A.~Esmaili, S.~K. Kang, and P.~D. Serpico, {\it {IceCube events and decaying
  dark matter: hints and constraints}},  {\em JCAP} {\bf 1412} (2014), no.~12
  054, [\href{http://arxiv.org/abs/1410.5979}{{\tt arXiv:1410.5979}}].

\bibitem{Anchordoqui:2015lqa}
L.~A. Anchordoqui, V.~Barger, H.~Goldberg, X.~Huang, D.~Marfatia, L.~H.~M.
  da~Silva, and T.~J. Weiler, {\it {IceCube neutrinos, decaying dark matter,
  and the Hubble constant}},  {\em Phys. Rev.} {\bf D92} (2015), no.~6 061301,
  [\href{http://arxiv.org/abs/1506.08788}{{\tt arXiv:1506.08788}}]. [Erratum:
  Phys. Rev.D94,no.6,069901(2016)].

\bibitem{Boucenna:2015tra}
S.~M. Boucenna, M.~Chianese, G.~Mangano, G.~Miele, S.~Morisi, O.~Pisanti, and
  E.~Vitagliano, {\it {Decaying Leptophilic Dark Matter at IceCube}},  {\em
  JCAP} {\bf 1512} (2015), no.~12 055,
  [\href{http://arxiv.org/abs/1507.01000}{{\tt arXiv:1507.01000}}].

\bibitem{Ko:2015nma}
P.~Ko and Y.~Tang, {\it {IceCube Events from Heavy DM decays through the
  Right-handed Neutrino Portal}},  {\em Phys. Lett.} {\bf B751} (2015) 81--88,
  [\href{http://arxiv.org/abs/1508.02500}{{\tt arXiv:1508.02500}}].

\bibitem{Aisati:2015ova}
C.~El~Aisati, M.~Gustafsson, T.~Hambye, and T.~Scarna, {\it {Dark Matter Decay
  to a Photon and a Neutrino: the Double Monochromatic Smoking Gun Scenario}},
  {\em Phys. Rev.} {\bf D93} (2016), no.~4 043535,
  [\href{http://arxiv.org/abs/1510.05008}{{\tt arXiv:1510.05008}}].

\bibitem{Kistler:2015oae}
M.~D. Kistler, {\it {On TeV Gamma Rays and the Search for Galactic Neutrinos}},
   \href{http://arxiv.org/abs/1511.05199}{{\tt arXiv:1511.05199}}.

\bibitem{Chianese:2016opp}
M.~Chianese, G.~Miele, S.~Morisi, and E.~Vitagliano, {\it {Low energy IceCube
  data and a possible Dark Matter related excess}},  {\em Phys. Lett.} {\bf
  B757} (2016) 251--256, [\href{http://arxiv.org/abs/1601.02934}{{\tt
  arXiv:1601.02934}}].

\bibitem{Fiorentin:2016avj}
M.~Re~Fiorentin, V.~Niro, and N.~Fornengo, {\it {A consistent model for
  leptogenesis, dark matter and the IceCube signal}},  {\em JHEP} {\bf 11}
  (2016) 022, [\href{http://arxiv.org/abs/1606.04445}{{\tt arXiv:1606.04445}}].

\bibitem{Dev:2016qbd}
P.~S.~B. Dev, D.~Kazanas, R.~N. Mohapatra, V.~L. Teplitz, and Y.~Zhang, {\it
  {Heavy right-handed neutrino dark matter and PeV neutrinos at IceCube}},
  {\em JCAP} {\bf 1608} (2016), no.~08 034,
  [\href{http://arxiv.org/abs/1606.04517}{{\tt arXiv:1606.04517}}].

\bibitem{DiBari:2016guw}
P.~Di~Bari, P.~O. Ludl, and S.~Palomares-Ruiz, {\it {Unifying leptogenesis,
  dark matter and high-energy neutrinos with right-handed neutrino mixing via
  Higgs portal}},  {\em JCAP} {\bf 1611} (2016), no.~11 044,
  [\href{http://arxiv.org/abs/1606.06238}{{\tt arXiv:1606.06238}}].

\bibitem{Chianese:2016smc}
M.~Chianese and A.~Merle, {\it {A Consistent Theory of Decaying Dark Matter
  Connecting IceCube to the Sesame Street}},  {\em JCAP} {\bf 1704} (2017),
  no.~04 017, [\href{http://arxiv.org/abs/1607.05283}{{\tt arXiv:1607.05283}}].

\bibitem{Murase:2016gly}
K.~Murase and E.~Waxman, {\it {Constraining High-Energy Cosmic Neutrino
  Sources: Implications and Prospects}},  {\em Phys. Rev.} {\bf D94} (2016),
  no.~10 103006, [\href{http://arxiv.org/abs/1607.01601}{{\tt
  arXiv:1607.01601}}].

\bibitem{Murase:2015xka}
K.~Murase, D.~Guetta, and M.~Ahlers, {\it {Hidden Cosmic-Ray Accelerators as an
  Origin of TeV-PeV Cosmic Neutrinos}},  {\em Phys. Rev. Lett.} {\bf 116}
  (2016), no.~7 071101, [\href{http://arxiv.org/abs/1509.00805}{{\tt
  arXiv:1509.00805}}].

\bibitem{Palladino:2016xsy}
A.~Palladino, M.~Spurio, and F.~Vissani, {\it {On the IceCube spectral
  anomaly}},  {\em JCAP} {\bf 1612} (2016), no.~12 045,
  [\href{http://arxiv.org/abs/1610.07015}{{\tt arXiv:1610.07015}}].

\bibitem{Faraggi:2000pv}
A.~E. Faraggi and M.~Pospelov, {\it {Selfinteracting dark matter from the
  hidden heterotic string sector}},  {\em Astropart. Phys.} {\bf 16} (2002)
  451--461, [\href{http://arxiv.org/abs/hep-ph/0008223}{{\tt hep-ph/0008223}}].

\bibitem{Halverson:2016nfq}
J.~Halverson, B.~D. Nelson, and F.~Ruehle, {\it {String Theory and the Dark
  Glueball Problem}},  {\em Phys. Rev.} {\bf D95} (2017), no.~4 043527,
  [\href{http://arxiv.org/abs/1609.02151}{{\tt arXiv:1609.02151}}].

\bibitem{Forestell:2016qhc}
L.~Forestell, D.~E. Morrissey, and K.~Sigurdson, {\it {Non-Abelian Dark Forces
  and the Relic Densities of Dark Glueballs}},  {\em Phys. Rev.} {\bf D95}
  (2017), no.~1 015032, [\href{http://arxiv.org/abs/1605.08048}{{\tt
  arXiv:1605.08048}}].

\bibitem{Boddy:2014yra}
K.~K. Boddy, J.~L. Feng, M.~Kaplinghat, and T.~M.~P. Tait, {\it
  {Self-Interacting Dark Matter from a Non-Abelian Hidden Sector}},  {\em Phys.
  Rev.} {\bf D89} (2014), no.~11 115017,
  [\href{http://arxiv.org/abs/1402.3629}{{\tt arXiv:1402.3629}}].

\bibitem{Ibarra:2013cra}
A.~Ibarra, D.~Tran, and C.~Weniger, {\it {Indirect Searches for Decaying Dark
  Matter}},  {\em Int. J. Mod. Phys.} {\bf A28} (2013) 1330040,
  [\href{http://arxiv.org/abs/1307.6434}{{\tt arXiv:1307.6434}}].

\bibitem{Abdo:2010nc}
A.~A. Abdo et~al., {\it {Fermi LAT Search for Photon Lines from 30 to 200 GeV
  and Dark Matter Implications}},  {\em Phys. Rev. Lett.} {\bf 104} (2010)
  091302, [\href{http://arxiv.org/abs/1001.4836}{{\tt arXiv:1001.4836}}].

\bibitem{Vertongen:2011mu}
G.~Vertongen and C.~Weniger, {\it {Hunting Dark Matter Gamma-Ray Lines with the
  Fermi LAT}},  {\em JCAP} {\bf 1105} (2011) 027,
  [\href{http://arxiv.org/abs/1101.2610}{{\tt arXiv:1101.2610}}].

\bibitem{Ackermann:2012qk}
{\bf Fermi-LAT} Collaboration, M.~Ackermann et~al., {\it {Fermi LAT Search for
  Dark Matter in Gamma-ray Lines and the Inclusive Photon Spectrum}},  {\em
  Phys. Rev.} {\bf D86} (2012) 022002,
  [\href{http://arxiv.org/abs/1205.2739}{{\tt arXiv:1205.2739}}].

\bibitem{Esmaili:2012us}
A.~Esmaili, A.~Ibarra, and O.~L.~G. Peres, {\it {Probing the stability of
  superheavy dark matter particles with high-energy neutrinos}},  {\em JCAP}
  {\bf 1211} (2012) 034, [\href{http://arxiv.org/abs/1205.5281}{{\tt
  arXiv:1205.5281}}].

\bibitem{Dugger:2010ys}
L.~Dugger, T.~E. Jeltema, and S.~Profumo, {\it {Constraints on Decaying Dark
  Matter from Fermi Observations of Nearby Galaxies and Clusters}},  {\em JCAP}
  {\bf 1012} (2010) 015, [\href{http://arxiv.org/abs/1009.5988}{{\tt
  arXiv:1009.5988}}].

\bibitem{Cirelli:2009dv}
M.~Cirelli, P.~Panci, and P.~D. Serpico, {\it {Diffuse gamma ray constraints on
  annihilating or decaying Dark Matter after Fermi}},  {\em Nucl. Phys.} {\bf
  B840} (2010) 284--303, [\href{http://arxiv.org/abs/0912.0663}{{\tt
  arXiv:0912.0663}}].

\bibitem{Zhang:2009ut}
L.~Zhang, C.~Weniger, L.~Maccione, J.~Redondo, and G.~Sigl, {\it {Constraining
  Decaying Dark Matter with Fermi LAT Gamma-rays}},  {\em JCAP} {\bf 1006}
  (2010) 027, [\href{http://arxiv.org/abs/0912.4504}{{\tt arXiv:0912.4504}}].

\bibitem{Zaharijas:2010ca}
{\bf Fermi-LAT} Collaboration, G.~Zaharijas, A.~Cuoco, Z.~Yang, and J.~Conrad,
  {\it {Constraints on the Galactic Halo Dark Matter from Fermi-LAT Diffuse
  Measurements}},  {\em PoS} {\bf IDM2010} (2011) 111,
  [\href{http://arxiv.org/abs/1012.0588}{{\tt arXiv:1012.0588}}].

\bibitem{Zaharijas:2012dr}
{\bf Fermi-LAT} Collaboration, G.~Zaharijas, J.~Conrad, A.~Cuoco, and Z.~Yang,
  {\it {Fermi-LAT measurement of the diffuse gamma-ray emission and constraints
  on the Galactic Dark Matter signal}},  {\em Nucl. Phys. Proc. Suppl.} {\bf
  239-240} (2013) 88--93, [\href{http://arxiv.org/abs/1212.6755}{{\tt
  arXiv:1212.6755}}].

\bibitem{Aliu:2012ga}
{\bf VERITAS} Collaboration, E.~Aliu et~al., {\it {VERITAS Deep Observations of
  the Dwarf Spheroidal Galaxy Segue 1}},  {\em Phys. Rev.} {\bf D85} (2012)
  062001, [\href{http://arxiv.org/abs/1202.2144}{{\tt arXiv:1202.2144}}].
  [Erratum: Phys. Rev.D91,no.12,129903(2015)].

\bibitem{Baring:2015sza}
M.~G. Baring, T.~Ghosh, F.~S. Queiroz, and K.~Sinha, {\it {New Limits on the
  Dark Matter Lifetime from Dwarf Spheroidal Galaxies using Fermi-LAT}},  {\em
  Phys. Rev.} {\bf D93} (2016), no.~10 103009,
  [\href{http://arxiv.org/abs/1510.00389}{{\tt arXiv:1510.00389}}].

\bibitem{Consortium:2010bc}
{\bf CTA Consortium} Collaboration, M.~Actis et~al., {\it {Design concepts for
  the Cherenkov Telescope Array CTA: An advanced facility for ground-based
  high-energy gamma-ray astronomy}},  {\em Exper. Astron.} {\bf 32} (2011)
  193--316, [\href{http://arxiv.org/abs/1008.3703}{{\tt arXiv:1008.3703}}].

\bibitem{Pierre:2014tra}
M.~Pierre, J.~M. Siegal-Gaskins, and P.~Scott, {\it {Sensitivity of CTA to dark
  matter signals from the Galactic Center}},  {\em JCAP} {\bf 1406} (2014) 024,
  [\href{http://arxiv.org/abs/1401.7330}{{\tt arXiv:1401.7330}}]. [Erratum:
  JCAP1410,E01(2014)].

\bibitem{Abeysekara:2013tza}
A.~U. Abeysekara et~al., {\it {Sensitivity of the High Altitude Water Cherenkov
  Detector to Sources of Multi-TeV Gamma Rays}},  {\em Astropart. Phys.} {\bf
  50-52} (2013) 26--32, [\href{http://arxiv.org/abs/1306.5800}{{\tt
  arXiv:1306.5800}}].

\bibitem{Sako:2009xa}
T.~K. Sako, K.~Kawata, M.~Ohnishi, A.~Shiomi, M.~Takita, and H.~Tsuchiya, {\it
  {Exploration of a 100 TeV gamma-ray northern sky using the Tibet air-shower
  array combined with an underground water-Cherenkov muon-detector array}},
  {\em Astropart. Phys.} {\bf 32} (2009) 177--184,
  [\href{http://arxiv.org/abs/0907.4589}{{\tt arXiv:0907.4589}}].

\bibitem{Ahlers:2013xia}
M.~Ahlers and K.~Murase, {\it {Probing the Galactic Origin of the IceCube
  Excess with Gamma-Rays}},  {\em Phys. Rev.} {\bf D90} (2014), no.~2 023010,
  [\href{http://arxiv.org/abs/1309.4077}{{\tt arXiv:1309.4077}}].

\bibitem{Cheng:2016slx}
H.-C. Cheng, W.-C. Huang, X.~Huang, I.~Low, Y.-L.~S. Tsai, and Q.~Yuan, {\it
  {AMS-02 Positron Excess and Indirect Detection of Three-body Decaying Dark
  Matter}},  {\em JCAP} {\bf 1703} (2017), no.~03 041,
  [\href{http://arxiv.org/abs/1608.06382}{{\tt arXiv:1608.06382}}].

\bibitem{Ting:2238506}
S.~Ting, {\it {The First Five Years of the Alpha Magnetic Spectrometer on the
  International Space Station. The First Five Years of the Alpha Magnetic
  Spectrometer on the International Space Station}}, .

\bibitem{Petrovic:2014uda}
J.~Petrović, P.~D. Serpico, and G.~Zaharijaš, {\it {Galactic Center gamma-ray
  "excess" from an active past of the Galactic Centre?}},  {\em JCAP} {\bf
  1410} (2014), no.~10 052, [\href{http://arxiv.org/abs/1405.7928}{{\tt
  arXiv:1405.7928}}].

\bibitem{Carlson:2014cwa}
E.~Carlson and S.~Profumo, {\it {Cosmic Ray Protons in the Inner Galaxy and the
  Galactic Center Gamma-Ray Excess}},  {\em Phys. Rev.} {\bf D90} (2014), no.~2
  023015, [\href{http://arxiv.org/abs/1405.7685}{{\tt arXiv:1405.7685}}].

\bibitem{LINDENTALK}
T.~Linden, {\it {Talk given at COSMO-14, August 25-29}},  {\em {talk given at
  COSMO-14, August 25-29}} (2014).

\bibitem{Gordon:2014gya}
C.~Gordon and O.~Macias, {\it {Can Cosmic Rays Interacting With Molecular
  Clouds Explain The Galactic Center Gamma-Ray Excess?}},  {\em PoS} {\bf
  CRISM2014} (2015) 042, [\href{http://arxiv.org/abs/1410.7840}{{\tt
  arXiv:1410.7840}}].

\bibitem{Cholis:2014lta}
I.~Cholis, D.~Hooper, and T.~Linden, {\it {Challenges in Explaining the
  Galactic Center Gamma-Ray Excess with Millisecond Pulsars}},  {\em JCAP} {\bf
  1506} (2015), no.~06 043, [\href{http://arxiv.org/abs/1407.5625}{{\tt
  arXiv:1407.5625}}].

\bibitem{Alves:2014yha}
A.~Alves, S.~Profumo, F.~S. Queiroz, and W.~Shepherd, {\it {Effective field
  theory approach to the Galactic Center gamma-ray excess}},  {\em Phys. Rev.}
  {\bf D90} (2014), no.~11 115003, [\href{http://arxiv.org/abs/1403.5027}{{\tt
  arXiv:1403.5027}}].

\bibitem{Berlin:2014tja}
A.~Berlin, D.~Hooper, and S.~D. McDermott, {\it {Simplified Dark Matter Models
  for the Galactic Center Gamma-Ray Excess}},  {\em Phys. Rev.} {\bf D89}
  (2014), no.~11 115022, [\href{http://arxiv.org/abs/1404.0022}{{\tt
  arXiv:1404.0022}}].

\bibitem{Ipek:2014gua}
S.~Ipek, D.~McKeen, and A.~E. Nelson, {\it {A Renormalizable Model for the
  Galactic Center Gamma Ray Excess from Dark Matter Annihilation}},  {\em Phys.
  Rev.} {\bf D90} (2014), no.~5 055021,
  [\href{http://arxiv.org/abs/1404.3716}{{\tt arXiv:1404.3716}}].

\bibitem{Cheung:2014lqa}
C.~Cheung, M.~Papucci, D.~Sanford, N.~R. Shah, and K.~M. Zurek, {\it {NMSSM
  Interpretation of the Galactic Center Excess}},  {\em Phys. Rev.} {\bf D90}
  (2014), no.~7 075011, [\href{http://arxiv.org/abs/1406.6372}{{\tt
  arXiv:1406.6372}}].

\bibitem{Gherghetta:2015ysa}
T.~Gherghetta, B.~von Harling, A.~D. Medina, M.~A. Schmidt, and T.~Trott, {\it
  {SUSY implications from WIMP annihilation into scalars at the Galactic
  Center}},  {\em Phys. Rev.} {\bf D91} (2015) 105004,
  [\href{http://arxiv.org/abs/1502.07173}{{\tt arXiv:1502.07173}}].

\bibitem{Pospelov:2007mp}
M.~Pospelov, A.~Ritz, and M.~B. Voloshin, {\it {Secluded WIMP Dark Matter}},
  {\em Phys. Lett.} {\bf B662} (2008) 53--61,
  [\href{http://arxiv.org/abs/0711.4866}{{\tt arXiv:0711.4866}}].

\bibitem{Martin:2014sxa}
A.~Martin, J.~Shelton, and J.~Unwin, {\it {Fitting the Galactic Center
  Gamma-Ray Excess with Cascade Annihilations}},  {\em Phys. Rev.} {\bf D90}
  (2014), no.~10 103513, [\href{http://arxiv.org/abs/1405.0272}{{\tt
  arXiv:1405.0272}}].

\bibitem{Abdullah:2014lla}
M.~Abdullah, A.~DiFranzo, A.~Rajaraman, T.~M.~P. Tait, P.~Tanedo, and A.~M.
  Wijangco, {\it {Hidden on-shell mediators for the Galactic Center
  $\gamma$-ray excess}},  {\em Phys. Rev.} {\bf D90} (2014) 035004,
  [\href{http://arxiv.org/abs/1404.6528}{{\tt arXiv:1404.6528}}].

\bibitem{Ko:2014gha}
P.~Ko, W.-I. Park, and Y.~Tang, {\it {Higgs portal vector dark matter for
  $\mathinner{\mathrm{GeV}}$ scale $\gamma$-ray excess from galactic center}},
  {\em JCAP} {\bf 1409} (2014) 013, [\href{http://arxiv.org/abs/1404.5257}{{\tt
  arXiv:1404.5257}}].

\bibitem{Freytsis:2014sua}
M.~Freytsis, D.~J. Robinson, and Y.~Tsai, {\it {Galactic Center Gamma-Ray
  Excess through a Dark Shower}},  {\em Phys. Rev.} {\bf D91} (2015), no.~3
  035028, [\href{http://arxiv.org/abs/1410.3818}{{\tt arXiv:1410.3818}}].

\bibitem{Rothstein:2009pm}
I.~Z. Rothstein, T.~Schwetz, and J.~Zupan, {\it {Phenomenology of Dark Matter
  annihilation into a long-lived intermediate state}},  {\em JCAP} {\bf 0907}
  (2009) 018, [\href{http://arxiv.org/abs/0903.3116}{{\tt arXiv:0903.3116}}].

\bibitem{Baumgart:2009tn}
M.~Baumgart, C.~Cheung, J.~T. Ruderman, L.-T. Wang, and I.~Yavin, {\it
  {Non-Abelian Dark Sectors and Their Collider Signatures}},  {\em JHEP} {\bf
  04} (2009) 014, [\href{http://arxiv.org/abs/0901.0283}{{\tt
  arXiv:0901.0283}}].

\bibitem{Nomura:2008ru}
Y.~Nomura and J.~Thaler, {\it {Dark Matter through the Axion Portal}},  {\em
  Phys. Rev.} {\bf D79} (2009) 075008,
  [\href{http://arxiv.org/abs/0810.5397}{{\tt arXiv:0810.5397}}].

\bibitem{Mardon:2009rc}
J.~Mardon, Y.~Nomura, D.~Stolarski, and J.~Thaler, {\it {Dark Matter Signals
  from Cascade Annihilations}},  {\em JCAP} {\bf 0905} (2009) 016,
  [\href{http://arxiv.org/abs/0901.2926}{{\tt arXiv:0901.2926}}].

\bibitem{Bergstrom:2013jra}
L.~Bergstrom, T.~Bringmann, I.~Cholis, D.~Hooper, and C.~Weniger, {\it {New
  limits on dark matter annihilation from AMS cosmic ray positron data}},  {\em
  Phys. Rev. Lett.} {\bf 111} (2013) 171101,
  [\href{http://arxiv.org/abs/1306.3983}{{\tt arXiv:1306.3983}}].

\bibitem{Chen:2003gz}
X.-L. Chen and M.~Kamionkowski, {\it {Particle decays during the cosmic dark
  ages}},  {\em Phys. Rev.} {\bf D70} (2004) 043502,
  [\href{http://arxiv.org/abs/astro-ph/0310473}{{\tt astro-ph/0310473}}].

\bibitem{Padmanabhan:2005es}
N.~Padmanabhan and D.~P. Finkbeiner, {\it {Detecting dark matter annihilation
  with CMB polarization: Signatures and experimental prospects}},  {\em Phys.
  Rev.} {\bf D72} (2005) 023508,
  [\href{http://arxiv.org/abs/astro-ph/0503486}{{\tt astro-ph/0503486}}].

\bibitem{Slatyer:2009yq}
T.~R. Slatyer, N.~Padmanabhan, and D.~P. Finkbeiner, {\it {CMB Constraints on
  WIMP Annihilation: Energy Absorption During the Recombination Epoch}},  {\em
  Phys. Rev.} {\bf D80} (2009) 043526,
  [\href{http://arxiv.org/abs/0906.1197}{{\tt arXiv:0906.1197}}].

\bibitem{Madhavacheril:2013cna}
M.~S. Madhavacheril, N.~Sehgal, and T.~R. Slatyer, {\it {Current Dark Matter
  Annihilation Constraints from CMB and Low-Redshift Data}},  {\em Phys. Rev.}
  {\bf D89} (2014) 103508, [\href{http://arxiv.org/abs/1310.3815}{{\tt
  arXiv:1310.3815}}].

\bibitem{Fan:2012gr}
J.~Fan and M.~Reece, {\it {Simple dark matter recipe for the 111 and 128 GeV
  Fermi-LAT lines}},  {\em Phys. Rev.} {\bf D88} (2013), no.~3 035014,
  [\href{http://arxiv.org/abs/1209.1097}{{\tt arXiv:1209.1097}}].

\bibitem{Gao:2010qx}
Y.~Gao, A.~V. Gritsan, Z.~Guo, K.~Melnikov, M.~Schulze, and N.~V. Tran, {\it
  {Spin determination of single-produced resonances at hadron colliders}},
  {\em Phys. Rev.} {\bf D81} (2010) 075022,
  [\href{http://arxiv.org/abs/1001.3396}{{\tt arXiv:1001.3396}}].

\bibitem{ArkaniHamed:2008qp}
N.~Arkani-Hamed and N.~Weiner, {\it {LHC Signals for a SuperUnified Theory of
  Dark Matter}},  {\em JHEP} {\bf 12} (2008) 104,
  [\href{http://arxiv.org/abs/0810.0714}{{\tt arXiv:0810.0714}}].

\bibitem{Cheung:2009su}
C.~Cheung, J.~T. Ruderman, L.-T. Wang, and I.~Yavin, {\it {Lepton Jets in
  (Supersymmetric) Electroweak Processes}},  {\em JHEP} {\bf 04} (2010) 116,
  [\href{http://arxiv.org/abs/0909.0290}{{\tt arXiv:0909.0290}}].

\bibitem{Patt:2006fw}
B.~Patt and F.~Wilczek, {\it {Higgs-field portal into hidden sectors}},
  \href{http://arxiv.org/abs/hep-ph/0605188}{{\tt hep-ph/0605188}}.

\bibitem{MarchRussell:2008yu}
J.~March-Russell, S.~M. West, D.~Cumberbatch, and D.~Hooper, {\it {Heavy Dark
  Matter Through the Higgs Portal}},  {\em JHEP} {\bf 07} (2008) 058,
  [\href{http://arxiv.org/abs/0801.3440}{{\tt arXiv:0801.3440}}].

\bibitem{Dienes:1996zr}
K.~R. Dienes, C.~F. Kolda, and J.~March-Russell, {\it {Kinetic mixing and the
  supersymmetric gauge hierarchy}},  {\em Nucl. Phys.} {\bf B492} (1997)
  104--118, [\href{http://arxiv.org/abs/hep-ph/9610479}{{\tt hep-ph/9610479}}].

\bibitem{Falkowski:2009yz}
A.~Falkowski, J.~Juknevich, and J.~Shelton, {\it {Dark Matter Through the
  Neutrino Portal}},  \href{http://arxiv.org/abs/0908.1790}{{\tt
  arXiv:0908.1790}}.

\bibitem{Essig:2013lka}
{R.~Essig et al.}, {\it {Working Group Report: New Light Weakly Coupled
  Particles}},  \href{http://arxiv.org/abs/1311.0029}{{\tt arXiv:1311.0029}}.

\bibitem{Holdom:1985ag}
B.~Holdom, {\it {Two U(1)'s and Epsilon Charge Shifts}},  {\em Phys. Lett.}
  {\bf B166} (1986) 196.

\bibitem{Strassler:2006im}
M.~J. Strassler and K.~M. Zurek, {\it {Echoes of a hidden valley at hadron
  colliders}},  {\em Phys. Lett.} {\bf B651} (2007) 374--379,
  [\href{http://arxiv.org/abs/hep-ph/0604261}{{\tt hep-ph/0604261}}].

\bibitem{Han:2007ae}
T.~Han, Z.~Si, K.~M. Zurek, and M.~J. Strassler, {\it {Phenomenology of hidden
  valleys at hadron colliders}},  {\em JHEP} {\bf 07} (2008) 008,
  [\href{http://arxiv.org/abs/0712.2041}{{\tt arXiv:0712.2041}}].

\bibitem{Hooper:2012cw}
D.~Hooper, N.~Weiner, and W.~Xue, {\it {Dark Forces and Light Dark Matter}},
  {\em Phys. Rev.} {\bf D86} (2012) 056009,
  [\href{http://arxiv.org/abs/1206.2929}{{\tt arXiv:1206.2929}}].

\bibitem{Berlin:2014pya}
A.~Berlin, P.~Gratia, D.~Hooper, and S.~D. McDermott, {\it {Hidden Sector Dark
  Matter Models for the Galactic Center Gamma-Ray Excess}},  {\em Phys. Rev.}
  {\bf D90} (2014), no.~1 015032, [\href{http://arxiv.org/abs/1405.5204}{{\tt
  arXiv:1405.5204}}].

\bibitem{Cline:2014dwa}
J.~M. Cline, G.~Dupuis, Z.~Liu, and W.~Xue, {\it {The windows for kinetically
  mixed Z'-mediated dark matter and the galactic center gamma ray excess}},
  {\em JHEP} {\bf 08} (2014) 131, [\href{http://arxiv.org/abs/1405.7691}{{\tt
  arXiv:1405.7691}}].

\bibitem{Liu:2014cma}
J.~Liu, N.~Weiner, and W.~Xue, {\it {Signals of a Light Dark Force in the
  Galactic Center}},  {\em JHEP} {\bf 08} (2015) 050,
  [\href{http://arxiv.org/abs/1412.1485}{{\tt arXiv:1412.1485}}].

\bibitem{Cline:2015qha}
J.~M. Cline, G.~Dupuis, Z.~Liu, and W.~Xue, {\it {Multimediator models for the
  galactic center gamma ray excess}},  {\em Phys. Rev.} {\bf D91} (2015),
  no.~11 115010, [\href{http://arxiv.org/abs/1503.08213}{{\tt
  arXiv:1503.08213}}].

\bibitem{Aguilar:2014mma}
{\bf AMS} Collaboration, M.~Aguilar et~al., {\it {Electron and Positron Fluxes
  in Primary Cosmic Rays Measured with the Alpha Magnetic Spectrometer on the
  International Space Station}},  {\em Phys. Rev. Lett.} {\bf 113} (2014)
  121102.

\bibitem{Accardo:2014lma}
{\bf AMS} Collaboration, L.~Accardo et~al., {\it {High Statistics Measurement
  of the Positron Fraction in Primary Cosmic Rays of 0.5-500 GeV with the Alpha
  Magnetic Spectrometer on the International Space Station}},  {\em Phys. Rev.
  Lett.} {\bf 113} (2014) 121101.

\bibitem{Cirelli:2008pk}
M.~Cirelli, M.~Kadastik, M.~Raidal, and A.~Strumia, {\it {Model-Independent
  Implications of the E+-, Anti-Proton Cosmic Ray Spectra on Properties of Dark
  Matter}},  {\em Nucl. Phys.} {\bf B813} (2009) 1--21,
  [\href{http://arxiv.org/abs/0809.2409}{{\tt arXiv:0809.2409}}]. [Addendum:
  Nucl. Phys.B873,530(2013)].

\bibitem{Mardon:2009gw}
J.~Mardon, Y.~Nomura, and J.~Thaler, {\it {Cosmic Signals from the Hidden
  Sector}},  {\em Phys. Rev.} {\bf D80} (2009) 035013,
  [\href{http://arxiv.org/abs/0905.3749}{{\tt arXiv:0905.3749}}].

\bibitem{Abramowski:2014tra}
{\bf HESS} Collaboration, A.~Abramowski et~al., {\it {Search for dark matter
  annihilation signatures in H.E.S.S. observations of Dwarf Spheroidal
  Galaxies}},  {\em Phys. Rev.} {\bf D90} (2014) 112012,
  [\href{http://arxiv.org/abs/1410.2589}{{\tt arXiv:1410.2589}}].

\bibitem{Zitzer:2015eqa}
{\bf VERITAS} Collaboration, B.~Zitzer, {\it {Search for Dark Matter from Dwarf
  Galaxies using VERITAS}},  in {\em {Proceedings, 34th International Cosmic
  Ray Conference (ICRC 2015)}}, 2015.
\newblock \href{http://arxiv.org/abs/1509.01105}{{\tt arXiv:1509.01105}}.

\bibitem{Hooper:2012sr}
D.~Hooper, C.~Kelso, and F.~S. Queiroz, {\it {Stringent and Robust Constraints
  on the Dark Matter Annihilation Cross Section From the Region of the Galactic
  Center}},  {\em Astropart. Phys.} {\bf 46} (2013) 55--70,
  [\href{http://arxiv.org/abs/1209.3015}{{\tt arXiv:1209.3015}}].

\bibitem{Storm:2012ty}
E.~Storm, T.~E. Jeltema, S.~Profumo, and L.~Rudnick, {\it {Constraints on Dark
  Matter Annihilation in Clusters of Galaxies from Diffuse Radio Emission}},
  {\em Astrophys. J.} {\bf 768} (2013) 106,
  [\href{http://arxiv.org/abs/1210.0872}{{\tt arXiv:1210.0872}}].

\bibitem{vonHarling:2012sz}
B.~von Harling and K.~L. McDonald, {\it {Secluded Dark Matter Coupled to a
  Hidden CFT}},  {\em JHEP} {\bf 08} (2012) 048,
  [\href{http://arxiv.org/abs/1203.6646}{{\tt arXiv:1203.6646}}].

\bibitem{Kuno:1999jp}
Y.~Kuno and Y.~Okada, {\it {Muon decay and physics beyond the standard model}},
   {\em Rev. Mod. Phys.} {\bf 73} (2001) 151--202,
  [\href{http://arxiv.org/abs/hep-ph/9909265}{{\tt hep-ph/9909265}}].

\bibitem{Michel:1949qe}
L.~Michel, {\it {Interaction Between Four Half Spin Particles and the Decay of
  the $\mu$ Meson}},  {\em Proc. Phys. Soc.} {\bf A63} (1950) 514--531.

\bibitem{Finkbeiner:2011dx}
D.~P. Finkbeiner, S.~Galli, T.~Lin, and T.~R. Slatyer, {\it {Searching for Dark
  Matter in the CMB: A Compact Parameterization of Energy Injection from New
  Physics}},  {\em Phys. Rev.} {\bf D85} (2012) 043522,
  [\href{http://arxiv.org/abs/1109.6322}{{\tt arXiv:1109.6322}}].

\bibitem{Galli:2011rz}
S.~Galli, F.~Iocco, G.~Bertone, and A.~Melchiorri, {\it {Updated CMB
  constraints on Dark Matter annihilation cross-sections}},  {\em Phys. Rev.}
  {\bf D84} (2011) 027302, [\href{http://arxiv.org/abs/1106.1528}{{\tt
  arXiv:1106.1528}}].

\bibitem{Hutsi:2011vx}
G.~Hutsi, J.~Chluba, A.~Hektor, and M.~Raidal, {\it {WMAP7 and future CMB
  constraints on annihilating dark matter: implications on GeV-scale WIMPs}},
  {\em Astron. Astrophys.} {\bf 535} (2011) A26,
  [\href{http://arxiv.org/abs/1103.2766}{{\tt arXiv:1103.2766}}].

\bibitem{Giesen:2012rp}
G.~Giesen, J.~Lesgourgues, B.~Audren, and Y.~Ali-Haimoud, {\it {CMB photons
  shedding light on dark matter}},  {\em JCAP} {\bf 1212} (2012) 008,
  [\href{http://arxiv.org/abs/1209.0247}{{\tt arXiv:1209.0247}}].

\bibitem{Weniger:2013hja}
C.~Weniger, P.~D. Serpico, F.~Iocco, and G.~Bertone, {\it {CMB bounds on dark
  matter annihilation: Nucleon energy-losses after recombination}},  {\em
  Phys.Rev.} {\bf D87} (2013), no.~12 123008,
  [\href{http://arxiv.org/abs/1303.0942}{{\tt arXiv:1303.0942}}].

\bibitem{Bechtol:2015cbp}
{\bf DES} Collaboration, K.~Bechtol et~al., {\it {Eight New Milky Way
  Companions Discovered in First-Year Dark Energy Survey Data}},  {\em
  Astrophys. J.} {\bf 807} (2015), no.~1 50,
  [\href{http://arxiv.org/abs/1503.02584}{{\tt arXiv:1503.02584}}].

\bibitem{Koposov:2015cua}
S.~E. Koposov, V.~Belokurov, G.~Torrealba, and N.~W. Evans, {\it {Beasts of the
  Southern Wild: Discovery of Nine Ultra Faint Satellites in the Vicinity of
  the Magellanic Clouds}},  {\em Astrophys. J.} {\bf 805} (2015), no.~2 130,
  [\href{http://arxiv.org/abs/1503.02079}{{\tt arXiv:1503.02079}}].

\bibitem{Drlica-Wagner:2015xua}
{\bf DES, Fermi-LAT} Collaboration, A.~Drlica-Wagner et~al., {\it {Search for
  Gamma-Ray Emission from Des Dwarf Spheroidal Galaxy Candidates with Fermi-Lat
  Data}},  {\em Submitted to: Astrophys. J.} (2015)
  [\href{http://arxiv.org/abs/1503.02632}{{\tt arXiv:1503.02632}}].

\bibitem{Geringer-Sameth:2015lua}
A.~Geringer-Sameth, M.~G. Walker, S.~M. Koushiappas, S.~E. Koposov,
  V.~Belokurov, G.~Torrealba, and N.~W. Evans, {\it {Indication of Gamma-Ray
  Emission from the Newly Discovered Dwarf Galaxy Reticulum II}},
  \href{http://arxiv.org/abs/1503.02320}{{\tt arXiv:1503.02320}}.

\bibitem{Hooper:2015ula}
D.~Hooper and T.~Linden, {\it {On the Gamma-Ray Emission from Reticulum II and
  Other Dwarf Galaxies}},  \href{http://arxiv.org/abs/1503.06209}{{\tt
  arXiv:1503.06209}}.

\bibitem{Blasi:2009hv}
P.~Blasi, {\it {The Origin of the Positron Excess in Cosmic Rays}},  {\em Phys.
  Rev. Lett.} {\bf 103} (2009) 051104,
  [\href{http://arxiv.org/abs/0903.2794}{{\tt arXiv:0903.2794}}].

\bibitem{Gleeson:1968zza}
L.~J. Gleeson and W.~I. Axford, {\it {Solar Modulation of Galactic Cosmic
  Rays}},  {\em Astrophys. J.} {\bf 154} (1968) 1011.

\bibitem{Moskalenko:2001ya}
I.~V. Moskalenko, A.~W. Strong, J.~F. Ormes, and M.~S. Potgieter, {\it
  {Secondary Anti-Protons and Propagation of Cosmic Rays in the Galaxy and
  Heliosphere}},  {\em Astrophys. J.} {\bf 565} (2002) 280--296,
  [\href{http://arxiv.org/abs/astro-ph/0106567}{{\tt astro-ph/0106567}}].

\bibitem{Beischer:2009zz}
B.~Beischer, P.~von Doetinchem, H.~Gast, T.~Kirn, and S.~Schael, {\it
  {Perspectives for indirect dark matter search with AMS-2 using cosmic-ray
  electrons and positrons}},  {\em New J. Phys.} {\bf 11} (2009) 105021.

\bibitem{Hooper:2012gq}
D.~Hooper and W.~Xue, {\it {Possibility of Testing the Light Dark Matter
  Hypothesis with the Alpha Magnetic Spectrometer}},  {\em Phys. Rev. Lett.}
  {\bf 110} (2013), no.~4 041302, [\href{http://arxiv.org/abs/1210.1220}{{\tt
  arXiv:1210.1220}}].

\bibitem{Evoli:2008dv}
C.~Evoli, D.~Gaggero, D.~Grasso, and L.~Maccione, {\it {Cosmic-Ray Nuclei,
  Antiprotons and Gamma-Rays in the Galaxy: a New Diffusion Model}},  {\em
  JCAP} {\bf 0810} (2008) 018, [\href{http://arxiv.org/abs/0807.4730}{{\tt
  arXiv:0807.4730}}].

\bibitem{2011ascl.soft06011M}
L.~{Maccione}, C.~{Evoli}, D.~{Gaggero}, and D.~{Grasso}, ``{DRAGON: Galactic
  Cosmic Ray Diffusion Code}.'' Astrophysics Source Code Library, June, 2011.

\bibitem{Evoli:2011id}
C.~Evoli, I.~Cholis, D.~Grasso, L.~Maccione, and P.~Ullio, {\it {Antiprotons
  from Dark Matter Annihilation in the Galaxy: Astrophysical Uncertainties}},
  {\em Phys. Rev.} {\bf D85} (2012) 123511,
  [\href{http://arxiv.org/abs/1108.0664}{{\tt arXiv:1108.0664}}].

\bibitem{Pshirkov:2011um}
M.~S. Pshirkov, P.~G. Tinyakov, P.~P. Kronberg, and K.~J. Newton-McGee, {\it
  {Deriving Global Structure of the Galactic Magnetic Field from Faraday
  Rotation Measures of Extragalactic Sources}},  {\em Astrophys. J.} {\bf 738}
  (2011) 192, [\href{http://arxiv.org/abs/1103.0814}{{\tt arXiv:1103.0814}}].

\bibitem{DiBernardo:2012zu}
G.~Di~Bernardo, C.~Evoli, D.~Gaggero, D.~Grasso, and L.~Maccione, {\it {Cosmic
  Ray Electrons, Positrons and the Synchrotron Emission of the Galaxy:
  Consistent Analysis and Implications}},  {\em JCAP} {\bf 1303} (2013) 036,
  [\href{http://arxiv.org/abs/1210.4546}{{\tt arXiv:1210.4546}}].

\bibitem{Jaffe:2009hh}
T.~R. Jaffe, J.~P. Leahy, A.~J. Banday, S.~M. Leach, S.~R. Lowe, and
  A.~Wilkinson, {\it {Modelling the Galactic Magnetic Field on the Plane in
  2D}},  {\em Mon. Not. Roy. Astron. Soc.} {\bf 401} (2010) 1013,
  [\href{http://arxiv.org/abs/0907.3994}{{\tt arXiv:0907.3994}}].

\bibitem{Hinton:2004eu}
{\bf HESS} Collaboration, J.~A. Hinton, {\it {The Status of the H.E.S.S.
  project}},  {\em New Astron. Rev.} {\bf 48} (2004) 331--337,
  [\href{http://arxiv.org/abs/astro-ph/0403052}{{\tt astro-ph/0403052}}].

\bibitem{Abramowski:2013ax}
{\bf HESS} Collaboration, A.~Abramowski et~al., {\it {Search for
  Photon-Linelike Signatures from Dark Matter Annihilations with H.E.S.S.}},
  {\em Phys. Rev. Lett.} {\bf 110} (2013) 041301,
  [\href{http://arxiv.org/abs/1301.1173}{{\tt arXiv:1301.1173}}].

\bibitem{Sinnis:2004je}
G.~Sinnis, A.~Smith, and J.~E. McEnery, {\it {HAWC: A Next generation all - sky
  VHE gamma-ray telescope}},  in {\em {On recent developments in theoretical
  and experimental general relativity, gravitation, and relativistic field
  theories. Proceedings, 10th Marcel Grossmann Meeting, MG10, Rio de Janeiro,
  Brazil, July 20-26, 2003. Pt. A-C}}, pp.~1068--1088, 2004.
\newblock \href{http://arxiv.org/abs/astro-ph/0403096}{{\tt astro-ph/0403096}}.

\bibitem{Harding:2015bua}
{\bf HAWC} Collaboration, J.~P. Harding and B.~Dingus, {\it {Dark Matter
  Annihilation and Decay Searches with the High Altitude Water Cherenkov (HAWC)
  Observatory}},  in {\em {Proceedings, 34th International Cosmic Ray
  Conference (ICRC 2015)}}, 2015.
\newblock \href{http://arxiv.org/abs/1508.04352}{{\tt arXiv:1508.04352}}.

\bibitem{Pretz:2015zja}
{\bf HAWC} Collaboration, J.~Pretz, {\it {Highlights from the High Altitude
  Water Cherenkov Observatory}},  in {\em {Proceedings, 34th International
  Cosmic Ray Conference (ICRC 2015)}}, 2015.
\newblock \href{http://arxiv.org/abs/1509.07851}{{\tt arXiv:1509.07851}}.

\bibitem{Weekes:2001pd}
T.~C. Weekes et~al., {\it {VERITAS: The Very energetic radiation imaging
  telescope array system}},  {\em Astropart. Phys.} {\bf 17} (2002) 221--243,
  [\href{http://arxiv.org/abs/astro-ph/0108478}{{\tt astro-ph/0108478}}].

\bibitem{Holder:2006gi}
{\bf VERITAS} Collaboration, J.~Holder et~al., {\it {The first VERITAS
  telescope}},  {\em Astropart. Phys.} {\bf 25} (2006) 391--401,
  [\href{http://arxiv.org/abs/astro-ph/0604119}{{\tt astro-ph/0604119}}].

\bibitem{Geringer-Sameth:2013cxy}
{\bf VERITAS} Collaboration, A.~Geringer-Sameth, {\it {The VERITAS Dark Matter
  Program}},  in {\em {4th International Fermi Symposium Monterey, California,
  USA, October 28-November 2, 2012}}, 2013.
\newblock \href{http://arxiv.org/abs/1303.1406}{{\tt arXiv:1303.1406}}.

\bibitem{FlixMolina:2005hv}
{\bf MAGIC} Collaboration, J.~Flix~Molina, {\it {Planned dark matter searches
  with the MAGIC Telescope}},  in {\em {Proceedings, 40th Rencontres de Moriond
  on Very High Energy Phenomena in the Universe}}, pp.~421--424, 2005.
\newblock \href{http://arxiv.org/abs/astro-ph/0505313}{{\tt astro-ph/0505313}}.

\bibitem{Ahnen:2016qkx}
{\bf Fermi-LAT, MAGIC} Collaboration, M.~L. Ahnen et~al., {\it {Limits to dark
  matter annihilation cross-section from a combined analysis of MAGIC and
  Fermi-LAT observations of dwarf satellite galaxies}},  {\em JCAP} {\bf 1602}
  (2016), no.~02 039, [\href{http://arxiv.org/abs/1601.06590}{{\tt
  arXiv:1601.06590}}].

\bibitem{Ciafaloni:2000df}
M.~Ciafaloni, P.~Ciafaloni, and D.~Comelli, {\it {Bloch-Nordsieck violating
  electroweak corrections to inclusive TeV scale hard processes}},  {\em Phys.
  Rev. Lett.} {\bf 84} (2000) 4810--4813,
  [\href{http://arxiv.org/abs/hep-ph/0001142}{{\tt hep-ph/0001142}}].

\bibitem{Ciafaloni:1998xg}
P.~Ciafaloni and D.~Comelli, {\it {Sudakov enhancement of electroweak
  corrections}},  {\em Phys. Lett.} {\bf B446} (1999) 278--284,
  [\href{http://arxiv.org/abs/hep-ph/9809321}{{\tt hep-ph/9809321}}].

\bibitem{Ciafaloni:1999ub}
P.~Ciafaloni and D.~Comelli, {\it {Electroweak Sudakov form-factors and
  nonfactorizable soft QED effects at NLC energies}},  {\em Phys. Lett.} {\bf
  B476} (2000) 49--57, [\href{http://arxiv.org/abs/hep-ph/9910278}{{\tt
  hep-ph/9910278}}].

\bibitem{Chiu:2009mg}
J.-y. Chiu, A.~Fuhrer, R.~Kelley, and A.~V. Manohar, {\it {Factorization
  Structure of Gauge Theory Amplitudes and Application to Hard Scattering
  Processes at the LHC}},  {\em Phys. Rev.} {\bf D80} (2009) 094013,
  [\href{http://arxiv.org/abs/0909.0012}{{\tt arXiv:0909.0012}}].

\bibitem{Hryczuk:2011vi}
A.~Hryczuk and R.~Iengo, {\it {The one-loop and Sommerfeld electroweak
  corrections to the Wino dark matter annihilation}},  {\em JHEP} {\bf 01}
  (2012) 163, [\href{http://arxiv.org/abs/1111.2916}{{\tt arXiv:1111.2916}}].
  [Erratum: JHEP06,137(2012)].

\bibitem{Baumgart:2014vma}
M.~Baumgart, I.~Z. Rothstein, and V.~Vaidya, {\it {Calculating the Annihilation
  Rate of Weakly Interacting Massive Particles}},  {\em Phys. Rev. Lett.} {\bf
  114} (2015) 211301, [\href{http://arxiv.org/abs/1409.4415}{{\tt
  arXiv:1409.4415}}].

\bibitem{Bauer:2014ula}
M.~Bauer, T.~Cohen, R.~J. Hill, and M.~P. Solon, {\it {Soft Collinear Effective
  Theory for Heavy WIMP Annihilation}},  {\em JHEP} {\bf 01} (2015) 099,
  [\href{http://arxiv.org/abs/1409.7392}{{\tt arXiv:1409.7392}}].

\bibitem{Ovanesyan:2014fwa}
G.~Ovanesyan, T.~R. Slatyer, and I.~W. Stewart, {\it {Heavy Dark Matter
  Annihilation from Effective Field Theory}},  {\em Phys. Rev. Lett.} {\bf 114}
  (2015), no.~21 211302, [\href{http://arxiv.org/abs/1409.8294}{{\tt
  arXiv:1409.8294}}].

\bibitem{Baumgart:2014saa}
M.~Baumgart, I.~Z. Rothstein, and V.~Vaidya, {\it {Constraints on Galactic Wino
  Densities from Gamma Ray Lines}},  {\em JHEP} {\bf 04} (2015) 106,
  [\href{http://arxiv.org/abs/1412.8698}{{\tt arXiv:1412.8698}}].

\bibitem{Baumgart:2015bpa}
M.~Baumgart and V.~Vaidya, {\it {Semi-inclusive wino and higgsino annihilation
  to LL'}},  {\em JHEP} {\bf 03} (2016) 213,
  [\href{http://arxiv.org/abs/1510.02470}{{\tt arXiv:1510.02470}}].

\bibitem{Wells:2004di}
J.~D. Wells, {\it {PeV-scale supersymmetry}},  {\em Phys. Rev.} {\bf D71}
  (2005) 015013, [\href{http://arxiv.org/abs/hep-ph/0411041}{{\tt
  hep-ph/0411041}}].

\bibitem{ArkaniHamed:2004fb}
N.~Arkani-Hamed and S.~Dimopoulos, {\it {Supersymmetric unification without low
  energy supersymmetry and signatures for fine-tuning at the LHC}},  {\em JHEP}
  {\bf 06} (2005) 073, [\href{http://arxiv.org/abs/hep-th/0405159}{{\tt
  hep-th/0405159}}].

\bibitem{Giudice:2004tc}
G.~F. Giudice and A.~Romanino, {\it {Split supersymmetry}},  {\em Nucl. Phys.}
  {\bf B699} (2004) 65--89, [\href{http://arxiv.org/abs/hep-ph/0406088}{{\tt
  hep-ph/0406088}}]. [Erratum: Nucl. Phys.B706,487(2005)].

\bibitem{Arvanitaki:2012ps}
A.~Arvanitaki, N.~Craig, S.~Dimopoulos, and G.~Villadoro, {\it {Mini-Split}},
  {\em JHEP} {\bf 02} (2013) 126, [\href{http://arxiv.org/abs/1210.0555}{{\tt
  arXiv:1210.0555}}].

\bibitem{ArkaniHamed:2012gw}
N.~Arkani-Hamed, A.~Gupta, D.~E. Kaplan, N.~Weiner, and T.~Zorawski, {\it
  {Simply Unnatural Supersymmetry}},
  \href{http://arxiv.org/abs/1212.6971}{{\tt arXiv:1212.6971}}.

\bibitem{Hall:2012zp}
L.~J. Hall, Y.~Nomura, and S.~Shirai, {\it {Spread Supersymmetry with Wino LSP:
  Gluino and Dark Matter Signals}},  {\em JHEP} {\bf 01} (2013) 036,
  [\href{http://arxiv.org/abs/1210.2395}{{\tt arXiv:1210.2395}}].

\bibitem{Cohen:2013ama}
T.~Cohen, M.~Lisanti, A.~Pierce, and T.~R. Slatyer, {\it {Wino Dark Matter
  Under Siege}},  {\em JCAP} {\bf 1310} (2013) 061,
  [\href{http://arxiv.org/abs/1307.4082}{{\tt arXiv:1307.4082}}].

\bibitem{Ciafaloni:2012gs}
P.~Ciafaloni, D.~Comelli, A.~De~Simone, A.~Riotto, and A.~Urbano, {\it
  {Electroweak Bremsstrahlung for Wino-Like Dark Matter Annihilations}},  {\em
  JCAP} {\bf 1206} (2012) 016, [\href{http://arxiv.org/abs/1202.0692}{{\tt
  arXiv:1202.0692}}].

\bibitem{Bauer:2000ew}
C.~W. Bauer, S.~Fleming, and M.~E. Luke, {\it {Summing Sudakov logarithms in B
  ---> X(s gamma) in effective field theory}},  {\em Phys. Rev.} {\bf D63}
  (2000) 014006, [\href{http://arxiv.org/abs/hep-ph/0005275}{{\tt
  hep-ph/0005275}}].

\bibitem{Bauer:2000yr}
C.~W. Bauer, S.~Fleming, D.~Pirjol, and I.~W. Stewart, {\it {An Effective field
  theory for collinear and soft gluons: Heavy to light decays}},  {\em Phys.
  Rev.} {\bf D63} (2001) 114020,
  [\href{http://arxiv.org/abs/hep-ph/0011336}{{\tt hep-ph/0011336}}].

\bibitem{Bauer:2001ct}
C.~W. Bauer and I.~W. Stewart, {\it {Invariant operators in collinear effective
  theory}},  {\em Phys. Lett.} {\bf B516} (2001) 134--142,
  [\href{http://arxiv.org/abs/hep-ph/0107001}{{\tt hep-ph/0107001}}].

\bibitem{Bauer:2001yt}
C.~W. Bauer, D.~Pirjol, and I.~W. Stewart, {\it {Soft collinear factorization
  in effective field theory}},  {\em Phys. Rev.} {\bf D65} (2002) 054022,
  [\href{http://arxiv.org/abs/hep-ph/0109045}{{\tt hep-ph/0109045}}].

\bibitem{Chiu:2007yn}
J.-y. Chiu, F.~Golf, R.~Kelley, and A.~V. Manohar, {\it {Electroweak Sudakov
  corrections using effective field theory}},  {\em Phys. Rev. Lett.} {\bf 100}
  (2008) 021802, [\href{http://arxiv.org/abs/0709.2377}{{\tt
  arXiv:0709.2377}}].

\bibitem{Chiu:2007dg}
J.-y. Chiu, F.~Golf, R.~Kelley, and A.~V. Manohar, {\it {Electroweak
  Corrections in High Energy Processes using Effective Field Theory}},  {\em
  Phys. Rev.} {\bf D77} (2008) 053004,
  [\href{http://arxiv.org/abs/0712.0396}{{\tt arXiv:0712.0396}}].

\bibitem{Chiu:2008vv}
J.-y. Chiu, R.~Kelley, and A.~V. Manohar, {\it {Electroweak Corrections using
  Effective Field Theory: Applications to the LHC}},  {\em Phys. Rev.} {\bf
  D78} (2008) 073006, [\href{http://arxiv.org/abs/0806.1240}{{\tt
  arXiv:0806.1240}}].

\bibitem{Chiu:2009ft}
J.-y. Chiu, A.~Fuhrer, R.~Kelley, and A.~V. Manohar, {\it {Soft and Collinear
  Functions for the Standard Model}},  {\em Phys. Rev.} {\bf D81} (2010)
  014023, [\href{http://arxiv.org/abs/0909.0947}{{\tt arXiv:0909.0947}}].

\bibitem{Bauer:2002nz}
C.~W. Bauer, S.~Fleming, D.~Pirjol, I.~Z. Rothstein, and I.~W. Stewart, {\it
  {Hard scattering factorization from effective field theory}},  {\em Phys.
  Rev.} {\bf D66} (2002) 014017,
  [\href{http://arxiv.org/abs/hep-ph/0202088}{{\tt hep-ph/0202088}}].

\bibitem{Chiu:2012ir}
J.-Y. Chiu, A.~Jain, D.~Neill, and I.~Z. Rothstein, {\it {A Formalism for the
  Systematic Treatment of Rapidity Logarithms in Quantum Field Theory}},  {\em
  JHEP} {\bf 05} (2012) 084, [\href{http://arxiv.org/abs/1202.0814}{{\tt
  arXiv:1202.0814}}].

\bibitem{Becher:2010tm}
T.~Becher and M.~Neubert, {\it {{Drell-Yan} Production at Small $q_T$,
  Transverse Parton Distributions and the Collinear Anomaly}},  {\em Eur. Phys.
  J.} {\bf C71} (2011) 1665, [\href{http://arxiv.org/abs/1007.4005}{{\tt
  arXiv:1007.4005}}].

\bibitem{Denner:2016etu}
A.~Denner, L.~Jenniches, J.-N. Lang, and C.~Sturm, {\it {Gauge-independent
  $\overline{MS}$ renormalization in the 2HDM}},  {\em JHEP} {\bf 09} (2016)
  115, [\href{http://arxiv.org/abs/1607.07352}{{\tt arXiv:1607.07352}}].

\bibitem{Bergstrom:2012vd}
L.~Bergstrom, G.~Bertone, J.~Conrad, C.~Farnier, and C.~Weniger, {\it
  {Investigating Gamma-Ray Lines from Dark Matter with Future Observatories}},
  {\em JCAP} {\bf 1211} (2012) 025, [\href{http://arxiv.org/abs/1207.6773}{{\tt
  arXiv:1207.6773}}].

\bibitem{Schlegel:1997yv}
D.~J. Schlegel, D.~P. Finkbeiner, and M.~Davis, {\it {Maps of dust IR emission
  for use in estimation of reddening and CMBR foregrounds}},  {\em Astrophys.
  J.} {\bf 500} (1998) 525, [\href{http://arxiv.org/abs/astro-ph/9710327}{{\tt
  astro-ph/9710327}}].

\bibitem{Edwards:2017mnf}
T.~D.~P. Edwards and C.~Weniger, {\it {A Fresh Approach to Forecasting in
  Astroparticle Physics and Dark Matter Searches}},
  \href{http://arxiv.org/abs/1704.05458}{{\tt arXiv:1704.05458}}.

\bibitem{Burkert:1995yz}
A.~Burkert, {\it {The Structure of dark matter halos in dwarf galaxies}},  {\em
  IAU Symp.} {\bf 171} (1996) 175,
  [\href{http://arxiv.org/abs/astro-ph/9504041}{{\tt astro-ph/9504041}}].
  [Astrophys. J.447,L25(1995)].

\bibitem{Diemer:2014gba}
B.~Diemer and A.~V. Kravtsov, {\it {A universal model for halo
  concentrations}},  {\em Astrophys. J.} {\bf 799} (2015), no.~1 108,
  [\href{http://arxiv.org/abs/1407.4730}{{\tt arXiv:1407.4730}}].

\bibitem{Acero:2016qlg}
{\bf Fermi-LAT} Collaboration, F.~Acero et~al., {\it {Development of the Model
  of Galactic Interstellar Emission for Standard Point-Source Analysis of Fermi
  Large Area Telescope Data}},  {\em Astrophys. J. Suppl.} {\bf 223} (2016),
  no.~2 26, [\href{http://arxiv.org/abs/1602.07246}{{\tt arXiv:1602.07246}}].

\bibitem{Schaller:2014uwa}
M.~Schaller, C.~S. Frenk, R.~G. Bower, T.~Theuns, A.~Jenkins, J.~Schaye, R.~A.
  Crain, M.~Furlong, C.~D. Vecchia, and I.~G. McCarthy, {\it {Baryon effects on
  the internal structure of $\Lambda$CDM haloes in the EAGLE simulations}},
  {\em Mon. Not. Roy. Astron. Soc.} {\bf 451} (2015), no.~2 1247--1267,
  [\href{http://arxiv.org/abs/1409.8617}{{\tt arXiv:1409.8617}}].

\bibitem{Bullock:1999he}
J.~S. Bullock, T.~S. Kolatt, Y.~Sigad, R.~S. Somerville, A.~V. Kravtsov, A.~A.
  Klypin, J.~R. Primack, and A.~Dekel, {\it {Profiles of dark haloes.
  Evolution, scatter, and environment}},  {\em Mon. Not. Roy. Astron. Soc.}
  {\bf 321} (2001) 559--575, [\href{http://arxiv.org/abs/astro-ph/9908159}{{\tt
  astro-ph/9908159}}].

\bibitem{Nezri:2012tu}
E.~Nezri, R.~White, C.~Combet, D.~Maurin, E.~Pointecouteau, and J.~A. Hinton,
  {\it {gamma-rays from annihilating dark matter in galaxy clusters: stacking
  vs single source analysis}},  {\em Mon. Not. Roy. Astron. Soc.} {\bf 425}
  (2012) 477, [\href{http://arxiv.org/abs/1203.1165}{{\tt arXiv:1203.1165}}].

\bibitem{Sanchez-Conde:2013yxa}
M.~A. Sánchez-Conde and F.~Prada, {\it {The flattening of the
  concentration–mass relation towards low halo masses and its implications
  for the annihilation signal boost}},  {\em Mon. Not. Roy. Astron. Soc.} {\bf
  442} (2014), no.~3 2271--2277, [\href{http://arxiv.org/abs/1312.1729}{{\tt
  arXiv:1312.1729}}].

\bibitem{Moline:2016pbm}
A.~Molin\`e, M.~A. S\`anchez-Conde, S.~Palomares-Ruiz, and F.~Prada, {\it
  {Characterization of subhalo structural properties and implications for dark
  matter annihilation signals}},  {\em Mon. Not. Roy. Astron. Soc.} {\bf 466}
  (2017), no.~4 4974--4990, [\href{http://arxiv.org/abs/1603.04057}{{\tt
  arXiv:1603.04057}}].

\bibitem{Lu:2016vmu}
Y.~Lu, X.~Yang, F.~Shi, H.~J. Mo, D.~Tweed, H.~Wang, Y.~Zhang, S.~Li, and S.~H.
  Lim, {\it {Galaxy groups in the 2MASS Redshift Survey}},  {\em Astrophys. J.}
  {\bf 832} (2016), no.~1 39, [\href{http://arxiv.org/abs/1607.03982}{{\tt
  arXiv:1607.03982}}].

\bibitem{1996AJ....111..794K}
I.~D. {Karachentsev} and D.~A. {Makarov}, {\it {The Galaxy Motion Relative to
  Nearby Galaxies and the Local Velocity Field}},  {\em AJ} {\bf 111} (Feb.,
  1996) 794.

\bibitem{Karachentsev:2013jca}
I.~D. Karachentsev, R.~B. Tully, P.-F. Wu, E.~J. Shaya, and A.~E. Dolphin, {\it
  {Infall of nearby galaxies into the Virgo cluster as traced with HST}},  {\em
  Astrophys. J.} {\bf 782} (2014) 4,
  [\href{http://arxiv.org/abs/1312.6769}{{\tt arXiv:1312.6769}}].

\bibitem{1993AJ....105.2035D}
A.~{Diaferio}, M.~{Ramella}, M.~J. {Geller}, and A.~{Ferrari}, {\it {Are groups
  of galaxies virialized systems?}},  {\em Astrophys. J.} {\bf 105} (June,
  1993) 2035--2046.

\bibitem{Bhattacharjee:1998qc}
P.~Bhattacharjee and G.~Sigl, {\it {Origin and propagation of extremely
  high-energy cosmic rays}},  {\em Phys. Rept.} {\bf 327} (2000) 109--247,
  [\href{http://arxiv.org/abs/astro-ph/9811011}{{\tt astro-ph/9811011}}].

\bibitem{Ishiwata:2008cu}
K.~Ishiwata, S.~Matsumoto, and T.~Moroi, {\it {High Energy Cosmic Rays from the
  Decay of Gravitino Dark Matter}},  {\em Phys. Rev.} {\bf D78} (2008) 063505,
  [\href{http://arxiv.org/abs/0805.1133}{{\tt arXiv:0805.1133}}].

\bibitem{Grefe:2008zz}
M.~Grefe, {\em {Neutrino signals from gravitino dark matter with broken
  R-parity}}.
\newblock PhD thesis, Hamburg U., 2008.
\newblock \href{http://arxiv.org/abs/1111.6041}{{\tt arXiv:1111.6041}}.

\bibitem{Alloul:2013bka}
A.~Alloul, N.~D. Christensen, C.~Degrande, C.~Duhr, and B.~Fuks, {\it
  {FeynRules 2.0 - A complete toolbox for tree-level phenomenology}},  {\em
  Comput. Phys. Commun.} {\bf 185} (2014) 2250--2300,
  [\href{http://arxiv.org/abs/1310.1921}{{\tt arXiv:1310.1921}}].

\bibitem{Alwall:2011uj}
J.~Alwall, M.~Herquet, F.~Maltoni, O.~Mattelaer, and T.~Stelzer, {\it {MadGraph
  5 : Going Beyond}},  {\em JHEP} {\bf 06} (2011) 128,
  [\href{http://arxiv.org/abs/1106.0522}{{\tt arXiv:1106.0522}}].

\bibitem{Alwall:2014hca}
J.~Alwall, R.~Frederix, S.~Frixione, V.~Hirschi, F.~Maltoni, O.~Mattelaer,
  H.~S. Shao, T.~Stelzer, P.~Torrielli, and M.~Zaro, {\it {The automated
  computation of tree-level and next-to-leading order differential cross
  sections, and their matching to parton shower simulations}},  {\em JHEP} {\bf
  07} (2014) 079, [\href{http://arxiv.org/abs/1405.0301}{{\tt
  arXiv:1405.0301}}].

\bibitem{Fortin:2009rq}
J.-F. Fortin, J.~Shelton, S.~Thomas, and Y.~Zhao, {\it {Gamma Ray Spectra from
  Dark Matter Annihilation and Decay}},
  \href{http://arxiv.org/abs/0908.2258}{{\tt arXiv:0908.2258}}.

\bibitem{Finkbeiner:2014sja}
D.~P. Finkbeiner and N.~Weiner, {\it {X-ray line from exciting dark matter}},
  {\em Phys. Rev.} {\bf D94} (2016), no.~8 083002,
  [\href{http://arxiv.org/abs/1402.6671}{{\tt arXiv:1402.6671}}].

\bibitem{Sommerfeld:1938}
A.~Sommerfeld, {\it {{\"U}ber die Beugung und Bremsung der Elektronen}},  {\em
  Annalen der Physik} {\bf 403} (1931) 257--330.

\bibitem{Pospelov:2008jd}
M.~Pospelov and A.~Ritz, {\it {Astrophysical Signatures of Secluded Dark
  Matter}},  {\em Phys. Lett.} {\bf B671} (2009) 391--397,
  [\href{http://arxiv.org/abs/0810.1502}{{\tt arXiv:0810.1502}}].

\bibitem{Byckling:1969sx}
E.~Byckling and K.~Kajantie, {\it {N-particle phase space in terms of invariant
  momentum transfers}},  {\em Nucl. Phys.} {\bf B9} (1969) 568--576.

\bibitem{Kersevan:2004yh}
B.~P. Kersevan and E.~Richter-Was, {\it {Improved phase space treatment of
  massive multi-particle final states}},  {\em Eur. Phys. J.} {\bf C39} (2005)
  439--450, [\href{http://arxiv.org/abs/hep-ph/0405248}{{\tt hep-ph/0405248}}].

\bibitem{Boudjema:2005hb}
F.~Boudjema, A.~Semenov, and D.~Temes, {\it {Self-annihilation of the
  neutralino dark matter into two photons or a Z and a photon in the MSSM}},
  {\em Phys. Rev.} {\bf D72} (2005) 055024,
  [\href{http://arxiv.org/abs/hep-ph/0507127}{{\tt hep-ph/0507127}}].

\bibitem{Baro:2007em}
N.~Baro, F.~Boudjema, and A.~Semenov, {\it {Full one-loop corrections to the
  relic density in the MSSM: A Few examples}},  {\em Phys. Lett.} {\bf B660}
  (2008) 550--560, [\href{http://arxiv.org/abs/0710.1821}{{\tt
  arXiv:0710.1821}}].

\bibitem{Baro:2009na}
N.~Baro, F.~Boudjema, G.~Chalons, and S.~Hao, {\it {Relic density at one-loop
  with gauge boson pair production}},  {\em Phys. Rev.} {\bf D81} (2010)
  015005, [\href{http://arxiv.org/abs/0910.3293}{{\tt arXiv:0910.3293}}].

\bibitem{Manohar:2006nz}
A.~V. Manohar and I.~W. Stewart, {\it {The Zero-Bin and Mode Factorization in
  Quantum Field Theory}},  {\em Phys. Rev.} {\bf D76} (2007) 074002,
  [\href{http://arxiv.org/abs/hep-ph/0605001}{{\tt hep-ph/0605001}}].

\bibitem{Fuhrer:2010eu}
A.~Fuhrer, A.~V. Manohar, J.-y. Chiu, and R.~Kelley, {\it {Radiative
  Corrections to Longitudinal and Transverse Gauge Boson and Higgs
  Production}},  {\em Phys. Rev.} {\bf D81} (2010) 093005,
  [\href{http://arxiv.org/abs/1003.0025}{{\tt arXiv:1003.0025}}].

\bibitem{Passarino:1978jh}
G.~Passarino and M.~J.~G. Veltman, {\it {One Loop Corrections for e+ e-
  Annihilation Into mu+ mu- in the Weinberg Model}},  {\em Nucl. Phys.} {\bf
  B160} (1979) 151--207.

\bibitem{Denner:1992vza}
A.~Denner, H.~Eck, O.~Hahn, and J.~Kublbeck, {\it {Feynman rules for fermion
  number violating interactions}},  {\em Nucl. Phys.} {\bf B387} (1992)
  467--484.

\bibitem{Ellis:2007qk}
R.~K. Ellis and G.~Zanderighi, {\it {Scalar one-loop integrals for QCD}},  {\em
  JHEP} {\bf 02} (2008) 002, [\href{http://arxiv.org/abs/0712.1851}{{\tt
  arXiv:0712.1851}}].

\bibitem{'tHooft:1978xw}
G.~'t~Hooft and M.~J.~G. Veltman, {\it {Scalar One Loop Integrals}},  {\em
  Nucl. Phys.} {\bf B153} (1979) 365--401.

\bibitem{Ellis:2011cr}
R.~K. Ellis, Z.~Kunszt, K.~Melnikov, and G.~Zanderighi, {\it {One-loop
  calculations in quantum field theory: from Feynman diagrams to unitarity
  cuts}},  {\em Phys. Rept.} {\bf 518} (2012) 141--250,
  [\href{http://arxiv.org/abs/1105.4319}{{\tt arXiv:1105.4319}}].

\bibitem{Mertig:1990an}
R.~Mertig, M.~Bohm, and A.~Denner, {\it {FEYN CALC: Computer algebraic
  calculation of Feynman amplitudes}},  {\em Comput. Phys. Commun.} {\bf 64}
  (1991) 345--359.

\bibitem{Shtabovenko:2016sxi}
V.~Shtabovenko, R.~Mertig, and F.~Orellana, {\it {New Developments in FeynCalc
  9.0}},  \href{http://arxiv.org/abs/1601.01167}{{\tt arXiv:1601.01167}}.

\bibitem{Patel:2015tea}
H.~H. Patel, {\it {Package-X: A Mathematica package for the analytic
  calculation of one-loop integrals}},  {\em Comput. Phys. Commun.} {\bf 197}
  (2015) 276--290, [\href{http://arxiv.org/abs/1503.01469}{{\tt
  arXiv:1503.01469}}].

\bibitem{Bardin:1999ak}
D.~{\relax Yu}. Bardin and G.~Passarino, {\em {The Standard Model in the
  Making: Precision Study of the Electroweak Interactions}}.
\newblock 1999.

\bibitem{Stewart:1998ke}
I.~W. Stewart, {\it {Extraction of the D* D pi coupling from D* decays}},  {\em
  Nucl. Phys.} {\bf B529} (1998) 62--80,
  [\href{http://arxiv.org/abs/hep-ph/9803227}{{\tt hep-ph/9803227}}].

\bibitem{Manohar:2000dt}
A.~V. Manohar and M.~B. Wise, {\it {Heavy quark physics}},  {\em Camb. Monogr.
  Part. Phys. Nucl. Phys. Cosmol.} {\bf 10} (2000) 1--191.

\bibitem{Catani:1996jh}
S.~Catani and M.~H. Seymour, {\it {The Dipole formalism for the calculation of
  QCD jet cross-sections at next-to-leading order}},  {\em Phys. Lett.} {\bf
  B378} (1996) 287--301, [\href{http://arxiv.org/abs/hep-ph/9602277}{{\tt
  hep-ph/9602277}}].

\bibitem{Catani:1996vz}
S.~Catani and M.~H. Seymour, {\it {A General algorithm for calculating jet
  cross-sections in NLO QCD}},  {\em Nucl. Phys.} {\bf B485} (1997) 291--419,
  [\href{http://arxiv.org/abs/hep-ph/9605323}{{\tt hep-ph/9605323}}]. [Erratum:
  Nucl. Phys.B510,503(1998)].

\bibitem{Moult:2015aoa}
I.~Moult, I.~W. Stewart, F.~J. Tackmann, and W.~J. Waalewijn, {\it {Employing
  Helicity Amplitudes for Resummation}},  {\em Phys. Rev.} {\bf D93} (2016),
  no.~9 094003, [\href{http://arxiv.org/abs/1508.02397}{{\tt
  arXiv:1508.02397}}].

\bibitem{Chiu:2009yx}
J.-y. Chiu, A.~Fuhrer, A.~H. Hoang, R.~Kelley, and A.~V. Manohar, {\it
  {Soft-Collinear Factorization and Zero-Bin Subtractions}},  {\em Phys. Rev.}
  {\bf D79} (2009) 053007, [\href{http://arxiv.org/abs/0901.1332}{{\tt
  arXiv:0901.1332}}].

\bibitem{Denner:1991kt}
A.~Denner, {\it {Techniques for calculation of electroweak radiative
  corrections at the one loop level and results for W physics at LEP-200}},
  {\em Fortsch. Phys.} {\bf 41} (1993) 307--420,
  [\href{http://arxiv.org/abs/0709.1075}{{\tt arXiv:0709.1075}}].

\end{thebibliography}\endgroup
